\documentclass[11pt,twoside,notitlepage]{book}

\RequirePackage{lineno}
\usepackage{import}
\usepackage{titlesec}
\usepackage{enumitem}
\usepackage{amssymb, amsmath}
\usepackage{mathpazo}
\usepackage{float,graphicx}
\graphicspath{{./}{ePHENIXLOI/}}
\usepackage[figuresright]{rotating}
\usepackage[usenames,dvipsnames]{color}
\usepackage{microtype}
\usepackage{tabulary}
\usepackage{booktabs,multirow,dcolumn,bigdelim}

\usepackage{xr-hyper}
\usepackage[pdftex,plainpages=false,pdfpagelabels,backref,pdfborder={0 0 0}]{hyperref}
\usepackage{url}
\usepackage[toc,page,titletoc]{appendix}
\usepackage{afterpage}
\usepackage[font=small,labelfont=bf]{caption}
\usepackage[scaled]{helvet}
\usepackage{sectsty}
\allsectionsfont{\normalfont\sffamily}
\usepackage{titletoc}
\titlecontents{chapter}
  [1.5em]
  {\linespread{0.9}\normalfont\sffamily\bfseries}
  {\contentslabel{1em}} 
  {\hspace*{-2.3em}} 
  {\mdseries\titlerule*[1pc]{.}\contentspage} 
\titlecontents{section}
  [3.5em]
  {\linespread{0.9}\normalfont\sffamily}
  {\contentslabel{2.3em}} 
  {\hspace*{-2.3em}} 
  {\titlerule*[1pc]{.}\contentspage} 
\titlecontents{subsection}
  [4.5em]
  {\linespread{0.9}\normalfont\sffamily}
  {\contentslabel{2.3em}} 
  {\hspace*{-2.3em}} 
  {\titlerule*[1pc]{.}\contentspage} 
\titlecontents{subsubsection}
  [5.5em]
  {\linespread{0.9}\normalfont\sffamily}
  {\contentslabel{2.3em}} 
  {\hspace*{-2.3em}} 
  {\titlerule*[1pc]{.}\contentspage} 
\usepackage[text={6.5in,8.75in},headheight=15pt,centering]{geometry}
\usepackage{xspace}

\DeclareGraphicsExtensions{.pdf,.png}

\usepackage{fancyhdr}
\fancypagestyle{plain}{  \fancyhf{}
  \fancyfoot[RO,LE]{\sffamily \thepage}
}

\usepackage{eurosym}
\usepackage{xcolor}
\usepackage{framed}
\colorlet{shadecolor}{blue!10}

\newcommand{\hic}{\mbox{$A$$+$$A$}\xspace}
\newcommand{\pA}{\mbox{$p$$+$$A$}\xspace}
\newcommand{\AuAu}{\mbox{Au$+$Au}\xspace}
\newcommand{\auau}{\mbox{Au$+$Au}\xspace} 
\newcommand{\raa}{\mbox{$R_\mathrm{AA}$}\xspace}
\newcommand{\pbpb}{\mbox{Pb$+$Pb}\xspace} \newcommand
{\pdau}{\mbox{$p(d)$$+$Au}\xspace} \newcommand
{\pau}{\mbox{$p$$+$Au}\xspace} \newcommand
{\aj}{\mbox{$A_J$}\xspace} \newcommand {\pp}{\mbox{$p$$+$$p$}\xspace}
\newcommand {\ppbar}{\mbox{$p$$+$$\overline{p}$}\xspace}
\newcommand{\pT}{\mbox{${p_T}$}\xspace}
\newcommand{\jpsi}{\mbox{$J/\psi$}}
\newcommand{\sqrtsnn}{\mbox{$\sqrt{s_{\scriptscriptstyle NN}}$}}

\newcommand{\ncoll}{$N_\mathrm{coll}$}
\newcommand{\qgp}{\mbox{quark-gluon plasma}\xspace}

\newcommand{\PbPb}{\mbox{Pb$+$Pb}} \newcommand{\pPb}{\mbox{$p$$+$Pb}}

 \newcommand{\ET}{\mbox{$E_T$}}
 
\newcommand{\RAA}{\mbox{$R_{AA}$}\xspace} 
\newcommand{\pt}{\mbox{${p_T}$}\xspace}

\newcommand{\dAu}{\mbox{$d$$+$Au}\xspace}
\newcommand{\pAu}{\mbox{$p$$+$Au}\xspace}

\newcommand{\fastjet}{\mbox{\sc FastJet}\xspace}
\newcommand{\geant}{\mbox{\sc Geant4}\xspace}
\newcommand{\antikt}{\mbox{anti-$k_T$}\xspace}
\newcommand{\pythia}{\mbox{\sc Pythia}\xspace}
\newcommand{\rapgap}{\mbox{\sc Rapgap}\xspace}
\newcommand{\milou}{\mbox{\sc Milou}\xspace}
\newcommand{\pyquen}{\mbox{\sc Pyquen}\xspace}
\newcommand{\hijing}{\mbox{\sc Hijing}\xspace}

\newcommand{\roounfold}{\mbox{\sc RooUnfold}\xspace}

\newcommand{\gj}{\mbox{$\gamma$+jet}\xspace}
\newcommand{\gh}{\mbox{$\gamma$+hadron}\xspace}
\newcommand{\martinimusic}{\mbox{\sc Martini+Music}\xspace}
\newcommand{\martini}{\mbox{\sc Martini}\xspace}
\newcommand{\music}{\mbox{\sc Music}\xspace}
\newcommand{\Ephenix}{Electron-Ion Collider (EIC) detector built
  around the BaBar magnet and sPHENIX calorimetry\xspace}
\newcommand{\ephenix}{EIC detector built around the BaBar magnet and
  sPHENIX calorimetry\xspace} 
\newcommand{\refdesign}{reference design\xspace}
\newcommand{\refconfig}{reference configuration\xspace}
\newcommand{\dijet}{\mbox{dijet}\xspace}
\newcommand{\fake}{\mbox{fake}\xspace}
\newcommand{\fast}{\mbox{fast}\xspace}
\newcommand{\veryfast}{\mbox{very fast}\xspace}
\newcommand{\epem}{\mbox{$e^+e^-$}\xspace}
\newcommand{\onewidth}{0.6\linewidth}
\newcommand{\twowidth}{0.48\linewidth}

\newcommand{\egoing}{\mbox{electron-going}\xspace}
\newcommand{\hgoing}{\mbox{hadron-going}\xspace}
\newcommand{\egodir}{electron-going direction\xspace}
\newcommand{\hgodir}{hadron-going direction\xspace}

\usepackage{subfig}

\begin{document}

\frontmatter

\pagestyle{empty}
\makeatletter{}\renewcommand*\familydefault{\sfdefault}
{\sffamily
\vfill
\vspace{4cm}
\begin{figure}[H]
  \begin{center}
  \includegraphics[width=0.6\linewidth]{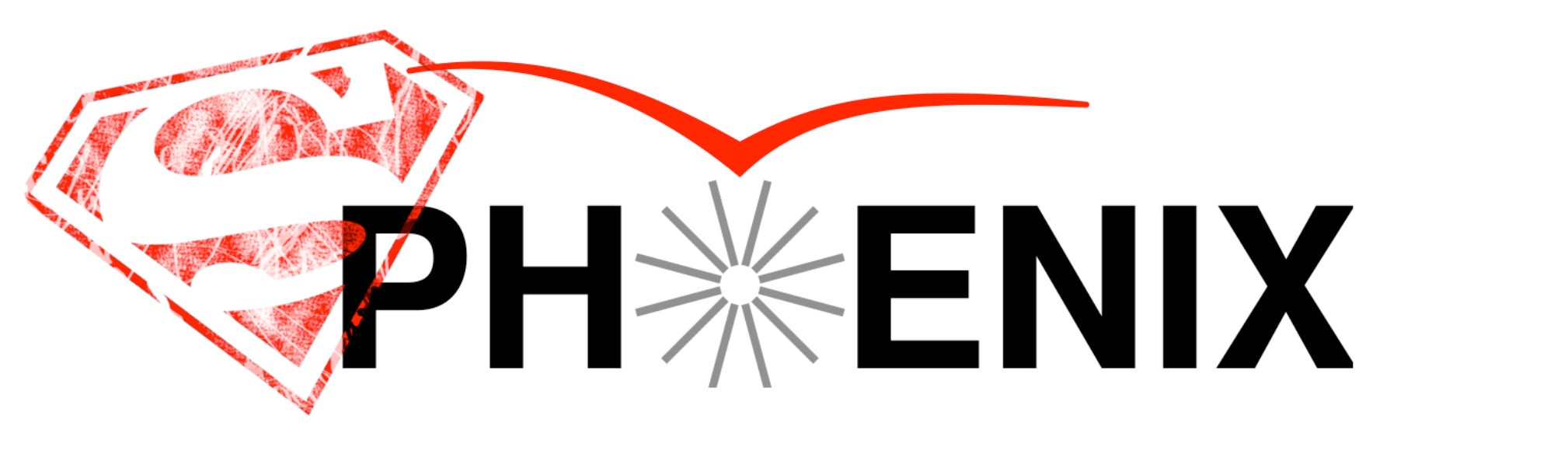}
  \end{center}
\end{figure}

\begin{center}
  \large
  \emph{\Large{An Upgrade Proposal from the PHENIX Collaboration}}

  \begin{tabular}{rl}
  &November 19, 2014 \\
  \end{tabular}
\end{center}

\vspace{2cm}

\begin{figure}[H]
  \begin{center}
    \includegraphics[width=0.7\linewidth]{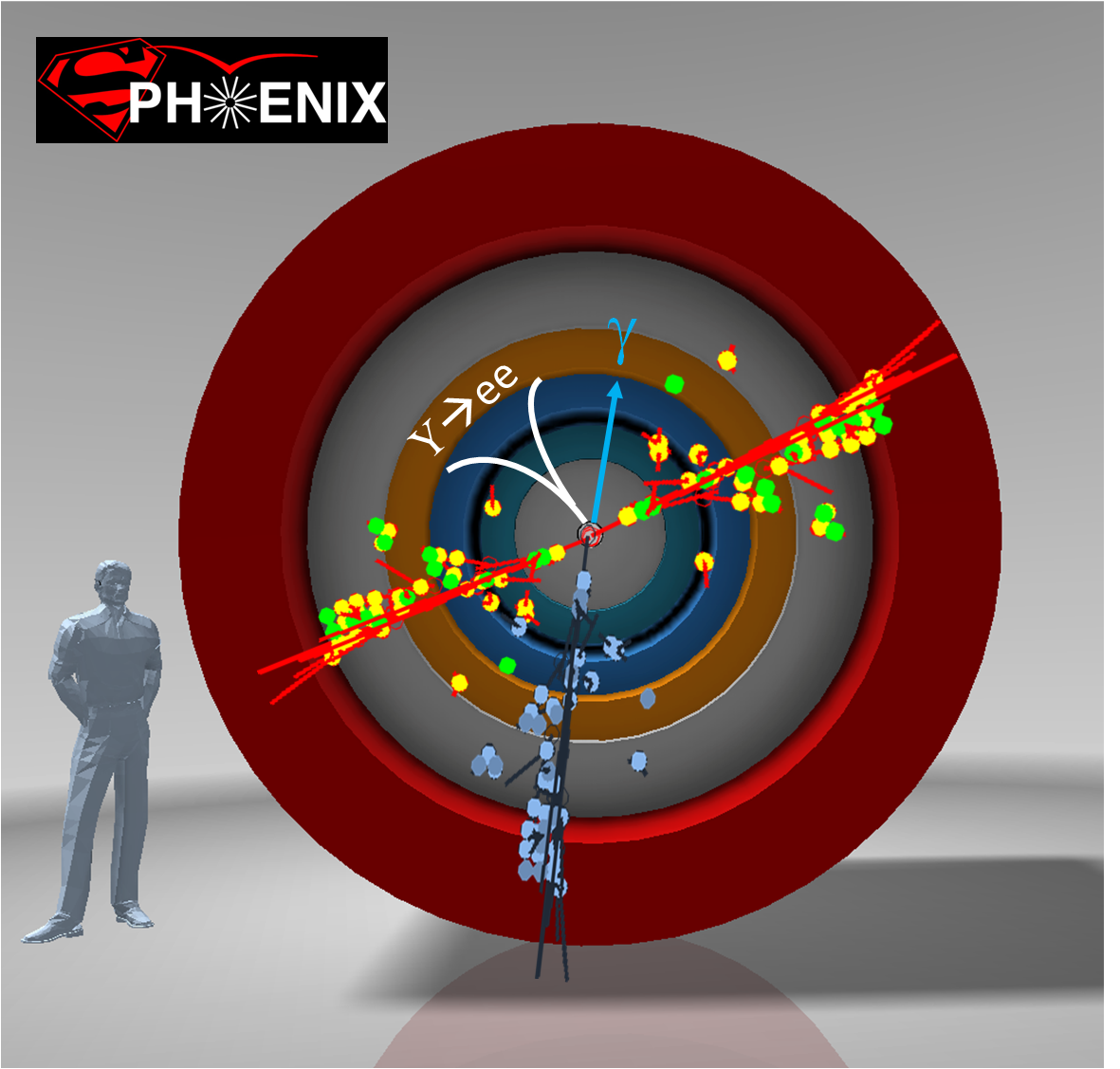}
  \end{center}
\end{figure}
}

\vfill
\renewcommand*\familydefault{\rmdefault}

\cleardoublepage
\pagestyle{fancy}
\makeatletter{}\chapter*{Executive Summary}
\label{executive_summary}
\setcounter{page}{1}

In this document the PHENIX collaboration proposes a major upgrade to
the PHENIX detector at the Relativistic Heavy Ion Collider.  This
upgrade, sPHENIX, enables an extremely rich jet and beauty quarkonia
physics program addressing fundamental questions about the nature of
the strongly coupled \qgp, discovered experimentally at RHIC to be a
perfect fluid.  The startling dynamics of the QGP on fluid-like length
scales is an emergent property of QCD, seemingly implicit in the
Lagrangian but stubbornly hidden from view.  QCD is an asymptotically
free theory, but how QCD manifests as a strongly coupled fluid with
specific shear viscosity near $T_C$ as low as allowed by the
uncertainty principle is as fundamental an issue as that of how
confinement itself arises.

Questions such as this can only be fully addressed with jet, dijet,
$\gamma$-jet, fragmentation function, and Upsilon observables at RHIC
energies, which probe the medium over a variety of length
scales. Comparing these measurements with ones at the Large Hadron
Collider will yield important insights into the thermodynamics of QCD,
and these issues have acquired fresh new importance as recent analyses
of data from $p(d)$$+$A collisions have raised questions regarding the
minimum size, shape, and temperature needed for the formation of
droplets of \qgp.  Finally, beyond the physics program described here,
sPHENIX provides an excellent foundation for a possible future
detector able to exploit the novel physics opportunities of an
electron-ion collider at RHIC.

The sPHENIX upgrade addresses specific questions whose answers are
necessary to advance our understanding of the \qgp:
\begin{itemize}[topsep=2pt, partopsep=2pt, parsep=4pt, itemsep=4pt]
\item How does a partonic shower develop and propagate in the \qgp? 
\item How does one reconcile the observed strongly coupled \qgp with
  the asymptotically free theory of quarks and gluons?
\item What are the dynamical changes in the \qgp in terms of
  quasiparticles and excitations as a function of temperature?
\item How sharp is the transition of the \qgp from the most strongly
  coupled regime near $T_c$ to a weakly coupled system of partons
  known to emerge at asymptotically high temperatures?
\end{itemize}

The development of the sPHENIX physics program has benefited from very
active engagement with the theory community.  For current-day
questions regarding the perfect fluidity of the \qgp, engagement
between theorists and experimentalists, fed by increasingly
comprehensive data from RHIC and the LHC, has moved the physics
discussion beyond merely constraining $\eta/s$ to exploring its
temperature dependence and other properties.  In an analogous manner,
there is great progress in the theoretical understanding of jet
quenching --- see Ref.~\cite{JETCollaboration:QhatConstraint} from the
JET Collaboration, for example.  We foresee that truly comprehensive
jet data from RHIC and the LHC --- to which sPHENIX contributes
crucially --- will move the physics discussion beyond merely
constraining the single transport property $\hat{q}$ to a deeper
understanding of the dynamics of the \qgp.

\begin{figure}[hbt!]
  \centering
  \includegraphics[width=\linewidth]{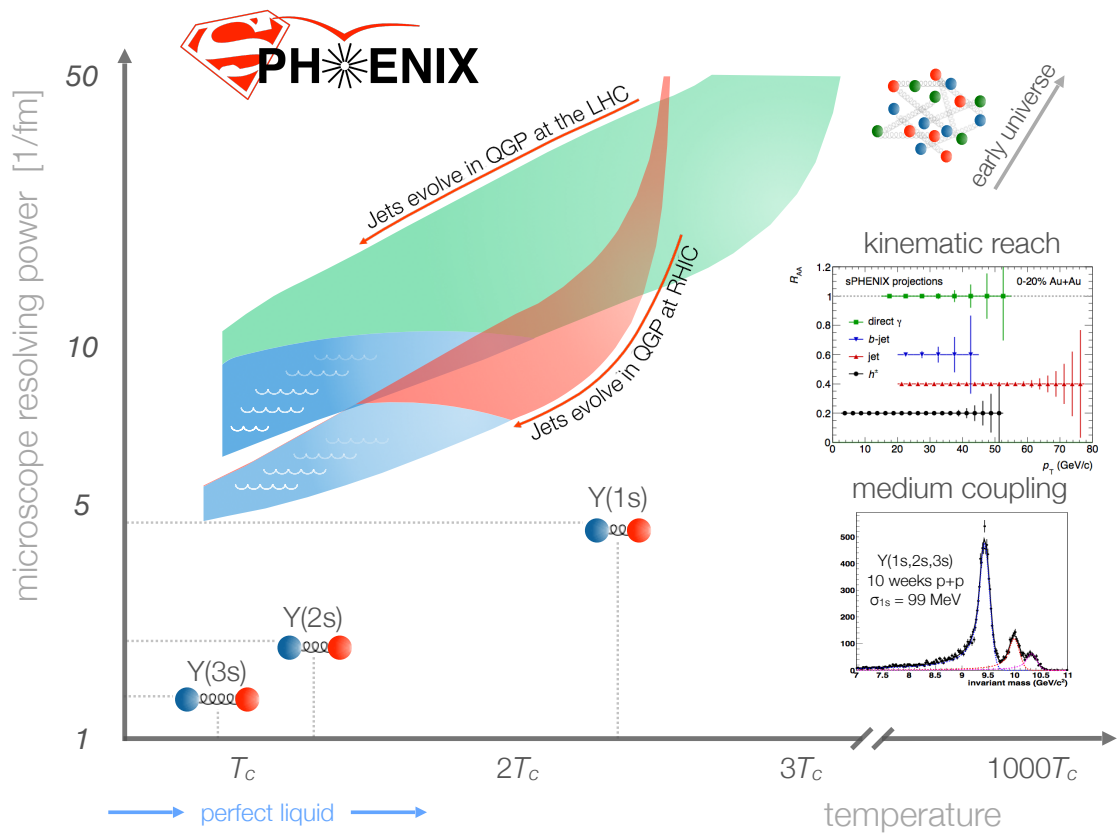}
  \caption[Conceptual diagram of sPHENIX physics]{ The physics goals of sPHENIX, graphically. Hard scattered
    partons at the LHC and at RHIC evolve through splittings and
    interaction with the medium, providing sensitivity to QGP dynamics
    over a wide range of length scales.  The heavy quarkonia states
    are well-localized in this space and provide uniquely valuable
    probes of the coupling strength of the medium.  Shown as inserts
    are projections of the capabilities of sPHENIX for measuring these
    key probes. }
  \label{fig:magic_plot} 
\end{figure}

Figure~\ref{fig:magic_plot} depicts the physics goals of sPHENIX. Hard
scattered partons at both the LHC and at RHIC begin with a very large
virtuality at the earliest, hottest stage of the collision.  These
highly virtual partons have very fine resolving power and probe the
medium on extremely small length scales.  The scattered partons
initially shed their virtuality, evolving downward in scale, through
splittings as though they were in vacuum.  At later times the momentum
scale of the developing partonic shower becomes comparable to that of
the hot QCD medium and the nascent jet becomes more and more sensitive
to mesoscopic, fluid-scale excitations in the medium.  At the same
time, the medium is populated with heavy quarkonia whose physical size
and temperature sensitive coupling to the medium provide precisely
locatable probes of the medium in this space.  At the longest scales,
one sees the well-established hydrodynamic behavior of the medium with
minimal specific shear viscosity, the so-called perfect liquid.  The
sPHENIX detector will be able to measure jets, $b$-tagged jets,
photons, charged hadrons and their correlations over a wide range of
energies, and it will also have mass resolution sufficient to
separately distinguish the three states of the Upsilon family.  These
capabilities will enable us to map out the dynamics of the QGP across
this space and address the fundamental questions posed above.

\begin{figure}[hbt!]
  \centering
  \includegraphics[trim = 300 0 300 0, clip, width=0.6\linewidth]{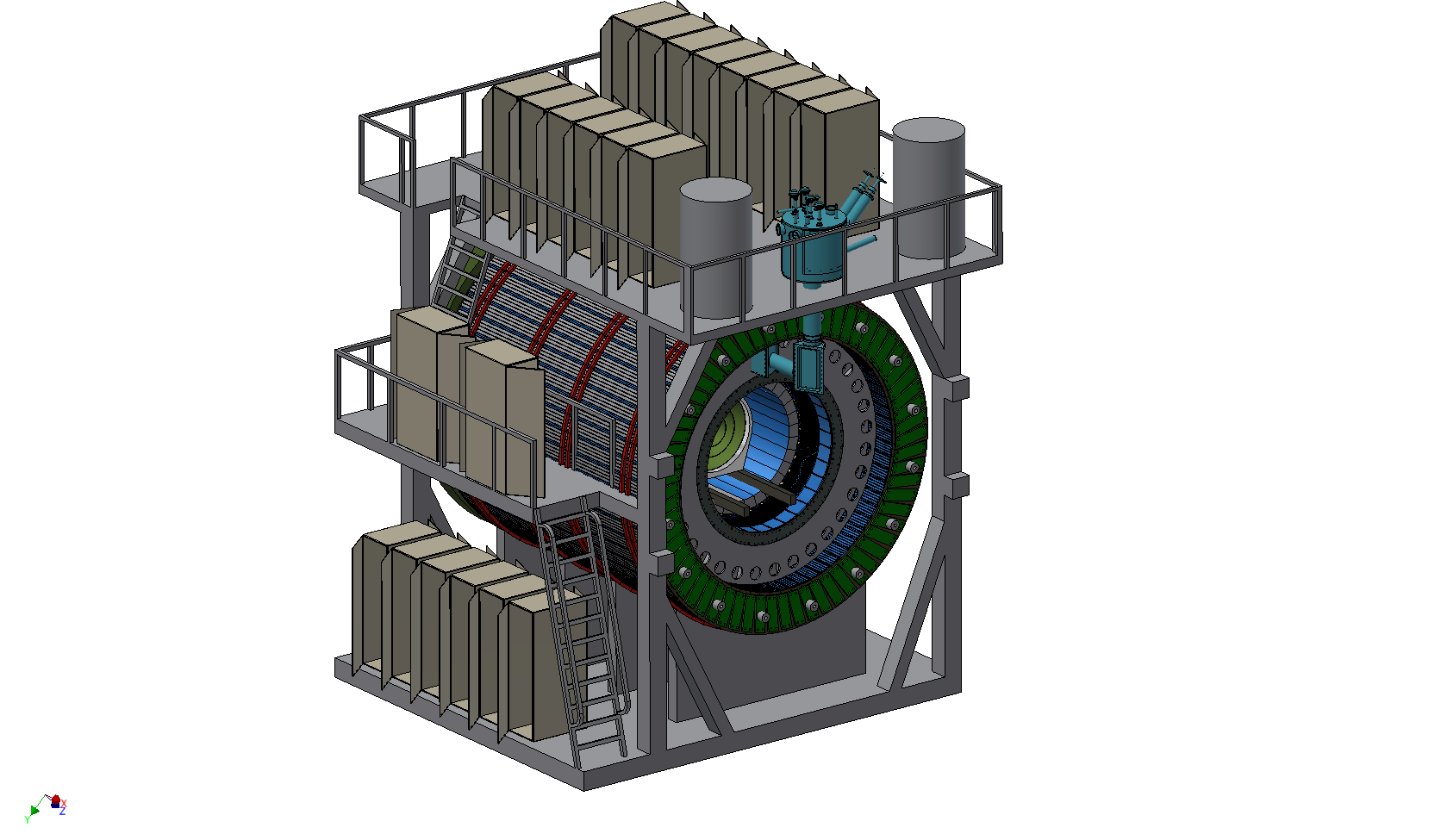}
  \caption[Engineering drawing of sPHENIX, showing the superconducting
  solenoid containing the electromagnetic calorimeter and surrounded
  by the hadronic calorimeter]{An engineering drawing of sPHENIX,
    showing the superconducting solenoid containing the
    electromagnetic calorimeter and surrounded by the hadronic
    calorimeter, with a model of the associated support structure, as
    it would sit in the PHENIX IR.}
  \label{fig:evolution}
\end{figure}

To pursue these physics questions we are proposing an upgrade
consisting of a 1.5\,T superconducting magnetic solenoid of inner
radius 140\,cm with silicon tracking, electromagnetic calorimetry, and
hadronic calorimetry providing uniform coverage for $|\eta|<1$.  The
sPHENIX solenoid is an existing magnet developed for the BaBar
experiment at SLAC, and recently ownership of this key component was
officially transferred to BNL.  An engineering drawing of the sPHENIX
detector and its incorporation into the PHENIX interaction region are
shown in Figure~\ref{fig:evolution}.

The sPHENIX plan has been developed in conjunction with the official
timeline from BNL management.  The expectation is for RHIC running
through 2016, a shutdown in 2017, RHIC running for the increased
luminosity beam energy scan in 2018--2019, a shutdown in 2020, and
RHIC running in 2021 and 2022. We anticipate installing the magnet,
the hadronic calorimeter and portions of the tracking system to enable
significant commissioning of sPHENIX during the 2019 running period.
The sPHENIX detector will be completely integrated during the 2020
shutdown and would be available for physics at the start of the 2021
run.  With the high luminosity available at RHIC and the high sPHENIX
data acquisition bandwidth, sPHENIX will record 100 billion and sample
over 2/3 of a trillion \auau collisions at $\sqrt{s_{NN}} = 200$\,GeV
in a 22 week physics run period.  The high rate capability of sPHENIX
will enable the recording of over 10 million dijet events with $\ET >
20$\,GeV, along with a correspondingly large $\gamma$$+$jet sample.
We envision a run plan for 2021--2022 consisting of two 30 week
physics runs allowing a period for final commissioning, 22 weeks of
Au$+$Au running, and extended periods of \pp and \pdau running.

The design of sPHENIX takes advantage of a number of technological
advances to enable a significantly lower cost per unit solid angle
than has been previously possible, and we have obtained budgetary
guidance from well-regarded vendors for the major components of
sPHENIX.  Further cost savings are achieved by reusing significant
elements of the existing PHENIX mechanical and electrical
infrastructure.  Thus sPHENIX physics will be delivered in a very cost
effective way.

We have designed sPHENIX so that it could serve as the foundation for
a detector intended to make physics measurements at a future electron
ion collider (EIC) at RHIC.  The BNL implementation of the EIC, eRHIC,
adds a 5--15\,GeV electron beam to the current hadron and nuclear beam
capabilities of RHIC.  The sPHENIX detector, when combined with future
upgrades in the backward ($\eta < -1$) and forward ($\eta > 1$)
regions enables a full suite of EIC physics measurements as described
in Appendix~\ref{chap:ePHENIX}.  There is also the potential, if one
can realize appropriate instrumentation in the hadron-going direction
while polarized \pp and \pA collisions are available at RHIC, to
pursue a rich program of forward physics
measurements~\cite{forwardWhitePaper}.

In Chapter~\ref{chap:physics_case}, we detail the physics accessible
via jet, dijet, $\gamma$$+$jet, fragmentation function, and Upsilon
measurements at RHIC to demonstrate mission need.  In
Chapter~\ref{chap:detector_requirements}, we detail the sPHENIX
detector and subsystem requirements needed to achieve the physics
goals.  In Chapter~\ref{chap:detector_concept}, we detail the specific
detector design and \geant simulation results.  In
Chapter~\ref{chap:jet_performance}, we detail the physics performance
with full detector simulations.  In Appendix~\ref{chap:PreshowerFHCal}
we describe two possible augmentations of the baseline sPHENIX
detector: one, a preshower for the electromagnetic calorimeter to
extend the reach of direct photon measurements; and two, a forward
calorimeter to extend the acceptance of sPHENIX and to provide
access to additional physics. Lastly, in Appendix~\ref{chap:ePHENIX}
we include a copy of a Letter of Intent for an EIC detector built
around the BaBar magnet and the sPHENIX calorimetry.

\cleardoublepage

\resetlinenumber

\tableofcontents
\cleardoublepage

\mainmatter

\renewcommand{\thepage}{\arabic{page}}
\setcounter{chapter}{0}
\setcounter{page}{1}

\makeatletter{}\chapter{The Physics Case for sPHENIX}
\label{chap:physics_case}

\begin{figure}[!hbt]
 \begin{center}
   \raisebox{4pt}{\includegraphics[trim = 2 2 2 2, clip, width=0.46\linewidth]{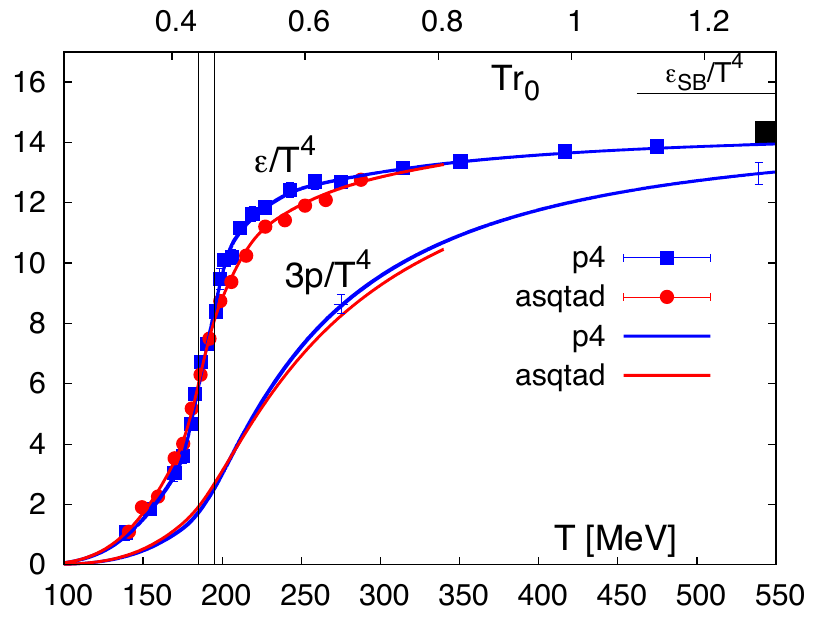}}
    \hfill
    \includegraphics[trim = 2 2 2 2, clip, width=0.52\linewidth]{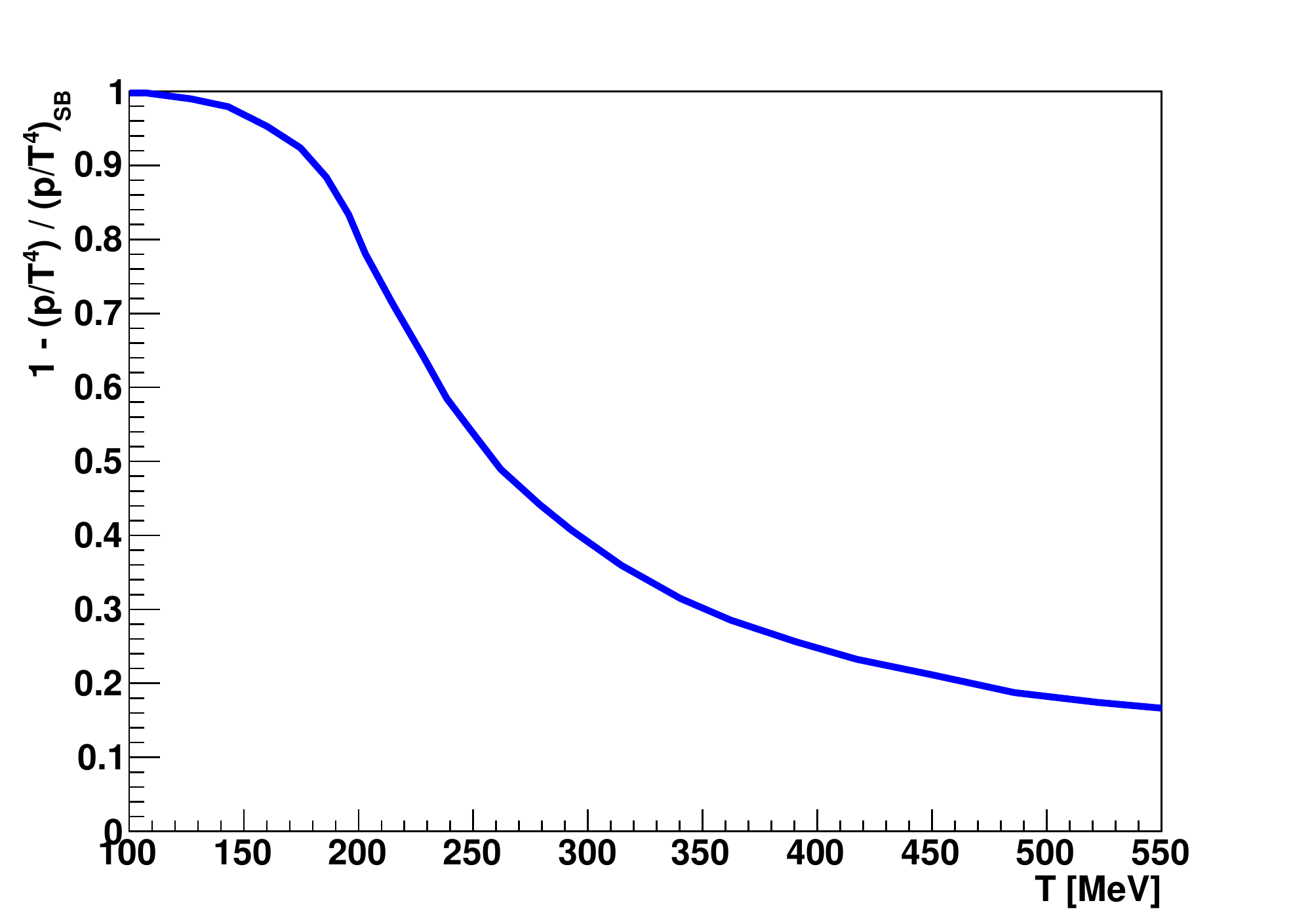}
    \caption[$E/3p$ and $p/T^{4}$ vs temperature]{(left) The energy density and three times the pressure
      normalized by $1/T^{4}$ as a function of
      temperature~\protect\cite{PhysRevD.80.014504}{}. (right) Deviation in 
      $p/T^{4}$ relative to the Stefan-Boltzmann value as a function
      of temperature.  The deviation from the Stefan-Boltzmann value is 23\%, 39\%, 53\%, and 80\% at
      temperatures of 420, 300, 250, and 200~MeV, respectively.
    \label{fig:lattice}}
 \end{center}
\end{figure}

\makeatletter{}Hadronic matter under conditions of extreme temperature or net baryon
density transitions to a new state of matter called the \qgp.  Lattice
QCD calculations at zero net baryon density indicate a smooth
crossover transition at $T_{c} \approx 170$~MeV, though with a rapid
change in properties at that temperature as shown in the left panel of
Figure~\ref{fig:lattice}~\cite{PhysRevD.80.014504}.  This \qgp
dominated the early universe for the first six microseconds of its
existence.  Collisions of heavy nuclei at the Relativistic Heavy Ion
Collider (RHIC) have sufficient initial kinetic energy that is then
converted into heat to create \qgp with an initial
temperature---measured via the spectrum of directly emitted
photons---of greater than 300~MeV~\cite{Adare:2008ab}. The higher
energy collisions at the Large Hadron Collider (LHC) produce an even
higher initial temperature $T > 420$~MeV~\cite{Luzum:2009sb}.

In materials where the dominant forces are electromagnetic, the
coupling $\alpha_{\mathrm{em}}$ is always much less than one.  Even so,
many-body collective effects can render perturbative calculations
non-convergent and result in systems with very strong
effective coupling~\cite{Adams:2012th}.  In cases where the nuclear force is
dominant, and at temperature scales of order 1--3\,$T_{c}$, the
coupling constant $\alpha_{s}$ is not much less than one and the
system is intrinsically non-perturbative.  In addition, the many-body
collective effects in the \qgp and their temperature dependence near
$T_{c}$ are not yet well understood.

The right panel of Figure~\ref{fig:lattice} shows the deviation from
the Stefan-Boltzmann limit of Lattice QCD results for the pressure
normalized by $1/T^{4}$.  The Stefan-Boltzmann limit holds for a
non-interacting gas of massless particles (i.e., the extreme of the
weakly coupled limit), and as attractive inter-particle interactions
grow stronger the pressure decreases.  Thus, one might expect that the
\qgp would transition from a weakly coupled system at high temperature
to a more strongly coupled system near $T_{c}$.  However, a direct
quantitative extraction of the coupling strength warrants caution as
string theory calculations provide an example where the coupling is
very strong and yet the deviation from the Stefan-Boltzmann limit is
only 25\%~\cite{Gubser:2009fc,Gubser:1996de}.  The change in initial
temperature between RHIC and LHC collisions is thus expected to be
associated with important changes in the nature of the
\qgp~\cite{Wiedemann:2009sa}.  If not, the question is why not.

The collisions at RHIC and the LHC involve a time evolution during
which the temperature drops as the \qgp expands.  The real constraint
on the temperature dependence of the \qgp properties will come from
calculations which simultaneously describe observables measured at both energies.
Since we are studying a phase transition, it is crucial to do
experiments near the phase transition and compare them with
experiments done further above $T_c$.  Typically, all the non-scaling
behavior is found near the transition.

For many systems the change in coupling strength is related to
quasiparticle excitations or strong coherent fields, and to study
these phenomena one needs to probe the medium at a variety of length
scales.  For example, in a superconductor probed at long length
scales, one scatters from Cooper pairs; in a superconductor probed at
short distance scales one observes the individual electrons.  Hard
scattered partons generated in heavy ion collisions that traverse the
\qgp serve as the probes of the medium.  Utilizing these partonic
probes, measured as reconstructed jets, over the broadest possible
energy scale is a key part of unraveling the quasiparticle puzzle in
the \qgp.  Jets at the LHC reach the highest energies, the largest initial
virtualities, and large total energy loss to probe the shortest distance scales. The lower
underlying event activity at RHIC will push the jet probes to lower energies and lower
initial virtualities thus probing the important longer distance scales in the medium.
Measurements of the three Upsilon states that span a large range in binding energy 
and size are an excellent complement to the jet program, with precision required at 
both RHIC and the LHC.

Continued developments in techniques for jet reconstruction in the
environment of a heavy ion collision have allowed the LHC experiments
to reliably recover jets down to
40~GeV~\cite{Aad:2013sla,Aad:2012vca}, which is well within the range
of reconstructed jet energies at RHIC.  This overlap opens the
possibility of studying the \qgp at the same scale but under different
conditions of temperature and coupling strength.

Apart from the temperature and coupling strength differences in the medium created
at RHIC and the LHC, the difference in the steepness of the hard scattering \pt spectrum
plays an important role.   The less steeply falling spectrum at the LHC has the benefit
of giving the larger reach in \pt with reconstructed jets expected up to 1 TeV.   At RHIC,
the advantage of the more steeply falling spectrum is the greater sensitivity to the medium
coupling and \qgp modifications of the parton shower.   This greater sensitivity may enable
true tomography in particular with engineering selections for quarks and/or gluons with longer
path length through the medium.   In addition, for correlations, once a clean direct photon or jet
tag is made, the underlying event is 2.5 times smaller at RHIC compared to the LHC thus giving
cleaner access to the low energy remnants of the parton shower and possible medium response.

This Chapter is organized into Sections as follows.  We first describe
the key ways of 'pushing' and 'probing' the \qgp to understand its
properties.  We then discuss three different aspects in which the RHIC
jet results are crucial in terms of (1) the temperature dependence of
the QGP, (2) the microscopic inner workings of the QGP, and (3) the
QGP time evolution along with the parton shower evolution.  We relate each
of these three aspects to specific observables measurable with sPHENIX.  We then
discuss the current state of jet probe measurements from RHIC and LHC
experiments, followed by a review of theoretical calculations for RHIC
jet observables.  We discuss the specific physics of heavy quark jets and open
heavy flavor in terms of Upsilon observables.   Finally, we review the rates available
that enable precision measurements across this comprehensive program.

\section{Pushing and probing the QGP}

Results from RHIC and LHC heavy ion experiments have provided a wealth
of data for understanding the physics of the \qgp.  One very
surprising result discovered at RHIC was the fluid-like flow of the
\qgp~\cite{Adcox:2004mh}, in stark contrast to some expectations that
the \qgp would behave as a weakly coupled gas of quarks and gluons.
It was originally thought that even at temperatures as low as
2--5\,$T_{c}$, the \qgp could be described with a weakly coupled
perturbative approach despite being quite far from energy scales
typically associated with asymptotic freedom.  

The \qgp created in heavy ion collisions expands and cools, eventually passing through the
phase transition to a state of hadrons, which are then measured by
experiment.  Extensive measurements of the radial and flow coefficients of
various hadrons, when compared to hydrodynamics calculations, imply a very
small ratio of shear viscosity to entropy density,
$\eta/s$~\cite{Luzum:2008cw}.  In the limit of very weak coupling
(i.e., a non-interacting gas), the shear viscosity is quite large as
particles can easily diffuse across a velocity gradient in the medium.
Stronger inter-particle interactions inhibit diffusion to the limit
where the strongest interactions result in a very short mean free path
and thus almost no momentum transfer across a velocity gradient,
resulting in almost no shear viscosity.  

The shortest possible mean
free path is of order the de~Broglie wavelength, which sets a lower
limit on $\eta/s$~\cite{Danielewicz:1984ww}.  A more rigorous
derivation of the limit $\eta/s \ge 1/4\pi$ has been calculated
within string theory for a broad class of strongly coupled gauge
theories by Kovtun, Son, and Starinets (KSS)~\cite{Kovtun:2004de}.
Viscous hydrodynamic calculations assuming $\eta/s$ to be temperature
independent through the heavy ion collision time evolution are
consistent with the experimental data where $\eta/s$ is within 50\% of
this lower bound for strongly coupled
matter~\cite{Luzum:2008cw,Song:2007ux,Alver:2010dn,Teaney:2009qa,Schenke:2011zz,Adare:2011tg}.
Even heavy quarks (i.e., charm and beauty) are swept up in the fluid
flow and theoretical extractions of the implied $\eta/s$ are equally
small~\cite{Adare:2006nq}.

\begin{figure}[t]
 \begin{center}
   \includegraphics[width=\onewidth]{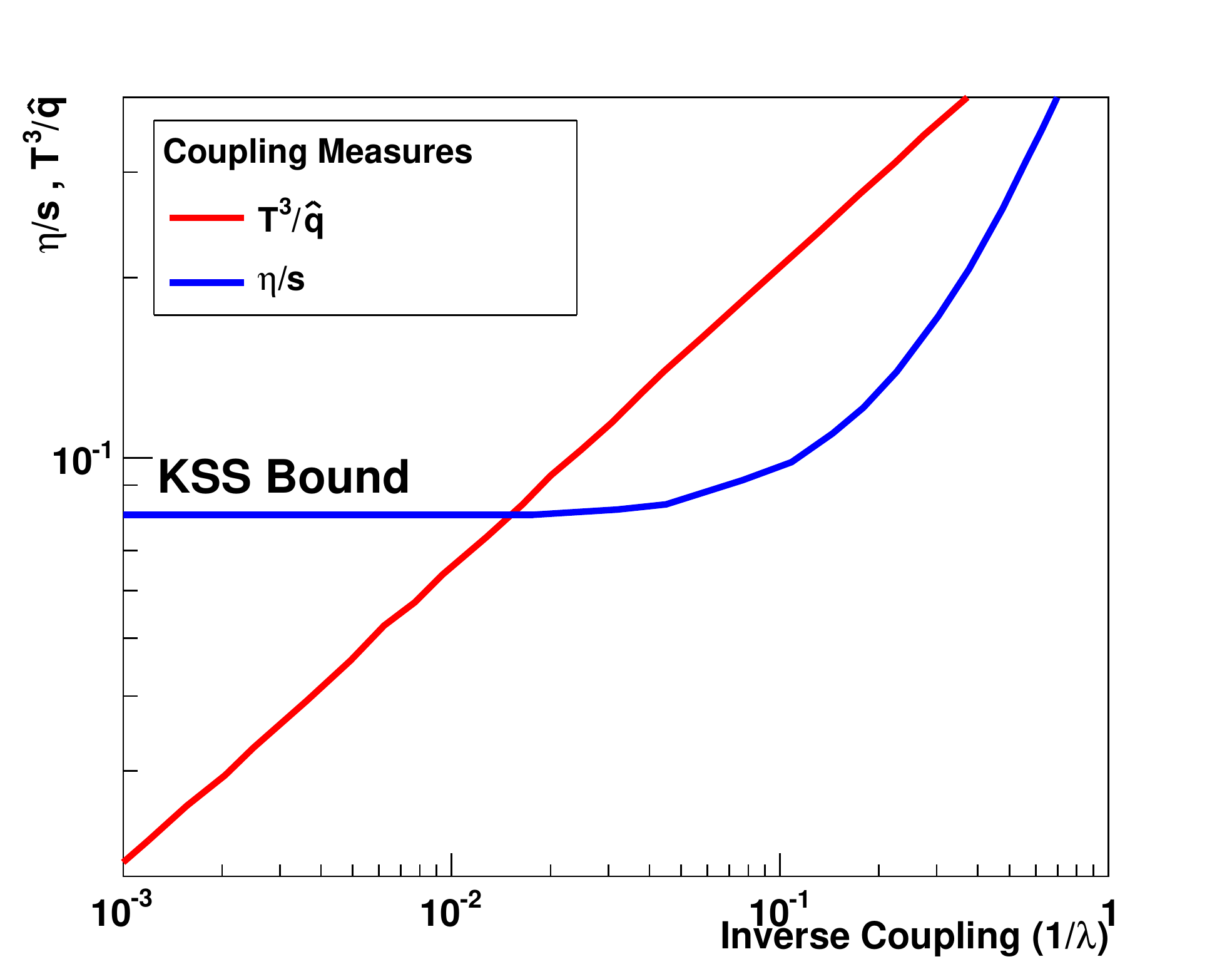} 
   \caption[$\eta/s$ and $T^3/\hat{q}$ vs inverse 't~Hooft coupling]{
     $\eta/s$ (blue) and $T^3/\hat{q}$ (red) as a function of the
     inverse of the 't~Hooft coupling\protect\cite{Majumder:2007zh}{}.
     For large $\lambda$ (i.e., small $1/\lambda$), $\eta/s$
     approaches the quantum lower bound asymptotically, losing its
     sensitivity to further changes in the coupling strength.}
    \label{fig:etas_versus_qhat} 
 \end{center}
\end{figure}

Other key measures of the coupling strength to the medium are found in
the passage of a hard scattered parton through the \qgp.  As the
parton traverses the medium it accumulates transverse momentum as
characterized by $\hat{q} = d(\Delta p_{T}^{2})/dt$ and transfers
energy to the medium via collisions as characterized by $\hat{e} =
dE/dt$.  Ref.~\cite{Liu:2006ug} has calculated $\hat{q}/T^3$ in
${\cal N} = 4$ supersymmetric Yang-Mills theory to be proportional to
the square root of the coupling strength whereas $\eta/s$
asymptotically approaches the quantum lower bound as the coupling
increases.  Both of these ratios are shown as a function of the
inverse coupling in Figure~\ref{fig:etas_versus_qhat}.  For large 't
Hooft coupling ($\lambda$), $\eta/s$ is already quite close to
$1/4\pi$, whereas $T^3/\hat{q}$ is still changing.  This behavior has
caused the authors of Ref.~\cite{Majumder:2007zh} to comment: ``The
ratio $T^{3}/\hat{q}$ is a more broadly valid measure of the coupling
strength of the medium than $\eta/s$.''

In vacuum, the hard scattered parton creates a shower of particles
that eventually form a cone of hadrons, referred to as a jet.  In the
\qgp, the lower energy portion of the shower may eventually be
equilibrated into the medium, thus giving a window on the rapid
thermalization process in heavy ion collisions.  This highlights part
of the reason for needing to measure the fully reconstructed jet
energy and the correlated particle emission with respect to the jet at
all energy scales.  In particular, coupling parameters such as
$\hat{q}$ and $\hat{e}$ are scale dependent and must take on weak
coupling values at high enough energies and strong
coupling values at thermal energies.

\begin{figure}[ht]
 \begin{center}
   \includegraphics[trim = 2 2 2 2, clip, width=\onewidth]{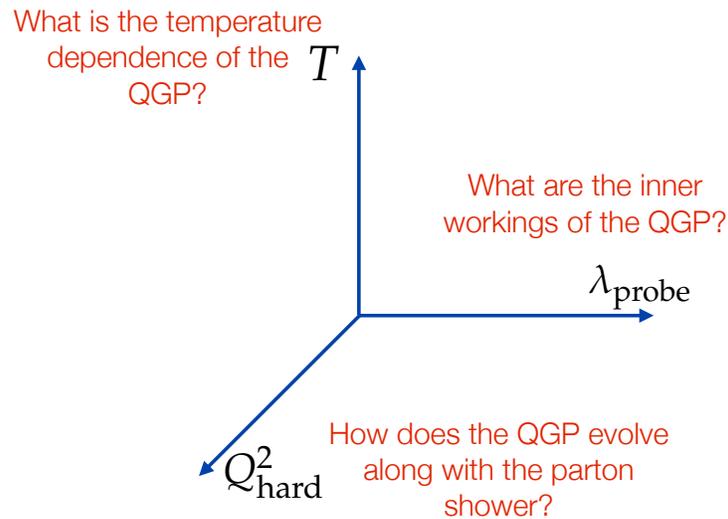}
   \caption[ Pushing and probing the \qgp along three axes]{Pushing
     Three illustrative axes along which the \qgp may be pushed and
     probed.  The axes are the temperature of the \qgp, the
     $Q^2_{\mathrm{hard}}$ of the hard process that sets of the scale
     for the virtuality evolution of the probe, and the wavelength
     with which the parton probes the medium
     $\lambda_{\mathrm{probe}}$.}
   \label{fig:threeaxes}
 \end{center}
\end{figure}

The focus of this proposal is the measurement of jet probes of the
medium as a way of understanding the coupling of the medium, the
origin of this coupling, and the mechanism of rapid equilibration.
The \qgp is one form of the ``condensed matter''
of QCD and in any rigorous investigation of condensed matter of any
type, it is critical to make measurements as one pushes the system
closer to and further from a phase transition and with probes at
different length scales.  
Substantially extending these scales with measurements at RHIC, particularly closer 
to the transition temperature and at longer distance scales, is the unique ability provided by this proposal.

The critical variables to manipulate for this program are the temperature of
the \qgp, the length scale probed in the medium, and the virtuality of
the hard process as shown schematically in Figure~\ref{fig:threeaxes}.
In the following three sections we detail the physics of each axis.

\section{What is the temperature dependence of the QGP?}
\label{sec:temperature_dependence}

\begin{figure}[t]
 \begin{center}
    \includegraphics[trim = 2 2 2 2, clip, width=\twowidth]{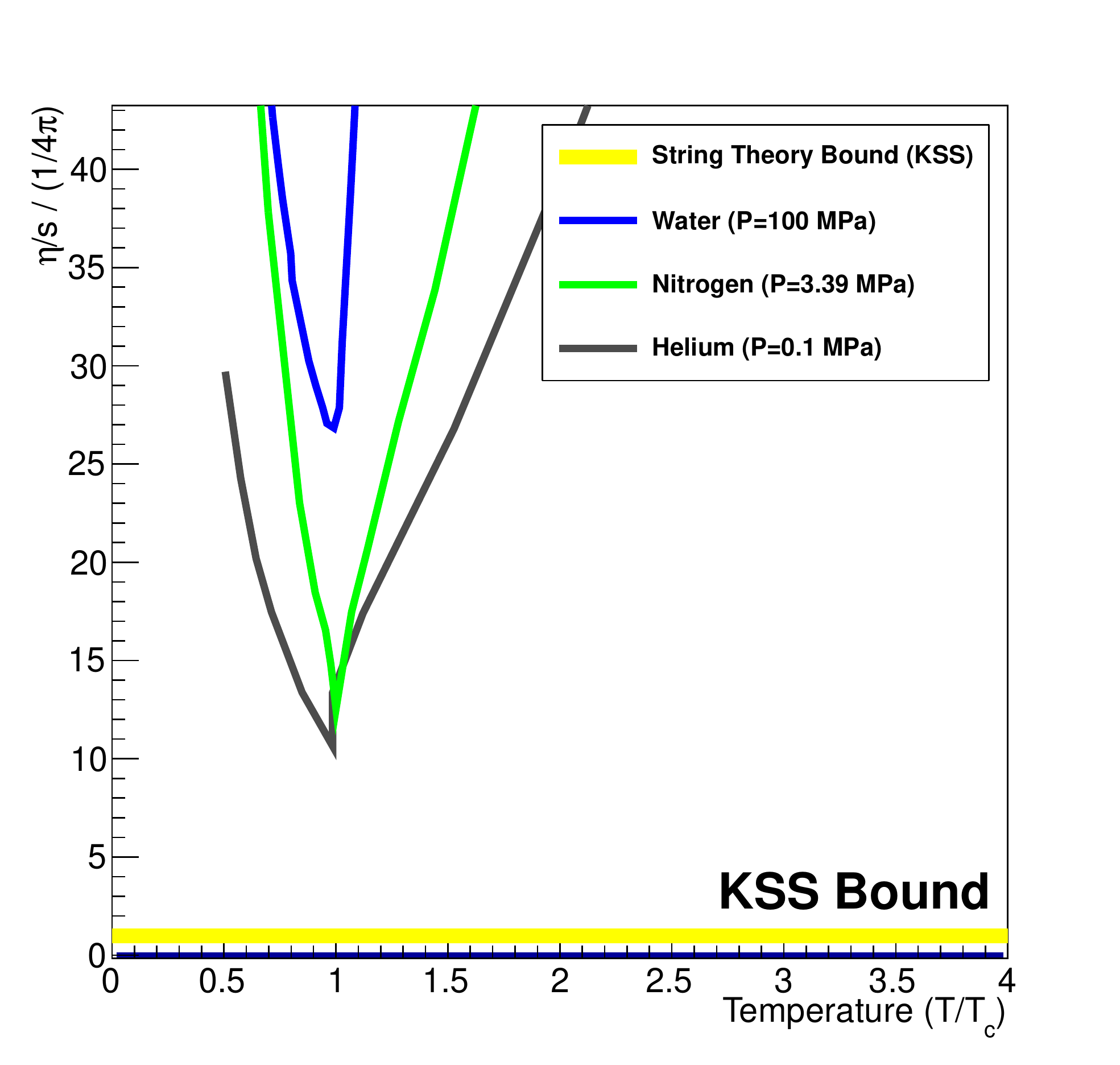}
    \hfill
    \includegraphics[trim = 2 2 2 2, clip, width=\twowidth]{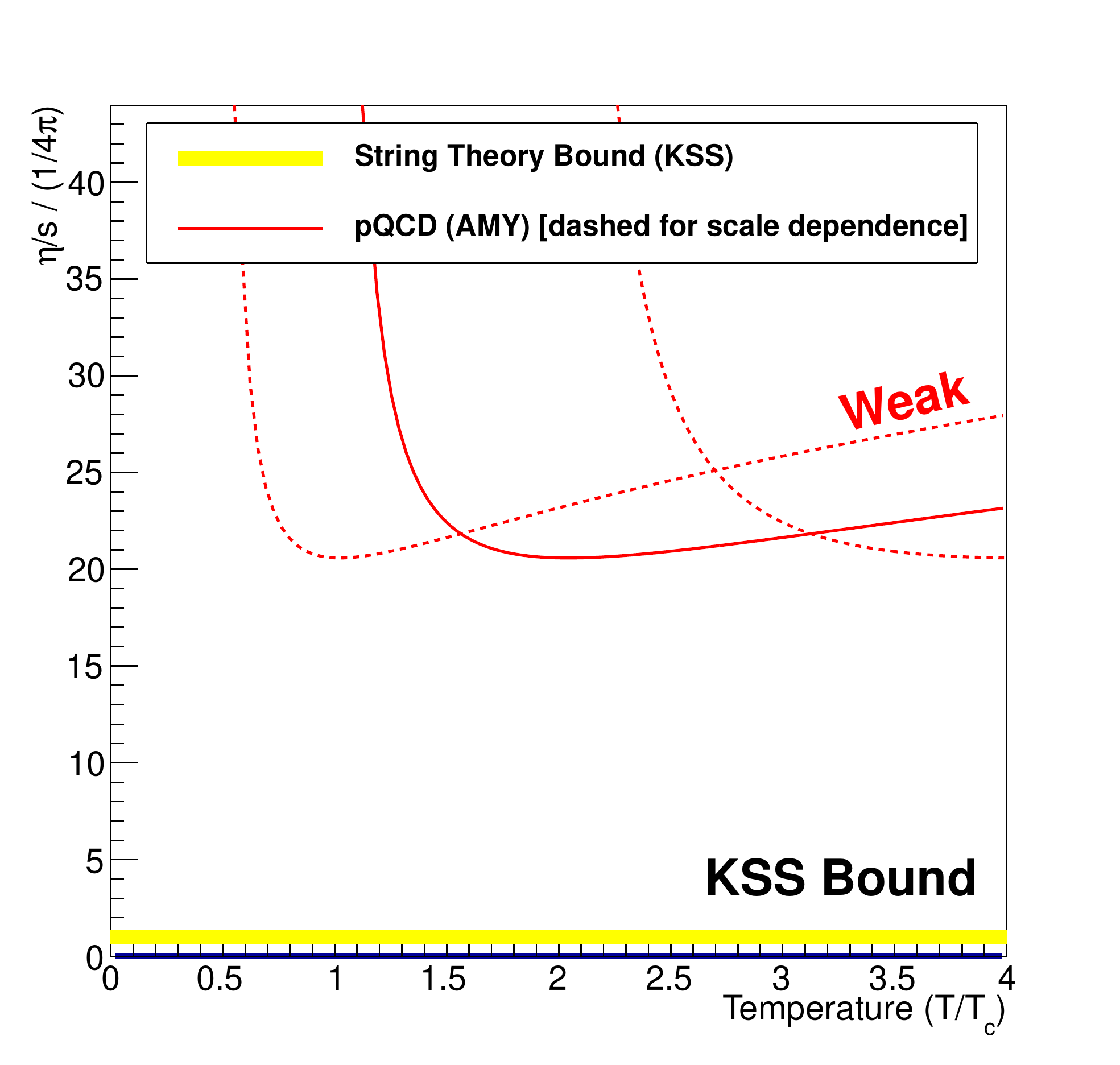}
    \caption[$\eta/s$ vs $T/T_{c}$ for water, nitrogen, and
    helium]{(left) The ratio of shear viscosity to entropy density,
      $\eta/s$, normalized by the conjectured KSS bound as a function
      of the reduced temperature, $T/T_{c}$, for water, nitrogen, and
      helium. The cusp for Helium corresponds to the case at the
      critical pressure.  (right) Calculation of hot QCD matter
      (quark-gluon plasma) for a weakly coupled system.  Dashed lines
      show the scale dependence of the perturbative calculation.}
    \label{fig:etaovers_1}
 \end{center}
\end{figure}

The internal dynamics of more familiar substances---the subjects of
study in conventional condensed matter and material physics---are
governed by quantum electrodynamics.  It is well known that near a
phase boundary they demonstrate interesting behaviors, such as the
rapid change in the shear viscosity to entropy density ratio,
$\eta/s$, near the critical temperature, $T_c$. This is shown in
Figure~\ref{fig:etaovers_1} for water, nitrogen, and
helium~\cite{Csernai:2006zz}.  Despite the eventual transition to
superfluidity at temperatures below $T_{c}$, $\eta/s$ for these
materials remains an order of magnitude above the conjectured quantum
bound of Kovtun, Son, and Starinets (KSS) derived from string
theory~\cite{Kovtun:2004de}.  These observations provide a deeper
understanding of the nature of these materials: for example the
coupling between the fundamental constituents, the degree to which a
description in terms of quasiparticles is important, and the
description in terms of normal and superfluid components.

The dynamics of the QGP are dominated by Quantum Chromodynamics and the
experimental characterization of the dependence of $\eta/s$ on
temperature will lead to a deeper understanding of strongly coupled
QCD near this fundamental phase transition.  Theoretically,
perturbative calculations in the weakly coupled limit indicate that
$\eta/s$ decreases slowly as one approaches $T_{c}$ from above, but
with a minimum still a factor of 20 above the KSS
bound~\cite{Arnold:2003zc} (as shown in the right panel of
Figure~\ref{fig:etaovers_1}).  However, as indicated by the dashed
lines in the figure, the perturbative calculation has a large
renormalization scale dependence and results for different values of
the scale parameter ($\mu, \mu/2, 2\mu$) diverge from each other near
$T_{c}$.

\begin{figure}[tp]
 \begin{center}
    \includegraphics[trim = 2 2 2 2, clip, width=\twowidth]{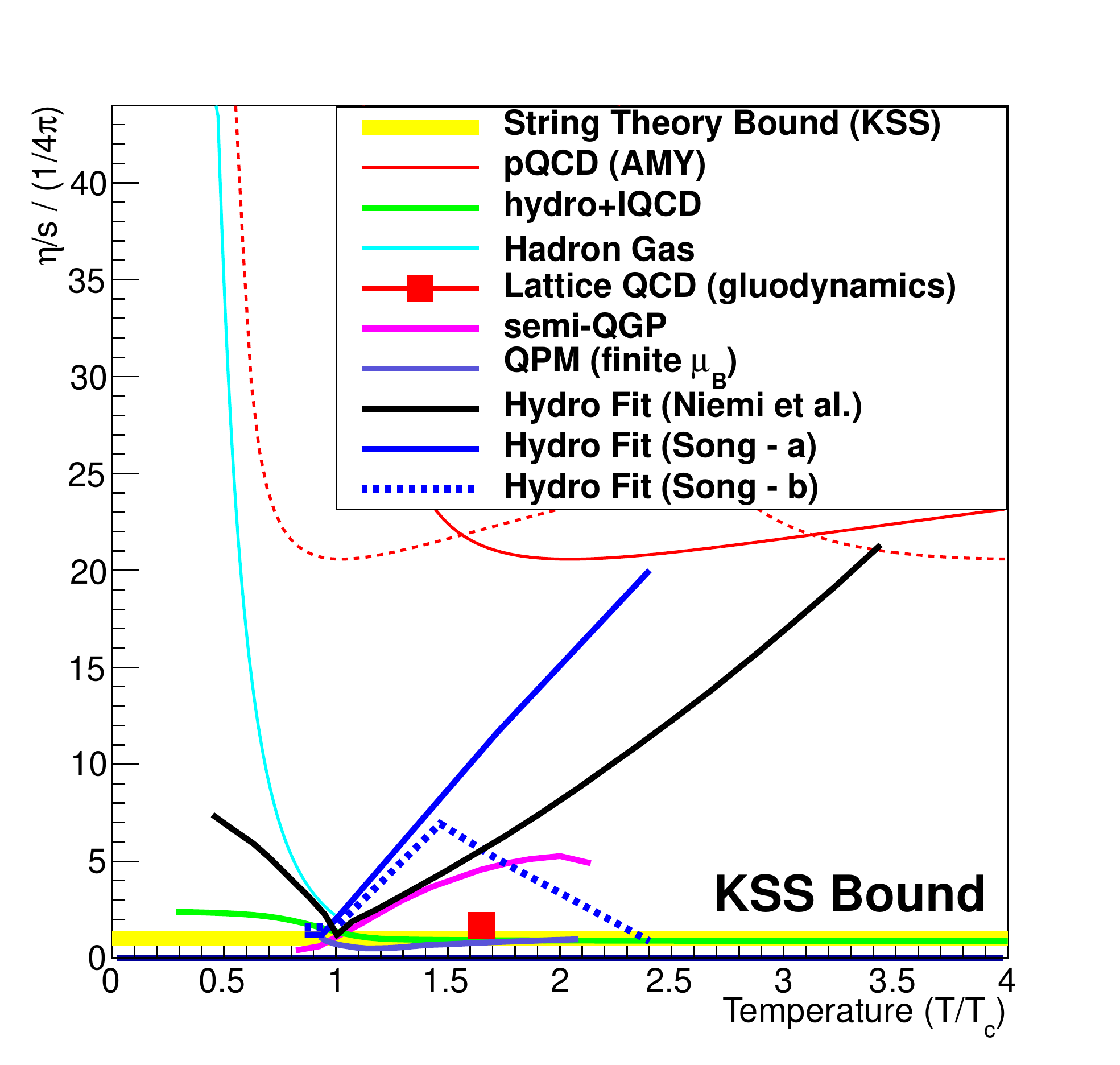}
    \hfill
    \includegraphics[trim = 2 2 2 2, clip, width=\twowidth]{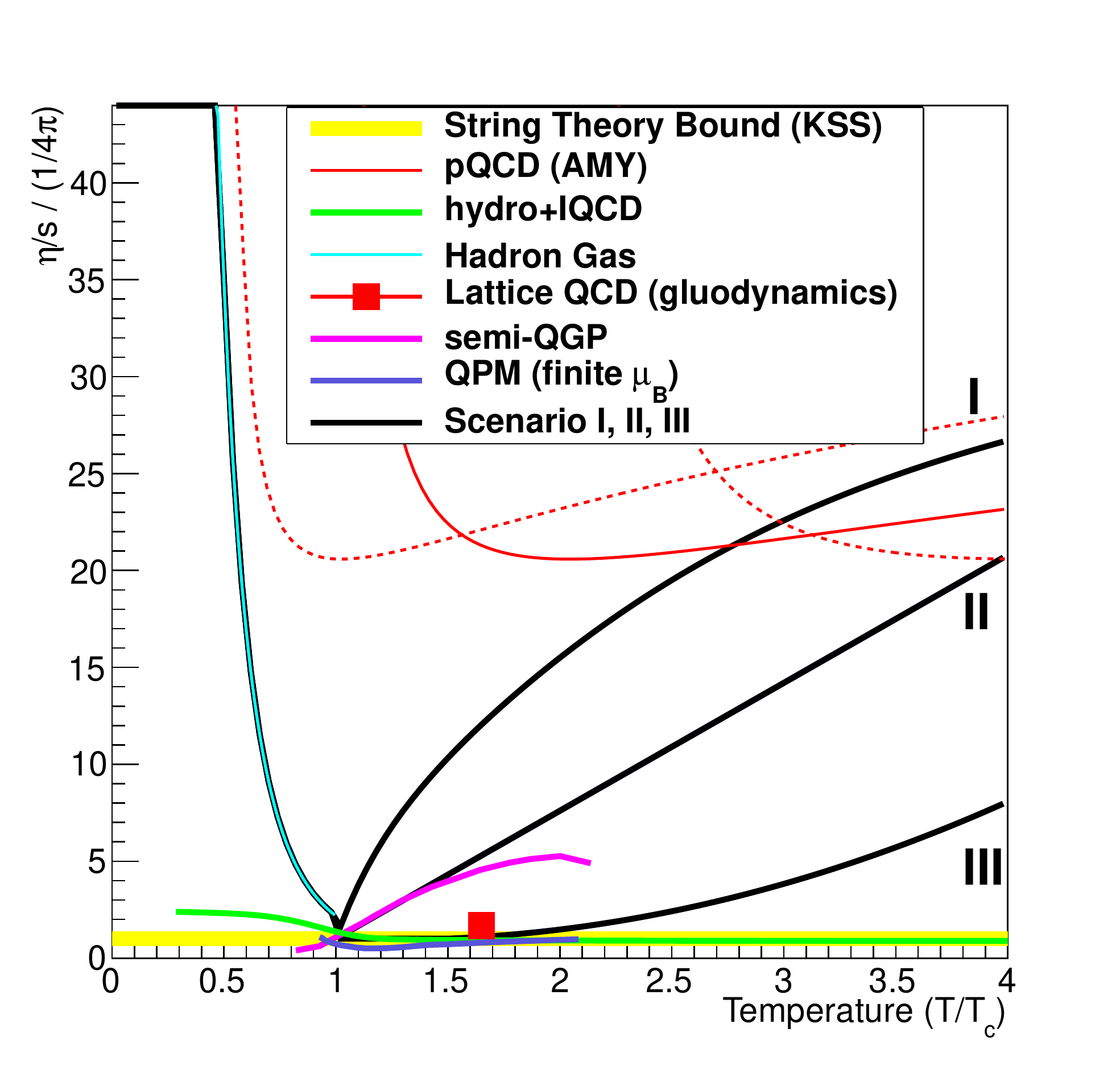}
    \caption[ $\eta/s$ vs $T/T_{c}$ for various \qgp
    calculations]{\label{fig:etaovers_2}(left) Shear viscosity divided
      by entropy density, $\eta/s$, renormalized by the conjectured
      KSS bound as a function of the reduced temperature, $T/T_{c}$,
      with various calculations for the quark-gluon plasma case. See
      text for discussion.  (right) Figure with three conjectured
      scenarios for the quark-gluon plasma transitioning from the
      strongly coupled bound (as a near-perfect fluid) to the weakly
      coupled case.}
 \end{center}
\end{figure}

Figure~\ref{fig:etaovers_2} (left panel) shows several
state-of-the-art calculations for $\eta/s$ as a function of
temperature. Hadron gas calculations show a steep increase in $\eta/s$
below $T_c$~\cite{Prakash:1993bt}, and similar results using the UrQMD
model have also been obtained~\cite{Demir:2008tr}.  Above $T_c$ there
is a lattice calculation in the SU(3) pure gauge
theory~\cite{Meyer:2007ic} resulting in a value near the KSS bound at
$T=1.65\,T_c$.  Calculations in the semi-QGP
model~\cite{Hidaka:2009ma}, in which color is not completely ionized,
have a factor of five increase in $\eta/s$ in the region of
1--2\,$T_c$.  Also shown are calculations from a quasiparticle model
(QPM) with finite $\mu_B$~\cite{Srivastava:2012sb} indicating little
change in $\eta/s$ up to 2\,$T_c$.  There is also an update on the
lower limit on $\eta/s$ from second order relativistic viscous
hydrodynamics~\cite{Kovtun:2011np}, with values remaining near
$1/4\pi$.  It is safe to say that little is known in a theoretically
reliable way about the nature of this transition or the approach to
weak-coupling.

Hydrodynamic modeling of the bulk medium does provide constraints on
$\eta/s$, and recent work has been done to understand the combined
constraints on $\eta/s$ as a function of temperature utilizing both
RHIC and LHC flow data sets~\cite{Gale:2012rq,Song:2011qa,Nagle:2011uz,Niemi:2011ix}.  
The results from~\cite{Niemi:2011ix} as
constrained by RHIC and LHC data on hadron transverse momentum spectra
and elliptic flow are shown in Figure~\ref{fig:etaovers_2} (left
panel). These reach the pQCD weak coupled value at $20 \times 1/4\pi$
for $T = 3.4 T_{c}$.  Also shown are two scenarios, labeled ``Song-a''
and ``Song-b'', for $\eta/s(T)$ in ~\cite{Song:2011qa} from which the 
authors conclude that ``one cannot unambiguously determine the
functional form of $\eta/s(T)$ and whether the QGP fluid is more
viscous or more perfect at LHC energy.''

Shown in Figure~\ref{fig:etaovers_2} (right panel) are three possible
scenarios for a more or less rapid modification of the medium from the
strong to the weak coupling limit.  Scenario I has the most rapid
change in $\eta/s(T)$ following the ``Song-a'' parametrization and
Scenario III has the least rapid change going through the lattice QCD
pure glue result~\cite{Meyer:2007ic}.  It is imperative to map out
this region in the `condensed matter' physics of QCD and extract the
underlying reason for the change.

The above discussion has focused on $\eta/s$ as the measure of the
coupling strength of the \qgp.  However, both $\eta/s$ and jet probe
parameters such as $\hat{q}$ and $\hat{e}$ are sensitive to the
underlying coupling of the matter, but in distinct ways. Establishing
for example the behavior of $\hat{q}$ around the critical temperature
is therefore essential to a deep understanding of the quark-gluon
plasma.  Hydrodynamic modeling may eventually constrain $\eta/s(T)$
very precisely, though it will not provide an answer to the question
of the microscopic origin of the strong coupling (something naturally
available with jet probes).

\begin{figure}[tp]
 \begin{center}
    \includegraphics[trim = 2 2 2 2, clip, width=\twowidth]{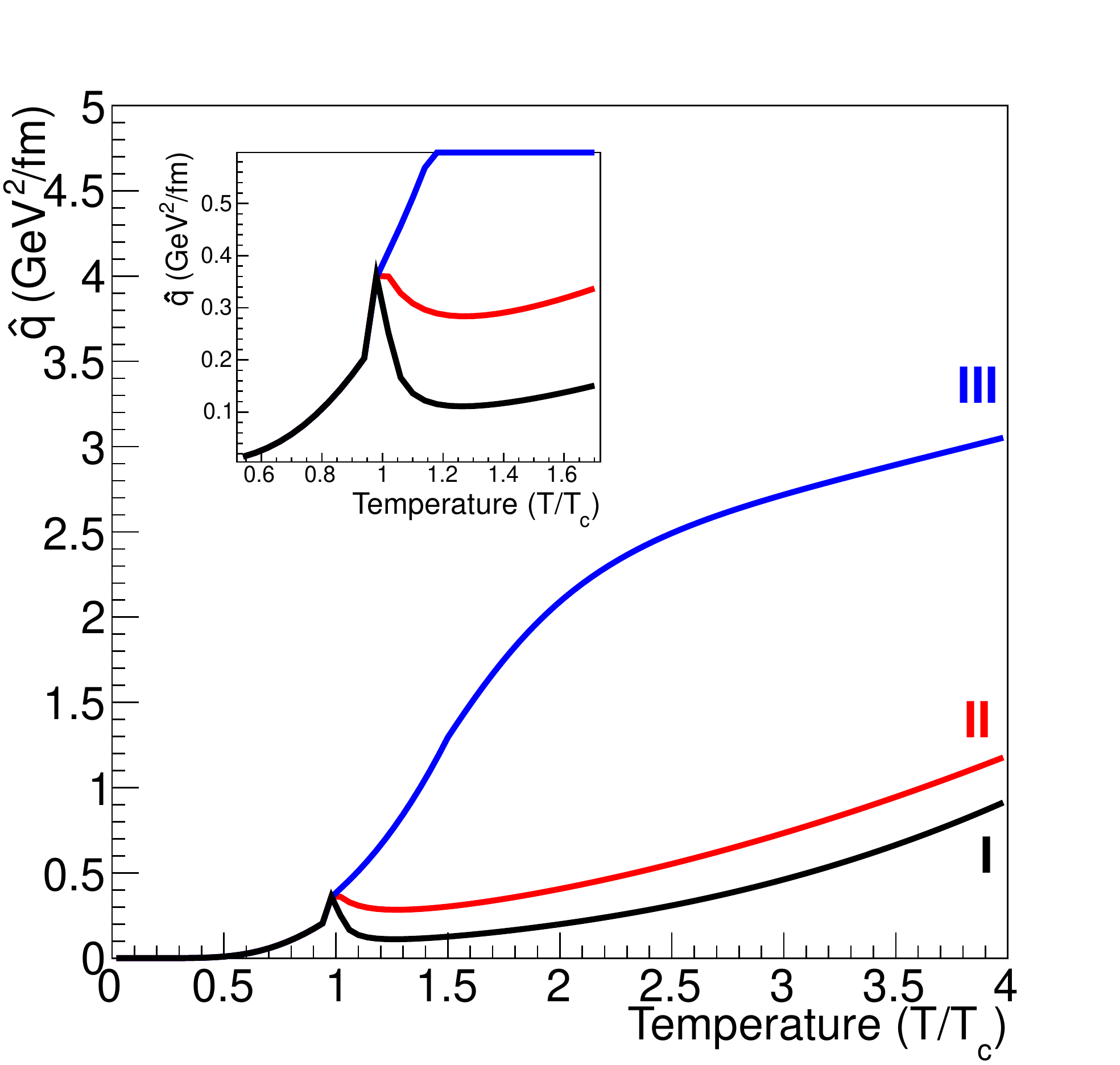}
    \hfill
    \includegraphics[trim = 2 2 2 2, clip, width=\twowidth]{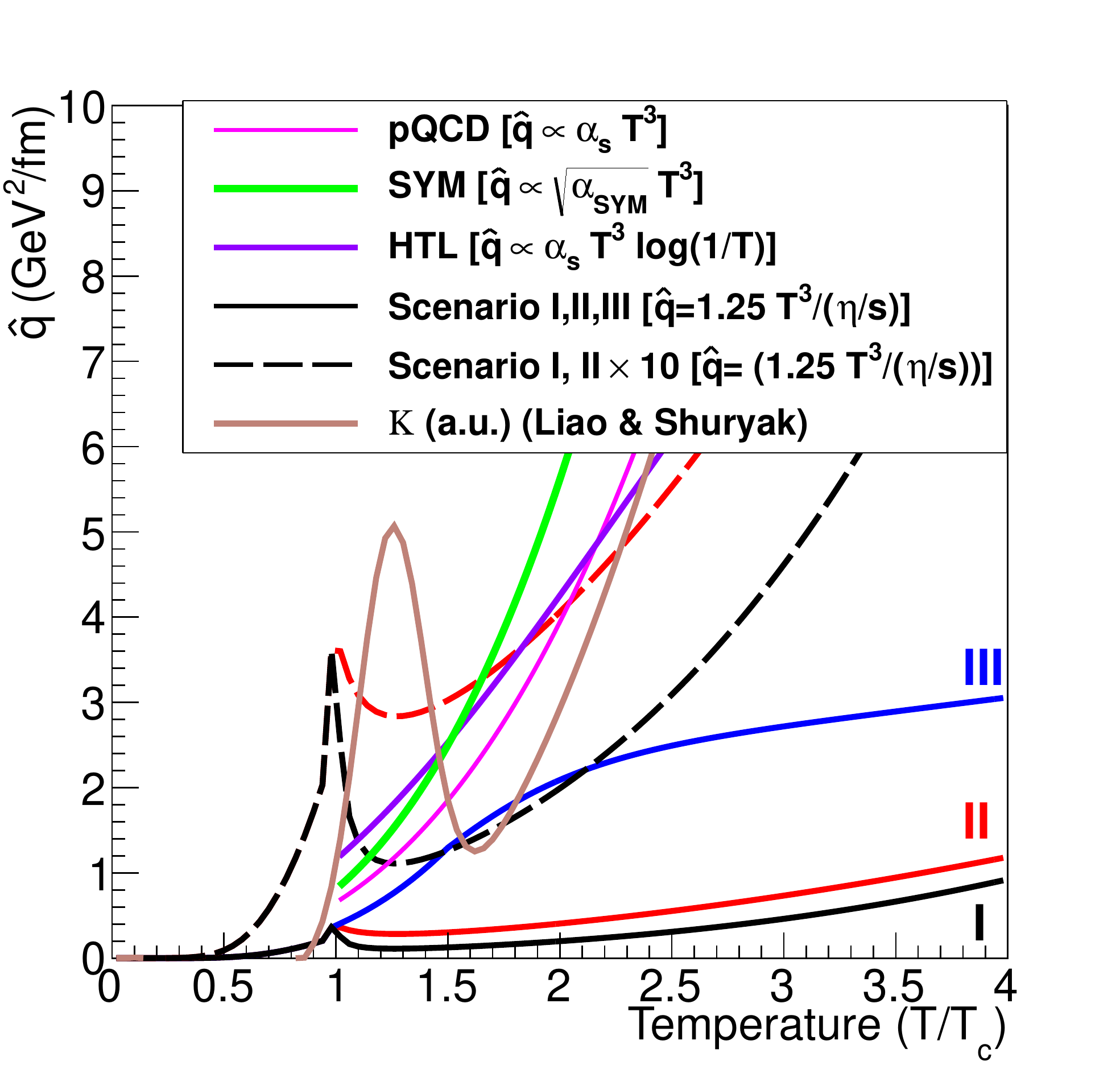}
    \caption[$\hat{q}$ vs $T/T_{c}$ in weak and strong coupling
    scenarios]{(left) $\hat{q}$ as a function of $T/T_{c}$ in the
      three scenarios as related with the weak-coupling
      calculation. (right) Different calculations for the scaling of
      $\hat{q}$ under weak and strong coupling assumptions.}
    \label{fig:qhatmap}
 \end{center}
\end{figure}

The authors of Ref~\cite{Majumder:2007zh} propose a test of the strong
coupling hypothesis by measuring both $\eta/s$ and $\hat{q}$.  They
derive a relation between the two quantities expected to hold in the
weak coupling limit:
\begin{equation}
\hat{q} \stackrel{?}{=} \frac{1.25 T^{3}}{\eta/s}.
\label{eq:qhat2etas}
\end{equation}
The authors conclude that ``an unambiguous determination of both sides
of [the equation] from experimental data would thus permit a model
independent, quantitative assessment of the strongly coupled nature of
the quark-gluon plasma produced in heavy ion collisions.''  For the
three scenarios of $\eta/s(T)$ shown in Figure~\ref{fig:etaovers_2}
(right panel), we calculate $\hat{q}$ as a function of temperature
assuming the equivalence case in Eqn.~\ref{eq:qhat2etas} and the
result is shown in Figure~\ref{fig:qhatmap} (left panel).  The inset in
Figure~\ref{fig:qhatmap} shows a magnified view of the region around
$T_c$ and a significant local maximum in $\hat{q}$ is observed in
scenarios I and II.

Figure~\ref{fig:qhatmap} (right panel) shows that for the equivalence
relation of Eqn.~\ref{eq:qhat2etas}, all three scenarios have a result
that differs significantly from the simple perturbative expectation of
$\alpha_{s} T^{3}$~\cite{Armesto:2011ht}.  Also shown in
Figure~\ref{fig:qhatmap} are the predicted temperature dependence of
$\hat{q}$ in the strongly coupled AdS/CFT (supersymmetric Yang-Mills)
case~\cite{Liu:2006ug} and the Hard Thermal Loop (HTL)
case~\cite{arnold:2000dr}.

Since the expected scaling of $\hat{q}$ with temperature is such a
strong function of temperature, jet quenching measurements should be
dominated by the earliest times and highest temperatures.  In order
to have sensitivity to temperatures around 1--2 $T_c$, measurements
at RHIC are needed in contrast to the LHC where larger initial
temperatures are produced.  In addition, the ability of RHIC to
provide high luminosity heavy-ion collisions at a variety of center
of mass energies can be exploited to probe the detailed temperature
dependence of quenching right in the vicinity of $T_c$.

\begin{figure}[ht]
  \centering
  \raisebox{0.2cm}{\includegraphics[width=0.49\linewidth]{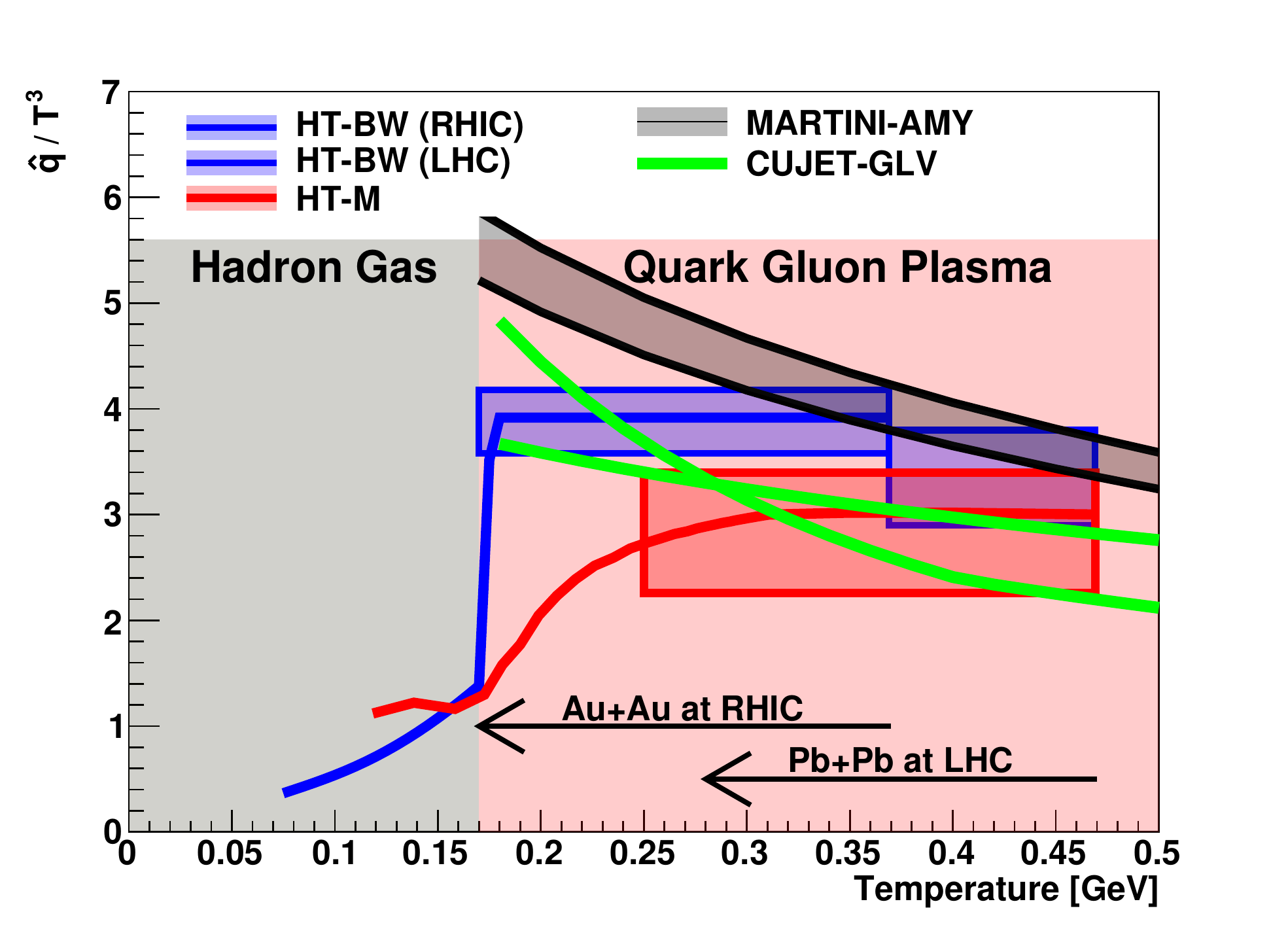}}
  \hfill
  \includegraphics[width=0.49\linewidth]{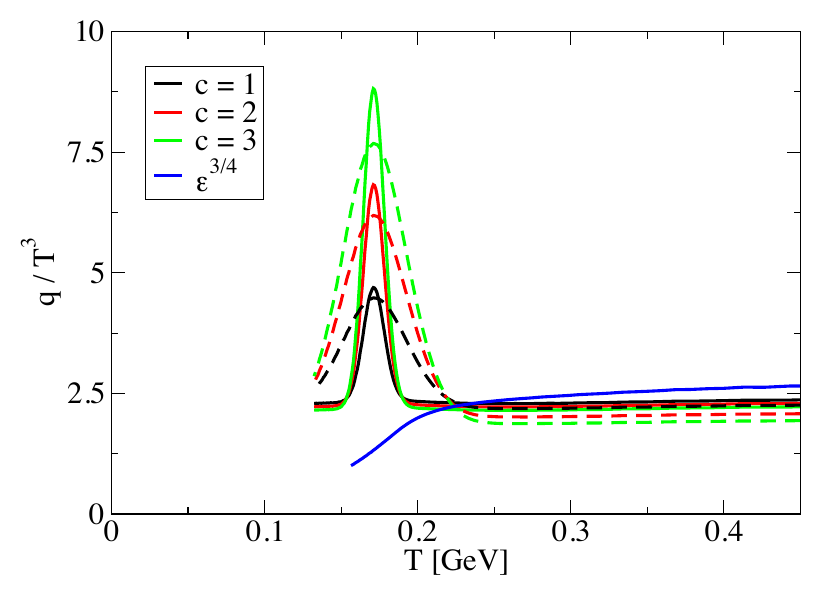}
  \caption[Calculations of $\hat{q}/T^{3}$ vs temperature, constrained
  by RHIC and LHC \raa data --- including near $T_C$ enhancement
  scenarios of $\hat{q}/T^{3}$]{(left) Calculations from four jet
    quenching frameworks constrained by RHIC and LHC \raa data
    with results for $\hat{q}/T^{3}$ as a function of temperature.
    Details of the calculation are given in
    Ref.~\cite{JETCollaboration:QhatConstraint}. (right) Near $T_C$
    enhancement scenarios of $\hat{q}/T^{3}$ considered in
    Ref.~~\cite{Renk:2014nwa}.}
  \label{fig:qhatconstraint}
\end{figure}

Theoretical developments constrained simultaneously by data from RHIC and the LHC 
have been important in discriminating against some models with very
large $\hat{q}$  -- see Figure~\ref{fig:cmsqhat} from Ref.~\cite{CMS:2012aa} and theory
references therein.  Models such as PQM and ASW with very large values of $\hat{q}$ have been ruled out
by the combined constraint.  Shown in the left panel of Figure~\ref{fig:qhatconstraint} is a recent compilation of four
theoretical calculations with a directly comparable extraction of $\hat{q}$.  Developments on the 
theory and experimental fronts have significantly narrowed the 
range of $\hat{q}$~\cite{JETCollaboration:QhatConstraint}.
This theoretical progress lends strength to the case that the tools will be
available on the same time scale as sPHENIX data to have precision determinations
of $\hat{q}$ and then ask deeper additional questions about the \qgp and its underlying properties.

It is notable that a number of calculations favor an increased coupling strength near the
transition temperature.  Shown in the right panel of Figure~\ref{fig:qhatconstraint} are a set of scenarios
considered by Renk in Ref.~\cite{Renk:2014nwa}.   This paper states that ``Comparing weak coupling scenarios with data, 
NTC [near $T_{C}$ enhancement] is favored.   An answer to this question will require a systematic picture 
across several different high \pt observables.''   

In Ref.~\cite{Liao:2008dk}, Liao and Shuryak use 
RHIC measurements of single hadron suppression and azimuthal
anisotropy to infer that ``the jet quenching is a few times stronger
near $T_c$ relative to the quark-gluon plasma at $T > T_c$.''  This
enhancement of $\hat{q}$ is shown in Figure~\ref{fig:qhatmap} (right panel)
and is the result of color magnetic monopole excitations in the plasma
near $T_{c}$.   

\begin{figure}[ht]
  \centering
  \raisebox{0.2cm}{\includegraphics[width=0.68\linewidth]{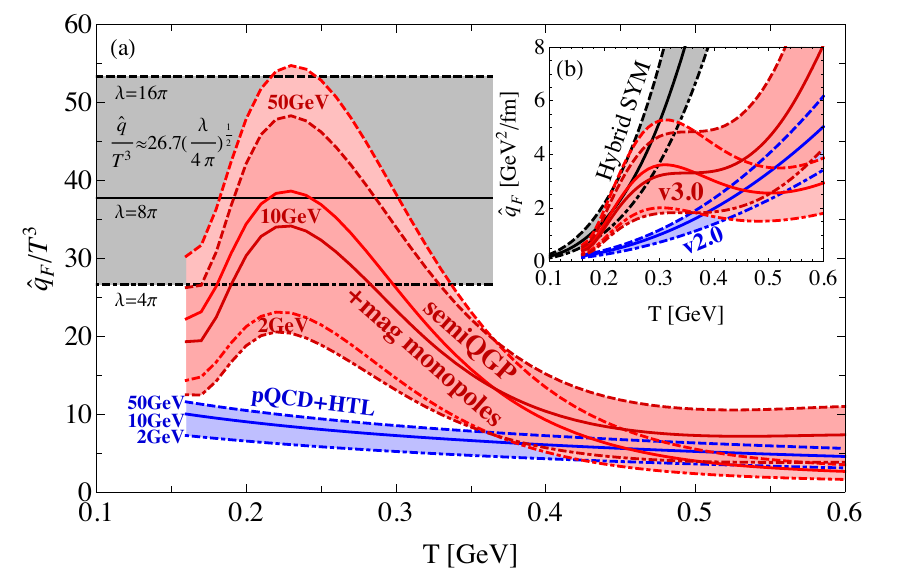}}
  \caption[Magnetic Monopole/CUJET 3.0 result for $\hat{q}/T^{3}$ vs temperature, constrained
  by RHIC and LHC \raa data]{Results from calculations within CUJET 3.0 with magnetic monopole
excitations that result in enhanced coupling near $T_{c}$.   Plotted are the constraints on $\hat{q}/T^{3}$ 
as a function of temperature as shown in Ref.~~\cite{Xu:2014tda}.}
  \label{fig:cujet_monopole}
\end{figure}

Most recently this strong coupling picture with color magnetic monopole excitations has been
implemented within CUJET 3.0 for a broader comparison with experimental observables and
previous theory calculations~\cite{Xu:2014tda}.   Shown in Figure~\ref{fig:cujet_monopole} are
results from their constrained RHIC and LHC data fit for the temperature dependence of the scaled
quenching power $\hat{q}/T^{3}$.

Within the jet quenching model WHDG~\cite{Horowitz:2011gd}, the authors 
constrain $\hat{q}$ by the PHENIX $\pi^{0}$ nuclear modification factor.  
They find the prediction scaled by the expected increase in the color charge 
density created in higher energy LHC collisions when compared to the ALICE
results~\cite{Aamodt:2010jd} over-predicts the suppression.
This over-prediction based on the assumption of an unchanging probe-medium
coupling strength led to title of Ref.~\cite{Horowitz:2011gd}: ``The
surprisingly transparent sQGP at the LHC.''  They state that ``one
possibility is the sQGP produced at the LHC is in fact more
transparent than predicted.''  Similar conclusions have been reached
by other authors~\cite{Chen:2011vt,Zakharov:2011ws,Buzzatti:2011vt}.
Recently work has been done to incorporate the running of the QCD coupling
constant~\cite{Buzzatti:2012dy}.  

It is important to note that most all calculations predict a stronger
coupling near the transition, even if just from the running of the
coupling constant $\alpha_{s}$, and the goal is to experimentally
determine the degree of the effect.  Lower energy data at RHIC also
provides important constraints -- see for example
Refs.~\cite{Adare:2012uk,Schmah:2013vea}.  The full set of
experimental observables need to be considered spanning the largest
range of collision energy, system size, and engineering path length.

\begin{figure}[!hbt]
 \begin{center}
    \includegraphics[trim = 2 2 2 2, clip, width=0.6\linewidth]{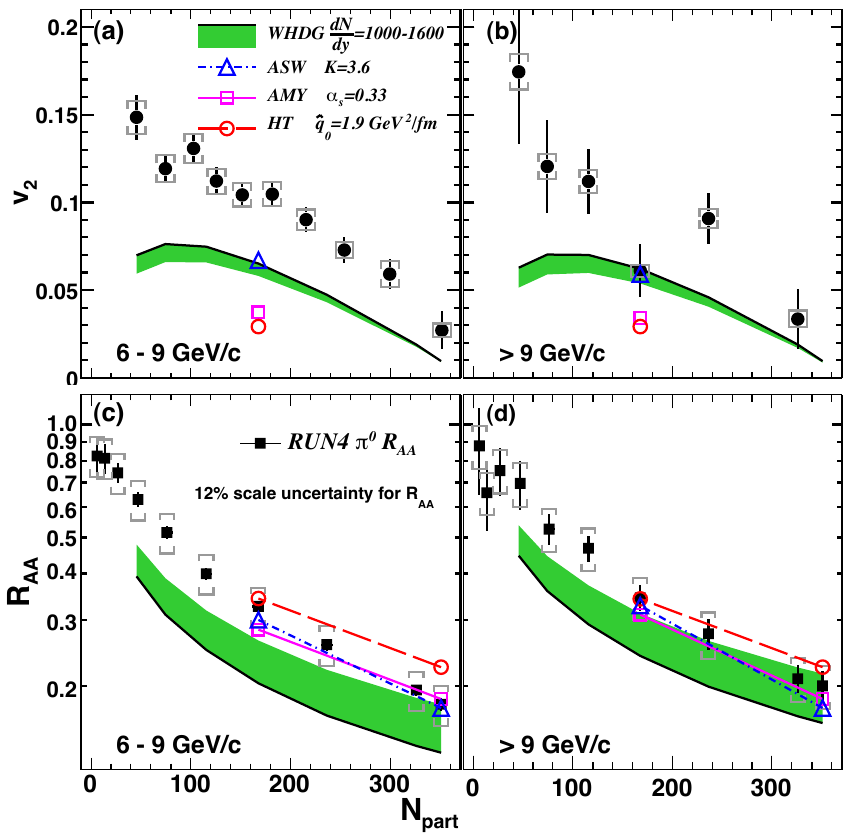}
    \caption[Experimental $\pi^0$ $v_2$ and \raa along with
    calculations from four weakly coupled energy loss
    models]{\label{fig:phenixpi0v2} $\pi^0$ $v_2$
      (top panels) and \raa (bottom panels) for $6<p_T<9$~GeV/c
      (left panels) and $p_T>9$~GeV/c (right panels).  Calculations
      from four weakly coupled energy loss models are shown as
      well~\cite{Bass:2008rv,Wicks:2005gt}.  From
      Ref.~\protect\cite{Adare:2010sp}{}.}
  \end{center}
\end{figure}

One observable that has been particularly challenging for energy loss
models to reproduce is the azimuthal anisotropy of $\pi^0$ production
with respect to the reaction plane.  A weak dependence on the path
length in the medium is expected from radiative energy loss.  This
translates into a small $v_2$ for high $p_T$ particles (i.e., only a
modest difference in parton energy loss when going through a short
versus long path through the QGP).  Results of $\pi^0$ $v_2$ are shown
in Figure~\ref{fig:phenixpi0v2}~\cite{Adare:2010sp}.  Weakly coupled
radiative energy loss models are compared to the \raa (bottom
panels) and $v_2$ (top panels) data.  These models reproduce the
\raa, but they fall far short of the $v_2$ data in both $p_T$
ranges measured (6--9~GeV/c and $>9$~GeV/c).  This large path length
dependence is naturally described by strongly coupled energy loss
models~\cite{Marquet:2009eq,Adare:2010sp}.  Note that one can match
the $v_2$ by using a stronger coupling, larger $\hat{q}$, but at the
expense of over-predicting the average level of suppression.  New
strong coupling
models~\cite{Casalderrey-Solana:2014wca,Casalderrey-Solana:2014bpa}
also need to confront the full data set available at RHIC.

\begin{figure}[t]
 \begin{center}     \includegraphics[width=0.8\linewidth]{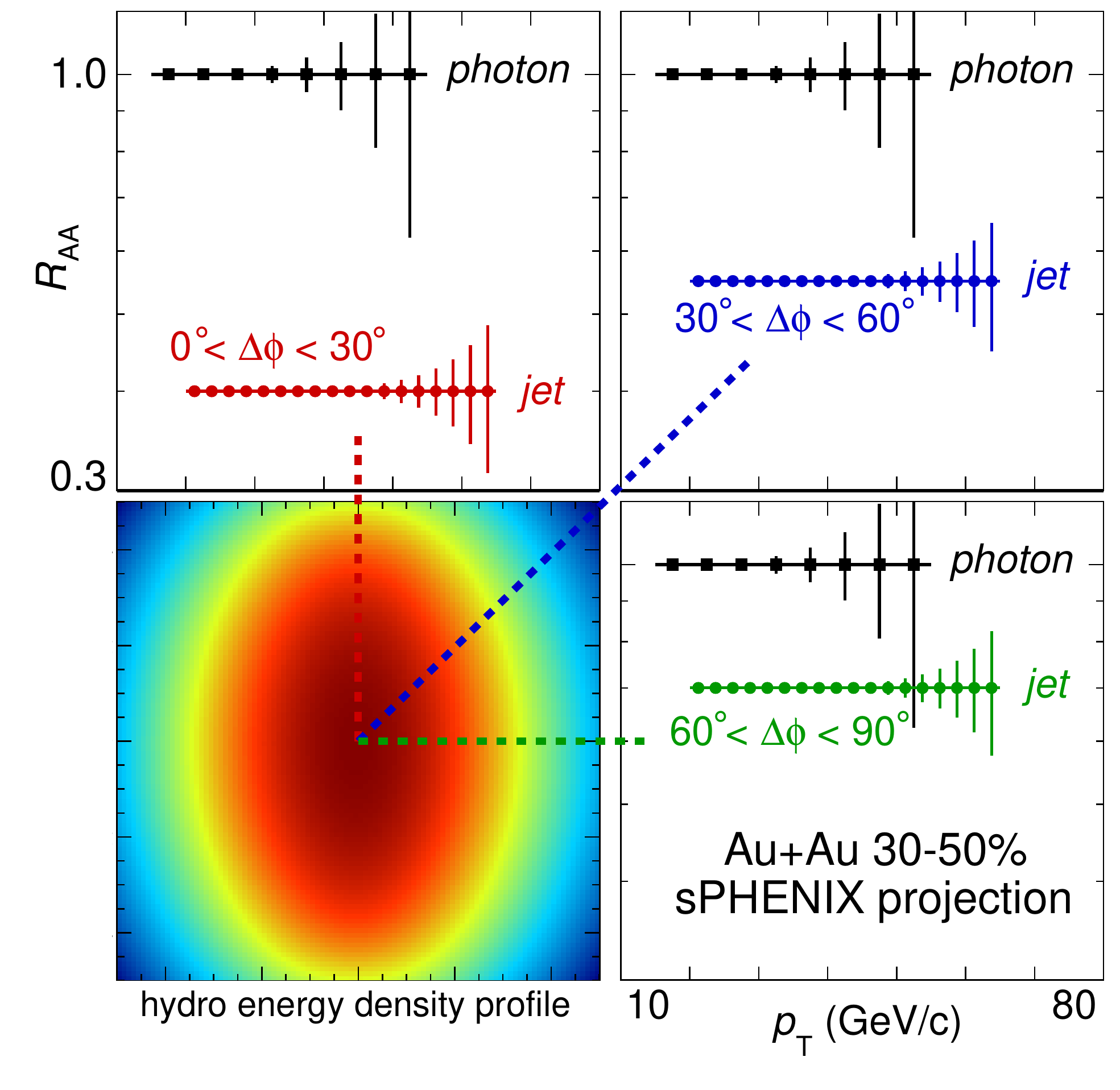}
    \caption[Statistical reach of azimuthally-sensitive hard probes in
    sPHENIX]{Demonstration of the statistical reach for
      azimuthally-sensitive hard probes measurements in sPHENIX. Each
      panel shows the projected statistical uncertainty for the
      $R_\mathrm{AA}$ of inclusive jets and photons, with each a panel
      a different $\Delta\phi$ range with respect to the reaction
      plane in $30$--$50$\% Au+Au events. sPHENIX would additionally
      have tremendous statistical reach in the analogous charged
      hadron
      $R_\mathrm{AA}$. \label{fig:sPHENIX_AuAu_ReactionPlane_figure} }
    \end{center}
\end{figure}

The measurement of jet quenching observables as a detailed function of
orientation with respect to the reaction plane is directly sensitive
to the coupling strength and the path length dependence of the
modification to the parton shower.  In addition, medium response may
be optimally measured in mid-central collisions with a lower
underlying event and where the medium excitations are not damped out
over a longer time evolution.  Shown in
Figure~\ref{fig:sPHENIX_AuAu_ReactionPlane_figure} are projected
uncertainties from sPHENIX --- detailed in the chapter on physics
performance --- for the direct photon and reconstructed jet observables
in three orientation selections.  One expects no orientation
dependence for the direct photons and the question is whether the
unexpectedly large dependence for charged hadrons persists in
reconstructed jets up to the highest \pt.  Note that the same
measurements can be made for beauty tagged jets, charged hadrons up to
50 GeV/c, and a full suite of correlation measurements including
jet-jet, hadron-jet, $\gamma$-jet.

All measurements in heavy ion collisions are the result of emitted
particles integrated over the entire time evolution of the reaction,
covering a range of temperatures.  Similar to the hydrodynamic model constraints, the
theory modeling for jet probes requires a consistent temperature and scale dependent
model of the \qgp and is only well constrained by precision data through
different temperature evolutions, as measured at RHIC and the LHC.

\clearpage

\section{What are the inner workings of the QGP?}
\label{Section:InnerWorkings}

A second axis along which one can investigate the underlying
structure of the \qgp concerns the question of what length scale of
the medium is being probed by jet quenching processes.  In
electron scattering, the scale is set by the virtuality of the
exchanged photon, $Q^{2}$. By varying this virtuality one can
obtain information over an enormous range of scales: from
pictures of viruses at length scales of $10^{-5}$ meters, to the
partonic make-up of the proton in deep inelastic electron scattering
at length scales of less than $10^{-18}$ meters. 

\begin{figure}[!hbt]
 \begin{center}
   \raisebox{0.05in}{\includegraphics[trim = 2 2 2 2, clip,width=0.57\linewidth]{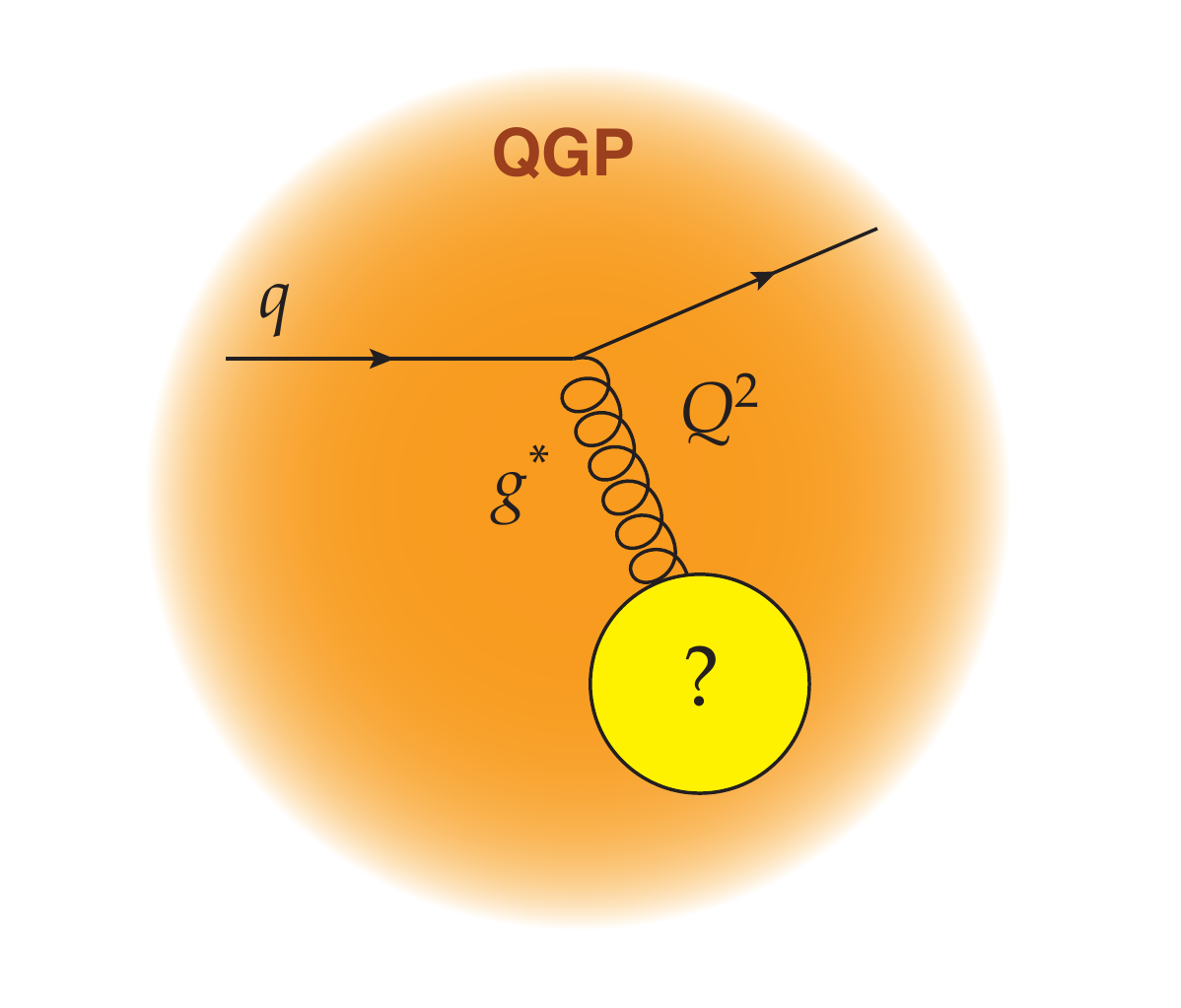}}
    \hfill
    \includegraphics[trim = 2 2 2 50, clip,width=0.4\linewidth]{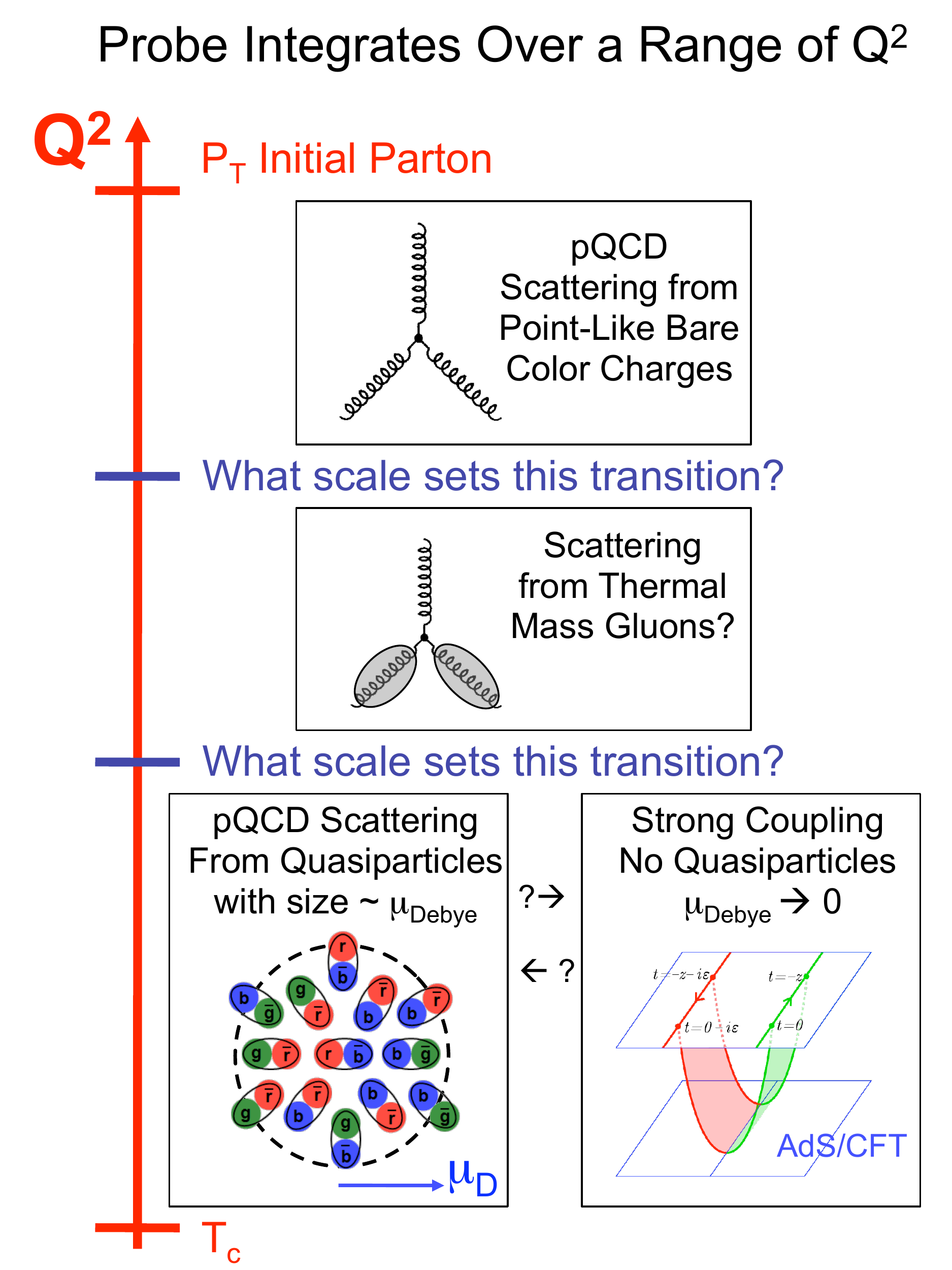}
    \caption[Interaction scale for the interaction of partons with the
    \qgp and possibilities for the recoil objects]{(left) Diagram of a
      quark exchanging a virtual gluon with an unknown object in the
      QGP.  This highlights the uncertainty for what sets the scale of
      the interaction and what objects or quasiparticles are
      recoiling.  (right) Diagram as a function of the $Q^{2}$ for the
      net interaction of the parton with the medium and the range of
      possibilities for the recoil objects.\label{fig:probescale}}
 \end{center}
\end{figure}

For the case of hard scattered partons in the quark-gluon plasma, the length scale probed
is initially set by the virtuality of the hard scattering process.  Thus, at the
highest LHC jet energies, the parton initially probes a very short length scale.  
Then as the evolution proceeds, the length scale is set by
the virtuality of the gluon exchanged with the
color charges in the medium, as shown in the left panel of
Figure~\ref{fig:probescale}.  However, if the exchanges are coherent, 
the total coherent energy loss through
the medium may set the length scale.

Figure~\ref{fig:probescale} (right panel) shows that if the length
scale probed is very small then one expects scattering directly from
point-like bare color charges, most likely without any influence from
quasiparticles or deconfinement.  As one probes longer length
scales, the scattering may be from thermal mass gluons and eventually
from possible quasiparticles with size of order the Debye screening
length.  In Ref.~\cite{krishna}, Rajagopal states that ``at some length scale, a
quasiparticulate picture of the QGP must be valid, even though on its
natural length scale it is a strongly coupled fluid.  It will be a
challenge to see and understand how the liquid QGP emerges from
short-distance quark and gluon quasiparticles.''

The extension of jet measurements over a wide range of energies and
with different medium temperatures again gives one the largest span
along this axis.  What the parton is scattering from in the medium is
tied directly to the balance between radiative energy loss and
inelastic collisional energy loss in the medium (encoded in $\hat{q}$ and $\hat{e}$). 
In the limit that the scattering centers in the medium are infinitely massive, one only
has radiative energy loss---as was assumed for nearly 10 years to be
the dominant parton energy loss effect.  In the model of Liao and
Shuryak~\cite{Liao:2008dk}, the strong coupling near the quark-gluon
plasma transition is due to the excitation of color magnetic
monopoles, and this should have a significant influence on the
collisional energy loss and equilibration of soft partons into the
medium.

\begin{figure}[ht]
  \centering
  \includegraphics[width=0.6\linewidth]{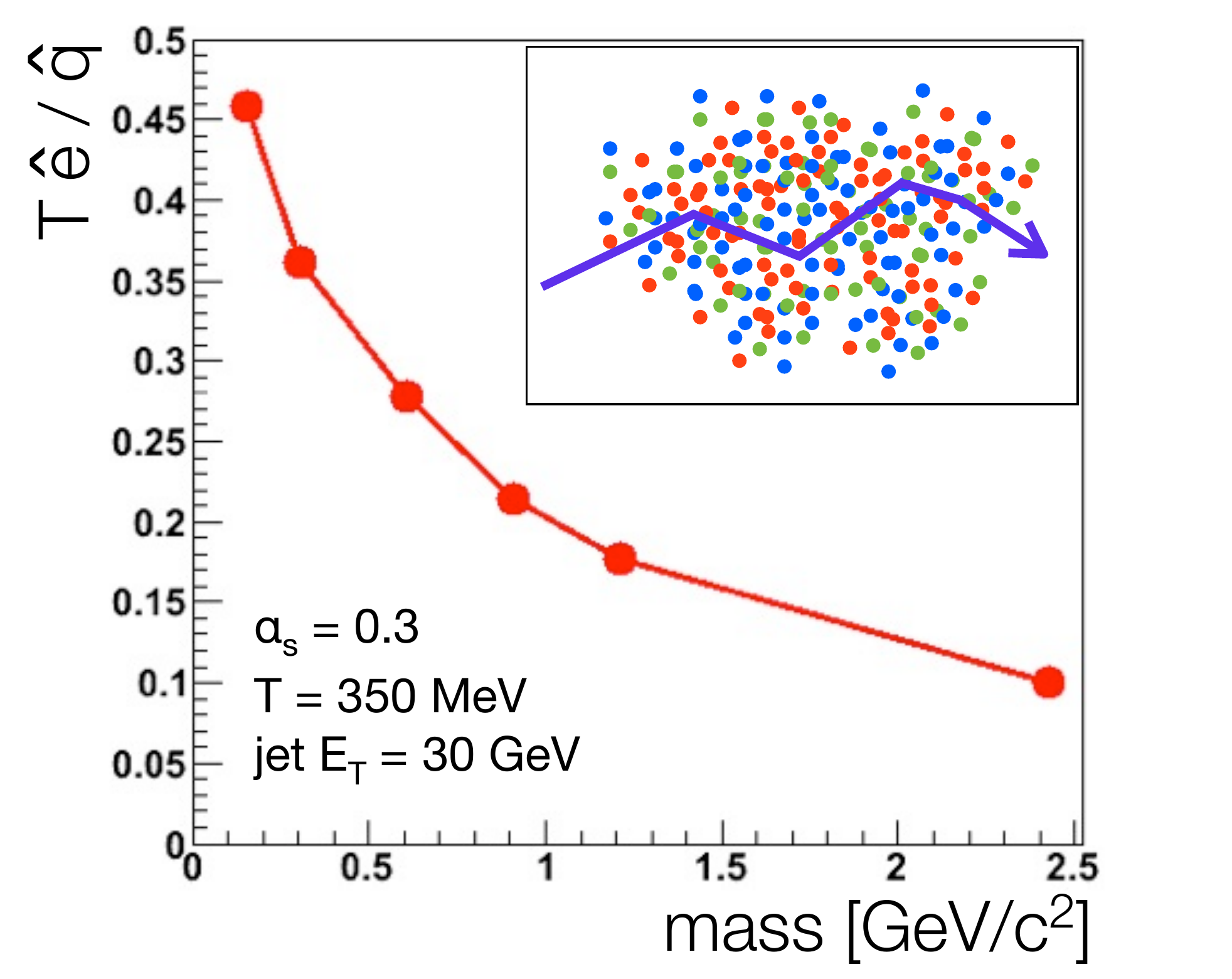}
  \caption[$T\hat{e}/\hat{q}$ as a function of the mass of the
  effective scattering centers in the medium]{$T\hat{e}/\hat{q}$ as a
    function of the mass of the effective scattering centers in the
    medium.  As the mass increases, the parton is less able to
    transfer energy to the medium and the ratio drops.}
  \label{fig:ehat_qhat}
\end{figure}

In a model by Coleman-Smith~\cite{ColemanSmith:2011rw,ColemanSmith:2011wd}
consisting of parton showers propagating in a medium of deconfined
quarks and gluons, one can directly vary the mass of the effective scattering centers and extract
the resulting values for $\hat{e}$ and $\hat{q}$.
Figure~\ref{fig:ehat_qhat} shows $T\hat{e}/\hat{q}$ as a function of
the mass of the effective scattering centers in the medium in this
model.  In the limit of infinitely massive scattering centers, the
interactions are elastic and no energy is transferred to the medium.

\begin{figure}[!hbt]
 \begin{center}
   \begin{minipage}[c]{0.27\textwidth}
     \includegraphics[trim = 505 2 15 2, clip,width=\textwidth]{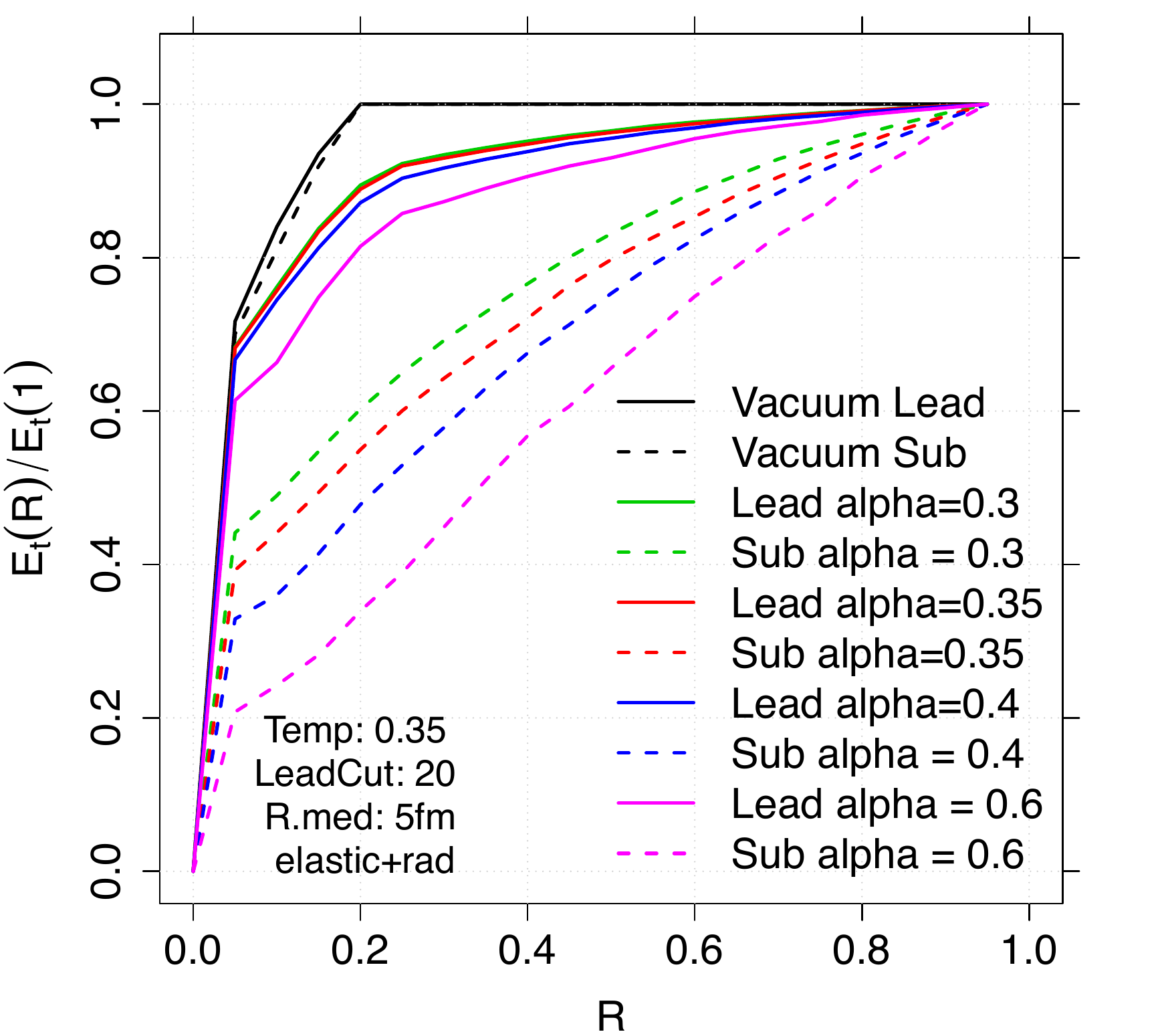}
     \includegraphics[trim = 2 2 520 2, clip,width=\textwidth]{figs/figure_physicscase_colemansmith_jetprofile}
   \end{minipage}
   \hfill
   \raisebox{-0.3cm}{
     \begin{minipage}[c]{0.71\textwidth}
       \includegraphics[width=\textwidth]{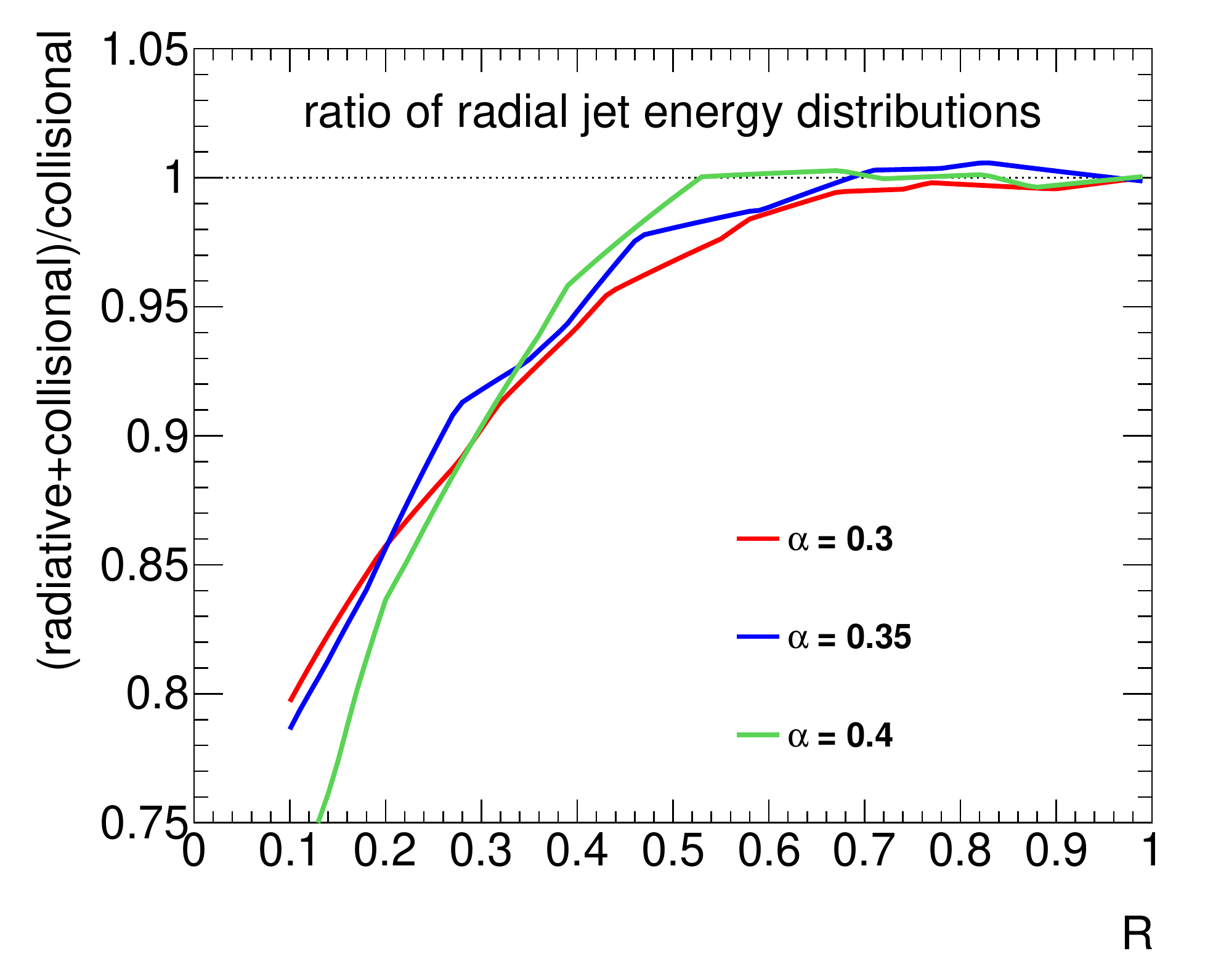}
     \end{minipage}
   }
   \caption[Calculations by Coleman-Smith of the jet energy profile as
   a function of radius for leading and sub-leading
   jets]{\label{fig:csjetprofile} (left) Calculations from
     Coleman-Smith~\protect\cite{ColemanSmith:2012vr}{} showing the
     jet energy profile as a function of radius for leading (solid
     lines) and sub-leading (dashed lines) jets.  Leading jets have
     $E_T>20$~GeV and sub-leading jets have $E_T>5$~GeV.  The medium
     temperature is 350~MeV. (right) Ratio the radial distribution of
     energy in sub-leading jets in a medium with radiative and elastic
     energy loss to the distribution in a medium with elastic energy
     loss only.  In these calculations, $\alpha$ serves as a proxy for
     the effective medium coupling.  }
 \end{center}
\end{figure}

Many observables are sensitive to the balance of $\hat{e}$ and $\hat{q}$, and thus sensitive to what is being
scattered from in the medium.   For example, in the same calculation by Coleman-Smith~\cite{ColemanSmith:2012vr}, 
the transverse radial jet energy profile is significantly modified by the balance of collisional and radiative energy loss.   Shown 
in the left panels Figure~\ref{fig:csjetprofile} are the vacuum and medium modified fractional energy distribution as a function
of distance $R$ from the jet axis.   The upper left panel is including both elastic and inelastic processes and the 
lower left panel with only elastic processes.    In the right panel we show the ratio of the profiles and for three different
effective medium coupling parameters. The sub-leading jet profiles are
dramatically modified compared to the vacuum and leading jet profiles. The elastic and radiative profiles clearly
separate, the radiative sub-leading jets become broader and softer than the elastic only. Both sets of sub-leading jets
become much broader and softer compared to the leading jets.

\begin{figure}[!hbt]
 \begin{center}
   \raisebox{0.55cm}{\includegraphics[trim = 2 2 2 2, clip, width=0.35\textwidth]{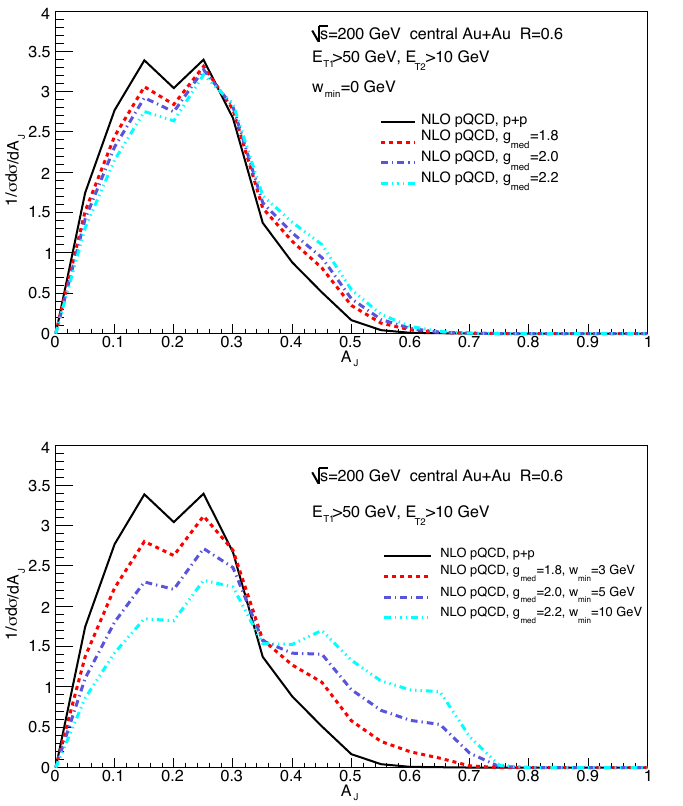}}
   \hfill
   \includegraphics[width=0.62\textwidth]{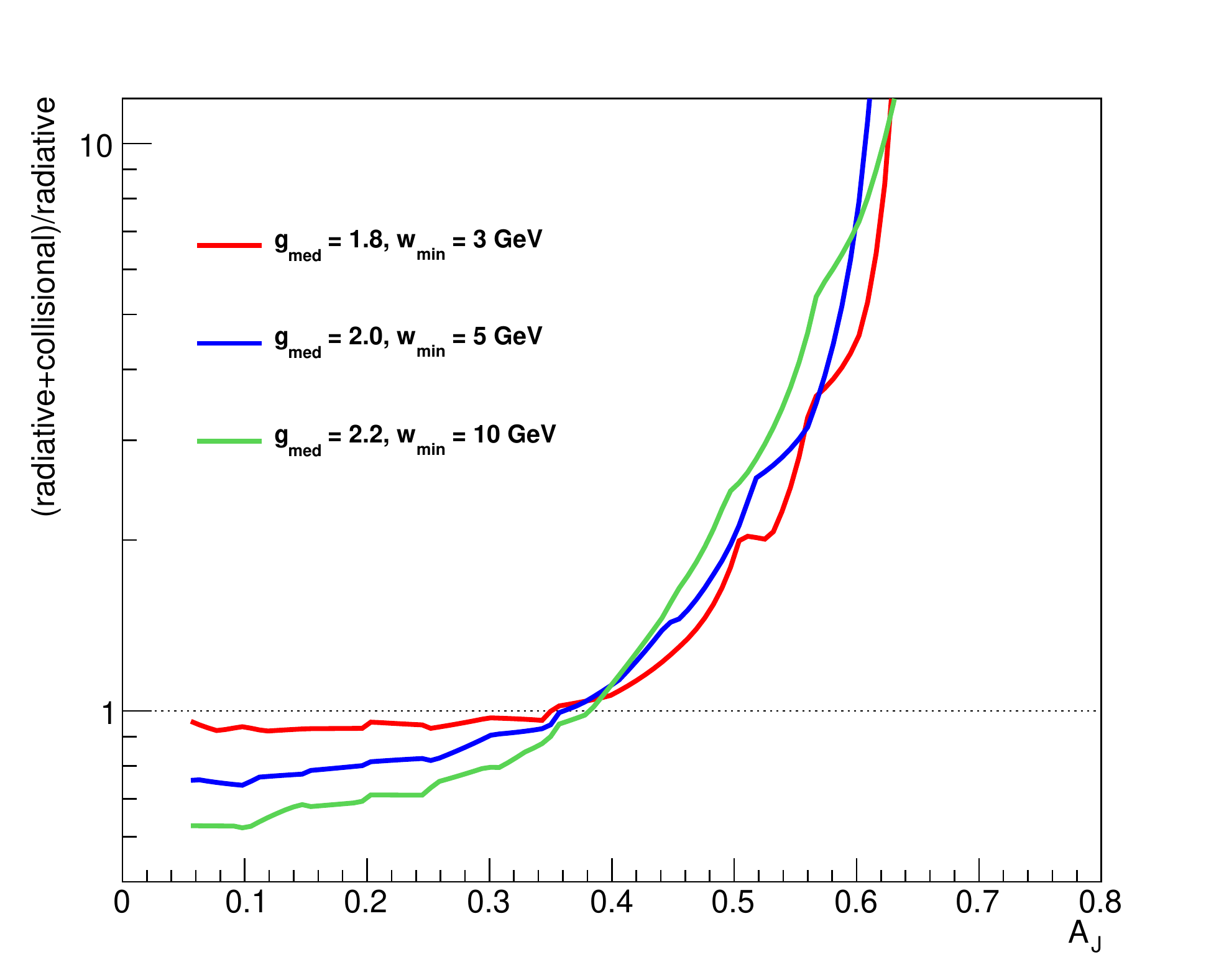}
   \caption[$A_J$ distributions by Vitev et al. for leading jet $E_T >
   50$~GeV, jet cone radius, $R = 0.6$, and different medium coupling
   strengths]{\label{fig:vitevaj} (left) $A_J$ distributions
     calculated by Vitev et
     al.~\protect\cite{He:2011pd,Neufeld:2011yh,Vitev:2009rd}{} for
     leading jet $E_T > 50$~GeV, jet cone radius, $R = 0.6$ and
     different medium coupling strengths.  The upper plot shows
     results for radiative energy loss only, and the lower plot
     includes collisional energy loss as well. (right) Ratio of $A_J$
     distributions with radiative and collisional energy loss to those
     with radiative energy loss only.}
 \end{center}
\end{figure}

In the calculation by Vitev et
al.~\cite{He:2011pd,Neufeld:2011yh,Vitev:2009rd}, the inclusion of
collisional energy loss results in a substantial shift in the dijet
asymmetry as shown comparing the top left and the bottom left of
Figure~\ref{fig:vitevaj}.  The right panel of Figure~\ref{fig:vitevaj}
shows the $A_{J}$ ratio with and without collisional energy loss.
There is a significant additional suppression of back-to-back matched
jets at low $A_{J}$ and a much larger number of very asymmetric jet
pairs.  Detailed measurements as a function of jet energy, jet radius,
and collision geometry are needed to map out the magnitude of the
collisional component, and thus $\hat{e}$ and its related effective
mass of the scattering centers.

One of the most sensitive observables to collisional energy loss is the modification of high \pt charm
and beauty heavy quarks in the medium.   We detail this physics in the later section specifically
on heavy quarks -- Section~\ref{sec:heavyquark}.

\clearpage

\section{How does the QGP evolve along with the parton shower?}
\label{sec:medium_evolution}

\begin{figure}[!hb]
 \begin{center}
    \includegraphics[trim = 2 2 2 2, clip, width=0.95\linewidth]{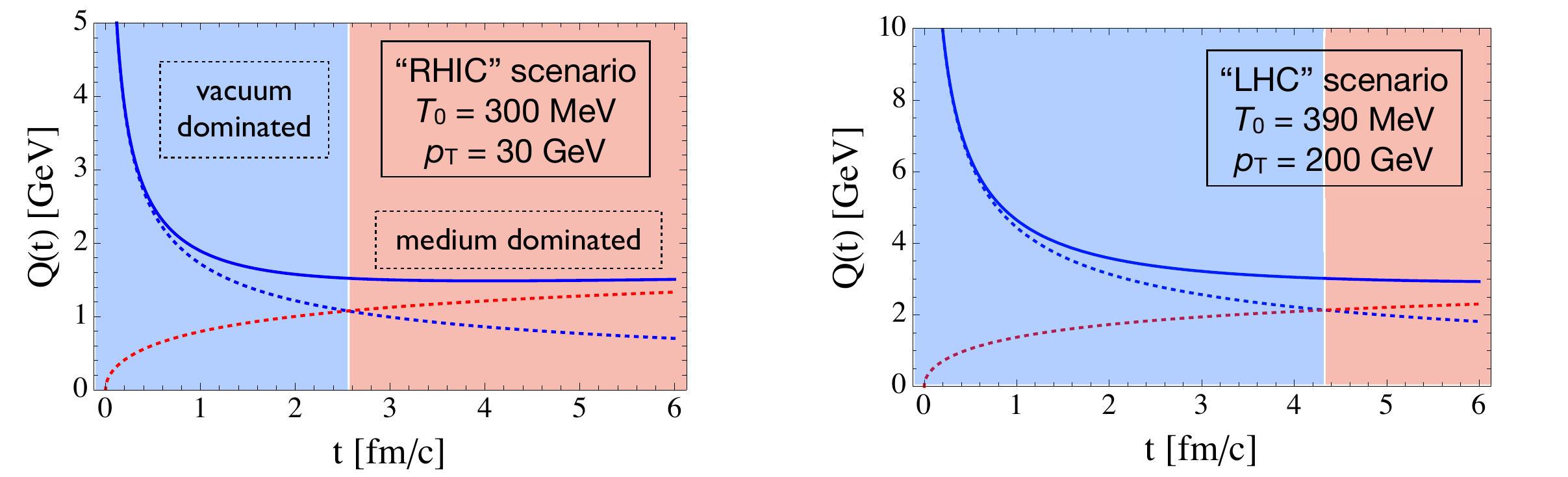}
    \caption[Jet virtuality evolution in medium at RHIC and
    LHC]{\label{fig:virtualityevolution}Jet virtuality evolution in
      medium at RHIC (left) and LHC (right).  Vacuum contributions to
      virtuality (blue dashed lines) decrease with time and medium
      induced contributions (red dashed lines) increase as the parton
      scatters in the medium.  The total virtuality (blue solid lines)
      is the quadrature sum of the two contributions.  At RHIC the
      medium induced virtuality dominates by 2.5\,fm/c while at the
      LHC the medium term does not dominate until 4.5\,fm/c.  From
      Ref.~\protect\cite{Muller:usersmeeting}{}.}
 \end{center}
\end{figure}

The initial hard scattered parton starts out very far off-shell and in
$e^{+}e^{-}$, \pp or \ppbar~collisions the virtuality evolves in
vacuum through gluon splitting down to the scale of hadronization.  In
heavy ion collisions, the vacuum virtuality evolution is interrupted
at some scale by scattering with the medium partons which increase the
virtuality with respect to the vacuum evolution.
Figure~\ref{fig:virtualityevolution} shows the expected evolution of
virtuality in vacuum, from medium contributions, and combined for a
quark-gluon plasma at $T_0 = 300$~MeV with the traversal of a 30~GeV
parton (left) and at $T_{0}=390$~MeV with the traversal of a 200~GeV
parton (right)~\cite{Muller:usersmeeting,Muller:2010pm}.  If this
picture is borne out, it ``means that the very energetic parton [in
the right picture] hardly notices the medium for the first 3--4 fm of
its path length~\cite{Muller:2010pm}.'' Spanning the largest possible
range of virtuality (initial hard process $Q^{2}$) is very important,
but complementary measurements at both RHIC and LHC of produced jets
at the same virtuality (around 50~GeV) will test the interplay between
the vacuum shower and medium scattering contributions.

In some theoretical frameworks --- for example
Refs~\cite{Renk:2013rla,Majumder:2011uk,Majumder:2014gda} --- the
parton splitting is simply dictated by the virtuality and in vacuum
this evolves relatively quickly from large to small scales as shown
above.  The $Q$ evolution means that the jet starts out being
considerably off mass shell when produced, and this off-shellness is
reduced by successive splits to less virtual partons.  In these
calculations, the scattering with the medium modifies this process of
parton splitting.  The scale of the medium as it relates to a
particular parton is $\hat{q}$ times the parton lifetime (this is the
mean transverse momentum that the medium may impart to the parent and
daughter partons during the splitting process). When the parton's
off-shellness is much larger than this scale, the effect of the medium
on this splitting process is minimal. As the parton drops down to a
lower scale, the medium begins to affect the parton splitting more
strongly.

\begin{figure}[!hbt]
 \begin{center}
   \includegraphics[width=0.43\linewidth]{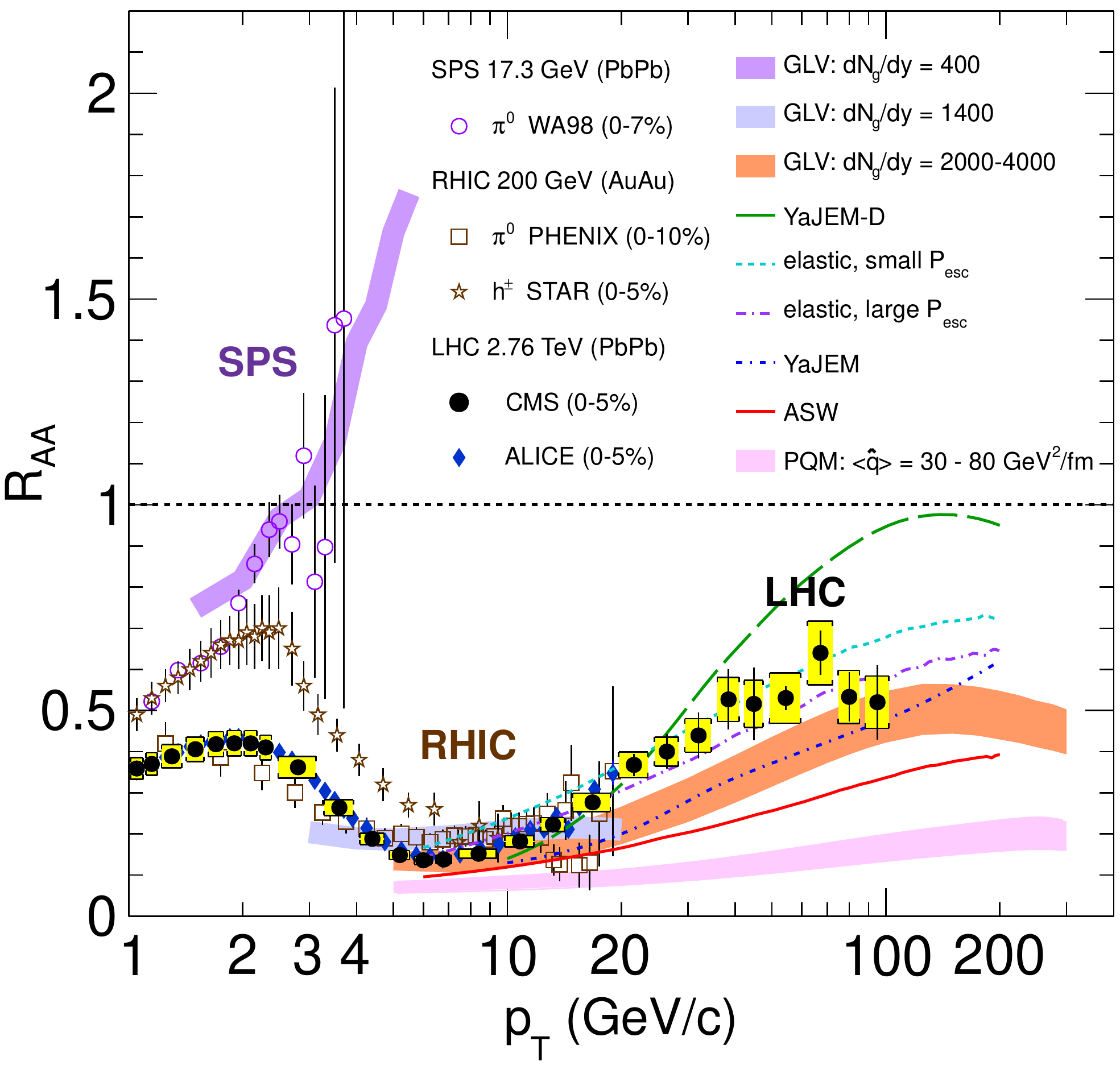}
   \raisebox{0.33cm}{\includegraphics[width=0.55\linewidth]{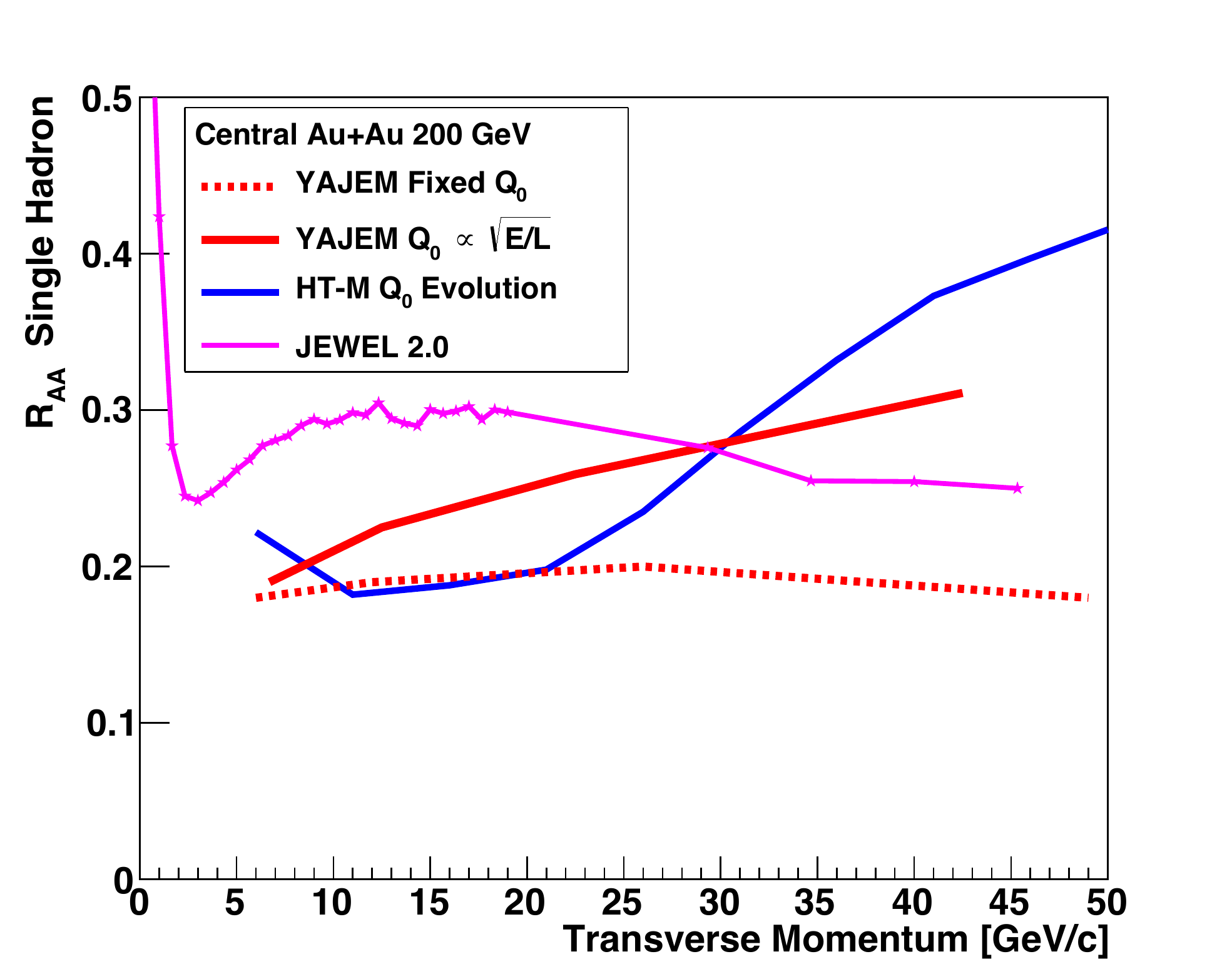}}
   \caption[ The nuclear modification factor \raa vs transverse
   momentum at the SPS, RHIC, and LHC, compared to various jet
   quenching calculations]{(left) The nuclear modification factor \raa
     as a function of transverse momentum in A$+$A collisions at the
     SPS, RHIC, and LHC.  Comparisons with various jet quenching
     calculations as detailed in Ref.~\cite{CMS:2012aa} and references
     therein are shown.  The simultaneous constraint of RHIC and LHC
     data is a powerful discriminator. (right) Predictions for single
     hadrons \raa to $\pT\sim50$~GeV/$c$ in central \auau at 200~GeV.}
  \label{fig:virtualityevolution_singlehadron}
  \label{fig:cmsqhat}
 \end{center}
\end{figure}

Shown in Figure~\ref{fig:virtualityevolution_singlehadron} is the
single hadron \raa in central \auau collisions at 200~GeV along
side measurements at other beam energies.  One specifically notes that
for the YAJEM calculation, inclusion of the virtuality evolution leads
to a factor of 50\% rise in \raa from 20--40~GeV/c, and in the HT-M
calculation a 100\% rise. A strong rise in \raa measured at
higher \pt at the LHC has been observed, and measurement of the
consistent effect within the same framework at RHIC is a key test of
this virtuality evolution description. It is notable that the JEWEL
calculation which describes the rising \raa at the
LHC~\cite{Zapp:2013vla}, results in a nearly flat \raa over the entire
\pt range at RHIC. sPHENIX can perform precision measurements of
charged hadrons over this \pt range.

\begin{figure}[!hbt]
 \begin{center}
   \includegraphics[width=0.95\linewidth]{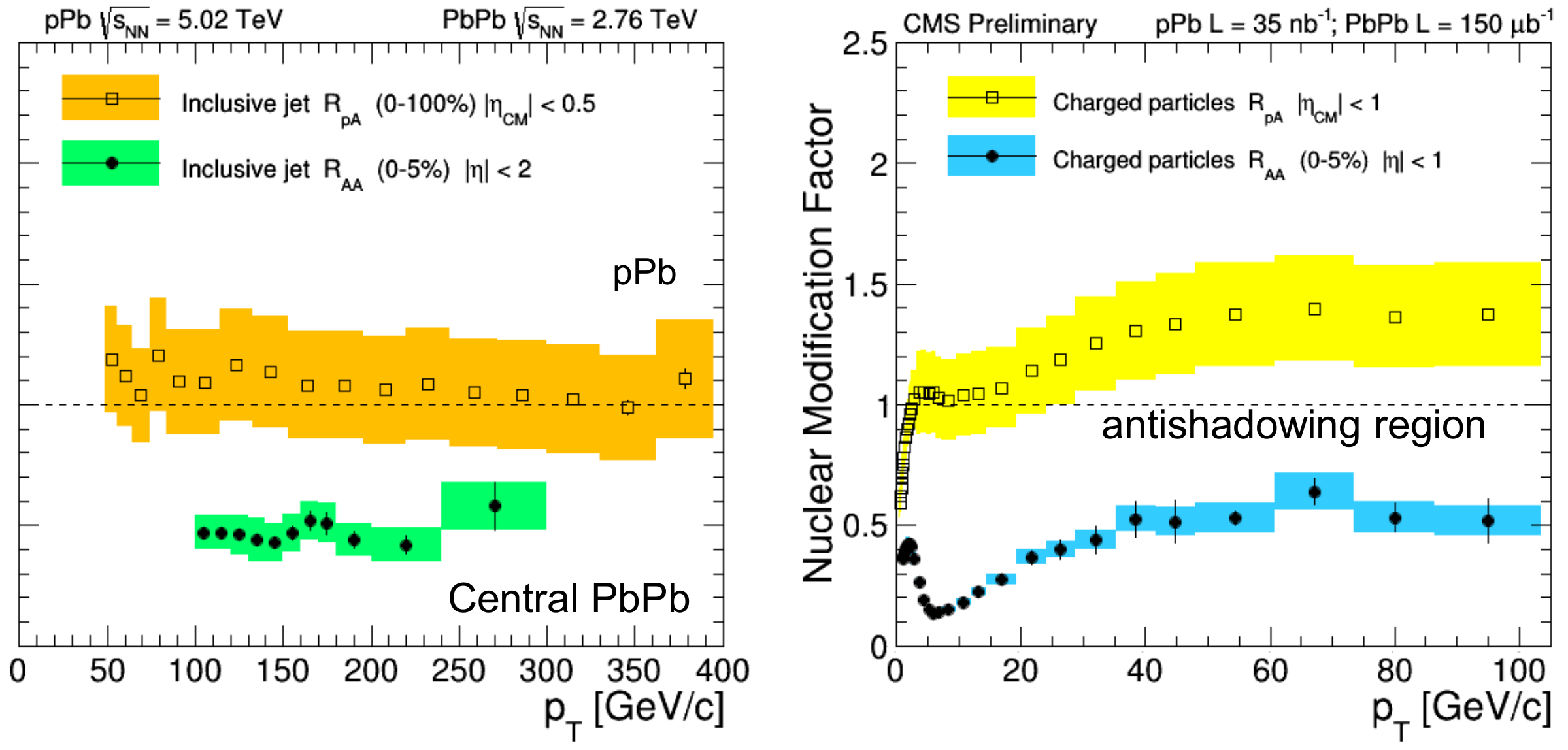}
   \caption[Preliminary CMS results of jet and charged particle \raa
   in \pA and \pbpb]{(left) Preliminary results from CMS showing the
     jet \raa in \pA near 1 and in \pbpb near 0.2. (right) Preliminary
     results from CMS showing the charged particle \raa in \pA and in
     \pbpb.  In contrast to the jet \raa results which are essentially
     flat with \pT, the charged particle \raa shows a rise with \pT.}
  \label{fig:cms_raa}
 \end{center}
\end{figure}

Further emphasizing the importance of having such measurements over the maximum kinematic reach at RHIC
and the LHC are the recent jet and charged hadron $R_{pA}$ measurements shown in Figure~\ref{fig:cms_raa}.  
It is quite striking that the flat reconstructed jet \raa and rising charged hadron \raa are mimicked already
in proton-nucleus collisions.   This may hint that all the physics of the rising hadron \raa does not fully
constrain the virtuality evolution and various calculations thus predict quite different effects at RHIC.

\begin{figure}[!hbt]
 \begin{center}
   \includegraphics[width=0.95\linewidth]{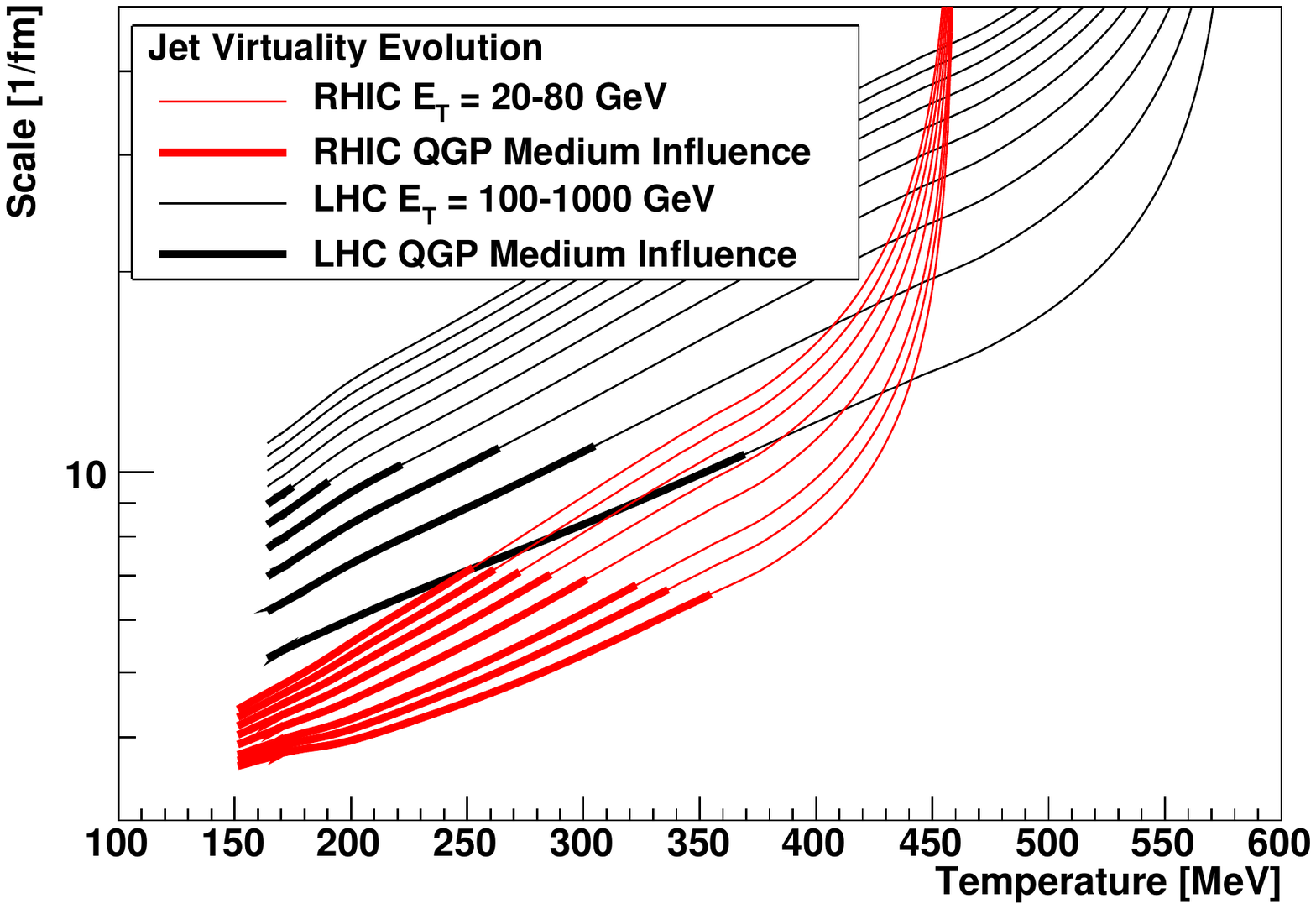}
   \caption[Scale probed in the medium via high energy partons as a
   function of the local temperature in the medium]{Scale probed in
     the medium in [1/fm] via high energy partons as a function of the
     local temperature in the medium.  The red (black) curves are for
     different initial parton energies in the RHIC (LHC) medium.}
  \label{fig:magic2}
 \end{center}
\end{figure}

To convey the scale probed and virtuality evolution differences at
RHIC and the LHC, we show the off-shellness of the initial hard
scattered parton virtuality in units of 1/fm as a function of the local temperature of
the QGP medium where the parton resides in Figure~\ref{fig:magic2}.  The calculation incorporates
the vacuum virtuality evolution which falls off quickly with time and
the medium scattering contribution that kicks the virtuality back up.
We incorporate the full time evolution of pre-equilibrium dynamics,
viscous hydrodynamics, and hadron cascade from
Ref.~\cite{Habich:2014jna} to map the time of the parton evolution to
the local temperature.  The medium virtuality contribution also scales
with the local temperature. The red (black) curves are for
different initial parton energies in the RHIC (LHC) medium.  The
thicker line regions highlight where the medium virtuality has a
substantial influence on the parton splitting.  It is notable that
highest energy partons at the LHC, of order 1 TeV, are always
dominated by the initial vacuum virtuality evolution (for more than 10
fm/c).  In contrast, the lower energy jets and the RHIC medium
evolution have the largest influence and map out a unique part of this
microscope resolving power and temperature of the \qgp.

\clearpage

\section{Current jet probe measurements}
 \label{currentjetdata}

 Jet quenching (i.e., the significant loss of energy for partons
 traversing the QGP) was discovered via measurements at RHIC of the
 suppression of single hadron yields compared to expectations from \pp
 collisions~\cite{Adcox:2001jp,Adler:2002xw}.  Since the time of that
 discovery there has been an enormous growth in jet quenching
 observables that have also pushed forward a next generation of
 analytic and Monte Carlo theoretical calculation to confront the
 data.

As detailed in Ref.~\cite{Adare:2008cg,Bass:2008rv}, many
formalisms assuming weakly coupled parton probes are able to achieve an equally
good description of the single inclusive hadron data at RHIC and the LHC.
The single high $p_T$ hadron suppression constrains
the $\hat{q}$ value within a model, but is not able to discriminate
between different energy loss mechanisms and formalisms for the
calculation.  Two-hadron correlations measure the correlated
fragmentation between hadrons from within the shower of one parton and
also between the hadrons from opposing scattered partons.  These
measurements, often quantified in terms of a nuclear modification
$I_{AA}$~\cite{Adare:2010ry,Adare:2010mq,Adams:2005ph}, are a
challenge for models to describe simultaneously~\cite{Nagle:2009wr}.

The total energy loss of the leading parton provides information on
one part of the parton-medium interaction.  Key information on the
nature of the particles in the medium being scattered from is
contained in how the soft (lower momentum) part of the parton shower
approaches equilibrium in the \qgp.  This information is 
accessible through full jet reconstruction, jet-hadron correlation,
and $\gamma$-jet correlation observables. 

The measurements of fully reconstructed jets and the particles correlated
with the jet (both inside the jet and outside) are crucial to testing
these pictures.  Not only does the strong coupling influence the
induced radiation from the hard parton (gluon bremsstrahlung)
and its inelastic collisions with the medium, but it also influences the way
soft partons are transported by the medium outside of the jet cone
as they fall into equilibrium with the medium.  Thus, the jet
observables combined with correlations get directly at the coupling of
the hard parton to the medium and the parton-parton coupling for the
medium partons themselves.

\begin{figure}[t]
 \begin{center}
    \includegraphics[trim = 2 2 2 2, clip, width=0.9\linewidth]{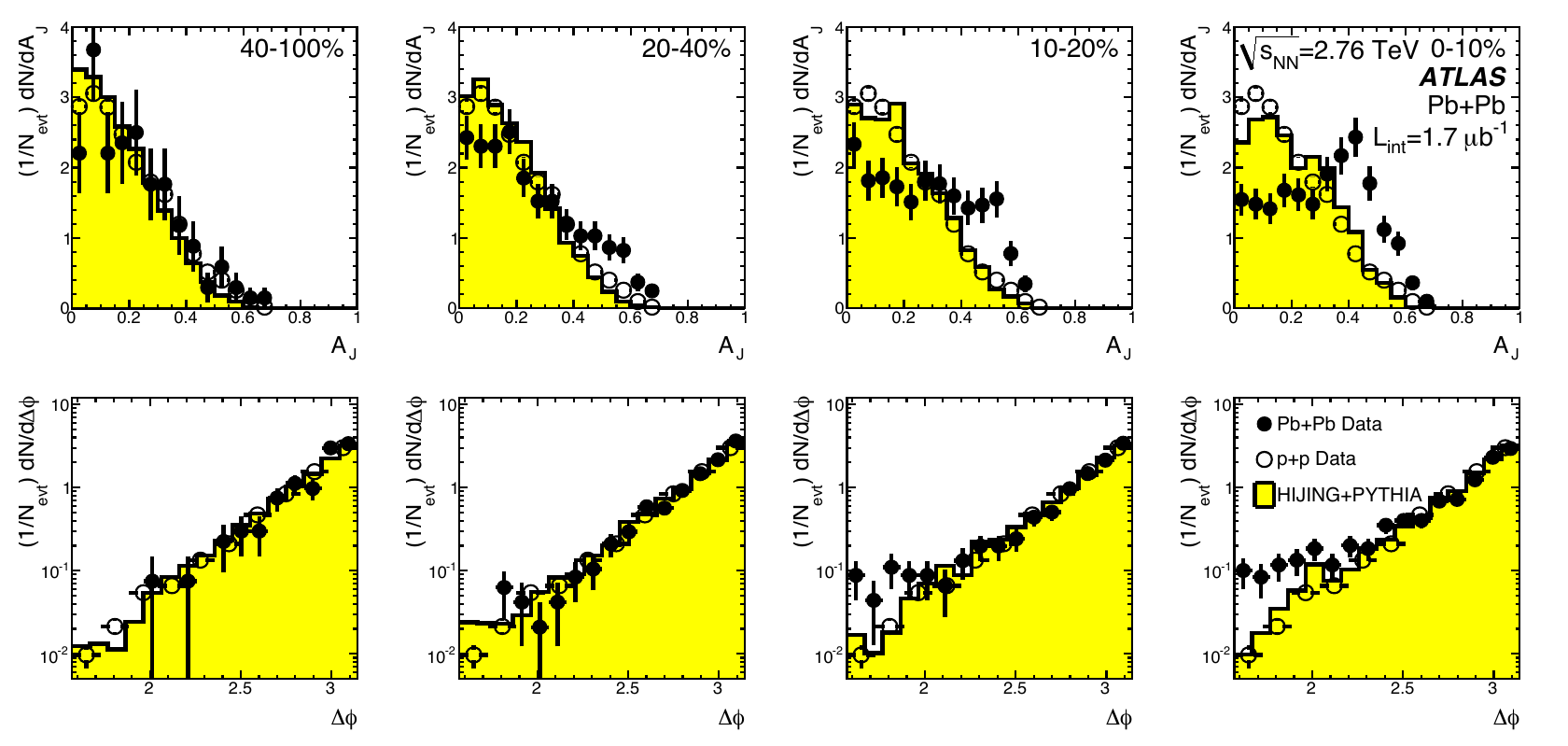}
    \caption[ATLAS $A_J$ and dijet $\Delta\phi$
    distributions]{\label{fig:atlasdijet}$A_J$ (top row) and dijet
      $\Delta\phi$ distribution from
      ATLAS~\protect\cite{Aad:2010bu}{}.  Jets are reconstructed with
      the anti-$k_T$ algorithm with $R=0.4$.  The leading jet has
      $E_T>100$~GeV and the associated jet has $E_T>25$~GeV.  \PbPb~
      data (solid points), \pp~data at 7~TeV (open points) and
      \pythia embedded in \hijing events and run through the ATLAS
      Monte Carlo (yellow histograms) are shown.  From
      Ref.~\protect\cite{Aad:2010bu}{}.}
 \end{center}
\end{figure}

\begin{figure}[t]
 \begin{center}
    \includegraphics[trim = 2 2 2 2, clip, width=\onewidth]{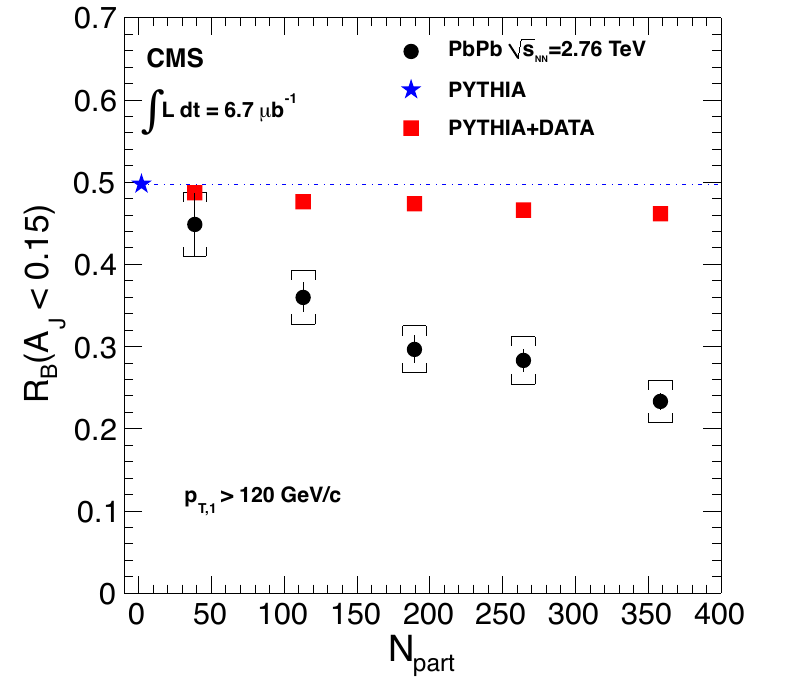}
    \caption[Fraction of dijets with $A_J<0.15$ in \PbPb~collisions as
    a function of centrality from CMS]{\label{fig:cmsdijetsupp}
      Fraction of dijets which have $A_J<0.15$ in \PbPb~collisions as
      a function of centrality.  Jets are reconstructed with an
      iterative cone algorithm with a radius of 0.5.  The leading jet
      is required to have an $E_T>120$~GeV and the associated jet has
      $E_T>50$~GeV.  Results are shown for \PbPb~data (circles),
      \pythia (star) and \pythia jets embedded into real data
      (squares).  From Ref.~\protect\cite{Chatrchyan:2011sx}{}.}
 \end{center}
\end{figure}

These jet observables are now available at the LHC.  The first results
based on reconstructed jets in heavy ion collisions were the
centrality dependent dijet asymmetries measured by
ATLAS~\cite{Aad:2010bu}. These results, shown in
Figure~\ref{fig:atlasdijet}, indicate a substantial broadening of
dijet asymmetry $A_{J} = (E_{1}-E_{2})/(E_{1}+E_{2})$ distribution for
increasingly central \PbPb~collisions and the lack of modification to
the dijet azimuthal correlations.  The broadening of the $A_J$
distribution points to substantial energy loss for jets and the
unmodified azimuthal distribution shows that the opposing jet
$\Delta\phi$ distribution is not broadened as it traverses the matter.
Figure~\ref{fig:cmsdijetsupp} shows CMS
results~\cite{Chatrchyan:2011sx} quantifying the fraction of dijets
which are balanced (with $A_J<0.15$) decreases with increasing
centrality.

\begin{figure}[ht]
  \centering
  \includegraphics[width=0.50\linewidth]{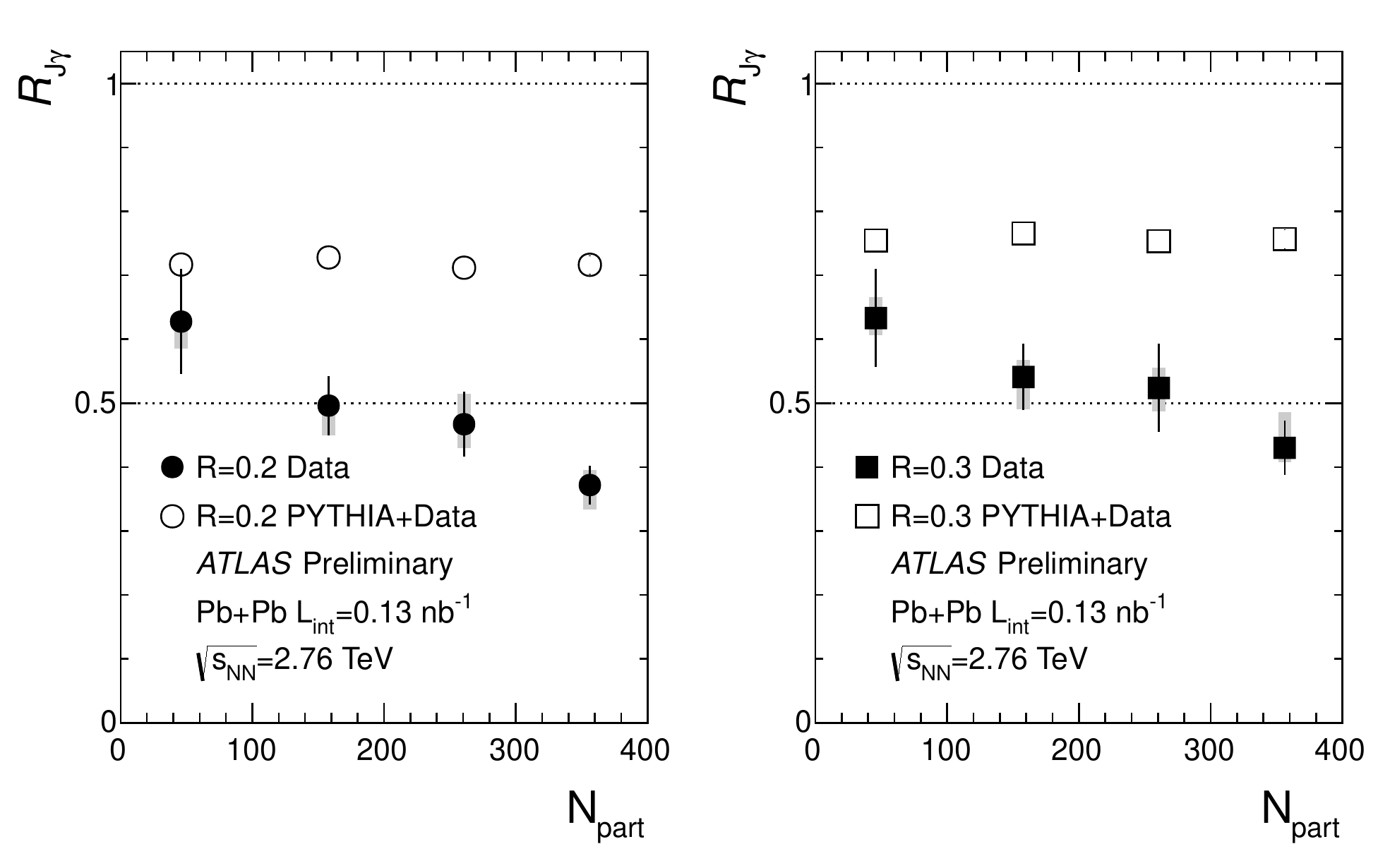}
  \hfill
  \raisebox{0.12cm}{\includegraphics[width=0.43\linewidth]{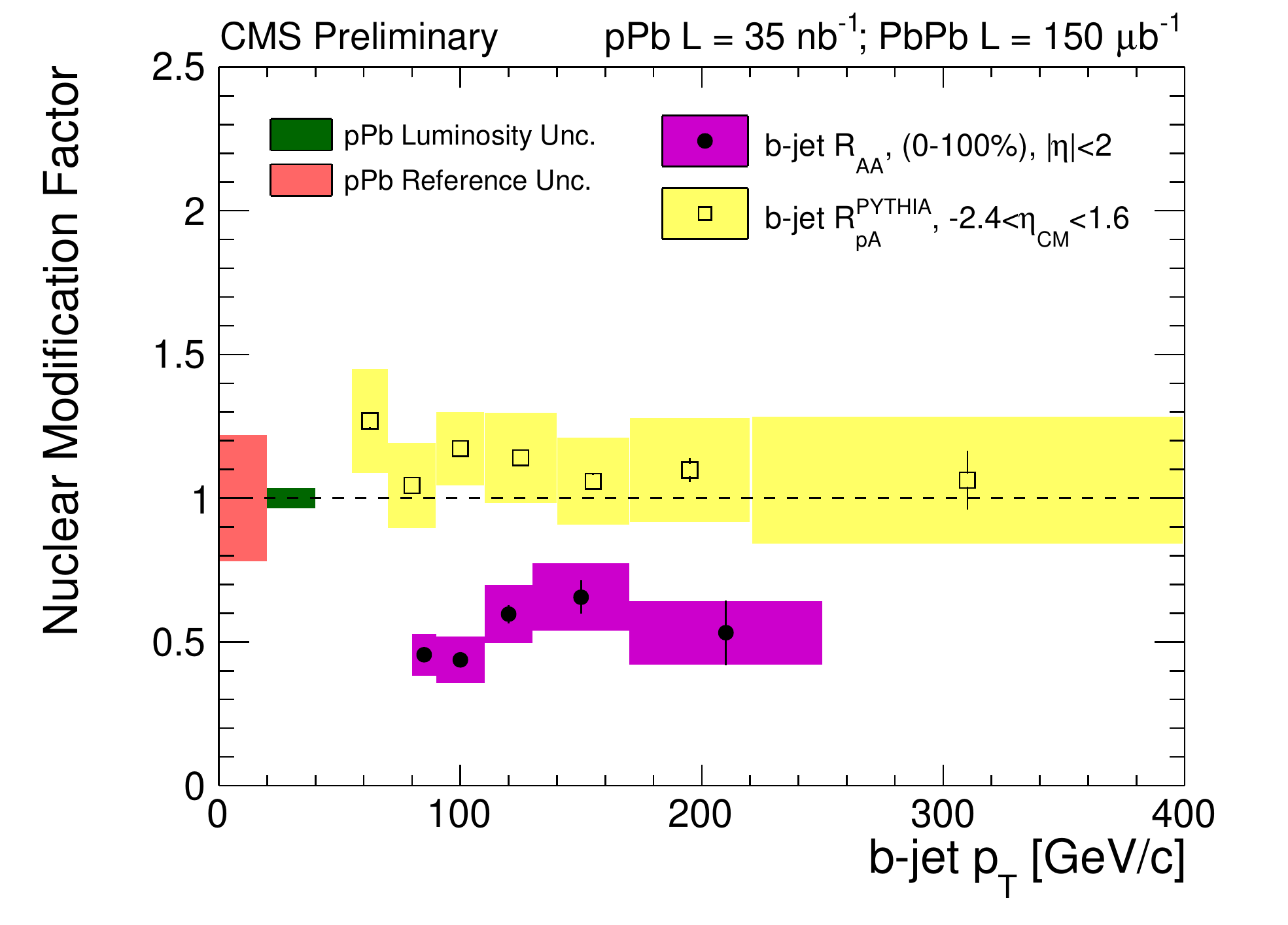}}
  \caption[ATLAS results on the change in balance of direct photons
  and jets and CMS results on the \raa for beauty tagged jets in \pbpb
  collisions at the LHC]{(left) ATLAS results on the change in balance
    of direct photons and jets in \pbpb collisions at the LHC.
    (right) CMS results on the \raa for beauty tagged jets in \pbpb
    collisions at the LHC.  }
  \label{fig:gamma_bjet}
\end{figure}

Direct photon-jet measurements are also a powerful tool to study jet
quenching. Unlike dijet measurements the photon passes through the
matter without losing energy, providing a cleaner handle on the
expected jet $p_T$~\cite{Wang:1996yh}.  CMS has results for
photons with $p_T>60$~GeV/c correlated with jets with
$p_T>30$~GeV/c~\cite{Chatrchyan:2012}.  Though with modest
statistical precision, the measurements indicate energy transported
outside the $R=0.3$ jet cone through medium interactions.  Similar results
from the ATLAS experiment are shown in the left panel of Figure~\ref{fig:gamma_bjet}, 
again indicating a shifting of energy outside the opposing jet radius.

Reconstructed jets have
significantly extended the kinematic range for jet quenching studies
at the LHC, and quenching effects are observed up to the highest
reconstructed jet energies ($>300$~GeV)~\cite{Chatrchyan:2012ni}.
They also provide constraints on the jet modification that are not
possible with particle based measurements.  For example, measurements from ATLAS
constrain jet fragmentation modification from vacuum fragmentation to
be small~\cite{Steinberg:2011qq} and  CMS results on jet-hadron
correlations have shown that the lost energy is recovered in low $p_T$
particles far from the jet cone~\cite{Chatrchyan:2011sx}.  
The lost energy is transported to very large angles
and the remaining jet fragments as it would in the vacuum.

Detector upgrades to PHENIX and STAR at RHIC with micro-vertex detectors will
allow the separate study of $c$ and $b$ quark probes of the medium, as tagged
via displaced vertex single electrons and reconstructed $D$ and $\Lambda_{c}$ hadrons.   Similar 
measurements at the LHC provide tagging of heavy flavor probes as well -- initial
results on beauty tagged jets from CMS are shown in the right panel of Figure~\ref{fig:gamma_bjet}.  
These measurements also provide insight on the different energy loss mechanisms,
in particular because initial measurements of non-photonic electrons from RHIC
challenge the radiative energy loss models.

There are other preliminary results on fully reconstructed jets from
both STAR~\cite{Caines:2011ew,Putschke:2011zz,Putschke:2008wn,Jacobs:2010wq}
and PHENIX experiments~\cite{Lai:2011zzb,Lai:2009zq}.  However, these
results have not yet proceeded to publication in part due to limitations
in the measurement capabilities.  In this proposal we demonstrate that
a comprehensive jet detector (sPHENIX) with large, uniform acceptance
and high rate capability, combined with the now completed RHIC
luminosity upgrade can perform these measurements to access this key
physics.

\begin{figure}[ht]
  \centering
  \includegraphics[width=0.7\linewidth]{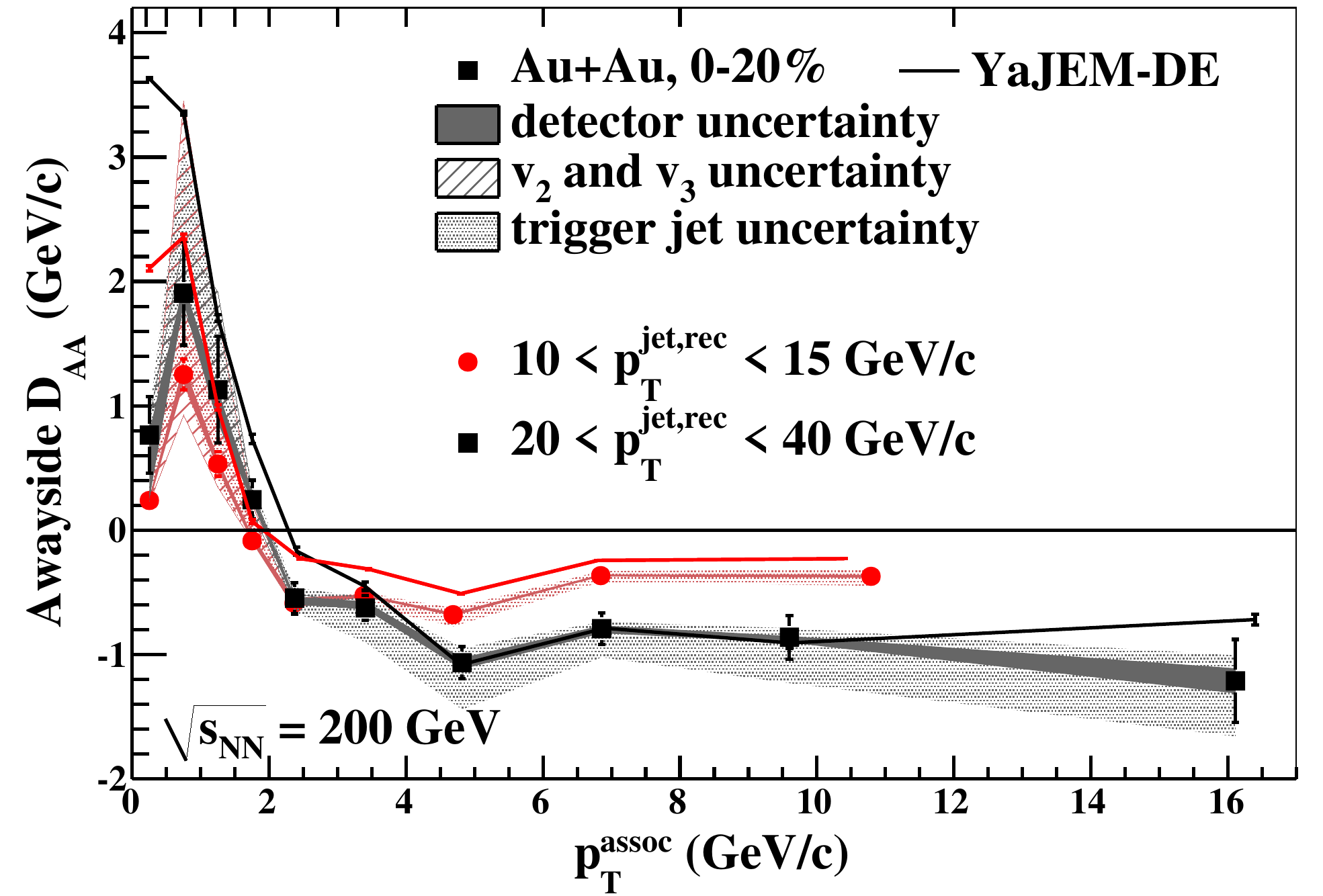}
  \caption[The away-side momentum difference, $D_{AA}$, of hadrons
  between \auau and \pp events, as measured by STAR]{The away-side
    momentum difference, $D_{AA}$, of hadrons between \auau and \pp
    events, as measured by STAR~\cite{Adamczyk:2013jei}, showing
    medium modification of jet fragmentation.}
  \label{fig:star_hjet}
\end{figure}

Figure~\ref{fig:star_hjet}
shows results from the STAR collaboration~\cite{Adamczyk:2013jei} on
correlations between reconstructed trigger jets and single charged
hadrons. The experimental results show the difference in the away-side
momentum of hadrons between Au$+$Au and \pp events.  The extent to
which this value differs from zero is an indication of the strength of
the medium modification of the fragmentation process.  The figure also
compares these results to calculations obtained using the YAJEM-DE
model that qualitatively reproduces the data.

\begin{figure}[t]
 \begin{center}
   \fbox{\includegraphics[angle=90,width=\linewidth]{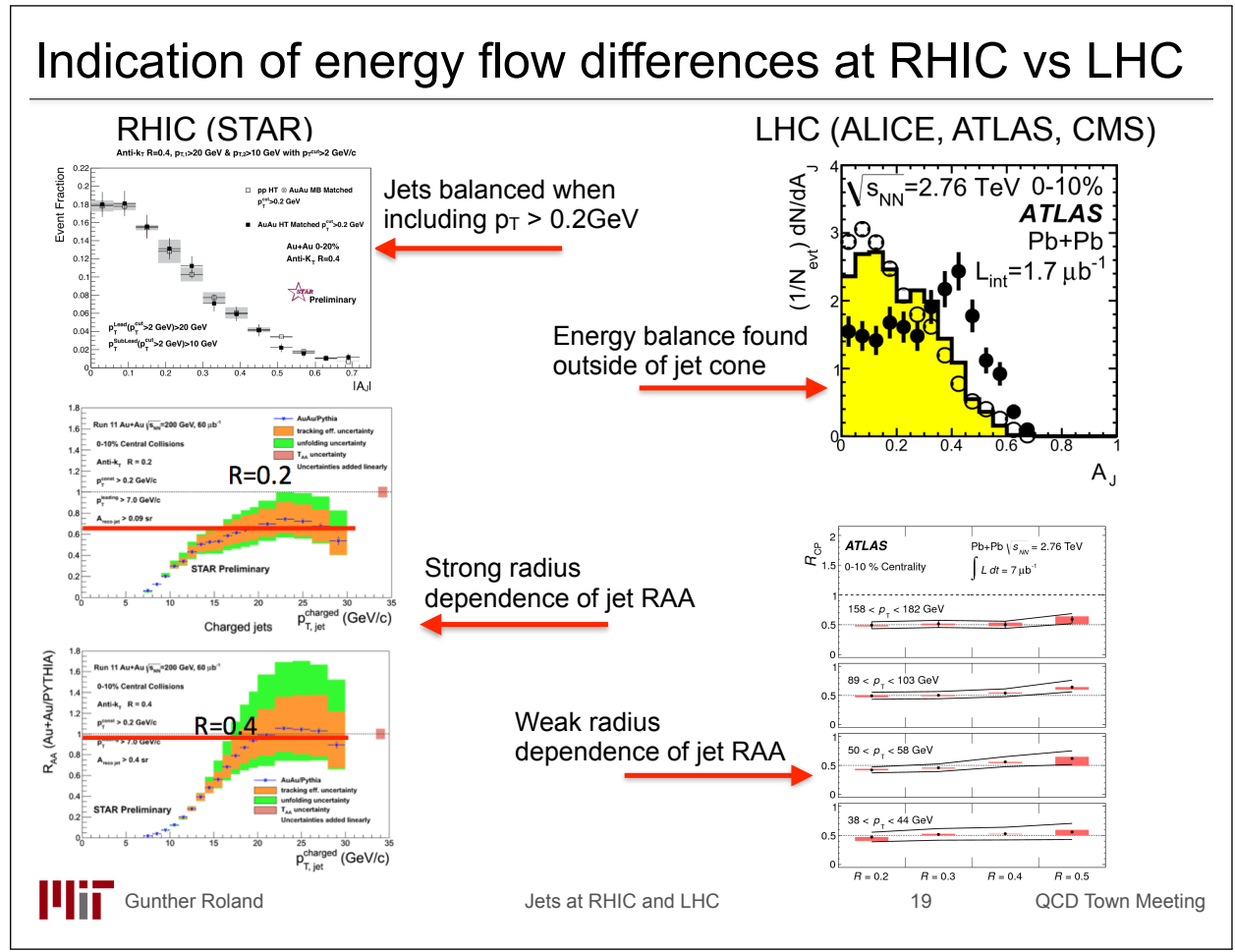}}
   \caption[Slide from G. Roland's talk at the QCD Town Meeting
   (September 2014) contrasting preliminary RHIC results from STAR for
   jet \raa and dijet asymmetry $A_J$  with LHC results]{ Slide from G. Roland's talk at the QCD Town Meeting
     (September 2014).  Shown are preliminary RHIC results from STAR
     for jet \raa and dijet asymmetry $A_J$ in comparison with LHC
     results.  The initial observation is for quite different trends.
     Data with overlapping energy ranges and comparable jet algorithms
     and jet bias selections from sPHENIX will shed significant light
     on the underlying physics differences.
    \label{fig:guntheruber}
  }
 \end{center}
\end{figure}

Figure~\ref{fig:guntheruber} shows a compilation panel with results
from RHIC preliminary jet results and LHC jet results.  They indicate
that with this set of observables, the behavior is quite different at
RHIC and the LHC.  Whether the significant radius $R$ dependence of
jet suppression \raa at RHIC, not observed at the LHC, is the result
of engineered bias selections on the STAR results remains to be
tested.  In addition, the recovery of most energy within $R=0.4$ is an
exciting result from STAR which could potentially indicate a different
redistribution of energy in the RHIC created \qgp.

It is clear that in addition to extending the RHIC observables to
include fully reconstructed jets and $\gamma$-jet correlations,
theoretical development work is required for converging to a coherent
'standard model' of the medium coupling strength and the nature of the
probe-medium interaction.  In the next section, we detail positive
steps in this direction.

\clearpage

\section{Theoretical calculations of jets at RHIC}
\label{sec:jetcalculations}

Motivated in part by the new information provided by LHC jet results
and the comparison of RHIC and LHC single and di-hadron results, the
theoretical community is actively working to understand the detailed
probe-medium interactions.  The challenge is to understand not only
the energy loss of the leading parton, but how the parton shower
evolves in medium and how much of the lost energy is re-distributed in
the \qgp.  Theoretical calculations attempting to describe the wealth
of new data from RHIC and the LHC have not yet reconciled some of the
basic features, with some models including large energy transfer to
the medium as heat (for example~\cite{CasalderreySolana:2011rq}) and
others with mostly radiative energy loss (for
example~\cite{Renk:2012cb,Renk:2011wb}).  None of the current
calculations available has been confronted with the full set of jet
probe observables from RHIC and the LHC.  Measurements of jets at RHIC
energies and with jets over a different kinematic range allow for
specific tests of these varying pictures.  In this section, we give a
brief review of a subset of calculations for jet observables at RHIC
enabled by the sPHENIX upgrade and highlight the sensitivity of these
observables to the underlying physics.

Much of this work has been carried out under the auspices of the
Department of Energy Topical Collaboration on Jet and Electromagnetic
Tomography of Extreme Phases of Matter in Heavy-ion Collisions
~\cite{jetcollaboration}.  Workshops held by the JET Collaboration at
Duke University in March 2012 and Wayne State University in August
2013 and 2014 have been dedicated to the topic of jet measurements at RHIC.
These workshops were attended by theorists as well as experimentalists
from both RHIC and the LHC.  This is an active collaborative effort.

In order to overcome specific theoretical hurdles regarding analytic
parton energy loss calculations and to couple these calculations with
realistic models of the QGP space-time evolution, Monte Carlo
approaches have been developed (as
examples~\cite{Zapp:2009pu,Renk:2010zx,Young:2011va,ColemanSmith:2011wd,Lokhtin:2011qq,Armesto:2009zc}).
Here we describe RHIC energy jet probe results from specific theory
groups utilizing different techniques for calculating the jet-medium
interactions.  These efforts indicate a strong theoretical interest
and the potential constraining power of a comprehensive jet physics
program at RHIC.

\begin{figure}[ht]
  \centering
  \includegraphics[trim = 30 0 30 30, clip, width=\linewidth]{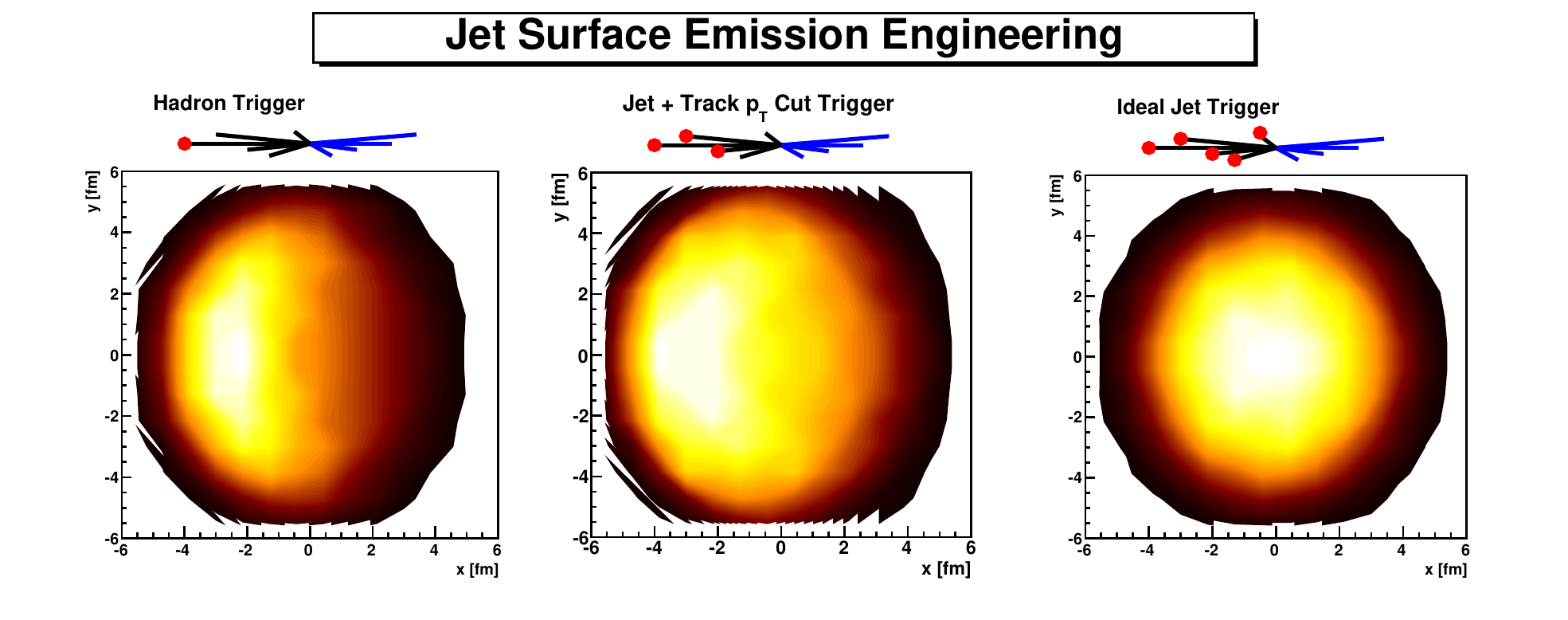}
  \caption[Dijet surface bias in YaJEM for various trigger
  definitions]{Dijet surface bias in YaJEM for various trigger
    definitions. As the trigger is changed from a single hadron (left)
    to a reconstructed jet with a minimum $p_T$ selection on charged
    tracks and electromagnetic clusters (middle) to an ideally
    reconstructed jet (right), the surface bias in the production
    point becomes less pronounced. sPHENIX is capable of all three
    types of measurements.  (Based on figure taken
    from~\cite{Renk:2012ve}.)}
  \label{fig:renk_jet_engineering}
\end{figure}

Jets provide a very rich spectrum of physics observables, ranging from
single jet observables such as \raa, to correlations of jets with
single particles, to correlations of trigger jets with other jets in
the event.  An example of how one can exploit this variety can be
found in recent calculations by Renk~\cite{Renk:2012ve}.
Figure~\ref{fig:renk_jet_engineering} is based on calculations using
the YaJEM model to illustrate what could be called ``jet surface
engineering''.  Triggers ranging from single hadrons on up to ideally
reconstructed jets are used to form correlations with another jet in
the event.  The different triggers demonstrate different degrees of
surface bias in the production point of the ``dijet'' and this bias
itself can be used as a lever to investigate properties of the medium.

\begin{figure}[ht]
 \begin{center}
    \includegraphics[trim = 2 2 2 2, clip, width=\twowidth]{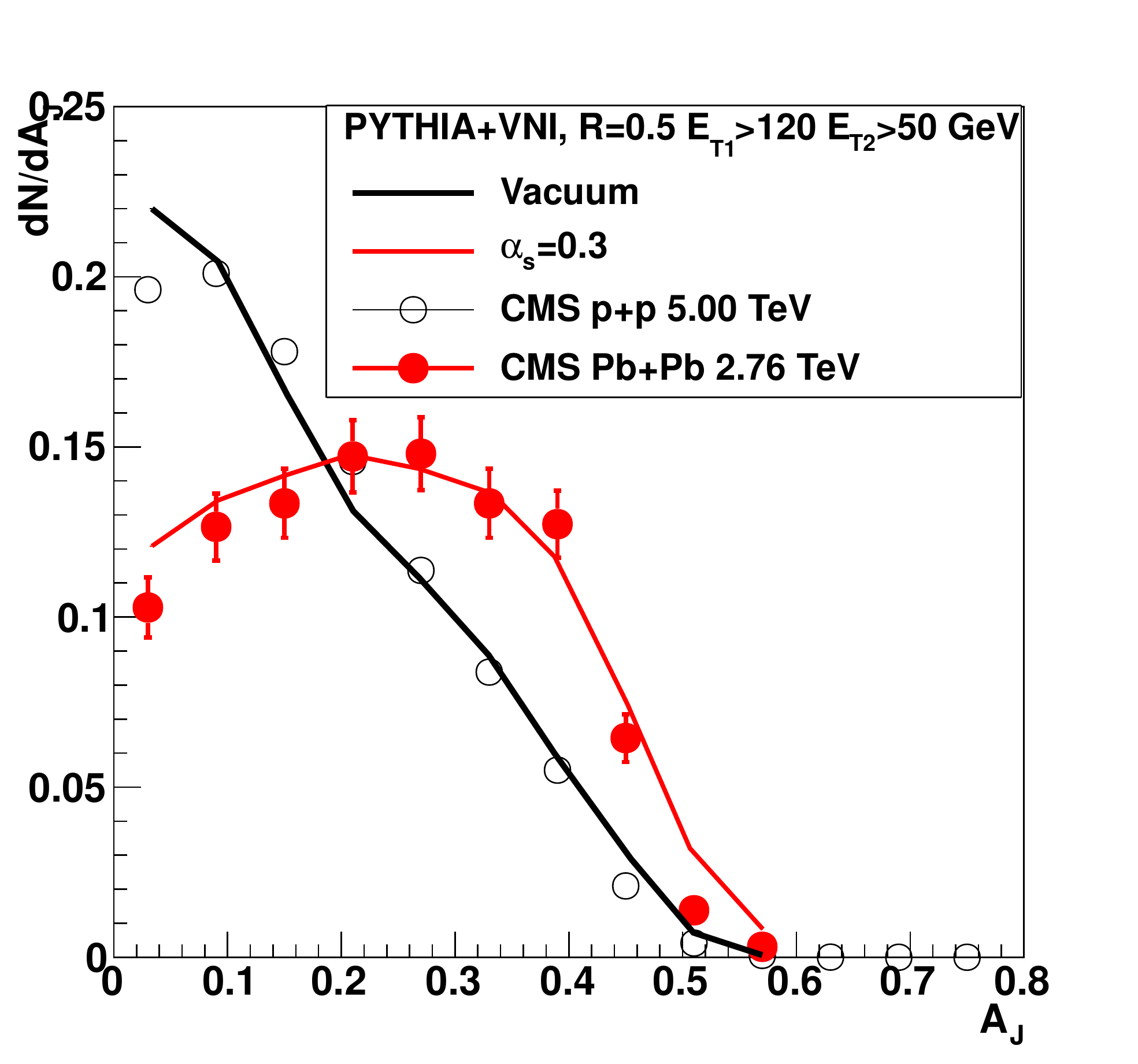}
    \hfill
    \includegraphics[trim = 2 2 2 2, clip,width=\twowidth]{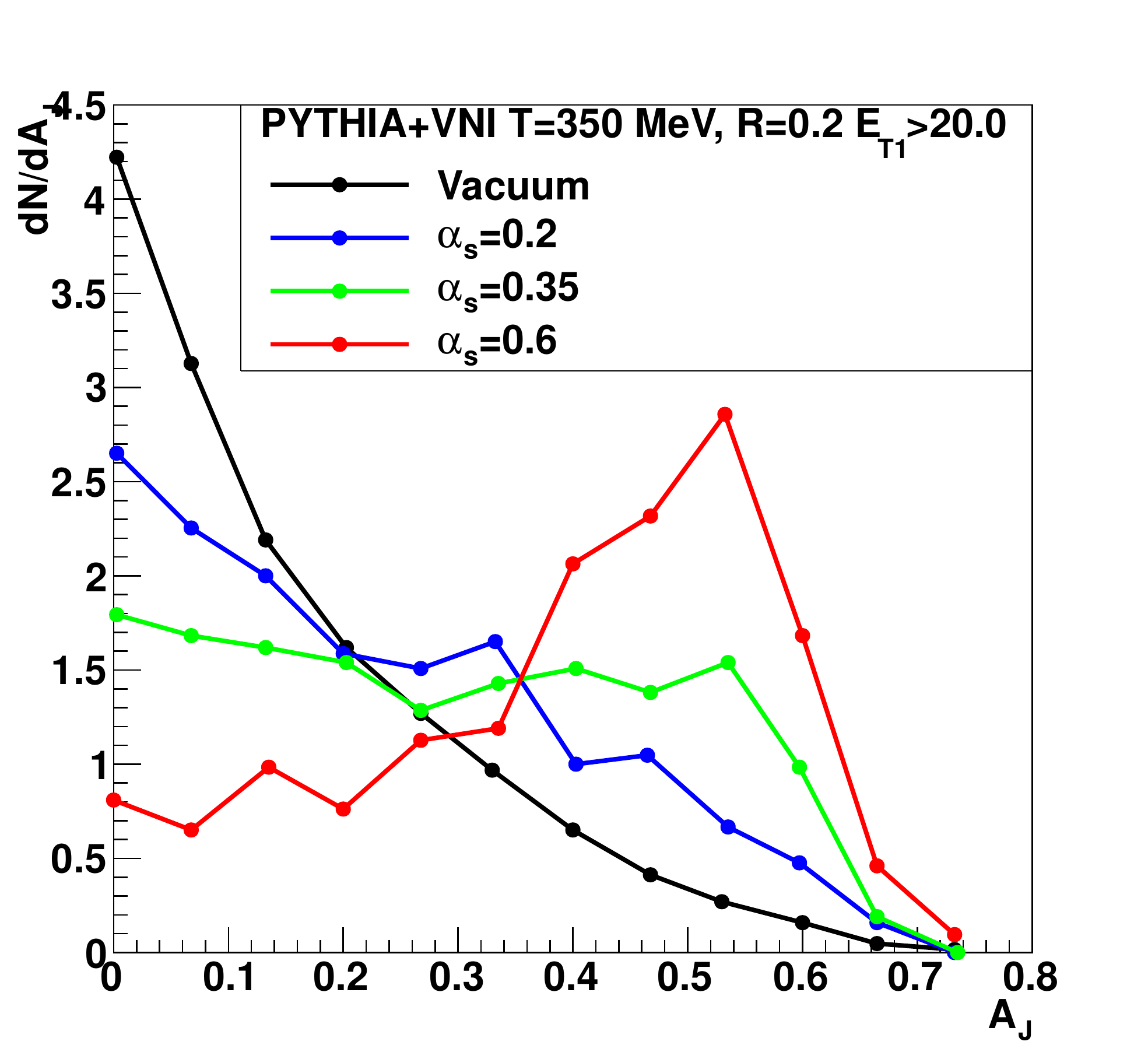}
    \caption[Dijet $A_J$ in VNI parton cascade compared to the CMS
    data and calculation for RHIC energies of $A_J$ for different
    values of $\alpha_s$]{\label{fig:colemansmith1}(left) Calculation
      in VNI parton cascade of dijet $A_J$ with $T=0.35$~GeV and
      $\alpha_s=0.3$ compared to the CMS
      data~\protect\cite{ColemanSmith:2011rw}{}.  (right) Calculation
      for RHIC jet energies, $E_{T,1}>20$~GeV, for a circular
      geometry of radius 5\,fm of $A_J$ for different values of
      $\alpha_s$ increasing to $\alpha_s=0.6$ (red
      line)~\protect\cite{ColemanSmith:2012vr}{}.}
 \end{center}
\end{figure}

We show results are from Coleman-Smith and
collaborators~\cite{ColemanSmith:2011rw,ColemanSmith:2011wd} where
they extract jet parton showers from \pythia (turning off
hadronization) and then embed the partons into a deconfined \qgp,
modeled with the VNI parton cascade~\cite{Geiger:1991nj}.  For the
calculations shown here, the background medium consists of a cylinder
of deconfined quarks and gluons at a uniform temperature.  One
excellent feature of the calculation is that it provides the ability
to track each individual parton and thus not only look at the full
time evolution of scattered partons from the shower, but also medium
partons that are kicked up and can contribute particles to the
reconstructed jets.

Calculation results for the dijet asymmetry $A_{J} =
(E_{1}-E_{2})/(E_{1}+E_{2})$ in a QGP with a temperature appropriate
for LHC collisions and fixed $\alpha_{s}=0.3$ are shown in
Figure~\ref{fig:colemansmith1} (left
panel)~\cite{ColemanSmith:2011rw}.  The jets in the calculation are
reconstructed with the anti-$k_T$ algorithm with radius parameter $R =
0.5$ and then smeared by a simulated jet resolution of
100\%/$\sqrt{E}$, and with requirements of $E_{T1}>120$~GeV and
$E_{T2}>50$~GeV on the leading and sub-leading jet, respectively.
The calculated $A_J$ distributions reproduce the CMS experimental
data~\cite{Chatrchyan:2011sx}.

In Figure~\ref{fig:colemansmith1} (right panel) the calculation is
repeated with a medium temperature appropriate for RHIC collisions and
with RHIC observable jet energies, $E_{T1} > 20$~GeV and $R = 0.2$.
The calculation is carried out for different coupling strengths
$\alpha_{s}$ between partons in the medium themselves and the parton
probe and medium partons.  The variation in the value of $\alpha_{s}$
should be viewed as changing the effective coupling in the many-body
environment of the QGP.  It is interesting to note that in the parton
cascade BAMPS, the authors find a coupling of $\alpha_{s} \approx 0.6$
is required to describe the bulk medium flow~\cite{Wesp:2011yy}.
These results indicate sizable modification to the dijet asymmetry and
thus excellent sensitivity to the effective coupling to the medium at
RHIC energies.  

\begin{figure}[!hbt]
  \centering
    \includegraphics[trim = 10 0 10 0, clip, width=0.98\linewidth]{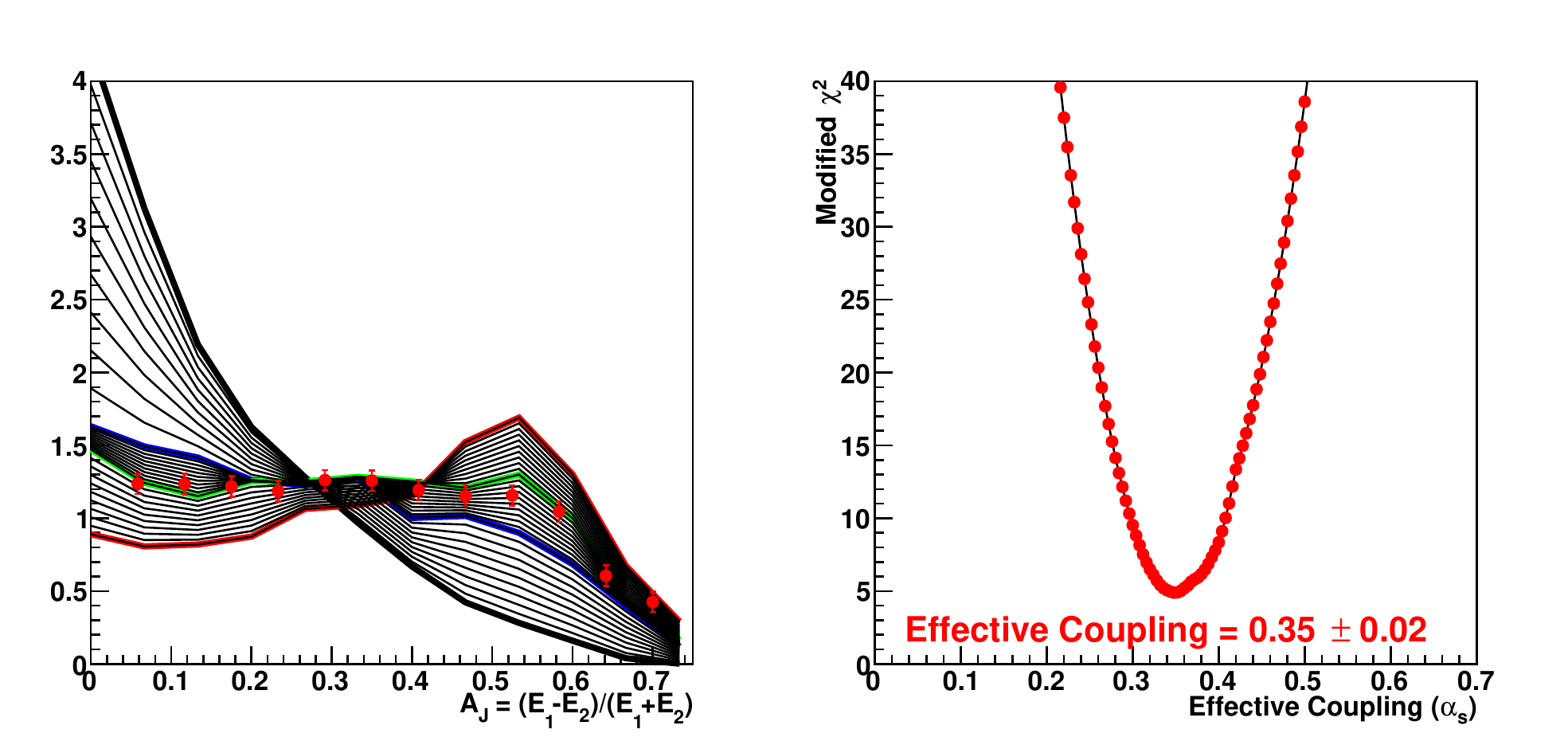}
    \caption{Determination of effective coupling strength in the model
      of Coleman-Smith.}
    \label{fig:coupling_determination}
\end{figure}

Figure~\ref{fig:coupling_determination} demonstrates the determination
of the effective coupling in the model of Coleman-Smith.  The
different curves in the left panel show the distribution of dijet
asymmetry for different values of the effective coupling.  The data
points are generated for a particular value of the coupling strength
and the uncertainties are representative of those that sPHENIX would
record.  By performing a modified $\chi_2$ comparison of the model to
the data, one obtains the curve in the right panel. From that curve,
one is able to determine the coupling with an uncertainty of about
5\%.

\begin{figure}[!hbt]
  \begin{center}
    \includegraphics[trim = 2 7 2 2, clip, width=\twowidth]{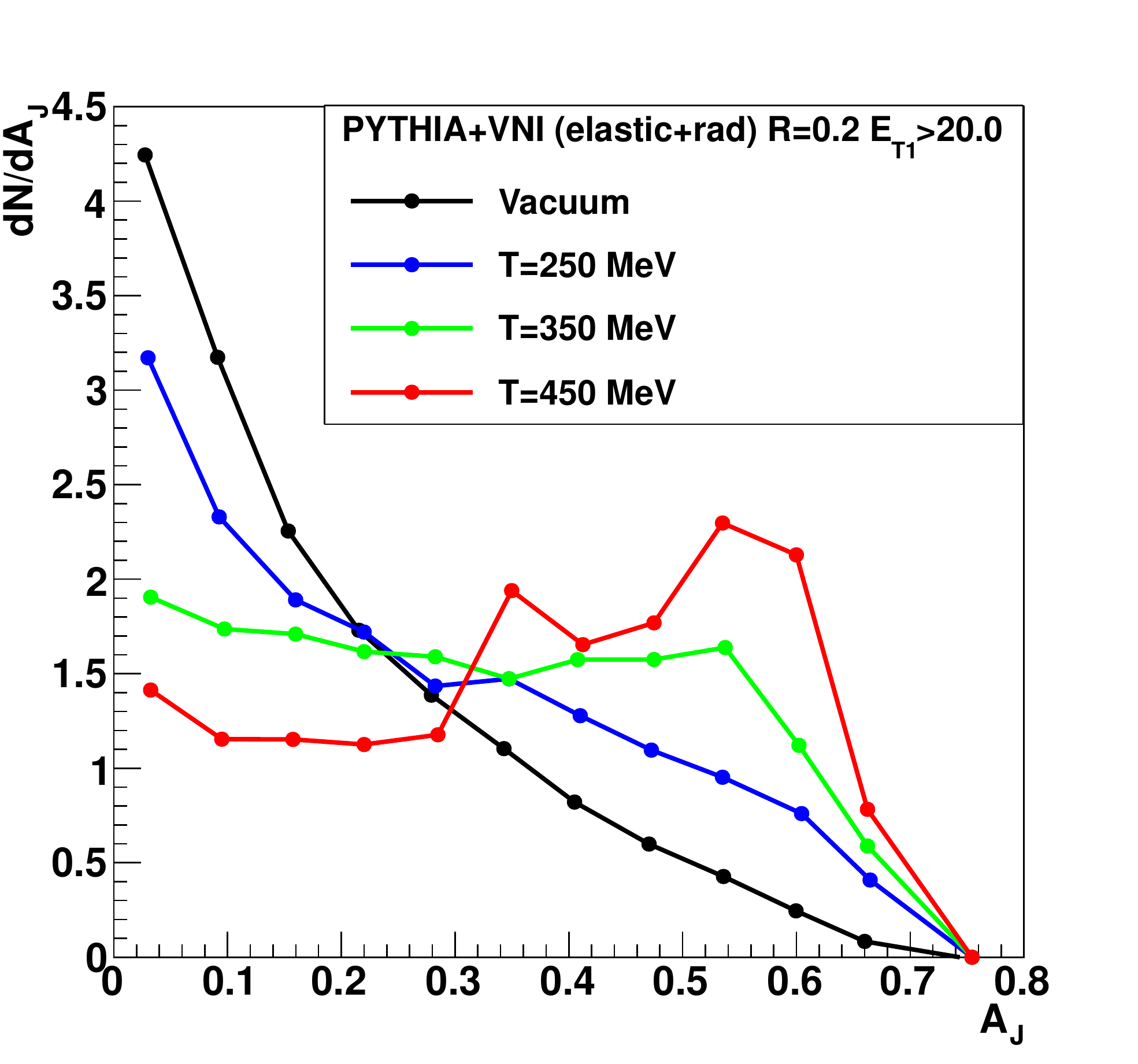}
    \hfill
    \includegraphics[trim = 2 2 2 2, clip, width=\twowidth]{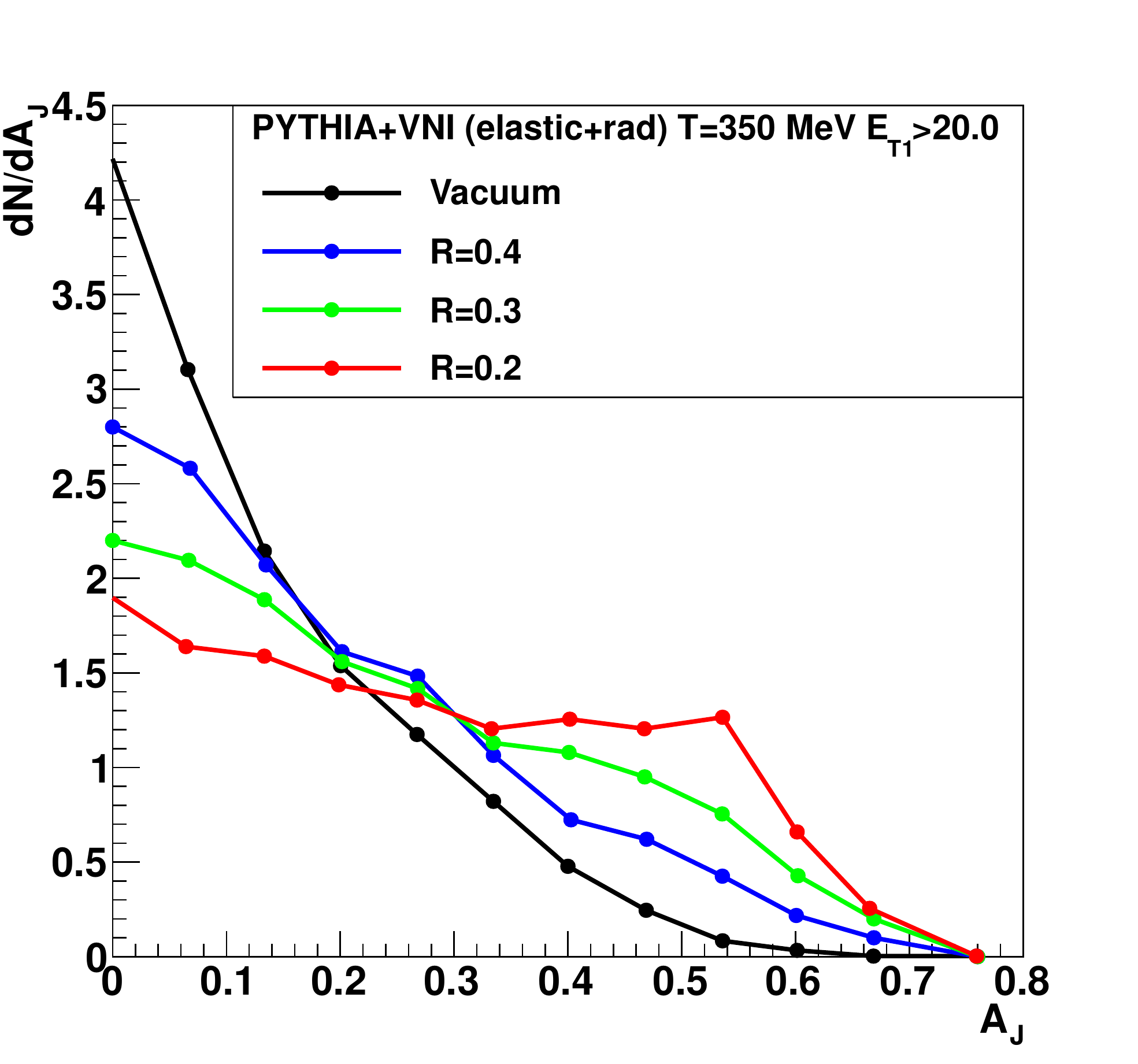}

    \caption[Calculation by Coleman-Smith of dijet asymmetry $A_J$ for
    leading jets with $E_T>20$~GeV as the medium temperature is
    varied and as the jet cone radius is varied at fixed
    temperature]{\label{fig:csprobescale}Calculations from
      Coleman-Smith~\protect\cite{ColemanSmith:2012vr}{} for dijets
      embedded into the VNI parton cascade.  The dijet asymmetry $A_J$
      for leading jets with $E_T>20$~GeV is shown as the medium
      temperature is varied (left panel) and as the jet cone radius is
      varied with fixed temperature $T=350$~MeV (right panel).}
 \end{center}
\end{figure}

Figure~\ref{fig:csprobescale} (left panel) shows the temperature
dependence of the dijet asymmetry, now keeping the coupling
$\alpha_{s}$ fixed.  One observes a similar sharp drop in the fraction
of energy balanced dijets with increasing temperature to that seen for
increasing the effective coupling, and so combining these observations
with constrained hydrodynamic models and direct photon emission
measurements is important.  Given that the initial temperatures of the
QGP formed at RHIC and the LHC should be significantly different, this
plot shows that if RHIC and LHC measure the $A_J$ distribution at the
same jet energy there should still be a sensitivity to the temperature
which will lead to an observable difference.  Thus, having overlap in
the measured jet energy range at RHIC and the LHC is important, and
this should be available for jet energies of 40--70~GeV.
Figure~\ref{fig:csprobescale} (right panel) shows the jet cone size,
$R$, dependence of $A_J$ at a fixed temperature.  The narrowest jet
cone $R=0.2$ has the most modified $A_J$ distribution, as partons are
being scattered away by the medium to larger angles.

\begin{figure}[!hbt]
 \begin{center}
    \includegraphics[trim = 2 2 2 2, clip, width=0.8\linewidth]{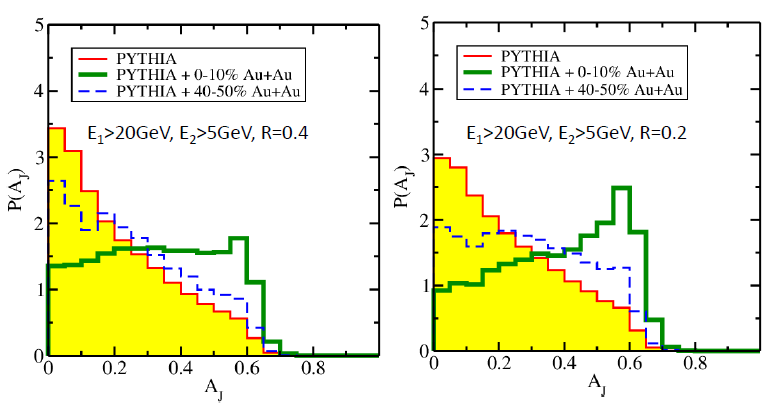}
    \caption[Calculations by Qin et al. of dijet $A_{J}$ for
    $E_{T,1}>20$~GeV and $E_{T,2}>5$~GeV for $R=0.2, 0.4$
    jets]{\label{fig:qinaj1} Calculations from Qin et
      al.~\protect\cite{qin_privatecomm}{} of dijet $A_{J}$ for
      $E_{T,1}>20$~GeV and $E_{T,2}>5$~GeV for $R=0.4$ jets (left)
      and $R=0.2$ jets (right).  Central (green) and mid-central
      (blue) distributions are shown along with the initial \pythia
      distributions (red).}
 \end{center}
\end{figure}

The second results are from Qin and
collaborators~\cite{Qin:2010mn,qin_privatecomm} where they solve a
differential equation that governs the evolution of the radiated gluon
distribution as the jet propagates through the medium.  Energy
contained inside the jet cone is lost by dissipation through elastic
collisions and by scattering of shower partons to larger angles.
Their calculation is able to describe the LHC measured dijet
asymmetry~\cite{Qin:2010mn}.  Figure~\ref{fig:qinaj1} shows the
predicted dijet asymmetry at RHIC for mid-central and central \auau
collisions for leading jets $E_{T1} > 20$~GeV and jet radius
parameter $R=0.4$ and $R=0.2$ in the left and right panels,
respectively.  Despite the calculation including a rather modest value
of $\hat{q}$ and $\hat{e}$, the modification for $R=0.2$ is as strong
as the result with $\alpha_{s} = 0.6$ from Coleman-Smith and
collaborators shown above in the right panel of
Figure~\ref{fig:colemansmith1}.  Calculations of $\gamma$-jet
correlations indicate similar level modifications.  It is also notable
that Qin and collaborators have calculated the reaction plane
dependence of the dijet $A_J$ distribution and find negligible
differences.  This observable will be particularly interesting to
measure at RHIC since these calculations have difficultly reproducing
the high $p_T$ $\pi^{0}$ reaction plane dependence ($v_2$) as
discussed in the previous section.

\begin{figure}[t]
 \begin{center}
    \includegraphics[trim = 2 2 2 2, clip, width=0.5\linewidth]{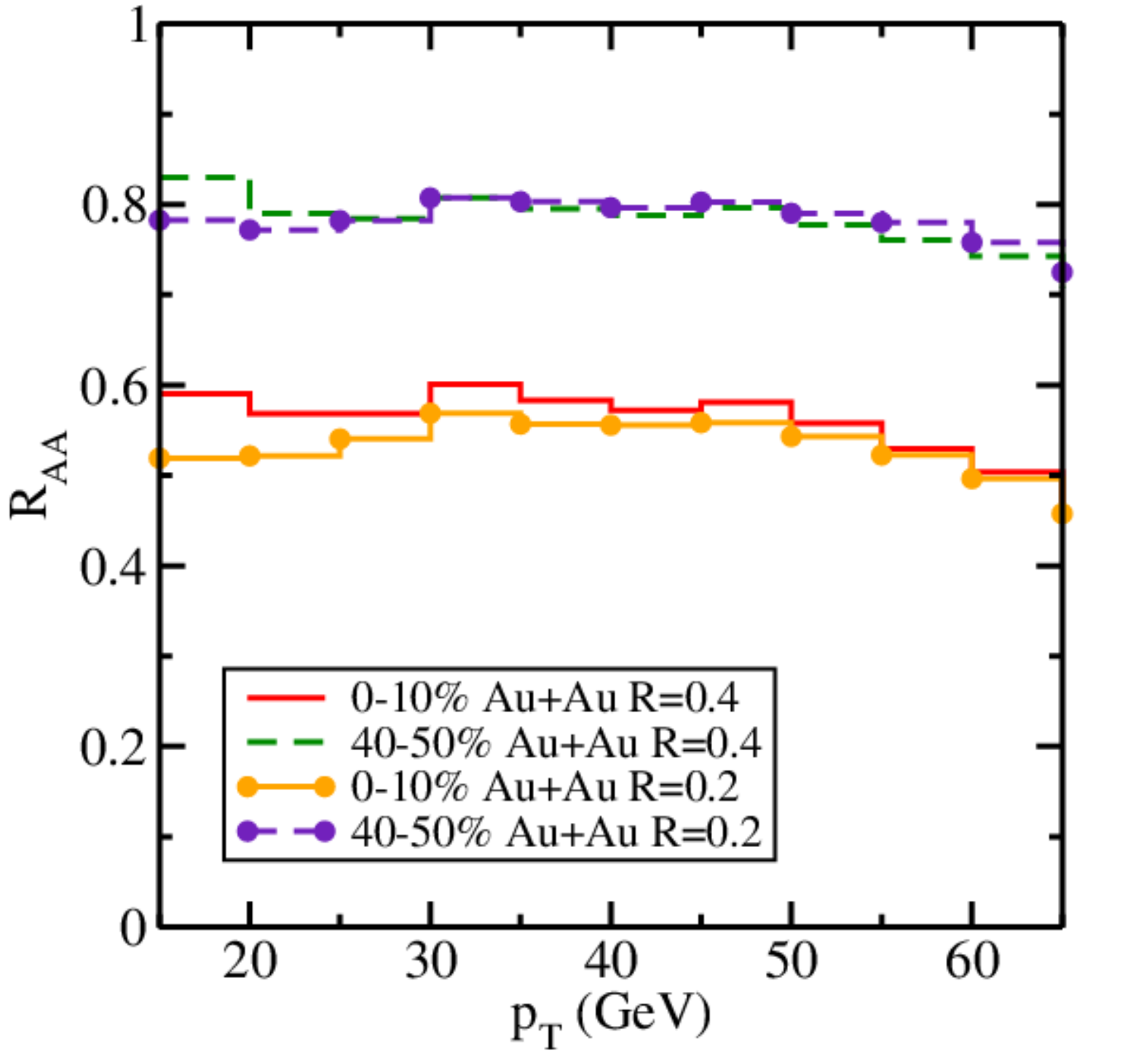}
    \caption[Calculations from Qin et al. of jet \raa for central
    and mid-central collisions for $R=0.2, 0.4$
    jets]{\label{fig:qinraa} Calculations from Qin et
      al.~\protect\cite{qin_privatecomm}{} for jet \raa for
      central (solid lines) and mid-central collisions (dashed lines)
      for $R=0.2$ and 0.4 jets.}
 \end{center}
\end{figure}

Figure~\ref{fig:qinraa} shows results for the inclusive jet \raa
as a function of $p_T$ for jet radius parameters $R=0.2$ and $R=0.4$.
It is striking that the modification is almost independent of $p_T$ of
the jet and there is very little jet radius dependence.  The modest
suppression, of order 20\%, in mid-central \auau collisions is of
great interest as previous measurements indicate modification of
single hadrons and dihadron correlations for this centrality category.
Measurements of jets with a broad range of radius parameters are
easier in the lower multiplicity mid-central collisions.

\begin{figure}[t]
 \begin{center}
    \raisebox{0.08in}{\includegraphics[trim = 0 0 0 0, width=0.48\linewidth]{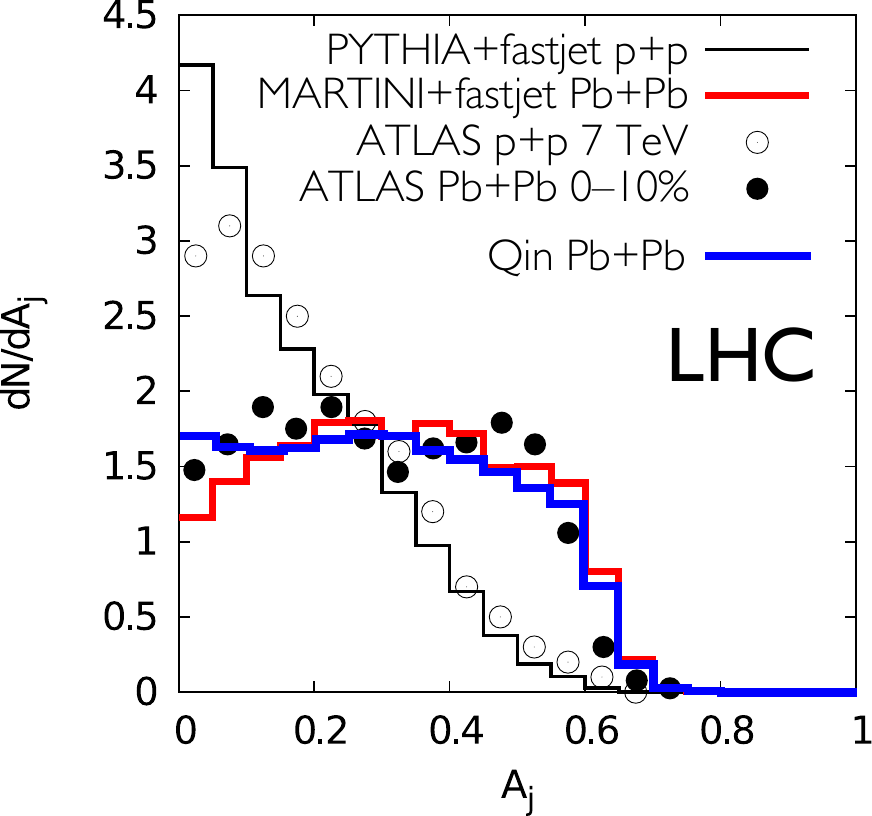}}
    \hfill
    \includegraphics[trim = 0 0 0 0, clip, width=0.47\linewidth]{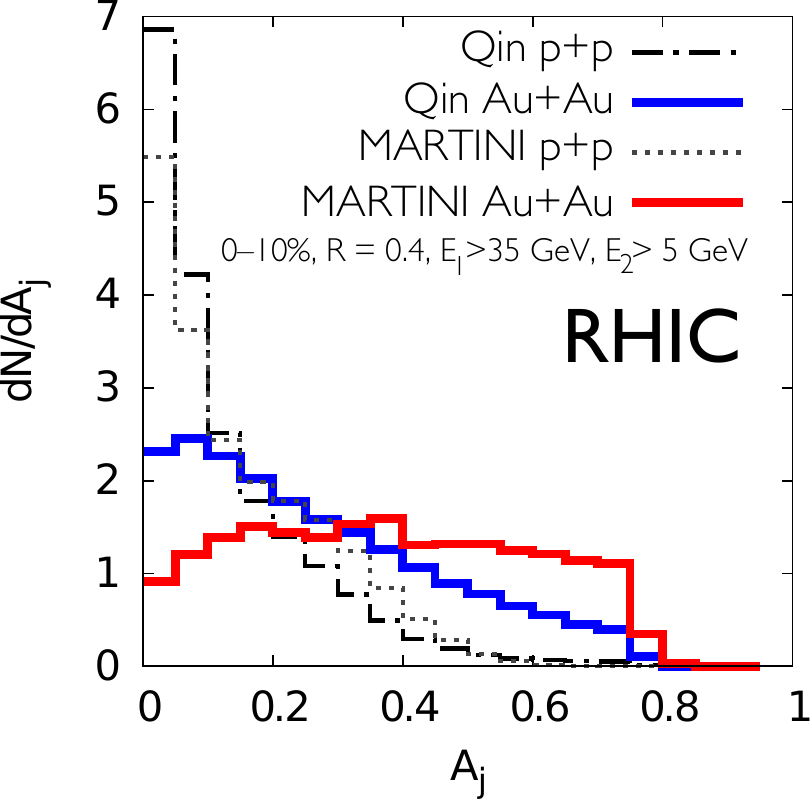}
    \caption[$A_J$ distributions in \martinimusic and in the model of
    Qin et al. at LHC and RHIC energies]{\label{fig:martiniaj} $A_J$
      distributions in
      \martinimusic~\protect\cite{Young_privatecomm}{} and the model
      of Qin et al.~\protect\cite{qin_privatecomm}{}.  (left)
      Comparison of $A_J$ calculations in \martinimusic and by Qin et
      al for \pbpb collisions at 2.76~TeV (red line, Qin et al; blue
      line, \martinimusic).  Both calculations show a similar broad
      $A_J$ distribution.  (right) Same as left panel, but for \auau
      collisions at 200~GeV (with leading jet $E_T>35$~GeV).  Here a
      difference in shape is observed between the two models with the
      Qin et al. model developing a peak at small $A_J$ while the
      \martinimusic calculation retains a shape in the calculation at
      the higher energy.}
 \end{center}
\end{figure}

The third results are from Young and Schenke and
collaborators~\cite{Young:2011va}.  These calculations utilize a jet
shower Monte Carlo, referred to as \martini~\cite{Schenke:2009vr},
and embed the shower on top of a hydrodynamic space-time background,
using the model referred to as \music~\cite{Schenke:2010nt}.
Figure~\ref{fig:martiniaj} shows the jet energy dependence of $A_J$
for RHIC energy dijets, $E_{T1}>25$~GeV and $E_{T1}>35$~GeV in the
left and right panels, respectively.  These results are directly
compared to the calculations from Qin and collaborators and indicate
a substantially different modification for the higher energy dijets.
Interestingly, both of these approaches, when applied at the higher
collision energies of the LHC, each reproduce the measured data quite
well~\cite{Young:2011qx,Qin:2010mn}.

Our next set of illustrative theory calculations come from Vitev and
collaborators~\cite{He:2011pd,Neufeld:2011yh,Vitev:2009rd} where they
utilize a Next-to-Leading-Order (NLO) calculation and consider not only
final-state inelastic parton interactions in the QGP, but also initial-state cold
nuclear matter effects.  Figure~\ref{fig:vitevaj} shown earlier plots the dijet asymmetry $A_J$ for jets 
with $E_{T1} > 50$~GeV and $R=0.6$. 
The plots are for cases of radiative energy loss only and including collisional energy loss as well,
and then the different colors are varying the probe-medium coupling by $\pm$10\%.   There is 
sensitivity even to these 10\% coupling modifications, and for the higher energy jets there is a
dramatic difference predicted from the inclusion of collisional energy loss.

\begin{figure}[t]
 \begin{center}
    \includegraphics[width=0.6\linewidth]{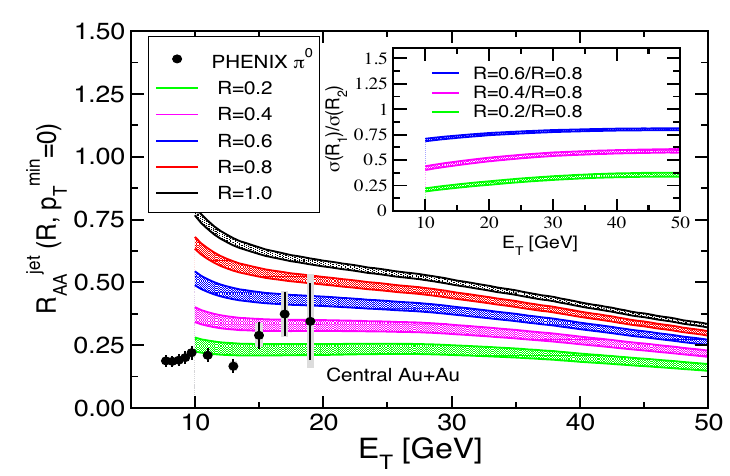}
    \caption[Calculations by Vitev et al. for the inclusive jet \raa
    vs jet energy and radius]{\label{fig:vitev_raasize} Calculations
      from Vitev et al. for the inclusive jet \raa as a function of
      the jet energy and radius.  }
 \end{center}
\end{figure}

For the inclusive jet suppression, these calculations predict a
significant jet radius $R$ dependence to the modification, in contrast
to the result from Qin and collaborators.
Figure~\ref{fig:vitev_raasize} shows the significant radius
dependence.  In addition, Vitev and collaborators hypothesize a
substantial cold nuclear matter effect of initial state parton energy
loss.  Because the high energy jets originate from hard scattering of
high Bjorken $x$ partons, a modest energy loss of these partons
results in a reduction in the inclusive jet yields.  At RHIC with
\dAu~running we will make cold nuclear matter measurements at the same
collision energy and determine the strength of these effects as a
baseline to heavy ion measurements.

\begin{figure}[t]
 \begin{center}
    \includegraphics[width=0.95\linewidth]{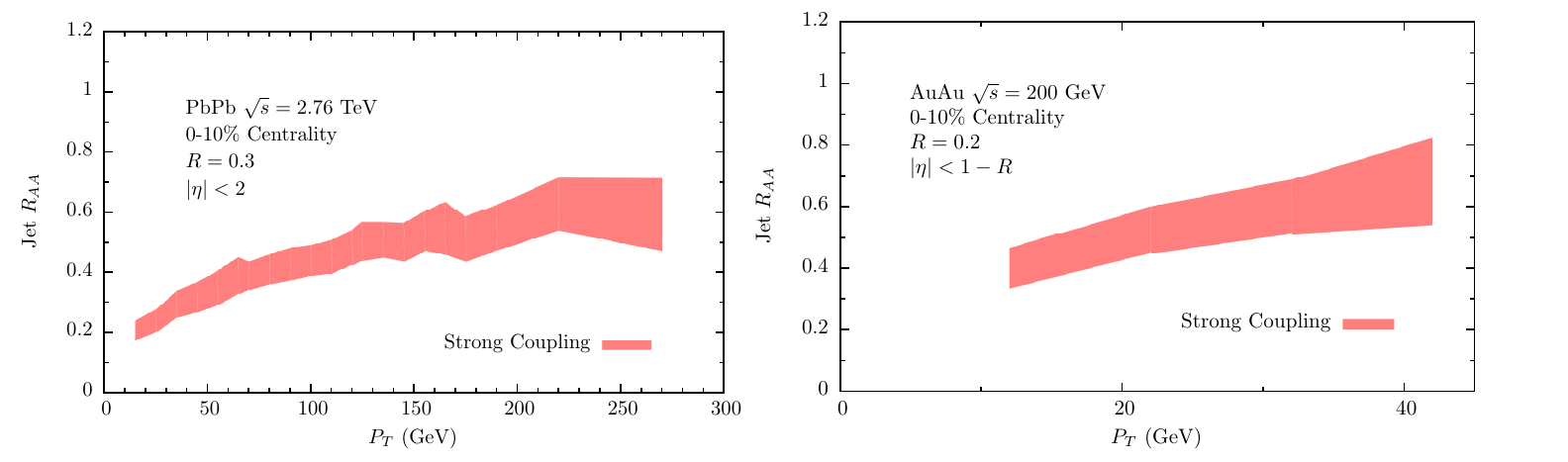}
    \caption[Calculations of jet \raa vs \pt for central collisions at
    the LHC and at RHIC using the hybrid strong coupling approach of
    Casalderrey-Solana et al]{\label{fig:uberstrong} Calculations from
      a hybrid strong coupling approach with predictions for
      reconstructed jet \raa as a function of \pt appropriate for
      central collisions at the LHC (left) and RHIC
      (right)~\cite{Casalderrey-Solana:2014wca,Casalderrey-Solana:2014bpa}.
    }
 \end{center}
\end{figure}

Recently a framework with a hybrid strong coupling approach has been implemented with initial success at describing
specific jet quenching observables~\cite{Casalderrey-Solana:2014wca,Casalderrey-Solana:2014bpa}.   Shown in 
Figure~\ref{fig:uberstrong} are the predicted \raa for reconstructed jets at the LHC (left) and at RHIC (right).   
The jet \raa shows a rise as a function of \pt at both energies, in contrast to calculations as shown in Figure~\ref{fig:qinraa} for example.
This framework enables an alternate set of  predictions for a host of observables sensitive to the redistribution
of energy within the parton shower at RHIC and the LHC.  The steeper spectrum at RHIC and the lower energy jets should
make them more sensitive to the details of the hybrid calculations.

\begin{figure}[t]
 \begin{center}
    \includegraphics[height=0.9\textheight]{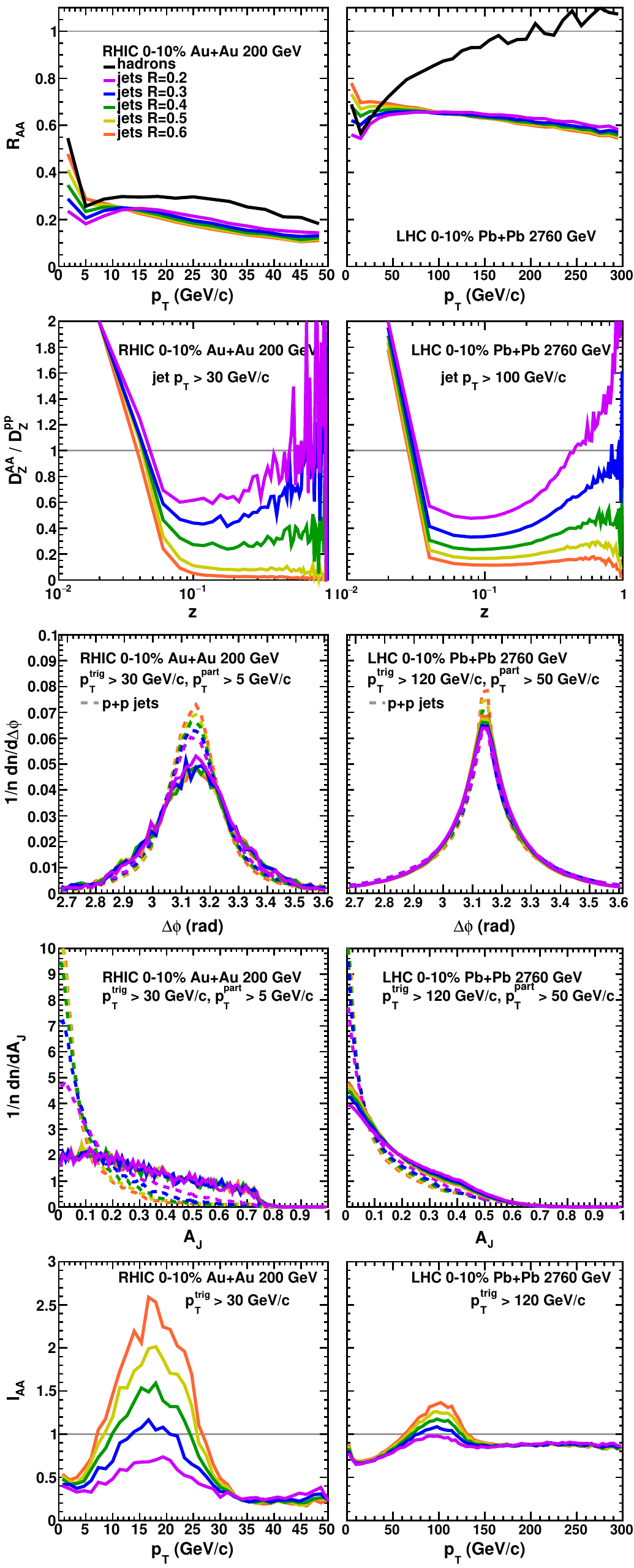}
    \caption[Compilation of results for jet and hadron observables at
    RHIC and the LHC using the JEWEL parton shower Monte
    Carlo]{\label{fig:uberjewel} JEWEL 2.0 parton shower Monte Carlo
      results for jet and hadron observables at RHIC and the LHC using
      the publicly available code~\cite{Zapp:2013vla,Zapp:2008gi}.
      See text for the detailed description of the panels.  }
 \end{center}
\end{figure}

The simultaneous development of parton shower Monte Carlo codes -- for
example see
Refs.~\cite{Zapp:2009pu,Renk:2010zx,Young:2011va,ColemanSmith:2011wd,Lokhtin:2011qq,Armesto:2009zc}
-- and in some cases their public availability allows the community to
explore a full range of experimental observables.  We have run the
JEWEL 2.0 code~\cite{Zapp:2013vla} at both RHIC and LHC kinematics and
medium parameters and then run the HEPMC output through the FASTJET
reconstruction code.  Results of a suite of observables for both
energies are shown in Figure~\ref{fig:uberjewel}.  The top panel shows
the jet \raa for different jet radii and charged hadron \raa.  It is
striking as pointed out earlier that the charged hadron \raa is quite
flat at RHIC and at the same time has the characteristic rise at the
LHC as observed in data.  The next panel shows the modified
fragmentation function from inclusive jets, where the observable
reflects both the modification in the parton shower and the
potentially reduced fraction of energy captured within the
reconstructed jet.  The next panel shows the dijet azimuthal asymmetry
with the dashed lines in \pp collisions and the solid lines in central
heavy ion collisions.  The RHIC predictions show a measurable
broadening of the azimuthal distribution.  The next panel shows the
$A_J$ dijet asymmetry distribution.  The steeper falling spectrum at
RHIC leads to a more significant depletion of balanced jets (i.e. $A_J
\approx 0$) and a larger shifts in the $\left<A_{J}\right>$.  The
bottom panel shows the related $I_{AA}$ for dijets, in this case with
a narrow trigger jet with $R=0.2$ and varying the away-side jet
radius.  Again, a dramatic modification is expected at RHIC from the
interplay of both larger surface bias from the trigger jet and a bias
for the away-side parton to be a gluon opposed a quark trigger parton.
sPHENIX will have excellent statistics across this breadth of
observables and more.

It is notable that in the recommended running mode for JEWEL, one
retains recoil partons for hadron reconstruction and not for jet
reconstruction.  JEWEL treats the medium partons as a gas of nearly
free quarks and gluons and thus it is relatively easy to transfer
energy to these partons, which then recoil.  In the case where recoil
partons are included in the jet reconstruction they have also their
initial thermal energy and thus one gets large jet \raa enhancements.
When excluding them, some energy is lost and \raa for large radius
jets appears smaller than expected.  We are working on a running mode
to only include the transferred energy and thus test the sensitivity
to the model of the medium partons.  We continue to directly engage
the theory community for development of these Monte Carlo codes for
optimal comparison between data and theory.

\clearpage

\section{Direct Photons and Fragmentation Functions}
\label{sec:direct_photons_and_ff}

Ideally, one would like to understand how a quark or gluon of
perfectly known energy interacts traversing the \qgp and the
redistribution of energy and particles both longitudinal and
transverse to the initial parton direction.  The \emph{golden channel}
for the calibration of initial quark energy is to tag them via an
opposing direct photon~\cite{Wang:1996yh}.  One can measure fully
reconstructed jets opposite the photon with different jet radii to
parse out the transverse energy redistribution.

\begin{figure}[!hbt]
 \begin{center}
   \raisebox{0.03cm}{\includegraphics[width=0.50\linewidth]{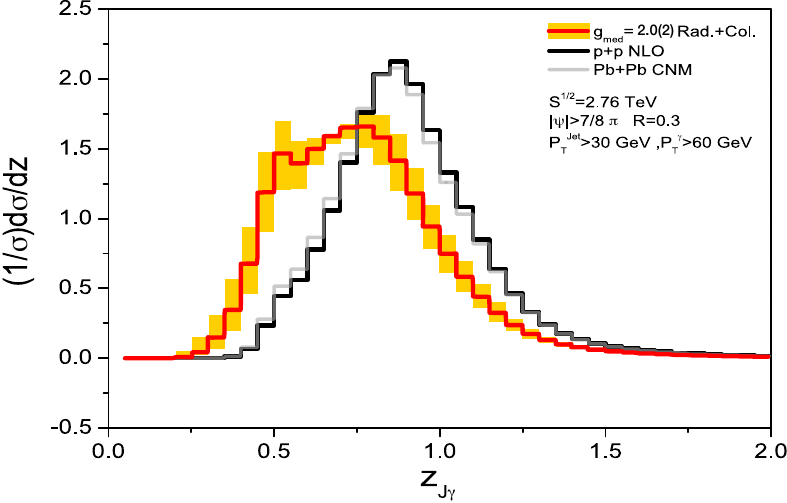}}
   \hfill
   \includegraphics[width=0.48\linewidth]{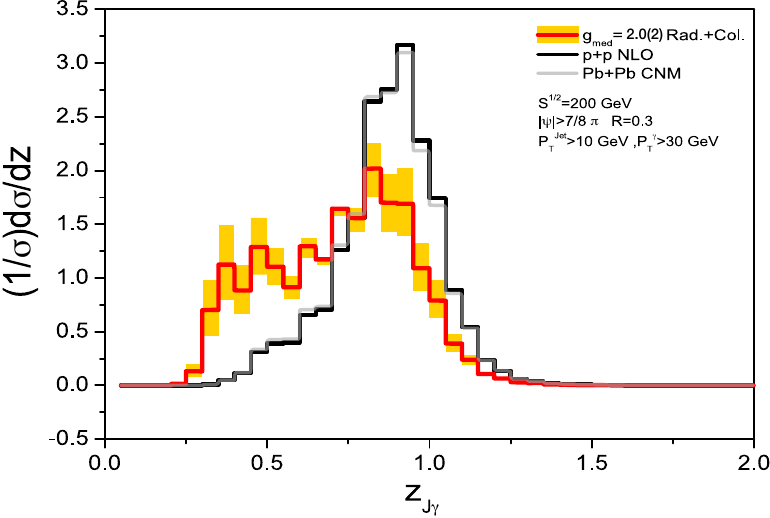}
   \caption[Calculations by Vitev et al. of the vacuum and medium
   modified distribution for direct photon triggered reconstructed jet
   events at LHC and RHIC energies]{\label{fig:vitevgammajet}
     Calculation results for the vacuum and medium modified
     distribution for direct photon --- reconstructed jet events at
     LHC collision energy (left) and RHIC collision energy
     (right)~\cite{Dai:2012am}.  }
 \end{center}
\end{figure}

\begin{figure}[!hbt]
 \begin{center}
   \includegraphics[width=0.6\linewidth]{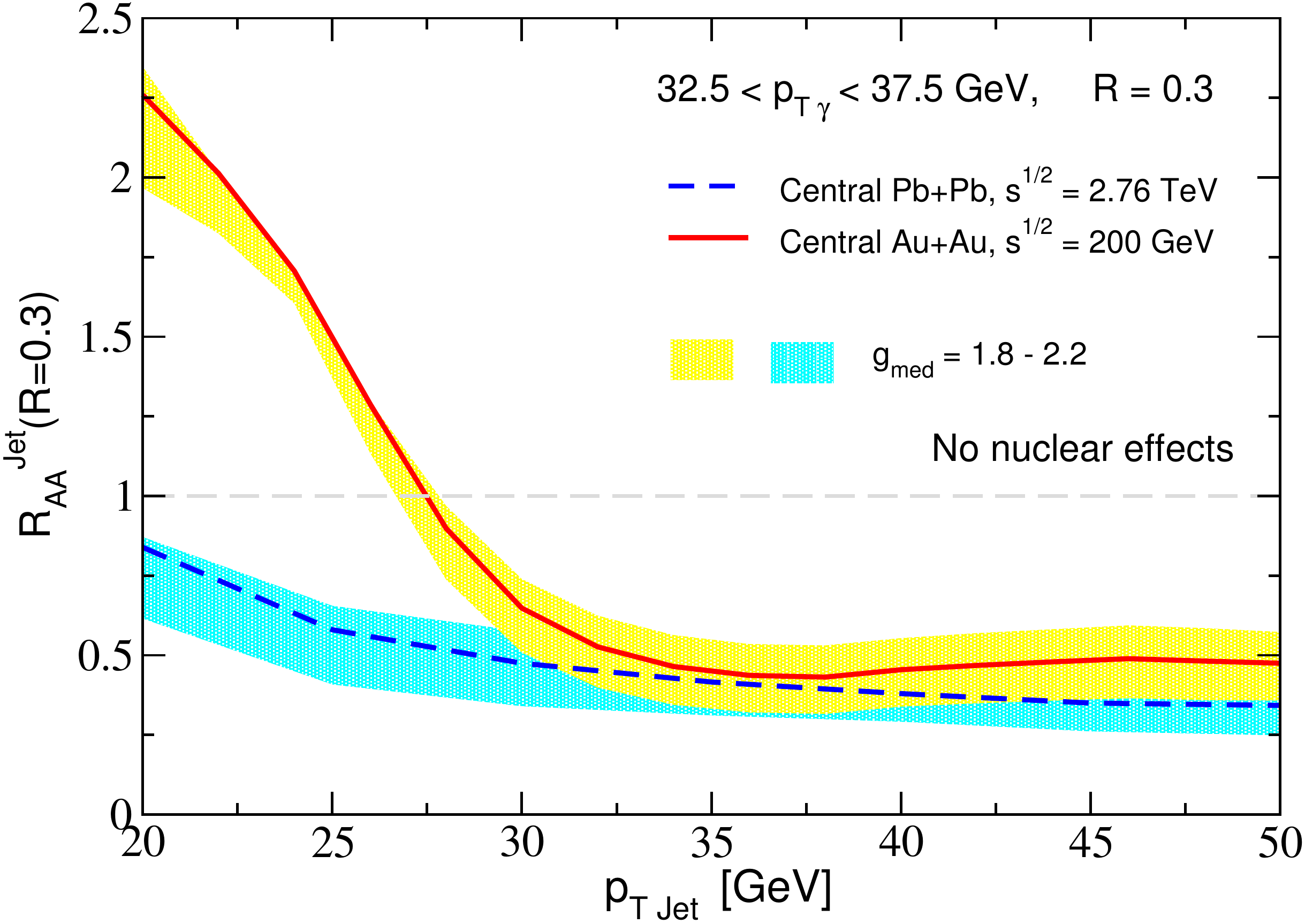}
   \caption[Calculations by Vitev et al. of the jet \raa opposite to a tagged
     direct photon in \auau collisions at 200~GeV and \pbpb collisions
     at 2.76~TeV]{Calculation results for the jet \raa opposite to a tagged
     direct photon in \auau collisions at 200~GeV and \pbpb collisions
     at 2.76~TeV~\cite{Dai:2012am}.}
  \label{fig:vitev_gammajet_raa}
 \end{center}
\end{figure}

Figure~\ref{fig:vitevgammajet} shows the event distribution for the
ratio of the reconstructed jet energy with $R=0.3$ relative to the
direct photon energy~\cite{Dai:2012am}.  As the authors note, ``The
steeper falling cross sections at RHIC energies lead not only to a
narrow $z_{J_{\gamma}}$ distribution in \pp collisions but also to a
larger broadening end shift in $\left<z_{J_{\gamma}}\right>$ in A+A
collisions.''  This results in a greater sensitivity to the
redistribution of energy, which is again sensitive to the balance of
processes including radiative and collisional energy loss.
Figure~\ref{fig:vitev_gammajet_raa} shows the jet \raa opposite a
35~GeV direct photon~\cite{Dai:2012am}.  There is a dramatic
difference between the RHIC and LHC result, where one expects a factor
of two enhancement in jets near 20~GeV in these collision systems.  As
detailed in the sPHENIX performance section in
Figure~\ref{fig:g4_photon_iaa}, with an underlying event energy a
factor of 2.5 lower at RHIC compared to the LHC, sPHENIX can
reconstruct jets over a very broad range of radii and energies
opposite these direct photons.

\begin{figure}[!hbt]
 \begin{center}
    \includegraphics[width=0.85\linewidth]{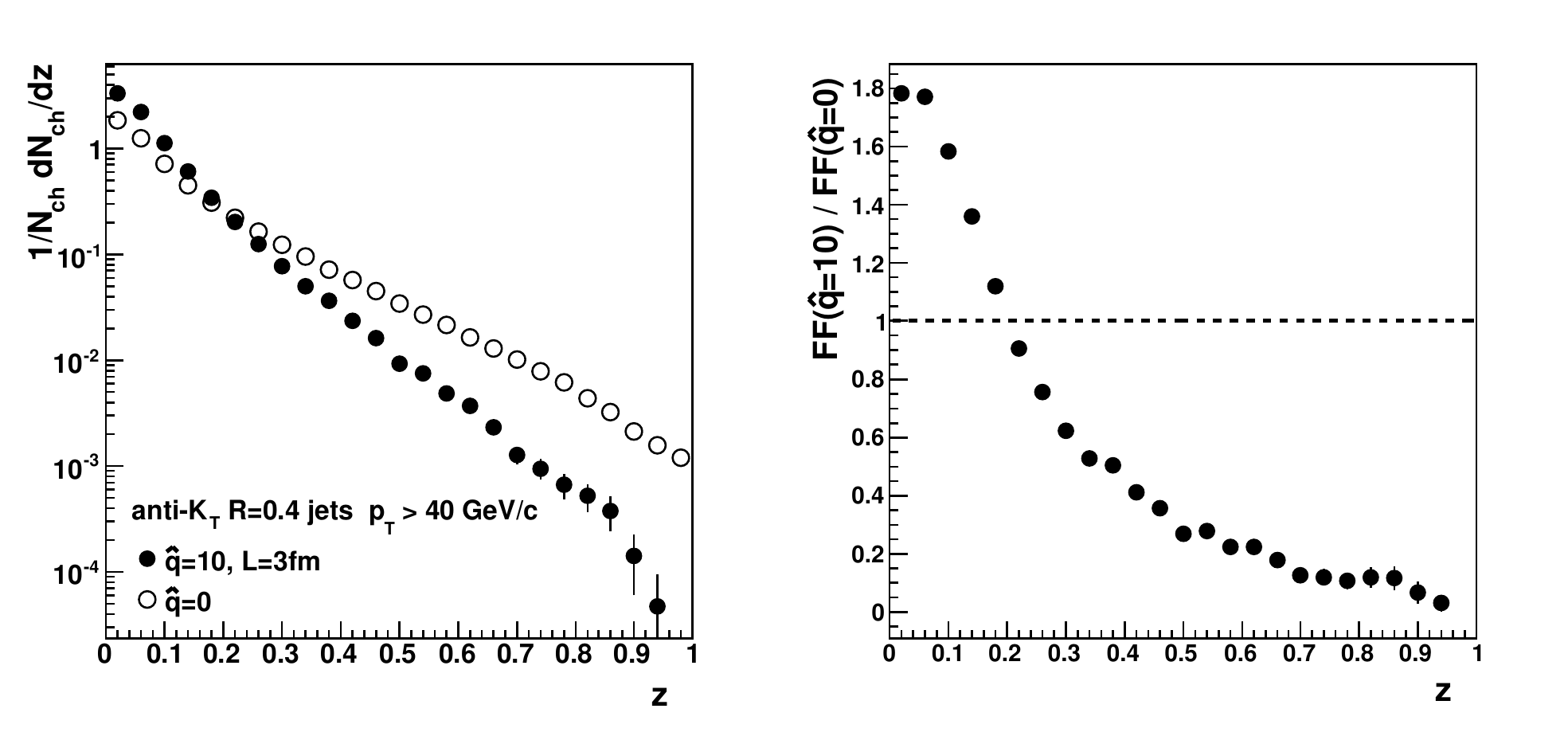}
    \caption[Q-\pythia simulation with quenching parameter $\hat{q} =
    0$ and $\hat{q} = 10$~GeV/c$^{2}$ for the fragmentation function
    of light quark and gluon jets as a function of $z$.
    ]{\label{fig:fragz}Q-\pythia simulation with quenching parameter
      $\hat{q} = 0$ (i.e., in vacuum) and $\hat{q} = 10$~GeV/c$^{2}$
      for the fragmentation function of light quark and gluon jets as
      a function of $z$.  }
 \end{center}
\end{figure}

With charged particle tracking one can also measure the longitudinal
redistribution of hadrons opposite the direct photon.  sPHENIX will
have excellent statistical reach for such direct photon measurements.
At the same time, it is advantageous to measure modified fragmentation
functions within inclusive reconstructed jets and via correlations as
well.  The original predictions of jet quenching in terms of induced
forward radiation had the strongest modification in the longitudinal
distribution of hadrons from the shower (i.e., a substantial softening
of the fragmentation function).  One may infer from the nuclear
suppression of $\pi^{0}$ in central \auau collisions $R_{AA} \approx
0.2$ that the high $z$ (large momentum fraction carried by the hadron)
showers are suppressed.  Shown in Figure~\ref{fig:fragz} is the
fragmentation function for 40~GeV jets in vacuum (\pythia) compared
with the case of substantial jet quenching (Q-\pythia with a quenching
factor used to match RHIC single hadron suppression observables).  In
the sPHENIX upgrade, fragmentation functions via precision charged
track measurements are available from high-$z$ where the effects are
predicted to be largest to low-$z$ where medium response and
equilibration effects are relevant.  The independent measurement of
jet energy (via calorimetry) and the hadron $p_T$ via tracking is
crucial.  This independent determination also dramatically reduces the
\fake track contribution by the required coincidence with a high
energy jet.

\begin{figure}[!hbt]
 \begin{center}
   \raisebox{3.5ex}{\includegraphics[width=0.45\linewidth]{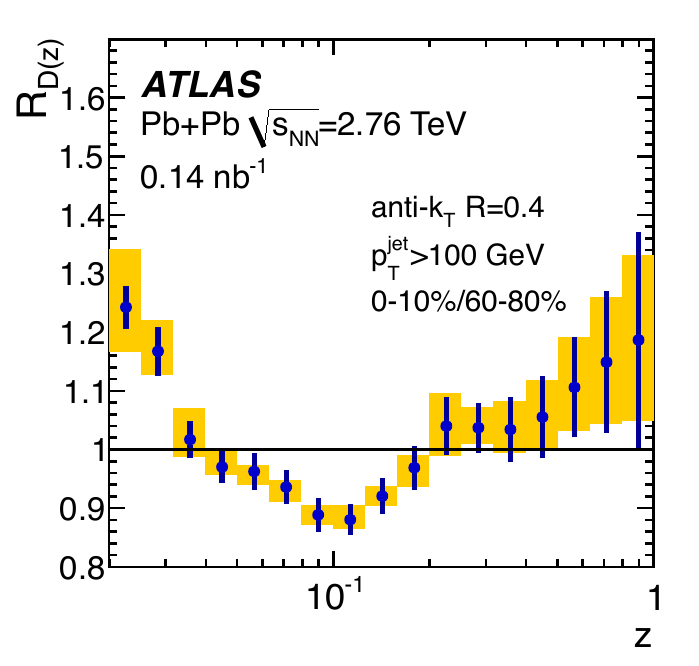}}
   \includegraphics[width=0.51\linewidth]{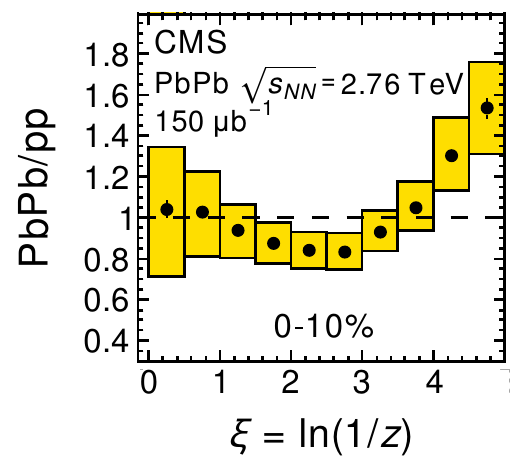} 
   \caption[The modified fragmentation function for $p_\mathrm{T} >
   100$~GeV jets in central Pb+Pb collisions vs $z$ from ATLAS and vs
   $\xi$ from CMS]{\label{fig:LHC_FF_measurements} Measurements of the
     modified fragmentation function for $p_\mathrm{T} > 100$~GeV jets
     in central Pb+Pb collisions from ATLAS~\cite{Aad:2014wha} (left
     plot, as a function of $z = \vec{p}_\mathrm{T}^\mathrm{\
       hadron}\cdot\vec{p}_\mathrm{T}^\mathrm{\ jet} /
     \left|\vec{p}_\mathrm{T}^\mathrm{\ jet}\right|^{2}$) and
     CMS~\cite{Chatrchyan:2014ava} (right plot, as a function of $\xi
     = 1/z$).  } \end{center}
\end{figure}

Measurements at the LHC reveal a very different behavior as shown in Figure~\ref{fig:LHC_FF_measurements} where
a slight enhancement is hinted at for large $z$, rather than a large suppression.
Measurements of fragmentation functions within reconstructed jets from the CMS and
ATLAS experiments in \pbpb collisions show very modest modification within
uncertainties. Although one explanation is that the jets that are
reconstructed are from near the surface and thus not modified, with a
nuclear modification factor for inclusive jets $R_{AA} \approx 0.5$
that explanation is challenged.   Similar measurements at RHIC energies
are crucial to fully map out the re-distribution of energy in the shower
and medium response.   An example of the sPHENIX precision for such measurements is shown 
later in Figure~\ref{fig:fastsim_ff}.

\begin{figure}[!hbt]
 \begin{center}
    \includegraphics[width=0.7\linewidth]{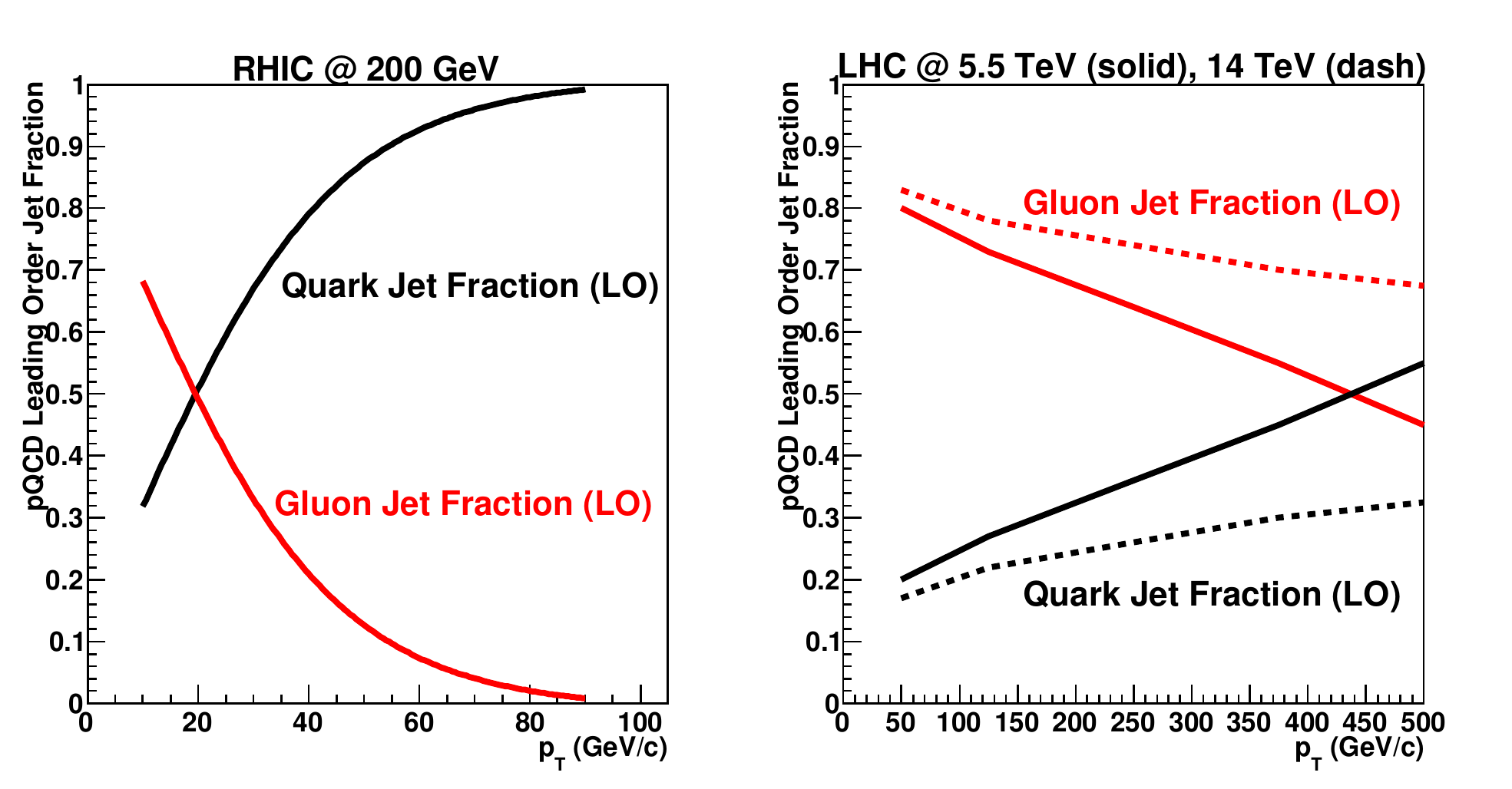}
\caption{Comparison of the fraction of quark and gluon jets from leading order pQCD calculations for RHIC and LHC energies.
\label{fig:partonfrac}}
\end{center}
\end{figure}

One can also access less directly this transverse and longitudinal redistribution of energy and particles
via trigger high \pt hadrons and narrow reconstructed jets.    Similar measurements have been carried out by
the STAR experiments, as discussed earlier in the context of Figure~\ref{fig:star_hjet}.   
With the large kinematic reach of sPHENIX, 
one can have very high statistics observables that span a reach where the opposing parton is mostly a gluon near 20~GeV and then
increases in quark fraction for higher energy triggers.   This is another complement between the kinematics at RHIC
and the LHC as shown in Figure~\ref{fig:partonfrac} comparing the quark-quark, quark-gluon, gluon-gluon relative contributions
as a function of \pt.

\begin{figure}[!hbt]
 \begin{center}
    \includegraphics[width=0.7\linewidth]{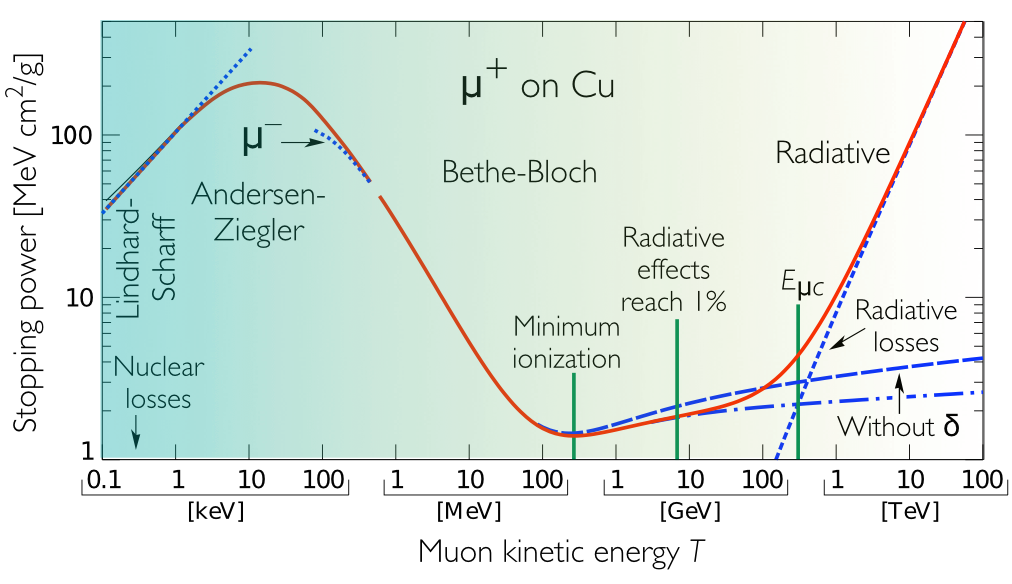}
    \caption[The muon stopping power in copper]{The muon stopping
      power in copper demonstrates a comprehensive understanding of
      the interaction of a fundamental particle with matter over an
      enormous range of scales.\label{fig:mucu}}
\end{center}
\end{figure}

Combining high statistics results on this full set of observables from
RHIC and the LHC can lead to a detailed description of the quark and
gluon interaction in the \qgp as a function of parton energy analogous
to that from the Particle Data Group for the muon in copper as shown
in Figure~\ref{fig:mucu}.

\clearpage

\section{Heavy Quark Jets}
\label{sec:heavyquark}

A main motivation for studying heavy flavor jets in heavy ion
collisions is to understand the mechanism for parton-medium
interactions and to further explore the issue of {\it strong versus weak} coupling~\cite{horowitz}.  
As detailed in Section~\ref{Section:InnerWorkings}, a major goal is understanding the constituents of the medium
and how fast partons transfer energy to the medium.   Heavy quarks have gathered special attention as they are 
particularly sensitive to the contribution of collisional energy loss, due to suppressed radiative energy loss from the
``dead cone'' effect~\cite{dk_dead_cone}.   Measurements of beauty-tagged jets and reconstructed $D$ mesons over the broadest
kinematic reach will enable the disentangling of $\hat{q}$ and $\hat{e}$.   

\begin{figure}[!hbt]
 \begin{center}
    \includegraphics[width=0.7\linewidth]{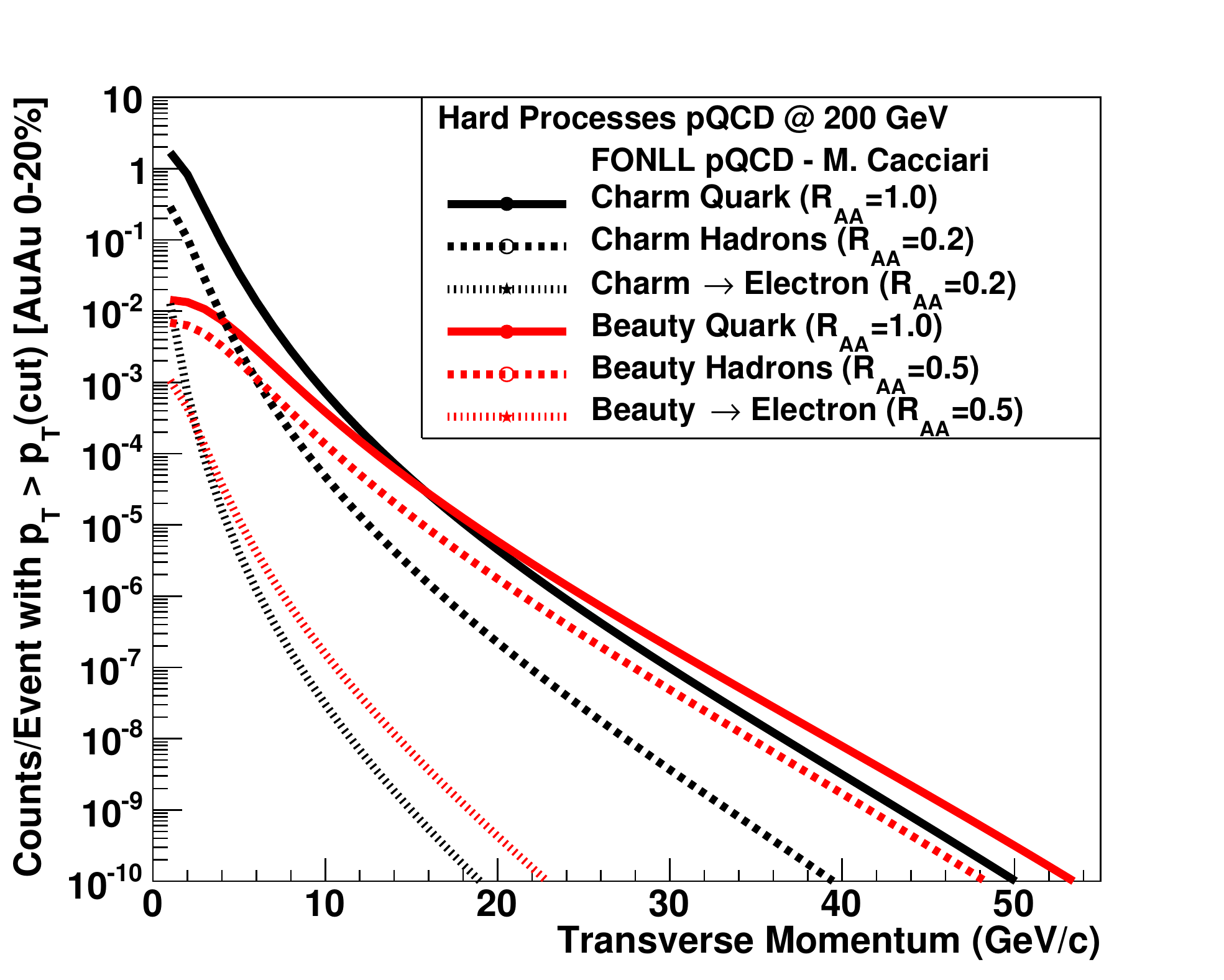}
    \caption[FONLL calculations of heavy flavor jets, fragmentation
    hadrons, and decay electrons vs \pt]{\label{fig:heavyrates}FONLL
      calculations~\cite{Cacciariprivate} for heavy flavor (charm and
      beauty) jets, fragmentation hadrons ($D, B$ mesons primarily),
      and decay electrons as a function of transverse momentum.  The
      rates have been scaled to correspond to counts with $p_{T} >
      p_{T}(cut)$ for \auau 0--20\% central collisions.}
 \end{center}
\end{figure}

There are important measurements currently being made of single electrons from semileptonic $D$ and $B$ decays and direct $D$ meson 
reconstruction with the current PHENIX VTX and STAR Heavy Flavor Tracker (HFT).  The sPHENIX program can significantly expand 
the experimental acceptance and physics reach by having the ability to reconstruct full jets with a heavy flavor tag.
The rates for heavy flavor production from perturbative QCD calculations~\cite{Cacciariprivate} are shown in 
Figure~\ref{fig:heavyrates}.

\begin{figure}[!hbt]
 \begin{center} \includegraphics[width=0.95\linewidth]{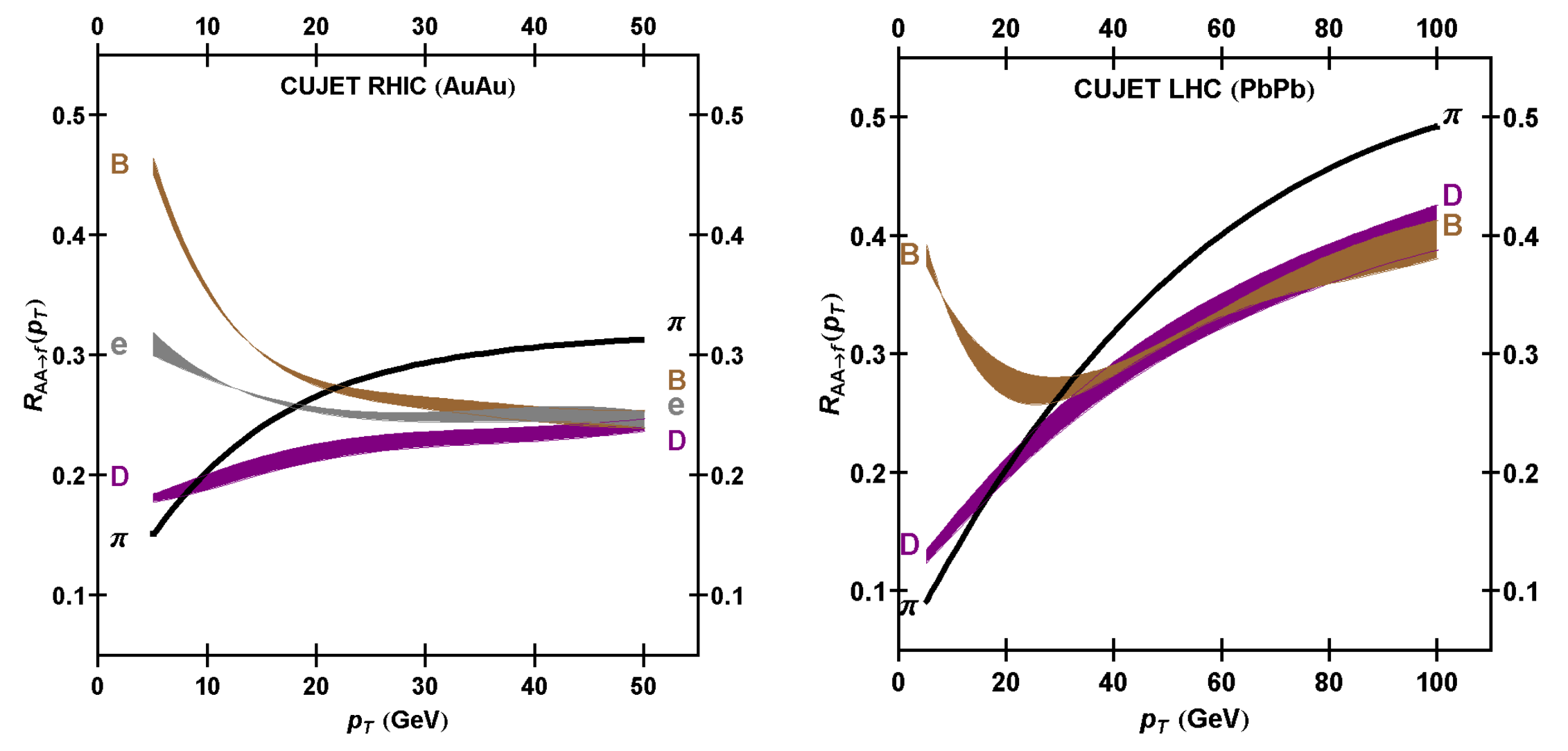} 
   \caption[CUJET calculations of $R_\mathrm{AA}$ in central \auau
   collisions at RHIC and in Pb+Pb collisions at the LHC, with light,
   charm and beauty hadrons and electrons shown as separate
   curves.]{Calculations within the CUJET 2.0~\cite{Xu:2014ica}
     framework of the $R_\mathrm{AA}$ in central Au+Au collisions at
     RHIC (left panel) and Pb+Pb collisions at the LHC (right panel),
     with light, charm and beauty hadrons and electrons shown as
     separate curves.  }
  \label{fig:cujet}
 \end{center}
\end{figure}

Calculations including both radiative and collisional energy loss for
light quark and gluon jets, charm jets, and beauty jets have been
carried out within the CUJET 2.0 framework~\cite{Xu:2014ica}.  The
resulting \raa values in central \auau at RHIC and \pbpb at the LHC
for $\pi, D, B$ mesons are shown as a function of \pt in
Figure~\ref{fig:cujet}.  The mass orderings are a convolution of
different initial spectra steepness, different energy loss mechanisms,
and final fragmentation.  Measurements of $D$ mesons to high \pt and
reconstructed beauty-tagged jets at RHIC will provide particularly
sensitive constraints in a range where, due to their large masses, the
charm and beauty quark velocities are not near the speed of light.

\begin{figure}[!hbt]
 \begin{center} \includegraphics[width=0.55\linewidth]{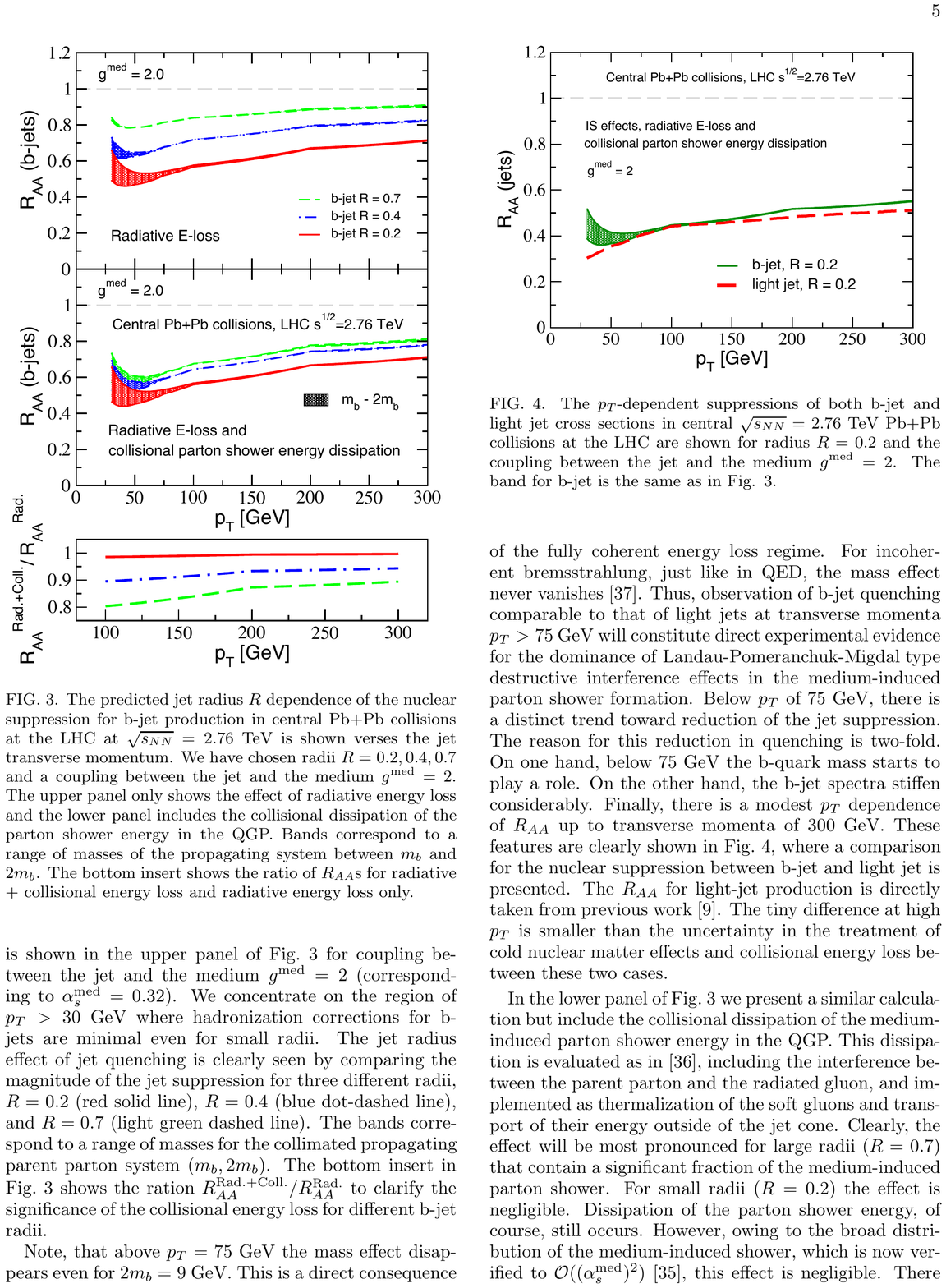}
   \caption[Calculations by Vitev et al. of beauty tagged jets showing
   the sensitivity to radiative and collisional energy loss
   contributions]{Calculations from Ref.~\cite{Huang:2013vaa} are
     shown for beauty tagged jets showing the sensitivity to radiative
     and collisional energy loss contributions.  }
  \label{fig:bsup}
 \end{center}
\end{figure}

Shown in Figure~\ref{fig:bsup} are calculations from
Ref.~\cite{Huang:2013vaa} that highlight the sensitivity of beauty
quark jets to collisional energy loss mechanisms.  Initial
measurements from the LHC show first indications of this mass
ordering, and precision data from higher statistics in future LHC
running and at RHIC are needed.  One expects larger effects at RHIC
where radiative energy loss contributions for the lower \pt beauty
quarks are suppressed.  Another promising tool is the study of heavy
flavor jet-shape modification in \auau relative to \pp collisions.
Different mechanisms of energy loss (radiative versus collisional)
predict different re-distributions of the jet fragments both inside
and outside the jet cone.  There are also scenarios where the heavy
meson forms inside the medium and is dissociated in the
matter~\cite{Adil:2006ra,Sharma:2009hn}.  This would lead to a nearly
unmodified jet shape relative to \pp collisions and a much softer
fragmentation function for the leading heavy meson.

\begin{figure}[!hbt]
 \begin{center}
    \includegraphics[width=0.7\linewidth]{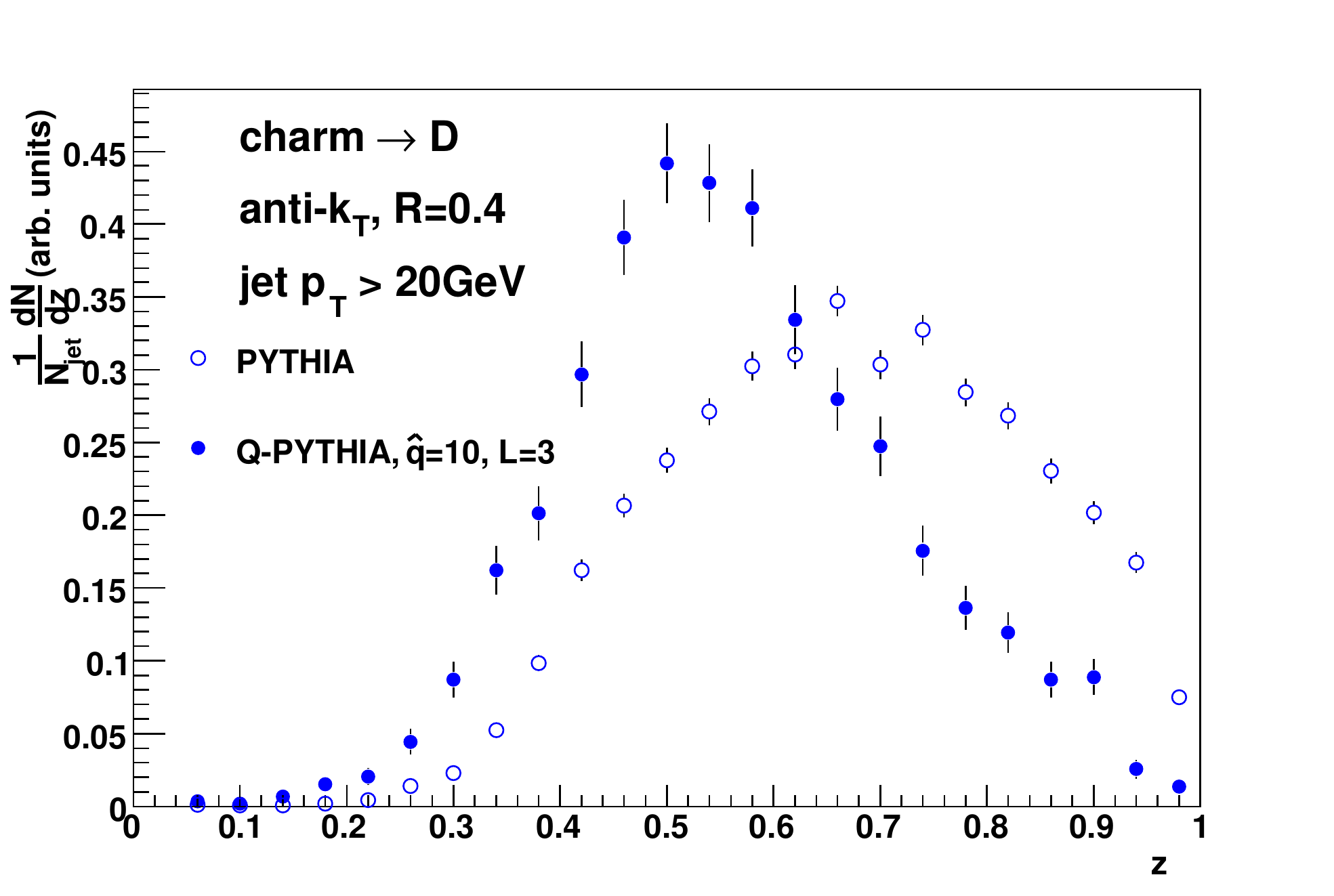}
    \caption[$D$ meson fragmentation function in \pythia and Q-\pythia
    for anti-$k_{T}$ jets with $R = 0.4$ and $E_{T}^\mathrm{jet} > 20$~GeV as
    a function of $z$]{\label{fig:charmfrag} $D$ meson fragmentation
      function in \pythia (open points) and Q-\pythia (solid points)
      for anti-$k_{T}$ jets with $R = 0.4$ and $E_{T}(jet) > 20$~GeV
      as a function of $z$, the fractional momentum of the $D$ meson
      relative to the charm quark.  }
 \end{center}
\end{figure}

Figure~\ref{fig:charmfrag} shows the D meson fragmentation function in
\pythia and Q-\pythia for 20~GeV charm jets.  The peak of the
fragmentation function is shifted in Q-\pythia from $z \approx 0.7$ to
$z \approx 0.5$.  Thus, for a given $p_{T}$, $D$ mesons are more
suppressed than charm jets. Measurement of D mesons within a reconstructed jet
will provide access to fragmentation function modifications with
emphasis on effects at large z.  

The tagging of charm and beauty jets has an extensive history in
particle physics experiments.  There are multiple ways to tag heavy
flavor jets.  First is the method of tagging via the selection of a
high $p_{T}$ electron with a displaced vertex inside the jet.  In
minimum bias \auau collisions at $\sqrt{s_{NN}} = 200$~GeV, the
fraction of inclusive electrons from $D$ and $B$ meson decays is
already greater than 50\% for $p_{T} > 2$~GeV/c.  The sPHENIX tracking
can confirm the displaced vertex of the electron from the collision
point, further enhancing the signal.  Since the semileptonic branching
fraction of $D$ and $B$ mesons is approximately 10\%, this method
provides a reasonable tagging efficiency.  Also, the relative angle of
the lepton with respect to the jet axis provides a useful
discriminator for beauty jets as well, due to the decay kinematics.
Second, the direct reconstruction of $D$ mesons is possible within
sPHENIX as detailed in the performance section.  The third method
utilizes jets with many tracks that do not point back to the primary
vertex.  This technique is detailed by the $D0$ collaboration to
identify beauty jets at the Tevatron~\cite{d0_nim}, and employed with
variations by ATLAS and CMS at the LHC.  This method exploits the fact
that most hadrons with a beauty quark decay into multiple charged
particles all with a displaced vertex.  The detailed performance
metrics for tagged beauty jets are given in Section~\ref{sec:hqjets}.

\begin{figure}[hbt!]
  \centering \includegraphics[width=0.6\textwidth]{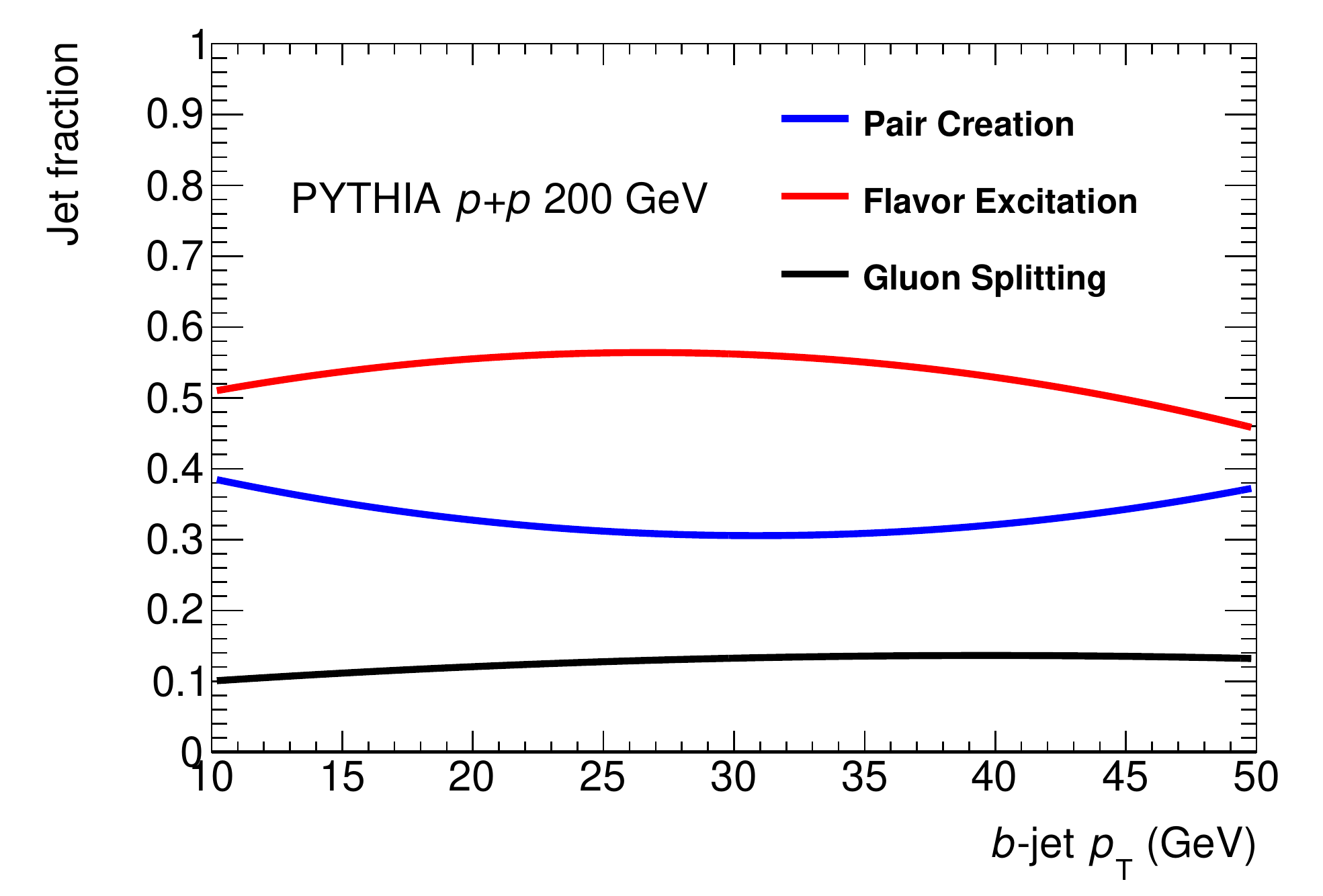}
  \caption[Fraction of inclusive $b$-jets, as a function of jet
  $p_\mathrm{T}$, originating from the pair creation, flavor
  excitation and gluon splitting modes in $\sqrt{s} = 200$~GeV \pythia
  events]{Fraction of inclusive $b$-jets, as a function of jet
    $p_\mathrm{T}$, originating from the pair creation (black), flavor
    excitation (red) and gluon splitting (blue) modes in $\sqrt{s} =
    200$~GeV \pythia events.  \label{fig:bjet_process_fraction}}
\end{figure}

As detailed in Ref.~\cite{Huang:2013vaa}, beauty tagged jets at the
LHC come from a variety of initial processes.  In fact, most often a
tagged beauty jet does not have a back-to-back partner beauty jet.  As
shown in Figure~\ref{fig:bjet_process_fraction}, at RHIC energies the
pair creation process represents $\sim35\%$ of the beauty jet
cross-section, which is a larger fractional contribution than at the
LHC, though flavor excitation still produces $\sim50\%$ of all $b$-jets
at RHIC.
Measurements at RHIC offer a different mixture of initial processes,
and thus kinematics, when looking at correlated back-to-back jets
including heavy flavor tags.

\clearpage

\section{Beauty Quarkonia in the QGP}
\label{sec:quarkonia_introduction}

An extensive program of $J/\psi$ measurements in A$+$A collisions has
been carried out at the SPS ($\sqrt{s_{NN}} = 17.3$~GeV) and RHIC
($\sqrt{s_{NN}} = 200$~GeV) and the LHC ($\sqrt{s_{NN}} =
2.76$~TeV). These measurements were motivated by a desire to observe
the suppression of $J/\psi$ production by color screening in the
QGP. In fact, strong suppression is observed at all three energies,
but it has become clear that the contribution of color screening to
the observed modification can not be uniquely determined without a
good understanding of two strong competing effects.

The first of these, the modification of the $J/\psi$ production cross
section in a nuclear target, has been addressed at RHIC using $d+$Au collisions and 
at the SPS using $p+$Pb collisions, and is being addressed at the LHC using $p+$Pb collisions.
The second complicating effect arises from
the possibility that previously unbound heavy quark pairs could
coalesce into bound states due to interactions with the medium.  This
opens up the possibility that if a high enough density of heavy quark
pairs is produced in a single collision, coalescence of heavy quarks
formed in different hard interactions might actually increase the
production cross section beyond the initial population of bound
pairs~\cite{Zhao:2011cv}.

\begin{figure}
  \begin{center}
    \includegraphics[width=0.6\textwidth]{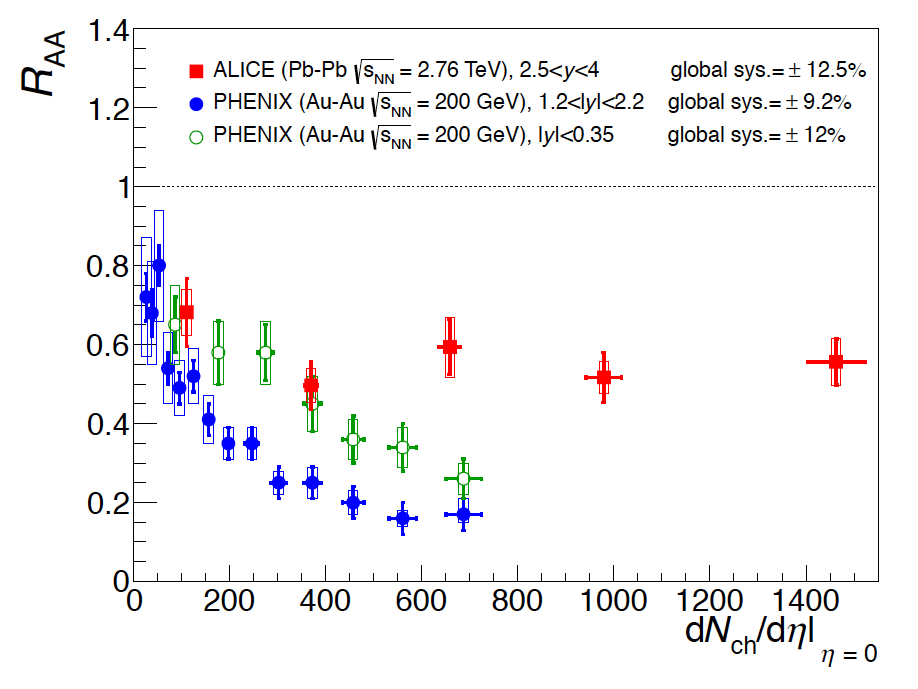}
  \end{center}
  \caption[Comparison of nuclear modification measured by PHENIX and
  ALICE, showing that suppression is much stronger at the lower
  energy]{\label{fig:quarkonia_phenix_alice_comparison} Comparison of
    nuclear modification measured by PHENIX and ALICE, showing that
    suppression is much stronger at the lower
    energy~\cite{Abelev:2012rv}. The modification measured by NA50 at
    low energy is similar to the PHENIX midrapidity result.  }
\end{figure}

Using $p$$+$Pb and $d$$+$Au data as a baseline, and under the
assumption that cold nuclear matter (CNM) effects can be factorized
from hot matter effects, the suppression in central collisions due to
the presence of hot matter in the final state has been estimated to be
about 25\% for \pbpb at the SPS~\cite{Arnaldi:2009ph}, and about 50\%
for \auau at RHIC~\cite{Brambilla:2010cs}, both measured at
midrapidity. The first $J/\psi$ data in \pbpb collisions at
$\sqrt{s_{NN}} = 2.76$~TeV from ALICE~\cite{Abelev:2012rv}, measured
at forward rapidity, are shown alongside PHENIX data in
Figure~\ref{fig:quarkonia_phenix_alice_comparison}. Interestingly, the
suppression in central collisions is far greater at RHIC than at the
LHC. This is qualitatively consistent with a
predicted~\cite{Zhao:2011cv} strong coalescence component due to the
very high $c \overline{c}$ production rate in a central collision at LHC.
There is great promise that, with CNM effects estimated from
$p$$+$Pb data, comparison of these data at widely spaced collision
energies will lead to an understanding of the role of coalescence.

Upsilon measurements have a distinct advantage over charmonium
measurements as a probe of deconfinement in the \qgp. 
The $\Upsilon$(1S), $\Upsilon$(2S) and $\Upsilon$(3S) states can all be
observed with comparable yields via their dilepton decays. Therefore
it is possible to compare the effect of the medium simultaneously on
three bottomonium states---all of which have quite different radii and
binding energies.

At the LHC, CMS has measured Upsilon modification data at midrapidity in \pbpb collisions at 2.76~GeV
that show strong differential suppression of the 2S and 3S states
relative to the 1S state~\cite{Chatrchyan:2012lxa}. ALICE has measured the $\Upsilon(1S)$
modification at forward rapidity in \pbpb collisions at 2.76~GeV~\cite{Abelev:2014nua}, and in $p+$Pb collisions
at 5.02~TeV~\cite{Abelev:2014oea}. With longer \pbpb
runs, and corresponding $p$$+$Pb modification data to establish a CNM baseline, the LHC measurements
will provide an excellent data set within which the suppression of the
three upsilon states relative to $p$$+$Pb can be measured
simultaneously at LHC energies.

At RHIC, upsilon measurements have been hampered by a combination
of low cross sections and acceptance, and insufficient momentum
resolution to resolve the three states. So far, there are 
measurements of the modification of the three states combined in \auau by
PHENIX~\cite{Adare:2014hje} and STAR~\cite{Adamczyk:2013poh}.  However a mass-resolved
measurement of the modifications of the three upsilon states at
$\sqrt{s_{NN}} = 200$~GeV would be extremely valuable for several
reasons.

\begin{figure}
  \begin{center}
    \includegraphics[width=0.6\textwidth]{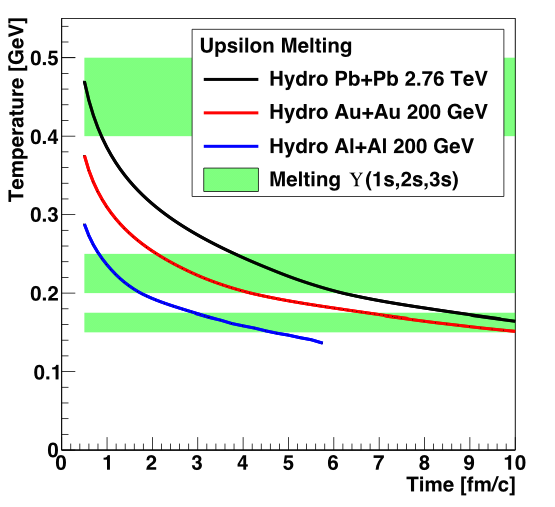}
  \end{center}
  \caption[Hydrodynamic simulations by Habich et al. of temperature vs
  time in \auau and Al$+$Al collisions at 200~GeV and \pbpb collisions
  at 2.76~TeV]{\label{fig:upsilon_time} Temperature as a function of
    time for the central cell in \auau and Al$+$Al collisions at 200
   ~GeV and \pbpb collisions at 2.76 TeV from hydrodynamic simulations
    that include earlier pre-equilibrium dynamics and post hadronic
    cascade~\cite{Habich:2014jna}.  Superimposed are the lattice
    expected dissociation temperatures with uncertainties for the
    three upsilon states.  }
\end{figure}

 First, the core QGP temperature is approximately $2 T_c$ at RHIC at
 1~fm/$c$ and is at least 30\% higher at the LHC (not including the
 fact that the system may thermalize faster)~\cite{muller:2012zq}.
 This temperature difference results in a different color screening
 environment.   Figure~\ref{fig:upsilon_time} shows the temperature as a function
of time for the central cell in \auau and Al$+$Al collisions at 200~GeV
and \pbpb collisions at 2.76 TeV from hydrodynamic simulations that include
earlier pre-equilibrium dynamics and post hadronic cascade~\cite{Habich:2014jna}.  
Superimposed are the lattice expected dissociation temperatures with uncertainties
for the three upsilon states.   The significant lever arm in temperature between
RHIC and LHC, and the use of either centrality or system size, allow one to
bracket the expected screening behavior.

Second, the bottomonium production rate at RHIC is lower
 than that at the LHC by $\sim 100$~\cite{Brambilla:2010cs}.  As a
 result, the average number of $b \overline{b}$ pairs in a central \auau
 collision at RHIC is $\sim 0.05$ versus $\sim 5$ in central \pbpb at
 the LHC. Qualitatively, one would expect this to effectively remove
 at RHIC any contributions from coalescence of bottom quarks from
 different hard processes, making the upsilon suppression at RHIC
 dependent primarily on color screening and CNM effects. This seems to
 be supported by recent theoretical calculations~\cite{Emerick:2011xu}
 where, in the favored scenario, coalescence for the upsilon is
 predicted to be significant at the LHC and small at RHIC.

Finally, it is of interest at RHIC energy to directly compare the modifications of the 
\jpsi~and the $\Upsilon(2S)$ states as a way of constraining the effects of
coalescence by studying two states - in the same temperature environment - that have very similar binding energies and 
radii, but quite different underlying heavy quark populations. 

\begin{figure}
  \begin{center}
    \includegraphics[width=0.45\textwidth, height=0.38\textwidth]{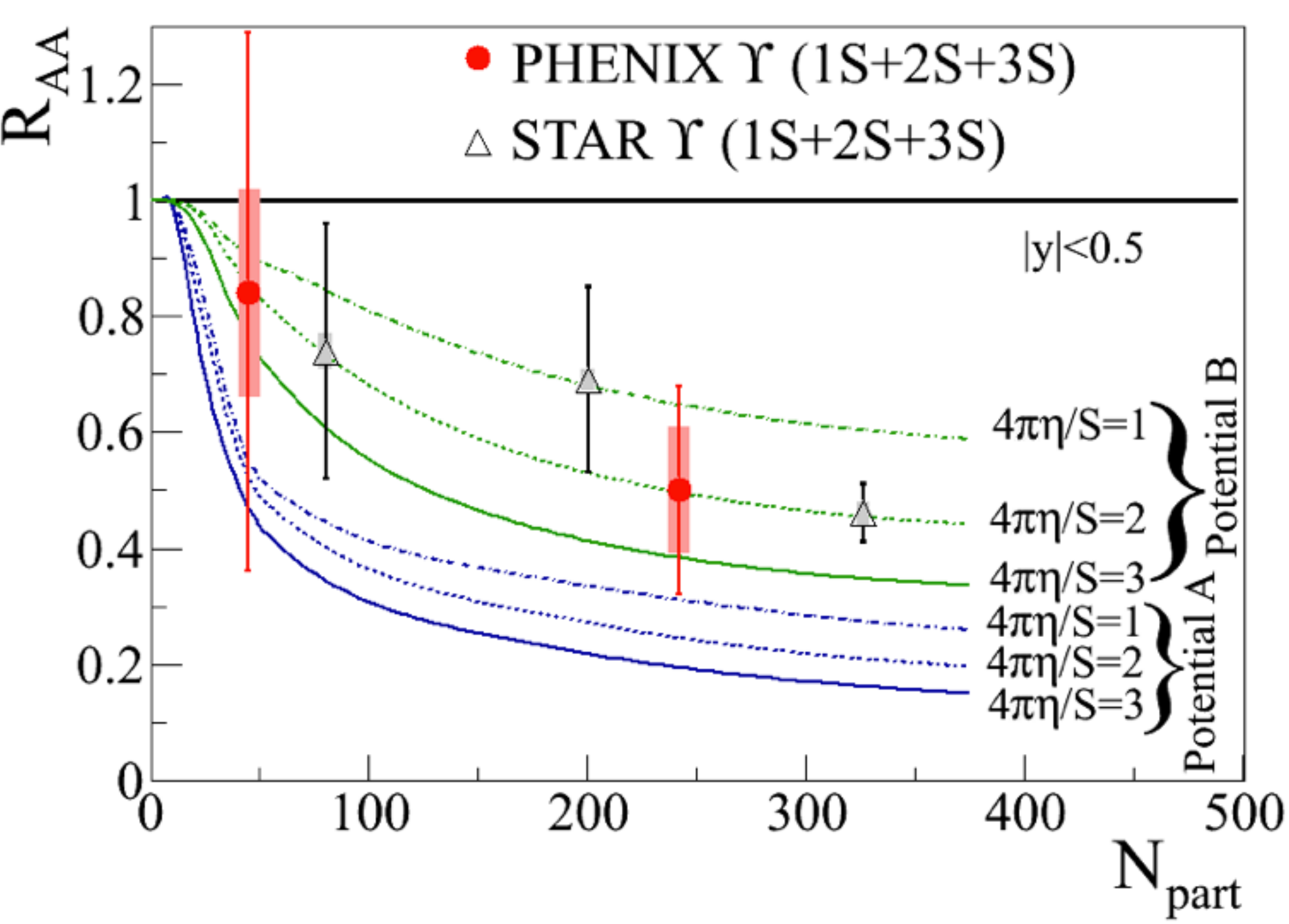}
    \hfill
    \includegraphics[width=0.45\textwidth, height=0.4\textwidth]{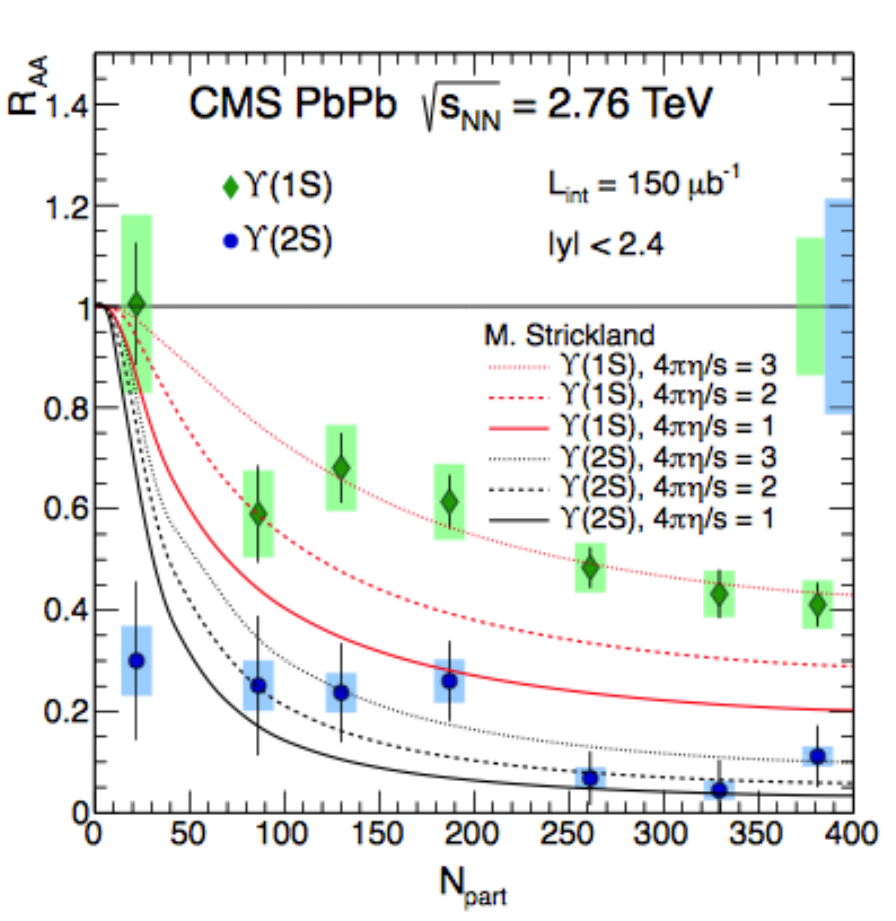}
  \end{center}
  \caption[ Calculations for Upsilon state suppression at RHIC and LHC
  energies vs collision centrality]{\label{fig:upsilon_theory}
    Calculations for Upsilon state suppression at RHIC and LHC
    energies as a function of collision centrality.  The current state
    of measurements are also shown from PHENIX and CMS.}
\end{figure}

 An example theoretical calculation for both RHIC and the LHC is shown
in Figure~\ref{fig:upsilon_theory} indicating the need for substantially
improved precision and separation of states in the temperature range probed at RHIC.

STAR has constructed a Muon Telescope Detector (MTD) to measure muons
at midrapidity~\cite{Ruan:2009ug}. The MTD 
has coverage over $|\eta| < 0.5$, with about 45\% effective
azimuthal coverage. The MTD will have a muon to pion enhancement
factor of 50--100, and the mass resolution will provide a clean
separation of the $\Upsilon$(1S) from the $\Upsilon$(2S+3S), and
likely the ability to separate the $\Upsilon$(2S) and $\Upsilon$(3S)
by fitting. While STAR has already taken data in the 2014 run with the MTD
installed, the upgrade to sPHENIX will provide better mass
resolution and approximately 10 times higher yields per run for
upsilon measurements.   In concert with the expected higher statistics results 
from the LHC experiments, sPHENIX data will provide the required precision to
discriminate models of breakup in the dense matter and the length scale
probed in the medium.

\section{Beauty Quarkonia in proton-nucleus collisions}
\label{sec:pA_quarkonia}

Measurements of quarkonia production in proton-nucleus collisions have
long been considered necessary to establish a cold nuclear matter
baseline for trying to understand hot matter effects in nuclear
collisions. It has become clear, however, that the physics of $p+$A
collisions is interesting in its own
right~\cite{Brambilla:2010cs}. Modification of quarkonia production in
a nuclear target has been described by models that include gluon
saturation effects (see for example~\cite{Eskola:2009uj}), breakup of
the forming quarkonia by collisions with nucleons in the
target~\cite{Arleo:1999af,McGlinchey:2012bp}, and partonic energy loss
in cold nuclear matter~\cite{Arleo:2012rs}.  These mechanisms, which
are all strongly rapidity and collision energy dependent, have been
used, in combination, to successfully describe \jpsi~ and
$\Upsilon(1S)$ data in $p(d)+$A collisions.

The observation of what appears to be hydrodynamic effects in $p$$+$Pb
collisions at the
LHC~\cite{Abelev:2012ola,Aad:2013fja,Chatrchyan:2013nka} and $d$$+$Au
collisions at RHIC~\cite{Adare:2013piz} has raised questions about the
longstanding assumption that $p(d)$$+$A collisions are dominated by cold
nuclear matter effects. For quarkonia, it raises the obvious question:
does the small hot spot produced in the $p(d)$$+$A collision affect the
quarkonia yield?

\begin{figure}
  \begin{center}
    \includegraphics[width=0.6\textwidth]{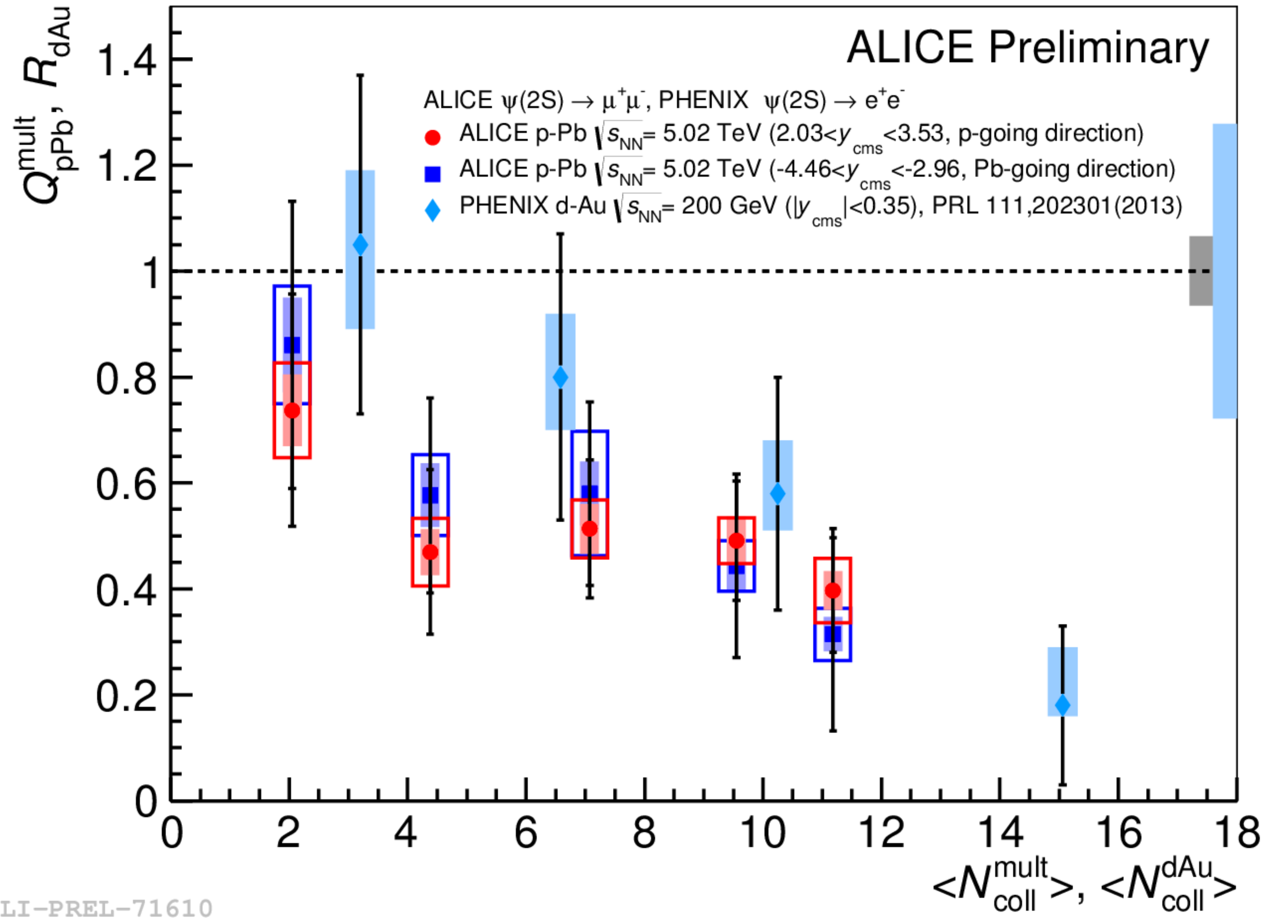}
  \end{center}
  \caption{\label{fig:psiprime_rhic_lhc} Comparison of $\psi'$ nuclear
    modification measured by PHENIX and ALICE, showing very similar
    suppression at RHIC and LHC energies. }
\end{figure}

Recent measurements of the modification of quarkonia excited states in
$p(d)$$+$A collisions have produced unexpected and puzzling
results. An example is shown in Figure~\ref{fig:psiprime_rhic_lhc},
where the centrality dependence of the $\psi^{\prime}$ modification in
$p(d)$$+$A collisions is shown for data measured at midrapidity at
$\sqrt{s_{NN}}=200$~GeV by PHENIX~\cite{Adare:2013ezl}, and
preliminary data at forward and backward rapidity at
$\sqrt{s_{NN}}=5.02$~TeV from ALICE. The suppression versus \ncoll~is
strikingly similar in all three cases, despite the large difference in
collision energy between the PHENIX and ALICE data, and the large
range of rapidities spanned by the three data sets. In two of the
cases --- PHENIX at midrapidity and ALICE at backward rapidity --- the
$\psi^{\prime}$ is much more strongly suppressed than the \jpsi. In
the third case --- ALICE at forward rapidity --- the \jpsi~ and
$\psi^{\prime}$ suppressions are much closer to each other.

The strong differential suppression between the $\psi^{\prime}$ and
\jpsi~ can not be understood as an effect of breakup by collisions
with nucleons in the target, because the time scale of the nuclear
crossing in all of these collisions is too short for the size
difference between the fully formed mesons to become
important. Similarly, shadowing and current models of energy loss in
cold nuclear matter lead to the expectation of similar modification in
$p(d)+$A collisions for the \jpsi~ and $\psi^{\prime}$ (see for
example the detailed discussion in~\cite{Ferreiro:2012mm}).  So
despite the fact that models which combine those effects have been
reasonably successful in describing \jpsi~ data, one must look
elsewhere for an explanation of the strong $\psi^{\prime}$
suppression. A possibility is breakup of the mesons by interactions
with comoving matter (which could be partonic or hadronic) produced in
the collision~\cite{Capella:1996va}. Since the time scale for
interactions with comoving matter is longer than the meson formation
time, this might produce stronger suppression of larger, more weakly
bound states.

The situation has become more interesting with the release of data from CMS on production of 
Upsilon excited states in $p+$Pb collisions. They find that the $\Upsilon(2S)$ to $\Upsilon(1S)$ ratio
is suppressed by about 20\% in minimum bias $p+$Pb collisions, while for the $\Upsilon(3S)$ the differential
suppression in minimum bias  collisions is about 30\%. The effect will be considerably 
larger in the most central collisions, but data showing the 
centrality dependence are not released yet.

A comprehensive $p+$A collision program with sPHENIX will provide Upsilon measurements in $p+$Au
collisions at RHIC energy with all three states resolved from each other. This data set will 
constrain theoretical efforts to understand the physics of $p+$A collisions in the following ways:

\begin{itemize}
\item Provide very precise measurements of the $\Upsilon(1S)$ modification at RHIC energies over 2 units
of rapidity and a broad \pt range that would complement the very precise data at LHC energies that will be 
available by 2023. 
These data for the $\Upsilon(1S)$ (binding energy 1.1~GeV) will, with the LHC data, constrain models of 
shadowing and partonic energy loss in cold nuclear matter.
\item Provide precise measurements at RHIC energies of the modification for the $\Upsilon(2S)$ and 
$\Upsilon(3S)$ states (binding 
energies of 540 and 200 MeV, respectively). Combined with precise data at LHC energies, these data 
will constrain models that 
attempt to explain the differential suppression of these excited states, as well as that of the $\psi^\prime$. 
\end{itemize}

\section{Rates and Physics Reach}
\label{sec:physicscasesummary}

Detailed information about the \qgp properties,
dynamics, time evolution, and structure at 1--2 $T_{c}$ is accessible 
at RHIC through the extensive set of reconstructed jet measurements
proposed here. The theoretical bridgework needed to connect these
measurements to the interesting and unknown medium characteristics 
of deconfined color charges is under active construction by many 
theorists. Combining this work with the flexible and high luminosity
RHIC accelerator facility can produce new discoveries in heavy ion 
collisions with an appropriate set of baseline measurements 
provided a suitable detector apparatus is constructed. Our proposed
design for a jet detector at RHIC that is best able to make use of these
opportunities is given in the following chapter.
Here we highlight the large rate of such events available at RHIC energies.

In order to realize this comprehensive program of jet probes, direct photon
tagged jets, Upsilons and more, one requires very high luminosities and the ability to sample that full 
physics without selection biases.

\begin{figure}[!hbt]
 \begin{center}
    \includegraphics[width=0.7\linewidth]{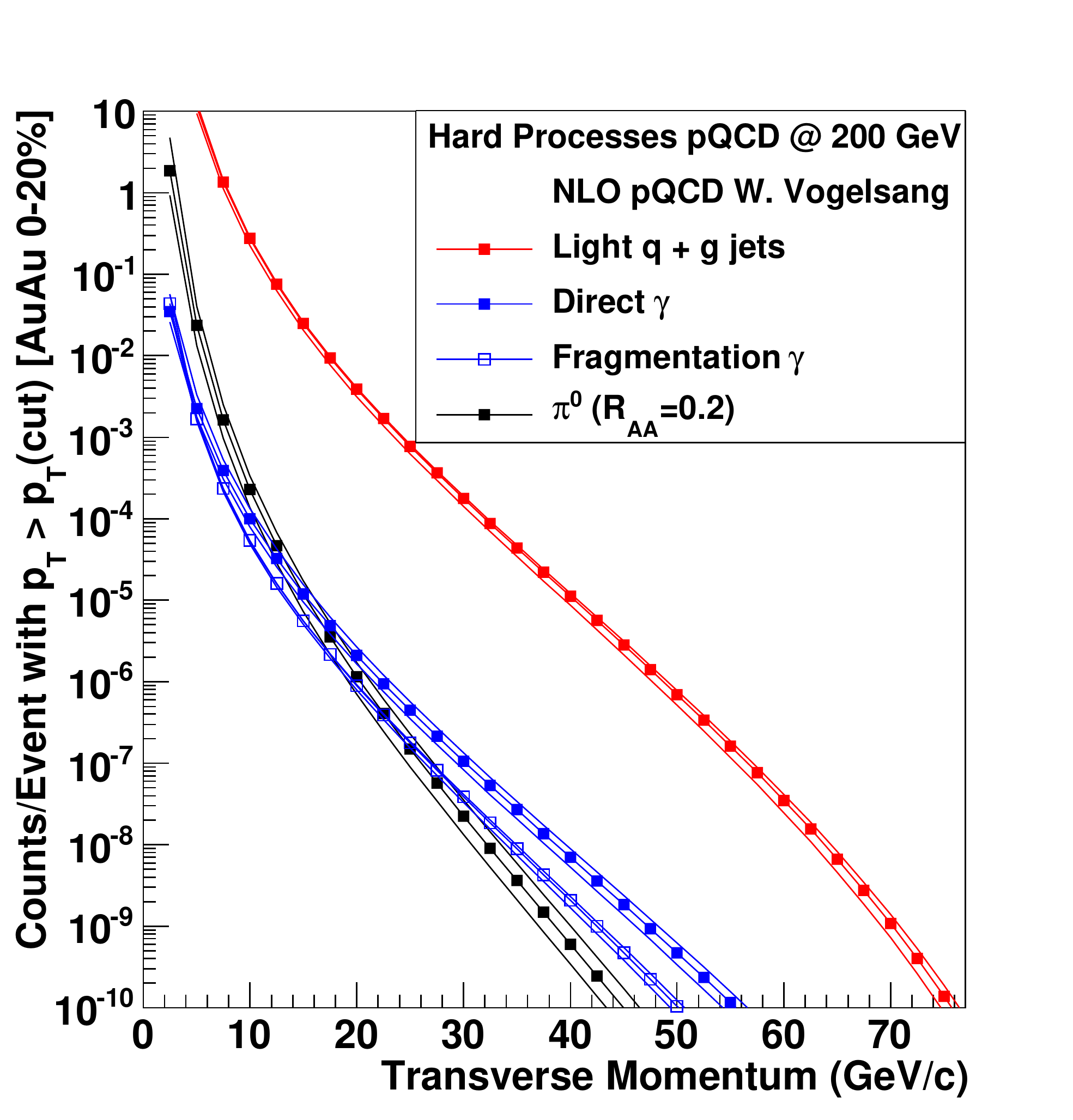} 
    \caption[Jet, photon and $\pi^{0}$ rates for $|\eta|<1.0$ from NLO
    pQCD calculations scaled to \AuAu central collisions for
    $\sqrt{s_{NN}} = 200$~GeV]{\label{fig:nlo_jetrates}Jet, photon and
      $\pi^{0}$ rates for $|\eta|<1.0$ from NLO
      pQCD~\protect\cite{Vogelsang:NLO}{} calculations scaled to
      \AuAu~ central collisions for $\sqrt{s_{NN}} = 200$~GeV .  The
      scale uncertainties on the pQCD calculations are shown as
      additional lines.  Ten billion \AuAu~ central collisions
      correspond to one count at $10^{-10}$ at the bottom of the
      y-axis range. A nominal 22 week RHIC run corresponds to 20
      billion central \auau events.}
 \end{center}
\end{figure}

The inclusive jet yield within $|\eta| < 1.0$ in 0--20\% central \auau
collisions at 200~GeV has been calculated for \pp collisions by
Vogelsang in a Next-to-Leading-Order (NLO) perturbative QCD
formalism~\cite{Vogelsang:NLO} and then scaled up by the expected
number of binary collisions, as shown in
Figure~\ref{fig:nlo_jetrates}.  Also shown are calculation results for
$\pi^{0}$ and direct and fragmentation photon yields.  The bands
correspond to the renormalization scale uncertainty in the calculation
(i.e., $\mu, \mu/2, 2\mu$).

The effect of the completed stochastic cooling upgrade to the RHIC
accelerator~\cite{Fischer:2010zzd} has been incorporated into the RHIC
beam projections~\cite{RHICBeam}.  Utilizing these numbers and
accounting for accelerator and experiment uptime and the fraction of
collisions within $|z| < 10$ cm, the nominal full acceptance range for
the detector, the sPHENIX detector can record 100 billion \auau
minimum bias collisions in a one-year 22 week run.  In fact, with the
latest luminosity projections, for the purely calorimetric jet and
$\gamma$-jet observables with modest trigger requirements, one can
sample 0.6 trillion \auau minimum bias collisions -- see details in
Section~\ref{Section:Rates}.  Note that the PHENIX experiment has a
nearly dead-timeless high-speed data acquisition and trigger system
that has already sampled tens of billions of \auau minimum bias
collisions, and maintaining this high rate performance with the
additional sPHENIX components is an essential design feature.

Figure~\ref{fig:nlo_jetrates} shows the counts per event with $p_T$
larger than the value on the x-axis for the most central 20\% \auau
collisions at $\sqrt{s_{NN}} = 200$~GeV.  With 20 billion events per
RHIC year for this centrality selection, this translates into jet
samples from 20--80~GeV and direct photon statistics out beyond 40~GeV.
It is notable that within the acceptance of the sPHENIX detector, over 80\%
of the inclusive jets will also be accepted dijet events.   The
necessary comparable statistics are available with 10 weeks of \pp and 10 weeks of $p$+$Au$ running.

\begin{figure}[!hbt]
 \centering
 \begin{minipage}[c]{0.70\linewidth}
   \includegraphics[trim = 20 0 45 40, clip, width=\linewidth]{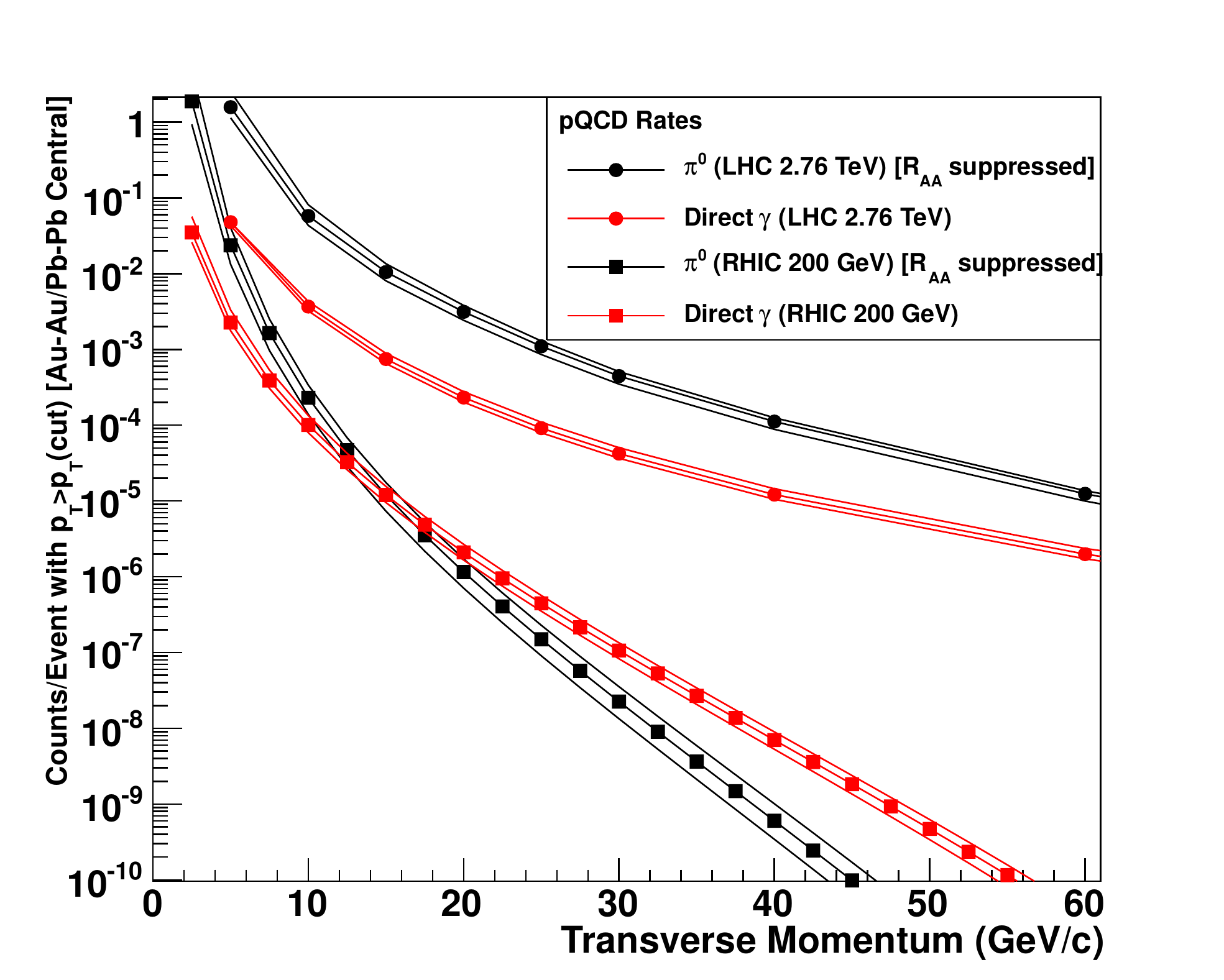}
 \end{minipage}
 \hfill
 \raisebox{0.32cm}{
 \begin{minipage}[c]{0.28\linewidth}
   \includegraphics[trim = 8 0 45 40, clip, width=\linewidth]{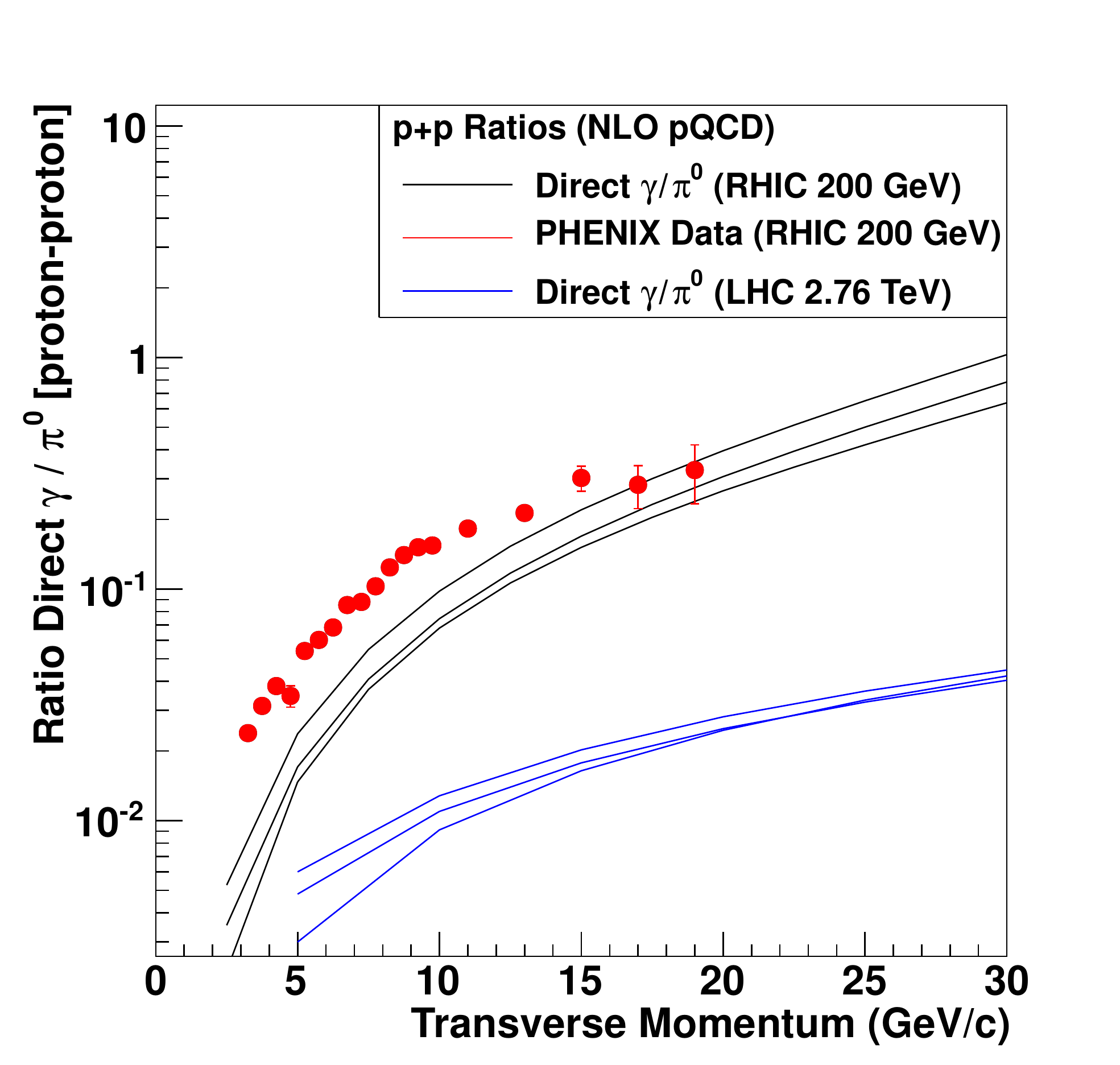}
   \\
   \includegraphics[trim = 8 0 45 40, clip, width=\linewidth]{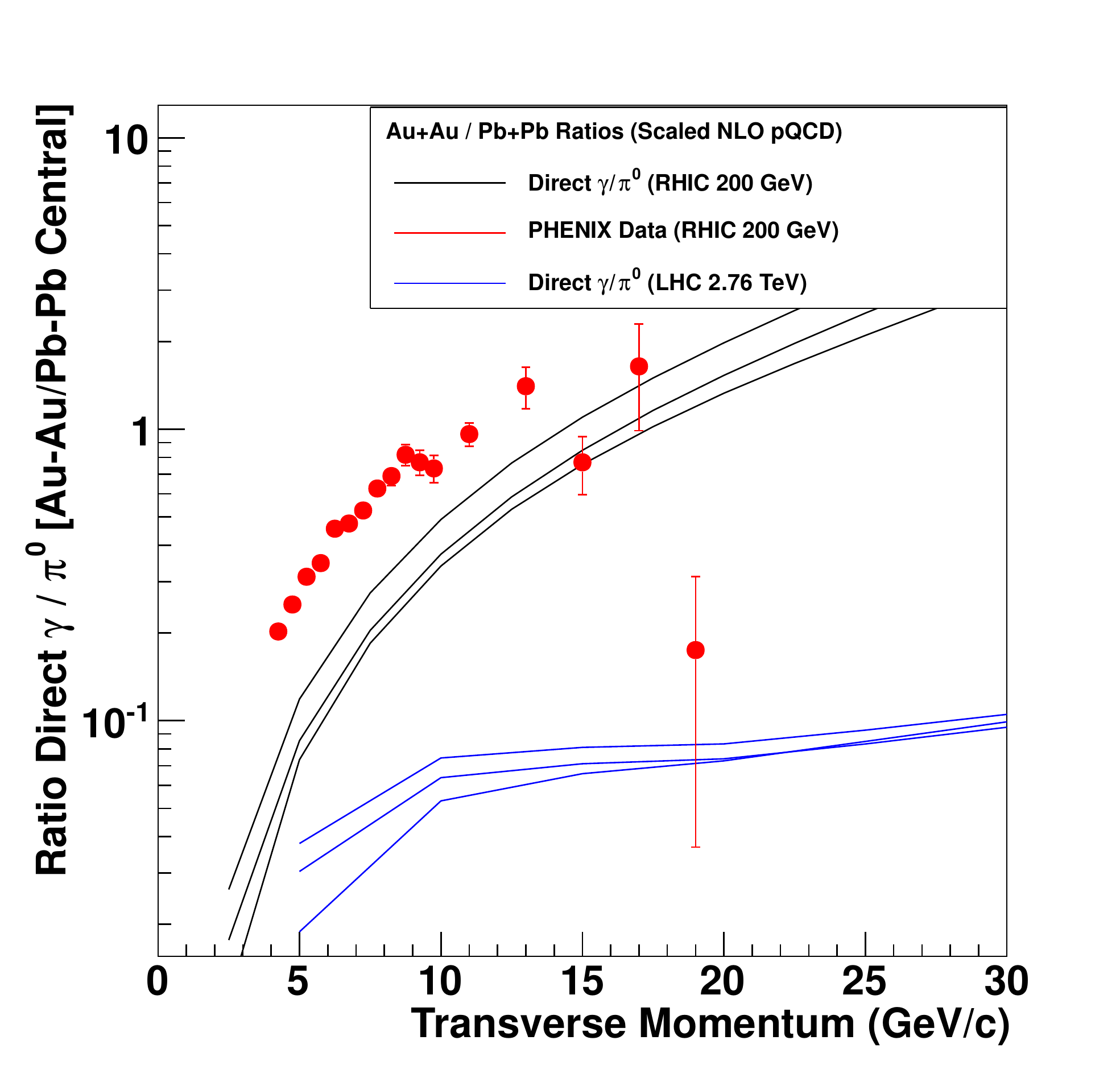}
 \end{minipage}
}
\caption[NLO pQCD calculations of direct photons and $\pi^{0}$ for
RHIC and LHC, compared to PHENIX measurements of direct $\gamma$ to
$\pi^{0}$ ratio in \pp (\auau or \pbpb)
collisions]{\label{fig:nlo_gammarates}NLO pQCD calculations of direct
  photons and $\pi^{0}$ for RHIC and LHC.  The plot on the left shows
  the counts per event in \auau or \pbpb collisions (including the
  measured \raa suppression factor for $\pi^{0}$).  The upper
  (lower) panel on the right shows the direct $\gamma$ to $\pi^{0}$
  ratio in \pp (\auau or \pbpb) collisions, in comparison with
  measurements from the PHENIX experiment at
  RHIC~\cite{Afanasiev:2012dg,Adare:2012yt}.}
\end{figure}    

Measurement of direct photons requires them to be separated from the
other sources of inclusive photons, largely those from $\pi^0$ and
$\eta$ meson decay.  The left panel of Figure~\ref{fig:nlo_gammarates}
shows the direct photon and $\pi^0$ spectra as a function of
transverse momentum for both $\sqrt{s}=200$~GeV and 2.76~TeV \pp
collisions.  The right panels show the $\gamma/\pi^0$ ratio as a
function of $p_T$ for these energies with comparison PHENIX
measurements at RHIC.  At the LHC, the ratio remains below 10\% for
$p_T < 50$~GeV while at RHIC the ratio rises sharply and exceeds one
at $p_T\approx 30$~GeV/c.  In heavy ion collisions the ratio is
further enhanced because the $\pi^0$s are significantly suppressed.
Taking the suppression into account, the $\gamma/\pi^0$ ratio at RHIC
exceeds one for $\pT > 15$~GeV/c.  The large signal to background
means that it will be possible to measure direct photons with the
sPHENIX calorimeter alone, even before applying isolation cuts.
Beyond measurements of inclusive direct photons, this enables
measurements of $\gamma$-jet correlations and $\gamma$-hadron
correlations.

\begin{figure}[p]
  \begin{center}

    \includegraphics[width=0.8\textwidth]{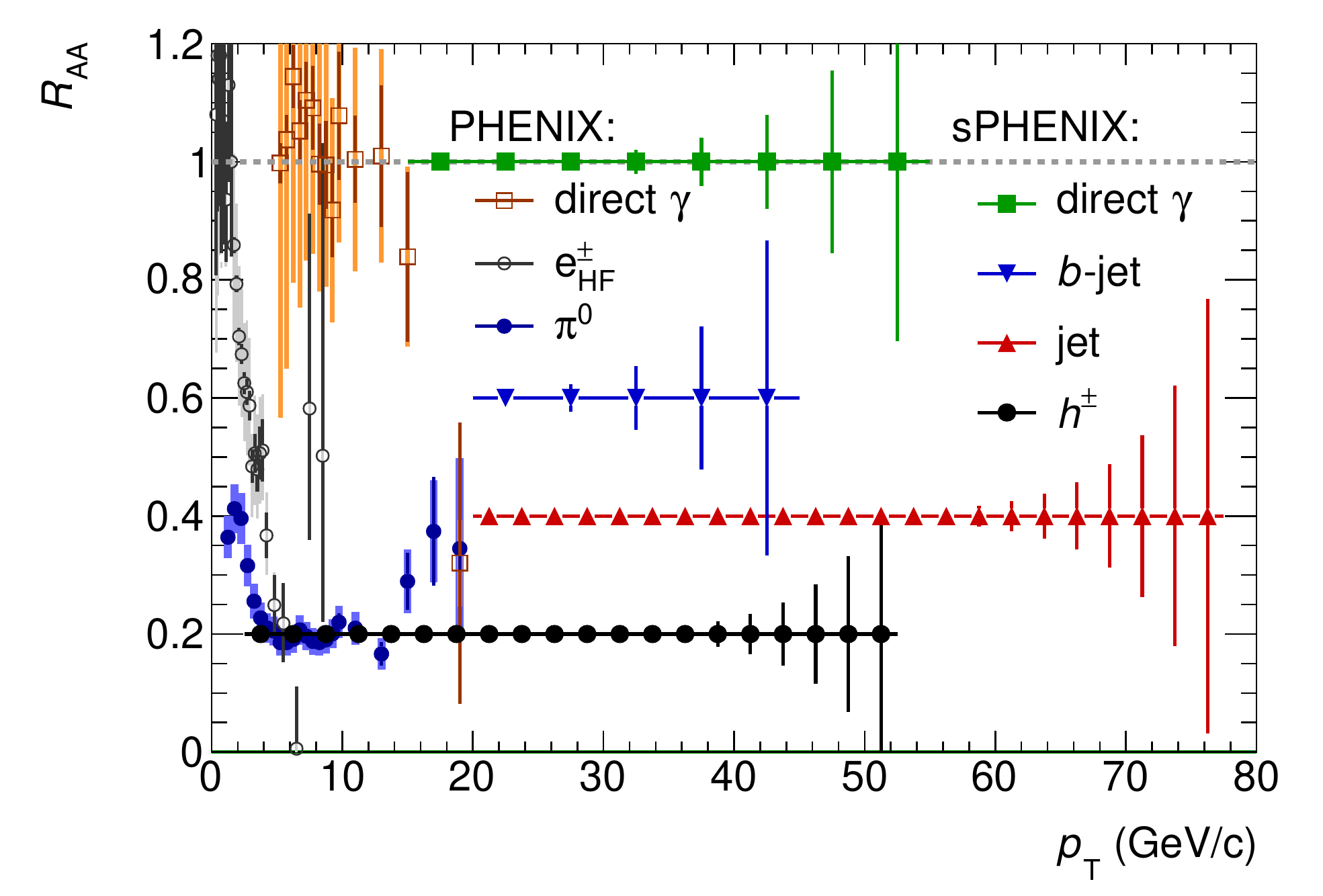}
    \hfill
    \includegraphics[width=0.8\textwidth]{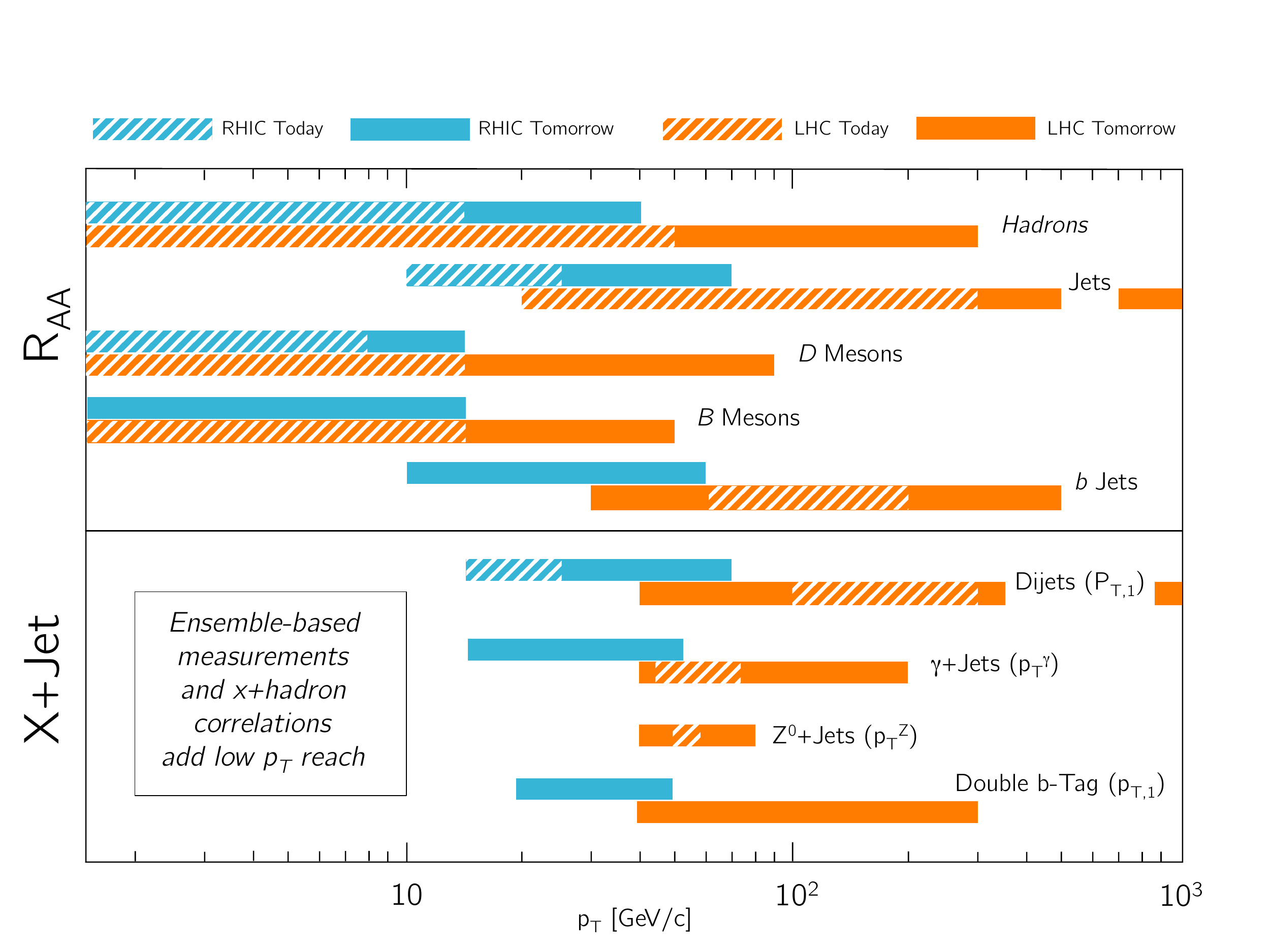}
  \end{center}

  \caption[Statistical projections of $R_\mathrm{AA}$ for hard probes
  in central Au+Au events with the sPHENIX detector after two years of
  data-taking, and kinematic reach of various jet quenching
  observables from previous and future RHIC and LHC
  data-taking]{\label{fig:guntherplot} (Top) Statistical projections
    for the $R_\mathrm{AA}$ of various hard probes vs $p_\mathrm{T}$
    in $0$--$20$\% Au+Au events with the sPHENIX detector after two
    years of data-taking, compared with a selection of current hard
    probes data from PHENIX.  (Bottom) Kinematic reach of various jet
    quenching observables from previous and future RHIC and LHC
    data-taking. Adapted from slides by G. Roland at the QCD Town
    Meeting at Temple University.  }
\end{figure}

Figure~\ref{fig:guntherplot} summarizes the current and future state
of hard probes measurements in A+A collisions in terms of their
statistical reach. The top panel shows the most up to date $R_\mathrm{AA}$ measurements of hard probes in central
Au+Au events by the PHENIX Collaboration (sometimes called the
``T-shirt plot'') plotted against statistical projections for sPHENIX
channels measured after the first two years of data-taking. 
While these existing measurements have greatly expanded our
knowledge of the QGP created at RHIC, the overall kinematic reach is
constrained to $< 20$~GeV even for the highest statistics
measurements. Due to the superior acceptance, detector capability and
collider performance, sPHENIX will greatly expand the previous
kinematic range studied at RHIC energies (in the case of inclusive
jets, the data could extend to $80$~GeV/c, four times the range of the
current PHENIX $\pi^0$ measurements) and will allow access to new
measurements entirely (such as fully reconstructed $b$-tagged jets).

The bottom panel of Figure~\ref{fig:guntherplot}, adapted from slides
shown by G. Roland at the QCD Town Meeting in September 2014, shows
the statistical reach in $p_\mathrm{T}$ for single inclusive measurements
(i.e. the $R_\mathrm{AA}$) and for ``jet+$X$'' correlation
measurements. Although there are some $p_\mathrm{T}$ ranges in common 
between present day measurements at RHIC and the LHC, it can be seen
that the higher kinematic ranges accessed by sPHENIX (referred to in
the figure as ``RHIC Tomorrow'') will have substantially more overlap
with current and future LHC data in a wide variety of channels. Thus
sPHENIX in tandem with the LHC experiments will allow for a detailed
set of measurements of the same observables within the same kinematic
ranges.

\makeatletter{}\chapter{Physics-Driven Detector Requirements}
\label{chap:detector_requirements}

\begin{figure}[hbt!]
 \begin{center}
  \includegraphics[width=0.8\linewidth]{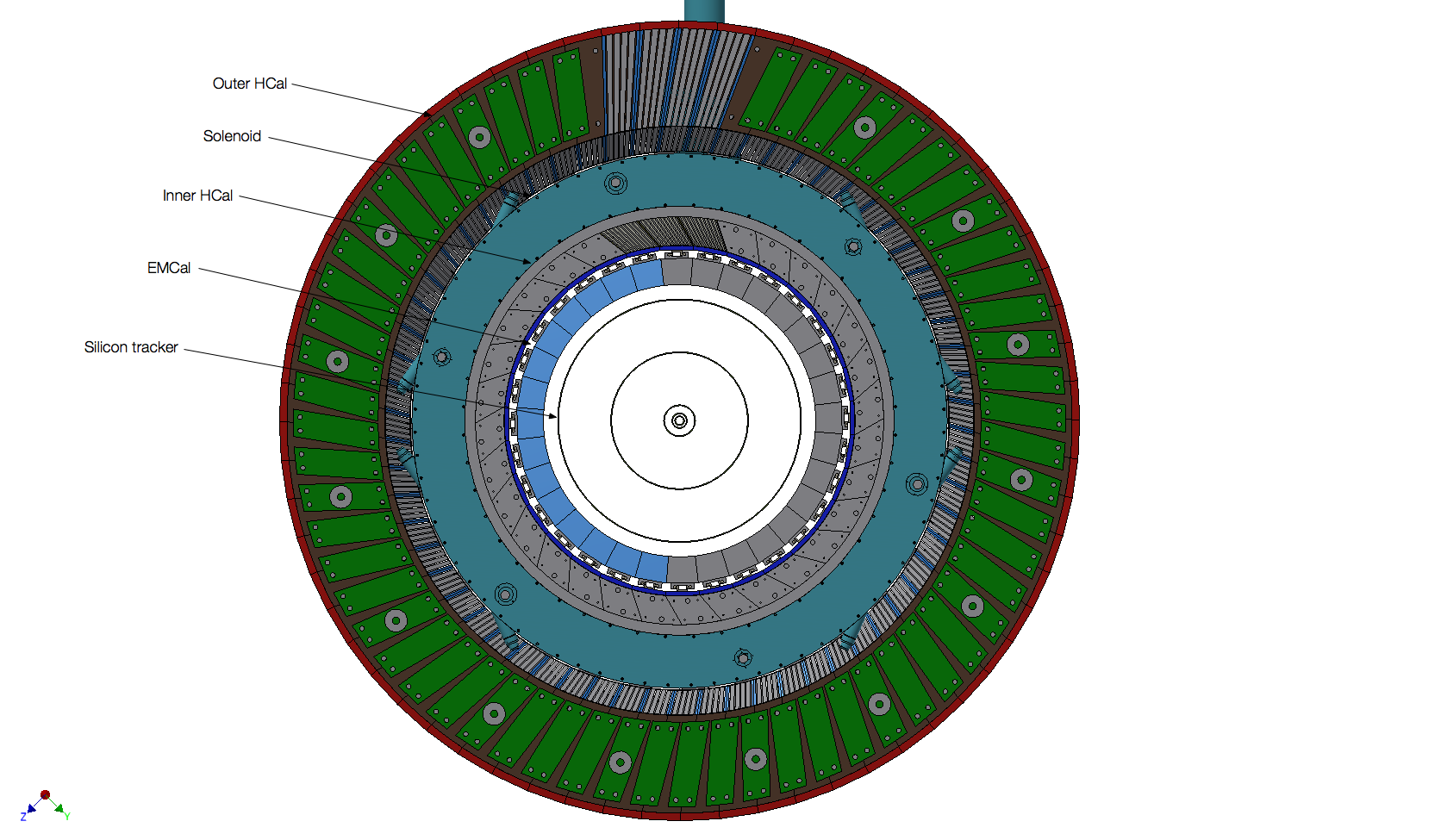}
  \caption[End view of the sPHENIX detector with its component
  subdetectors]{\label{fig:OverviewCutaway} End view of the sPHENIX
    detector with its component subdetectors.}
 \end{center}
\end{figure}

In order to perform the physics measurements outlined in
Chapter~\ref{chap:physics_case}, sPHENIX must satisfy a set of
detector requirements.  In this Chapter we discuss the physics-driven
requirements on the performance of the sPHENIX detector.  In addition,
as outlined in the Executive Summary, this sPHENIX upgrade serves as
the foundation for a future upgrade to a world class \Ephenix, and
those requirements are taken into account.  The details of specific
detector and \geant simulations regarding the physics capability of
the sPHENIX reference design are given in
Chapter~\ref{chap:jet_performance}.  The sPHENIX physics program rests
on several key measurements, and the requirements that drive any
particular aspect of the detector performance come from a broad range
of considerations related to those measurements.  A consideration of
the physics requirements has led to the development of the reference
design shown in Figure~\ref{fig:OverviewCutaway} and this will be described
in detail in Chapter~\ref{chap:detector_concept}.

The primary components of the sPHENIX reference design are as follows.

\begin{description}

\item[Magnetic Solenoid] solenoid built for the BaBar experiment at
  SLAC which became available after the termination of the BaBar
  program.  The cryostat has an inner radius of 140~cm and is 33~cm
  thick, and can produce a central field of 1.5~T.

\item[Silicon Tracking] seven layers of silicon tracking for charged
  track reconstruction and momentum determination.

\item[Electromagnetic Calorimeter] tungsten-scintillating fiber
  sampling calorimeter inside the magnet bore read out with silicon
  photo-multipliers.  The calorimeter has a small Moli\`ere radius and
  short radiation length. allowing for a compact design.

\item[Inner Hadronic Calorimeter] sampling calorimeter of non-magnetic
  metal and scintillator located inside the magnet bore. 

\item[Outer Hadronic Calorimeter] sampling calorimeter of steel
  scintillator located outside the cryostat which doubles as the flux
  return for the solenoid.  

\end{description}

In the following list we provide a high-level mapping between physics
aims and various detector requirements.  The justification for these
requirements is then discussed in more detail in subsequent sections.

\begin{description}

\item[Upsilons] The key to the physics is high statistics \pp, \pA,
  and \hic data sets, with mass resolution and signal-to-background
  sufficient to separate the three states of the $\Upsilon$ family.
  \begin{itemize}
  \item large acceptance ($\Delta\phi = 2\pi$ and $|\eta| < 1$)
  \item high rate data acquisition (15~kHz)
  \item trigger for electrons from $\Upsilon\rightarrow\epem$ ($>90$\% efficiency)
    in \pp and \pA
  \item track reconstruction efficiency $>90$\% and purity $>90$\% for
    $\pT > 3$~GeV/c
  \item momentum resolution of 1.2\% for $\pT$ in the range 4-10~GeV/c.
  \item electron identification with efficiency $>70$\% and charged
    pion rejection of 90:1 or better in central \auau at $\pT =
    4$~GeV/c.
  \end{itemize}

  \hrule

\item[Jets] The key to the physics is to cover jet energies of
  20--70~GeV, for all centralities, for a range of jet sizes, with
  high statistics and performance insensitive to the details of jet
  fragmentation.

  \begin{itemize}
  \item energy resolution $<120\%/\sqrt{E_{\mathrm{jet}}}$ in \pp
    for $R = 0.2$--0.4 jets
  \item energy resolution $<150\%/\sqrt{E_{\mathrm{jet}}}$ in
    central \auau for $R = 0.2$ jets
  \item energy scale uncertainty $<3$\% for inclusive jets
  \item energy resolution, including effect of underlying event, such
    that scale of unfolding on raw yields is less than a factor of
    three
  \item measure jets down to $R=0.2$ (segmentation no coarser than
    $\Delta\eta \times\Delta\phi \sim 0.1 \times 0.1$)
  \item underlying event influence event-by-event (large coverage
    HCal/EMCal) (ATLAS method)
  \item energy measurement insensitive to softness of fragmentation (quarks or gluons) --- HCal + EMCal
  \item jet trigger capability in \pp and \pA without jet bias (HCal
    and EMCal)
  \item rejection ($>95$\%) of high $p_T$ charged track backgrounds (HCal)
  \end{itemize}

\hrule

\item[Dijets] The key to the physics is large acceptance in
  conjunction with the general requirements for jets as above
  \begin{itemize}
  \item $>80$\% containment of opposing jet axis
  \item $>70$\% full containment for $R=0.2$ dijets
  \item \RAA and \aj measured with $<10$\% systematic uncertainty (also
    key in \pA, onset of effects)
  \end{itemize}

\hrule

\item[Fragmentation functions] The key to the physics is unbiased
  measurement of jet energy
  \begin{itemize}
  \item excellent tracking resolution out to $>40$~GeV/$c$ ($dp/p <
    0.2\% \times p$)
  \item independent measurement of $p$ and $E$ ($z=p/E$)
  \end{itemize}

\hrule

\item[Heavy quark jets] The key to the physics is tagging identified
  jets containing a displaced secondary vertex
  \begin{itemize}
  \item precision DCA ($< 100$ microns) for electron $\pT > 4$~GeV/$c$
  \item electron identification for high $\pT > 4$~GeV/$c$
  \end{itemize}

\hrule

\item[Direct photon] The key to the physics is identifying photons
  \begin{itemize}
  \item EMCal resolution for photon ID ($<15\%/\sqrt{E}$)
  \item EMCal cluster trigger capability in \pp and \pA with
    rejections $>100$ for $E_\gamma > 10$~GeV
  \end{itemize}

\hrule

\item[High statistics] Ability to sample high statistics for \pp, \pA,
  \hic at all centralities --- requires high rate, high throughput DAQ
  (10~kHz).

\end{description}

In the following sections, we detail the origin of key requirements.

\section{Acceptance}

The total acceptance of the detector is determined by the requirement
of high statistics jet measurements and the need to fully contain both
single jets and dijets.  To fully contain hadronic showers in the
detector requires both large solid angle coverage and a calorimeter
deep enough to fully absorb the energy of hadrons up to 70~GeV.

\begin{figure}[hbt!]
 \begin{center}
   \raisebox{0.5mm}{\includegraphics[trim = 2 2 2 2, clip, width=0.53\linewidth]{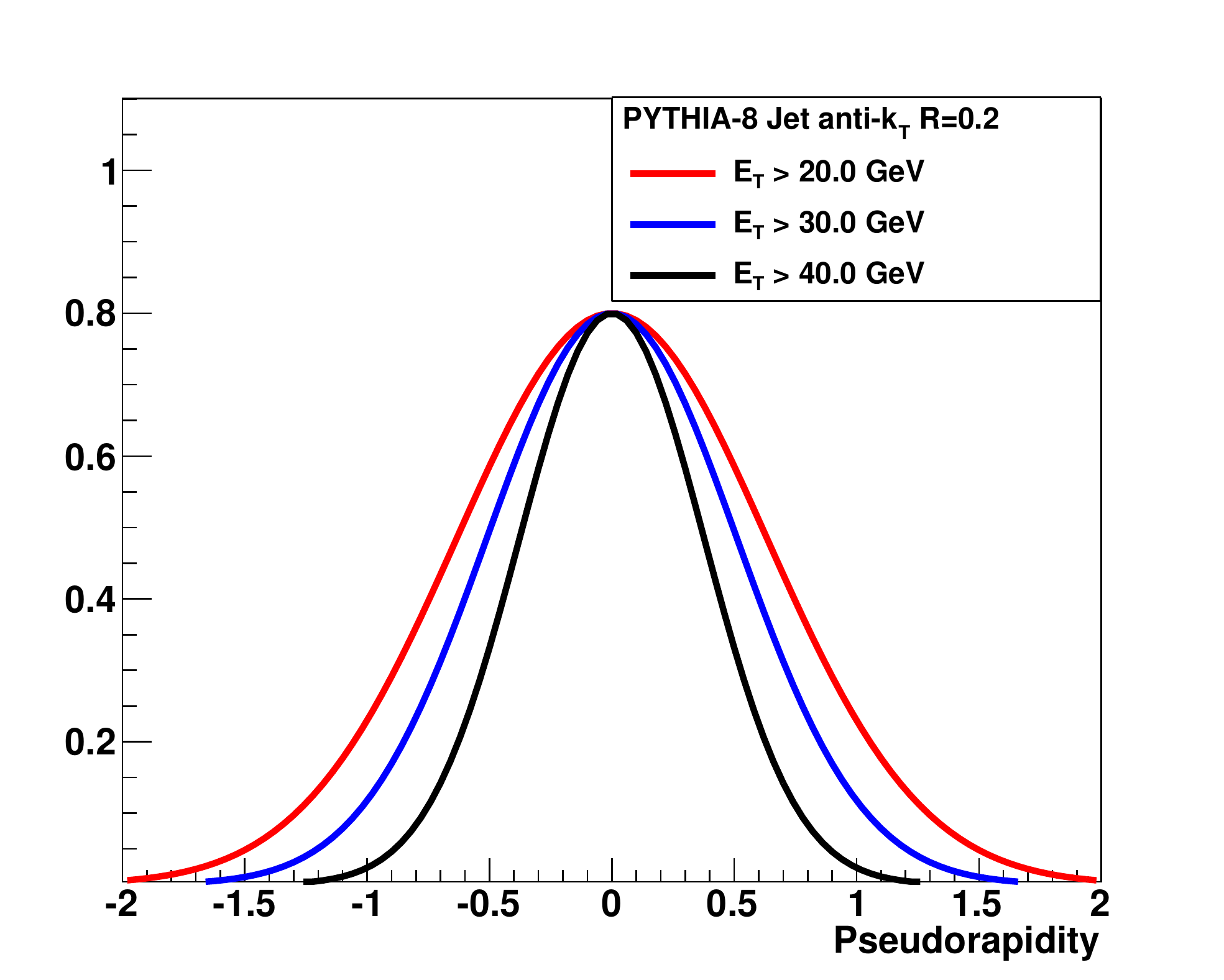}}
  \hfill
  \includegraphics[trim = 2 2 2 2, clip, width=0.45\linewidth]{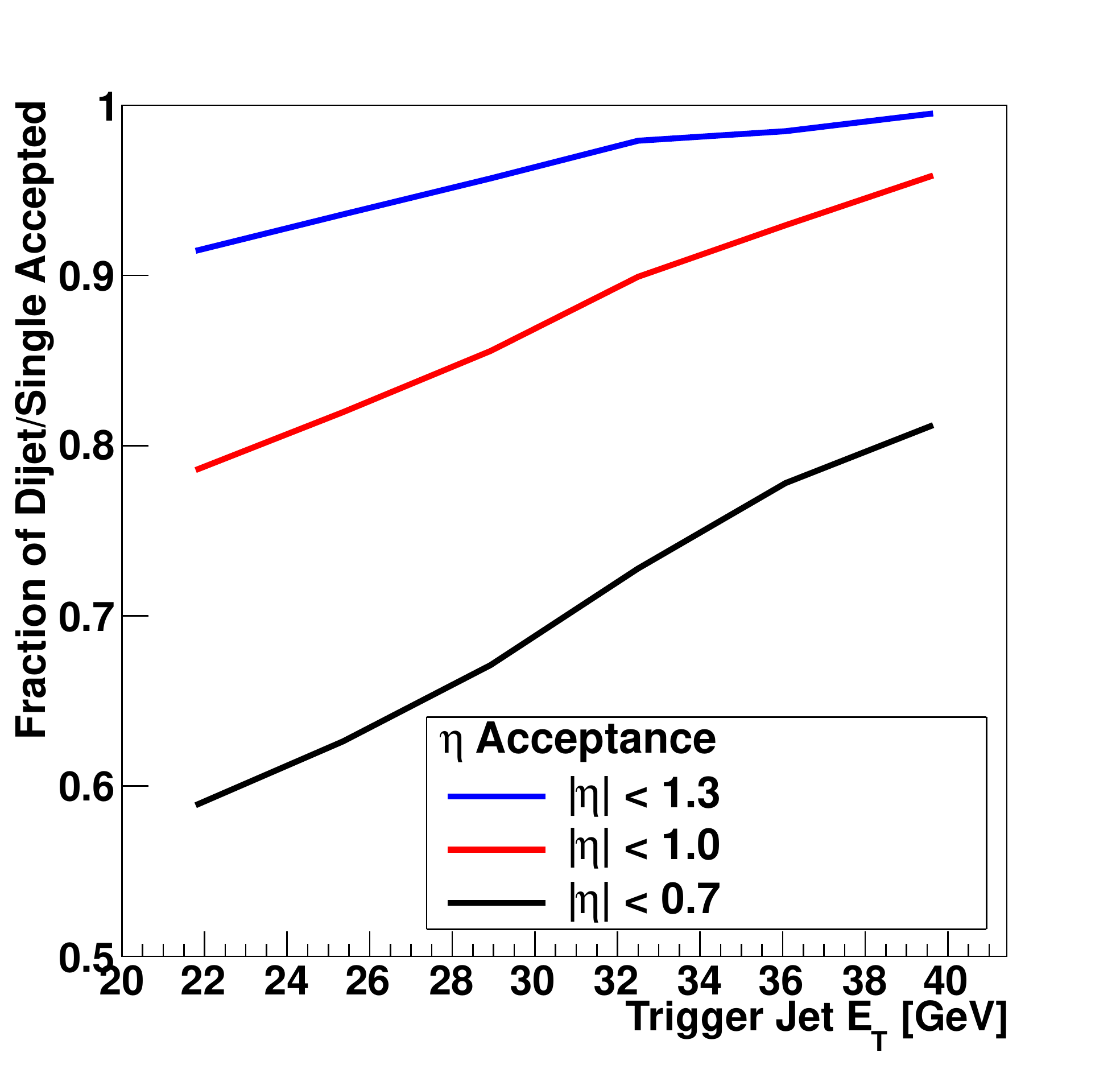}
  \caption[Pseudorapidity distribution of \pythia jets reconstructed
  with the \fastjet anti-k$_{T}$ and the fraction of events in which
  the leading and subleading jet are in the specified
  acceptance]{\label{fig:pythia_dijet_accept}(Left) Pseudorapidity
    distribution of \pythia jets reconstructed with the \fastjet
    anti-k$_{T}$ and R=0.2 for different transverse energy selections.
    (Right) The fraction of \pythia events where the leading jet is
    accepted into a given pseudorapidity range where the opposite side
    jet is also within the acceptance.  Note that the current PHENIX
    acceptance of $|\eta|<0.35$ corresponds to a fraction below 30\%.}
 \end{center}
\end{figure}

\begin{figure}[hbt!]
 \begin{center}
   \includegraphics[width=0.6\linewidth]{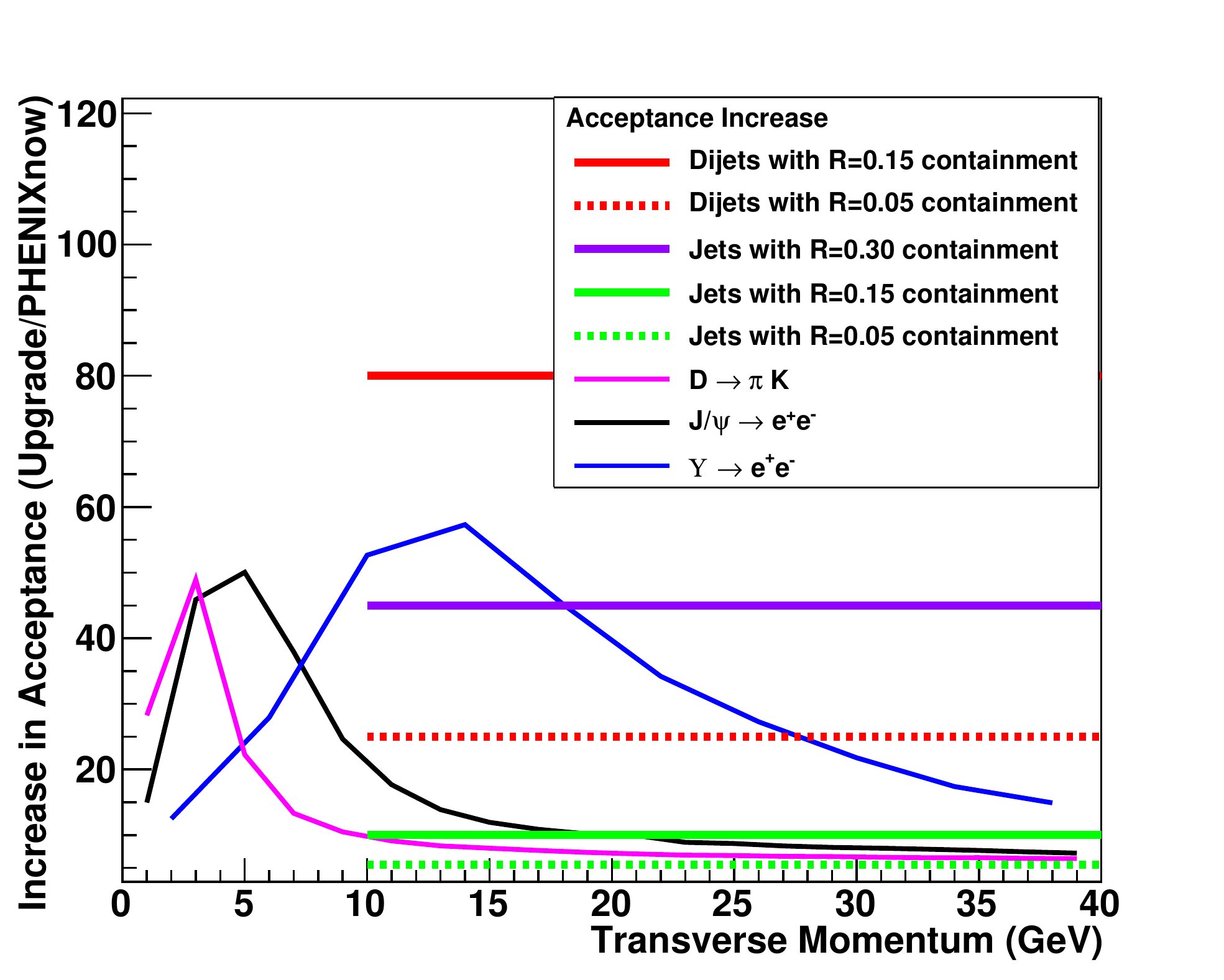}
   \caption[Acceptance increase for various processes for the proposed
   sPHENIX barrel detector compared with the current PHENIX central
   arm spectrometers]{\label{fig:acceptincrease}Acceptance increase
     for various processes (as modeled using the \pythia event
     generator) for the proposed sPHENIX barrel detector compared with
     the current PHENIX central arm spectrometers.}
 \end{center}
\end{figure}

The \pythia event generator has been used to generate a sample of \pp
at 200~GeV events which can be used to demonstrate the pseudorapidity
distribution of jets.  The left panel in
Figure~\ref{fig:pythia_dijet_accept} shows the pseudorapidity
distribution of jets with $E_{T}$ above 20, 30, and 40~GeV.  The right
panel in Figure~\ref{fig:pythia_dijet_accept} shows the fraction of
events where a trigger jet with $E_{T}$ greater than a given value
within a pseudorapidity range has an away side jet with $E_{T} >
5$~GeV accepted within the same coverage.  In order to efficiently
capture the away side jet, the detector should cover $|\eta|<1$, and
in order to fully contain hadronic showers within this fiducial
volume, the calorimetry should cover slightly more than that.  Given
the segmentation to be discussed below, the calorimeters are required
to cover $|\eta|<1.1$.

It should be noted that reduced acceptance for the away-side jet
relative to the trigger suffers not only a reduction in statistics for
the dijet asymmetry and $\gamma$-jet measurements but also results in a
higher contribution of low energy \fake jets (upward fluctuations
in the background) in those events where the away side jet is out of
the acceptance.   For the latter effect, the key is that both jet axes
are contained within the acceptance, and then events can be rejected
where the jets are at the edge of the detector and might have partial
energy capture.  

Compared to the current PHENIX acceptance (the central arms cover
$|\eta|<0.35$ and $\Delta \phi = \pi$), full azimuthal coverage with
$|\eta|<1.1$ results in a very substantial increase in the acceptance
of single jets and an even larger increase in the acceptance of dijets
for other observables including heavy quarkonia states, as shown in
Figure~\ref{fig:acceptincrease}.  The large acceptance and high rate
are key enablers of the physics program detailed in
Chapter~\ref{chap:physics_case}.

\section{Segmentation}

Jets are reconstructed from the four-vectors of the particles or
measured energies in the event via different algorithms (as described
in Chapter~\ref{chap:jet_performance}), and with a typical size $R =
\sqrt{\Delta\phi^{2} + \Delta\eta^{2}}$.  In order to reconstruct jets
down to radius parameters of $R = 0.2$ a segmentation in the hadronic
calorimeter of $\Delta \eta \times \Delta \phi = 0.1 \times 0.1$ is
required.  The electromagnetic calorimeter segmentation should be
finer as driven by the measurement of direct photons for
$\gamma$-jet correlation observables.  The compact electromagnetic
calorimeter design being considered for sPHENIX has a Moli\`ere
radius of $\sim 15$~mm, and with a calorimeter at a radius of about
100~cm, this leads to an optimal segmentation of $\Delta \eta \times
\Delta \phi = 0.024 \times 0.024$ in the electromagnetic section.

\section{Energy Resolution}

The requirements on the jet energy resolution are driven by
considerations of the ability to reconstruct the inclusive jet spectra
and dijet asymmetries and the fluctuations on the \fake jet background
(as detailed in Chapter~\ref{chap:jet_performance}.  The total jet
energy resolution is typically driven by the hadronic calorimeter
resolution and many other effects including the bending of charged
particles in the magnetic field out of the jet radius.  Expectations
of jet resolutions approximately 1.2 times worse than the hadronic
calorimeter resolution alone are typical (see a more detailed
discussion in Chapter~\ref{chap:jet_performance}).

In a central \AuAu event, the average energy within a jet cone of
radius $R = 0.2$ ($R = 0.4$) is approximately 10~GeV (40~GeV)
resulting in an typical RMS fluctuation of 3.5~GeV (7~GeV).  This sets
the scale for the required reconstructed jet energy resolution, as a
much better resolution would be dominated by the underlying event
fluctuations regardless.  A measurement of the jet energy for $E =
20$~GeV with $\sigma_{E} = 120\% \times \sqrt{E} = 5.4$~GeV gives a
comparable contribution to the underlying event fluctuation.  A full
study of the jet energy resolution with a \geant simulation of the
detector configuration is required and is presented in
Chapter~\ref{chap:jet_performance}.

Different considerations set the scale of the energy resolution
requirement for the EMCal. The jet physics requirement is easily met
by many EMCal designs.  For the direct $\gamma$-jet physics, the
photon energies being considered are $E_{\gamma} > 10$~GeV where even
a modest $\sigma_{E}/E = 12\%/\sqrt{E}$ represents only a blurring of
400~MeV.  In \auau central events, the typical energy in a $3 \times
3$ tower array is also approximately 400~MeV.  These values represent a
negligible performance degradation for these rather clean photon
showers even in central \auau events.

Most of these physics measurements require complete coverage over a
large range of rapidity and azimuthal angle ($\Delta\eta\le$ 1.1 and
$\Delta\phi$ = 2$\pi$) with good uniformity and minimal dead area. The
calorimeter should be projective (at least approximately) in $\eta$.
For a compact detector design there is a trade-off in terms of
thickness of the calorimeter and Moli\`ere radius versus the sampling
fraction and, therefore, the energy resolution of the device.  Further
optimization if these effects will be required as we work towards a
final design.

\section{Tracking}

The requirements on tracking capabilities are tied to three particular
elements of the sPHENIX physics program: fragmentation functions at
high and at low $z$, heavy flavor tagged jets, and the measurement of
the upsilon family of quarkonia states.

In order to utilize the available luminosity fully, the tracking
systems should have large, uniform acceptance and be capable of fast
readout.  Measuring fragmentation functions at low $z$ means looking
for possibly wide angle correlations between a trigger jet and a
charged hadron.  This places only moderate requirements on the
momentum resolution ($\Delta p/p \simeq 1\%\cdot p$), but reinforces
the requirement of large acceptance.  

Fragmentation functions at high $z$ place more stringent requirements
on momentum resolution.  In order to unfold the full fragmentation
function, $f(z)$, the smearing due to momentum uncertainty should
be very small compared to the corresponding smearing due to the
calorimetric jet measurement for a cleanly identified jet.  For a 40~GeV
jet this condition is satisfied by a tracking momentum resolution of
$\Delta p/p \simeq 0.2\%\cdot p$ or better.

The measurement of the $\Upsilon$ family places the most stringent
requirement on momentum resolution.  The large mass of the upsilon
means that one can focus primarily on electrons with momenta of
$\sim4-10$~GeV/c.  The $\Upsilon(3S)$ has about 3\% higher mass than the
$\Upsilon(2S)$ state; to distinguish them clearly one needs invariant
mass resolution of $\sim$100 MeV, or $\sim1$\%.  This translates into a momentum resolution
for the daughter $e^\pm$ of $\sim$ 1.2\% in the range $4-10$~GeV/c.  

The $\Upsilon$ measurement also generates requirements on the purity
and efficiency of electron identification.  The identification needs
to be efficient because of the low cross section for $\Upsilon$
production at RHIC, and it needs to have high purity against the
charged pion background to maintain a good signal to background ratio.
Generally speaking, this requires minimizing track ambiguities by 
optimizing the number of tracking layers, their spacing, and the segmentation 
of the strip layers. 
Translating this need into a detector requirement can be done only by performing
detailed simulations with a specific tracking configuration, 
followed by evaluation of the tracking performance.
This will be discussed in detail in
Section~\ref{sec:tracking}.

Tagging heavy-flavor jets introduces the additional tracking
requirement of being able to measure the displaced vertex of a D or B
meson decay, as described in Section~\ref{sec:hqjets}.  The $c\tau$
for D and B decays is 123~$\mu$m and 457~$\mu$m, respectively, and
the displaced vertex would need to be identified with a resolution
sufficient to distinguish these decays against backgrounds.

\section{Triggering}

The jet energy should be available at the Level-1 trigger as a
standard part of the PHENIX dead-timeless Data Acquisition and Trigger
system.  This triggering ability is important as one requires high
statistics measurements in proton-proton, proton-nucleus, light
nucleus-light nucleus, and heavy nucleus-heavy nucleus collisions with
a wide range of luminosities.  It is important to have combined EMCal
and HCal information available so as to avoid a specific bias on the
triggered jet sample.

\makeatletter{}\chapter{Detector Concept}
\label{chap:detector_concept}

\begin{figure}[hbt!]
  \centering
     \includegraphics[trim = 190 100 150 100, clip, width=0.8\linewidth]{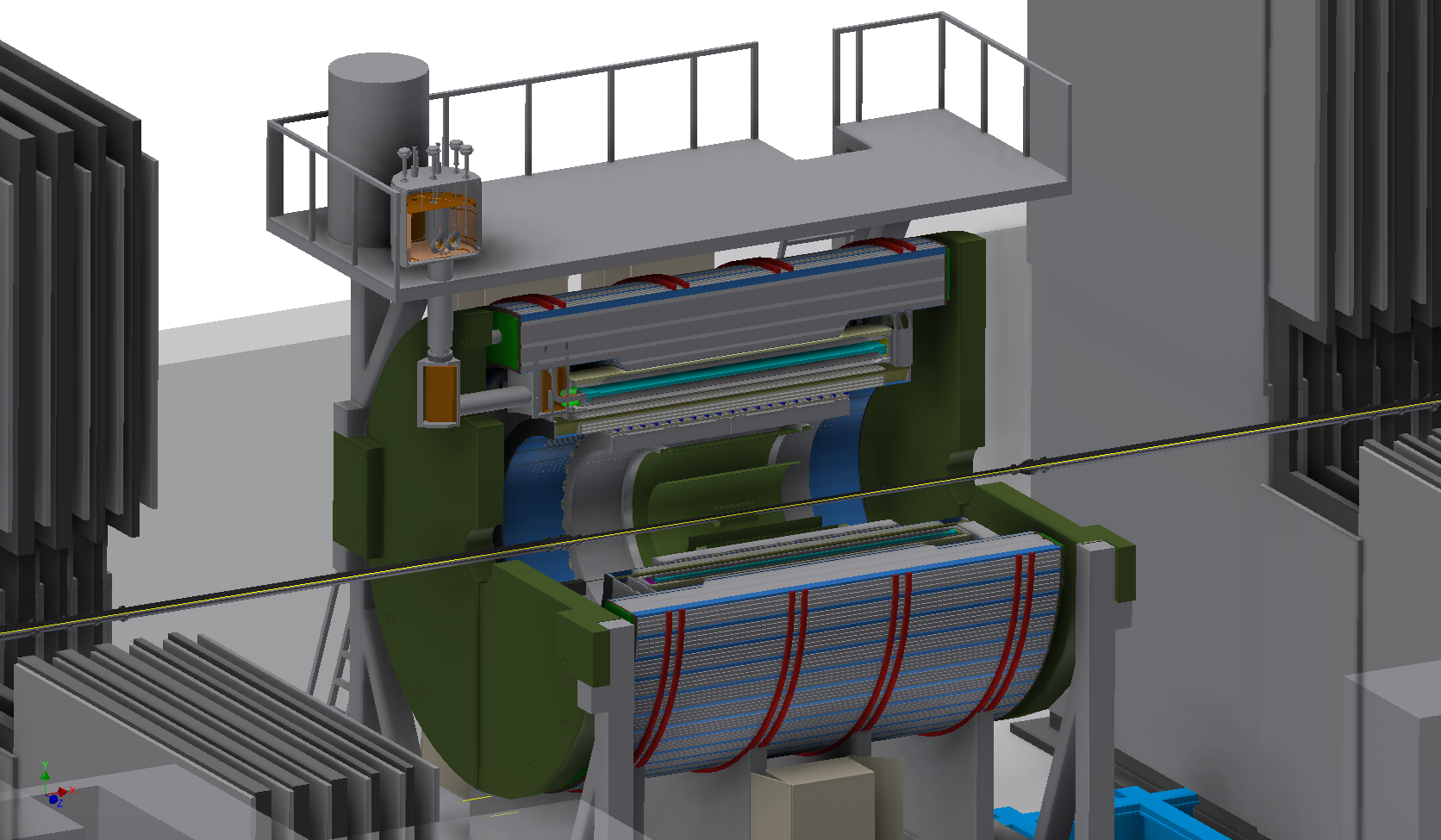}
         \caption[Engineering drawing of sPHENIX, showing the
     superconducting solenoid containing the electromagnetic
     calorimeter and surrounded by the hadronic calorimeter, with a
     model of the associated support structure]{An engineering drawing
       of sPHENIX, showing the superconducting solenoid containing the
       electromagnetic calorimeter and surrounded by the hadronic
       calorimeter, with a model of the associated support structure,
       as it would sit in the PHENIX IR.}
  \label{fig:fsphenix_big}
\end{figure}

In this Chapter we detail the sPHENIX detector design including the
magnetic solenoid, electromagnetic and hadronic calorimeters, silicon tracking, and
readout electronics.  Detector performance specifications are checked
using a full \geant simulation of the detector.  Full physics
performance measures are detailed in
Chapter~\ref{chap:jet_performance}.

The sPHENIX detector concept takes advantage of technological
developments to enable a compact design with excellent performance.  A
tungsten-scintillator electromagnetic calorimeter read out with
silicon photomultipliers (SiPMs) or avalanche photodiodes (APDs)
allows for a physically thin device which can operate in a magnetic
field, without the bulk of photomultiplier tubes and the need for high
voltage distribution.  The thinner electromagnetic calorimeter allows
space for the first longitudinal segment of the hadronic calorimeter
to sit inside the bore of the solenoid, with positive implications for
electron identification and reducing the overall size of the
calorimeter system.  The use of solid-state sensors for the hadronic
calorimeter allows for nearly identical electronic readout for the two
major systems.  A superconducting magnet coupled with high resolution
tracking detectors provides excellent momentum resolution inside the
solenoid.  The detector has been designed from the beginning to
minimize the number of distinct parts to be simpler to manufacture and
assemble.  The use of components insensitive to magnetic fields
enables the hadronic calorimeter to double as the flux return for the
solenoid, reducing both mass and cost.  Adapting existing electronic
designs for the readout allows for reduced development cost and risk,
and leverages a decade and a half of experience at PHENIX.  We now
detail each subsystem in the following Sections.

A number of alternative designs of both the electromagnetic and
hadronic calorimeter have been investigated by means of simulation as
well as construction of prototype devices which have demonstrated the
feasibility of the approach.  Work continues to optimize and simplify
the design and manufacture of the calorimeter, but we have chosen a
reference design of the technologies used in the calorimeters which
has been used to confirm that the design can achieve the physics goals
of the experiment.  The design discussed in this chapter is identical
to the concept used in the simulations shown in this proposal.

\makeatletter{}\section{Magnet}
\label{sec:magnet}

The magnet and tracking system should ultimately be capable of order 1\%
momentum resolution at 10\,GeV/c, cover the full $2\pi$ in azimuth and
$\left| \eta \right| < 1.1$.  
The BaBar solenoid is a good match to the requirements, became available 
in late 2012, and measures were taken to transfer ownership of the coil
and related equipment to Brookhaven in early 2013.

\renewcommand{\arraystretch}{1.9}
\addtolength{\tabcolsep}{-0.5pt}
\begin{table}[hbt!]
\caption{Key characteristics of the BaBar solenoid and cryostat.}
\centering
\begin{tabular}{lr}
\toprule
Central field in BaBar & 1.5 T \\
Cryostat inner radius & 140 cm \\
Cryostat outer radius & 173 cm \\
Cryostat length & 385 cm \\
Mean radius of windings & 153 cm \\
Coil length & 351 cm \\
Material thickness at normal incidence & $\sim 126$ mm Al \\
Operating current & 4596 A \\
Manufacturer & Ansaldo Energia (now ASG) \\
\bottomrule
\end{tabular}
\label{babar-tbl}
\end{table}

The main features of the BaBar solenoid are shown in Table~\ref{babar-tbl}.
The solenoid fits well into the mechanical infrastructure of the existing PHENIX 
interaction region (IR).
The RHIC beamline is 444.8\,cm
above the tracks that are used to move detectors into the collision
hall and 523.2\,cm above the floor, and we propose to keep the track
system in place for maneuvering detectors in and out.  
The hadron calorimeter which serves as the flux return for the magnet 
is about 100\,cm thick, so the outer radius of the 
hadronic calorimeter is about 150 cm above the tracks which provides
adequate clearance for support.
Instrumentation in the forward and
backward direction is not part of this proposal
but the space
available is approximately the same as the present muon tracker
systems.  

The BaBar magnet and related equipment, including the power supply, 
the quench protection electronics,
the dump resistor, rigging fixtures, and some cryogenic components have been removed
from the decommissioned BaBar detector and are in staging areas at SLAC.
The coil in its transfer frame have been surveyed for residual radiation and have been
found to be acceptable to move to Brookhaven.
\begin{figure}[htb!]
 \begin{center}
    \includegraphics[width=0.6\linewidth]{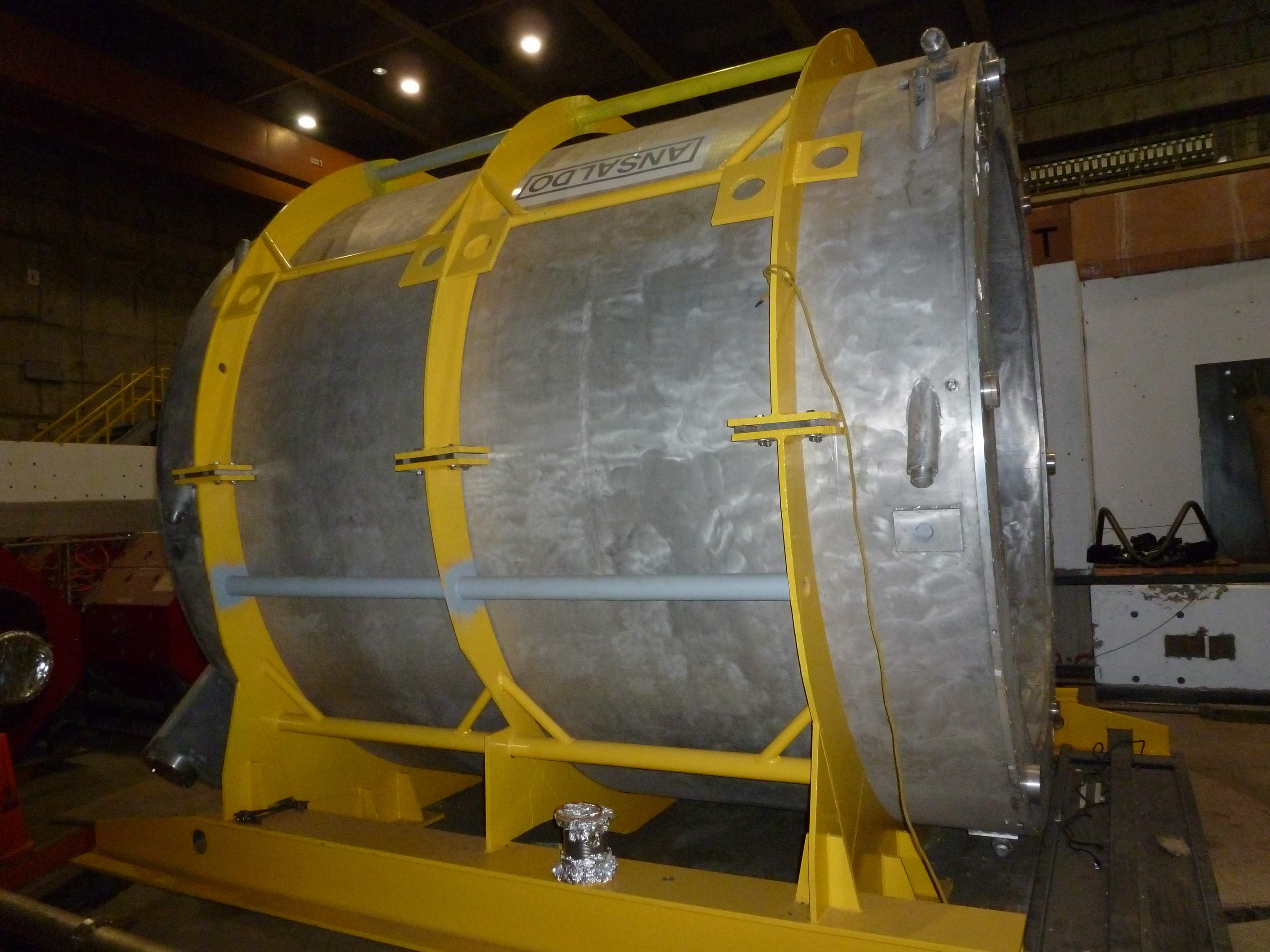}
    \caption{The BaBar solenoid in its transfer frame for shipping at SLAC in May, 2013.}
    \label{fig:babar-endstationa} 
 \end{center}
\end{figure}
The BaBar solenoid has been prepared for shipping, and is shown in its transfer frame
in Figure~\ref{fig:babar-endstationa}.

\subsection{Magnetic Field Calculations}

\begin{figure}[htb!]
 \begin{center}
    \includegraphics[width=0.6\linewidth]{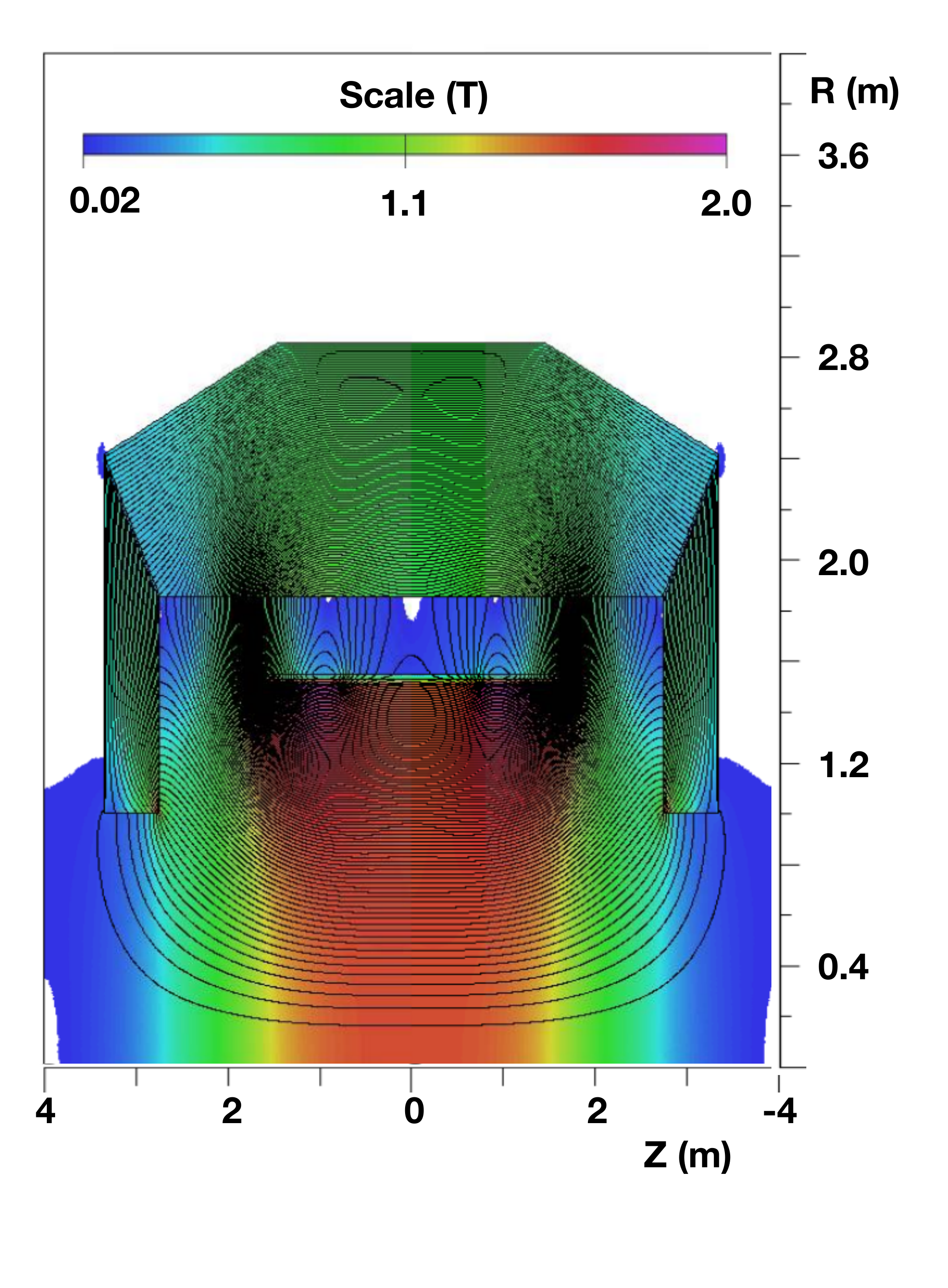}
    \caption{Calculation of the magnetic field from the solenoid with
      the flux returned by the hadronic calorimeter.}  
    \label{fig:field-map} 
 \end{center}
\end{figure}

Magnetic field calculations of the solenoid coil and a model of the
return steel were carried out with the OPERA magnetic field simulation
software package.  A field map is shown in Figure \ref{fig:field-map}.
Tools are under development for complete three dimensional field
calculations and calculations of the forces on the detector and flux
return.

\makeatletter{}\section{The Electromagnetic Calorimeter}
\label{sec:emcal}

The concept for the sPHENIX electromagnetic calorimeter follows from
the physics requirements outlined earlier in this proposal.  These
requirements lead to a calorimeter design that is compact (i.e. has a
small Moli\`{e}re radius and short radiation length), has a high
degree of segmentation ($0.024\times 0.024$ in $\eta$ and $\phi$), 
has small dead area,
and can be built at a reasonable cost. Since the calorimeter will be
located inside the solenoid cryostat, it will also have to
operate in a high magnetic field.  
A number of alternative designs
have been investigated and work continues to optimize and simplify the design
and manufacture of the calorimeter, but we have 
chosen a reference design.

\subsection{Segmentation and readout}

The segmentation of the calorimeter is determined by a number of
different requirements. One primary factor is the occupancy of the
individual readout towers in heavy ion collisions, which determines
the ability to resolve neighboring showers and to measure the energy
in the underlying event.  In addition, the degree of segmentation also
determines the ability to measure the transverse shower shape, which
is used in separating electrons from hadrons (e/$\pi$ rejection). 
The segmentation chosen
for the reference design of the  detector will provide the 
capability to perform the
physics program of this proposal.

The calorimeter will be divided into individual towers corresponding
to a segmentation in $\eta$ and $\phi$ of approximately
$0.024\times0.024$ and would result in about 25,000 readout channels 
(256 in $\phi \times$ 96 in $\eta$). 
The light can be collected at the front or the back of the calorimeter with 
short light guides forming 
towers measuring $\sim$ 2$\times$2 cm.
A design goal is to 
to minimize the radial space required by the light guide, SiPM,
readout electronics, and cables.

\begin{figure}[htb!]
 \begin{center}
    \includegraphics[width=0.4\linewidth]{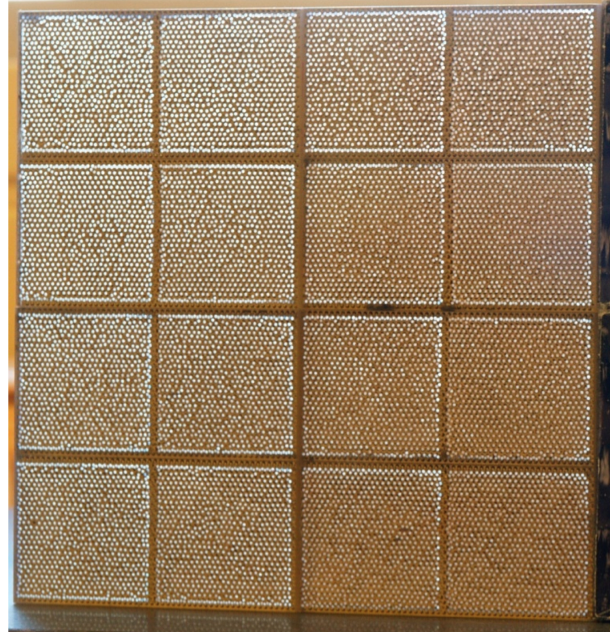}
    \caption[View of a prototype electromagnetic calorimeter module
    with fibers embedded before light guides are installed]{View of a
      prototype calorimeter module with fibers embedded before light
      guides are installed. (Figure courtesy of Oleg Tsai, EIC-RD1)}
    \label{fig:spacal_endview} 
 \end{center}
\end{figure}

The reference design for the electromagnetic calorimeter, which
satisfies the physics requirements of sPHENIX and the requirements of
an experiment at an electron-ion collider is a sampling calorimeter
with tungsten powder absorber and scintillating fibers constructed
with techniques developed at UCLA~\cite{Tsai:2012cpa}. A calorimeter
with 0.47 mm diameter fibers on 1 mm centers has a final density of
10.2 g/cm$^3$ and a radiation length of 7 mm which implies a
Moli\`{e}re radius of about 2.3 cm.  A calorimeter 18 radiation
lengths thick occupies 12.6 cm in radius, and with light collection,
sensors, preamps, and cables, the calorimeter is expected to occupy
radial space of about 25 cm.  Figure~\ref{fig:spacal_endview} shows an
end view of $4 \times 4$ towers with the end of the scintillating
fibers visible, before light guides are installed.

\begin{figure}[htb!]
 \begin{center}
    \includegraphics[width=0.7\linewidth]{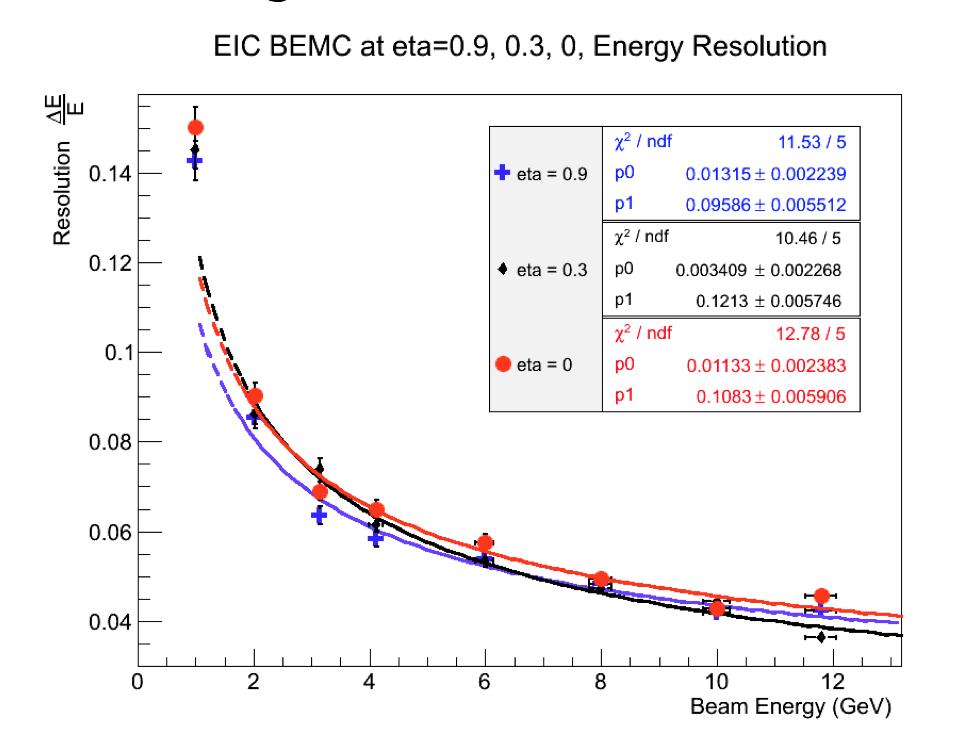}
    \caption[Energy resolution achieved in beam tests of a prototype
    electromagnetic calorimeter compared with \geant
    simulation]{Energy resolution achieved in beam tests of a
      prototype electromagnetic calorimeter compared with \geant
      simulation. (Figure courtesy of Oleg Tsai, EIC-RD1)}
    \label{fig:spacal_res} 
 \end{center}
\end{figure}

A key element of the design of the calorimeter is the light output of the scintillator available
to the photodetectors. 
There must be sufficient light produced by the scintillator at all energies of interest so that
photostatistics do not degrade the resolution of the calorimeter.
Measured light yields of $\sim$ 500 photoelectrons/GeV of incident energy
have been demonstrated with SiPM readout in beam tests.
The resulting energy resolution is less than about 12\%/$\sqrt{E}$ at the energies and angles
relevant to the calorimeter.
Figure~\ref{fig:spacal_res} shows the measured resolution of a prototype calorimeter compared with simulation
at three incident angles.

The reference design of the electromagnetic calorimeter is projective only in the azimuthal
direction; the calorimeter modules are expected to be wedges in $\phi$.
There are ongoing Monte Carlo simulations in conjunction with manufacturing feasibility studies
to study the costs and benefits of projectivity
at large pseudorapidity in jet reconstruction.

The PHENIX collaboration has worked with the company Tungsten Heavy Powder~\cite{THP}
on the design and fabrication of actual calorimeter components
with funding from a Phase I Small Business Innovation Research (SBIR) grant to study
and develop materials and components for compact tungsten based calorimeters for nuclear physics
applications.  Research and development, as part of a broader collaboration, has also been
supported by
a ``Joint Proposal to Develop Calorimeters for the Electron Ion Collider'' for EIC 
research and development funds (EIC-RD1).

\makeatletter{}\section{The Hadronic Calorimeter}
\label{sec:hcal}

The hadronic calorimeter is a key element of sPHENIX and many of the
overall performance requirements are directly tied to performance
requirements of the HCal itself.  The focus on measuring jets and
dijets in sPHENIX leads to a requirement on the energy resolution of
the calorimeter system as a whole---the particular requirement on the
HCal is that it have a single particle energy resolution better than
$\sigma_E/E = 100\%/\sqrt{E}$.

The jet measurement requirements also lead to a transverse
segmentation requirement of $\Delta \eta \times \Delta \phi \approx
0.1 \times 0.1$ over a rapidity range of $|\eta|<1.1$ with minimal
dead area and a longitudinal segmentation satisfactory for electron
identification and measurements of the structure of energy flow in the
underlying event.

\begin{figure}[htb!]
 \begin{center}
    \includegraphics[width=0.7\linewidth]{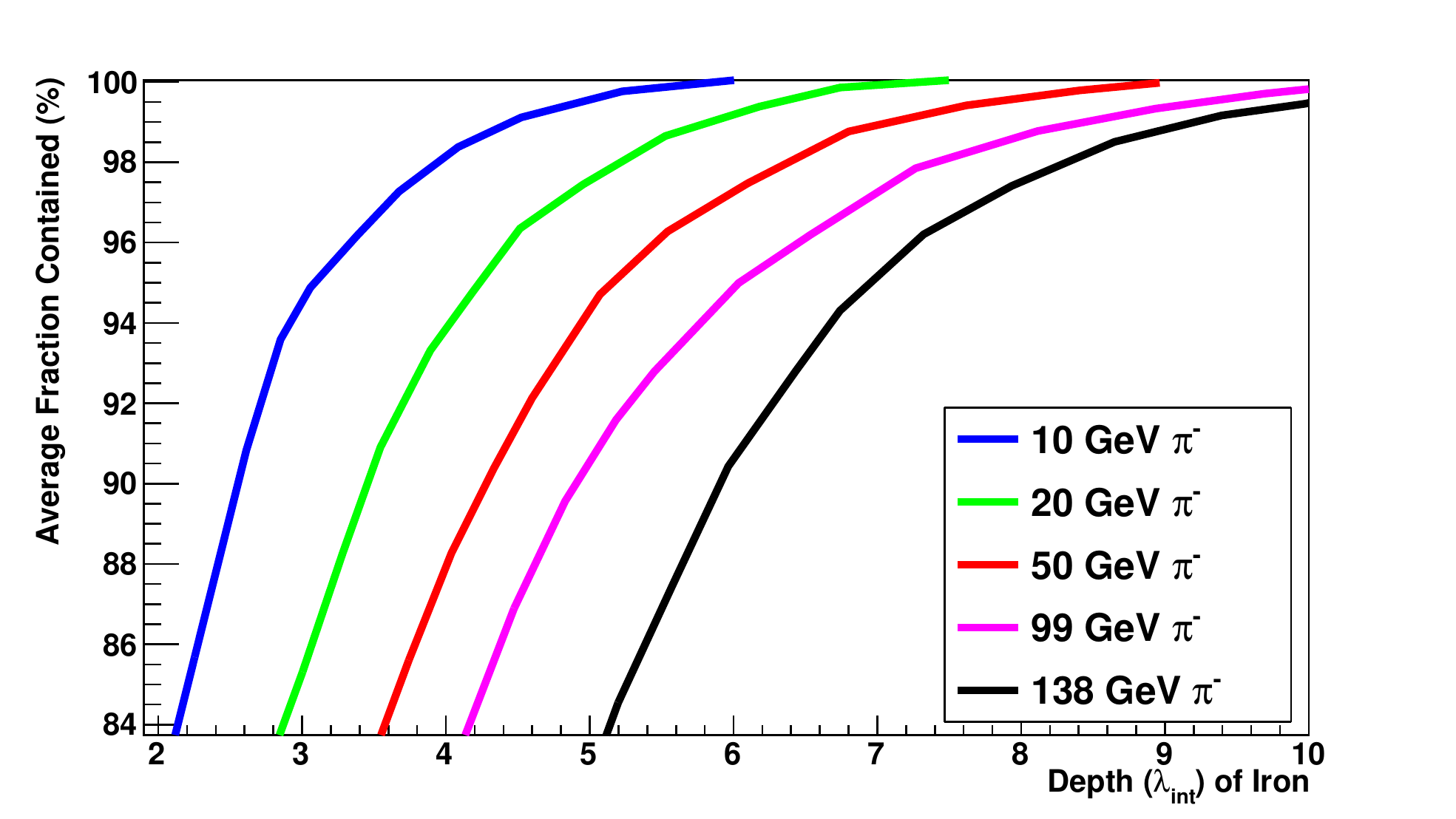}
    \caption[Average fractional energy contained in a block of iron vs
    block thickness]{Average energy fraction contained in a block of
      iron with infinite transverse dimensions, as a function of the
      thickness of the block.  Figure adapted from
      Ref.~\protect\cite{Wigmans:2000vf}{}.}

    \label{fig:HC-containment} 

 \end{center}
\end{figure}

The combination of the EMCal and the HCal needs to be at least
$\sim6\lambda_\mathrm{int}$ deep---sufficient to absorb $\sim 97\%$ of
the energy of impinging hadrons with momenta below 50~GeV/c, as shown
in Figure~\ref{fig:HC-containment}.  The electromagnetic calorimeter
is ~$\sim 1\lambda_\mathrm{int}$ thick, so the hadronic calorimeter
should be~$\sim5\lambda_\mathrm{int}$ deep.  Although the thickness of
the HCal is driven by physics needs, building it of iron plates and
scintillating tiles insures that the hadronic calorimeter can also
serve as the return yoke for the solenoid.

The HCal is divided into electromagnetic leakage section integrated
with the EMCal inside the solenoid with the bulk of the hadronic
calorimeter per se outside of the magnet.  This design minimizes the
overall size of the detector and minimizes the spread of the hadronic
shower in the radial space occupied by the cryostat.  A detailed
mechanical design is needed to determine exactly how much absorber
fits in the available space, but about one interaction length is
expected to be feasible with the space needed for support, light
collection, electronics, and cables.

\begin{figure}[htb!]
  \begin{center}
    \includegraphics[width=0.8\linewidth]{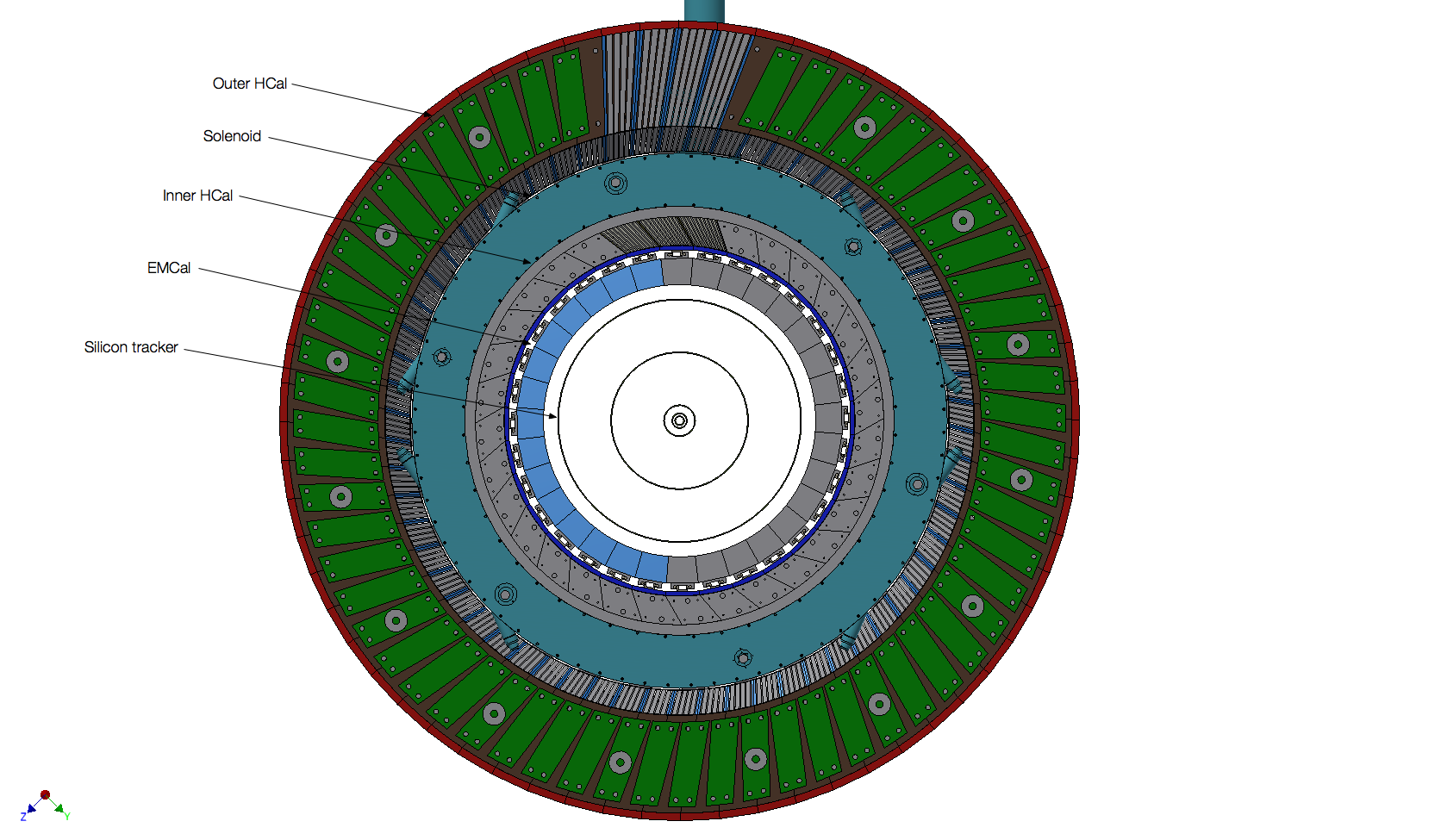}
    \caption{Cross section of sPHENIX. The outer hadronic calorimeter
      surrounds the solenoid cryostat.}
    \label{fig:sPHENIX-quadrant}

  \end{center}
\end{figure}

The outer hadronic calorimeter as shown in
Figure~\ref{fig:sPHENIX-quadrant} surrounds the cryostat in an
envelope which extends from just outside the cryostat at a radius of
about 180~cm to 264.5~cm.

\begin{figure}[htb!]
 \begin{center}
   \includegraphics[trim = 20 0 0 0, clip, width=0.7\linewidth]{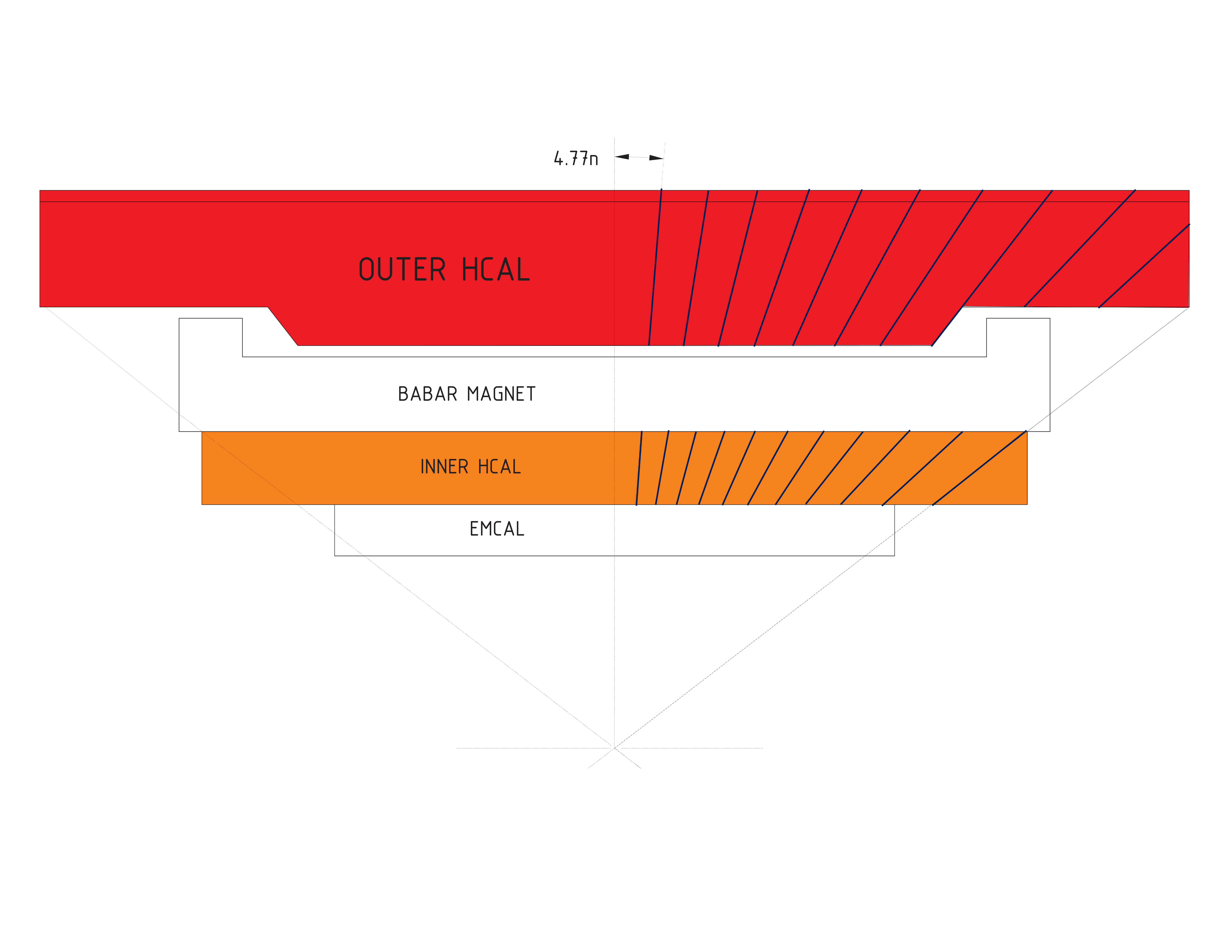}
   \caption{Scintillating tiles in the sampling gap of sPHENIX
     hadronic calorimeter, showing the transverse segmentation into
     elements 0.1 units of pseudorapidity wide.}
   \label{fig:HC-TiledLayer} 
 \end{center}
\end{figure}

Both the inner and outer longitudinal segments of the calorimeter are
constructed of tapered absorber plates, creating a finned structure,
with each fin oriented at an angle of $\sim \pm 10^\circ$ with respect
to a radius vector perpendicular to the beam axis, with the angle
chosen so that a radial ray from the interaction point crosses four
scintillator layers.  There are 384 tapered plates in each of the
inner and outer segments.  The plates in the inner and outer segments
are radially tilted in opposite directions resulting in a $\sim
20^\circ$ angle with respect to each other.  They are also staggered
by half a fin thickness.  The gaps between the iron plates are 8~mm
wide and contain individually wrapped 7~mm thick scintillating tiles
with a diffuse reflective coating and an embedded wavelength shifting
fiber which traverses the entire tile, entering and exiting on the
same edge.  The slight tilt and the azimuthal staggering of steel fins
and scintillating tiles prevents particles from traversing the depth
of the calorimeter without encountering the steel absorber
(channeling).  The benefits of two longitudinal segments include a
further reduction in the channeling of particles in the scintillator,
shorter scintillators with embedded fibers for collecting the light,
shower depth information, more symmetric response for particles of
opposite charge, and less variation in sampling fraction with depth.

With plates oriented as described, particles striking the calorimeter
at normal incidence will, on average, traverse about 90 cm of absorber
resulting in a probability for the punch through of particles with
momenta above $\sim 2$~GeV/c of only 1\%.  This design has a very
small number of distinct components which makes it simple to
fabricate, assemble, and to model.

Within each gap, there are 22 separately wrapped scintillator tiles of 11
different shapes, corresponding to a detector segmentation in
pseudorapidity of $\Delta\eta\sim0.1$ (see
Figure~\ref{fig:HC-TiledLayer}).  Azimuthally, the hadronic
calorimeter is divided into 64 wedges ($\Delta\phi\simeq 0.1$).  Each
wedge is composed of six sampling cells (steel plate and
scintillating tile) with the scintillating tile edges pointing towards
the origin.  The 22 pseudorapidity slices result in towers about
10~cm$\times 10$~cm in size at the inner surface of the
calorimeter. The total channel count in the calorimeter is
$1408 \times 2$.

\begin{figure}[htb!]
  \begin{center}
    \includegraphics[trim=0 250 0 250, clip, width=0.8\linewidth]{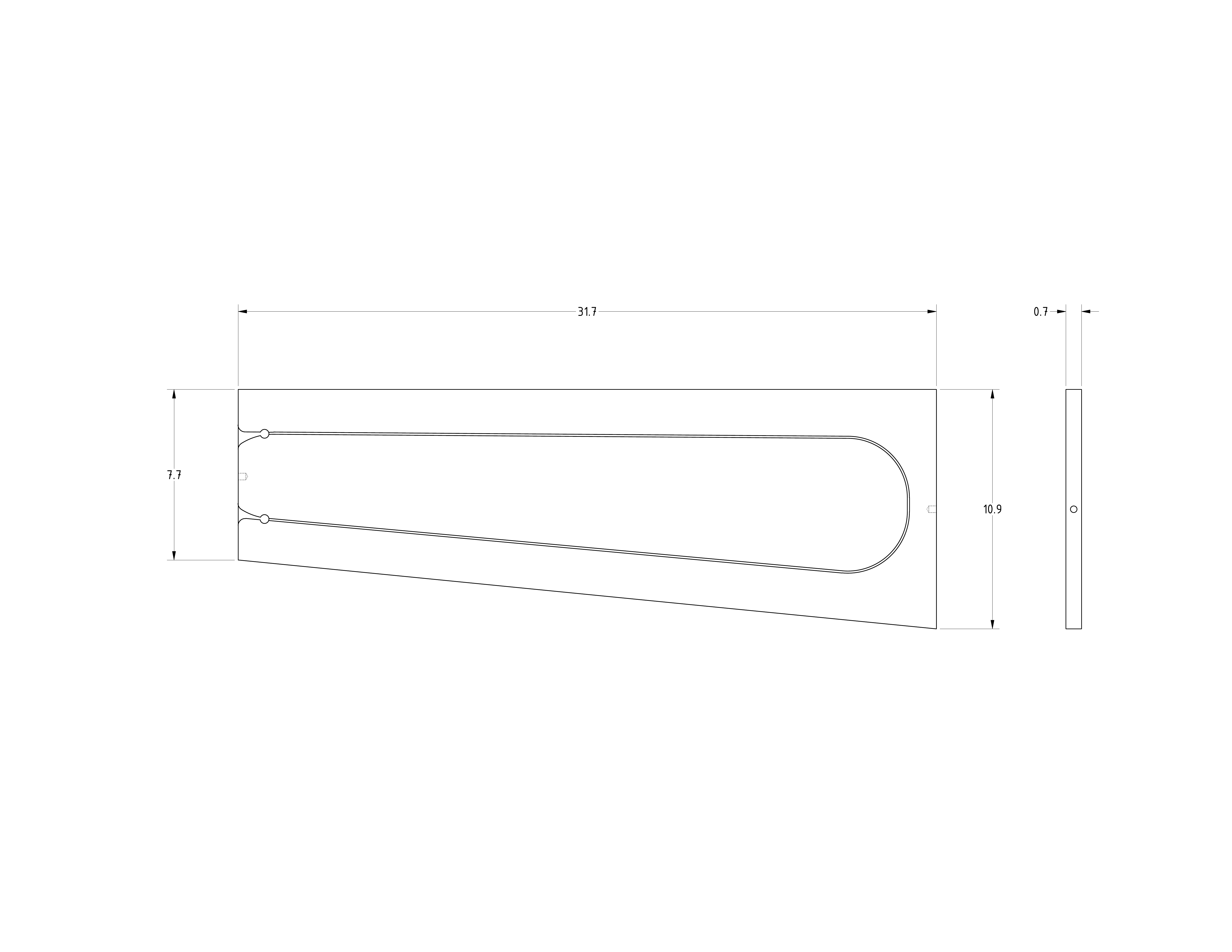}
    \caption{Grooved scintillating tiles for inner HCal section,
      showing the path of the fiber and the uniform thickness of the
      tiles.  This was the design of the tile used in the prototype.}
    \label{fig:HC-Tiles} 
  \end{center}
\end{figure}

The light from the scintillating tiles between the steel fins is
collected using wavelength shifting fibers laid along a 
path as shown in Figure~\ref{fig:HC-Tiles}.  This arrangement provides
relatively uniform light collection efficiency over the whole tile.
We have considered two fiber manufacturers: (1) Saint-Gobain (formerly
BICRON), product brand name BCF91A ~\cite{Saint-Gobain} and (2)
Kuraray, product name Y11~\cite{Kuraray}.  Both vendors offer single
and double clad fibers.

The calorimeter performance is determined by the sampling fraction and
the light collection and readout efficiency.  The readout contributes
mostly to the stochastic term in calorimeter resolution through
Poisson fluctuations in the number of photoelectrons on the input to
the analog signal processing.  Factors contributing to those
fluctuations are luminous properties of the scintillator, efficiency of
the light collection and transmission, and the photon detection
efficiency of the photon detector.

The scintillating tiles are based on the design of scintillators for
the T2K experiment by the INR group (Troitzk, Russia) who designed and
built 875~mm long scintillation tiles with a serpentine wavelength
shifting fiber readout~\cite{Izmaylov:2009jq}.  The T2K tiles are
injection molded polystyrene tiles of a geometry similar to those
envisioned for sPHENIX, read out with a single serpentine fiber, with
each fiber viewed by an SiPM on each end.  The measured light yield
value was 12 to 20 photoelectrons/MIP at
20$^\circ$C~\cite{Mineev:2006ek}.  
With 12~p.e./MIP measured by T2K
for 7~mm thick tiles (deposited energy $\sim1.4$~MeV) and an average
sampling fraction of 4\% estimated for the sPHENIX HCal we expect the
light yield from the HCal to be about 400~p.e./GeV.  A 40~GeV hadron
will share its energy nearly equally between the inner and outer HCal
segments so the upper limit of the dynamic range of the HCal can be
safely set to $\sim 30$~GeV which corresponds to a yield of
12000~p.e. on the SiPM.  To avoid signal saturation and ensure
uniform light collection, care will be required to both calibrate the
light yield (possibly requiring some attenuation) and randomize it.

\begin{figure}[tb]
  \begin{center}
    \includegraphics[width=0.95\linewidth]{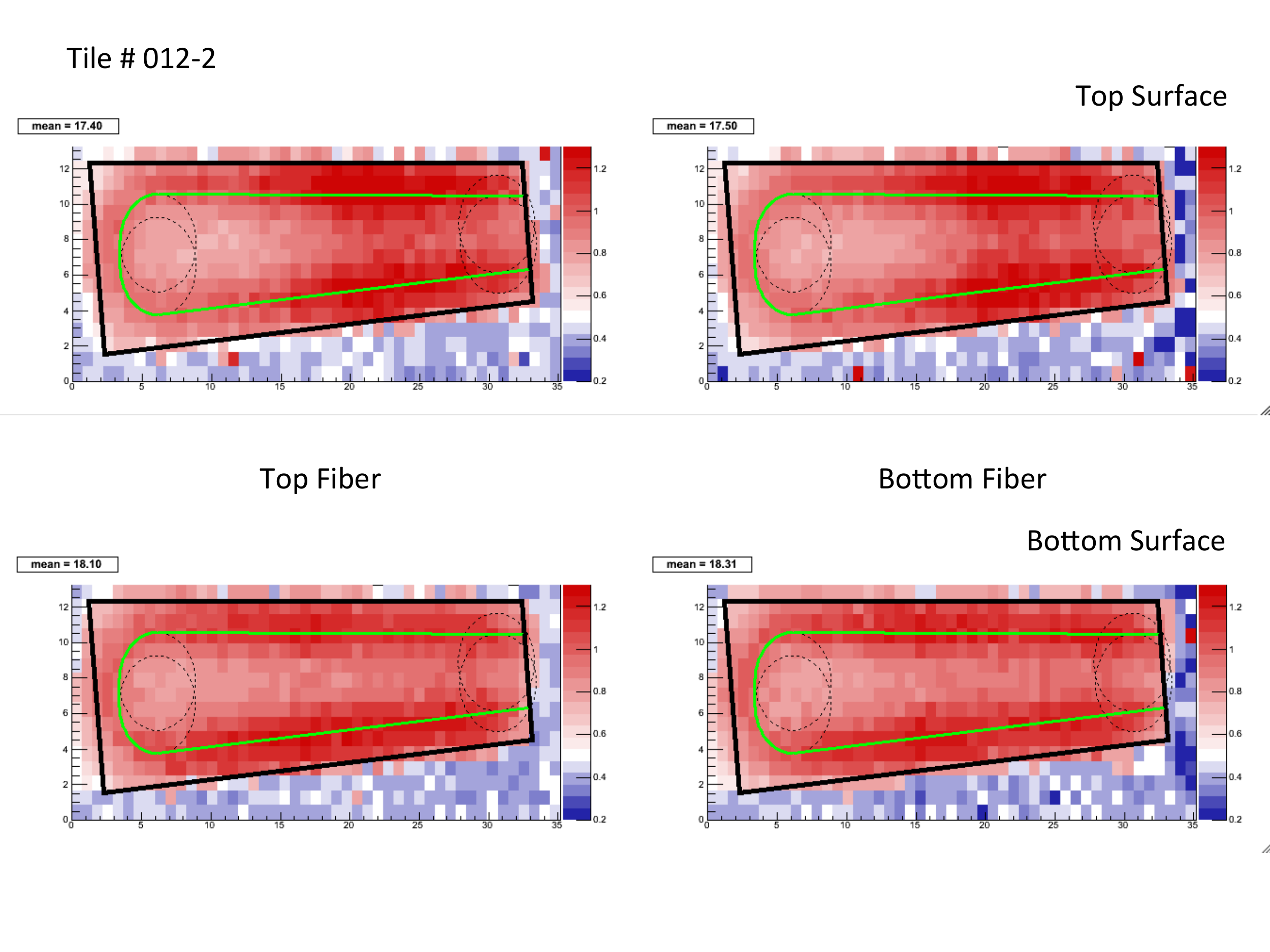}
    \caption{Measurement of uniformity of light collection in an
      sPHENIX prototype tile measured with a $^{90}$Sr source at the
      University of Colorado.}
    \label{fig:HC-LYProfile}
  \end{center}
\end{figure}

The uniformity of light collection from prototype tiles constructed
for the sPHENIX prototype arrangement can be judged from
Figure~\ref{fig:HC-LYProfile} with measurements made by scanning a
$^{90}$Sr source over the surface of a tile.  The largest drop in the
light yield is along the tile edges and in the corners farthest from
the fibers.

We note that this design is optimized for simplicity of manufacturing,
good light yield, and to serve as the flux return for the magnet.
As such, it has a manifestly non-uniform sampling fraction as a
function of depth and is not highly compensated.  However, the
performance specifications are quite different from particle physics
hadronic calorimeters, particularly with a limited energy measurement
range (0--60~GeV).  \geant simulations described in the next section
indicate a performance better than the physics requirements, and
test beam results which validate the design.

\makeatletter{}\section{Calorimeter Simulations\label{sec:g4sim}}

We have employed the \geant simulation toolkit
\cite{Agostinelli:2002hh} for our full detector simulations. It
provides collections of physics processes suitable for different
applications. We selected the QGSP\_BERT list which is recommended for
high energy detector simulations like the LHC experiments.  We have
also run additional tests with different physics lists in detailed
comparison with test beam data.  We have integrated the sPHENIX
simulations with the PHENIX software framework, enabling us to use
other analysis tools we have developed for PHENIX.

The detectors and readout electronics and support structures are
highly configurable in our \geant framework, making it easy to test
various geometries and detector concepts.  Magnetic field maps for the
BaBar magnet have been imported from OPERA calculations.  We keep
track of each particle and its descendants so every energy deposition
can be traced back to the original particle from the event generator.
The detector geometry can be easily configured when events are
generated from a number of libraries which simulate concentric
cylinders (the simplest idealized geometry), tilted plates, and
spaghetti fiber geometries for the electromagnetic and hadronic
calorimeters.

The superconducting magnet is simulated with the proper location of
the material thickness in the cryostat.

All tracks which reach a layer 10~cm behind the HCal are aborted to
prevent particles which are curled up by the field from re-entering
the detector.  Adding up the energy of those aborted tracks yields an
estimate of the energy which leaks from the back of the hadronic
calorimeter.

\subsection{Electromagnetic Calorimeter Simulation}

\begin{figure}[htb!]
 \begin{center}
    \includegraphics[width=0.7\linewidth]{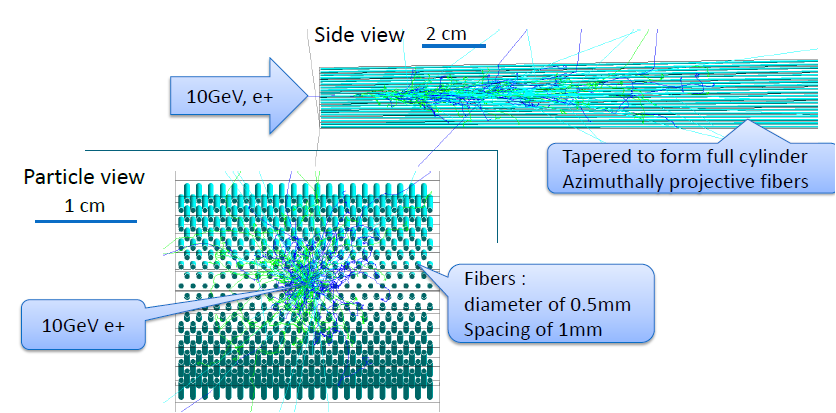}
    \caption{\label{fig:spacal}\geant event display showing the fiber
      matrix and an electron shower development in the calorimeter.  }
 \end{center}
\end{figure}

\begin{figure}[htb!]
 \begin{center}
    \includegraphics[width=0.6\linewidth]{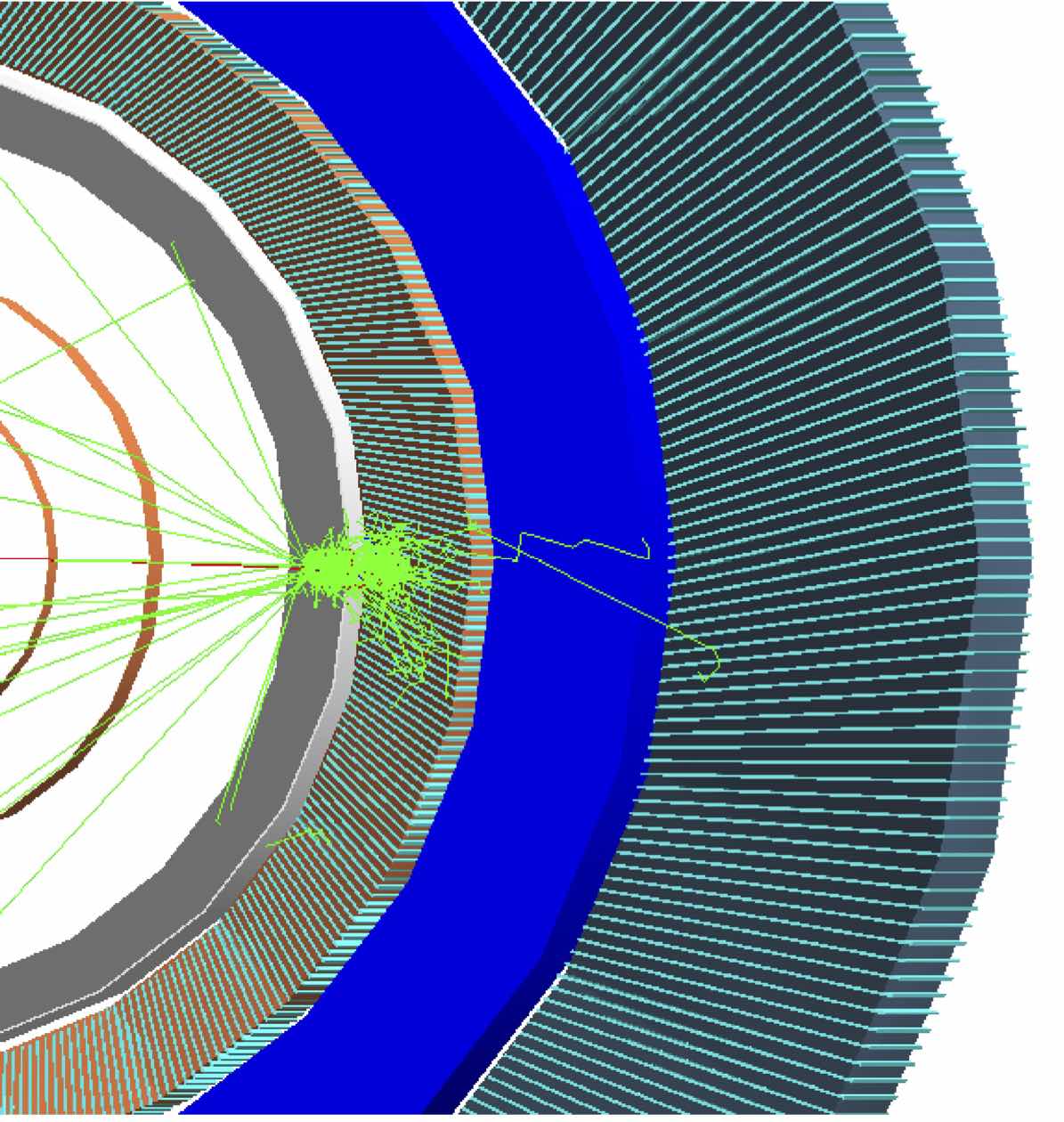}
    \caption[Transverse view of a 10~GeV/$c$ electron in sPHENIX,
    showing it showering mainly in the EMCal]{\label{fig:elec}
      Transverse view of a 10~GeV/$c$ electron in sPHENIX, showing it
      showering mainly in the EMCal.}
 \end{center}
\end{figure}

The electromagnetic calorimeter has been simulated using the \geant tools
described above.  A detailed description of the SPACAL design is included with
fiber locations and orientations as shown in Figure~\ref{fig:spacal}.   This
geometry is based on the \geant description developed by Alexander Kiselev for the
EIC research and development calorimeter project.

Figure~\ref{fig:elec} shows a typical \geant event in which a
10~GeV/$c$ electron hits the calorimeter.  Most of the shower develops
in the EMCal.
The resolution of the electromagnetic
calorimeter for electrons at normal incidence was shown earlier in comparison with the
team beam data in Figure~\ref{fig:spacal_res}.

\begin{figure}[htb!]
 \begin{center}
    \includegraphics[trim = 15 0 30 0, clip, width=0.8\linewidth]{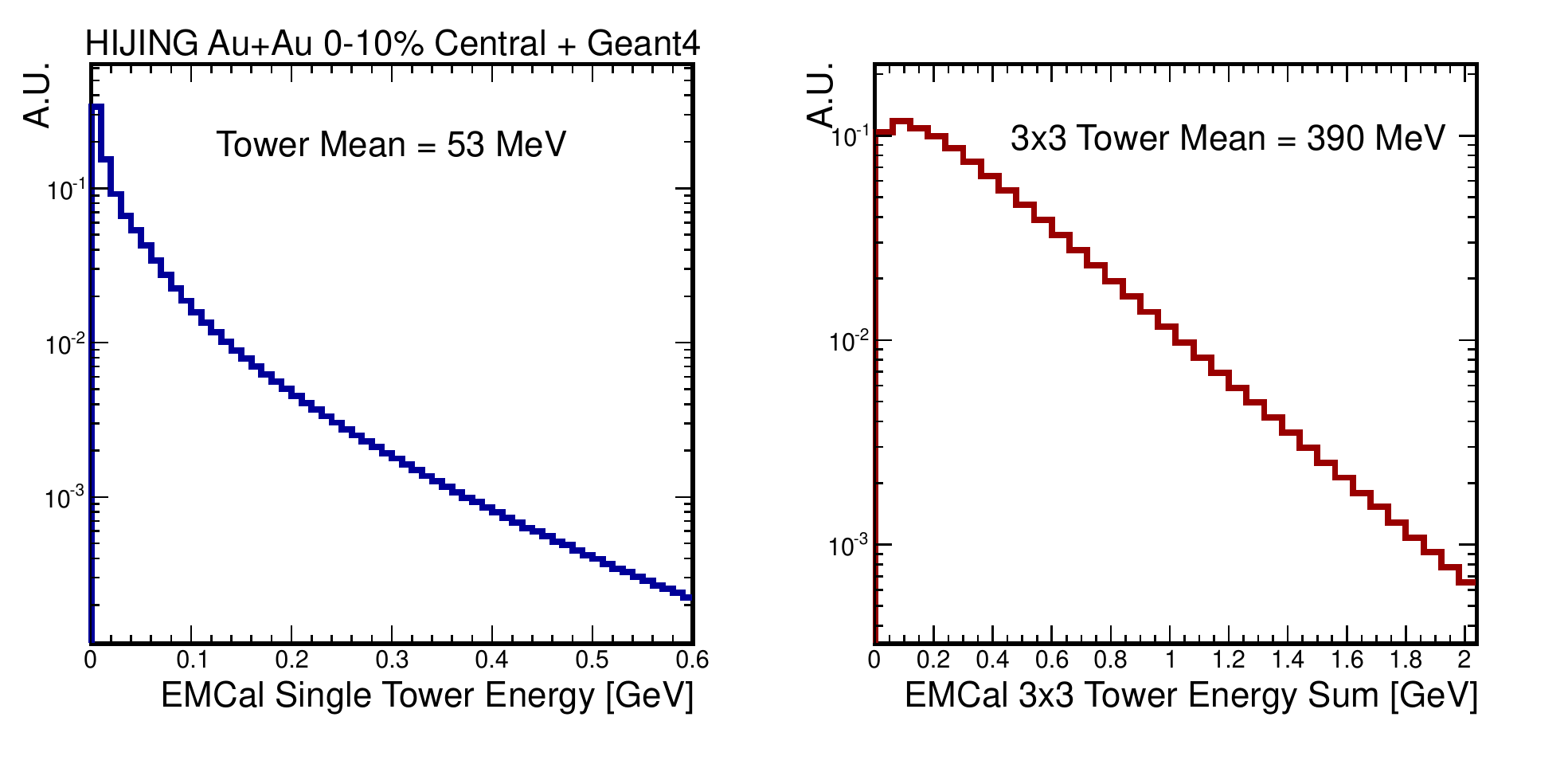}
    \caption[Distribution of energy deposited in the electromagnetic
    calorimeter for single towers and in $3\times3$ arrays of towers
    in central \hijing events]{\label{fig:g4emcal_hijingoccup}
      Distribution of energy deposited in the electromagnetic
      calorimeter for single towers (left panel) and in $3\times3$
      arrays of towers (right panel) in central \hijing events.}
 \end{center}
\end{figure}

The energy deposited in the electromagnetic calorimeter in central
\hijing events is shown in Figure~\ref{fig:g4emcal_hijingoccup}.
The mean energy deposited in any single tower is estimated to be 
approximately 50~MeV.  The existing PHENIX electromagnetic calorimeter cluster finding
algorithm has been adapted for the sPHENIX EMCal specifications.  Initial
results indicate that for a 10~GeV photon there is an extra 4\% of 
underlying event energy in the cluster and a degradation of less than 10\% in the energy resolution 
when embedded in a central \auau event (simulated with the \hijing event generator).

\subsection{Hadronic Calorimeter Simulation}

\begin{figure}[tb!]
 \begin{center}
  \includegraphics[width=0.6\linewidth]{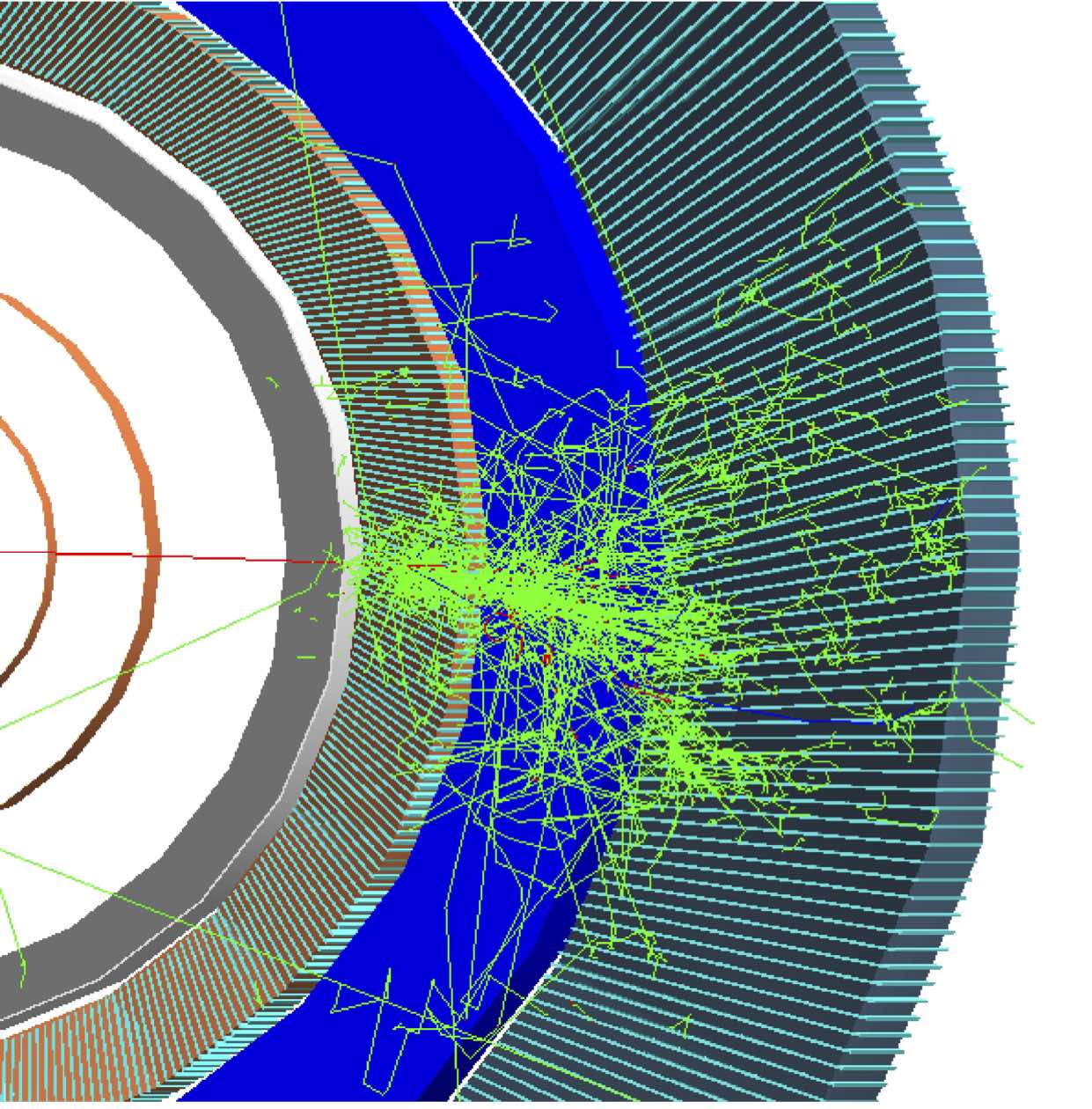}
  \caption{\label{fig:piplus} Transverse view of a 10~GeV/$c$ $\pi^-$
    in sPHENIX. It penetrates the EMCAL and magnet and showers in the
    first segment of the HCal.}
 \end{center}
\end{figure}

The hadronic calorimeter has been simulated using the \geant tools
described above.  The simulation includes the detailed geometry of the
steel plates and interleaved scintillator tiles.
Figure~\ref{fig:piplus} shows a typical \geant event in which a
10~GeV/$c$ $\pi^+$ incident on the calorimeter showers in the Hcal.

\begin{figure}[tb!]
  \begin{center}
    \includegraphics[width=0.8\linewidth]{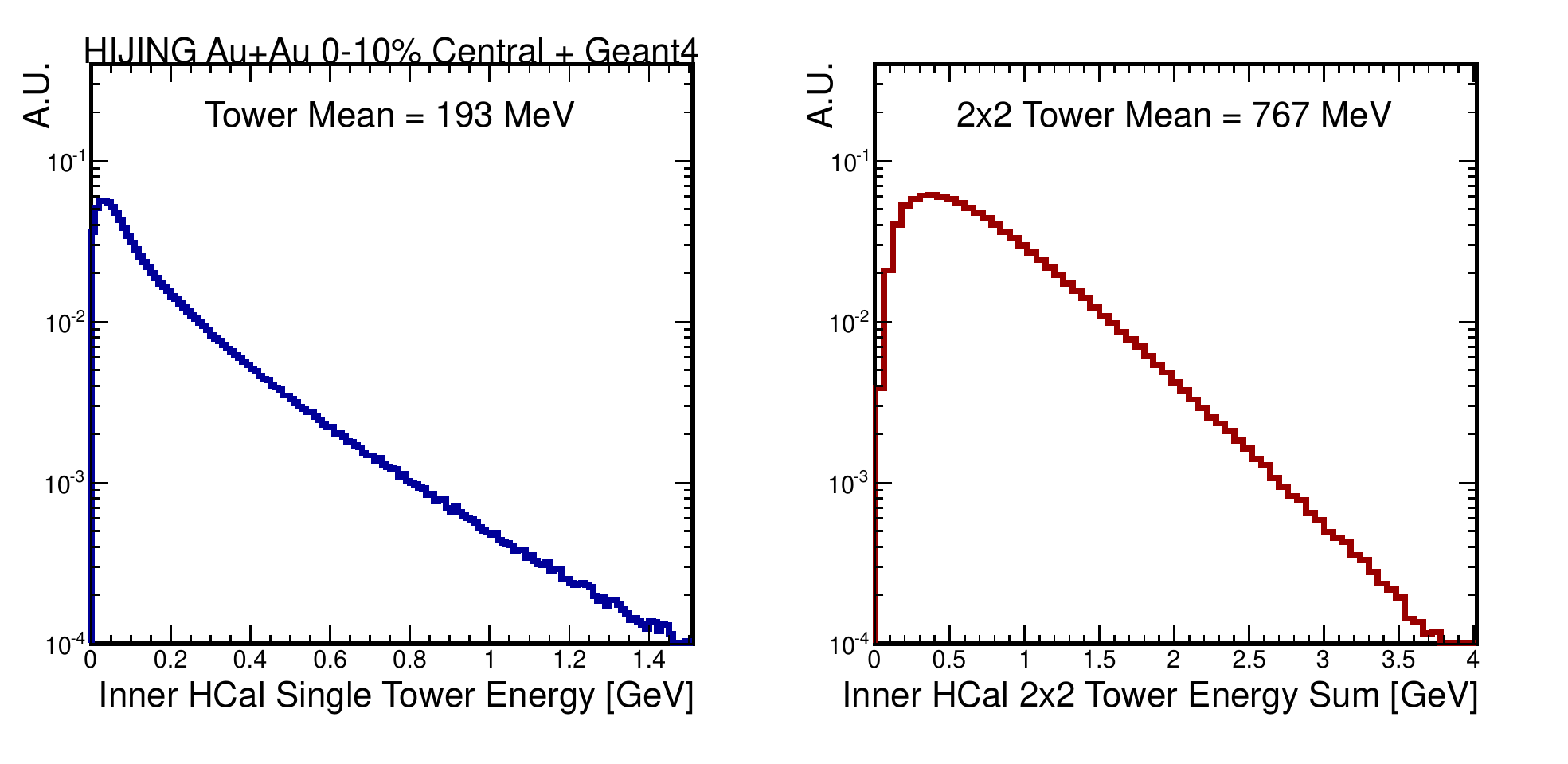}
    \hfill
    \includegraphics[width=0.8\linewidth]{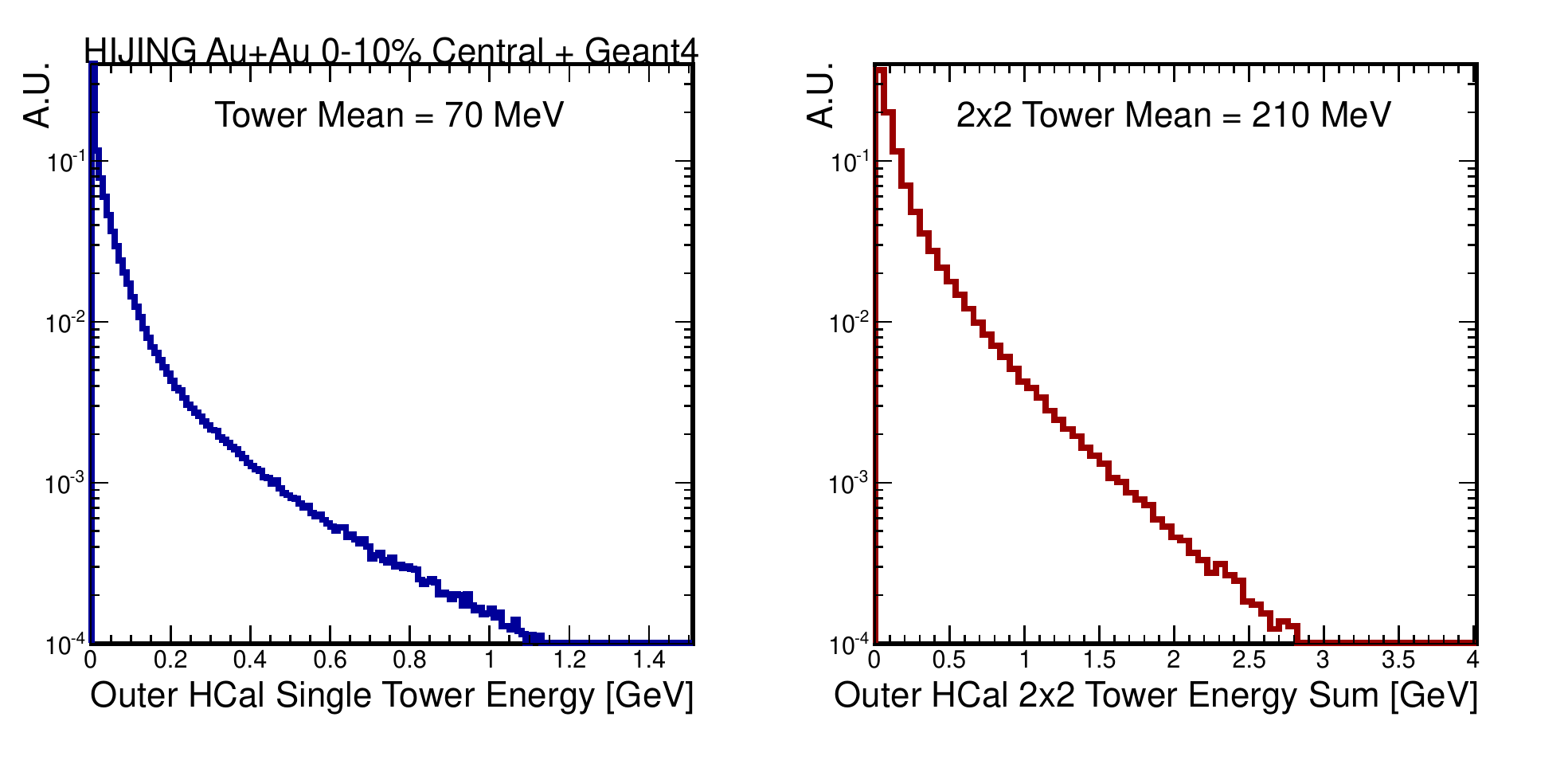}
    \caption{Energy occupancy distribution for the inner and outer
      hadronic calorimeter sections in 10\% central \auau \hijing
      events run through the full \geant simulation.}
    \label{fig:hcal_occup} 
  \end{center}
\end{figure}

The average energy deposition in the hadronic calorimeter inner and outer longitudinal segments in a \hijing 10\% central
\auau event is shown in Figure~\ref{fig:hcal_occup}.

\begin{figure}[tb!]
  \begin{center}
    \includegraphics[width=0.47\linewidth]{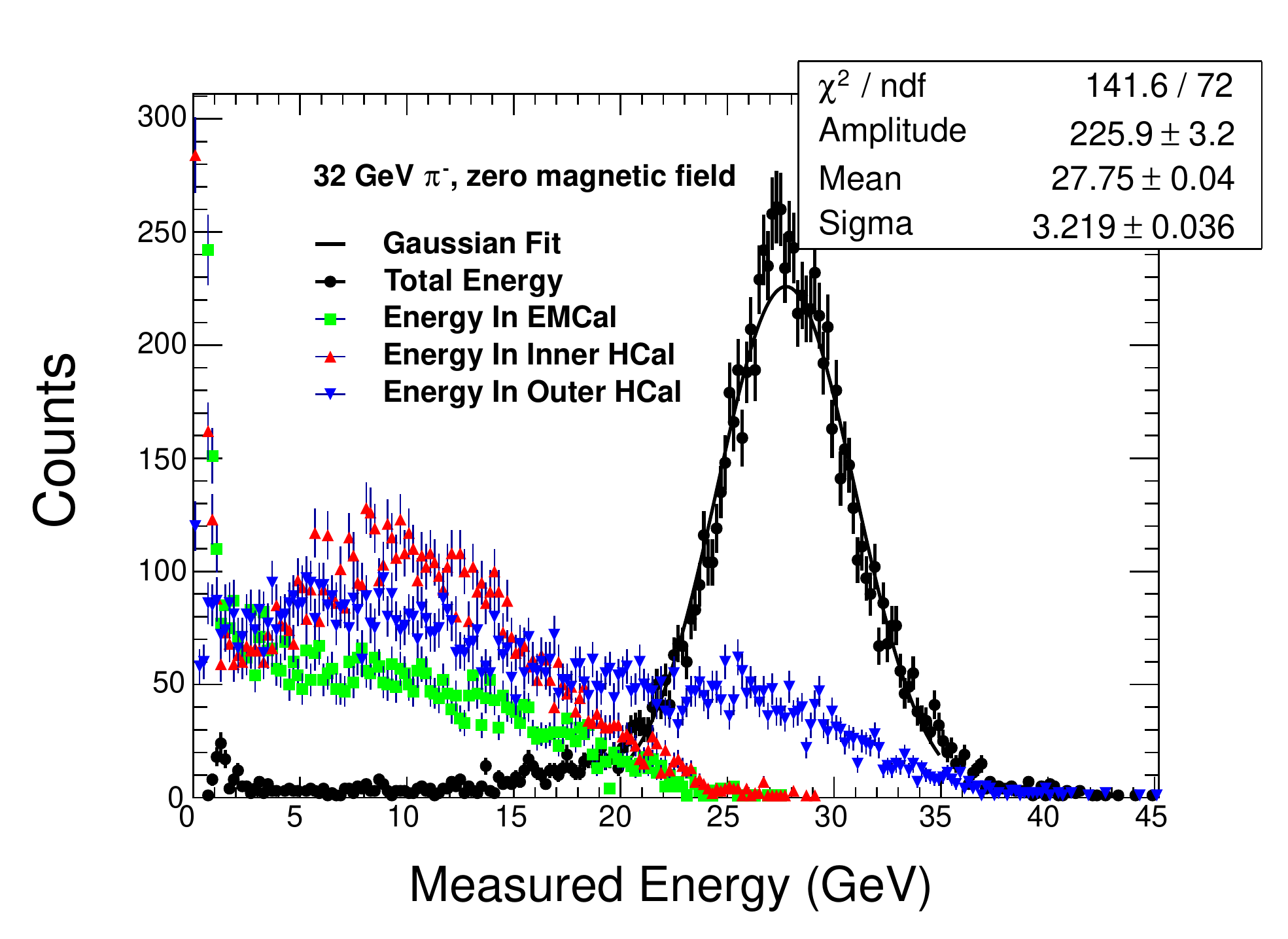}
    \hfill
    \includegraphics[width=0.47\linewidth]{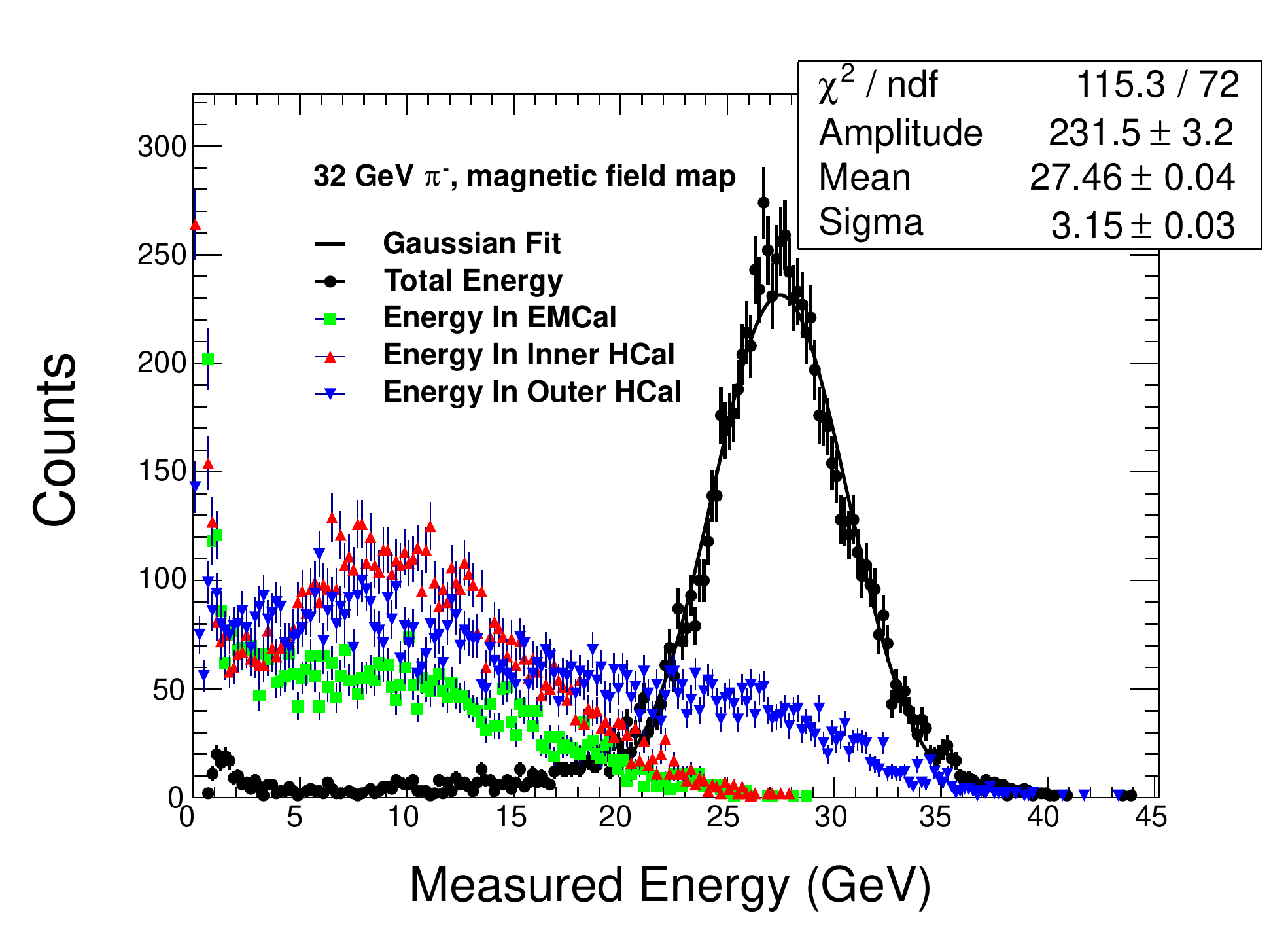}
    \caption[Energy deposited in the three longitudinal segments of
    the calorimetery by 32~GeV/$c$ charged pions, showing the good
    containment and Gaussian response of the calorimeter]{Energy
      deposited in the three longitudinal segments of the calorimetery
      by 32~GeV/$c$ charged pions, showing the good containment and
      Gaussian response of the calorimeter (shown with and without the
      magnetic field turned on).}
    \label{fig:hcal_proton} 
  \end{center}
\end{figure}

The single particle energy resolution in the HCal has been determined
using a full \geant description of the calorimeters.  The energy
deposition in the scintillator is corrected for the average sampling
fraction of the inner and outer sections and electromagnetic calorimeter separately.

The full calorimeter response to single protons and charged pions is shown
in Figure~\ref{fig:hcal_proton}.
The mean and standard deviation from a Gaussian fit to the measured
energy distribution are used to calculate the nominal detector
resolution.  These indicate a \geant performance level better
than the physics requirements.
Note that for jets in the energy range 20--70 GeV, the constant term is 
not a dominant effect.
Detailed comparisons with test beam data are necessary for final optimization
of the design.

\begin{figure}[tb!]
  \begin{center}
    \includegraphics[width=0.45\linewidth]{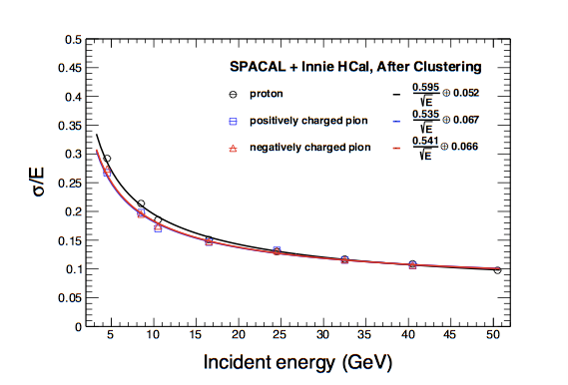}
    \hfill
    \includegraphics[width=0.45\linewidth]{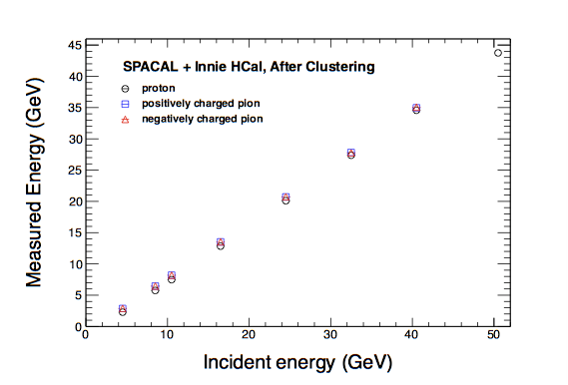}
    \caption[Energy resolution and linearity of the full calorimeter
    as a function of incident hadron energy]{Energy resolution of the full calorimeter as a function
      of incident hadron energy.  The left panel shows the Gaussian
      resolution width and the right panel shows the linearity of the
      energy response.   
      \label{fig:hcal_testbeamG4_resolution} }
  \end{center}
\end{figure}

As mentioned above, the proposed sPHENIX calorimeter system is about
$6\lambda_{\mathrm int}$ deep, and one expects some leakage of energy
out of calorimeter. The amount of this leakage and its energy
dependence can be estimated from literature,
Figure~\ref{fig:HC-containment} above, or from simulation which is
tuned to available experimental data.  The probability for a proton to
go through the whole depth of the calorimeter without an hadronic
interaction is about 0.6\% (verified with full \geant simulations).
Therefore, energy leakage out the back of the calorimeter is not
expected to be a serious problem.

\makeatletter{}\section{Electronics}
\label{sec:electronics}

For the readout of both the EMCal and HCal a common electronics design will 
be used to reduce the overall cost and minimize the design time. 
The \refdesign approach is based on electronics developed
for the PHENIX Hadron Blind Detector (HBD) and Resistive Plate Chambers (RPC), 
and uses the current PHENIX DAQ as the backend readout,
although there are alternatives which have been examined and could still be viable.

\subsection{Sensors}

For both the electromagnetic and hadronic calorimeters, we are
currently considering as sensors 3~mm$\times$3~mm silicon
photomultipliers (SiPMs), such as the Hamamatsu S10362-33-25C
MultiPixel Photon Counters (MPPC).  These devices have 14,400 pixels,
each 25~$\mu$m$\times$ 25~$\mu$m. Any SiPM device will have an
intrinsic limitation on its dynamic range due to the finite number of
pixels, and with over 14K pixels, this device has a useful dynamic
range of over 10$^4$.  The saturation at the upper end of the range is
correctable up to the point where all pixels have fired.  The photon
detection efficiency is $\sim36\%$ and it should therefore be possible
to adjust the light level to the SiPM using a mixer to place the full
energy range for each tower ($\sim$ 5~MeV--50~GeV) in its useful
operating range. For example, if the light levels were adjusted to
give 10,000 photoelectrons for 50~GeV, this would require only 200
photoelectrons/GeV, which should be easily achieved given the light
level from the fibers entering the mixer.

A number of sample devices, all 3~mm$\times$3~mm, from AdvansID,
Excelitas, SensL, and R\&D devices from RMD have been characterized
for use in sPHENIX, in addition to a suite of new sensors from
Hamamatsu.  Cost, photon detection efficiency, gain, number of
micro-pixels, and dark current have been compared for a wide variety
of devices.  Concern about radiation damage to SiPMs resulted in a
test in PHENIX in Run 14 in which two Hamamatsu SiPMs were placed at
approximately the location of the sPHENIX EMCal, and while the leakage
current was monitored during \auau operation, the thermal neutron flux
was measured with a $^{3}$He proportional counter.  The devices are
thought to be damaged by neutrons with an energy of a few MeV which
result from secondary neutrons produced by collision products, and so
simulation of the neutron background in sPHENIX will be necessary.
Radiation damage has also been measured on a variety of devices with a
14.1~MeV neutron generator at the BNL Instrumentation Solid State
Gamma Irradiation Facility, and these studies will allow selection of
the most rad-hard device.  KETEK, working with the CMS experiment, has
been working on devices which may be more immune to radiation, and
samples of those devices will be tested as well.  The result of these
and future studies should allow us to select the most appropriate
readout device for sPHENIX that will meet all of its requirements.

While we believe that the SiPMs are  likely the most suitable
sensor for the calorimeters, we are also considering avalanche
photodiodes (APDs) as an alternative. They have much lower gain
($\sim$50--100 compared to $\sim 10^5$ for SiPMs), and therefore would
require lower noise and more demanding readout electronics, but they
do provide better linearity over a larger dynamic range. In addition,
while the gain of both SiPMs and APDs depend on temperature, SiPMs
have a stronger gain variation than APDs (typically 10\%/$^\circ$C for
SiPMs vs 2\%/$^\circ$C for APDs). Thus, we are considering APDs as an
alternative solution as readout devices pending further tests with
SiPMs and our light mixing scheme.

\subsection{Digital and Analog Electronics}

\subsubsection {SiPM Preamplifier Circuitry}

The requirements of the sPHENIX calorimeter preamplifier circuit board
are to provide localized bias/gain control, temperature compensation,
signal wave shaping and differential drive of the SiPM signal to an
ADC for acquisition. Gain adjustment and temperature compensation are
performed as part of the same control circuit. Signal wave shaping is
performed by the differential driver to satisfy the sampling
requirements of the ADC.

\subsubsection{Temperature Compensation}

\begin{figure}[htb]
  \centering
  \includegraphics[width=0.7\linewidth]{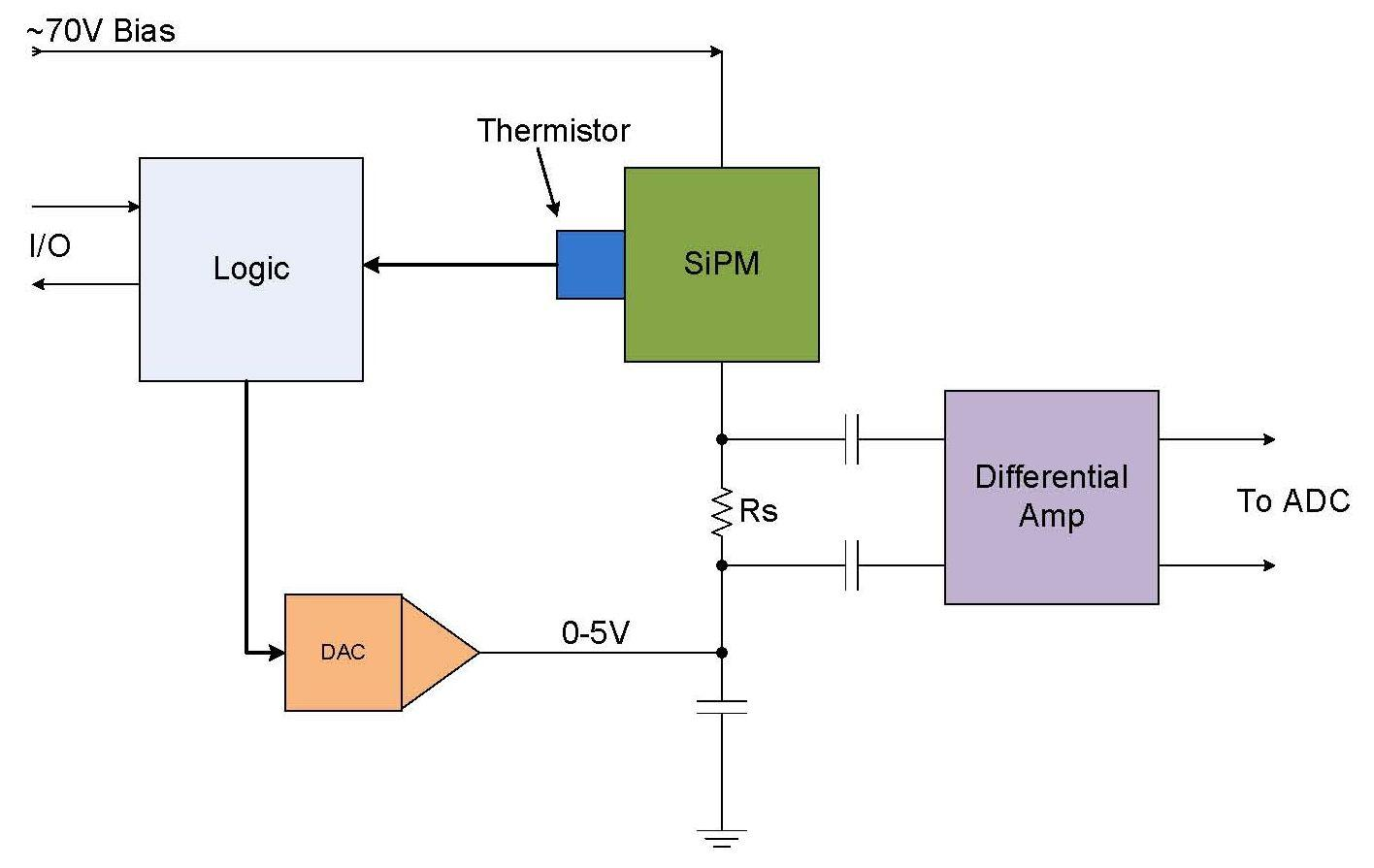}
  \caption{Block diagram of a temperature compensating circuit for SiPMs}
  \label{Fig:TempComp}
\end{figure}

The reverse breakdown voltage V$_{\mathrm br}$ for the Hamamatsu
S10362-33-25C device is nominally 70 Volts. As the bias is increased
over the value of V$_{\mathrm br}$ the SiPM begins to operate in
Geiger mode with a gain of up to $2.75 \times 10^{5}$. The range of this
over-voltage (V$_{\mathrm ov}$) is typically 1--2 Volts and represents
the useful gain range of the device.  The V$_{\mathrm br}$ increases
by 56~mV/$^\circ$C linearly with temperature and must be compensated to
achieve stable gain. This compensation is achieved using a closed
feedback loop circuit consisting of a thermistor, ADC, logic and DAC
voltage control as shown in Figure~\ref{Fig:TempComp}.

The thermistor is fixed to the back of the SiPM and provides a
significant voltage variation over temperature when used as part of a
voltage divider, thereby easing temperature measurement over a length
of cable. The bias supply for an array of SiPMs is fixed nominally at
V$_{\mathrm br} + 2.5$V. The DAC in each SiPM circuit then outputs a
subtraction voltage of~~0~V to 5~V to provide a full range of gain
control over the device temperature range. The SiPM gain may then be
adjusted externally through an interface to the logic.

\subsubsection{Preamplifier-Shaper-Driver}

\begin{figure}[htb]
  \centering
  \includegraphics[width=0.7\linewidth]{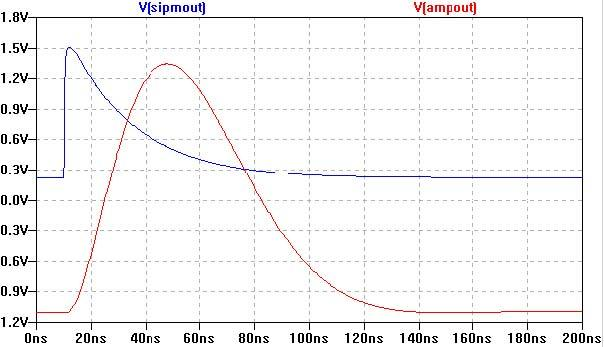}
  \caption{SPICE simulation of a prototype temperature compensating circuit 
for SiPM readout of the sPHENIX EMCal and HCal.}
  \label{Fig:Shaper}
\end{figure}

The SiPM current develops a voltage across the load resistor $R_s$
proportional to the number of pixels fired. To avoid the region of
greatest non-linearity due to saturation of the SiPM, the maximum
signal level is optically adjusted to 10K out of 14.4K pixels fired.
Simulations of the SiPM indicate that the current could be as much as
several tenths of an ampere at this maximum level. Results of a SPICE 
simulation are shown in Figure~\ref{Fig:Shaper}. Such a large current
affords the use of a small value for $R_s$ which virtually eliminates the
contribution of $R_s$ to non-linearity. This signal voltage is sensed
differentially, amplified and filtered by a low power, fully
differential amplifier. For sampling by a 65MSPS ADC, a peaking time
of approximately 35~ns is achieved through the use of a second order
Butterworth filter implemented in the differential driver circuit.

\subsubsection{Signal Digitization}
One solution for the readout of the EMCal and HCal detectors for sPHENIX is the
direct digitization of the SiPM signal. The signals from the SiPM are
shaped to match the sampling frequency, and digitized using a flash
ADC. The data are stored in local memory pending a Level-1 (L1) trigger
decision. After receiving an L1 trigger decision, the data are read out
to PHENIX Data Collection Modules (DCM II).  These second generation Data Collection
Modules would be the identical design as those developed and implemented for reading
out the current PHENIX silicon detectors.
One advantage of direct
digitization is the ability to do data processing prior to sending
trigger primitives to the L1 trigger system. The data processing can
include channel by channel gain and offset corrections, tower sums,
etc. This provides trigger primitives that will have near offline
quality, improved trigger efficiency, and provide better trigger
selection.

\begin{figure}[htb]
  \centering
  \includegraphics[width=0.9\linewidth]{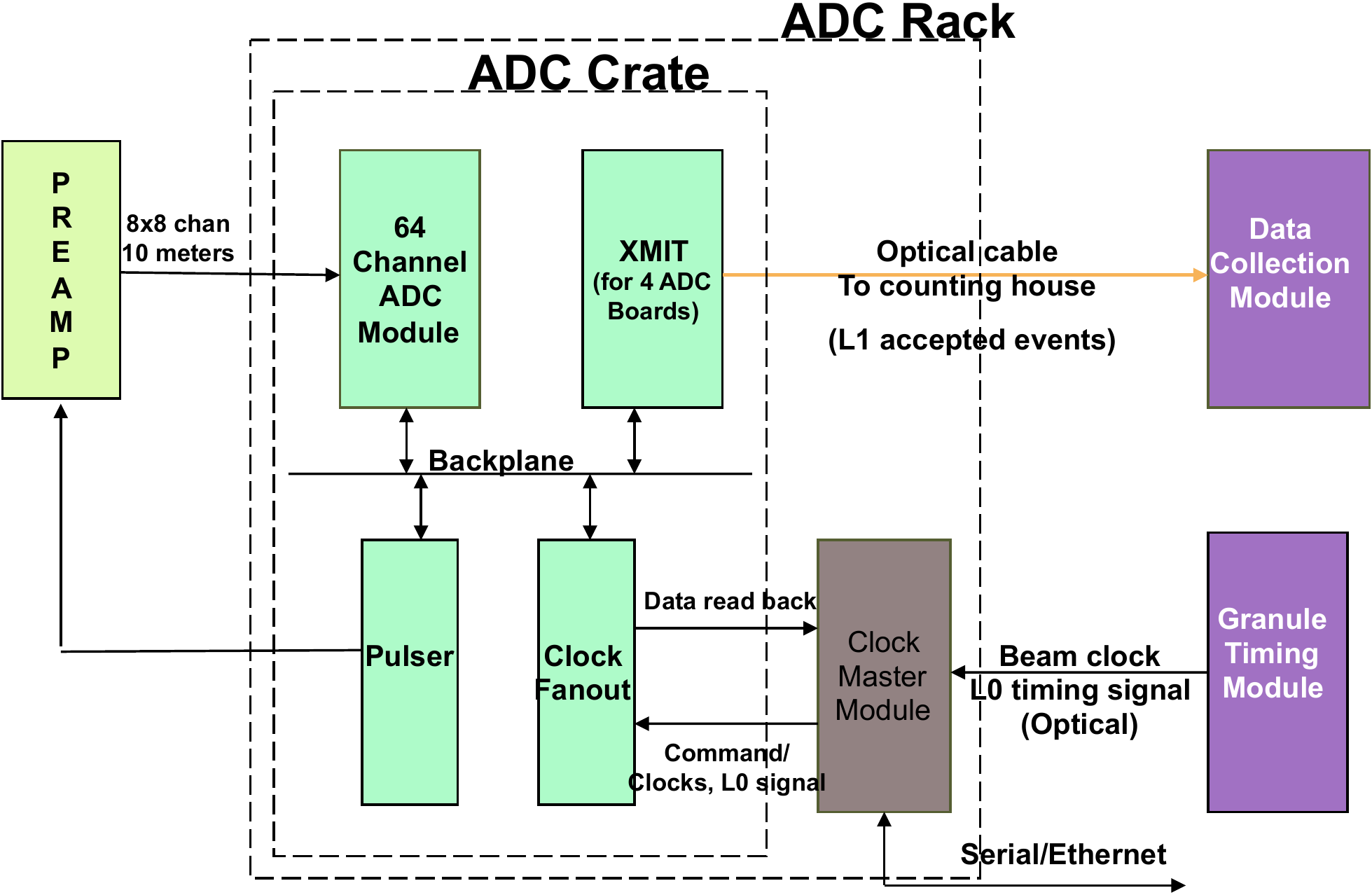}
  \caption{Block diagram of read out electronics based on electronics
    designed for the HBD.}
  \label{Fig:HBD}
\end{figure}

\begin{figure}[htb]
  \centering
  \includegraphics[width=0.9\linewidth]{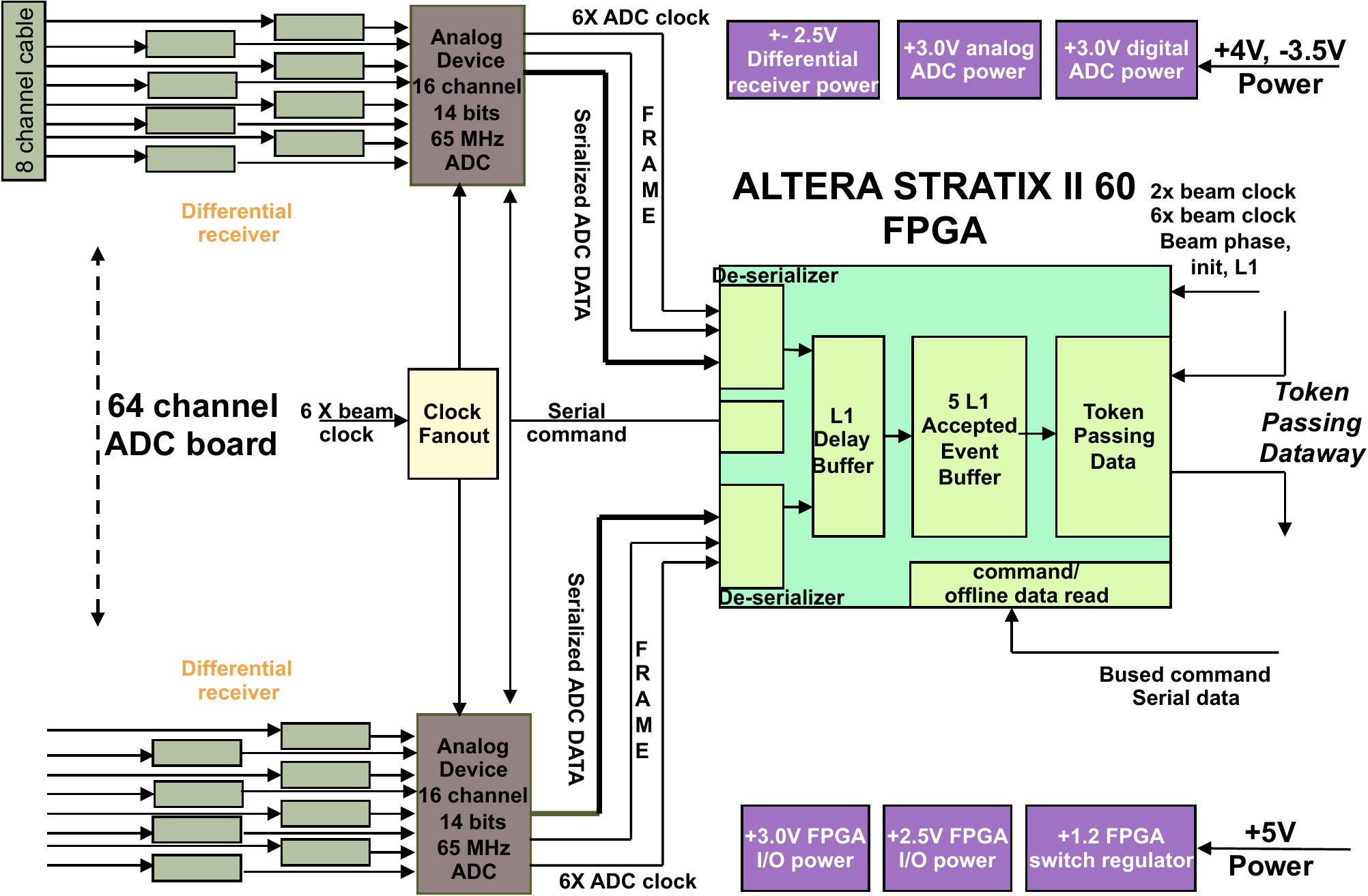}
  \caption{Block diagram of 64 channel ADC board based on design of
    the HBD system.}
  \label{Fig:HBD_ADC}
\end{figure}

A readout system based on this concept was implemented for the Hadron
Blind Detector (HBD) for the PHENIX experiment as shown in
Figure~\ref{Fig:HBD} and subsequently modified for the PHENIX
Resistive Plate Chamber (RPC) system.  The block diagram of the
Front-End Module (FEM) is showed in Figure~\ref{Fig:HBD_ADC}.  In the
HBD system, the discrete preamplifier-shaper is mounted on the
detector and the signals are driven out differentially on a 10 meter
Hard Metric cable. The signals are received by Analog Device AD8031
differential receivers which also serves as the ADC drivers. Texas
Instruments ADS5272 8 channel 12 bit ADCs receive the differential
signals from 8 channels and digitize them at 6x the beam crossing
clock . The 8 channels of digitized data are received differentially
by an Altera Stratix II 60 FPGA which provides a 40 beam crossing L1
delay and a 5 event L1 triggered event buffer.

The L1 triggered data from 4 FEMs is received by an XMIT board using
token passing to control the readout of the FEMs. The data is then
sent by 1.6~GBit optical links to the PHENIX DAQ.  A ClockMaster
module interfaces to the PHENIX Granule Timing Manager (GTM) system
and fans out the clocks, L1 triggers and test enable signals to the
FEMs and XMIT modules. The ClockMaster module also receives slow
control signals for configuring the readout.

Although not shown in the block diagram, the FEM has 4 LVDS outputs
that can be used to bring out L1 trigger primitives at 800~Mbits/sec.
This feature was not used for the HBD readout, however it has been
implemented for the RPC detector. A trigger module for the RPC system
based on the Altera Arria FPGA receives the trigger primitives from
the FEMs, combines them and sends them to the PHENIX L1 trigger system
through two 3.125~GBit optical links.

Distributing the analog and digital electronics directly on the
detector in close proximity to the sensors with all control and data
connections transmitted via high-speed optical fiber has been
considered, and has not been found to be feasible considering cost,
cooling, development time, and serviceability.  The approach chosen
for the reference design has the temperature compensating preamplifier
mounted on the detector which distributes the bias voltage to the
sensors and the drives the shaped and amplified signals differentially
to the digital modules located in racks near the detector using
shielded cables.  High speed fiber optic cables bring in all control
and clock signals to the digitizers and transmit Level 1 trigger
primitives and triggered data to the PHENIX DAQ.

\makeatletter{}\clearpage

\section{Charged Particle Tracking} 
\label{sec:tracking}

As discussed in Chapter~\ref{chap:detector_requirements}, the key
design requirements of the tracking system are precise momentum
resolution, high track reconstruction efficiency for the signals of
interest, good purity of the reconstructed tracks in central \auau
collisions, and precise measurement of displaced vertices. After
detailed \geant studies and extensive work on the tracking and pattern
recognition software, a \refdesign has been adopted that is capable of
meeting all of the key design requirements for the tracking
system. The \refdesign, which incorporates seven planes of silicon
detectors, is described and its performance detailed in this section.

\begin{figure}[htb!]
 \begin{center}
   \raisebox{0.88cm}{\includegraphics[trim = 2 2 2 50, clip, width=0.4\linewidth]{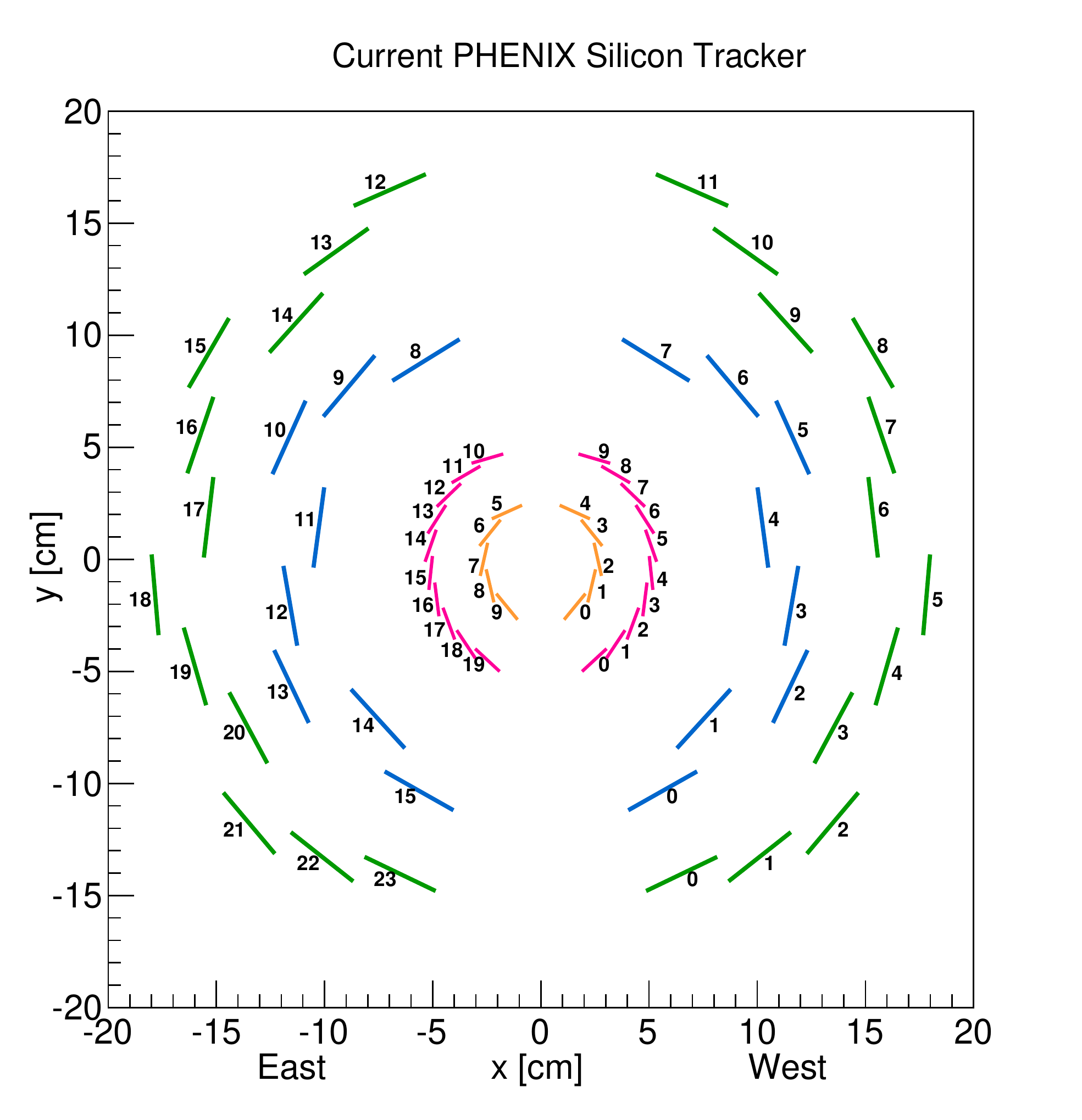}}
    \hfill
    \includegraphics[trim = 2 2 2 40, clip, width=0.55\linewidth]{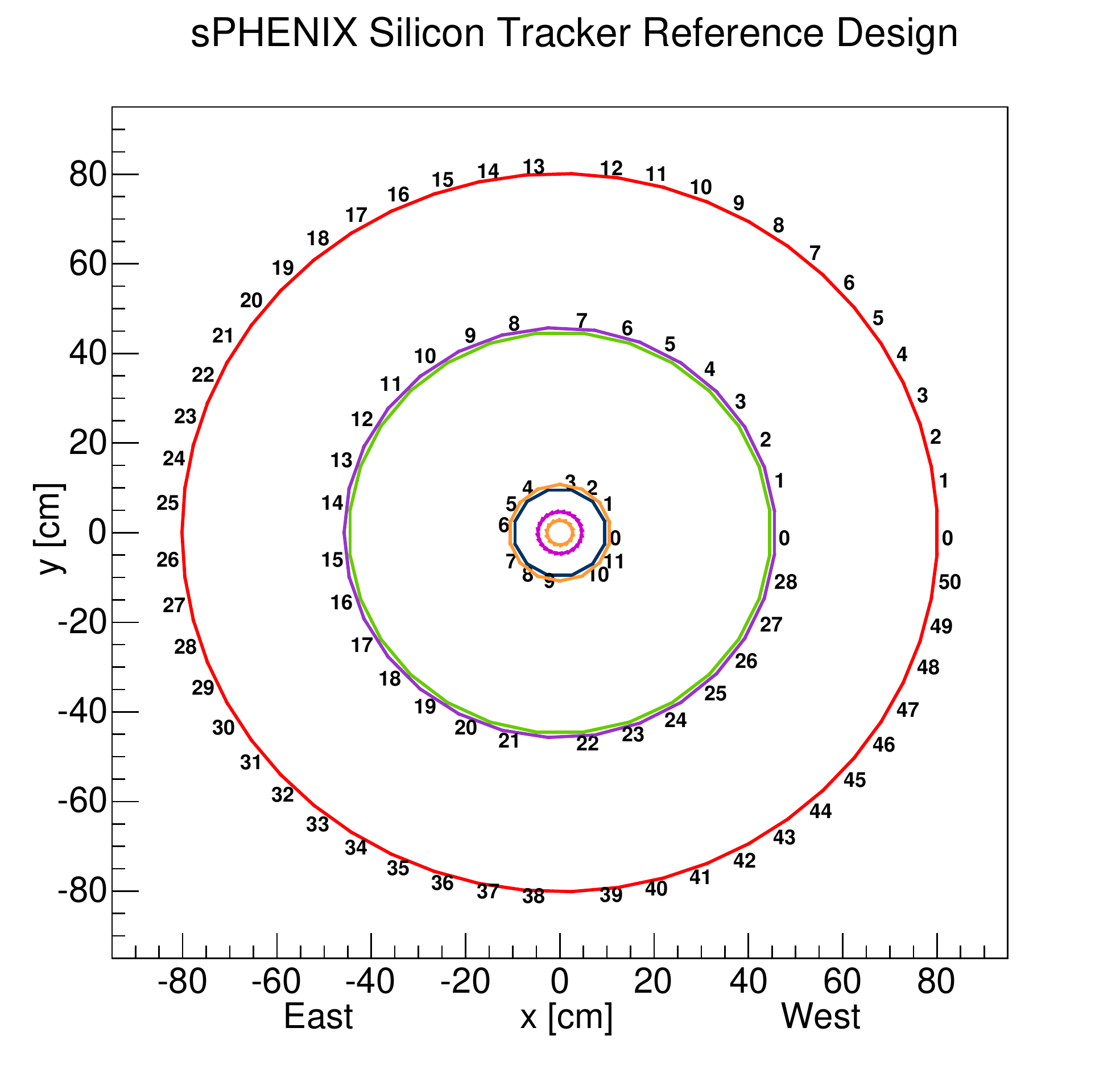}
    \caption[Current-day and reconfigured arrangement of silicon
    tracking layers in the PHENIX VTX]{\label{fig:tracking_evolve}
      (left) Present configuration of silicon tracking layers in the
      PHENIX VTX detector.  (right) Reconfiguration of the VTX inner
      two layers and additional tracking layers as described in the
      text.}
 \end{center}
\end{figure}

The current PHENIX silicon vertex tracker (VTX) consists of two inner
layers (pixels) at radii 2.5 and 5~cm from the beamline and two outer
layers (strip-pixels) at radii of 11.8 and 16.7~cm.  The ladders
comprising the current PHENIX VTX are shown in the left panel of
Figure~\ref{fig:tracking_evolve}.  The VTX, combined with the outer
PHENIX drift chambers (DC) and pad chambers (PC) provides good track
pattern recognition, high efficiency, and excellent displaced vertex
resolution with a specification for the distance of closest approach
resolution in the transverse plane of better than 100~$\mu$m for
$p_{T} > 1$~GeV/$c$.  This resolution is exceeded even in the high
occupancy \auau environment.

The \refconfig ~adds eight additional ladders to the two inner pixel
layers, thus completing azimuthal $2\pi$ coverage with the existing
$|\eta|<1.0$ coverage. In addition to the two inner pixel layers,
there will be five planes of strip detectors designed for precise
momentum measurement and pattern recognition in a high multiplicity
environment. Three of those layers will use strips of 60~$\mu$m pitch
and 8~mm length, and two will use strips of 240~$\mu$m pitch and 2~mm
length. The primary purpose of the latter two strip layers is pattern
recognition. Each of the two pattern recognition layers is mounted on
the same support and cooling structure as one of the longer strip
layers.  The lengths of the strips in the five outer layers represent
a compromise between cost and pattern recognition
performance. Table~\ref{table:tracker} summarizes the \refconfig
tracker layout.

\begin{table}[ht]
  \caption{The parameters of the \refconfig tracking layers.}
  \label{table:tracker}
  \centering
  \begin{tabular}{ccccccc} 
    \toprule
    Layer    & radius     & sensor pitch  & sensor length   & sensor depth & total thickness & area\\
             & (cm)       & ($\mu$m)     &  (mm)           & ($\mu$m)      &  \% $X_{0}$   & $m^2$    \\
    \midrule
    1        &  2.7       & 50            &   0.425         &  200           & 1.3           & 0.034    \\
    2        &  4.6       & 50            &   0.425         &  200           & 1.3           & 0.059    \\
    3        &  9.5       & 60            &   8             &  320           & 1.35          & 0.152    \\
    4        &  10.5      & 240           &   2             &  320           & 1.35          & 0.185    \\
    5        &  44.5      & 60            &   8             &  320           & 1             & 3.3      \\
    6        &  45.5      & 240           &   2             &  320           & 1             & 3.5      \\
    7        &  80.0      & 60            &   8             &  320           & 2             & 10.8     \\
    \bottomrule
  \end{tabular}
\end{table}

Charged particle tracks are reconstructed as follows. Modest
thresholds are applied on struck silicon cells. These thresholds
eliminate small energy deposits that are produced by low energy
spallation from passing particles, while preserving deposits from
lower momentum signal tracks that pass through the outer silicon
layers with large angles away from radial.  Adjacent hits are then
clustered, and the clusters are passed into the track pattern
recognition algorithm as a set of spatial points averaged from the
clustered hits. We employ a 5-dimensional Hough transform to locate
the helical hit patterns from tracks bending through the solenoid
field.  The large 5-d parameter space is spanned efficiently with low
memory overhead in the high occupancy of central heavy ion collisions
by a recursive search. The discovered track candidates are then passed
into a Kalman fitter assuming a constant magnetic field, and smoothing
is applied to measure the distance of closest approach (DCA) with
respect to the primary vertex. Some iterations are performed to
simultaneously determine the primary vertex position and the track
DCAs. Finally, a 1.6\% momentum recalibration is applied to account
for the small differences between the true field map of the BaBar
solenoid and the assumption of a constant field. We then select from
tracks sharing more than 3 hits the track with the best $\chi^{2}$ and
reject the others. This final rejection has minimal impact on the
track population for the \refdesign.
 
To evaluate how well the \refdesign and tracking software meet the key
requirements of the physics program, a full \geant simulation of the
tracking performance has been carried out using single particle events
and central \hijing \auau events --- with and without embedded single
particles.

\begin{figure}[!hbt]
 \begin{center}
    \includegraphics[width=0.6\linewidth]{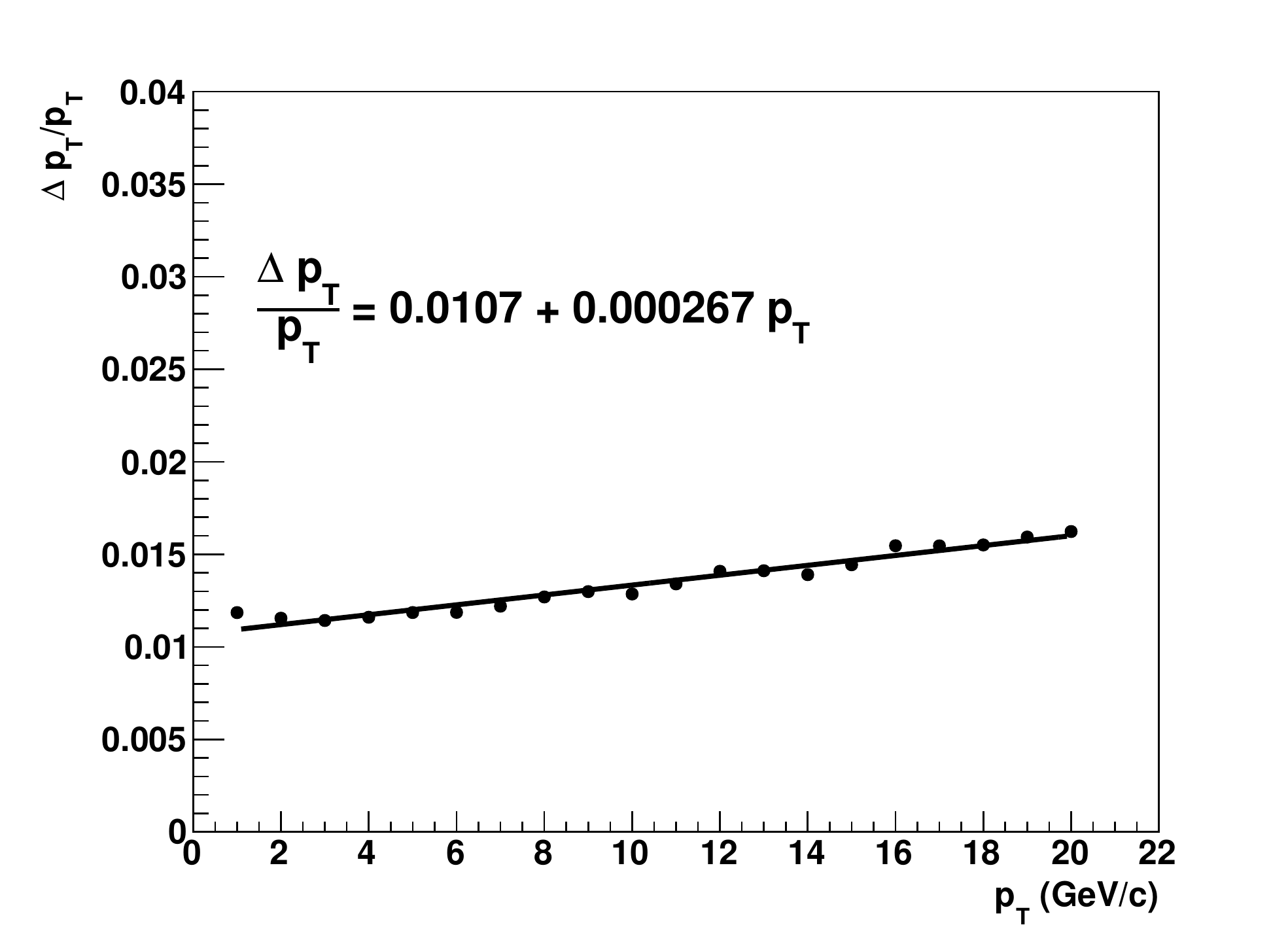}
    \caption[Full \geant simulation and track model evaluation of
    single pion \pt resolution]{\label{fig:mom_res}\geant and track
      model evaluation of single particle (pion) transverse momentum
      resolution. The fit consists of a term that is constant in \pt,
      and a term that is linear in \pt. The best fit parameters are
      shown on the plot.}
 \end{center}
\end{figure}

The \pt resolution for single pions is shown as a function of \pt in
Figure~\ref{fig:mom_res}. The constant term, which is due to multiple
scattering in the material of the tracker, is found to be 1.1\%.  The
linear term, determined by the position resolution of the tracker, is
2.7 x $10^{-4} (\mathrm{GeV}/c)^{-1}$.  This momentum resolution
leads to a mass resolution of just under 100~MeV for the
$\Upsilon(1S)$ state, which is sufficient to deliver the physics of
separate measurements of the Upsilon states. The momentum resolution
of the \refdesign is more than adequate for the less demanding (in
terms of momentum resolution) tasks of measuring heavy flavor tagged
jets and high-$z$ fragmentation functions.

The performance of the tracking system in high multiplicity events has
been investigated using a full \geant simulation of the tracker
response for 5000 \hijing \auau events with impact parameters in the
range 0-4 fm. This impact parameter range corresponds to about 0-10\%
collision centrality. For these studies only tracks that hit all seven
layers of the tracker were reconstructed.  To eliminate fake tracks,
cuts were made on the track quality ($\chi^2$ per degree of freedom)
and on the track distance of closest approach to the event vertex
(DCA). The track quality was required to satisfy quality $< 3$, and
the track DCA was required to satisfy DCA $< 1$~mm.

\begin{figure}[!hbt]
 \begin{center}
    \includegraphics[trim = 0 0 0 35, clip, width=0.5\linewidth]{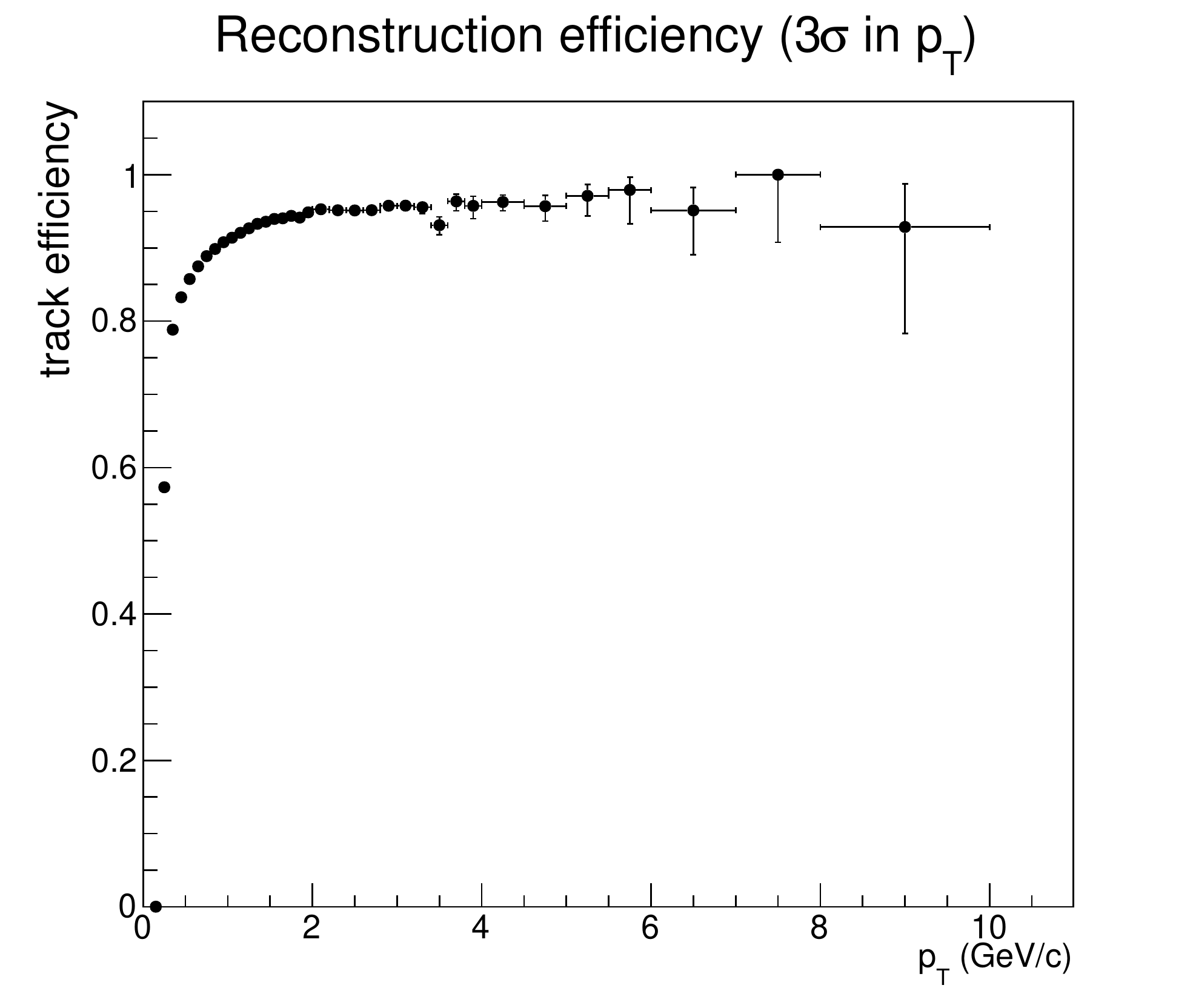}
    \hfill
    \includegraphics[width=0.44\linewidth]{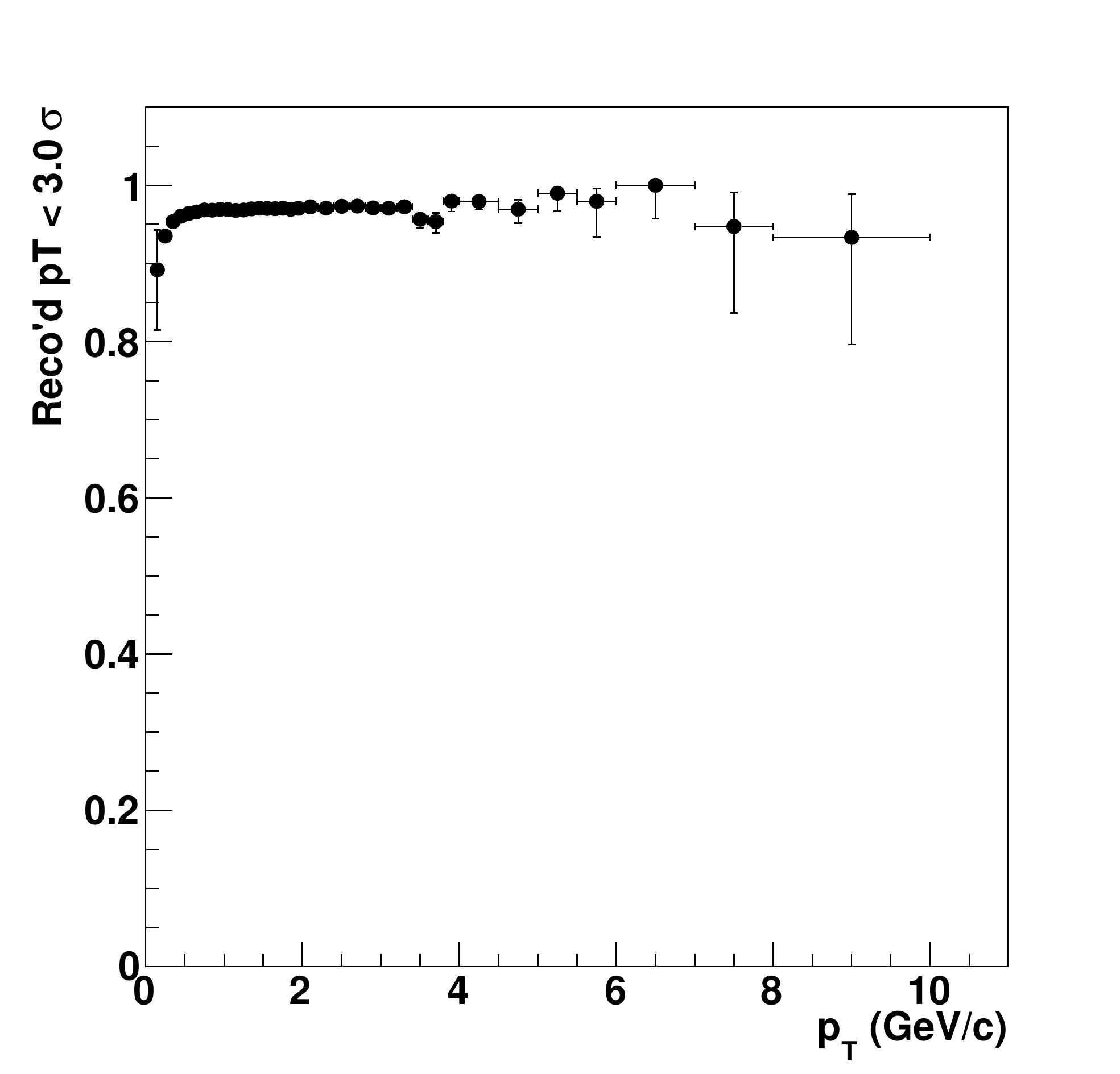}
    \caption[Fraction of primary tracks producing hits in all seven
    tracking layers that are reconstructed and fraction of
    reconstructed tracks whose reconstructed momentum lies within
    3$\sigma$ of the truth momentum of the associated \geant
    track]{\label{fig:track_reco_eff_purity} (left) The fraction of
      \geant tracks from the primary vertex with hits in all seven
      tracking layers that are reconstructed with quality $< 3$ and
      DCA $< 1$~mm, and whose momentum lies within 3$\sigma$ of the
      truth momentum. Only tracks that hit all seven layers were
      considered. (right) The fraction of all reconstructed tracks
      (passing cuts of quality $< 3$ and DCA $< 1$~mm) that also have
      reconstructed momentum within 3$\sigma$ of the truth momentum
      for the associated \geant track. }
 \end{center}
\end{figure}

To define the track reconstruction efficiency we start by counting all
truth tracks that originated at the primary vertex and deposited
energy in all seven layers. This is the denominator. The numerator is
then the number of reconstructed tracks that pass track cuts of
quality $< 3$ and DCA $< 1$~mm, and whose momentum lies within
$3\sigma$ of the truth momentum for the associated \geant track, The
resulting efficiency for 5000 \hijing events is shown in the left
panel of Figure~\ref{fig:track_reco_eff_purity}.  The efficiency is
found to be 88\% at 500~MeV/c, 92\% at 1 GeV/c and 97\% at high \pt.

Another way to look at the pattern recognition performance is to start
with all reconstructed tracks that have quality $< 3$ and DCA $<
1$~mm, and see what fraction of them satisfy the additional
requirement that their reconstructed momentum is within $3\sigma$ of
the truth momentum for the associated \geant track. The result from
5000 central \auau \hijing events is shown in the right panel of
Figure~\ref{fig:track_reco_eff_purity}.

\begin{figure}[!hbt]
 \begin{center}
    \includegraphics[width=0.9\linewidth]{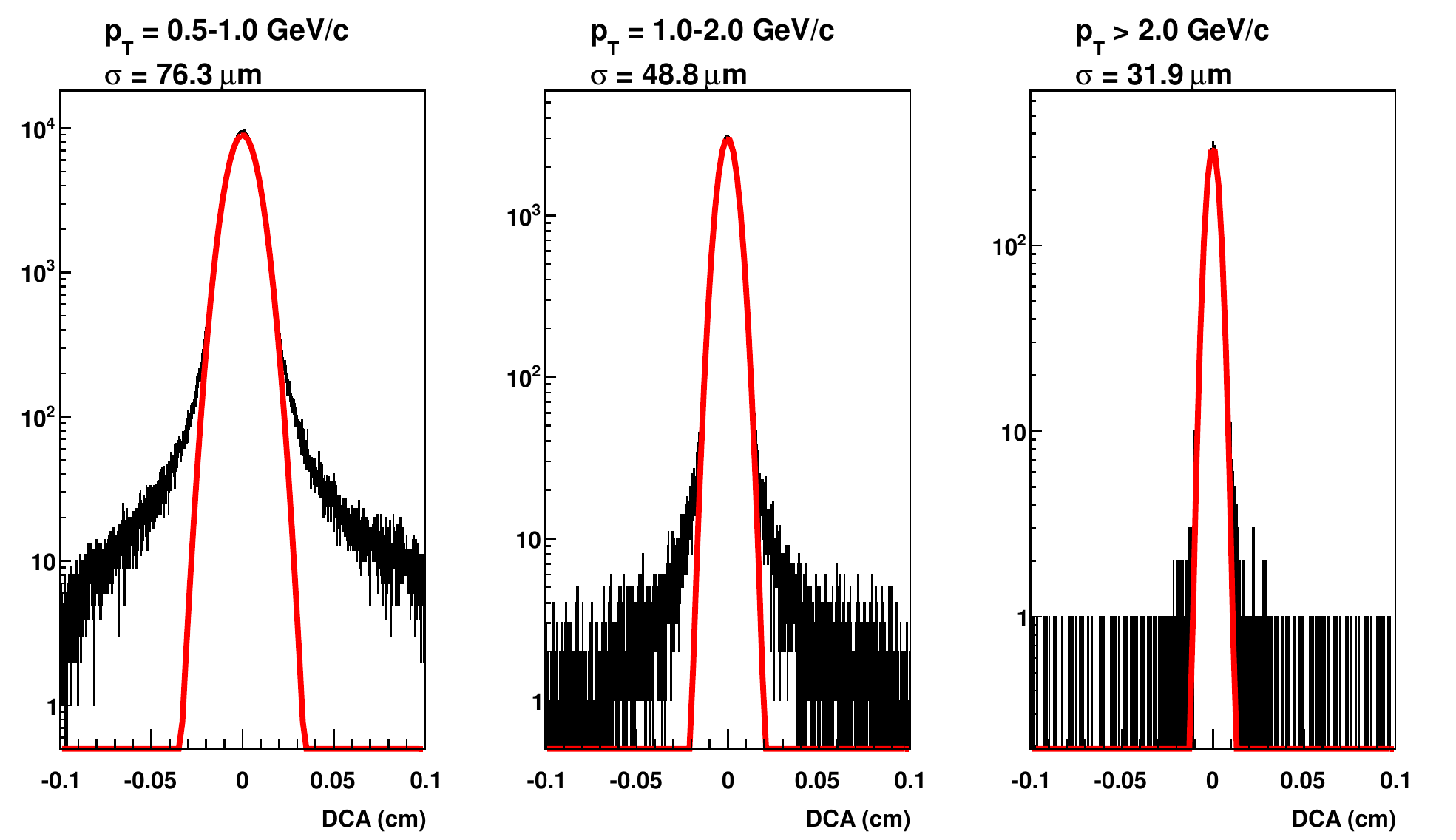}
    \caption{\label{fig:dca_hijing_3ptbins_svtxv1} DCA distributions
      in three \pt bins from reconstruction of 5000 central \hijing
      events.}
 \end{center}
\end{figure}

Heavy flavor tagged jet measurements rely critically on the DCA
resolution performance of the tracking system.
Figure~\ref{fig:dca_hijing_3ptbins_svtxv1} shows the DCA distribution
obtained from 5000 central \auau \hijing events in three \pt bins.
The distributions were made using all reconstructed tracks, with the
only track cut being quality $< 3$.

\begin{figure}[!hbt]
 \begin{center}
    \includegraphics[width=0.6\linewidth]{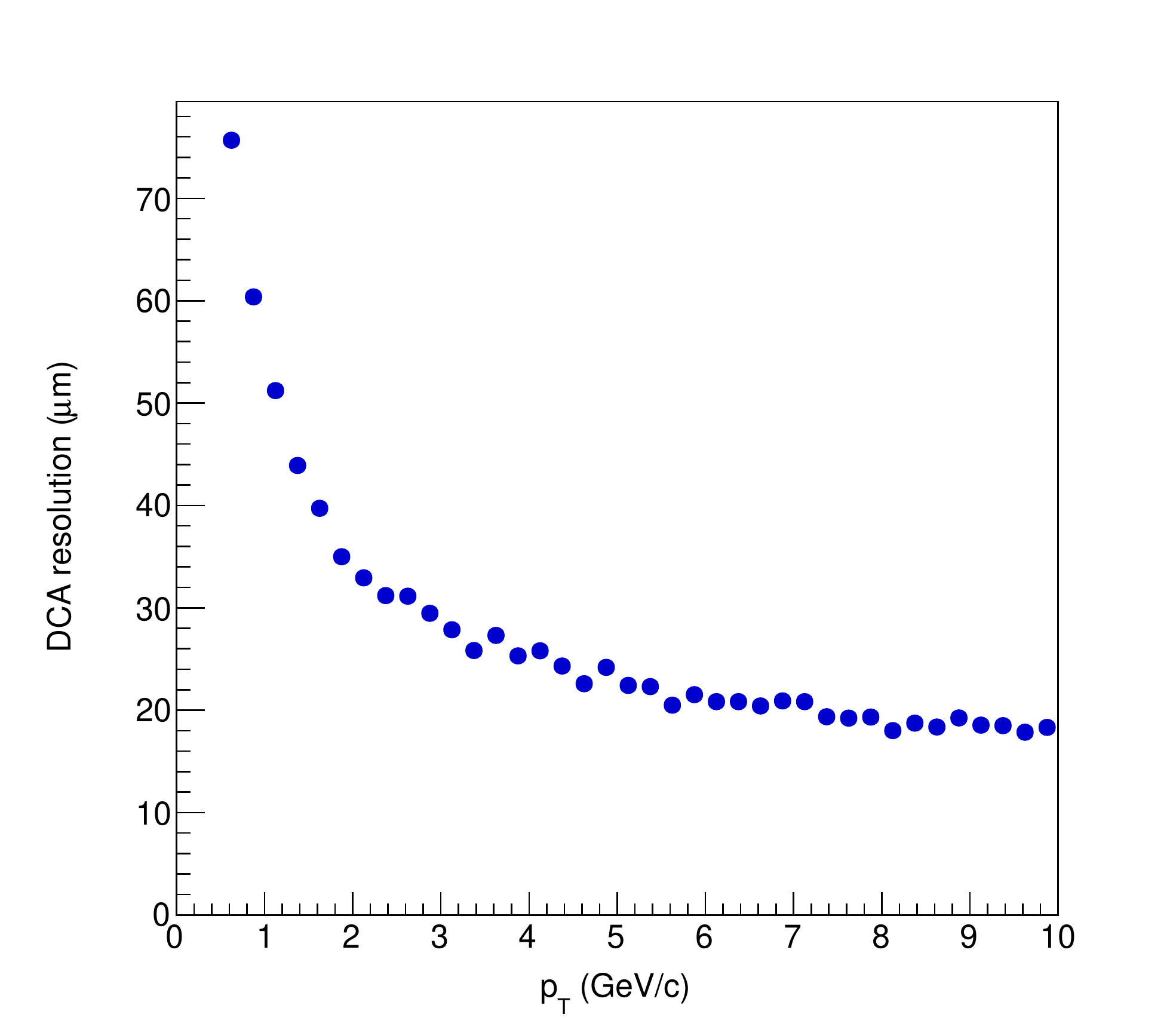}
    \caption{\label{fig:dca_ptbins} DCA resolution versus \pt from
      simulations with pions embedded in central \hijing events.}
 \end{center}
\end{figure}

We have also extracted the DCA resolution as a function of \pt for
single pions embedded in central \hijing events.  The result is shown
in Figure~\ref{fig:dca_ptbins}.  The standard track cuts of quality $<
3$ and DCA $< 1$~mm were applied. Because the embedded pions were
placed at the event vertex, this directly measures the DCA resolution
in each \pt bin.

These simulation results show that the \refdesign is capable of
delivering the momentum resolution, tracking efficiency, track purity
and displaced vertex resolution needed for the sPHENIX physics
program. We are considering possible modifications to the \refdesign
that would maintain the same performance but may reduce cost or add
redundancy. For example we are evaluating the possibility of using
pairs of stereo strips, inclined at a small angle, for layers 3/4 and
5/6. This would maintain the same tracking and pattern recognition
performance, but may preserve good track efficiency even if some
channels are lost. We are also looking critically at whether the
material budget can be reduced. Because the momentum resolution in the
range relevant for the Upsilon measurement is dominated by multiple
scattering in the tracker, reducing the tracker thickness would allow
us to reduce the radius of the outer tracking layer, translating to
lower cost for the same performance.  There is ongoing tracking R\&D,
particularly driven by interest in future use in an EIC detector, that
may inform our particular design choices.  We are also weighing the
cost and performance balance of other possible tracking options, such
as a potential time-projection chamber in place of the outer silicon
tracking layers in the reference design.

\clearpage 

\makeatletter{}\section{Electron Identification} 
\label{sec:electronid}

For the beauty quarkonia measurements (further discussed in
Section~\ref{sec:quarkonia_signal}), the electron track candidates
from the decayed Upsilon are identified using a combination of the
electromagnetic calorimeter (EMCal) and the inner hadron calorimeter
(Inner HCal).  The main backgrounds to reject are the hadron tracks,
which produces a continuous background under the Upsilon invariant
mass peaks (as simulated in Figure~\ref{fig:quarkonia_bg}).  To reject
this background, an EMCal energy matching with the track momentum and
a leaked energy veto in the Inner HCal are used.  By simulating the
full calorimeter system in \geant, the electron identification~(eID)
efficiency was studied against pion rejection for \pp and central
\auau events.

\begin{figure}[hbt!]
  \centering
  \includegraphics[clip,width=0.5\linewidth]{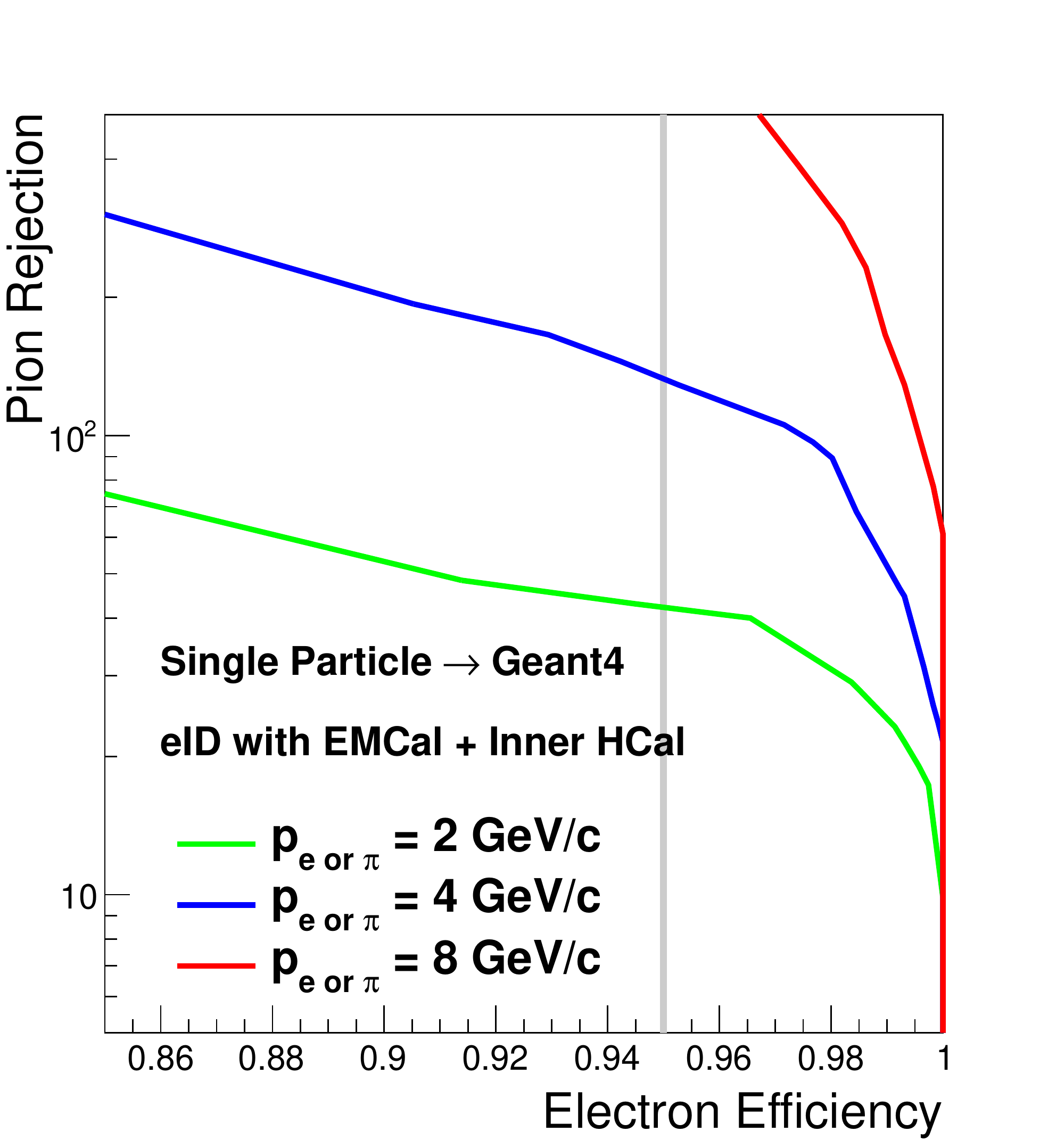}
  \caption[Electron ID efficiency versus pion rejection near central
  rapidity ($|\eta|<0.2$) based on a \geant simulation of single
  particles]{Electron ID efficiency versus pion rejection near central
    rapidity ($|\eta|<0.2$) for a \geant simulation of single
    particles of 2~(green), 4~(blue) and 8~(Red)~GeV$/c$ in the total
    momentum.  The vertical gray band highlights the proposed eID cut
    for \pp events, which corresponds to 95\% single electron ID
    efficiency.}
 \label{fig:electronid_pp}
\end{figure}

In \pp~collisions, the underlying event activity is low within the
shower size in the calorimeter.  Therefore, the eID performance is
studied using single track simulations as shown in
Figure~\ref{fig:electronid_pp}.  In this study, single events
containing an electron or negatively charged pion shower are simulated
in the full calorimeter system using \geant.  The cluster is built
around the initial track projection for each layer of the
calorimeters, which roughly corresponding to the size of $3\times3$
towers in that layer.

The electron track candidate is identified using a two-dimensional
likelihood analysis based on both EMCal and Inner HCal cluster
energies.  The usage of the Inner HCal information improves the
rejection by roughly a factor of two and the cut value is around the
level of 250~MeV (about 1~MIP) at 90\% eID efficiency.  The average
momentum for Upsilon-decayed electron is between the blue
($p=4$~GeV$/c$) and red curve ($p=8$~GeV$/c$), which corresponds to
better than 100:1 pion rejection for 95\% electron efficiency.

In central \auau~collisions, the underlying event fluctuation is quite
significant within the electron shower clusters.  Therefore, eID
becomes more challenging.  Nevertheless, the eID performance is
quantified for the most challenging environment of the central 0--10\%
\auau collisions, by embedding the above single-track candidates into
the full event \hijing and \geant simulations.  Comparing to the
EMCal, the Inner HCal picks up significant amount of background energy
due to its large cluster area size.  Therefore, eID in this study is
based on the EMCal cluster energy only, which is matched against the
sum of the expected electron track and the average background energy.

\begin{figure}[hbt!]
  \centering
  \includegraphics[clip,width=0.5\linewidth]{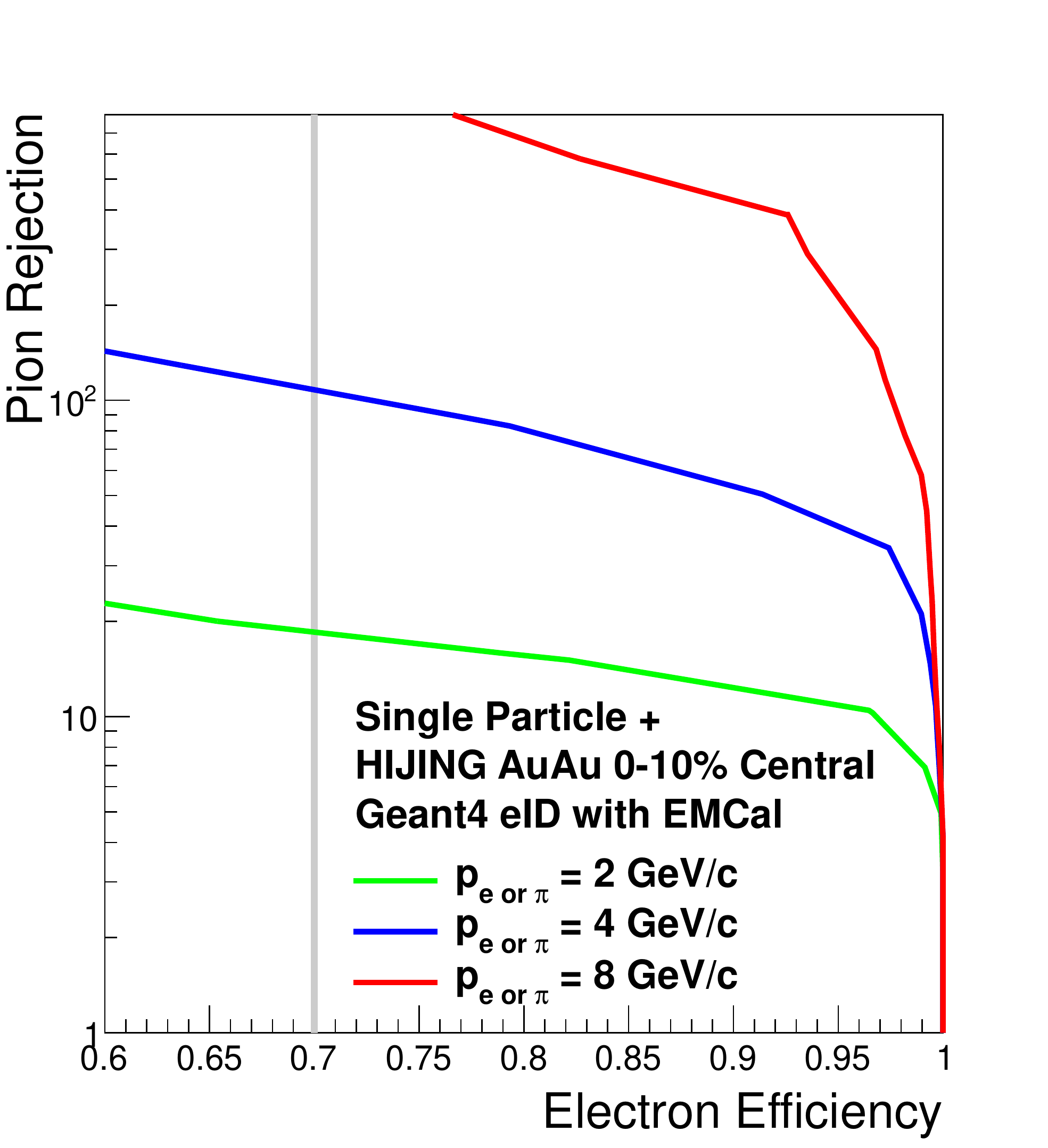}
  \caption[Electron ID efficiency versus pion rejection near central
  rapidity ($|\eta|<0.2$) based on a \geant simulation of single
  particles embedded in 0--10\% central \hijing events]{After embedding
    into 0--10\% central \hijing event, the electron ID efficiency
    versus pion rejection in the central rapidity ($|\eta|<0.2$) for
    particles of 2~(green), 4~(blue) and 8~(Red)~GeV$/c$ in the total
    momentum.  The vertical gray band highlights the proposed cut for
    the central \AuAu events, which corresponding to 70\% single
    electron ID efficiency.}
 \label{fig:electronid_AA}
\end{figure}

The efficiency-rejection curves for three typical momentum are
calculated again based on momentum-dependent likelihood analysis of
EMCal cluster energies, as shown in
Figure~\ref{fig:electronid_AA}. For $p=4$~GeV$/c$ tracks (blue
curves), as a conservative estimation for the average momentum tracks
for Upsilon-decayed electron candidates, the pion rejection is roughly
100:1 at 70\% of electron efficiency as highlighted by the gray
vertical line.

Our studies indicate that the electron ID performance is sufficient
for the beauty quarkonia measurements described in
Section~\ref{sec:quarkonia_signal}.  Meanwhile, the forward
pseudo-rapidity electron ID is still being optimized, as the details
for the calorimeter towering structure are being developed.

\makeatletter{}\section{Rates and DAQ}
\label{Section:Rates}

\begin{figure}[htb!]
 \begin{center}
   \includegraphics[width=0.45\linewidth]{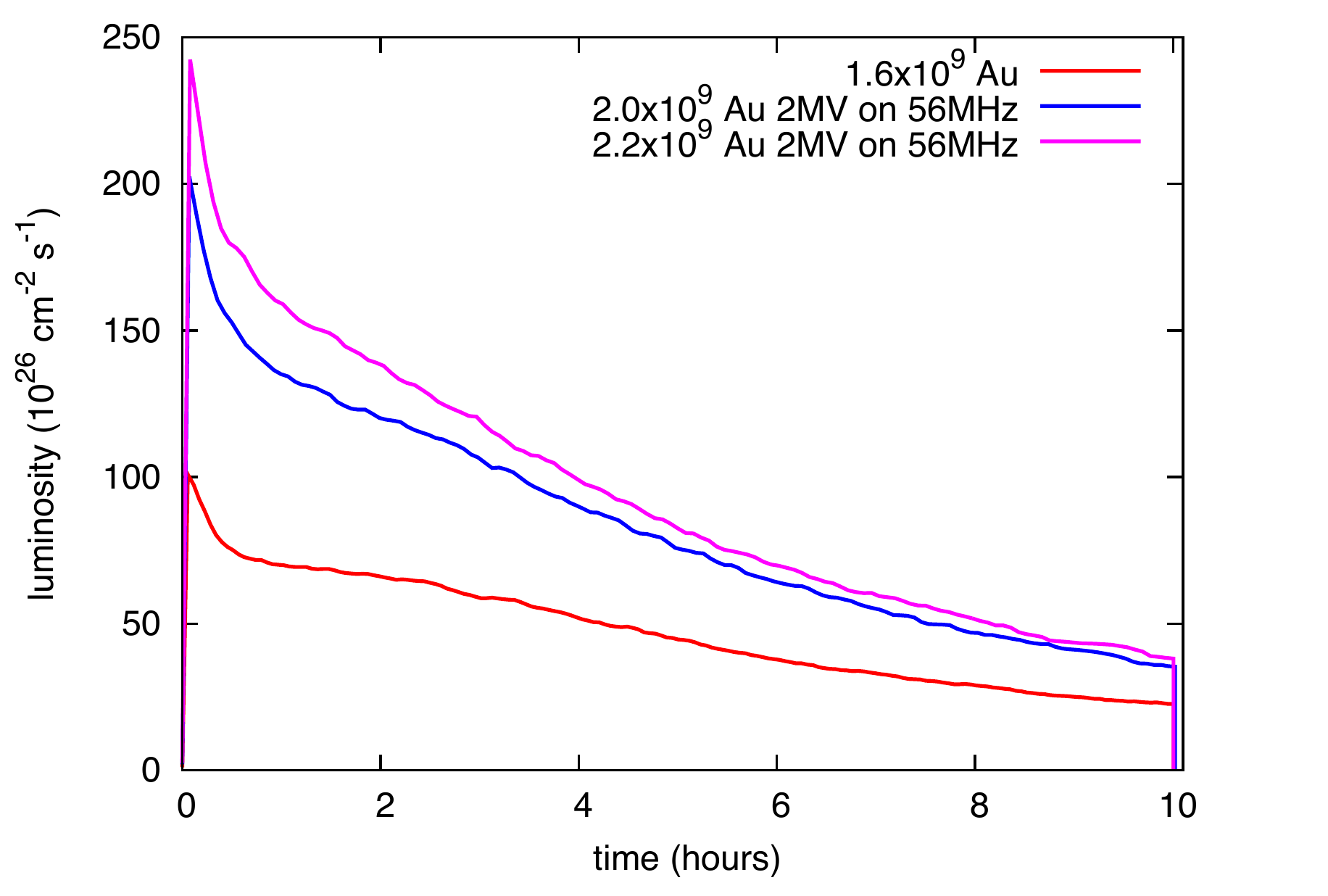}
   \includegraphics[width=0.45\linewidth]{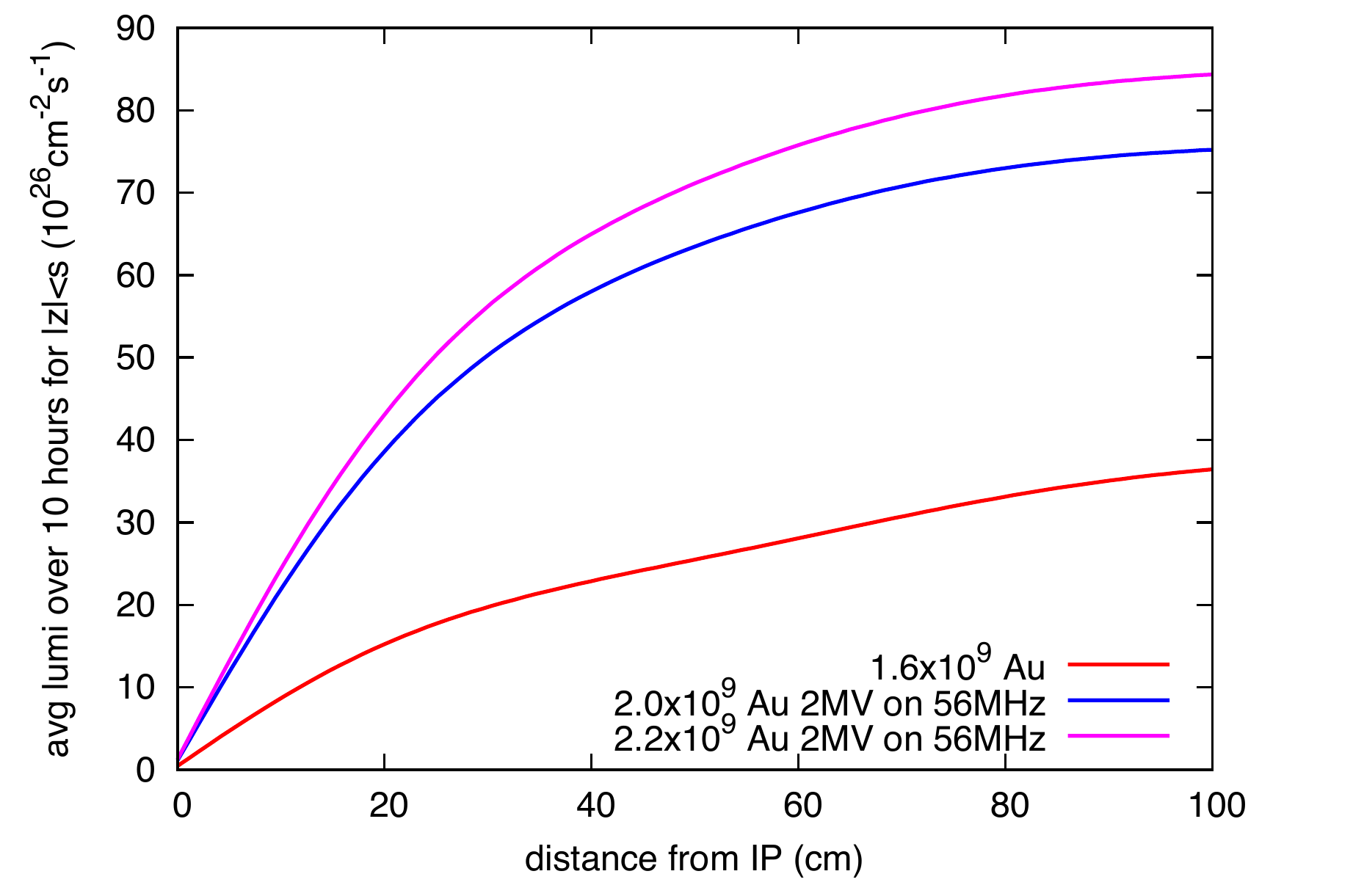}
   \caption[C-AD projections of instantaneous luminosity vs time in
   store and average luminosity as a function of the integration range
   around the nominal interaction point for \auau at 200~GeV]{(Left)
     Projections of instantaneous luminosity versus time in store for
     \auau at 200~GeV. (Right) Projections of average luminosity, as a
     function of the integration range around the nominal interaction
     point.  In both plots, the effect of the 56~MHz RF system is
     apparent.}
   \label{fig:timeinstore} 
 \end{center}
\end{figure}

A critical aspect of the sPHENIX detector is the ability to collect large data samples for high statistics jet and upsilon measurements.
The Collider-Accelerator Division (C-AD) has updated their RHIC Collider Projections as documented in Ref.~\cite{RHICBeam}.
For \auau collisions at 200 GeV, in the years 2021-2022, store luminosities in excess of $150 \times 10^{26}$ cm$^{2}$s$^{-1}$ are expected.   
They project that 35\% of those interactions will take place within the select z-vertex range $|z|<10$ cm.   
These projections represent an increase in delivered luminosity more than
a factor of two above the 2014 \auau achieved average numbers.   The interaction rate as a function of time-in-store from 
these projections is shown in Figure~\ref{fig:timeinstore}.  

The backbone of the PHENIX data acquisition system, which is the basis of the sPHENIX system, is the fully
pipelined and so-called ``dead-timeless'' Global Level-1 Trigger system and Granule Timing architecture.   The design limits the
maximum Level-1 Trigger accept rate to 25 kHz.    Currently the PHENIX silicon pixel layers (VTX) are planned for re-use in the 
inner sPHENIX tracker.   Tests show that rate above 15 kHz are achievable with the current VTX electronics with data
transmission to the Data Collection Module II boards with a modest growth in occupancy at higher luminosity.  
A Level-1 Trigger accept rate of 15 kHz for the reference design of the
entire system is a good match to the delivered luminosities from the C-AD projections.   
This would allow the recording without
any additional trigger bias of more than half of all collisions within $|z|<10$ cm at the very highest luminosities.    
At these highest luminosities,
many of the rarest probes can be sampled with more selective Level-1 triggers, as detailed in Section~\ref{Section:Trigger}.

Thus, for a 22 week physics running period of \auau at 200 GeV, sPHENIX with an uptime of 80\%, would record 100 billion
minimum bias events with $|z|<10$ cm.  More selective triggers could sample slightly less than a factor of two more events 
again within $|z|<10$ cm.   For observables not requiring the inner silicon tracking which has the more restrictive coverage,
one would utilize collisions over the much larger range $|z|<30$ cm, and sample with selective Level-1 triggers 0.6 trillion events.
As detailed in Section~\ref{Section:Trigger}, direct photons and purely calorimetric high energy jets would be able to utilize the
full 0.6 trillion events.  

The luminosity in \pp collisions is limited by the beam-beam tune shift, which will be reduced by fully operational electron
lenses.  This will bring the average store luminosity to $1.7 \times 10^{32}$ cm$^{2}$s$^{-1}$ at 200 GeV 
and $7.1 \times 10^{32}$ cm$^{2}$s$^{-1}$ at 510 GeV, a factor of two to three times the projected 2015 and 2016 average luminosity.   
C-AD projects for \pp collisions at 200 GeV delivering 63 pb$^{-1}$ per week over all vertices.   They project 35\% of these collisions to
be within $|z| < 10$ cm.   Accounting for sPHENIX uptime and projecting a 10 week physics data taking period, one can effectively
sample 500 pb$^{-1}$ over all z-vertices and 175 pb$^{-1}$ over $|z| <$ 10 cm.

For the $p$+$Au$ at 200 GeV case, the C-AD projection is delivering 400 nb$^{-1}$ per week over all vertices.  They project 30\% 
of these collisions within $|z| <$ 10 cm.    Again, accounting for sPHENIX uptime and projecting a 10 week physics data taking period,
one can effectively sample 3200 nb$^{-1}$ over all z-vertices and 960 nb$^{-1}$ over $|z| <$ 10 cm.
As detailed in the Physics Performance Chapter, the \pp and $p$+$Au$ data sets provide very robust baseline and cold-nuclear
matter statistics.

\makeatletter{}\section{sPHENIX Triggering}
\label{Section:Trigger}
Collider experiments typically require rather sophisticated trigger
capabilities to sample the rare physics of interest from the large
number of ``uninteresting'' collisions.  In the case of sPHENIX, for
many jet observables, selective triggering biases the physics of
interest and results in covering only a partial phase space of jets
(e.g. jets originating from partons emitted near the surface of the
medium).  For \auau collisions, the high bandwidth and deadtimeless
nature of the sPHENIX data acquisition system allows us to record
(with only a global \auau interaction or minimum bias trigger) a data
sample of 100 billion events within a z-vertex $|z|<10$ cm,
corresponding to the optimal acceptance of the silicon tracking
system.  During that same time period, a total of 0.6 trillion \auau
interactions over a wider z-vertex range will have taken place that
can be sampled with modestly selective triggering.  In addition,
critical trigger requirements are relevant for \pp and \pdau running,
and the lower occupancy environment simplifies the task at hand.

There are three systems where we plan for inputs to the sPHENIX Level-1 trigger system.   The current requirement is a modest 4.0 microsecond
(i.e. 40 beam clock) latency on the trigger decision.  The first two systems with trigger inputs are the electromagnetic and hadronic calorimeters.
Both systems utilize a common electronics that digitized the signals at full clock speed, thus removing any need for having separate trigger
thresholds applied to split analog signals.   The reference design has a set of full digitized energy values with a modest bit number reduction
collected into one module from the entire calorimeter systems (of order 25,000 channels).   Thus, one has full information in a set of FPGAs to apply a 
variety of trigger selections:

\begin{itemize}
\item Total electromagnetic energy, hadronic energy, and both
\item Jet patch energy sums including with average underlying event subtraction
\item Cluster energy in the electromagnetic calorimeter, or cluster pairs with geometric configurations
\end{itemize}

The third detector with input is envisioned utilizing the current PHENIX Beam-Beam Counters (BBC), described in detail in Ref.~\cite{Allen:2003zt}.
The BBC consists of 128 channels of quartz radiators with PMTs on each side in the z-direction of the collision point.  The detectors would be moved 
further back just outside the current design for the magnet flux return.   They would thus be moved about one unit further forward in
pseudorapidity from their current configuration.  In \auau collisions, these would provide a precision $\sigma < 1$ cm z-vertex resolution for
Level-1 triggering and an independent centrality and event plane determination.   

For the jet physics program with observables for single jets, high momentum photons, and high momentum hadrons, the electromagnetic calorimeter
cluster trigger inputs and jet patch capabilities are sufficient for sampling the full 600 billion events for the highest energies where the
increase in \auau statistics is particularly beneficially.  This triggering also works well in \pp and \pdau collisions, with the interaction
rates projected.

\begin{figure}[!hbt]
 \begin{center}
    \includegraphics[width=0.60\linewidth]{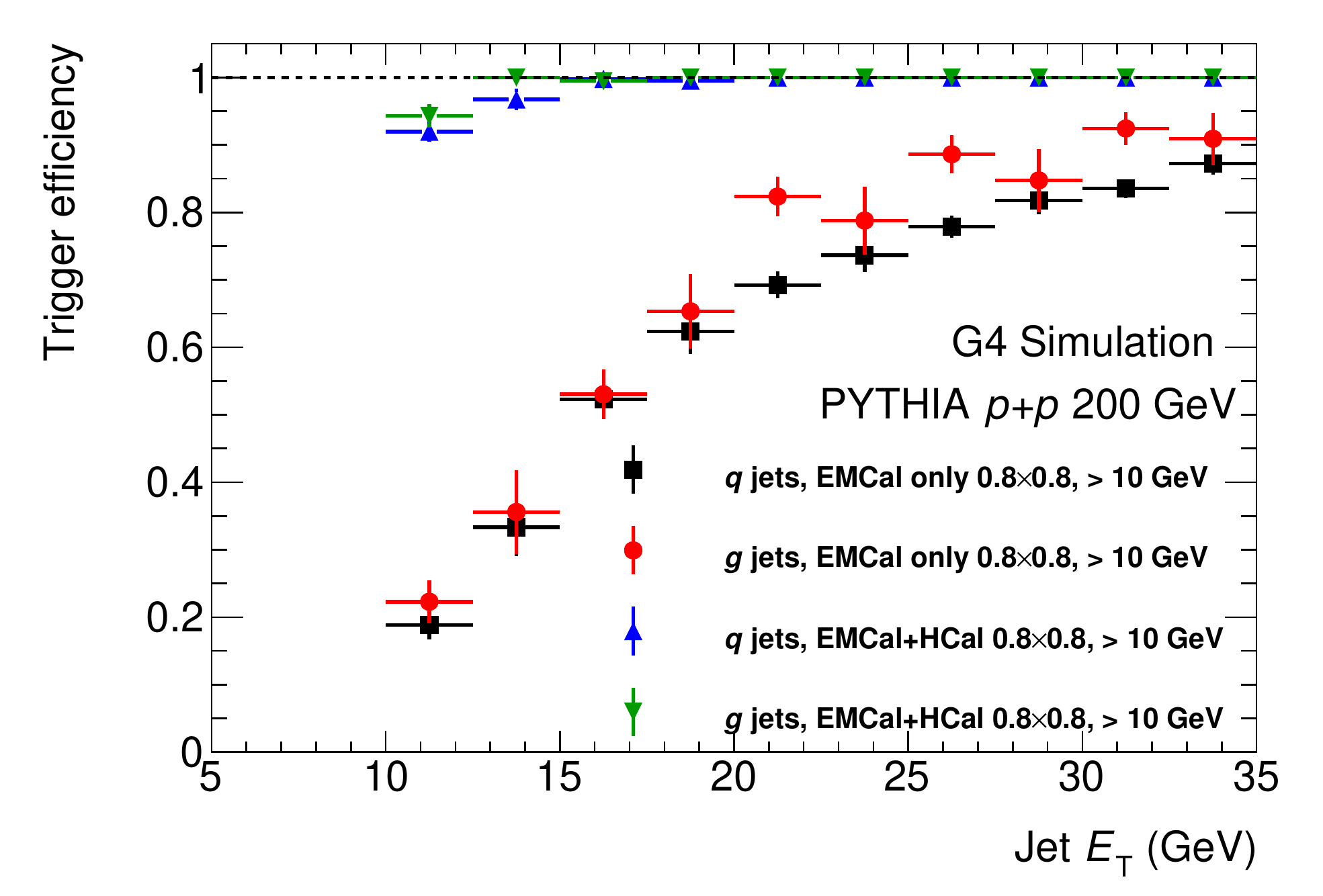}
    \caption[Trigger efficiency for jets using \geant simulated
    calorimeter-based triggers as a function of $R=0.4$ truth-level
    jet $E_\mathrm{T}$, with results for quark- and gluon-initiated
    jets shown separately]{\label{fig:sPHENIX_trigger_jet1} Trigger
      efficiency for jets using \geant simulated calorimeter-based
      triggers as a function of $R=0.4$ truth-level jet
      $E_\mathrm{T}$, with results for quark- and gluon-initiated jets
      shown separately. Results are shown for triggers requiring at
      least $E_\mathrm{T} > 10$~GeV in a
      $\Delta\eta\times\Delta\phi=0.8\times0.8$ calorimeter
      region. The efficiency using the electromagnetic calorimeter
      (EMCal) only is shown in black and red for quark- and gluon
      jets, respectively, and the efficiency using both calorimeters
      (EMCal+HCal) is shown in blue and green for quark- and
      gluon-jets, respectively.}
 \end{center}
\end{figure}

We have benchmarked the performance of possible ``jet patch'' triggers
in high-luminosity $p$+$p$ collisions implemented by examining the sum
of tower energies in the electromagnetic and hadronic calorimeters. In
this study, \pythia jet events with a hard scattering parameter
chosen to sample a wide kinematic range and minimum bias \pythia
events are examined under a \geant simulation of the calorimeter
response. The calorimeters are towerized into towers of size
$\Delta\eta\times\Delta\phi=0.1\times0.1$, and the total transverse
energy from both calorimeters is analyzed in sliding tower windows of
various sizes. For jet events, the highest energy window within
$\Delta{R}<0.4$ of the jet is considered for the purposes of
determining whether the jet fired the trigger. For minimum bias
events, the highest energy window anywhere in the event is considered.

Figure~\ref{fig:sPHENIX_trigger_jet1} illustrates the relevant results
for window sizes of $0.8\times0.8$, with a minimum transverse energy
requirement of 10 GeV. When both calorimeters are used for triggering,
the efficiency for $E_\mathrm{T} > 15$~GeV jets is unity, with no
dependence on the flavor of the jet. To demonstrate the importance of
using both calorimeters in the trigger, results are also shown for the
efficiency of an electromagnetic calorimeter-based trigger only. In
that case, it can be seen that the efficiency has a non-trivial jet
$E_\mathrm{T}$ dependence and reaches only $\epsilon\approx85$--$90\%$
even for $E_\mathrm{T}=35$~GeV jets. Furthermore, a systematic
difference can be observed between quark-- and gluon--initiated
jets. Thus, wide-area jet patch triggers utilizing both calorimeters
can most efficiently select an unbiased set of jets.

\begin{figure}[!hbt]
 \begin{center}
    \includegraphics[width=0.60\linewidth]{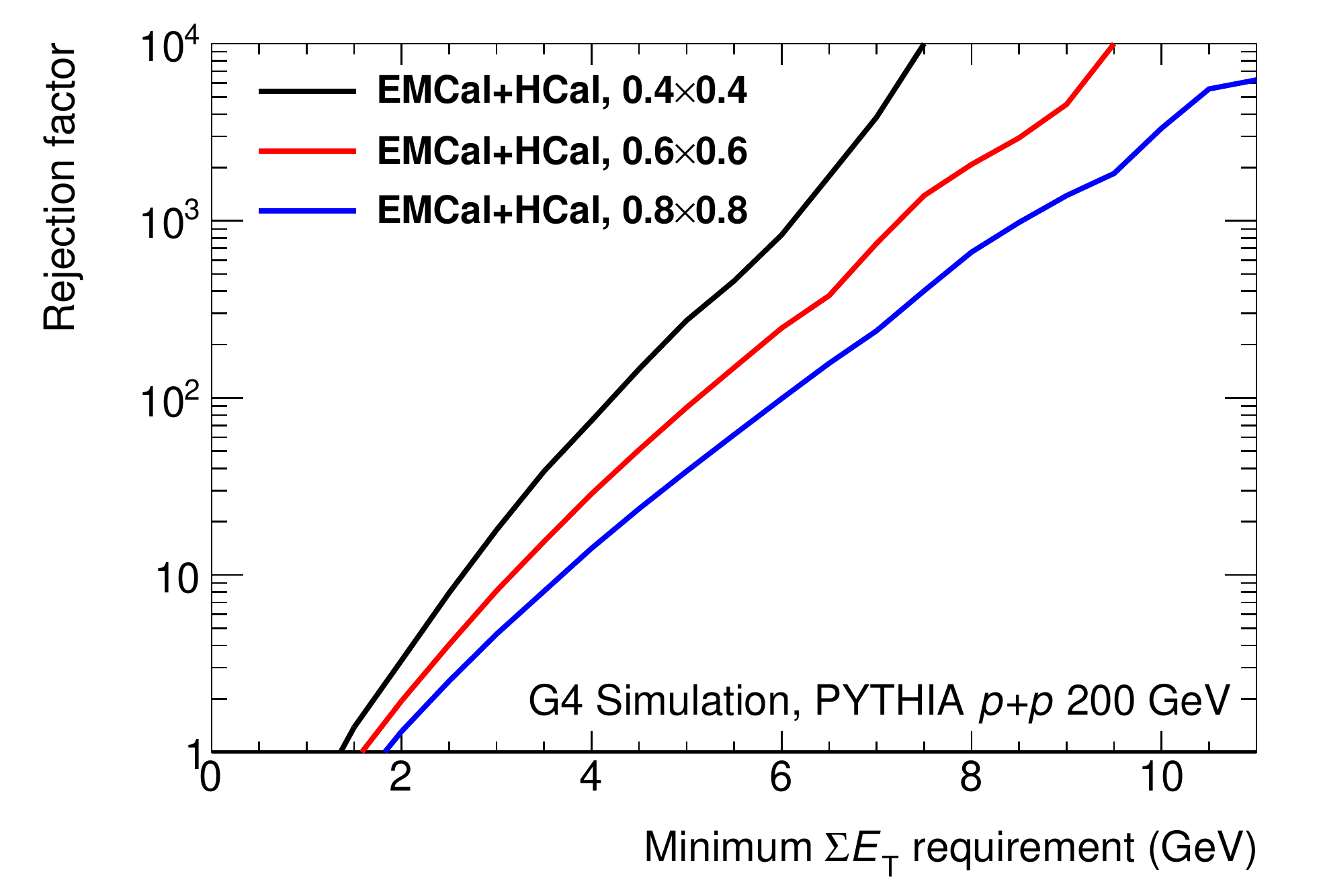}
    \caption[Rejection factor for minimum bias \pp events using \geant
    simulated calorimeter-based triggers, as a function of the minimum
    $E_\mathrm{T}$ trigger
    requirement]{\label{fig:sPHENIX_trigger_jet2} Rejection factor for
      minimum bias \pp events using \geant simulated calorimeter-based
      triggers, as a function of the minimum $E_\mathrm{T}$ trigger
      requirement. Results are shown for requiring this amount of
      energy in $\Delta\eta\times\Delta\phi=0.4\times0.4$ (black
      line), $0.6\times0.6$ (red line) and $0.8\times0.8$ (blue line)
      calorimeter regions.}
 \end{center}
\end{figure}

Figure~\ref{fig:sPHENIX_trigger_jet2} shows the rejection factor (the
inverse of the fraction of events which fire the trigger) for minimum
bias $p$+$p$ events of various electromagnetic and hadronic
calorimeter jet patch trigger schemes. The rejection factor is shown
for three choices of sliding window size and as a function of the
minimum required transverse energy. The figure demonstrates that a
minimum energy can be chosen to give rejection factors larger than
$10^{3}$, which will be necessary in high-luminosity $p$+$p$ and
$p(d)$+Au running.

Taken together, the results in
Figures~\ref{fig:sPHENIX_trigger_jet1}~and~\ref{fig:sPHENIX_trigger_jet2}
demonstrate that jet patch style triggers in sPHENIX will be
sufficient to sample an unbiased set of jets down to low
$E_\mathrm{T}$ while still providing the large rejections needed for
high-luminosity running.

\begin{figure}[!hbt]
 \begin{center}
    \includegraphics[width=0.60\linewidth]{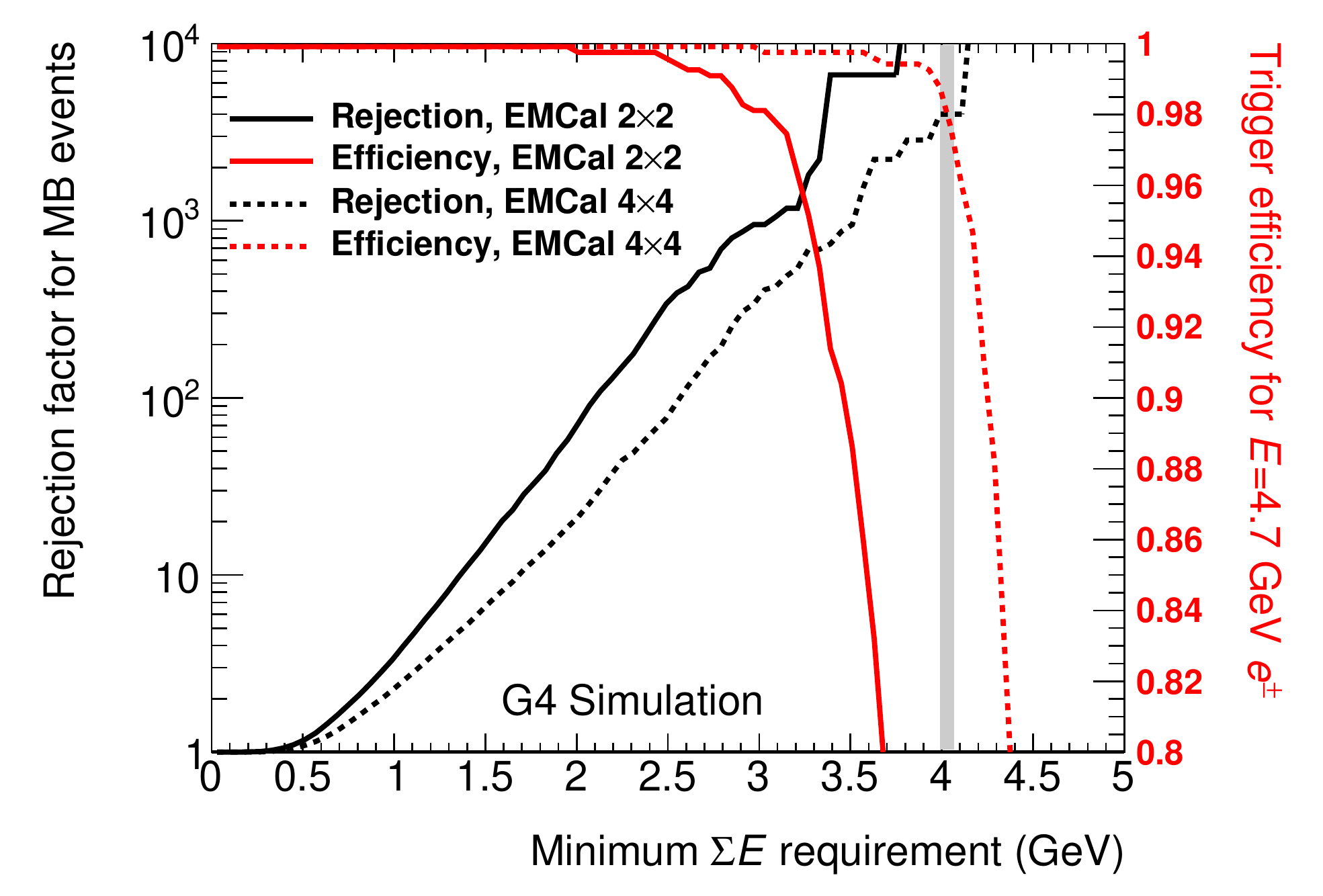}
    \caption[Rejection factor and efficiency for an electron trigger
    requiring a minimum amount of energy in a region of the
    electromagnetic calorimeter
    ($\Sigma{E}$)]{\label{fig:sPHENIX_trigger_electron} Rejection
      factor and efficiency for an electron trigger which requires
      some minimum amount of energy in a region of the electromagnetic
      calorimeter ($\Sigma{E}$). Results are shown for a full \geant
      simulation of the detector response. The rejection factor for
      minimum bias $p$+$p$ events (black lines) and the efficiency for
      $E_{e\pm}=4.7$ GeV electrons (red lines) are plotted as a
      function of the required energy $\Sigma{E}$. The solid and
      dashed lines show the results for trigger schemes in $2\times2$
      and $4\times4$ EMCal windows.}
 \end{center}
\end{figure}

The performance of possible electron triggers for selecting
di-electron $\Upsilon$ decays in high-luminosity $p$+$p$ running in
sPHENIX has also been investigated. These triggers are based on energy
sums in the electromagnetic calorimeter, and have been examined with a
\geant simulation of the calorimeter response. For this study,
electrons with an energy equal to half the nominal $\Upsilon$(1S)
mass, $E_{e\pm} = 4.7$~GeV, were generated, since this is the lowest
possible energy of the highest-energy electron in the decay of an at
rest ($p_\mathrm{T} = 0$) Upsilon. Thus, a successful trigger strategy
for $E_{e\pm} = 4.7$~GeV electrons is sufficient for all other
$\Upsilon$ decay topologies where both electrons are within the
sPHENIX acceptance. The electromagnetic calorimeter towers of size
$\Delta\phi\times\Delta\eta=0.025\times0.025$ were collected into
sliding tower windows made from $2\times2$ and $3\times3$ blocks of
these towers, and a $4\times4$ block made from sliding windows over
the $2\times2$ tower blocks. The total energy (not transverse energy)
in the electromagnetic calorimeter was considered. For each window
size, the distribution of largest energy sums in minimum bias \pythia
events were used to determine the rejection factors for the trigger.

Figure~\ref{fig:sPHENIX_trigger_electron} summarizes the performance
of such an electron trigger by simultaneously plotting the rejection
factor for minimum bias events and efficiency for $E_{e\pm} = 4.7$~GeV
electrons as a function of the minimum energy required in the
electromagnetic calorimeter tower windows. In particular, the vertical
gray band in the figure at $\Sigma{E} = 4$~GeV, gives an example of a
choice of minimum threshold energy in $4\times4$ windows for which the
rejection factor is $\approx5\times10^{3}$ while maintaining an
electron efficiency of 98\%. This demonstrates the feasibility of an
electron trigger for the Upsilon program in high-luminosity $p$+$p$
data-taking.

\begin{figure}[!hbt]
 \begin{center}
    \includegraphics[width=0.60\linewidth]{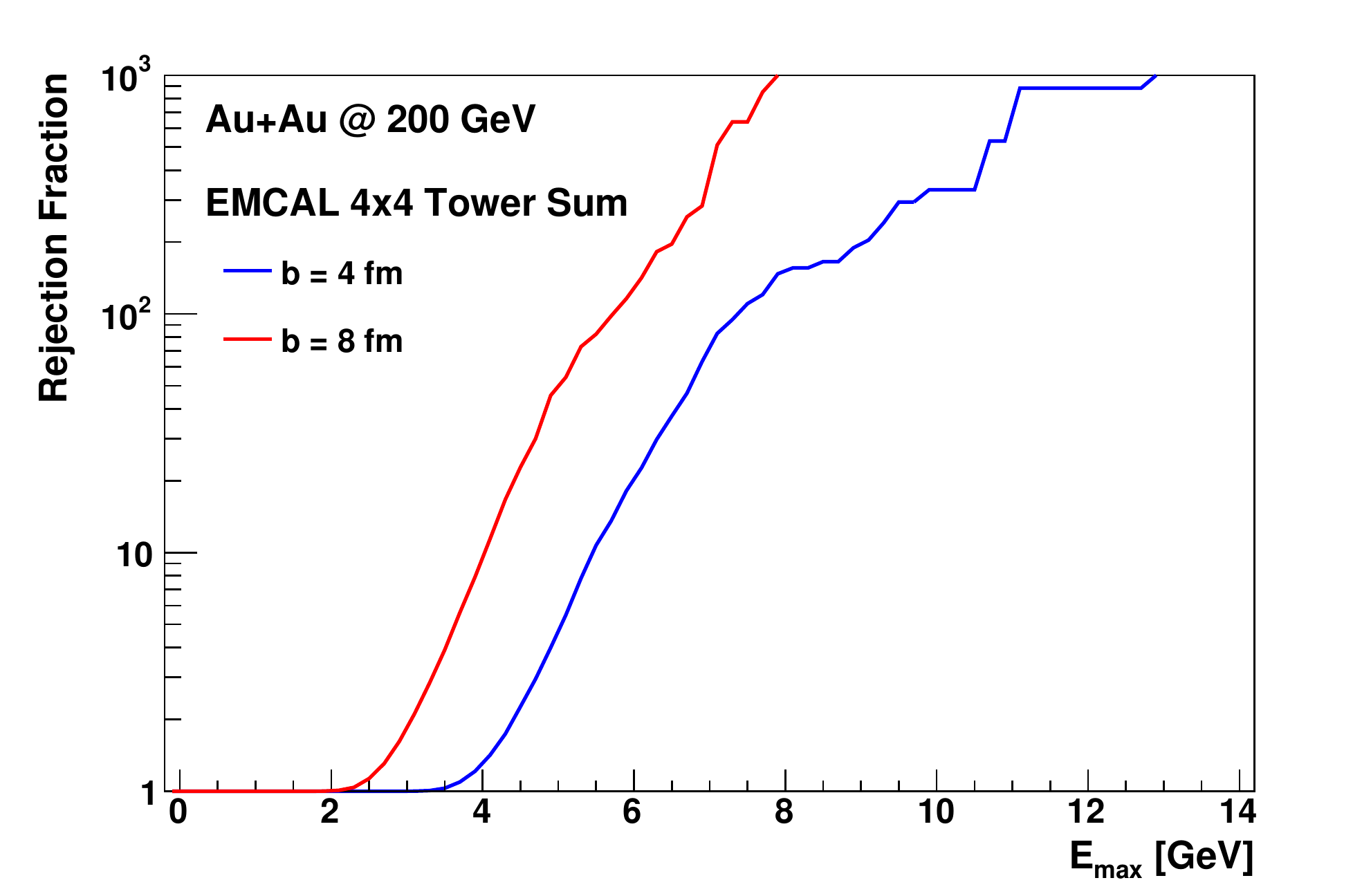}
    \caption{\label{fig:sPHENIX_emcaltrigger_auau} \hijing and \geant
      calculated EMCal trigger patch $4 \times 4$ rejections for
      central and mid-central events ($b=4$ and $b=8$~fm) as a
      function of threshold energy.  }
 \end{center}
\end{figure}

The reference design for the calorimeter digitizers have digitization available on
every crossing for triggering, and transmission of data to a Level 1 trigger board
capable of making trigger decisions such as shown in \ref{fig:sPHENIX_emcaltrigger_auau}
is being included from the beginning.
The digitizer electronics is being designed with the capability of transmitting data
from every channel with reduced precision, or $2\times2$ sums as trigger primitives which
can be used in more complex trigger algorithms running in FPGA-based trigger
boards similar to trigger boards developed for the PHENIX muon trigger.
The segmentation of the electronics and the detector matches well the need for 
$2\times2$ or $4\times4$ digital sums.
Cost and complexity, and the need for higher speed encoding and decoding of trigger
data are being considered in the overall system design.

\begin{figure}[!hbt]
 \begin{center}
    \includegraphics[width=0.48\linewidth]{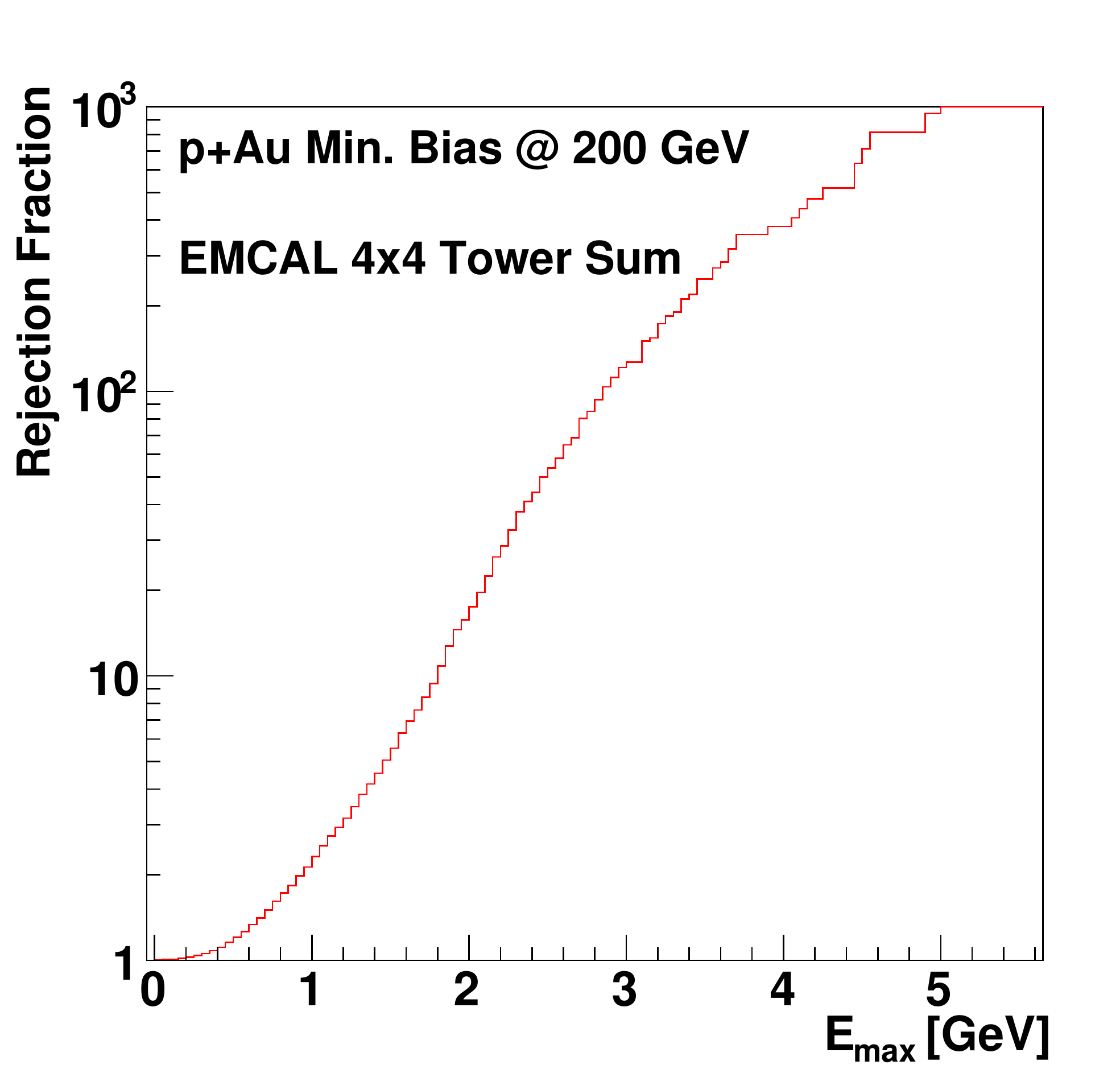}
    \hfill
    \includegraphics[width=0.48\linewidth]{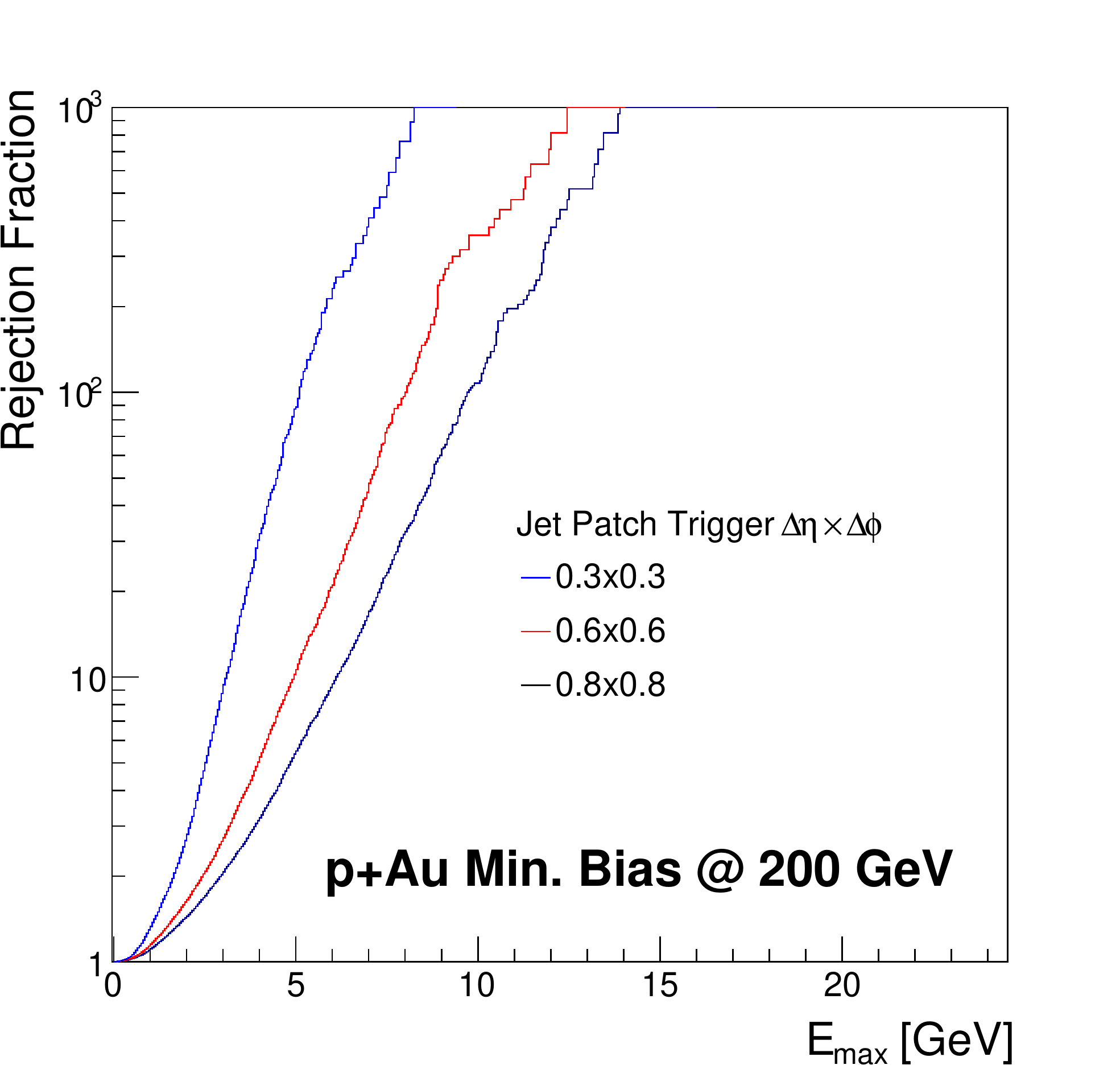}

    \caption{\label{fig:patrigger} (Left) \hijing and \geant
      calculated EMCal trigger patch $4 \times 4$ rejections for
      \pau minimum bias collisions.  (Right) \hijing and \geant calculated
      full calorimeter jet patch trigger rejections for \pau minimum bias collisions.}
 \end{center}
\end{figure}

We have extended these \pp trigger studies to minimum bias \pau collisions where we also
require selective physics based triggers.   Shown in Figure~\ref{fig:patrigger} are the rejection
factors as a function of EMCal trigger threshold (left) and the rejection factors as a function
of the total calorimeter jet patch trigger threshold (right) for different patch sizes.   The calculations
are carried out with \hijing simulated minimum bias \pau events run through the full \geant response chain.
The rejection factors are quite sufficient for sampling the full luminosity in \pau for unbiased jets and
Upsilons as in the \pp case.

We have also explored possibilities for rare probe triggers in \auau
events, where the high-multiplicity fluctuating background can
significantly affect the trigger performance. Though the performance
of the analogous \pp and \pdau triggers is worse due to the presence of the
underlying event, the required rejection factors are smaller. The
studies described below were performed using a full \geant simulation
of the electromagnetic and hadronic calorimeter in minimum bias $b =
4$~fm and $b = 8$~fm \hijing events.

For triggering on photons, we consider a trigger requiring some
minimum energy ($\Sigma{E}_\mathrm{T}$) in a $4\times4$ patch of EMCal
towers. Figure~\ref{fig:sPHENIX_emcaltrigger_auau} shows the rejection
factor for \hijing events of both $b$ values as a function of the
minimum energy required. It can be seen that for relatively modest
requirement of $\Sigma{E}_\mathrm{T} > 8$~GeV, the rejection factor
for minimum bias \hijing events is $>1000$, which is generously higher
than the rejection of a few hundred needed to sample the full rare
probe rate in the highest luminosity \auau data-taking. Thus, in such
a scenario, an unbiased sample of high-$p_\mathrm{T}$ photon events
can be recorded.

\begin{figure}[!hbt]
 \begin{center}
    \includegraphics[width=0.47\linewidth]{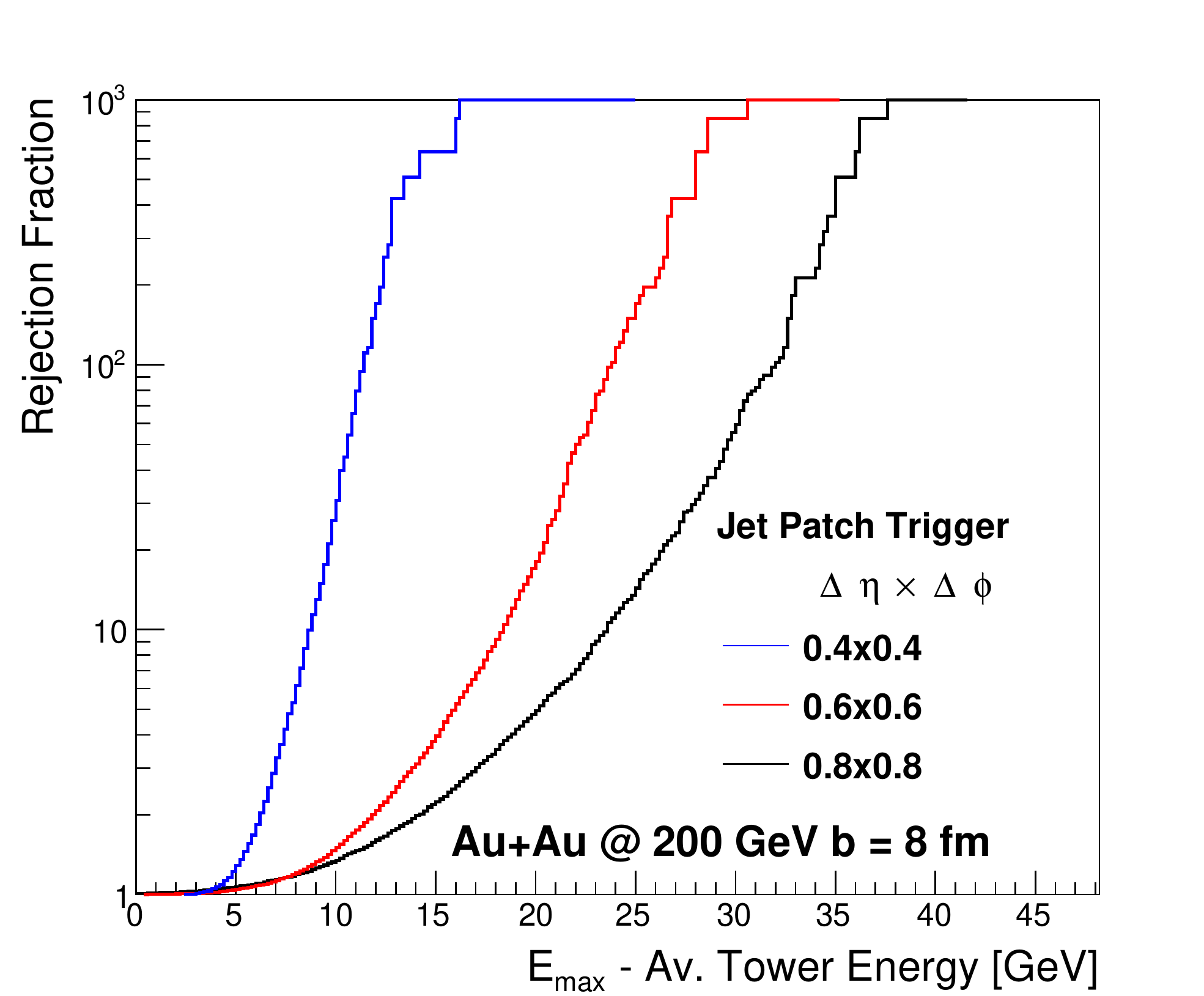}
    \hfill
    \includegraphics[width=0.47\linewidth]{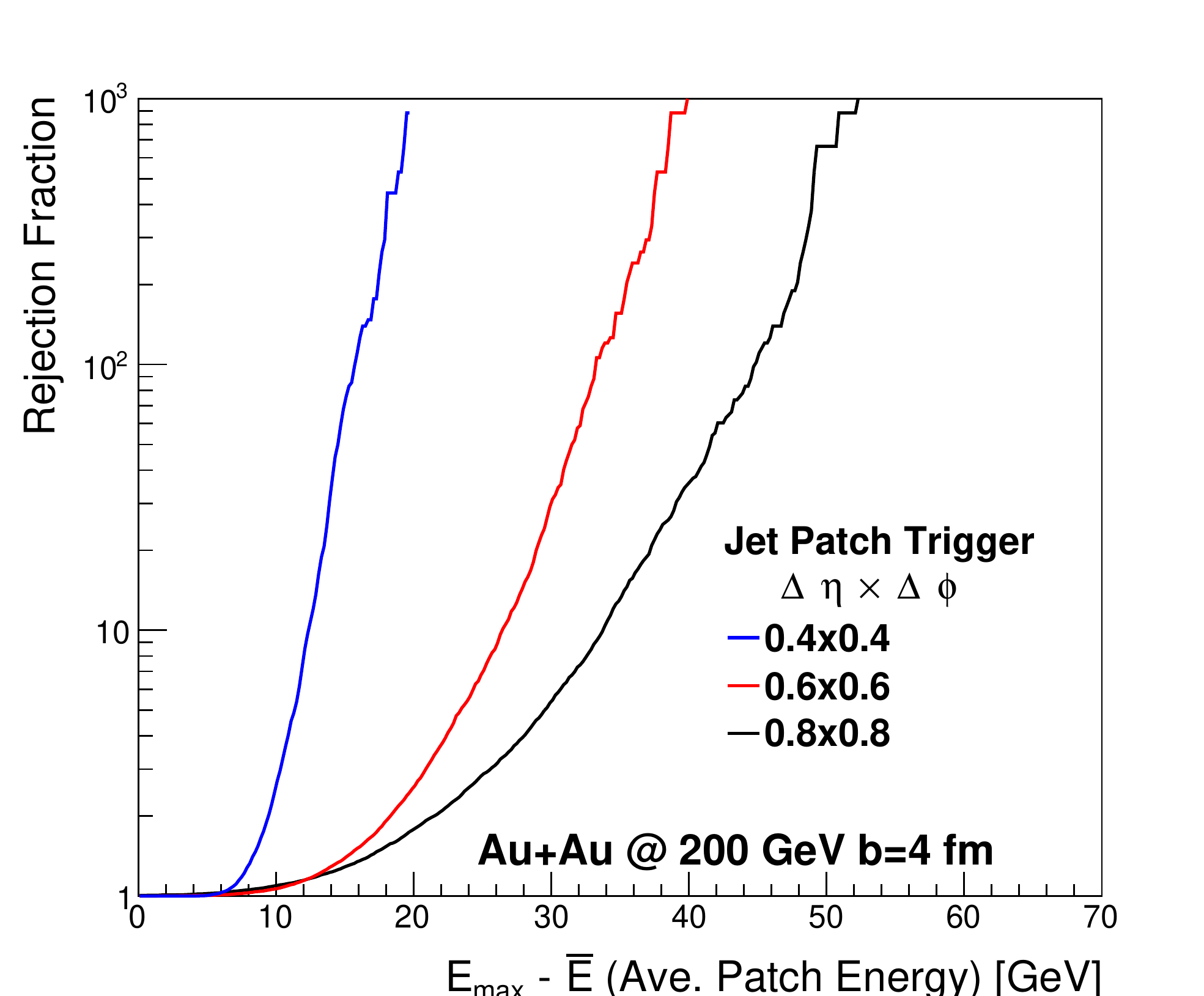}
    \caption[Full \hijing and \geant calculated calorimeter (EMCal and
    HCal) trigger patch rejections for central and mid-central events
    (b=4 and b=8 fm) as a function of threshold
    energy]{\label{fig:sPHENIX_emcaltrigger_auau2} Full \hijing and
      \geant calculated calorimeter (EMCal and HCal) trigger patch
      rejections for central and mid-central events (b=4 and b=8 fm)
      as a function of threshold energy.  The patch sizes considered
      are $\Delta \eta \times \Delta \phi = 0.4 \times 0.4$, $0.6
      \times 0.6$, and $0.8 \times 0.8$.  }
 \end{center}
\end{figure}

For triggering on jets, we consider instead the large area ``jet
patch'' triggers composed of sliding windows of EMCal and HCal towers
used above in studies of trigger in $p$+$p$ collisions. However, in
Au+Au events an underlying event subtraction is necessary at the
trigger level so that the jet patch trigger does not fire primarily on
the large $\Sigma{E}_\mathrm{T}$ of the underlying event
pedestal. This underlying event subtraction consists of subtracting
the mean energy density measured over the entire calorimeter, and is
kept simple to approximate what could be performed computationally in
a fast online trigger. Figure~\ref{fig:sPHENIX_emcaltrigger_auau2}
shows the rejection factors for jet patch triggers of various window
sizes and for $b = 4$~fm and $b = 8$~fm \hijing events, as a function
of the minimum window $\Sigma{E}_{T}$. It can be seen that even in the
central \hijing events, it will be possible to trigger on
high-$p_\mathrm{T}$ in a way that still maintains a rejection factor
of $100$--$200$ (for example, requiring $\Sigma{E}_\mathrm{T} >
30$~GeV in $0.6\times0.6$ windows). Thus, while the jet spectrum below
this cutoff would be measured using the minimum bias Au+Au event
sample, the full luminosity can be sampled to measure the
high-$p_\mathrm{T}$ end of the jet spectrum to its statistical limit.
Taken together, these figures demonstrate a triggering
strategy for rare probes in \auau collisions.

\makeatletter{}\section{Mechanical Design and Infrastructure Concept}
\label{sec:integration}

\begin{figure}[hbt!]
  \centering
  \includegraphics[trim = 300 0 300 0, clip, width=0.49\linewidth]{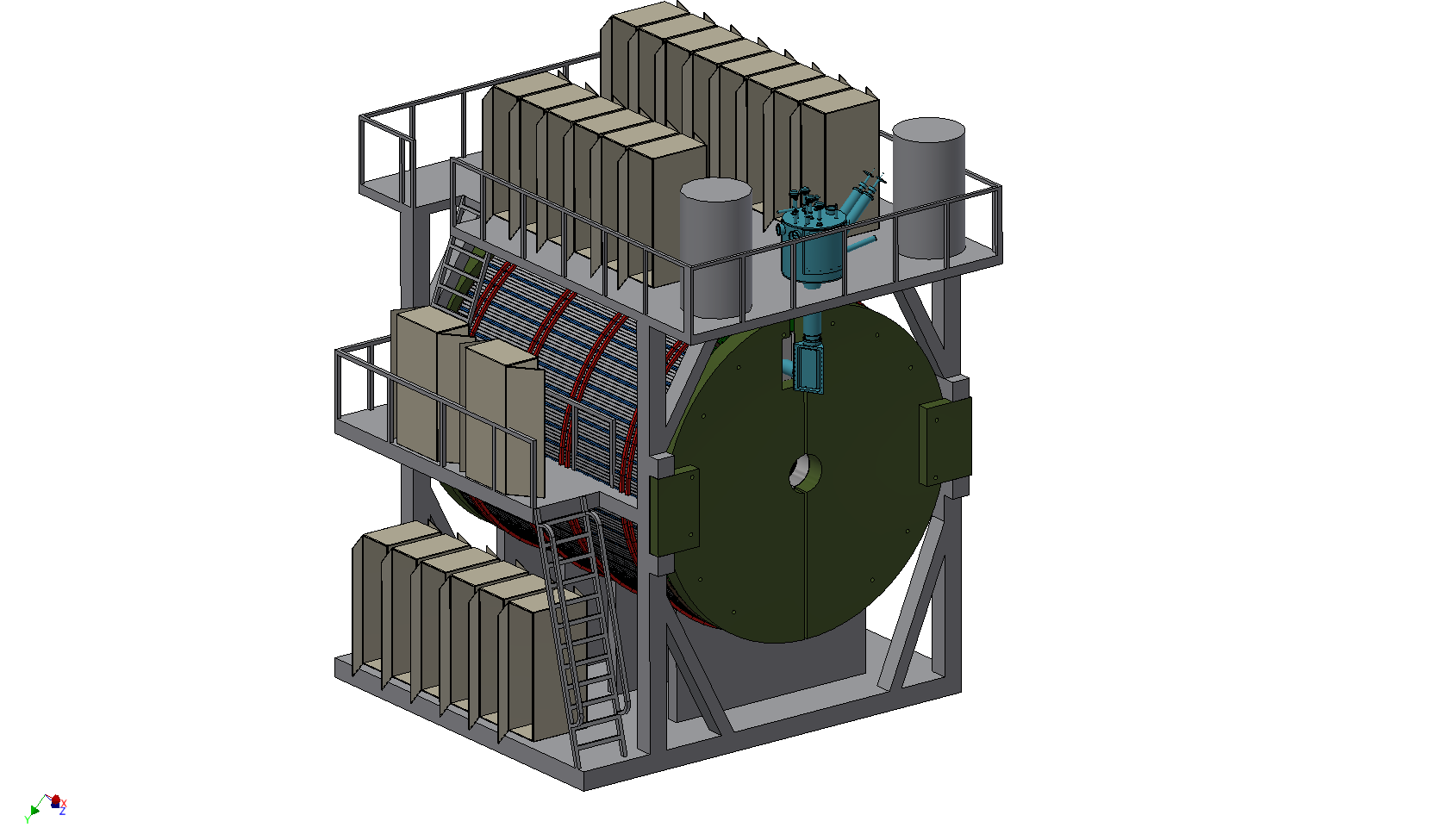}
  \hfill
  \includegraphics[trim = 300 0 300 0, clip, width=0.49\linewidth]{figs/sPhenix-Detector-SW-view-Cap-Removed.png}
  \caption{Illustration of sPHENIX underlying structural support,
    support equipment, overall assembly and maintenance concepts with
    and without endcaps.}
  \label{fig:sphenix_support_concept}
\end{figure}

sPHENIX has been designed to be straightforward to construct and
assemble, but it still requires significant infrastructure to support
and service it.  The overall concept for how sPHENIX will sit in the
existing PHENIX IR is shown in
Figure~\ref{fig:sphenix_support_concept}.  A set of envelope
dimensions for each of the major components of sPHENIX has been
established and is discussed below.

\subsection{Beampipe} The existing PHENIX beampipe will be used with
minimal modification. The current beampipe has a 40\,mm outside
diameter in the central area, and connected on either end with
transition pipe sections from 40\,mm to 75\,mm OD and 75\,mm OD to
125\,mm OD.  A new support structure to support the beampipe inside
the superconducting solenoid will need to be designed.

\subsection{Silicon Tracker} The support structure for the silicon tracker, utilities supply
and readout design will be designed to allow the tracker to be inserted into 
the superconducting solenoid cryostat. Existing VTX and
additional silicon layers will be integrated into a new
structural support design and mechanisms which will mount onto
rails allowing insertion and removal of the detector from within
the EMCal central bore.
The on-detector electronics and services inside the cryostat will generally not be
serviceable during runs.

\subsection{Superconducting solenoid magnet} The BaBar magnet has a 1.5 Tesla solenoid field,
140\,cm inner cryostat radius, 173\,cm outer cryostat
radius, 385\,cm cryostat length. 
The cryostat is not designed to be disassembled.
The cryostat will be supported by the hadronic calorimeter which also acts as the flux return.
The services stack will be modified to exit outside the acceptance
beyond the south end of the HCal
detector to carry cryogenic supply lines, power leads and monitoring cables.
The existing rigging fixtures from SLAC will be adapted 
for transport, lifting and installation whenever possible.
The Superconducting Magnet Division and Collider-Accelerator Department
have the technical expertise to integrate the solenoid into existing RHIC
infrastructure.

\subsection{Electromagnetic calorimeter} The EMCal will have a 13\,cm
radial thickness with electronics and services on the inner radius
and full $2\pi$ azimuthal coverage. 
The EMCal will be be supported by the Inner HCal, with 
provision for 
maintenance, assembly, disassembly and integration of component sectors.  
More detailed mechanical and structural design is ongoing, and assembly procedures
are being developed.

\subsection{Hadronic calorimeter} The HCal will have
full $2\pi$ azimuthal coverage, and the calorimeter is divided
into an inner radial section inside the solenoid and an outer radial section just outside the solenoid.
The inner radial section, which occupied 23\,cm in radial thickness in the simulation, 
will be designed to maximize the absorber inside the magnet while allowing
sufficient space for readout electronics and services.
The inner HCal was simulated with copper absorber, but non-magnetic stainless steel or brass
have almost the same interaction length and may be preferable mechanically.
The outer HCal was 67\,cm thick in the simulation, 
making the total HCal about 5.5 nuclear interaction lengths thick.
The Outer HCal will support the cryostat and the Inner HCal and EMCal assembly.
The HCal will also incorporate provision
for support of itself in the fully assembled configuration, any maintenance
configuration and for assembly/disassembly and integration of
component segments. 
The HCal will be constructed of 384 segments of 7\,mm
thick scintillator sections with embedded optical fibers which collect
the light.  
The scintillator sections will be sandwiched between
tapered steel plates tilted from the radial direction,
with the inner steel plates tilted in the opposite direction from
the outer steel and offset by a half a segment thickness. 

\subsection{Structural support apparatus} Structural support for the
sPHENIX major components will provide structural support
for all of the equipment with the following criteria:

\begin{itemize}

\item Appropriate structural support will be provided to all
  components, with integral connections and support interfaces and/or
  clearances for support structure designed into the comprising
  detector subassemblies and the superconducting solenoid.

\item Components will be able to be completely assembled in the PHENIX
  Assembly Hall (AH) utilizing existing cranes (40 ton load limit). The
  assembly will be mounted on the existing PHENIX rail system or a
  modification of the existing rail system.

\item Functional tests including pressure, and magnetic tests will be
  able to be performed in the AH.

\item The sPHENIX detector will be capable of disassembly
  to allow maintenance of electronics, support
  services and replaceable components. This capability will be
  available with the full assembly in the AH or the Interaction Region
  (IR), with full maintenance capabilities during shutdowns between
  runs and with as much maintenance capabilities during a run as
  possible.

\item The sPHENIX assembly will be relocatable from the AH to the IR
  using the existing rail system or a modification to the existing
  rail system. This relocation may be accomplished fully assembled or
  disassembled into subdivisions which are reassembled in the IR.
  Disassembly and re-assembly will use existing AH and IR cranes.
  
\item Support equipment for the above components and the utilities
  supplied to the above structure including provision for electronics
  racks, cooling services, cryogenics, power and signal cables, and
  monitoring and control equipment will be provided.

\item The assembled sPHENIX will allow partial disassembly during
  maintenance periods to provide access to all serviceable components,
  electronics and services. The assembled sPHENIX will provide for
  electronics racks and all other support components for operation and
  monitoring of the sPHENIX active components. Safe and efficient
  access to all service/monitoring components will be integrated into
  the design of the underlying structural support.

\item Infrastructure used successfully for the past fourteen years of
  of PHENIX operation will be adapted and expanded to support sPHENIX.
  This includes the chilled water system for electronic cooling, air
  conditioning, and safety systems.

\end{itemize}

\begin{figure}[hbt!]
  \centering
     \includegraphics[width=0.8\linewidth]{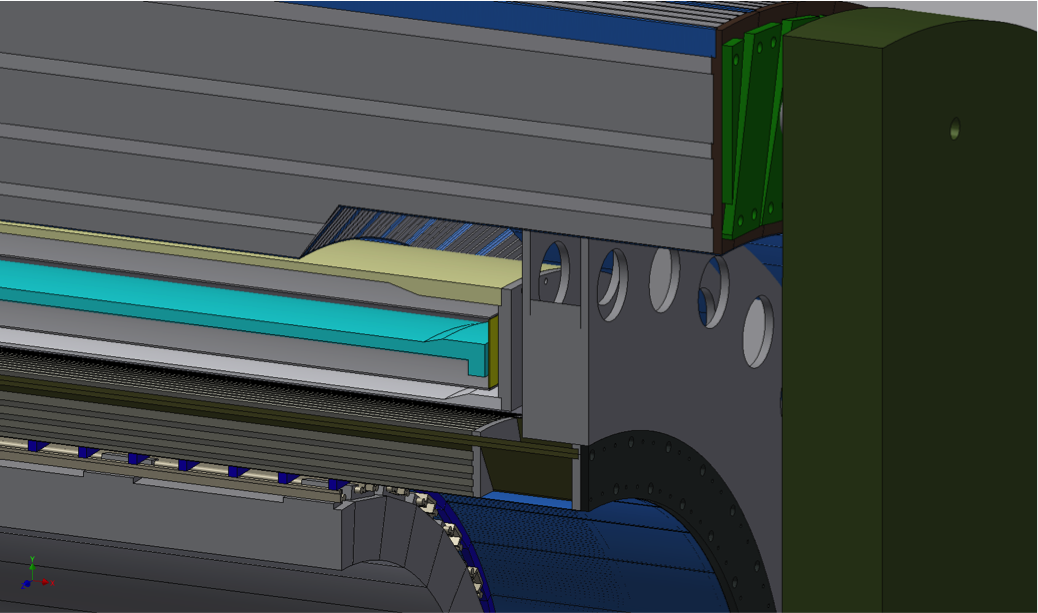}
     \caption{Closeup view of EMCal and HCal with the solenoid
       cryostat and services.}
  \label{fig:barrel_detail}
\end{figure}

Figure~\ref{fig:barrel_detail} shows a view of the HCal and EMCal
inside the solenoid cryostat with power and cryogenic services
provided through a modified chimney.

\makeatletter{}\section{Detector Development and Testing}
\label{sec:prototypes}

\begin{figure}[hbt!]
  \centering
  \includegraphics[width=0.8\linewidth]{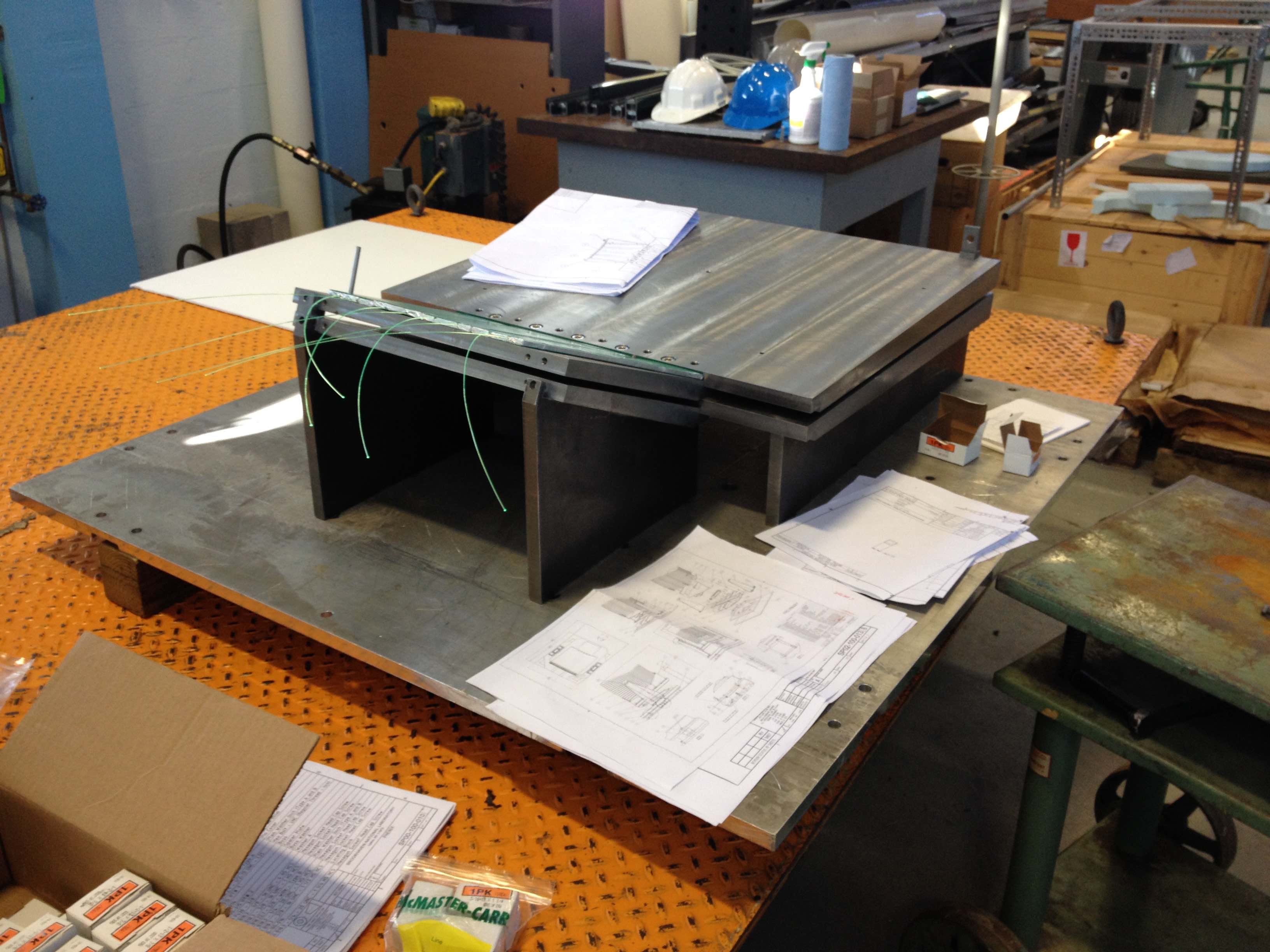}
  \\
  \includegraphics[width=0.8\linewidth]{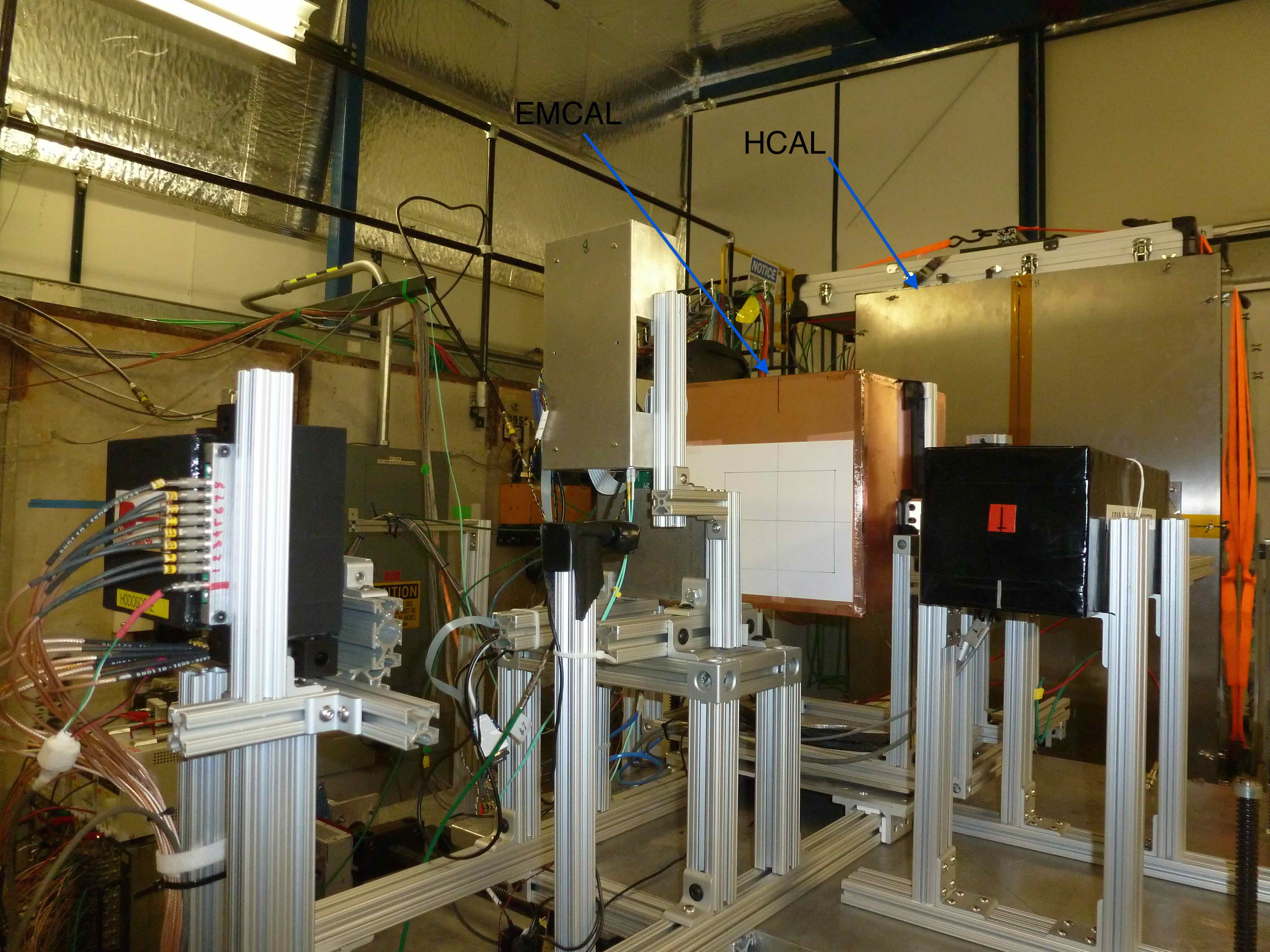}
  \caption[Pictures of the HCal prototype under construction and of
  the calorimeters in the MWEST beamline of the Fermilab Test Beam
  Facility]{Top: HCal prototype under construction.  The first layers
    of absorber are being stacked on the lift table for the beam
    test. Bottom: Calorimeters in MWEST beamline of the Fermilab Test
    Beam Facility.}
  \label{fig:hcal_prototype}
\end{figure}

Prototype electromagnetic and hadronic calorimeters have been
developed for beam tests to validate the design concepts and gain
experience with the readout and calibration of silicon
photomultipliers in an operating detector.  The first prototypes were
tested at the Fermilab Test Beam Facility as T-1044 February 5--25,
2014.  The EMCal prototype was a $7\times7$ device with 1 mm tungsten
absorber which can be rotated in the beam to study shower development
and energy resolution.  A beam test of the SPACAL electromagnetic
calorimeter was carried out by the UCLA group and collaborators
immediately following T-1044.  The HCal prototype consists of inner
and outer $4\times 4$ sections with machined tapered plates using a
mechanical design that is being evaluated for use in building the full
detector.  Both detectors are read out with Hamamatsu silicon
photomultipliers with signal conditioning that allows them to be flash
digitized at 60 MHz with existing PHENIX electronics.

\makeatletter{}\chapter{Physics Performance}
\label{chap:jet_performance}

In this Chapter we detail the expected sPHENIX physics performance.
The sPHENIX jet, dijet, $\gamma$-jet, fragmentation function, and
beauty quarkonia performance demonstrates the ability to measure key
observables that can test and discriminate different quenching
mechanisms, coupling strengths to the medium, and with sensitivity to different
length scales in the QGP.

The key aspects of jet performance are the ability to find jets with
high efficiency and purity, and to measure the kinematic properties of
jet observables with good resolution.  It is also necessary to
discriminate between jets from parton fragmentation and \fake jets
caused by fluctuations in the soft underlying event.  For the sPHENIX
physics program, there are four crucial observables that we have
simulated in detail to demonstrate the jet performance: single
inclusive jet yields, \dijet correlations, \gj correlations, and
modified fragmentation functions.  We also find that the combination
of full calorimetric reconstructed jets combined with track and
electromagnetic cluster jets allows one to engineer the surface
emission of the leading jet and thus the path of the partner
jet. Other significant observables such as the participant plane
dependence (e.g., $v_{2}$, $v_{3}$, etc.) of jets and jet-hadron
correlations are also enabled by this upgrade.

For beauty quarkonia decaying to \epem, the key aspects of performance
are electron identification (particularly in being able to
discriminate against charged pions), and good momentum resolution to
provide sufficient invariant mass resolution to distinguish clearly
the $\Upsilon(1s)$, $\Upsilon(2s)$, and $\Upsilon(3s)$  states.

An important focus will be to demonstrate the capabilities of sPHENIX
for central \auau collisions at $\sqrtsnn = 200$~GeV, where the
complications of the underlying event are the most severe.  We first
detail the physics performance for jet observables and then the
performance for the beauty quarkonia physics.

\section{Jet simulations}

It is not practical to simulate with \geant~\cite{Agostinelli:2002hh}
a sample of events equivalent to a full year of RHIC running.  We
therefore perform three different levels of simulations described in
detail below.  

The most sophisticated and computationally intensive are full \geant
simulations with \pythia~\cite{Sjostrand:2000wi} or
\hijing~\cite{Gyulassy:1994ew} events where all particles are traced
through the magnetic field, energy deposits in the calorimeters
recorded, clustering applied, and jets are reconstructed via the
\fastjet package~\cite{Cacciari:2005hq}.  We utilize this method to
determine the jet resolution in \pp and \auau collisions from the
combined electromagnetic and hadronic calorimeter information.  We
have also performed a full \geant study of the reconstruction of
\pythia jets embedded in central \auau \hijing events to gauge the
effect of the underlying event on jet observables.

For studies of \fake jets in \auau central collisions, one needs to
simulate hundreds of millions of events and for this we utilize a
\fast simulation where the particles from the event generator are
parsed by their particle type, smeared by the appropriate detector
resolution parametrization from \geant simulations, and segmented into
detector cells.  As described in detail below, a full underlying event
subtraction procedure is applied, and then jets are reconstructed
using \fastjet.  This method is also utilized for embedding events
from \pythia or \pyquen~\cite{Lokhtin:2005px} (a jet quenching parton
shower model) into \auau \hijing events to study \dijet and \gj
observables.  

Finally, in order to gain a more intuitive understanding of the
various effects, we run a \veryfast simulation where \pythia particles
are run directly through \fastjet and then the reconstructed jet
energies smeared by the parametrized resolutions and underlying event
fluctuations.

This section is organized as follows.  First we describe the jet
reconstruction and evaluate its performance in \pp collisions for both
an idealized detector as well as a fully simulated version.  Then we
describe our study of \fake jet contamination, which has been
published in Physical Review C~\cite{Hanks:2012wv}.  We show the
expected performance for sPHENIX measurements of inclusive single jet,
\dijet and \gj correlations, and modified fragmentation functions.

\section{Jet finding algorithms}
\label{sec:jet_finding_algorithms}

For all of the studies presented here we use the anti-$k_T$ jet
algorithm~\cite{Cacciari:2008gp} implemented as part of the \fastjet
package~\cite{Cacciari:2005hq}.  The \antikt algorithm is well suited
to heavy ion collisions and produces cone-like jets in an infrared and
collinear safe procedure.  The parameter that controls the size of the
jet in this algorithm is the jet radius, $R$.  While this is not
strictly a cone size it does specify the typical extent of the jet in
$\eta$-$\phi$ space.  High energy experiments typically use large $R$
values of 0.4--0.7 in order to come as close as possible to capturing
the initial parton energy.  In heavy ion collisions, the desire to
measure the quenching effects on the jet profile and to minimize the
effects of background fluctuations on jets has led to the use of a
range of $R$ values.  Values from 0.2 to 0.5 have been used to date in
\pbpb collisions at 2.76~TeV at the
LHC~\cite{Aad:2010bu,Chatrchyan:2012ni}.  We note that looking at the
jet properties as a function of the radius parameter is very
interesting and potentially sensitive to modifications to the jet
energy distribution in the medium.  

The excellent charged particle tracking capabilities of sPHENIX documented in
Section~\ref{sec:tracking}, there are a number of alternative jet reconstruction inputs
that are available.   These range from jet reconstruction with tracks only, as utilized 
recently by the STAR and ALICE experiments.   These inputs have the benefit of a well defined energy scale, though
with significant fluctuations due to non-charged track energy and track inefficiencies in central \auau or \pbpb
events.   There are results from the same experiments with charged tracks combined with electromagnetic energy.
There are also hybrid, particle flow type inputs as utilized to great benefit by the CMS 
experiment~\cite{Beaudette:2014cea,CMS-PAS-PFT-09-001,Nguyen:2011fg}.   We have implemented all of these
algorithms and have initial performance metrics with full \geant simulations in \pp \pythia reactions.

The particle flow algorithm in CMS results in a substantial improvement in the jet energy resolution, particularly
at lower jet energies, with contributions from multiple effects.  In \pp collisions, the benefits include (i) charged tracks
can be input to \fastjet with the momentum vector at the collision vertex rather than with calorimeter clusters where the
vector is modified as bent in the magnetic field, (ii) the charged tracking resolution is significantly better than the calorimeter
resolution for particles up to hundreds of GeV, (iii) charged tracks can be more easily associated with specific collision
points in the case of multiple interactions per bunch crossing.  The first two items are very significant for CMS since the
magnetic field strength at 4 Tesla really pulls the different charged constituents of the jet apart for easier unique association
with calorimeter clusters.

\begin{figure}[!hbt]
 \begin{center}
    \includegraphics[width=0.9\linewidth]{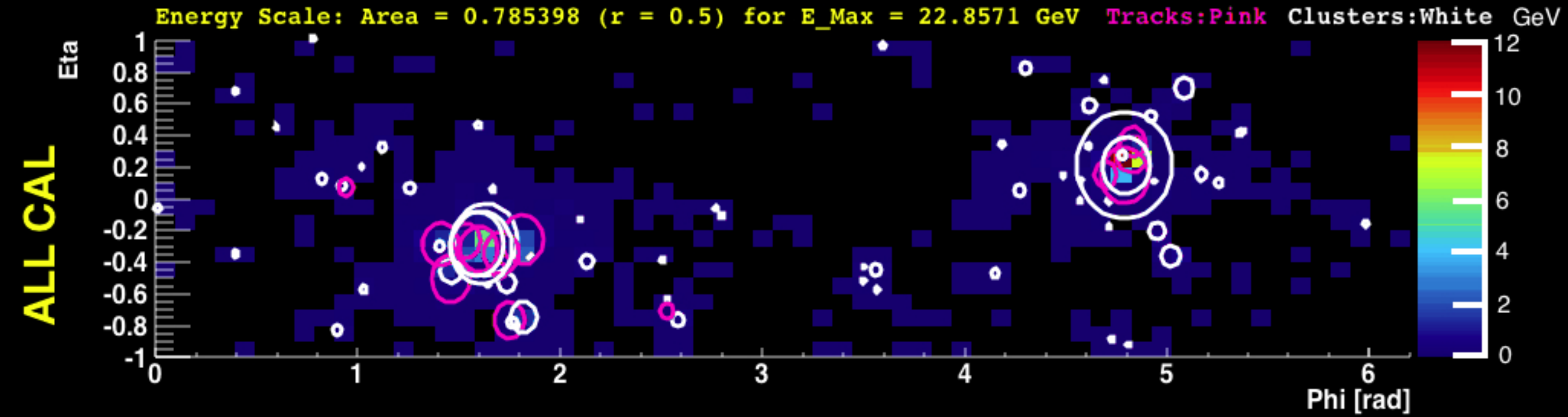}
    \vskip 1em
    \includegraphics[width=0.9\linewidth]{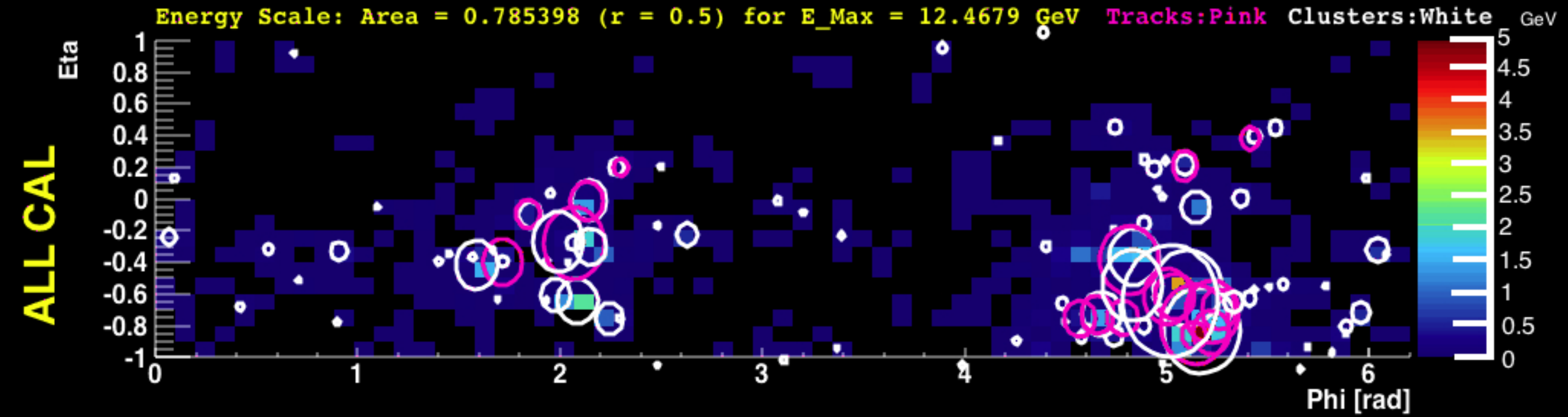}
    \vskip 1em
    \includegraphics[width=0.9\linewidth]{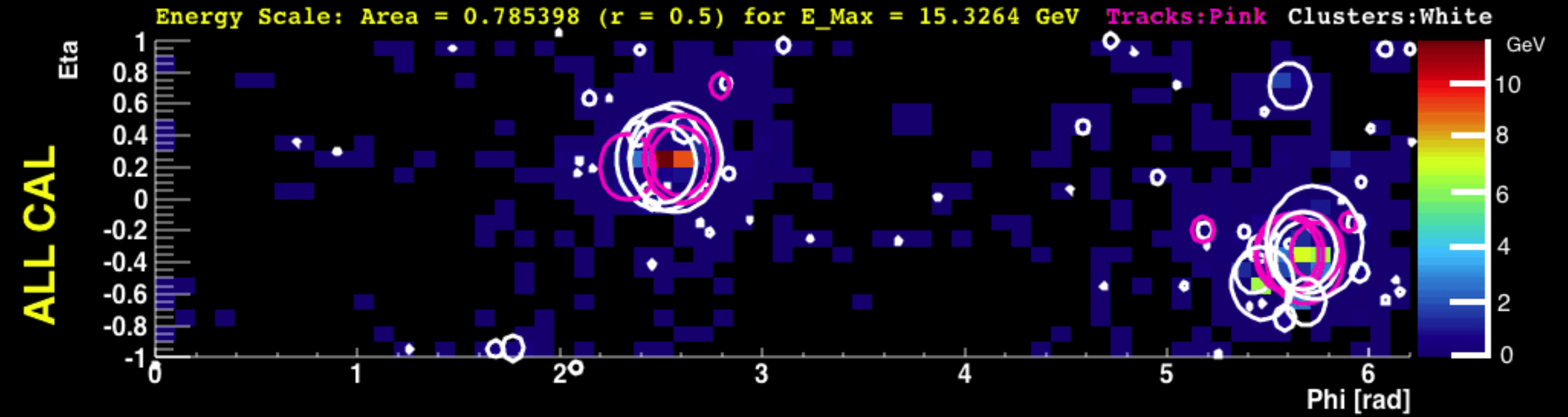}
    \vskip 1em
    \includegraphics[width=0.9\linewidth]{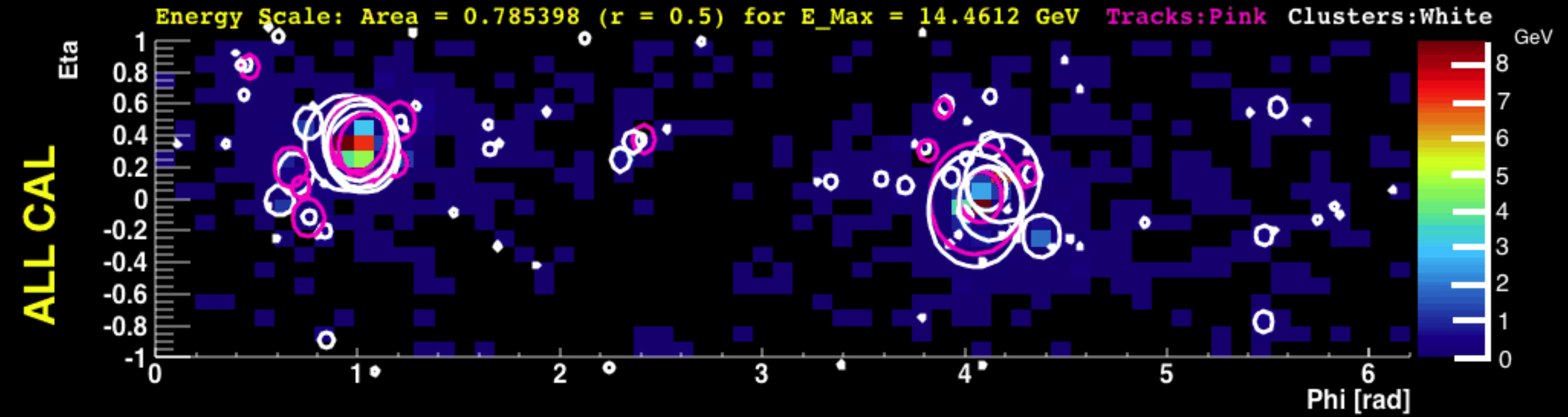}
    \caption[\geant event display of \pythia dijet events with
    indicating clusters found in the calorimeter and reconstructed
    charged tracks]{\label{fig:javier_display} \geant event display of
      \pythia dijet events.  Circles indicate clusters found in the
      calorimeter (white) and reconstructed charged tracks (pink).
      The area of all circles is proportional to the energy of the
      track or calorimeter cluster.  Thus, one can visually see
      closely matched tracks and clusters in position and energy.  }
 \end{center}
\end{figure}

We have implemented a first pass particle flow algorithm where charged
tracks are associated with energy clusters in the full calorimeter
system.  If there is a match within the 90\% confidence level for the
track energy (assuming it is a pion) and the calorimeter energy (as
determined with \geant single particle simulations), the cluster is
replaced by the track as an input to \fastjet.  If the track energy is
above this confidence interval, we do not include the track as there
is some probability for this to be a poor reconstruction or fake
track.  If the track energy is below this confidence interval, there
is a probability that the cluster has energy from additional sources
(neutrals or poor cluster splitting).  In this case, the track energy
is subtracted from the cluster energy and both are input to \fastjet.
Note that for this last scenario, the better tracking resolution does
not improve the jet resolution, since one also leaves any residual
from the poorer calorimeter resolution in the modified cluster.
Figure~\ref{fig:javier_display} shows four example \pythia dijet events
reconstructed through the sPHENIX \geant simulation.  The circles
represent reconstructed calorimeter clusters (white) and reconstructed
charged tracks (pink) with the area being proportional to the energy.

\begin{figure}[!hbt]
 \begin{center}
    \includegraphics[width=0.45\linewidth]{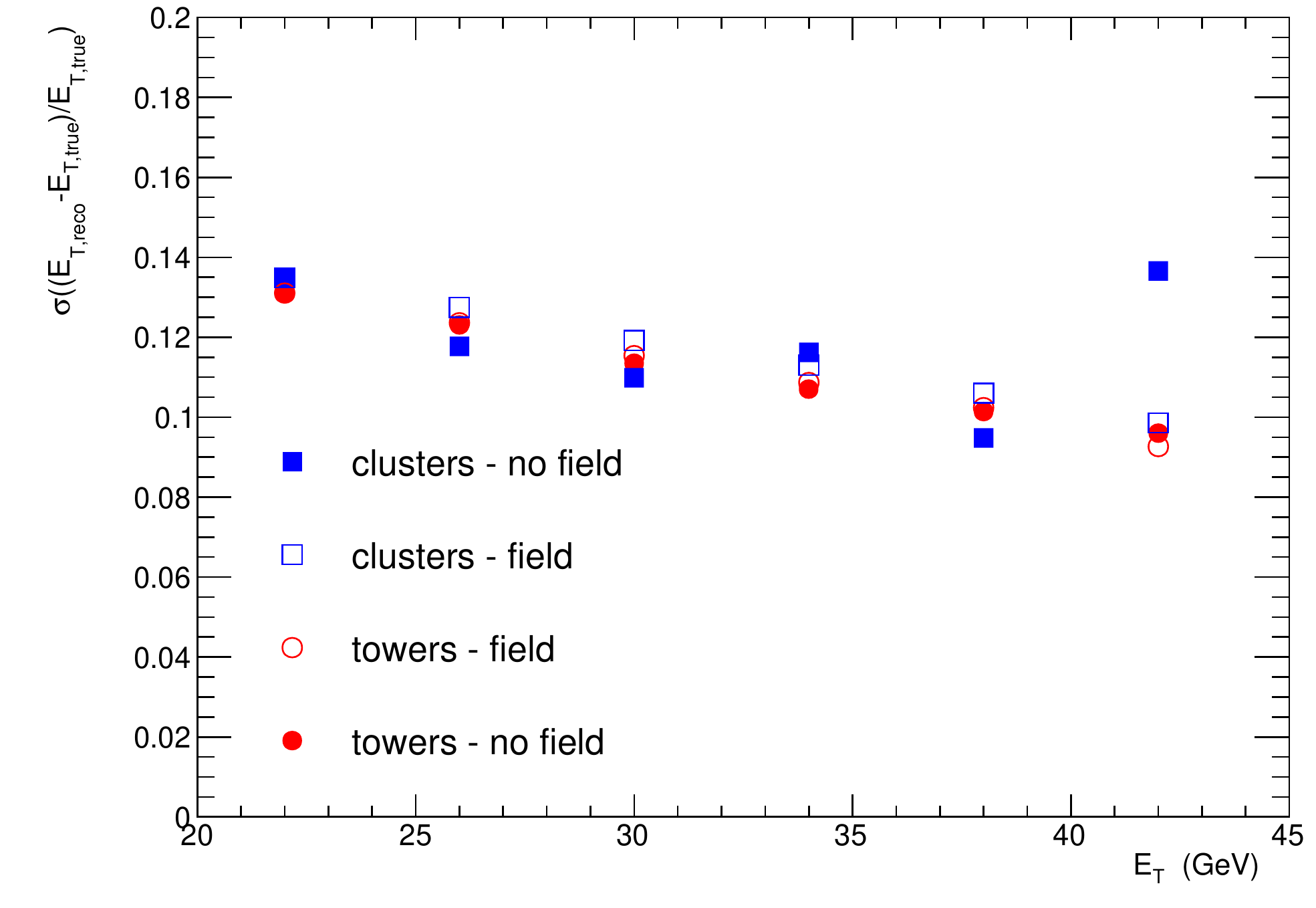}
    \hfill
    \includegraphics[width=0.45\linewidth]{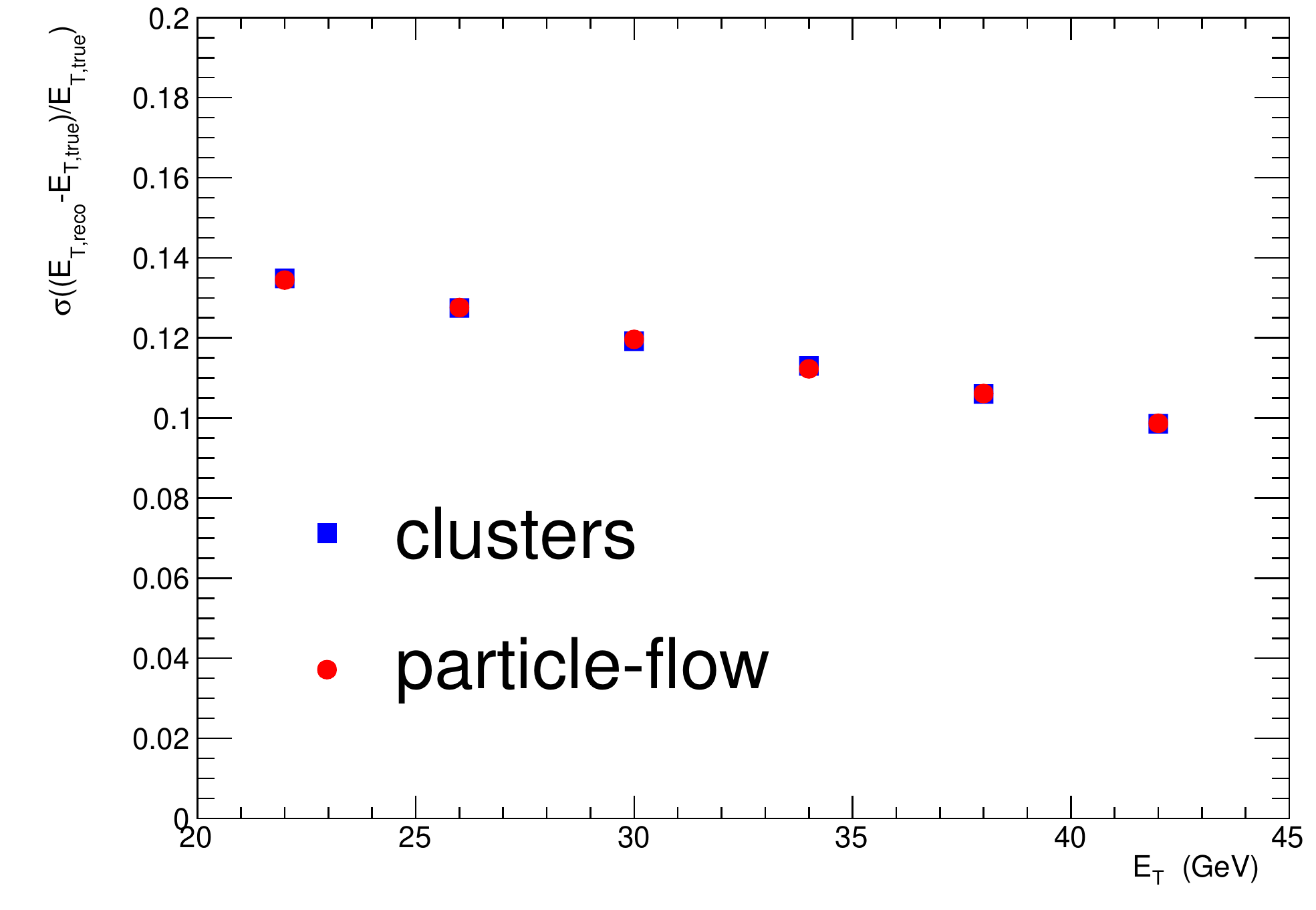}
    \caption[\geant simulations with \pythia dijet events and the
    resulting jet energy resolution for \antikt and $R=0.4$ with input
    calorimeter towers and clusters, and comparison of the jet energy
    resolution with pure calorimetric cluster input and particle flow
    jet algorithm]{\label{fig:ali_algorithm} (left) \geant
      simulations with \pythia dijet events and the resulting jet
      energy resolution for \antikt and $R=0.4$ with input calorimeter
      towers and clusters, with and without the magnetic field turned
      on.  (right) Comparison of the jet energy resolution with pure
      calorimetric cluster input and the first pass particle flow jet
      algorithm.  }
 \end{center}
\end{figure}

In order to gauge the benefit to the jet resolution of the particle
flow algorithm, we consider the three effects listed above.  Since the
luminosities at RHIC result in much lower numbers of collisions per
crossing in \pp and negligible in \auau, the third effect of pileup is
not a significant consideration.  To assess the possible benefit of
correcting the energy to the correct vector at the vertex, we first
compare the fully calorimetric results with the \geant magnetic field
turned off.  Shown in Shown in the left panel of
Figure~\ref{fig:ali_algorithm} are the \geant jet resolutions from
\pythia \pp events with the \antikt algorithm and $R=0.4$ when using
calorimeter towers or calorimeter reconstructed clusters as inputs,
with and without the magnetic field turned on.  The results all give
equivalent jet resolutions, which means that for $R=0.4$ jets the
moving calorimeter energies to the center of clusters and the bending
of soft charged particles in the magnetic field has minimal effect.
We do note that for $R=0.2$ jets, we observe a modest improvement in
the jet resolution with the magnetic field off as expected.  The right
panel of Figure~\ref{fig:ali_algorithm} compares the resolution with
calorimeter clusters to the first pass particle flow algorithm.  There
is only a very modest difference in the results.  This is not so
unexpected as detailed checks indicate that within jets, many of the
calorimeter clusters have multiple-particle energy contributions.  We
are exploring more sophisticated matching criteria that we expect to
yield some additional improvement in resolution.

Even with marginal jet resolution improvement, the particle flow algorithm allows one to make more detailed selections on
track constituents and individual calorimeter clusters.   The same applies for the the tracking only or tracking + EMCal
jet inputs.   The full suite of these algorithms will be further developed as the design of the overall system is
optimized.   These different algorithms have multiple benefits including very different systematics, including on the energy
scale, and will allow detailed comparisons with other experiments and their results.

\clearpage

\section{Jet performance in \pp collisions}
\label{sec:pp_jet_performance}

We begin by exploring the performance of the detector in \pp
collisions.  This allows us to investigate the effects of detector
resolution and how well the process of unfolding these
effects in simpler collisions works before considering the additional
effects of the underlying event and jet quenching in heavy-ion collisions.

The most realistic understanding of the sPHENIX jet reconstruction
performance comes from a full \geant simulation of the detector
response.  In this case, \pythia particles are run through a \geant
description of sPHENIX, the resulting energy deposition is corrected
for by the sampling fraction of the relevant calorimeter, binned in
cells of $\eta$-$\phi$ ($0.024\times 0.024$ for the EMCal and
$0.1\times 0.1$ for the HCal) and the resulting cells are used as
input to \fastjet.
Particles from \pythia events are put through \fastjet to determine
the truth jets.

We then calculate the difference between the energy of the
reconstructed calorimeter jets, $E_{\rm reco}$, and the particle-level
truth jets, $E_{\rm true}$.  The width of this distribution,
$\sigma(E)$, is fit with a functional form: $\sigma(E)/E = a/\sqrt{E}
+ b$.

\begin{figure}[hbt!]
  \centering
  \includegraphics[trim = 0 0 0 0,clip,width=\onewidth]{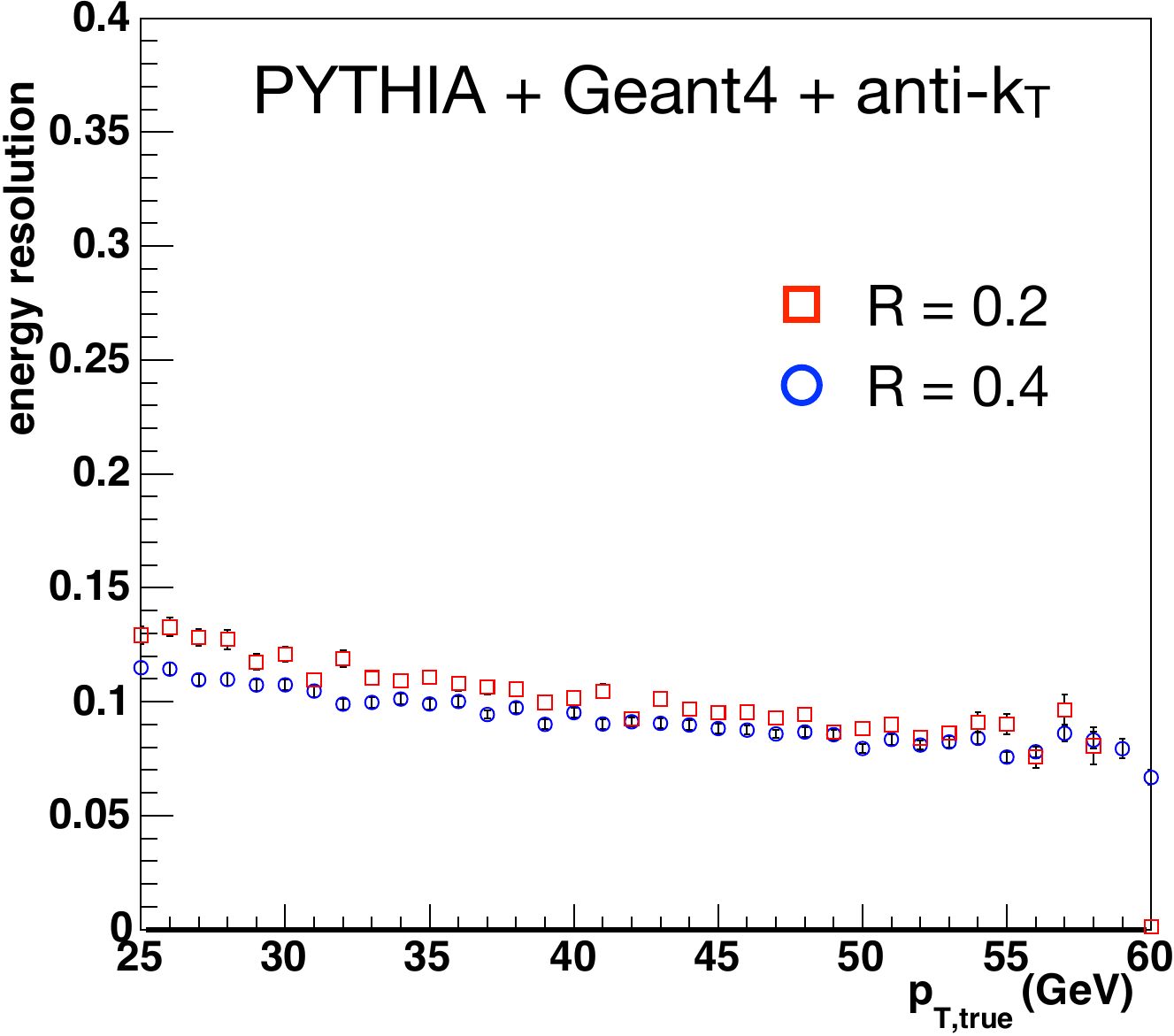}
  \caption{The \geant calculated energy resolution of single jets in
    \pp collisions reconstructed with the \fastjet \antikt algorithm
    with $R = 0.2$ and $R = 0.4$.}
 \label{fig:jet_energy_resolution}
\end{figure}

Full \geant calculations of the energy resolutions for jets in \pp collisions reconstructed with \antikt and
$R=0.2$ and $R=0.4$ are shown in Figure~\ref{fig:jet_energy_resolution}.
The resolutions are relatively independent of $R$ and in simulation are substantially better than 
the required specification detailed in Chapter~\ref{chap:detector_requirements}.

The jet energy resolution in collider experiments is often found to be
a factor of 1.2--1.3 worse than the quoted single particle resolution
of the hadronic calorimeter.  This factor is a balance of many effects
including the better resolution for the electromagnetic part of the
shower, soft particles that deflect out of the jet cone in the
magnetic field, some lost energy, etc.  The CMS quoted jet resolution
in \pp collisions at 7.0~TeV is approximately 120\%/$\sqrt{E}$ which
is roughly 1.2 times worse than the quoted single particle hadronic
calorimeter resolution~\cite{CMSJets:2010}.  The sPHENIX jet energy
resolution and hadronic calorimeter resolution from \geant are
consistent with this expectation, and both are within our performance
specifications.

\begin{figure}[hbt!]
  \centering
  \includegraphics[width=0.46\linewidth]{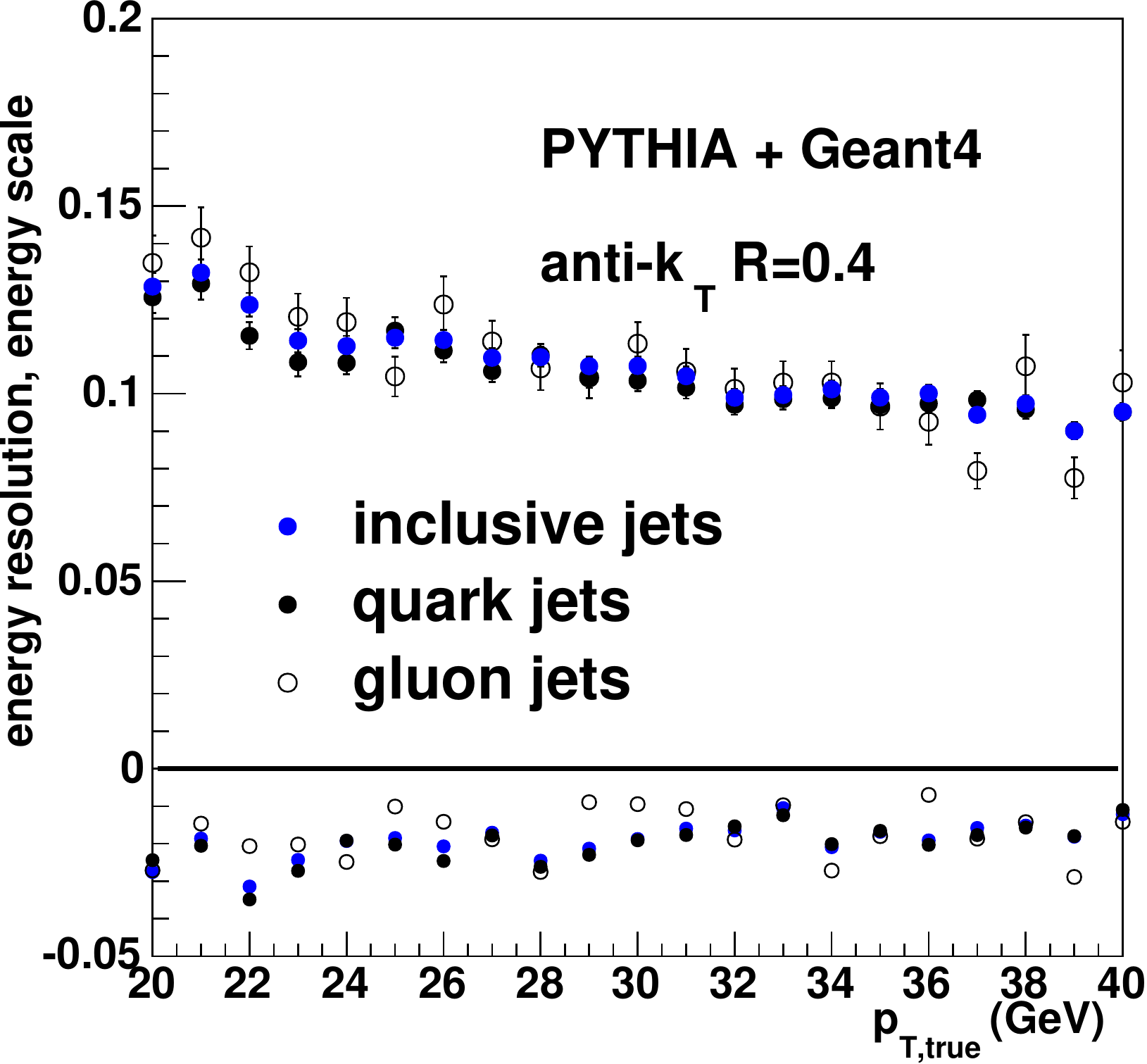}
  \hfill
  \raisebox{0.2cm}{\includegraphics[width=0.50\linewidth]{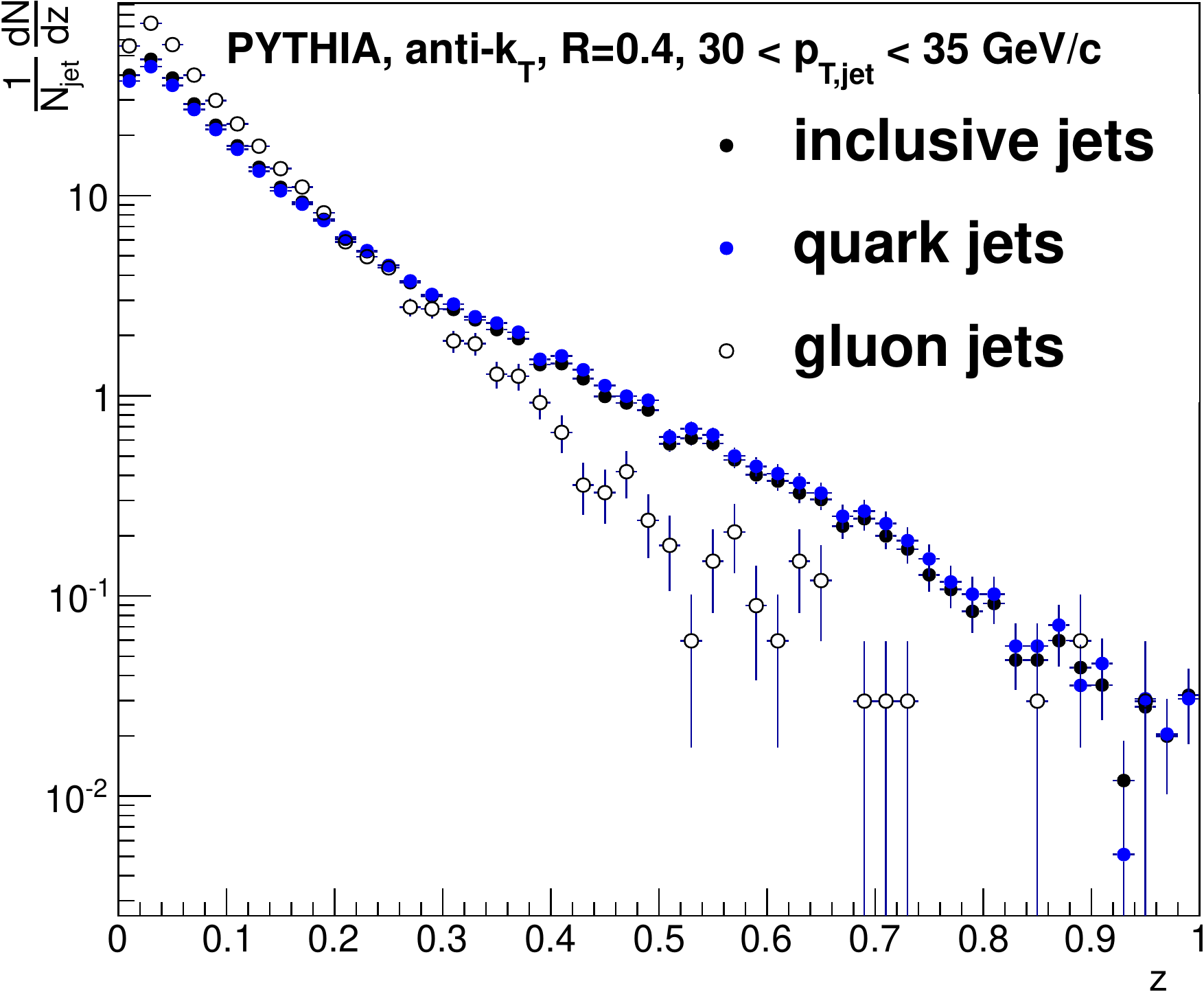}}
  \caption[\geant calculated energy resolution of single jets in \pp
  collisions separated into quark and gluon jets, and the \pythia
  calculated fragmentation function of quark and gluon jets
  separately]{(left) The \geant calculated energy resolution of single
    jets in \pp collisions separated into quark and gluon jets.
    (right) The \pythia calculated fragmentation function of quark and
    gluon jets separately.}
 \label{fig:jet_energy_resolution_quarkgluon}
\end{figure}

We also calculate the jet energy scale and resolution where we have
tagged from the truth information quark and gluon jets.  These results
are shown in Figure~\ref{fig:jet_energy_resolution_quarkgluon} (left)
and indicate no significant differences in jet energy scale and
resolution despite the significantly different fragmentation function
(right).

\subsection{\pp Inclusive Jet Spectra}
\label{sec:pp_very_fast_simulation}

In order to model the jet resolution effects described above on the
inclusive jet spectra in \pp collisions at $\sqrt{s_{NN}} = 200$~GeV,
we have used the \veryfast simulation.  This method entails running
\pythia, sending the resulting final state particles through \fastjet
to find jets, and then blurring the energy of the reconstructed jets
with values obtained from the full \geant simulation.

\begin{figure}[ht!]
  \centering
  \includegraphics[width=\onewidth]{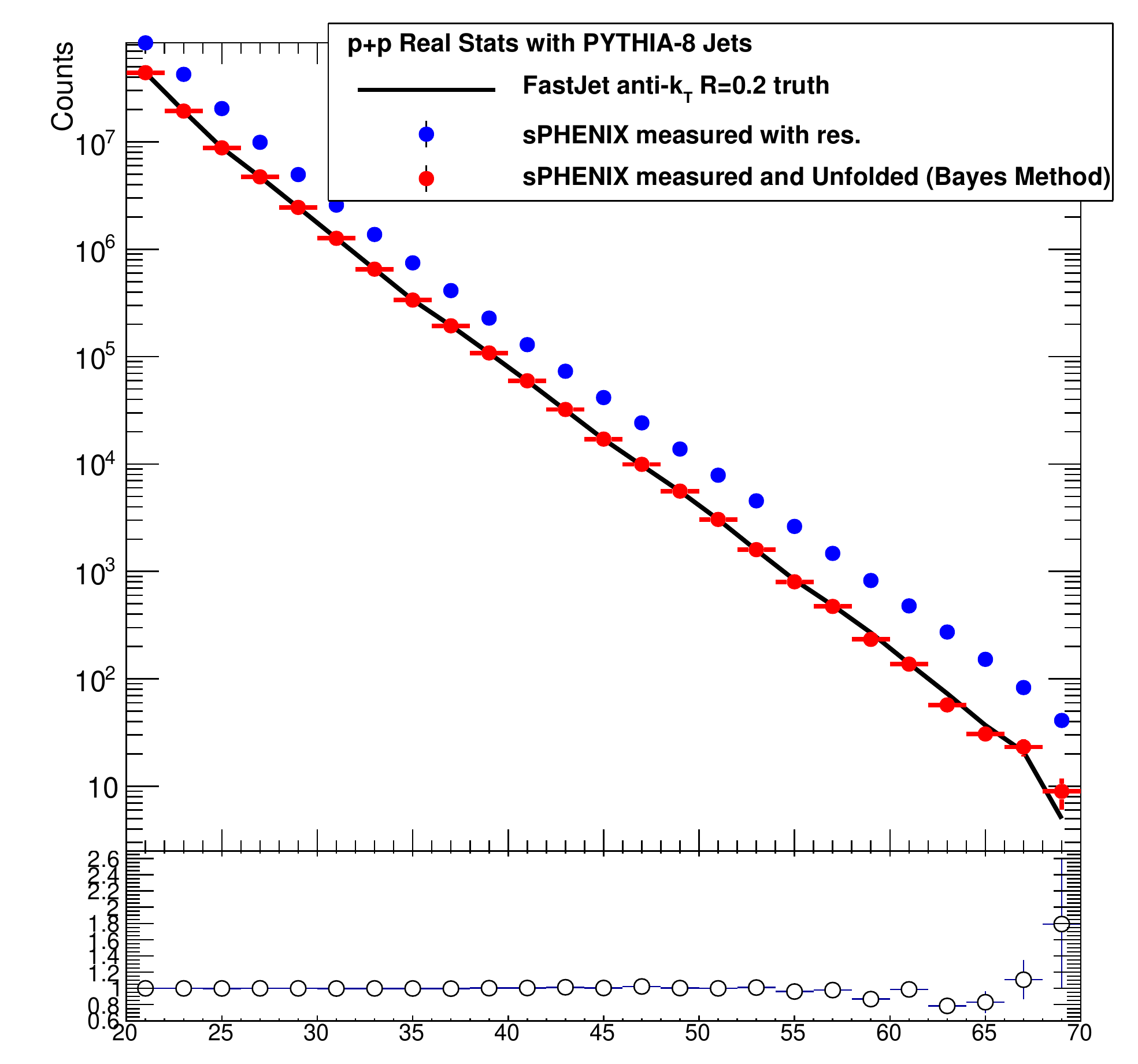}
  \caption[Unfolding the effect of finite detector resolution on jet
  reconstruction in \pp events]{Unfolding the effect of finite
    detector resolution on jet reconstruction in \pp events.  The
    black histogram is the truth spectrum of jets from \pythia, the
    blue dotted histogram is the spectrum after smearing by the jet
    energy resolution and the red histogram shows the result of using
    \roounfold Iterative Bayes method to unfold the detector effects.
    The lower panel shows the ratio of the unfolded to the true $E_T$
    spectrum.}
  \label{fig:pp_very_fast_inclusive}
\end{figure}

The truth spectrum of jets is obtained by using \fastjet to cluster
the \pythia~\cite{Sjostrand:2000wi} event with the \antikt algorithm.
Figure~\ref{fig:pp_very_fast_inclusive} shows the true jet \pt
spectrum as the solid histogram.  The convolution of the hard
parton-parton scattering cross section and the high-$x$ parton
distribution function results in a jet cross section that falls nearly
exponentially over the range 20--60~GeV, before turning steeply
downward as it approaches the kinematic limit, $x = 1$.

Figure~\ref{fig:pp_very_fast_inclusive} also shows the \veryfast
simulation result for the measured jet $E_T$ spectrum.  The main effects
of the jet resolution on the jet energy spectrum are to shift it to
higher energy and stiffen the slope slightly.  Both of these effects
can be undone reliably by a process of unfolding.  We have employed
the \roounfold~\cite{Adye:2011gm} package and for this demonstration
utilize the Iterative Bayes method with 4 iterations.  The results of
the unfolding are shown in
Figure~\ref{fig:pp_very_fast_inclusive}, along with the ratio of the
unfolded to the true $E_T$ spectrum, in the lower panel.  The ratio of
the two distributions demonstrates that the measurement provides an
accurate reproduction of the true jet energy spectrum.

\subsection{\pp Dijet Asymmetry}
\label{sec:dijet_pp}

\begin{figure}[hbt!]
  \centering
  \includegraphics[width=\onewidth]{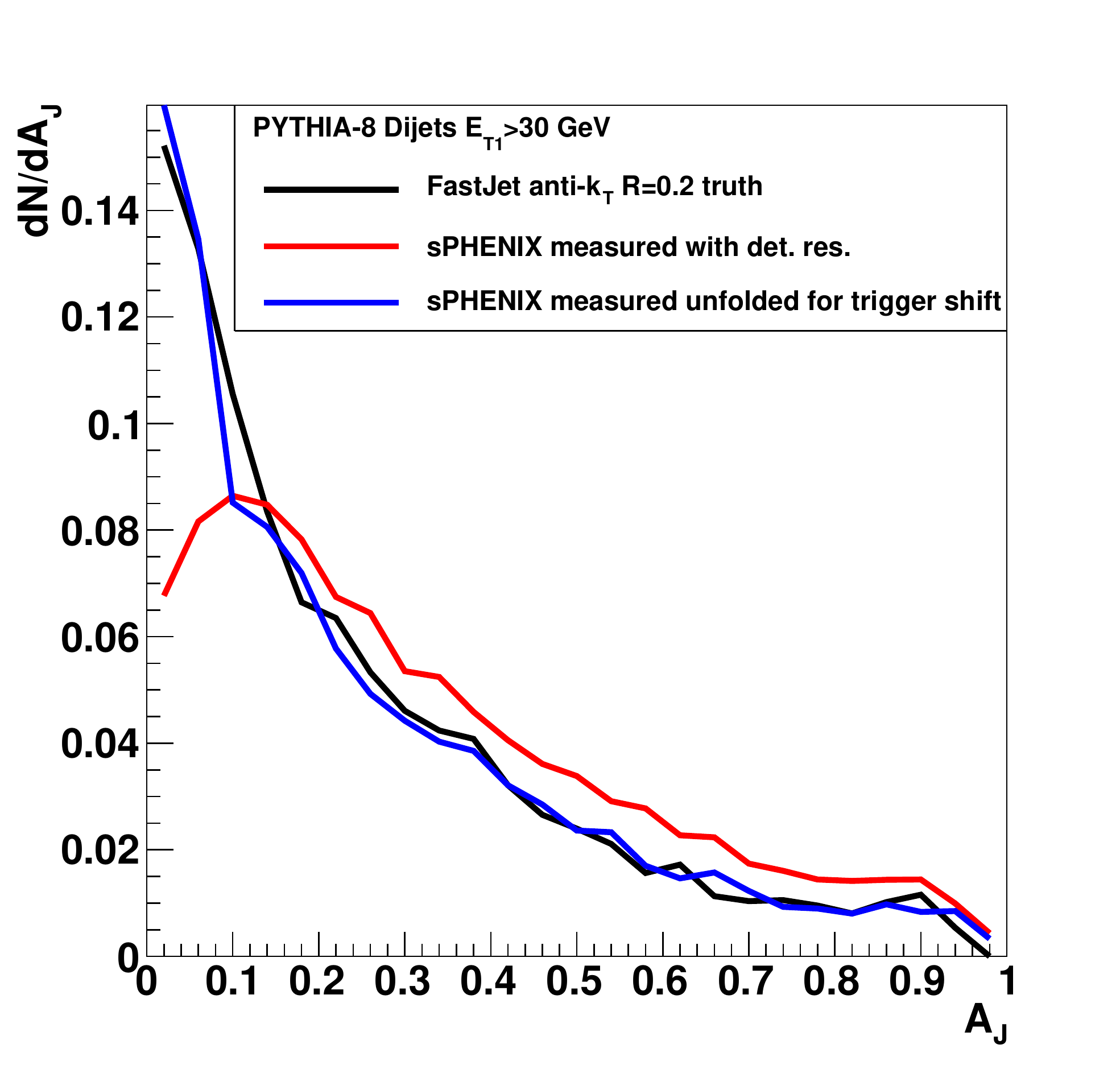}
  \caption[Dijet asymmetry, $A_J$, in \pp collisions]{Dijet asymmetry,
    $A_J$, in \pp collisions.  The truth spectrum is shown in black;
    the spectrum measured in \pythia and smeared by the jet energy
    resolution is shown in red.  The effect of the unfolding of the
    trigger jet bias is also shown in blue.}
  \label{fig:pp_very_fast_dijet}
\end{figure}

The \veryfast simulation is also used to establish expectations
for \dijet correlations.  Figure~\ref{fig:pp_very_fast_dijet}
shows the \dijet correlation for \pythia events reconstructed using the \antikt
algorithm with $R = 0.2$.  The highest energy jet in the event is taken
as the trigger jet and its transverse energy is compared to the transverse energy of the
highest energy jet in the opposite hemisphere.  

The jet asymmetry $A_{J} = (E_{T1} - E_{T2}) / (E_{T1} + E_{T2})$ for
the true jets, reconstructed at the particle level, is shown for
leading jets with $E_{T1} > 30$~GeV in
Figure~\ref{fig:pp_very_fast_dijet}.  Also shown is the simulated
measurement with the jet resolution included.  The resolution 
results in a reduction in the fraction of events observed with balanced jet energies (i.e. near
$A_{J} \approx 0$).  ATLAS and CMS \dijet asymmetries in
Pb+Pb collisions~\cite{Aad:2010bu,Chatrchyan:2011sx} are shown without unfolding for these
detector or underlying event effects.  A simultaneous
two-dimensional unfolding of both the jet energies (i.e., $E_{T1}({\rm
  meas}), E_{T2}({\rm meas}) \rightarrow E_{T1}({\rm true}),
E_{T2}({\rm true})$) is required in this case.  Both ATLAS and CMS
collaborations are actively working on this two-dimensional unfold,
and the sPHENIX group is as well.  At RHIC energies, the largest
effect is that the trigger jet is being selected from a steeply
falling spectrum and is biased by the resolution to be reconstructed
higher than the true energy.  If one simply shifts the trigger jet
down by this average bias (and inverts the identity of trigger and
associated jet if the trigger jet energy is then below that of the
associated jet), the original \dijet asymmetry distribution is
recovered, as shown in Figure~\ref{fig:pp_very_fast_dijet}.  This
procedure is not a replacement for the eventual two-dimensional
unfolding, but demonstrates the dominant effect.

\section{Jet performance in \AuAu collisions}
\label{sec:aa_jet_performance}

Here we simulate the performance of inclusive jet and \dijet
observables in heavy ion collisions.  The sPHENIX trigger and data
acquisition will sample jets from the full \auau minimum bias
centrality range, resulting in key measurements of the full centrality
dependence of jet quenching effects.  Finding jets and dealing with
the rate of \fake jets becomes much easier as the multiplicity due to
the underlying event drops, and so we have concentrated on showing
that we have excellent performance in central \auau collisions (i.e.,
in the most challenging case).

\begin{figure}[hbt!]
  \centering
  \includegraphics[trim = 0 0 0 0,clip,width=\onewidth]{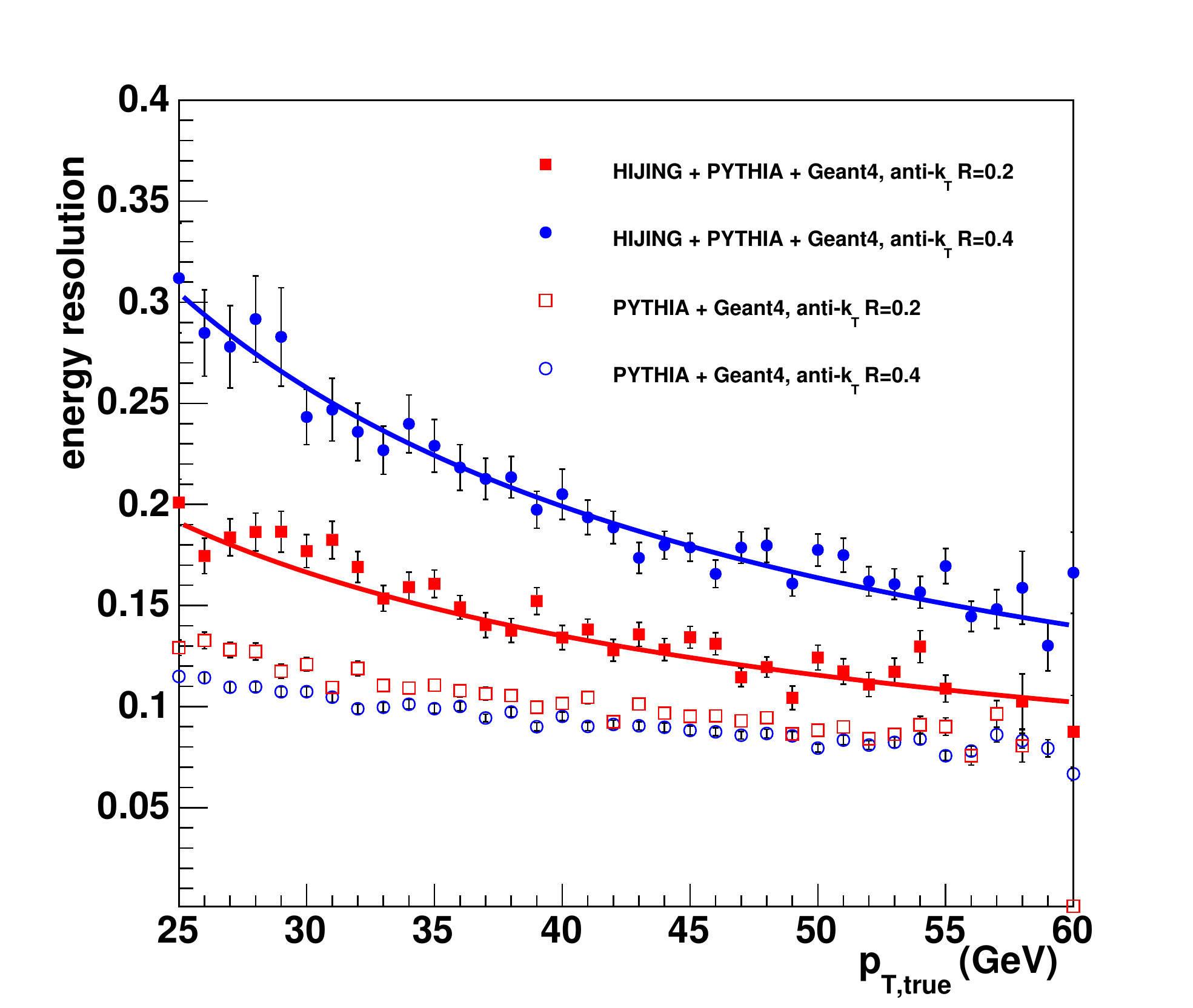}
  \caption[The \geant calculated energy resolution of PYTHIA jets
  embedded in a \auau HIJING event, reconstructed using the \antikt
  algorithm with $R = 0.2$ and $R = 0.4$]{The \geant calculated energy
    resolution of PYTHIA jets embedded in a \auau HIJING event,
    reconstructed using the \antikt algorithm with $R = 0.2$ and $R =
    0.4$. The points, showing the result of the full simulation, are
    compared to the dotted lines, showing the result obtained using
    the fast simulation.}
  \label{fig:jet_energy_resolution_AA}
\end{figure}

The effective jet resolution also has an important contribution from
fluctuations in the underlying event in the same angular space as the
reconstructed jet.  We have carried out a full \geant simulation embedding \pythia
jets into 0--10\% central \AuAu \hijing events.  The true \pythia reconstructed jets are
then compared with the \AuAu extracted jets (as detailed below) to determine the
jet energy resolution, as shown in Figure~\ref{fig:jet_energy_resolution_AA}.  Also shown
in the figure as dotted lines are the parametrized electromagnetic and hadronic calorimeter
resolution contributions used in the \fast simulation.  Again, the \geant resolutions are well
below our physics performance specifications.

In addition to the resolution effects, fluctuations in the underlying event
can create local maxima in energy that mimic jets, and are often
referred to as \fake jets.  While resolution effects can be accounted
for in a response matrix and unfolded, significant contributions of
\fake jets cannot be since they appear only in the measured
distribution and not in the distribution of jets from real hard
processes.  Thus, we first need to establish the range of jet
transverse energies and jet radius parameters for which \fake jet
contributions are minimal. Then within that range one can benchmark
measurements of the jet and \dijet physics observables.  

\subsection{Jet and Fake Jet Contributions}
\label{sec:fake_jets}

In this section we discuss both the performance for finding true jets
and estimations based on \hijing simulations for determining the contribution
from \fake jets.  
It is important to simulate very large event samples in order to evaluate the 
relative probabilities 
for reconstructing \fake jets compared to the rate of true high $E_T$ jets.  Thus, we employ 
the \fast simulation method and the \hijing simulation model for \auau collisions.
The ATLAS collaboration has found that the energy fluctuations in the heavy ion data are well matched by \hijing at 
$\sqrt{s_{NN}}=2.76$~TeV~\cite{Cole:2011zz}.
We have also added elliptic flow to the \hijing events used here.
The \fast simulation takes the particles from the event generator and parses them by their particle type.
 The calorimeter energies are summed into cells based on the detector segmentation 
and each tower is considered as a four-vector for input into \fastjet.

\begin{figure}[hbt!]
  \centering
  \includegraphics[width=\onewidth]{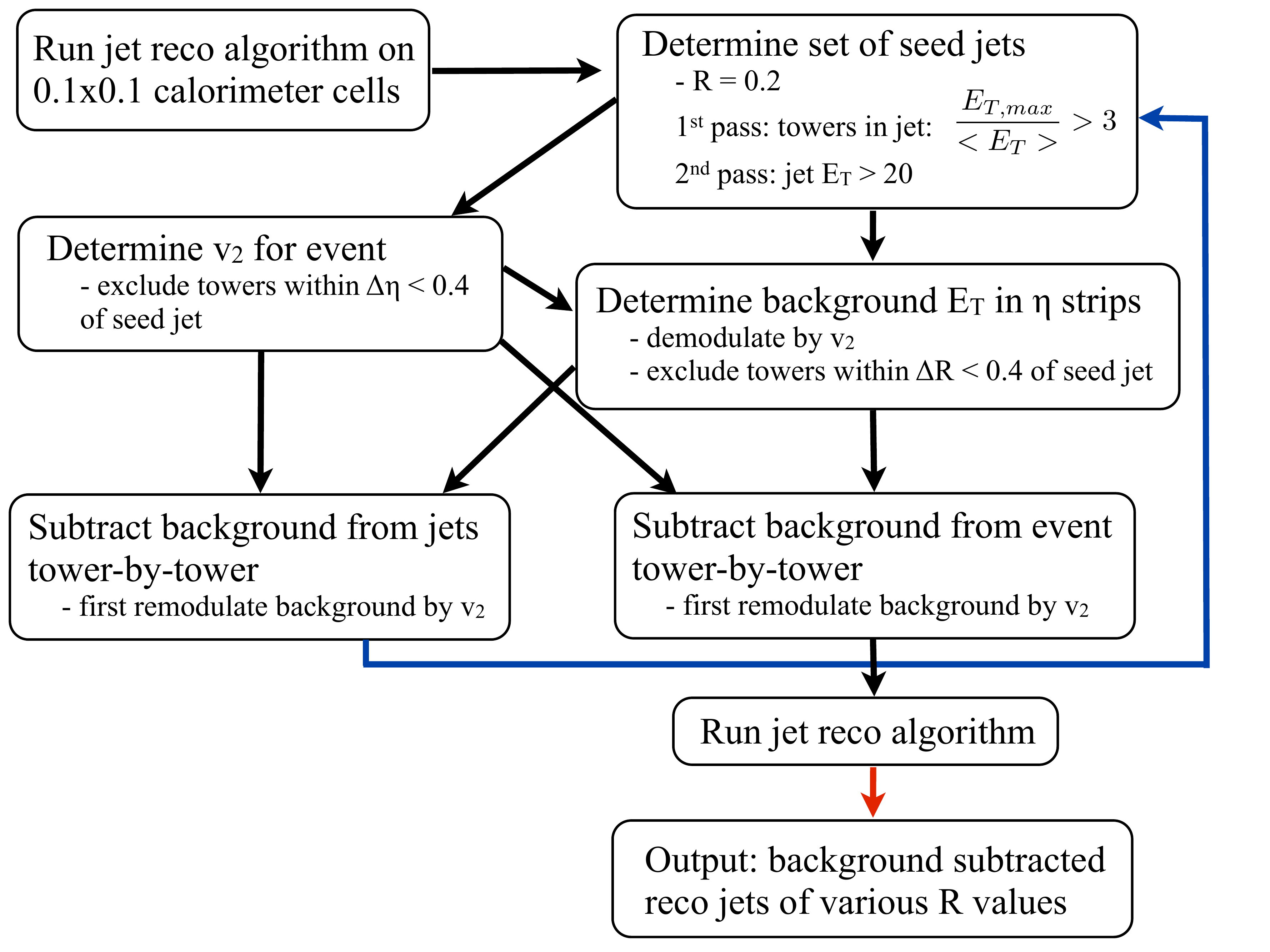}
  \caption{ Schematic diagram for the jet reconstruction algorithm.  }
  \label{fig:flowchart}
\end{figure}

Any jet measurements in heavy ion collisions must remove the
uncorrelated energy inside the jet cone from the underlying event.  
The approach developed in our studies is described in detail in Ref.~\cite{Hanks:2012wv}.
A schematic diagram of the algorithm (based on the ATLAS heavy ion method) is shown in Figure~\ref{fig:flowchart}.
Candidate jets are found and temporarily masked out of the event.  The
remaining event background is then characterized by the strength of
its $v_2$ and the overall background level in individual slices in pseudorapidity.  
Higher order flow harmonics were not included in this study.
New candidate jets are determined
and the background and $v_2$ are recalculated.  The jet finding algorithm
is then re-run on the background subtracted event to determine
the collection of final reconstructed jets.  
This process is then run iteratively to a convergent result.

In order to distinguish true jets from \fake jets we have augmented the \hijing code to run
the \fastjet \antikt algorithm with the output of each call to the fragmentation routine (HIJFRG).
In this way the true jets are identified from a single parton
fragmentation without contamination from the rest of the simulated event.
The reconstructed jets can then be compared to these true jets.  Reconstructed jets which are within
$\Delta R = \sqrt{\Delta\eta^2 + \Delta\phi^2} < 0.25$ of a true jet with $E_T>5$~GeV are considered to be matched
and those which are not are classified as \fake jets.  

Other estimates of \fake jet rates in heavy ion collisions have failed
to take into account how the structure of the background fluctuations
and the detector granularity affects the probability of any particular
fluctuation being reconstructed as a jet.  Note that simply blurring
individual particles by a Gaussian with an underlying event
fluctuation energy results in a substantial overestimate of the \fake
jet rate, and is not a replacement for a complete event simulation
incorporating \fastjet reconstruction with a full jet and underlying
event algorithm implementation.  Thus, we believe these studies
provide an accurate assessment of the effect of \fake jets.

\begin{figure}[hbt!]
  \centering
  \begin{minipage}[c]{0.47\linewidth}
    \includegraphics[width=\textwidth]{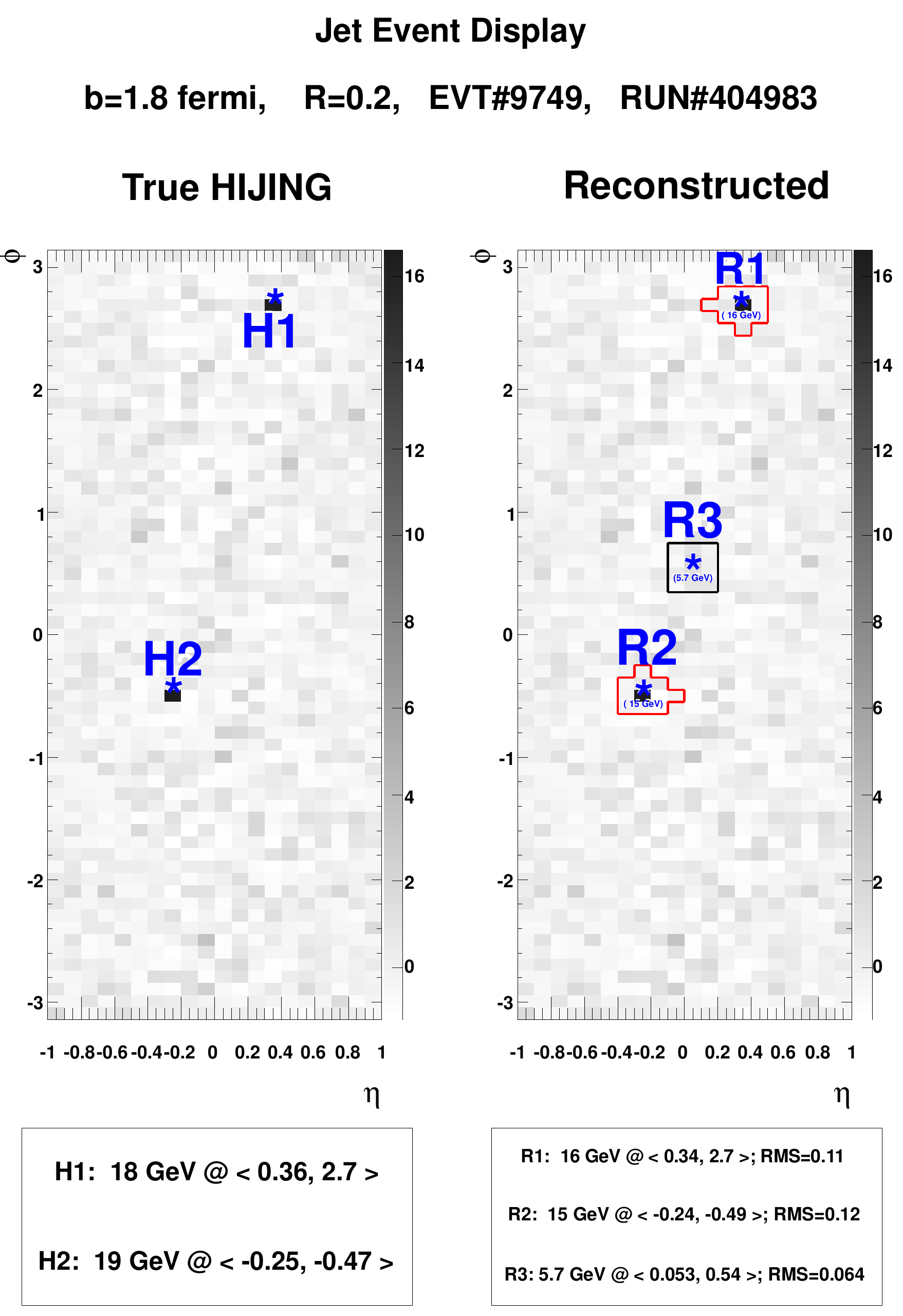} 
  \end{minipage}
  \begin{minipage}[c]{0.47\linewidth}
    \includegraphics[width=\textwidth]{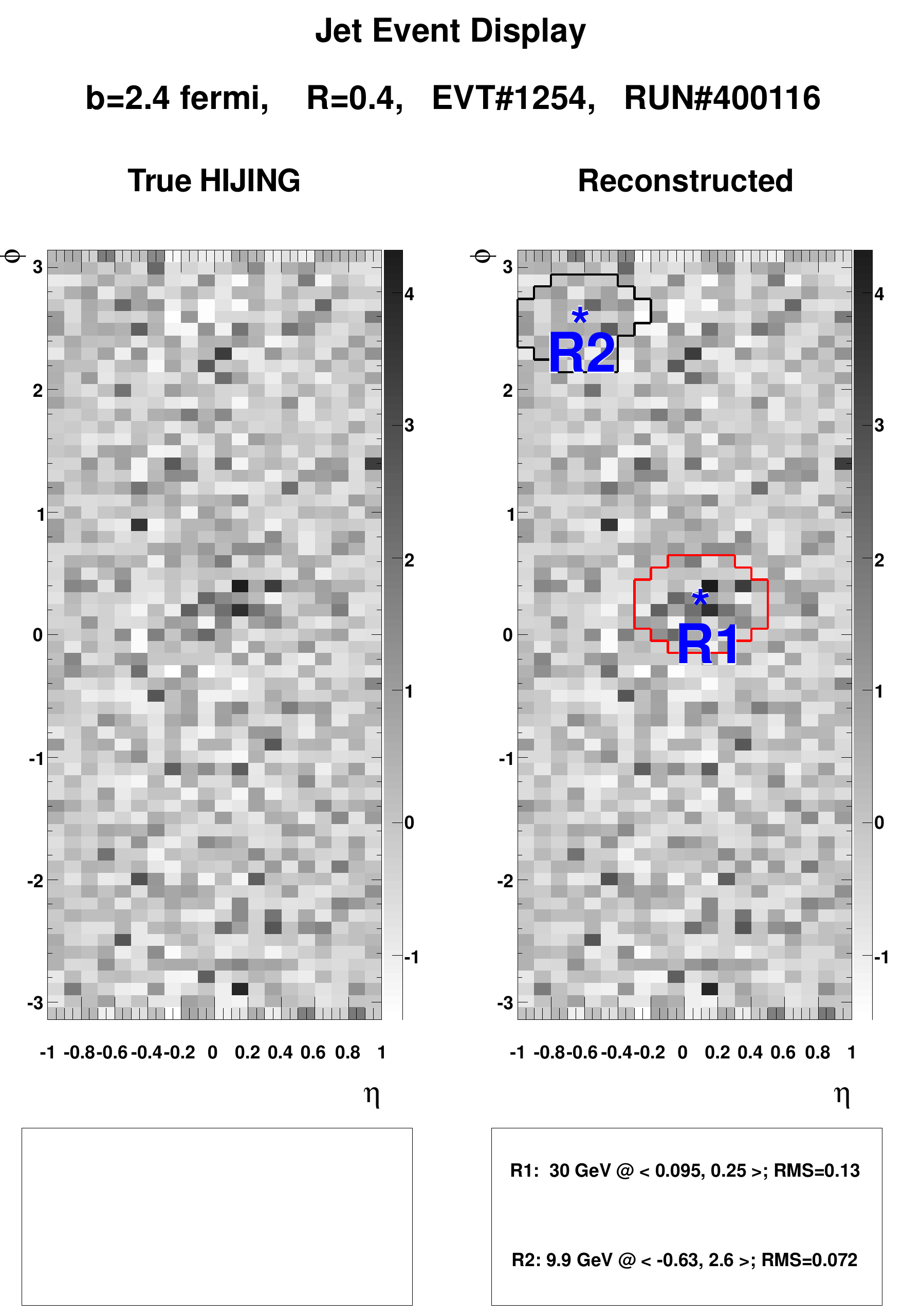}
   \end{minipage}
   \caption[Event displays of true and reconstructed jets shown
   overlaid on background subtracted calorimeter towers from fast
   simulation]{Event displays of true and reconstructed jets shown
     overlaid on background subtracted calorimeter towers from fast
     simulation.  The left event shows a \hijing \dijet event where
     both dijets (labeled H1 and H2) are reconstructed and matched (R1
     and R2).  A third jet, not matched to a true jet, is also
     reconstructed (R3).  The right event shows a \hijing event with
     no true jets with $E_T>5$~GeV.  Two \fake jets are
     reconstructed, one with $E_T=30$~GeV.}
  \label{fig:eventdisplays}
\end{figure}

As an illustration of true and \fake jets we show two calorimeter
event displays in Figure~\ref{fig:eventdisplays}.  True jets at high
$E_T$ are a rare occurrence.  A large energy background fluctuation at
high $E_T$ that mimics a jet is also a rare occurrence.  Thus the only
way to quantify the impact of \fake jets on the jet performance is to
run a large sample of untriggered simulated events and assess the
relative probability of true and \fake jets as a function of $E_T$ and
R.

A sample of over 750 million minimum bias \auau \hijing events at
$\sqrt{s_{NN}} = 200$~GeV with quenching turned off was used in these
studies.  The observable particles are binned in
$\eta$-$\phi$ cells of size $\Delta \eta \times \Delta \phi = 0.1
\times 0.1$.  In these studies, we have not
included smearing due to detector resolution as it is expected to be a
sub-dominant effect and we want to isolate the effects of the
underlying event.  At the end of this Section we present results
including detector resolution that do not change the key conclusions
of these studies.  

\begin{figure}[hbt!]
  \centering
  \begin{minipage}[c]{0.62\linewidth}
    \includegraphics[width=\textwidth]{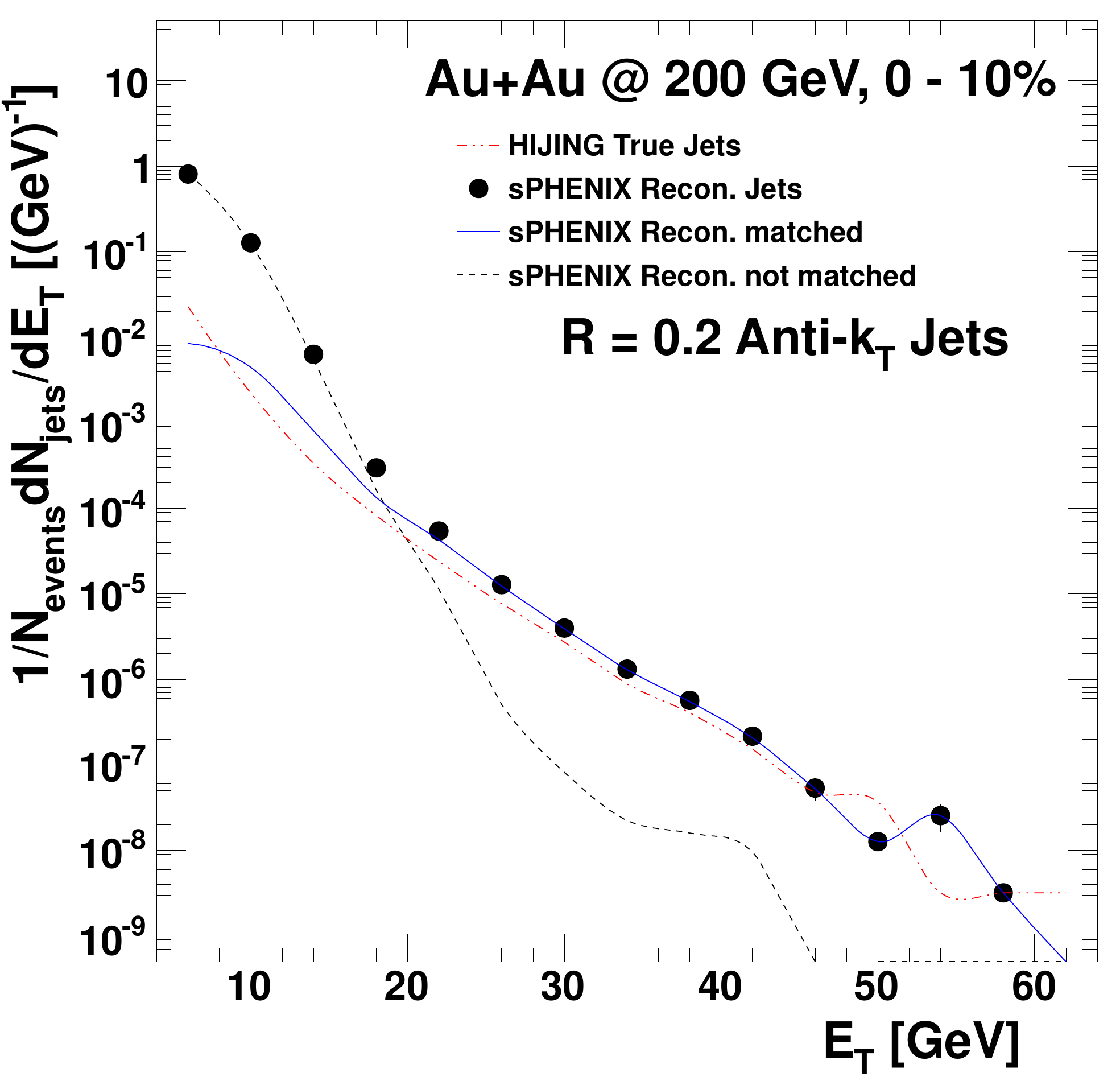} 
  \end{minipage}
  \raisebox{0.35cm}{\begin{minipage}[c]{0.36\linewidth}
    \includegraphics[trim=0 0 40 0,clip,width=\textwidth]{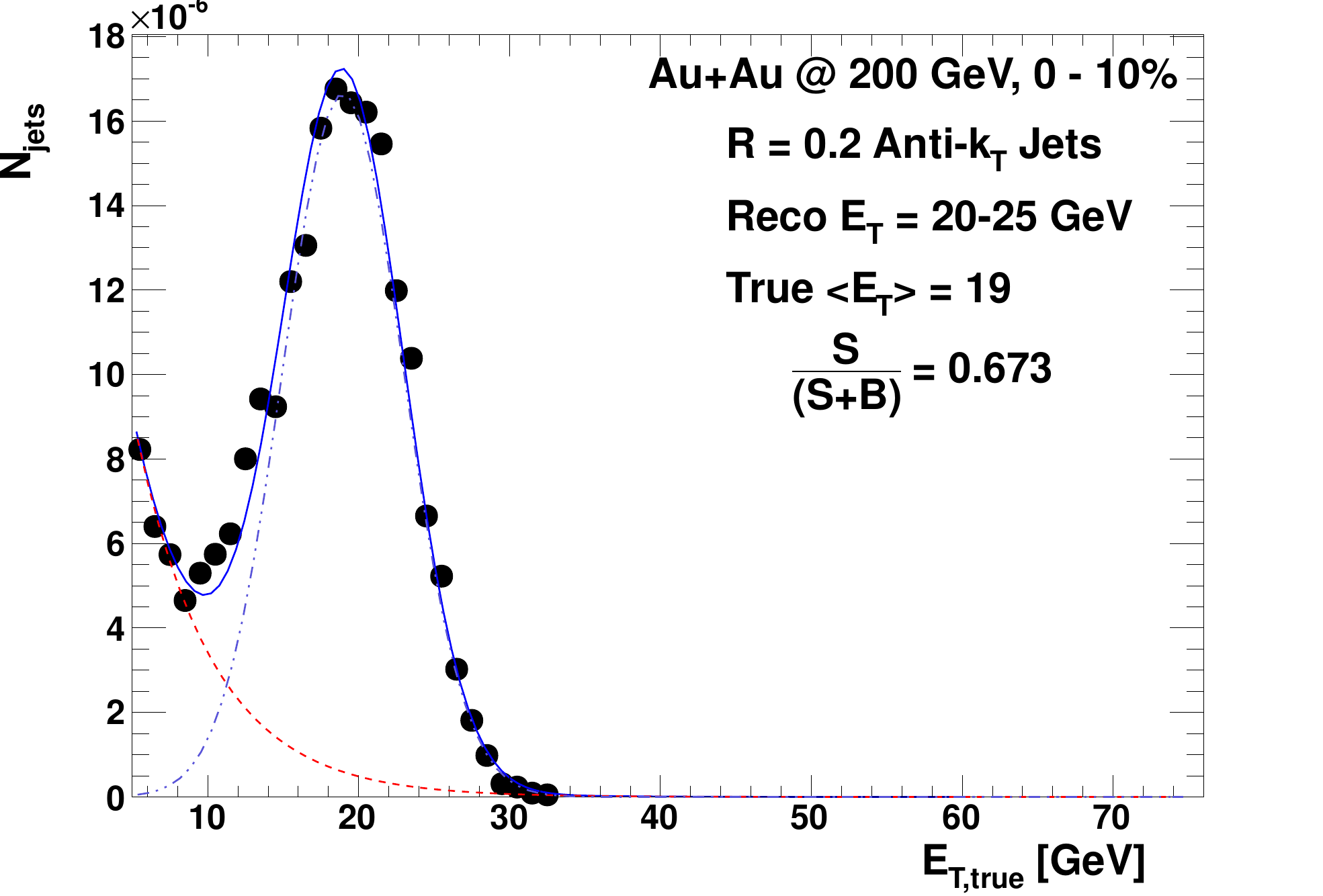}
    \\
    \\
    \includegraphics[trim=0 0 40 0,clip,width=\textwidth]{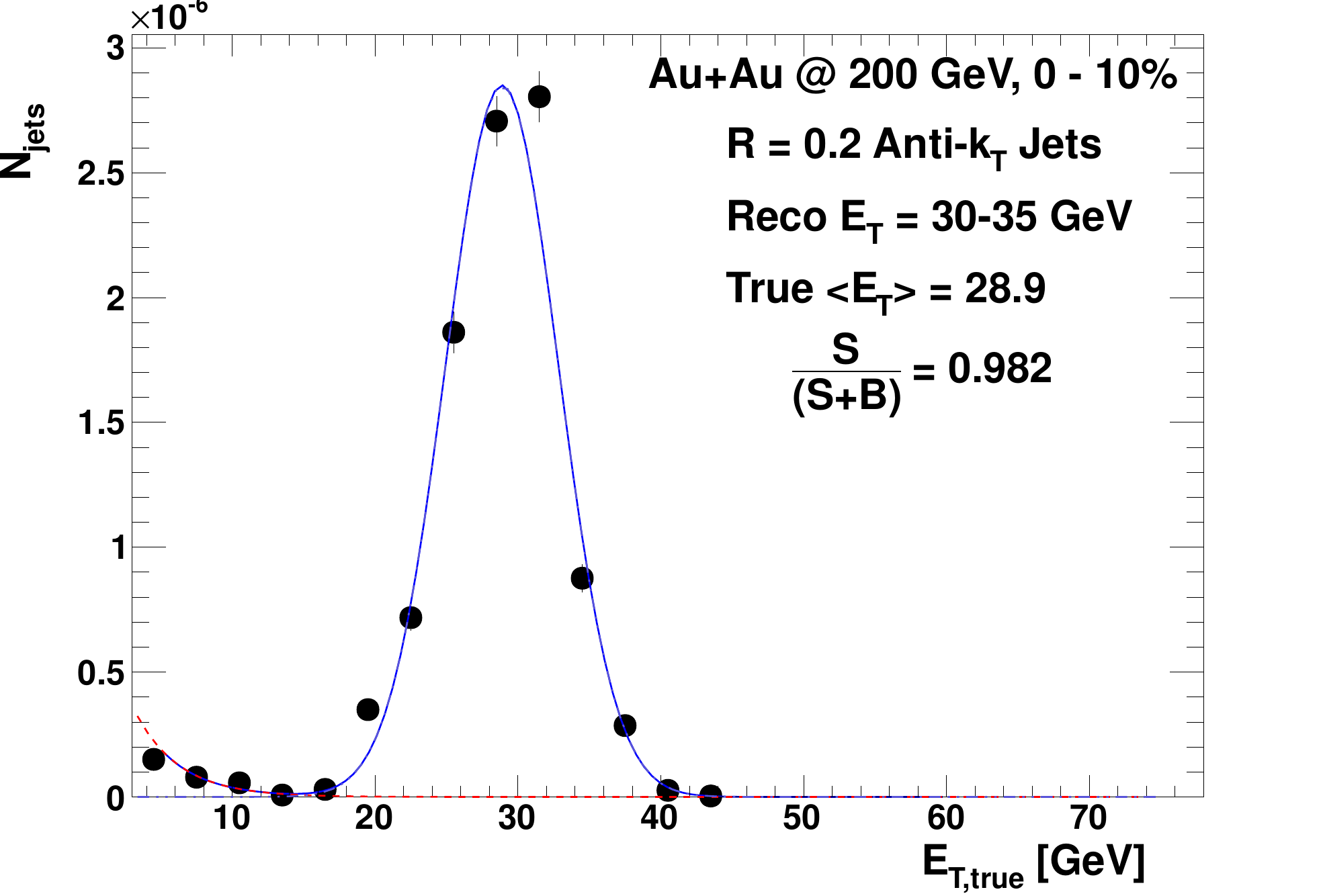}
  \end{minipage}}
\caption[The true/fake composition of the jet spectrum in central
0--10\% \auau based on 750M \hijing events]{The composition of the jet
  spectrum in central 0--10\% \auau based on 750M \hijing events.  The
  full spectrum is shown in the left plot as solid points.  The
  spectrum of those jets that are successfully matched to known real
  jets is shown as a blue curve.  That curve compares very well with
  the spectrum of true jets taken directly from \hijing.  The jets
  which are not matched with known jets are the \fake jets, and the
  spectrum of those jets is shown as the dashed curve. For $R = 0.2$,
  real jets begin to dominate over \fake jets above 20~GeV.  The
  panels on the right are slices in true jet energy showing the
  distribution and make up of the reconstructed jet energy.  At low
  $E_{\rm true}$, \fake jets can be seen encroaching on the low energy
  side of the distribution.  For higher $E_{\rm true}$ the \fake jets
  are negligible.}
  \label{fig:auau_fake_rate}
\end{figure}

\begin{figure}
\centering
\begin{minipage}[c]{0.49\linewidth}
\includegraphics[width=\textwidth]{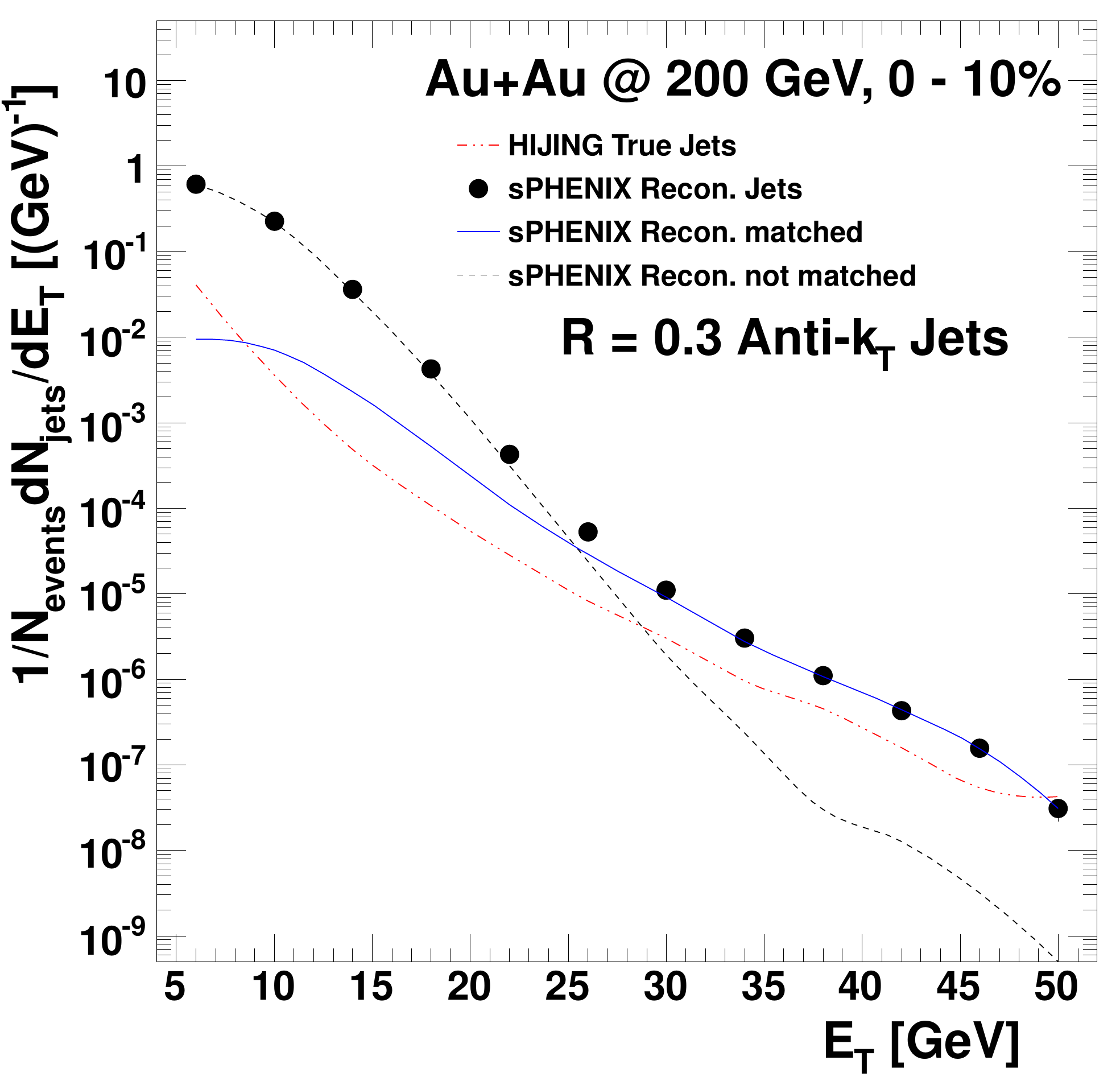}
\end{minipage}
\begin{minipage}[c]{0.49\linewidth}
\includegraphics[width=\textwidth]{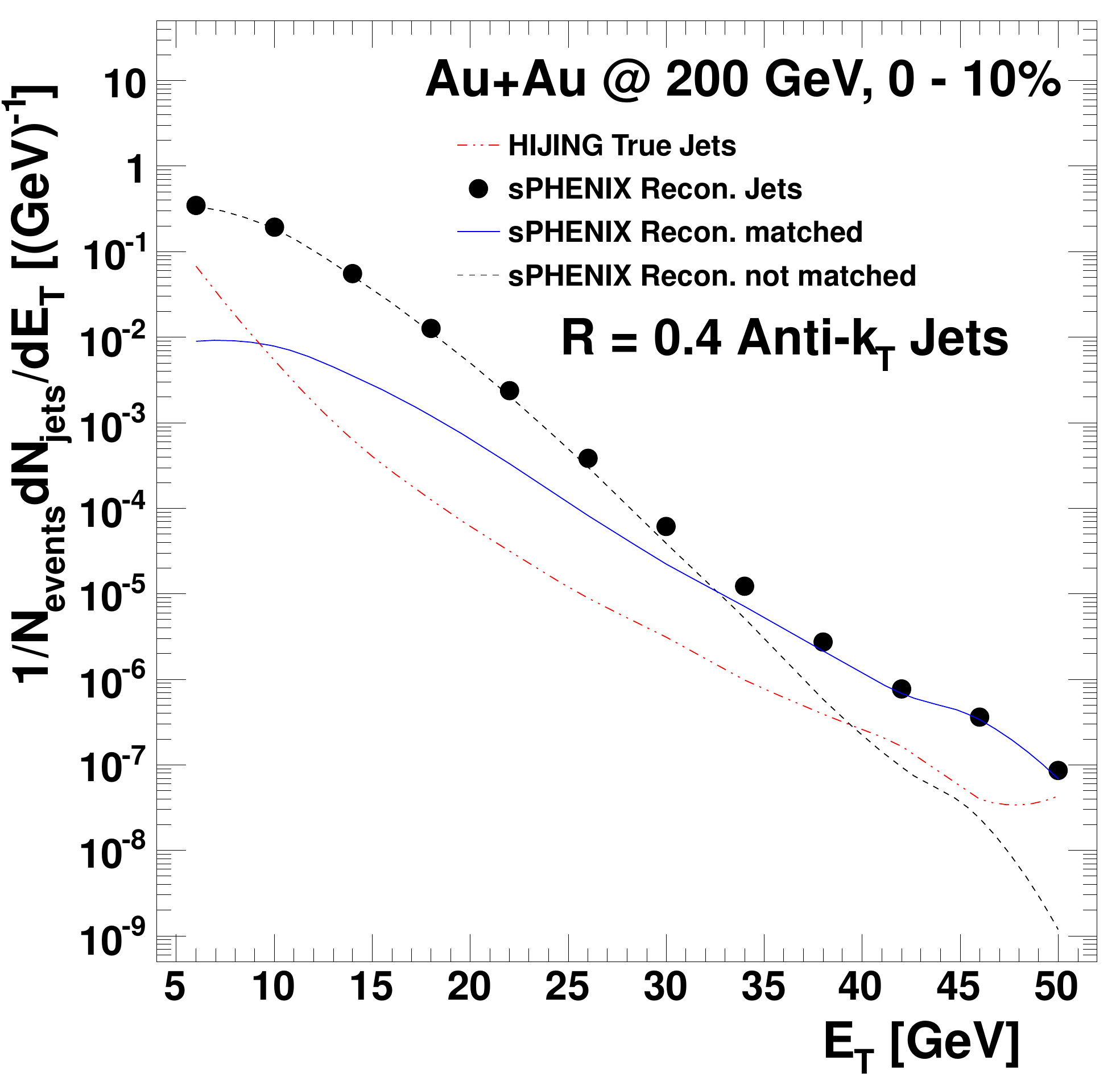}
\end{minipage}
\caption[The true/fake composition of the jet spectra in central
0--10\% \auau based on 750 million \hijing events for $R=0.3$ and
$R=0.4$ jets.]{Composition of the jet spectra in central 0--10\% \auau
  based on 750 million \hijing events for $R=0.3$ (left) and $R=0.4$
  (right) jets.}
\label{fig:auau_fake_rate_wide}
\end{figure}

The \fast simulation result for $R = 0.2$ jets without including
detector-level smearing of the jet energies is shown in
Figure~\ref{fig:auau_fake_rate}.  The full spectrum is shown on the
left as solid points.  The spectrum of those jets that are
successfully matched to true jets is shown as a blue curve.  That
curve compares very well with the spectrum of true jets taken directly
from \hijing.  The \fake jet, those not matched with a true jet,
spectrum is shown as the dashed curve. For $R = 0.2$, real jets begin
to dominate over \fake jets above 20~GeV.  The panels on the right of
Figure~\ref{fig:auau_fake_rate} are slices in reconstructed jet energy
showing the distribution and make up of the true jet energy.  For
reconstructed jets with $E_T = $25--30~GeV, a contribution of \fake
jets can be seen encroaching on the low energy side of the
distribution.  For $E_{\rm reco}>25$~GeV \fake jets are at the 10\%
level and for $E_{\rm reco}>30$~GeV \fake jets are negligible.
Contributions from \fake jets for larger jet cones are shown in
Figure~\ref{fig:auau_fake_rate_wide}.  The true jet rate becomes large
compared to the \fake jet rate at 30~GeV for $R=0.3$ and 40~GeV for
$R=0.4$.  We note that in one year of RHIC running, sPHENIX would
measure $10^{5}$ jets with $E_T > 30$~GeV and $10^{4}$ jets with $E_T
> 40$~GeV.

There are various algorithms for rejecting \fake jets based on the jet
profile or the particles within the jet.  These methods applied by the
ATLAS experiment significantly reduce the \fake rate by an order of
magnitude or more, increasing the energy and $R$ values over which it
is possible to measure jets~\cite{Aad:2012vca}. A detailed study of this \fake jet
rejection method and its utility is enabling new physics is discussed later
in Section~\ref{sec:jet_surface_emission_engineering}.

\begin{figure}[hbt!]
  \centering
  \includegraphics[angle=0,trim = 0 0 0 0,clip,width=\onewidth]{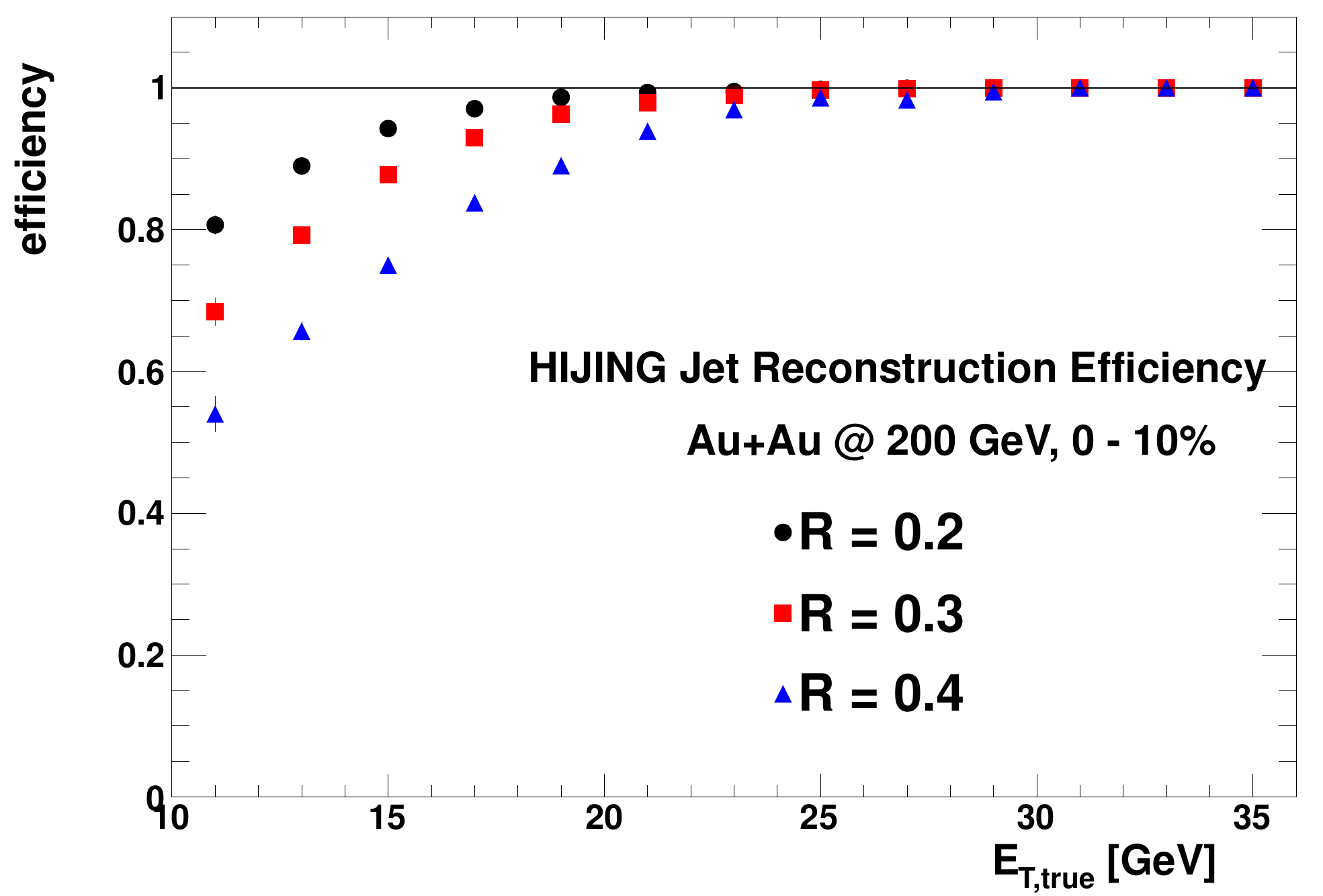}
  \caption{The efficiency for finding jets in central \auau collisions
    as a function of true jet energy and for $R = 0.2$, 0.3 and 0.4.}
  \label{fig:reconstruction_efficiency}
\end{figure}

The efficiency of finding true jets is shown in 
Figure~\ref{fig:reconstruction_efficiency}. We find $>95$\%
efficiency for finding jets above 20~GeV reconstructed with $R =
0.2$ or 0.3 and above 25~GeV for jets reconstructed using $R = 0.4$.

\begin{figure}[hbt!]
  \centering
  \includegraphics[width=0.8\textwidth]{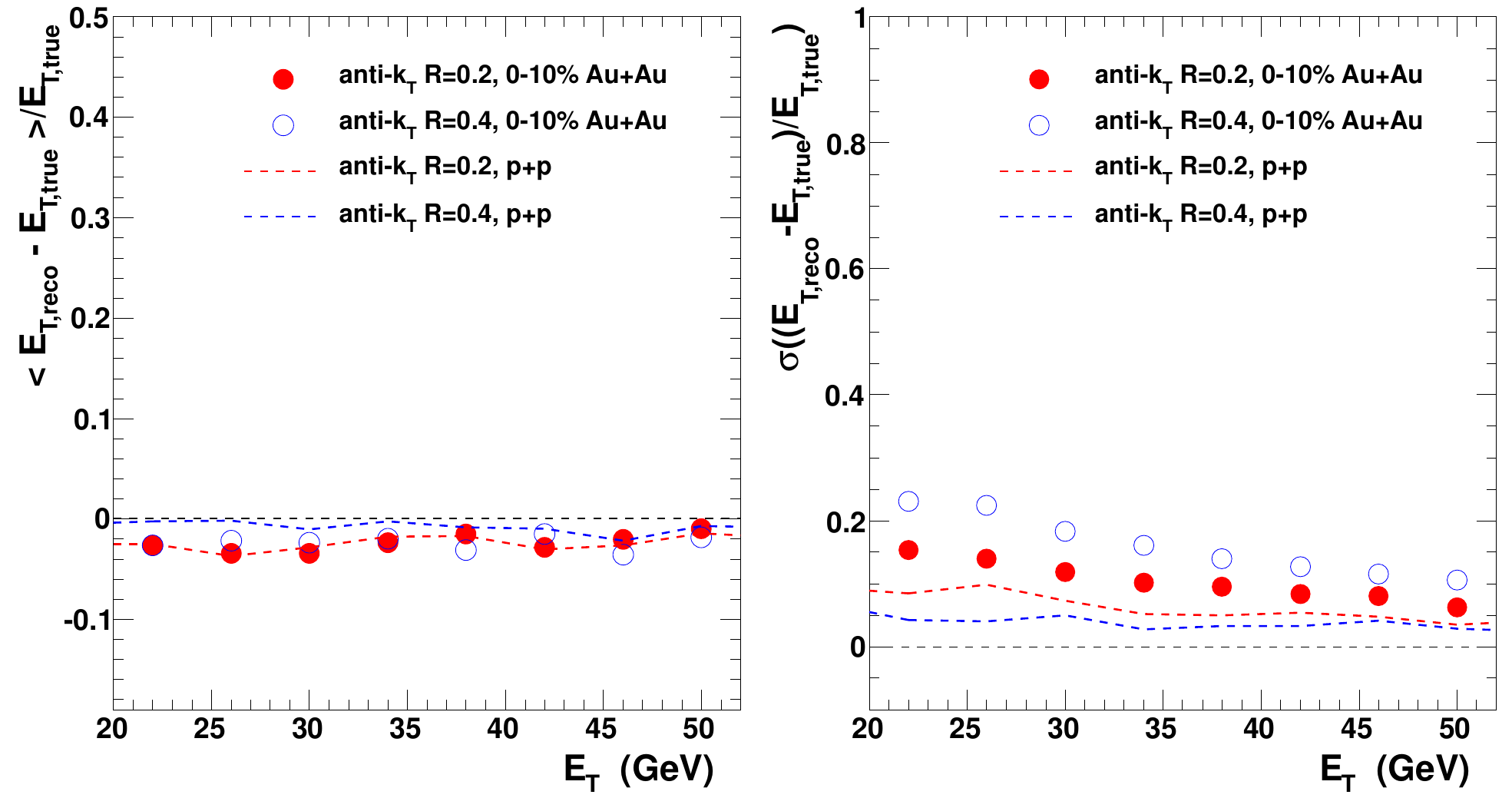}
  \caption[The jet energy scale and energy resolution of reconstructed
  jets in \auau collisions]{The jet energy scale and energy resolution
    of reconstructed jets in \auau collisions. The left plot shows the
    shift in the mean energy of the reconstructed jets compared to the
    true value.  There is only a few percent shift in the energy and
    no apparent dependence on jet cone size.  The right plot shows the
    jet energy resolution.}
  \label{fig:auau_jet_energy_scale}
\end{figure}

Having found the jets in \auau with good efficiency and having
established that the rate of \fake jets coming as a result of
background fluctuations are understood and under control, we also need
to show that we can reconstruct the kinematics of jets accurately and
precisely. This is quantified by the jet energy scale, the average shift
of the jet energy between reconstructed and true jets and the
jet energy resolution which shows the relative width of the 
difference between the true and reconstructed jet energies. 
Results from $R = 0.2$ and 0.4 are shown in Figure~\ref{fig:auau_jet_energy_scale}.  
For both jet radii the
jets are reconstructed within 4\% of the true energy over the 
measured range.  Note that this is just a first step energy scale determination.
The jet energy resolution shown in the right panel
only includes effects due to the detector segmentation applied
and the underlying event resolution.  In \pp collisions
the resolution for $R = 0.4$ jets is better than for $R = 0.2$ jets
because the segmentation can cause jet splitting with the smaller
jet cones.  In \auau collisions the order is swapped because the 
dominant effect is the additional smearing due to the underlying
event.

\begin{figure}[hbt!]
  \centering
  \includegraphics[trim=0 0 0 0,clip,width=\onewidth]{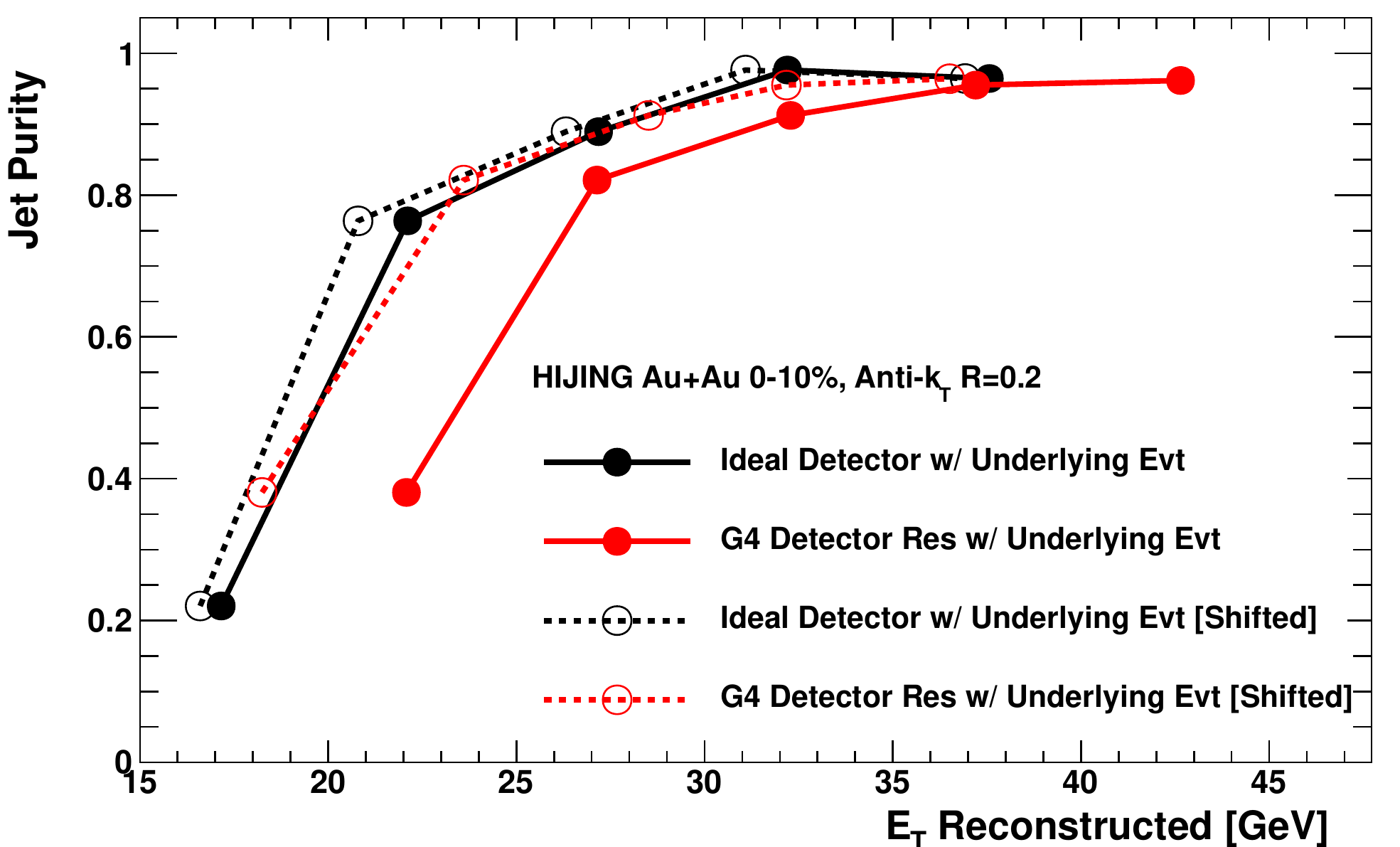}
  \caption[Jet purity ($S/(S+B)$) in 0--10\% \auau collisions from
  \hijing.  The purity values are for a ideal detector smeared by the  \geant parametrized
EMCal and HCal resolutions]{Results for the jet purity ($S/(S+B)$) in terms of matched
    true and \fake jets 
in 0--10\% \auau collisions from \hijing.  The purity values are for a ideal detector (i.e, 
sPHENIX segmentation with perfect resolution) and then including the \geant parametrized
EMCal and HCal resolutions.  Both results are then shifted down in $E_T$ by the 
reconstructed energy bias.
\label{fig:fakestudy_detres}}
\end{figure}
The \fast simulation results described above have been re-run with the
inclusion of the detector resolutions as parametrized from the single
particle \geant results --- detailed in Section~\ref{sec:g4sim}.  The
results shown in Figures~\ref{fig:auau_fake_rate} and
~\ref{fig:auau_fake_rate_wide} remain quite similar with the detector
resolution included, though with an overall shift of all the
distributions to higher $E_T$ due to the additional blurring on
falling spectra.  For $R=0.2$ jets, the smearing due to detector
resolution is comparable to the effect of the underlying event and for
larger jet cones the effect of the underlying event is found to be
much larger than detector resolution effects.
Figure~\ref{fig:fakestudy_detres} shows the jet purity
for $R=0.2$ jets as a function of reconstructed $E_T$.  The solid
black (red) points correspond to the cases without (with) detector
resolution effects.  Also shown as open points are both results
shifted down in energy by the average reconstructed energy bias as
determined from the reconstructed matched jet sample.  One observes
that the relative true and \fake jet contributions are the same for
the equivalent true jet energy ranges.

\clearpage

\subsection{Underlying Event and Detector Effects}
\label{sec:ue_and_detector}
To further evaluate possible differences between the fast
parameterized and full \geant simulations of the jet performance, a
study of the underlying event $E_\mathrm{T}$ distributions was
conducted. In this study, the total transverse energy
($\Sigma{E}_\mathrm{T}$) in fixed position windows with a large
acceptance in $\Delta\phi\times\Delta\eta$ was compared in \hijing
\auau $b = 4$~fm and $b = 8$~fm events under three different models of
the detector response: first, the truth transverse energy was summed
for all final-state, visible particles in the \hijing event record;
second, the $\Sigma{E}_\mathrm{T}$ was measured after a fast
parameterization of the detector response; third, the
$\Sigma{E}_\mathrm{T}$ was measured in calorimeter towers in the
window after a full \geant simulation. The $\Sigma{E}_\mathrm{T}$ thus
constructed was measured for the same events and in the same regions
for each model of the response.

\begin{figure}[!hbt]
 \begin{center}
    \includegraphics[width=0.48\linewidth]{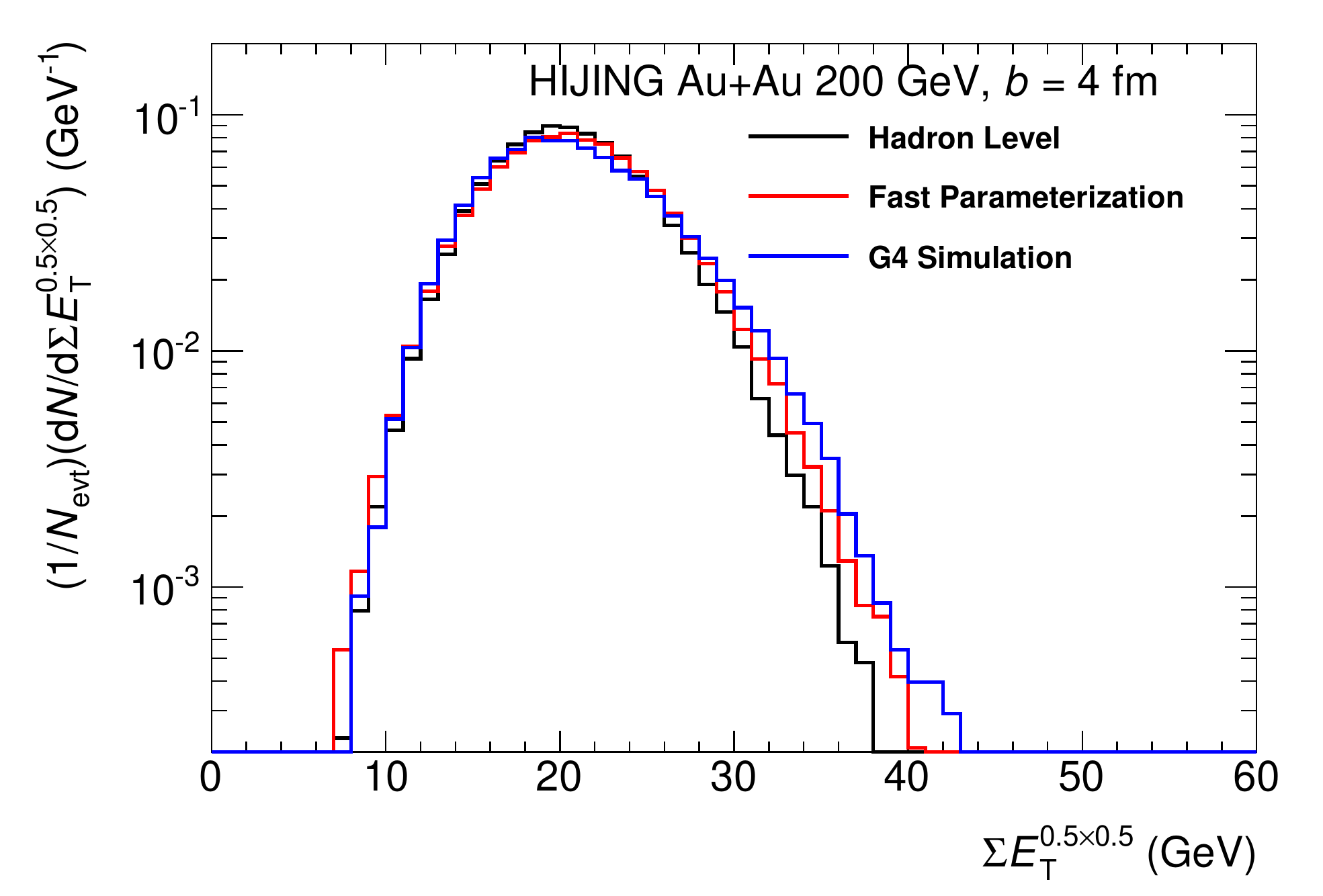}
    \includegraphics[width=0.48\linewidth]{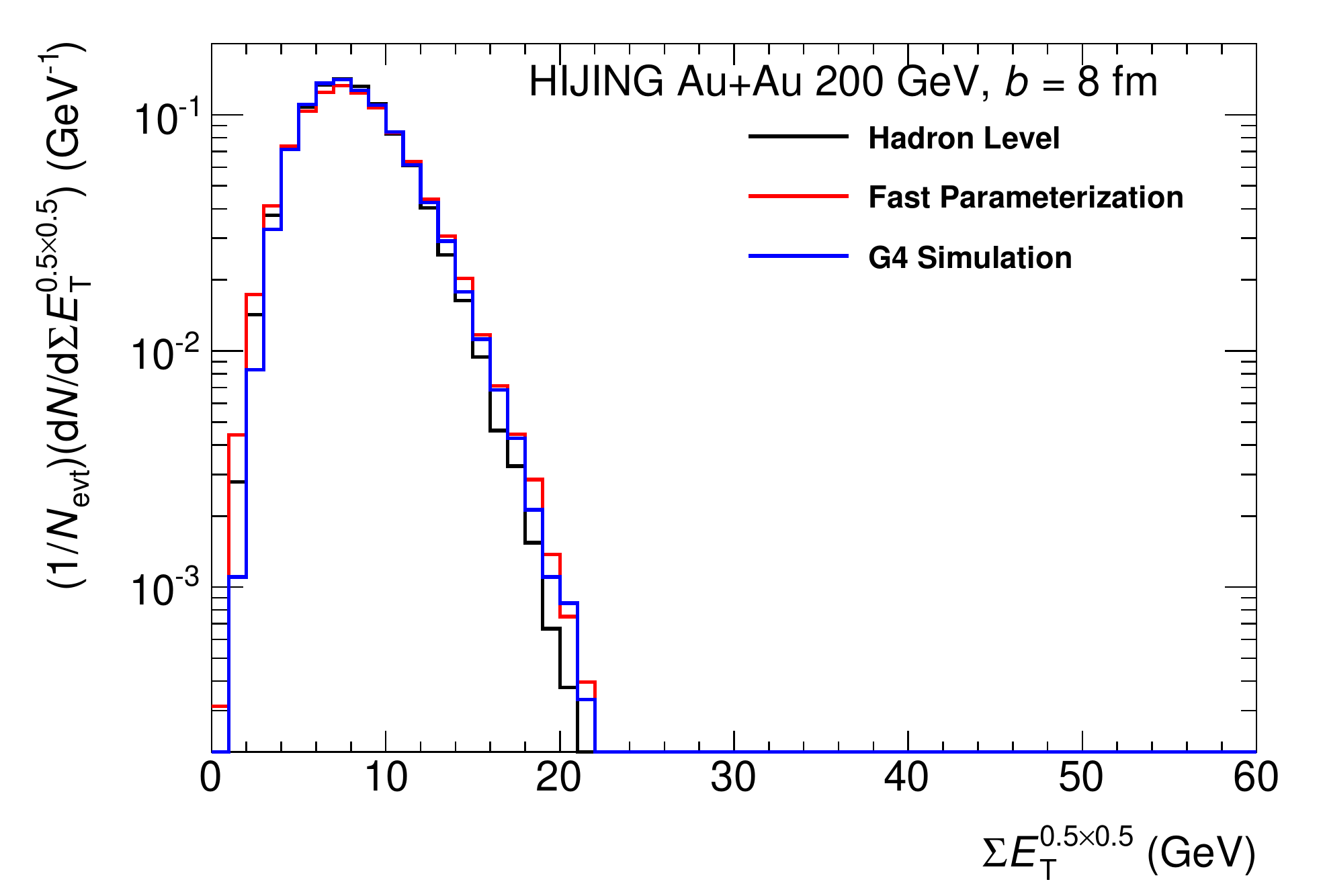}
    \caption[Total transverse energy in
    $\Delta\eta\times\Delta\phi=0.5\times0.5$ regions
    ($\Sigma{E}_\mathrm{T}^{0.5\times0.5}$) in \hijing \auau
    $\sqrt{s_\mathrm{NN}} = 200$~GeV events with $b = 4$~fm and $b =
    8$~fm]{\label{fig:sPHENIX_UEcompare_examples} Distributions of the
      total transverse energy in
      $\Delta\eta\times\Delta\phi=0.5\times0.5$ regions
      ($\Sigma{E}_\mathrm{T}^{0.5\times0.5}$) in \hijing \auau
      $\sqrt{s_\mathrm{NN}} = 200$~GeV events with $b = 4$~fm (left
      panel) and $b = 8$~fm (right panel). The total energy is shown at
      the final state hadron level (black lines), with a fast
      parameterization of the detector response (red lines) and with a
      full \geant-based simulation (blue lines). }
 \end{center}
\end{figure}

\begin{figure}[!hbt]
 \begin{center}
    \includegraphics[width=0.48\linewidth]{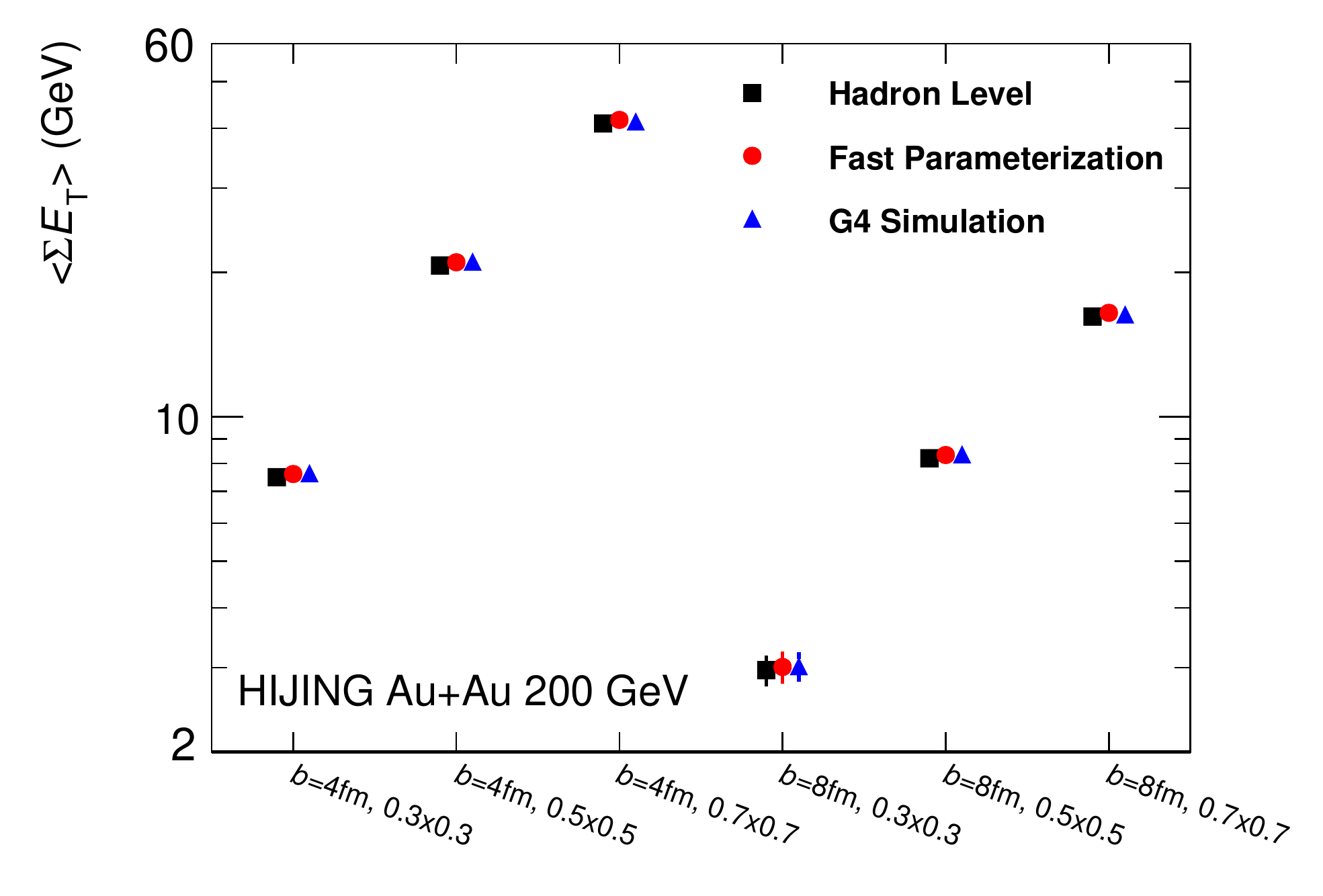}
    \includegraphics[width=0.48\linewidth]{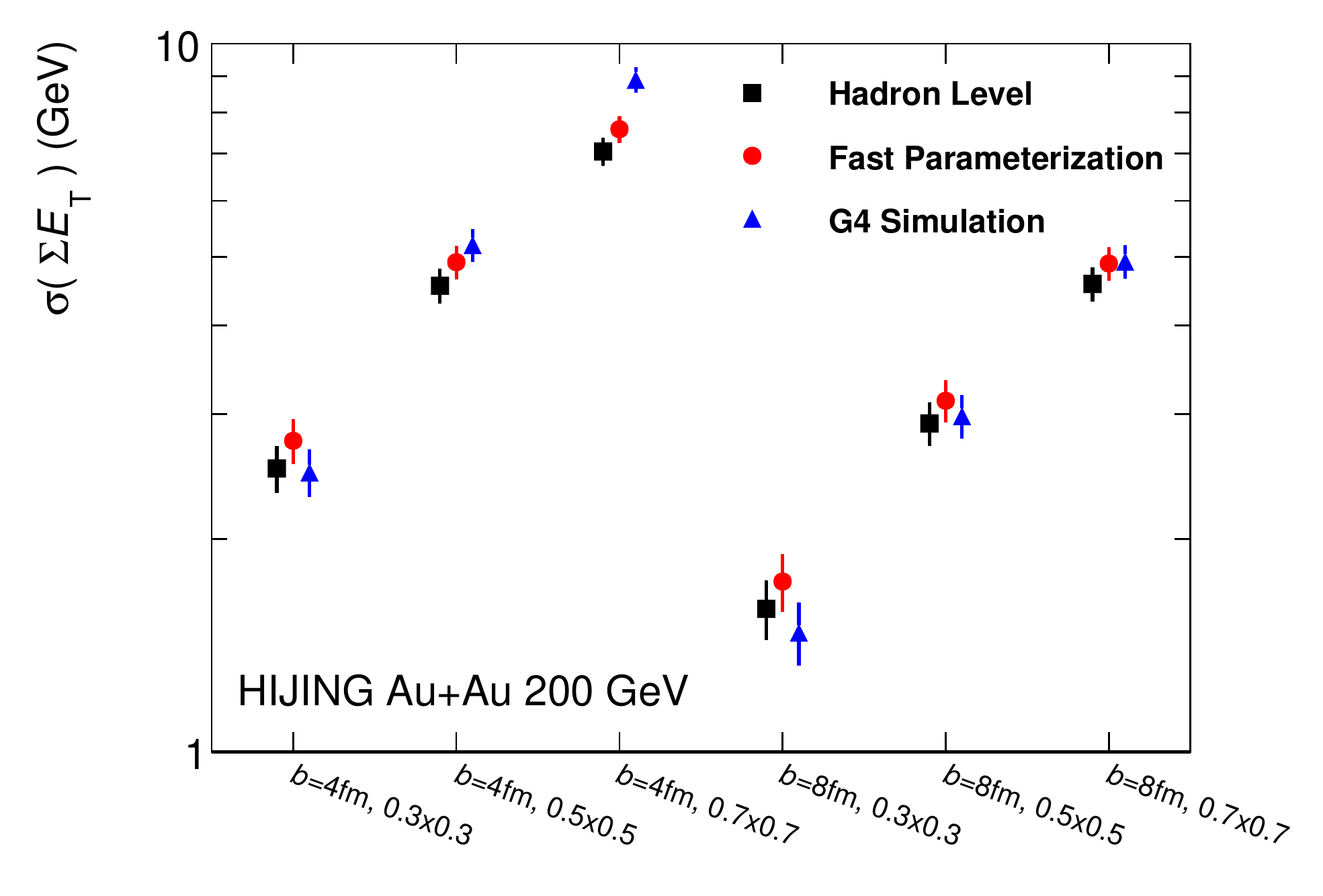}
    \caption[Mean and RMS of the total transverse energy in finite-sized
    regions ($\Sigma{E}_\mathrm{T}$), in \hijing \auau
    $\sqrt{s_\mathrm{NN}} = 200$~GeV
    events]{\label{fig:sPHENIX_UEcompare_quantified} Means (left panel)
      and RMS values (right panel) of the total transverse energy in
      finite-sized regions ($\Sigma{E}_\mathrm{T}$), in \hijing \auau
      $\sqrt{s_\mathrm{NN}} = 200$~GeV events. The horizontal axis shows
      the results for \hijing events at different impact parameter ($b =
      4$~fm and $b = 8$~fm) and for regions of different
      $\Delta\eta\times\Delta\phi$ size ($0.3\times0.3$, $0.5\times0.5$
      and $0.7\times0.7$). Results are shown for the total energy at the
      final state hadron level (black lines), with a fast
      parameterization of the detector response (red lines) and with a
      full \geant-based simulation (blue lines).  }
 \end{center}
\end{figure}

Figure~\ref{fig:sPHENIX_UEcompare_examples} shows an example of the
$\Sigma{E}_\mathrm{T}$ distributions for windows of size
$\Delta\eta\times\Delta\phi=0.5\times0.5$ (corresponding approximately
to the area under an $R=0.3$ jet), for the $b = 4$~fm and $b = 8$~fm
\hijing. The distributions are broadly similar, albeit with slight
differences in the shapes arising from the $E_\mathrm{T}$-dependent
resolution introduced by the fast parameterized and \geant
simulations. Figure~\ref{fig:sPHENIX_UEcompare_quantified} quantifies
the mean and root mean square values of the $\Sigma{E}_\mathrm{T}$
distributions for each model of the detector response. The panels show
these quantities for different choices of window size and separately
for $b = 4$~fm and $b = 8$~fm \hijing events. Generally, the fast
parameterized and \geant results reproduce the mean of the original
truth distributions well, but with slightly larger widths. These
initial studies demonstrate that while there are modest differences
between the different models of the detector response, the main
features of the $\Sigma{E}_\mathrm{T}$ distributions in these
high-multiplicity events are driven by the event to event fluctuations
of the soft particle production and not by the model of the detector
resolution.

Beyond effects due to various degrees of detector modeling realism,
one could be concerned that any particular method for dealing with the
underlying event could bias the obtained results.  In fact, there are
a number of alternate approaches in current to account for the effects
of the underlying event on jet observables.  The sPHENIX detector
has the capabilities to investigate multiple methods, not only to
gauge systematic uncertainties on a single result, but also to study
the physics issues highlighted by the different methods.

In Section~\ref{sec:fake_jets}, we described an approach for
subtracting contributions to the jet signal based on a method used by
the ATLAS experiment.  Here we show briefly the potential for a
different technique, used by both the STAR and ALICE experiments, in
which the background in A$+$+A and \pA events is calculated
event-by-event and then subtracted jet-by-jet using the formula
$p_{T,\mathrm{jet}} = p_{T,\mathrm{jet}}^\mathrm{rec} −
\rho_\mathrm{ch} \times A_\mathrm{jet}$, where the charged background
energy density $\rho_\mathrm{ch}$ is calculated as the median of
$p_T/A_\mathrm{jet}$ and $A_\mathrm{jet}$ is the jet area as
determined by the jet finder.

\begin{figure}[hbt!]
  \centering
  \includegraphics[width=0.8\linewidth]{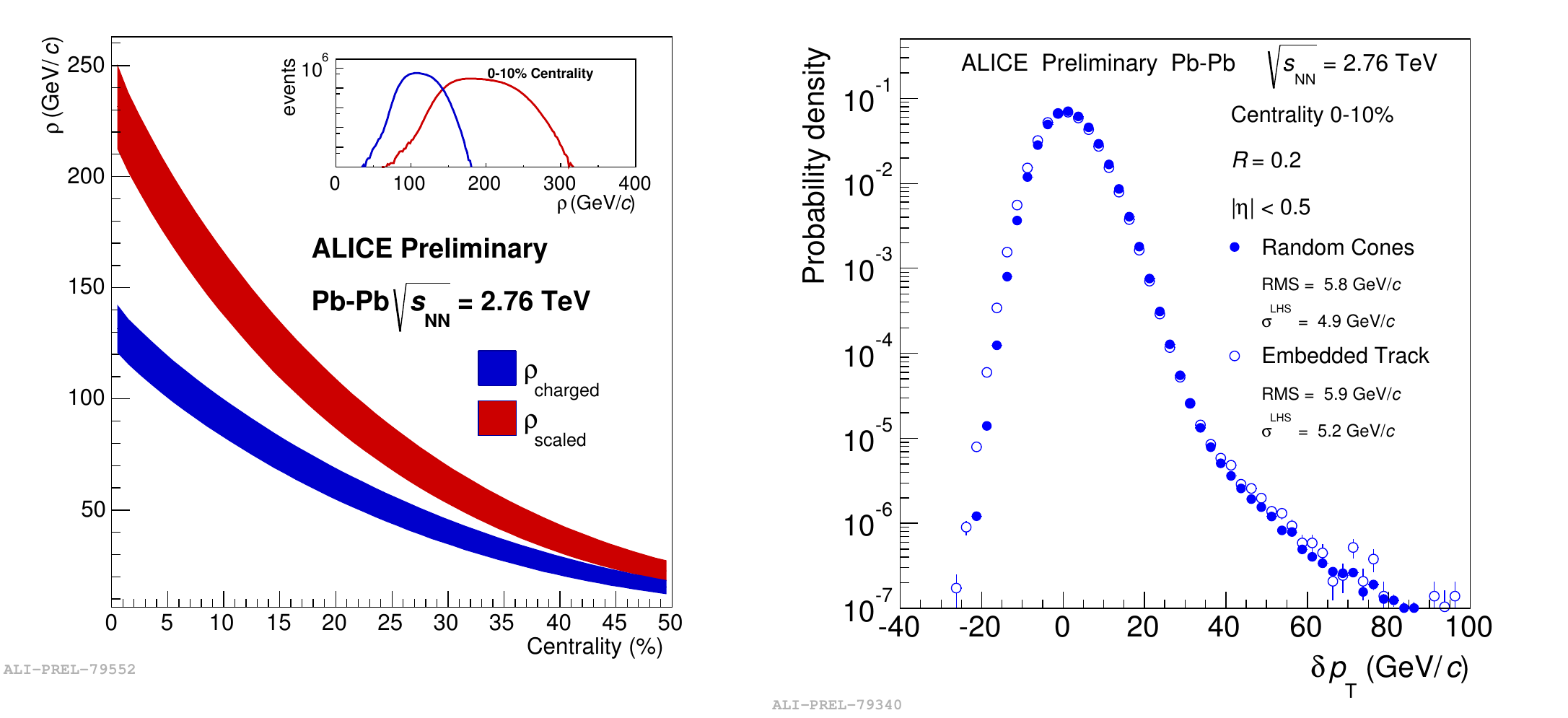}
  \caption[The charged background energy density, $\rho_\mathrm{ch}$,
  and the distribution of $\delta p_T$ for $R = 0.2$ jets with a $p_T$
  constituent cut of 150~MeV/$c$ for tracks and 350~MeV for clusters,
  determined using random cones and embedded tracks methods]{(left) The
    charged background energy density, $\rho_\mathrm{ch}$, is shown in
    blue and the total energy density is shown in red. (right) The
    distribution of $\delta p_T$ for $R = 0.2$ full jets with a $p_T$
    constituent cut of 150~MeV/$c$ for tracks and 350~MeV for clusters
    is shown as calculated by the random cones method in solid circles
    and the embedded track method in open circles.}
  \label{fig:alice_jet_energy_density}
\end{figure}

The philosophy behind this method is different from the ATLAS method
of subtracting the background event prior to reconstructing the
measured jets. This is a correction to the jet energy scale (JES), but
only for the effect of the average background density, and does not
correct for the additional smearing to the jet energy resolution (JER)
the fluctuations within the event cause. The fluctuations depend both
on the jet resolution parameter, the minimum $p_T$ of the jet
constituents and the centrality of the event. Additionally, the
fluctuations are smaller for the charged only background as any
fluctuations within the neutral sector are neglected. ALICE employs
two methods to calculate the background, but the default method is the
random cone method. In each event, a random cone is thrown with the
same R as the jet resolution parameter and the observable $\delta p_T$
is constructed by the following formula: $\delta p_T =
p_{T,\mathrm{jet}}^\mathrm{rec} − \rho_\mathrm{ch}\pi R^2$. This
distribution is shown for $R = 0.2$ full jets with the minimum
constituent cut in the 10\% most central events in
Figure~\ref{fig:alice_jet_energy_density}. This method additionally
quantifies the effect of jet overlap, as can be seen by the right hand
tail in this figure. 

\begin{figure}[!hbt]
 \begin{center}
    \includegraphics[width=0.8\linewidth]{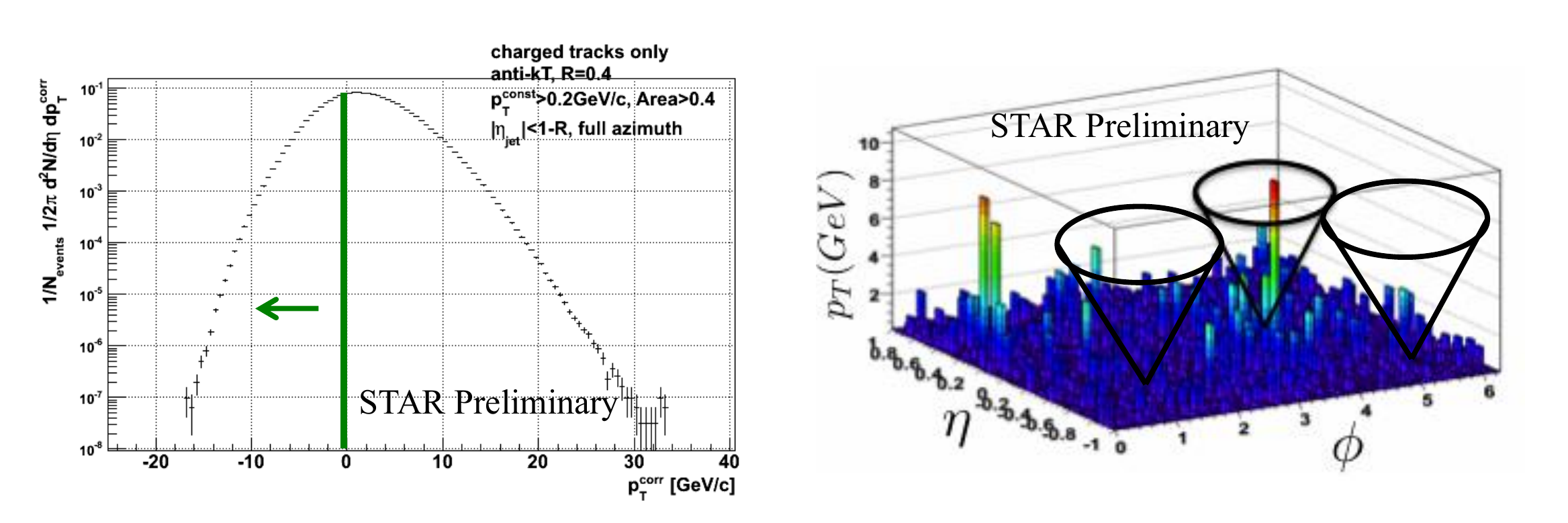}
    \caption[STAR preliminary result showing how jet candidates are
    found using the unmodified information in the
    event]{\label{fig:star_mean_algorithm} Preliminary result from STAR
      showing how jet candidates are found using the unmodified
      information in the event.  A distribution of jet energies relative
      to the ensemble median is then formed and is taken as the jet
      signal.}
 \end{center}
\end{figure}

The STAR experiment has used as approach very similar to that used by
the ALICE experiment.  Figure~\ref{fig:star_mean_algorithm} shows the
result of running the anti-$k_T$ jet finder on a raw, unmodified heavy
ion event.  The jets found by this procedure include some measure of
energy from the underlying event.  A distribution of jet energies
relative to the median for a sample of similar events is then formed
(left panel of Figure~\ref{fig:star_mean_algorithm}).  The ensemble
median energy is subtracted from the initial energy reconstructed for
each jet found.

sPHENIX will be able to reproduce these methods and study their
efficacy for producing physics observables and for constraining and
understanding systematic uncertainties.
\clearpage
\subsection{Inclusive Jet Yield in \AuAu Collisions}
\label{sec:auau_inclusive_jets}

\begin{figure}[hbt!]
  \centering
  \includegraphics[trim=0 0 0 10,clip,width=\onewidth]{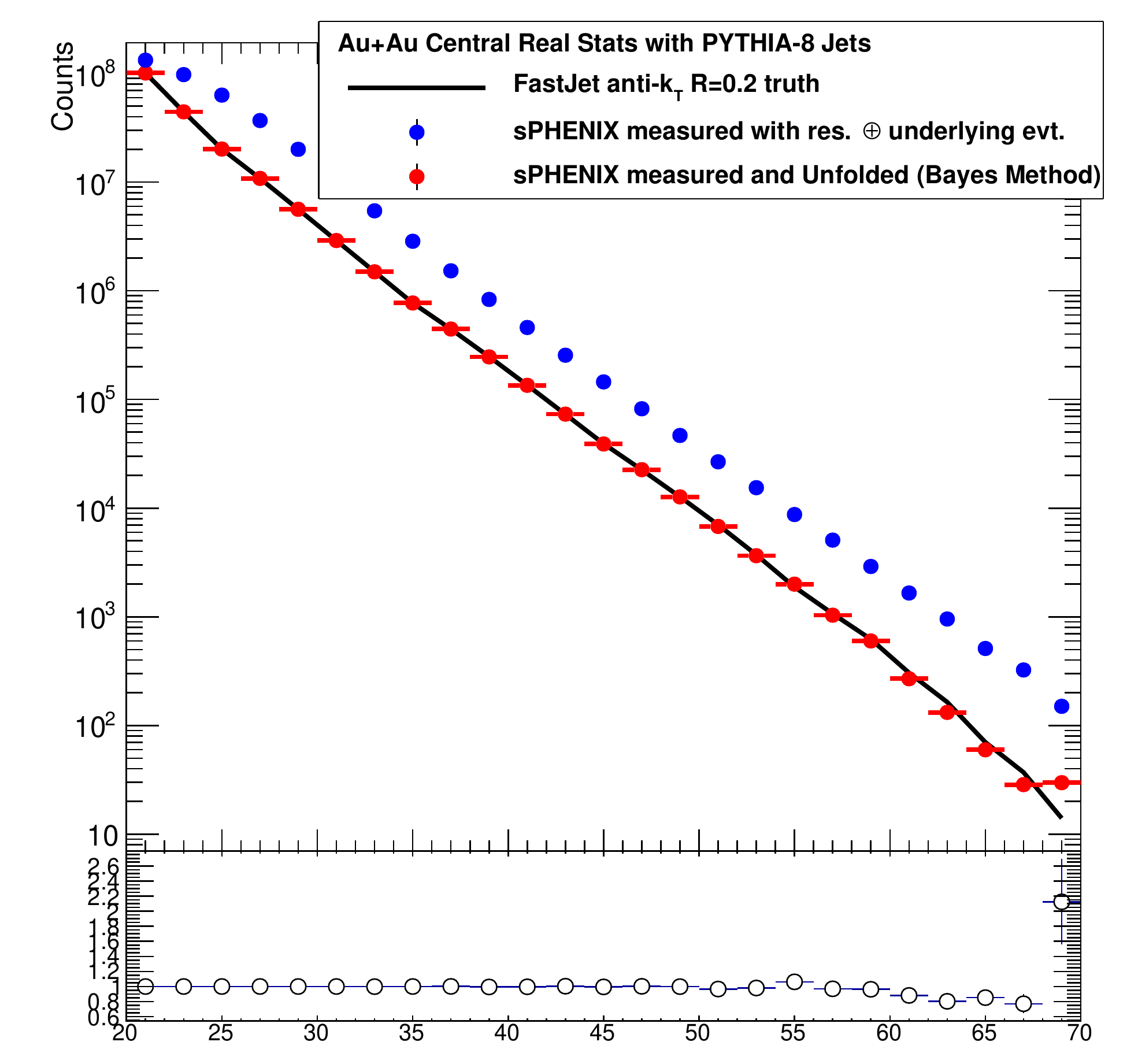}
  \caption[Effect of smearing the inclusive jet spectrum in \auau
  collisions]{Effect of smearing the inclusive jet spectrum in \auau
    collisions.  The jets found by \fastjet are smeared by the jet
    resolution contributions from the detector and the underlying event
    fluctuations.  The unfolded spectrum from the Iterative Bayes method
    is shown and the ratio of the unfolded to the true \pt spectrum
    (lower panel).}
  \label{fig:auau_smearing_and_unfolding}
\end{figure}

\begin{figure}[hbt!]
  \centering
  \includegraphics[width=\onewidth]{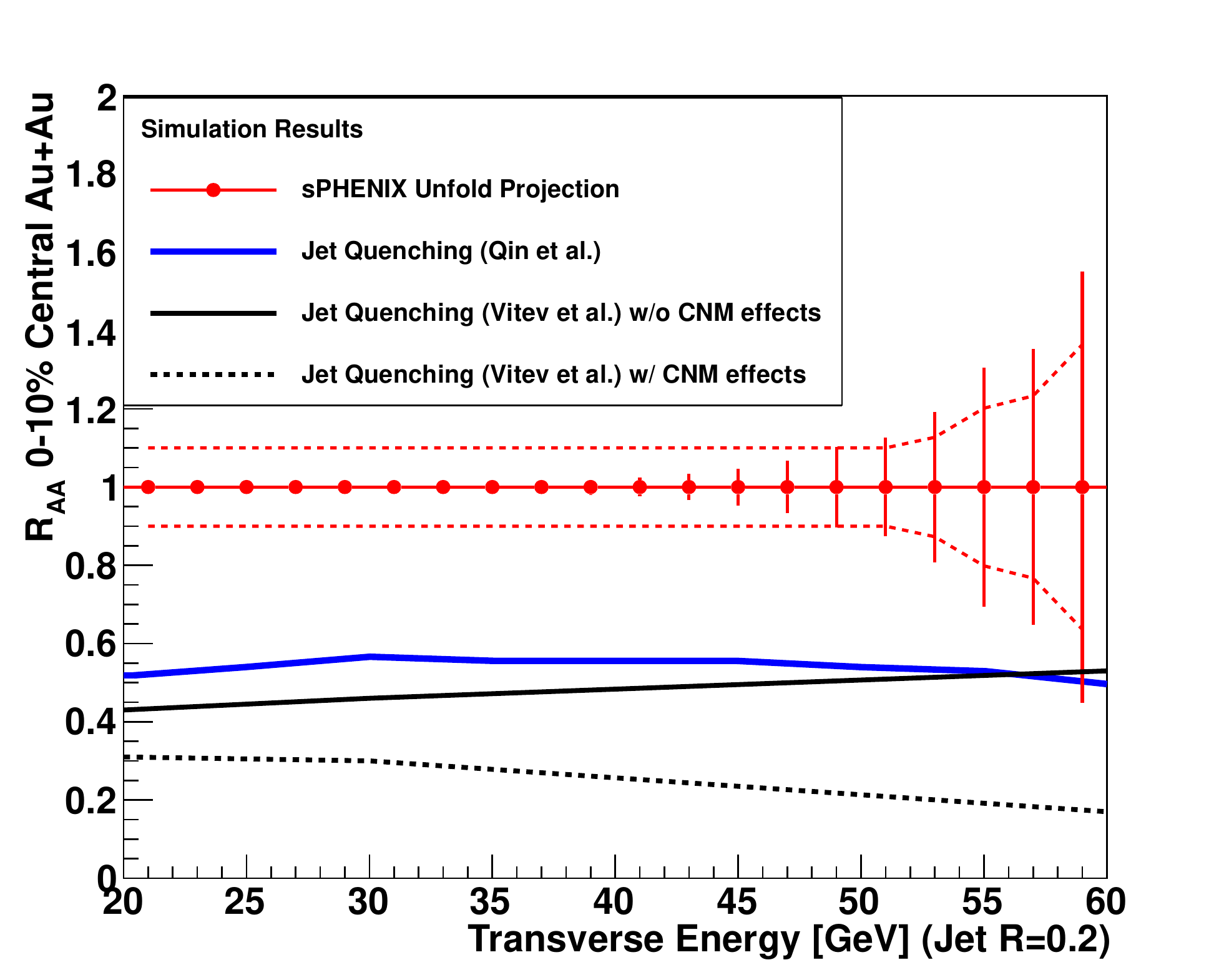}
  \caption[Single inclusive jet $R_{AA}$ with $R=0.2$ for \auau central
  events from the unfolding of the \pp and \auau spectra with an
  estimated systematic uncertainty as a multiplicative factor of
  approximately $\pm$ 10\%]{Single inclusive jet $R_{AA}$ with $R=0.2$
    for \auau central events from the unfolding of the \pp and \auau
    spectra with an estimated systematic uncertainty as a multiplicative
    factor of approximately $\pm$ 10\%.  Also shown are the predictions
    from a calculation including radiation and collisional energy loss
    and broadening~\cite{qin_privatecomm} and another with and without
    cold nuclear matter
    effects~\cite{He:2011pd,Neufeld:2011yh,Vitev:2009rd} (as discussed
    in Section~\ref{sec:jetcalculations}).}
  \label{fig:jet_aa}
\end{figure}

The inclusive jet spectrum is the most important first measurement to
assess the overall level of jet quenching in RHIC collisions.  The
results shown in Figure~\ref{fig:auau_smearing_and_unfolding} were
obtained by the \veryfast simulation approach described above.
\pythia was used to generate events and the final state particles were
sent to \fastjet in order to reconstruct jets.  The resulting jet
energy spectrum was smeared by the jet resolution determined for \pp
collisions from \geant, and an additional smearing by the underlying
event fluctuations (determined from the full 0--10\% central \hijing
\fast simulation).  Finally, an unfolding procedure was used to
recover the truth spectrum.  The ratio shown at the bottom of the plot
shows that the unfolding is very effective.

As an estimate of the uncertainties on a jet $R_{AA}$ measurement from one year of RHIC running, the uncertainties
from Figures~\ref{fig:pp_very_fast_dijet} and \ref{fig:auau_smearing_and_unfolding}
are propagated and shown in Figure~\ref{fig:jet_aa}.  For $E_T<50$~GeV the point to point
uncertainties are very small.   Also shown is an estimated systematic uncertainty including the effects
from unfolding.  All points are shown projected at $R_{AA} = 1$, and we show for comparison the predicted jet $R_{AA}$ including 
radiative and collisional energy loss and broadening from Ref.~\cite{qin_privatecomm}.

\subsection{Dijets in \AuAu collisions}
\label{sec:auau_dijets}

Fake jets contaminate \dijet observables much less than they do the
inclusive jet measurement.  In the case of inclusive jets, one is
working with a sample of $10^{10}$ central \auau events in a typical
RHIC year, so even if it is only a rare fluctuation in the background
that will be reconstructed as a real jet, there is a huge sample of
events in which to look for such fluctuations.

The case of \dijet correlations is very different.  There are $10^6$
clean trigger jets above $E_T = 30$~GeV in central \auau collisions
in a RHIC year --- detailed in Figure~\ref{fig:nlo_jetrates}.  This
means there is a factor of $10^4$ fewer chances to find the rare
background fluctuation that appears to be a true jet in the opposite
hemisphere.  Also, the presence of a high energy jet, for which the
fake rate is known to be low, tags the presence of a hard process
occurring in the event, and thus dramatically reduces the probability
of a jet in the opposite hemisphere being a fake.  Because of these
considerations, one can go to much lower \pt for the away side partner
of a \dijet pair.  Studies presented here include away side jets down
to 5~GeV, and we have found that the fake jet rate remains small for
the associated jets, even at these low jet energies.

\begin{figure}[hbt!]
  \centering
  \includegraphics[width=0.48\linewidth]{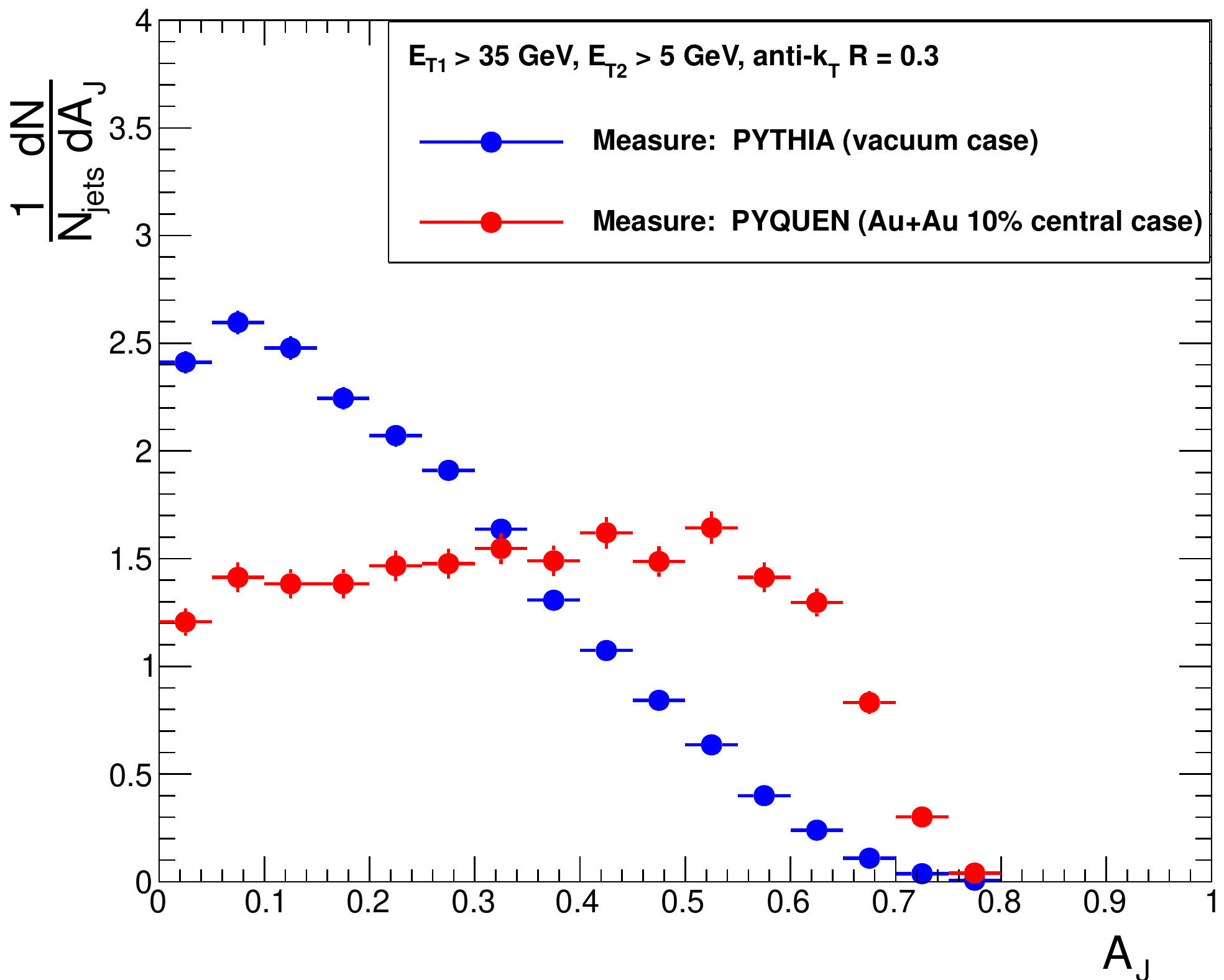}
  \hfill
  \includegraphics[width=0.48\linewidth]{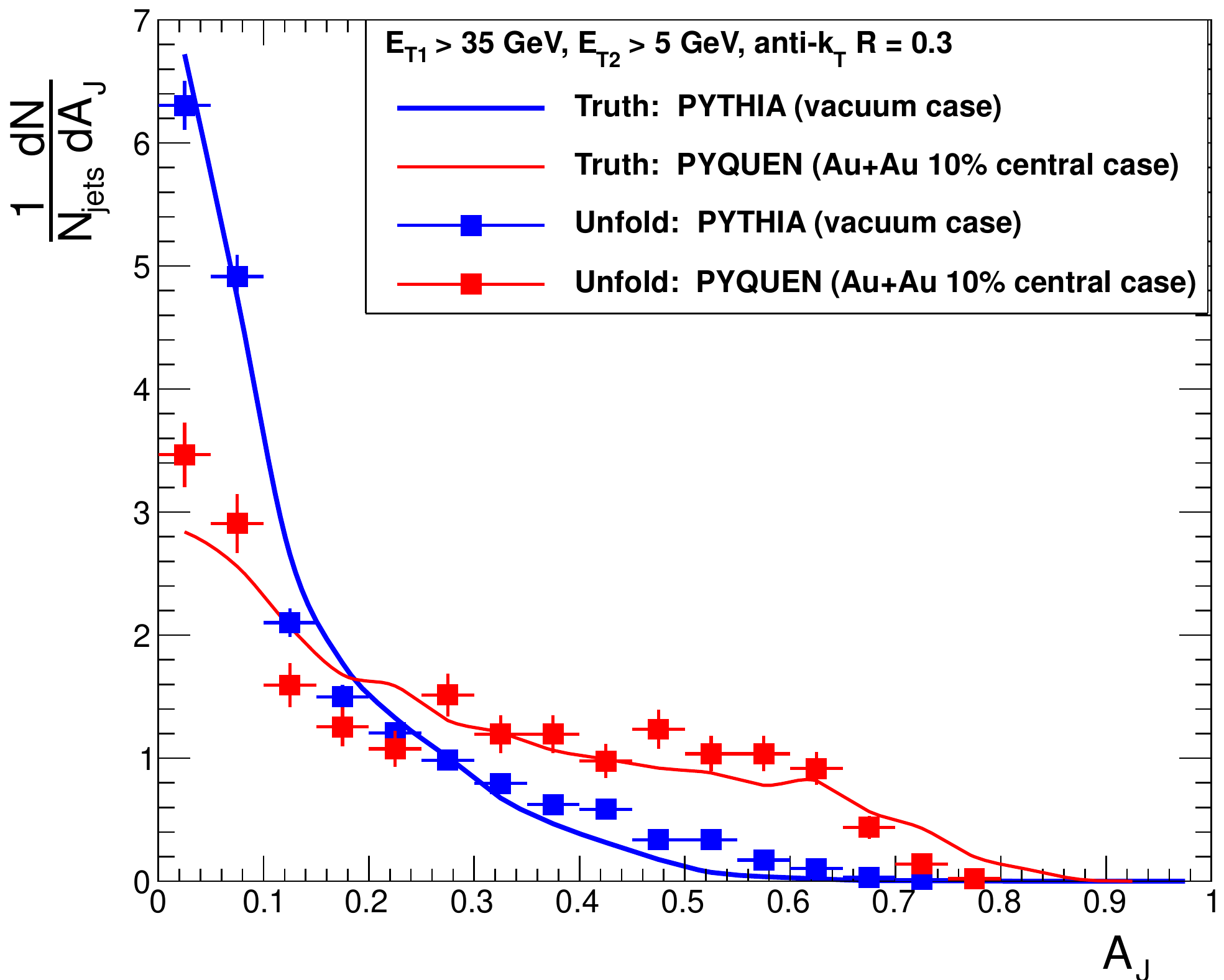}
  \caption[The effect of smearing on $A_J$ for $R = 0.3$ jets]{The
    effect of smearing on $A_J$ for $R = 0.3$ jets.  The left panel
    shows the effect of smearing on the ratio determined from jets
    reconstructed after embedding in \auau events.  Although smeared,
    the reconstructed data still show a distinct difference between the
    quenched and unquenched results.  The right panel shows the results
    of the ``unfolding'' procedure discussed in
    Section~\ref{sec:dijet_pp}.}
  \label{fig:auau_dijet_smearing}
\end{figure}

In order to address the sensitivity to modifications of the $A_J$
distributions that might be expected at RHIC here we compare \pythia
simulations with those from \pyquen~\cite{Lokhtin:2005px} (a jet
quenching parton shower model with parameters tuned to RHIC data).
All the \pyquen events generated are for central \auau events with
$b=2$~fm. Figure~\ref{fig:auau_dijet_smearing} shows the particle
level (i.e truth) $A_J$ distributions and how they are reconstructed
after being embedded in a central \auau event with a parametrized
detector smearing and segmentation applied.  As described above, the
full iterative underlying event subtraction method is applied.  The
simulated measured distributions (middle panel of
Figure~\ref{fig:auau_dijet_smearing}) show the effects of the
smearing; and the distinction between the \pythia and \pyquen
distributions remain large.  An unfolding procedure can be applied to
these embedded distributions to regain the true distributions.
However, as in the \pp case discussed in Section~\ref{sec:dijet_pp}
this should involve a full two-dimensional unfolding.  Applying the
same ``unfolding'' applied to the \pp case where the smearing of the
trigger jet is taken as the dominant effect recovers most of the
original distribution, as shown in the lower panel of
Figure~\ref{fig:auau_dijet_smearing}.  Again, this does not replace a
full unfolding procedure, but it does show that the reconstruction is
well under control and unfolding will be possible despite the presence
of a large fluctuations in the underlying event, after baseline and
flow subtraction.

\section{Extended kinematics and surface bias engineering}
\label{sec:jet_surface_emission_engineering}

Thus far we have documented a range of jet energy, radius, and
collision centralities over which inclusive jets dominate above
backgrounds and provide clean measurements of $R_{AA}$ and $A_{J}$
for example.  One can significantly extend the jet radius to larger
values and energies to lower values through various \fake jet
rejection methods including matching to track jets, identification of
individual particle energies in the jet (e.g. tracks or clusters) and
setting minimum energy thresholds, jet shape cuts, and more.  As we
demonstrate here, sPHENIX will have the full complement of these
methods available (thus having complementary overlap with existing
STAR jet observables).  All of these rejection methods present a bias
on the jet sample that often anti-correlates with the expected
modification in the \qgp medium.

Experiments have employed \fake jet rejection cuts to substantially
extend the high purity jet energy range accessible in central heavy
ion collisions --- for example see
Refs.~\cite{Aad:2012vca,Adamczyk:2013jei}.  With the sPHENIX detector
we can utilize track + electromagnetic jets matched to fully
calorimetric jets in a similar fashion.  In addition to extending the
measurable jet energy range to lower energies, for energies with high
purity without any selection one can turn this method into a powerful
tool to engineer the degree of jet surface emission.  For example, in
the sample of $10^5$ jets with $R=0.4$ and $E_{T} > 40$~GeV, we can
measure a high purity sample of reconstructed jets in central \AuAu
collisions.  We can then dial in the required track + electromagnetic
cluster jet characteristics to achieve a particular surface bias ---
as proposed by Renk~\cite{Renk:2012ve} and shown earlier in
Figure~\ref{fig:renk_jet_engineering}.

\begin{figure}[hbt!]
  \centering
  \includegraphics[width=0.7\linewidth]{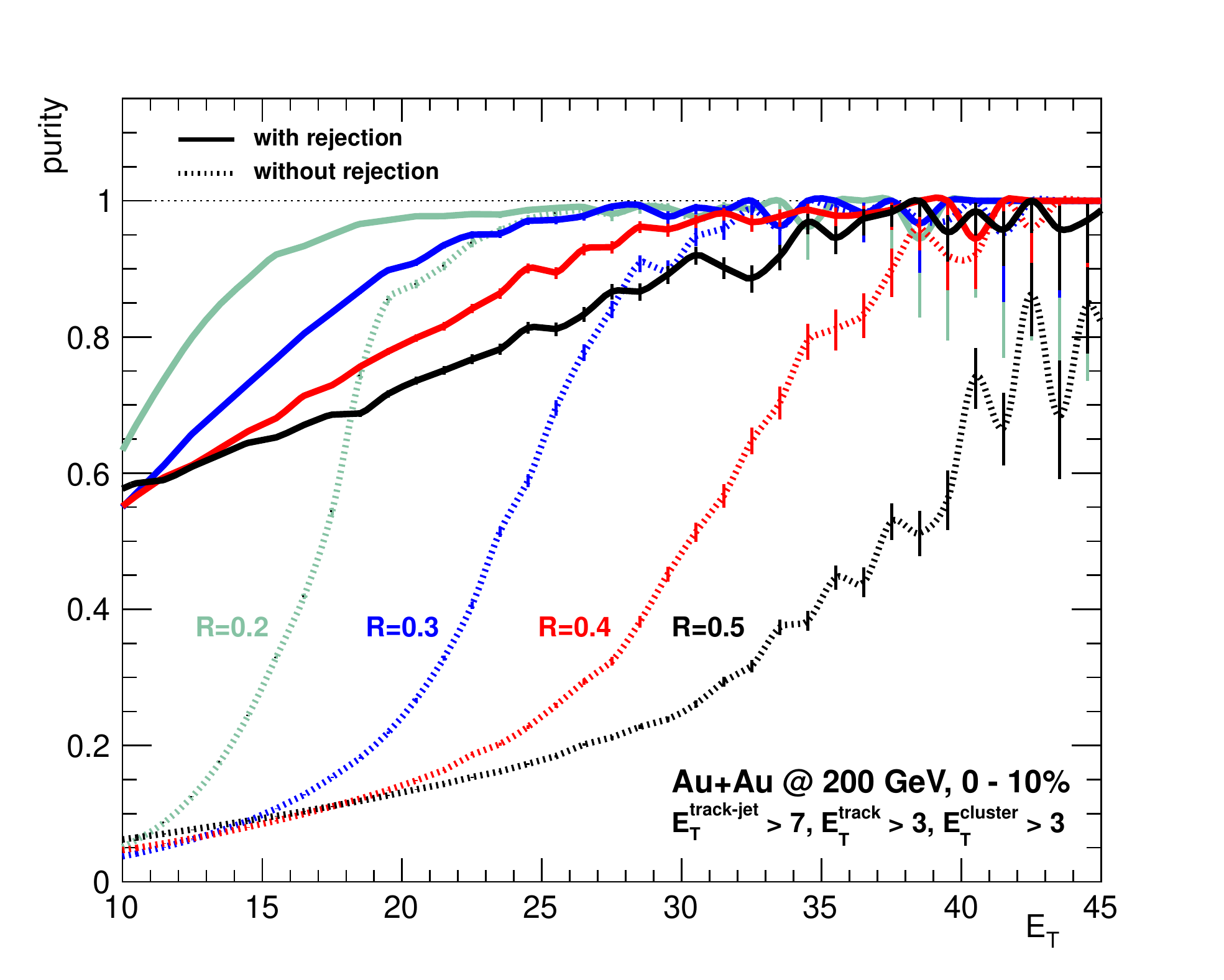}
  \caption[Purity results for $R=0.2, 0.3, 0.4, 0.5$ anti-$k_{T}$
  calorimetric reconstructed jets in 0--10\% central \AuAu \hijing
  events]{Purity results for $R=0.2, 0.3, 0.4, 0.5$ anti-$k_{T}$
    calorimetric reconstructed jets in 0--10\% central \AuAu \hijing
    events.  The dashed lines are without any track and electromagnetic
    cluster jet match requirement and the solid lines are with the match
    requirement.  The purities are significant higher for mid-central
    collision geometries.  }
  \label{fig:fakematch2}
\end{figure}

One can also incorporate electromagnetic clusters, which provide
additional input to the alternate jet reconstruction.  The
electromagnetic clusters and tracks have the same minimum energy cut
and are then input to the \fastjet algorithm.
Figure~\ref{fig:fakematch2} shows the jet purity for different jet
radii $R = 0.2, 0.3, 0.4, 0.5$ with a nominal track + electromagnetic
jet match requirement ($E_T > 7$~GeV for the match jet, $E_T > 3$~GeV
for the electromagnetic cluster and charged track) in central \auau
events.  The results are very good and indicate that even $R=0.5$ jets
can be reconstructed in the most central \AuAu events.  The effects of
the underlying event on jet observables are most severe in central
\auau events, and these results demonstrate the dramatically increased
range for jet reconstruction in mid-central \AuAu collisions, where
significant jet quenching effects have already been measured including
the theoretically challenging high \pt hadron azimuthal anisotropy.

sPHENIX will be also able to reproduce existing jet measurements at
RHIC, complete with the biases inherent in the various techniques used
to date.  However, the wider capabilities of sPHENIX will enable us
able to do more than merely confirm earlier results.  We will be able
to place those results along a spectrum of bias and to study the
effect on the jet observables of the alteration or removal of that
bias. 

\begin{figure}[hbt!]
  \centering
  \includegraphics[width=\linewidth]{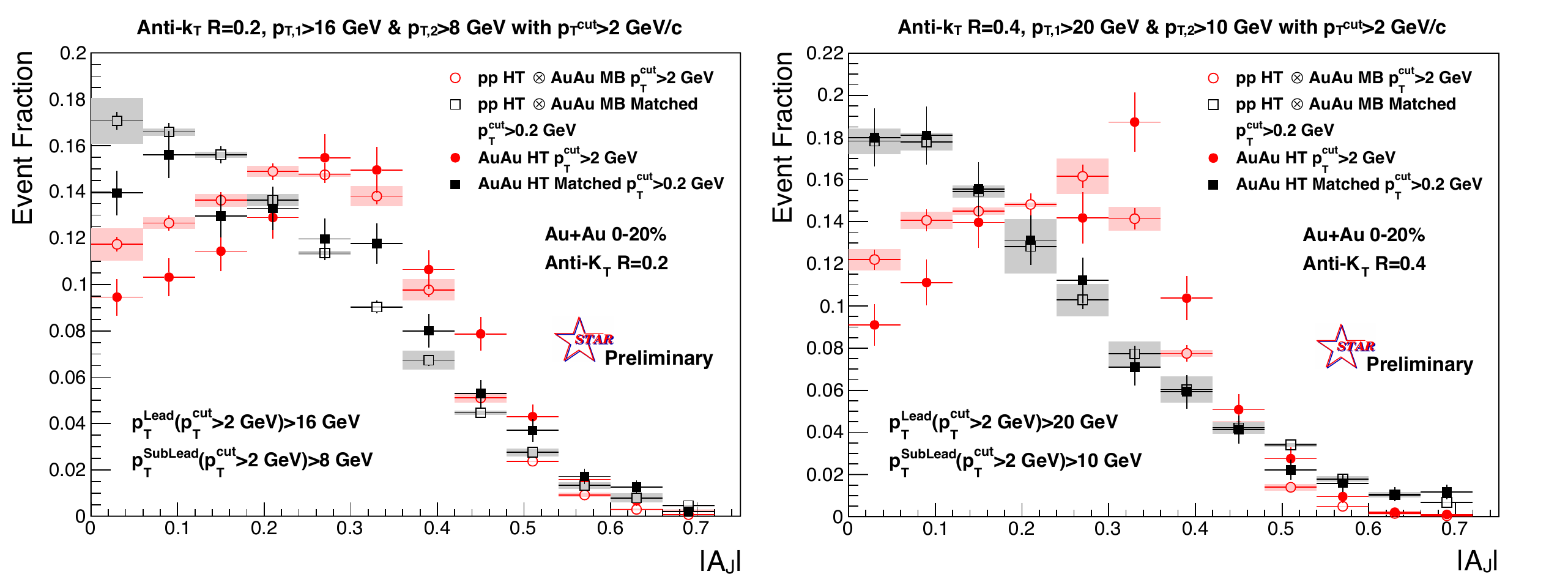}
  \caption[STAR result for $A_J$ for $R=0.2$ and $R=0.4$ jets, triggered
  on a high $E_T$ electromagnetic calorimeter tower]{STAR result for
    $A_J$ for $R=0.2$ and $R=0.4$ jets, triggered on a high $E_T$
    electromagnetic calorimeter tower.  When lower $p_T$ constituents
    are included in the $R=0.4$ jets, the shape of the $A_J$
    distribution is essentially identical for \pp and \auau.}
  \label{fig:star_jet_aj}
\end{figure}

Figure~\ref{fig:star_jet_aj} shows a preliminary result from the STAR
collaboration of $A_J$ for jets in events triggered on the presence of
a single EMCal tower above 5.4~GeV.  The left panel shows $A_J$ for
$R=0.2$ jets; the right for $R=0.4$ jets.  When a cut of
$p_T>2$~GeV/$c$ is placed on the constituents, there is a distinct
difference in the $A_J$ distributions for $R=0.2$ that disappears for
the larger $R=0.4$ jets, presumably because the larger radius recover
the full energy of the medium-modified jets.  sPHENIX will be able to
used jets triggered in a wide variety of ways as inputs to physics
analyses.

\clearpage

\section{Fragmentation Function and Photon-Jet Observables}
\label{sec:fragmentation_function}

Measurements that probe the redistribution of energy within the parton
shower directly are an important class of observables for
understanding the underlying dynamics of jet quenching. This
redistribution may take place both along the parton direction, such as
may be measured through the modification of longitudinal fragmentation
functions in \auau collisions, and transverse to the parton
direction. The capability of the sPHENIX detector and the high
statistics provided by RHIC will allow for measurements of the
two-dimensional distribution of energy to be measured and compared
between \pp, \pdau, and \auau collisions. Such measurements take
advantage of the fully calorimetric jet and photon observables in
tandem with the measurement of charged hadrons by the precision
tracking capabilities.

\begin{figure}[hbt!]
  \centering
  \includegraphics[width=0.7\textwidth]{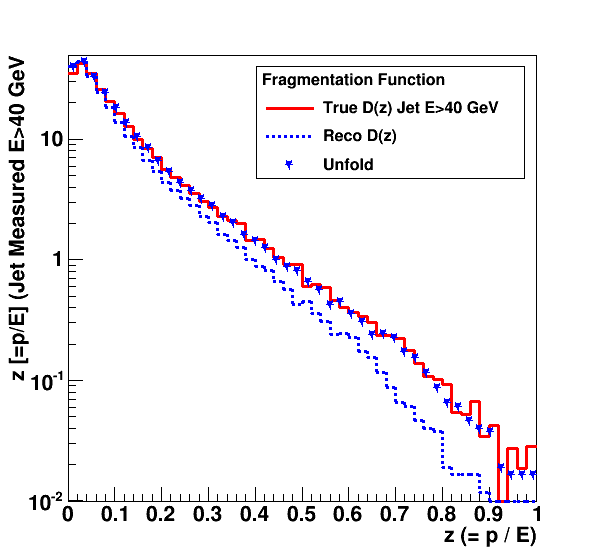}
  \caption[sPHENIX capabilities for measuring the inclusive jet
  fragmentation function, $D(z)$, for $p_\mathrm{T} > 40$~GeV/$c$
  jets]{Demonstration of the sPHENIX capabilities for measuring the
    inclusive jet fragmentation function, $D(z)$, for $p_\mathrm{T} >
    40$~GeV/$c$ jets. The original ``truth'' distribution is shown in red,
    while the reconstructed level distribution is shown in the blue
    dashed line. The reconstructed level distribution is corrected for
    detector effects to give the unfolded quantity shown in blue
    stars.  \label{fig:fastsim_ff} }
\end{figure}

Figure~\ref{fig:fastsim_ff} shows the results of a fast simulation of
the sPHENIX measurement capabilities for the inclusive jet
fragmentation function for high-$p_\mathrm{T}$ jets. This figure
compares the truth-level $D(z)$ quantity with the measured,
detector-level $D(z)$ which incorporates the effects of a fast
parameterized detector response on the jet and hadron
$p_\mathrm{T}$. Due to the finite momentum resolution in both cases
(and, in the case of the jet, the upfeeding from the underlying event
fluctuations), the reconstructed-level $D(z)$ is generally shifted
towards lower values at fixed $z$. The reconstructed level
distribution is unfolded to correct for these detector effects,
resulting in an unfolded $D(z)$ distribution that is able to
successfully recover the original truth-level $D(z)$. Furthermore, the
statistical uncertainties used in the plot are chosen to correspond to
that available in a 22 week \auau run. This figure demonstrates the
plausibility of measuring the fragmentation function within sPHENIX.

\begin{figure}[hbt!]
  \centering
  \includegraphics[width=0.7\textwidth]{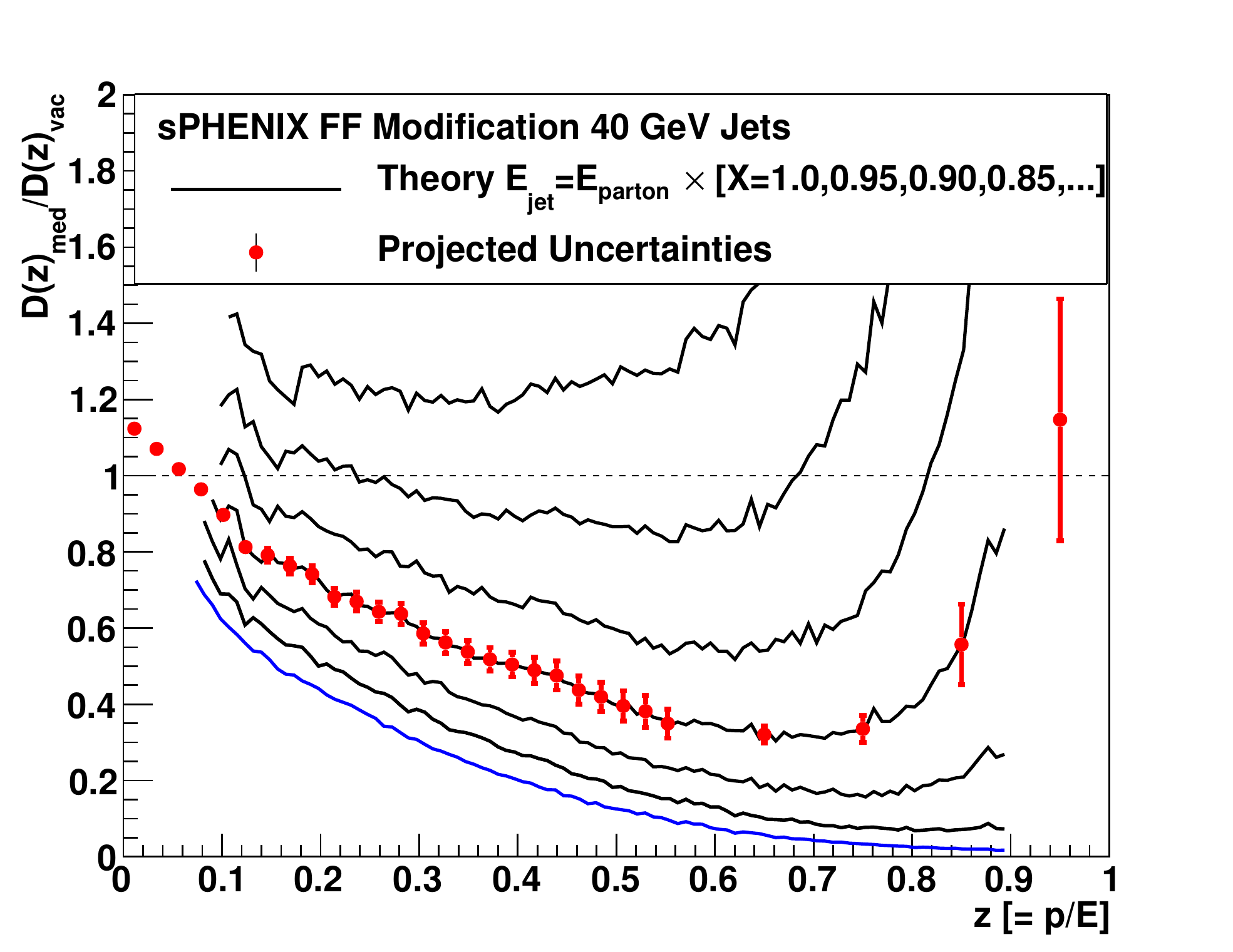}
  \caption[Modified fragmentation function $D(z)$ in the
    medium expressed as the ratio of the
    modified $D(z)$ to that assuming vacuum fragmentation]{Modified fragmentation function $D(z)$ in the
    medium~\cite{Armesto:2007dt} expressed as the ratio of the
    modified $D(z)$ to that assuming vacuum fragmentation. The
    different black curves show the results of different assumptions
    for how much of the parton energy is retained in the jet cone,
    $x$, with the original prediction corresponding to $x=1.0$ shown
    as the lower blue curve.  The projected statistical
    uncertainties achievable for 22 weeks and 10 weeks of \auau and
    $p$+$p$ data-taking are shown on top of the curve at $x=0.85$.
    \label{fig:fastsim_ff_mod} }
\end{figure}

When combined with a reference measurement of the unquenched
fragmentation function in $p$+$p$ collisions, sPHENIX will be able to
make detailed measurements of the medium modifications to the
longitudinal structure of the jet. Figure~\ref{fig:fastsim_ff_mod}
shows a calculation for the ratio of $D(z)$ in 200~GeV \auau to
$p$+$p$ collisions~\cite{Armesto:2007dt}, under different assumptions
of how much of the total parton energy is recovered in the jet
cone. The red points in the figure show the projected statistical
uncertainty for a particular value of the ratio, corresponding to the
first two years of sPHENIX data-taking.

In addition to the per-jet fragmentation function, the photon
capability in sPHENIX will allow for the measurement of photon-tagged
jets. Although lower in statistics than measurements of inclusive
jets, photon-jet events offer several crucial advantages. First,
photon performance is much less sensitive to the effects of the
underlying event in \auau collisions (i.e. there are no ``fake''
photons). Second, the resolution for high-$p_\mathrm{T}$ photons,
measured using only the EMCal, is better than that for full jets,
which include the resolution introduced by the HCal, resulting in more
modest corrections for detector effects. Third, photons provide an
independent, well-calibrated probe against which to measure jet
modification. Finally, unlike the case for inclusive jets, the
presence of a high-$p_\mathrm{T}$ photon in the event substantially
raises the probability that the balancing jet is a real jet, resulting
a much smaller fake rate and the ability to measure to lower
$E_\mathrm{T}$ and larger $R$ value. This feature is particularly
important because it allows for the possibility of exploring the cone
size dependence of jet modification down to much lower $E_\mathrm{T}$
than may be possible with inclusive jet measurements only.

\begin{figure}[hbt!]
  \centering
  \includegraphics[width=0.9\linewidth]{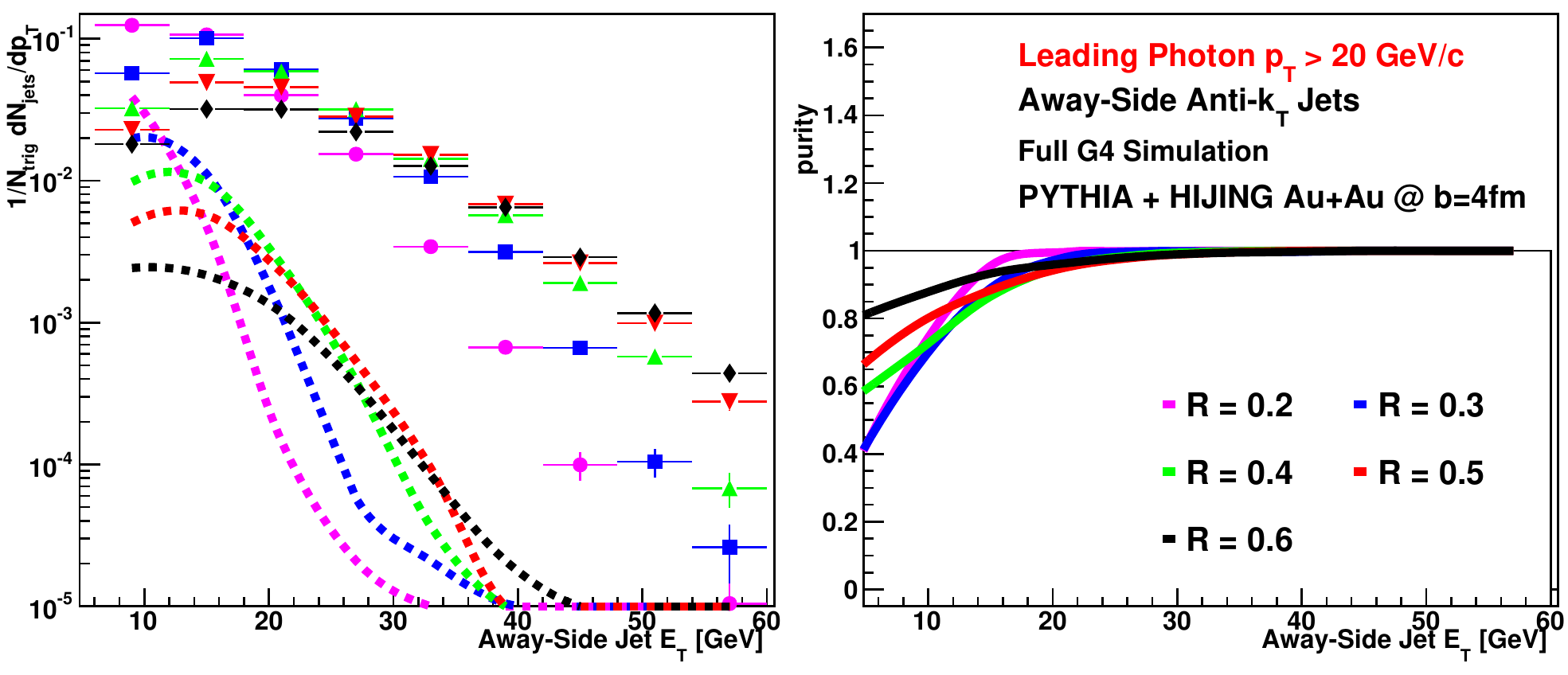}
  \caption[Photon-jet capability in sPHENIX, derived from a full \geant
  simulation of \pythia events with a $>20$~GeV photon embedded into
  central \auau \hijing events. ]{Demonstration of the photon-jet capability in sPHENIX, derived
    from a full \geant simulation of \pythia events with a $>20$~GeV
    photon embedded into central \auau \hijing events. (left)
    Distribution of jet $E_\mathrm{T}$ in these events for real jets
    (solid markers) and fakes (dashed lines). In both cases, different
    colors correspond to different $R$ values. (right) Purity of hard
    scattered jets as a function of jet $E_\mathrm{T}$, shown for
    different size jets.
    \label{fig:g4_photon_iaa} }
\end{figure}

Two key benchmarks of the simulated performance for photon-jet
measurements in sPHENIX are described
below. Figure~\ref{fig:g4_photon_iaa} quantitatively demonstrates,
using a full \geant description, the high jet purity for
photon-balancing jets. In events with a $p_\mathrm{T} > 20$~GeV
photon, the purity of away-side jets with $E_\mathrm{T} > 20$~GeV is
$>90$\% even for $R=0.6$ jets. 

\begin{figure}[hbt!]
  \centering
  \includegraphics[width=0.48\linewidth]{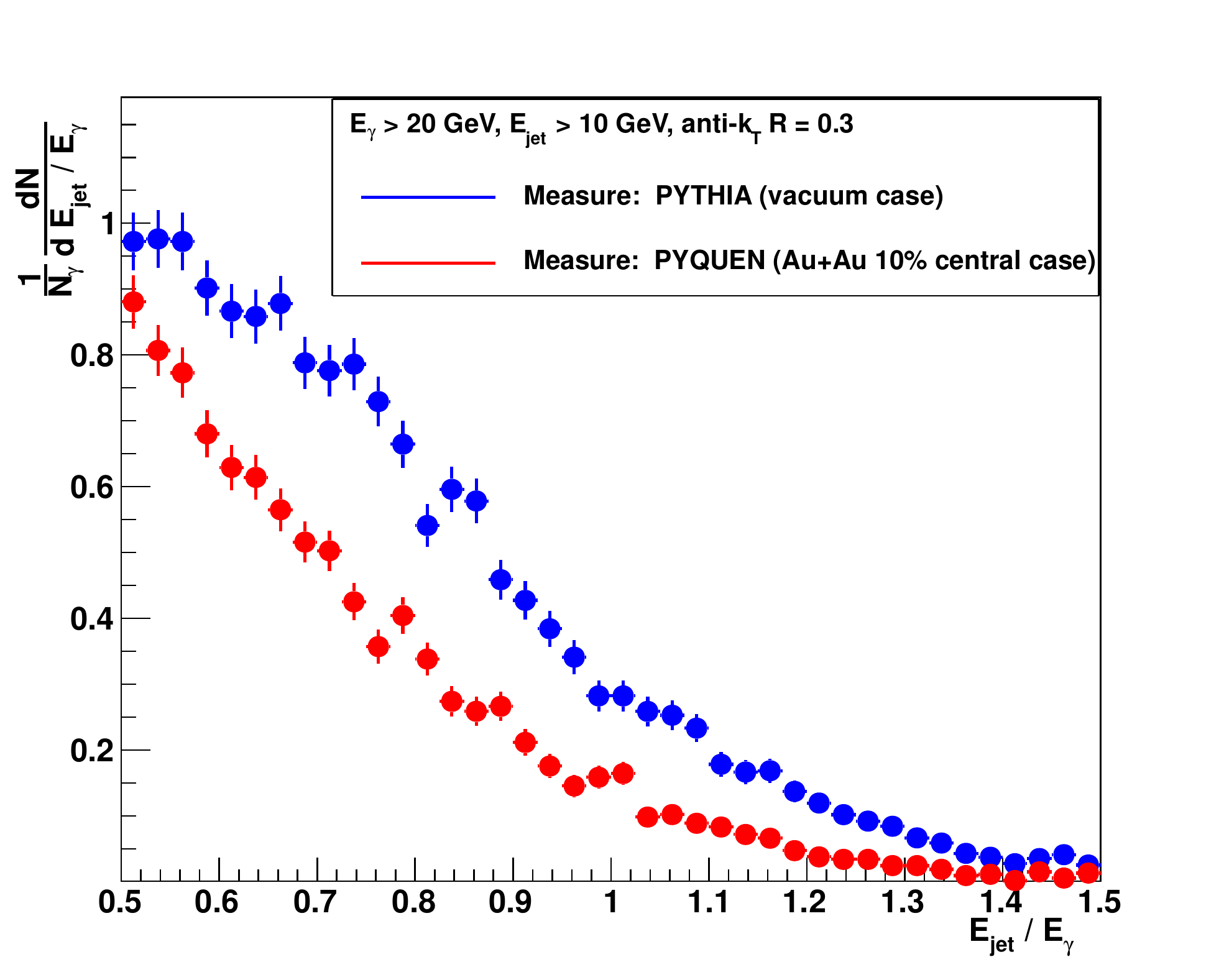}
  \hfill
  \includegraphics[width=0.48\linewidth]{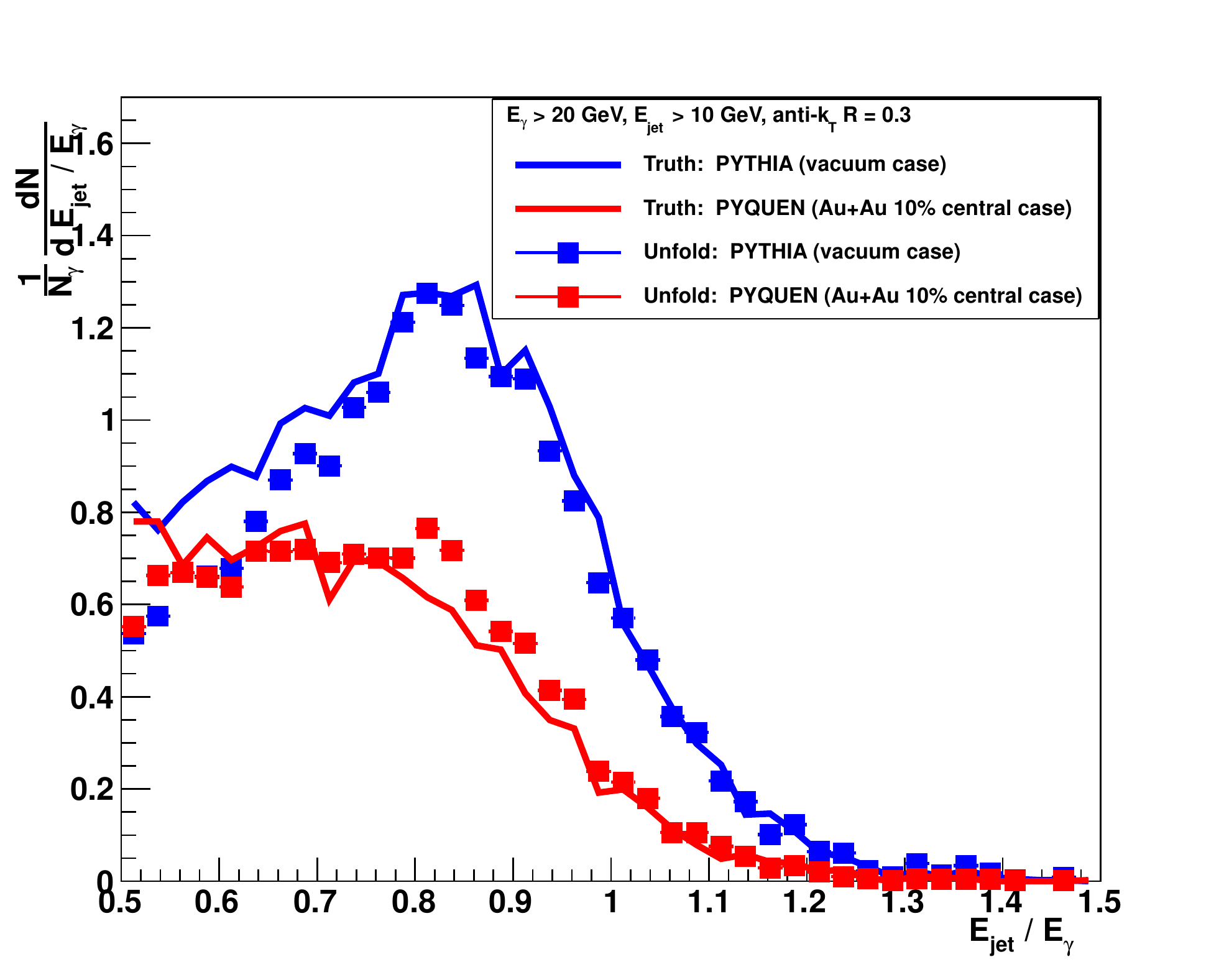}
  \caption[Effect of smearing on the energy ratio $E_{\rm jet}/E_\gamma$
  for $R = 0.3$ jets]{The effect of smearing on energy ratio $E_{\rm
      jet}/E_\gamma$ for $R = 0.3$ jets.  The left panel shows the
    effect of smearing on the ratio determined from jets reconstructed
    after embedding in \auau events.  Although smeared, the
    reconstructed data still show a distinct difference between the
    quenched and unquenched results.  Results of a one dimensional
    unfolding are compared with the truth particle level distributions
    in the right panel.}
  \label{fig:auau_gammajet_smearing}
\end{figure}

In contrast to the \dijet case studied above, the $\gamma$-jet
measurements do not compare two similar objects affected in a similar
way by the presence of the underlying event. In this case, the
quantity of interest is taken to be $x \equiv E_{\mathrm
  {jet}}/E_{\gamma}$ (rather than an asymmetry $A_J$). While in a
leading order QCD picture the $\gamma$ and recoiling parton should
exactly balance in energy, in reality this is not necessarily the
case, resulting in a vacuum distribution of $x$ that is peaked below
$1$. In particular, for small jet sizes there is a significant
probability that the away side parton shower is split into more than
one jet by the reconstruction procedure, with each carrying a fraction
of the energy needed to balance that of the $\gamma$.

Figure~\ref{fig:auau_gammajet_smearing} shows the distribution of $x$
values for $E_\mathrm{T}^\gamma > 20$~GeV events with an $R=0.3$ jet
with $E_\mathrm{T}^\mathrm{jet} > 10$~GeV, at the truth, reconstructed
and unfolded levels, for the vacuum fragmentation case as implemented
in \pythia and for the case of the \pyquen generator. Due to the
finite jet energy resolution and the effect of split jets, both
exacerbated by the underlying event in the embedded event, the
reconstructed level $x$ distribution is qualitatively different than
the truth distribution. However, the \pythia and \pyquen distributions
are still noticeably different. The figure also shows how well an
unfolding procedure can recover the original distributions. Although
nominally both the photon and the jet $E_\mathrm{T}$ must be unfolded,
In the $\gamma$-jet case the unfolding may be treated to good
approximation as being one-dimensional. This is because, as described
above, the dominant smearing effect is on the jet energy.  In the
figure above, an Iterative Bayesian unfolding is performed on the
detector-level $\gamma$-jet $x$ distributions for the $R=0.3$
jets. The unfolded results compare well with the particle level
distributions for both \pythia and \pyquen.

While the inclusive fragmentation function and photon-jet energy
balance measurements have been discussed here in detail, many
additional, potentially revealing measurements will be possible. For
example, measurements of the transverse momentum distribution of
hadrons with respect to the jet axis or opposite side photon axis may
provide additional information. Furthermore, one can use \gh
correlations to study the redistribution of energy lost by the
opposite going parton. Previous results from
CMS~\cite{Chatrchyan:2011sx} and STAR~\cite{Adamczyk:2013jei} on \gh
correlations indicate that this energy is spread over a wide angular
range. However, measurements at RHIC of \gh correlations have not had
the statistical precision or the acceptance necessary to make
comparable statements about the modification to jet fragmentation. The
combined jet and hadron capabilities of sPHENIX will provide useful
data.

\begin{figure}[hbt!]
  \centering
  \includegraphics[width=0.65\linewidth]{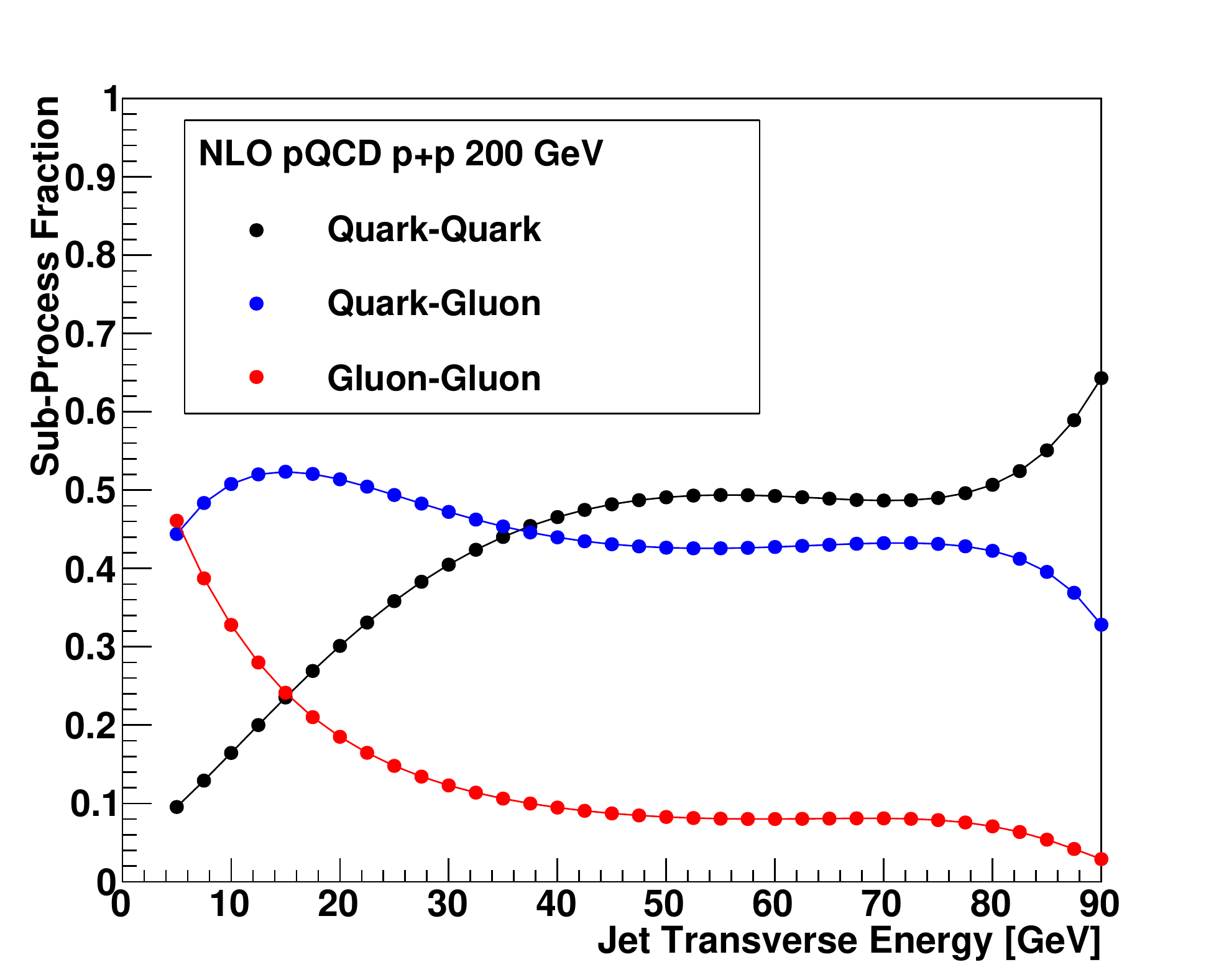}
  \caption[NLO pQCD calculation for the subprocess contributions as a
  function of jet transverse energy at midrapidity in \pp collisions at
  200~GeV]{NLO pQCD calculation for the subprocess contributions as a
    function of jet transverse energy at midrapidity in \pp collisions
    at 200~GeV.  At the lowest jet $E_{T}$ gluon-gluon processes
    dominate, transitioning to a mixture of quark-gluon and quark-quark
    processes, and finally dominated by quark-quark processes near the
    kinematic limit of $E_{T}$.  }
  \label{fig:subprocess}
\end{figure}

Photon-tagged jets offer another well-known experimental handle: the
away-side jet is dominantly a quark-initiated one. Thus, the
measurements above chiefly measure the medium modification of
quark-initiated parton showers. However, at RHIC energies, an
experimental handle on gluon jets is also
available. Figure~\ref{fig:subprocess} shows, in a leading order
picture, the breakdown of the inclusive jet cross-section by outgoing
parton flavor. It can be seen that, in events with a $20$--$30$~GeV
quark jet, the away-side jet is a gluon-initiated jet approximately
$2/3^\mathrm{rds}$ of the time. Thus, the presence of a quark jet,
which may be indicated by a particularly harder fragmentation pattern
or tighter energy profile, or by a more modest medium modification, is
a good way to select gluon jets on the away side. 

\begin{figure}[hbt!]
  \centering
  \includegraphics[width=0.9\linewidth]{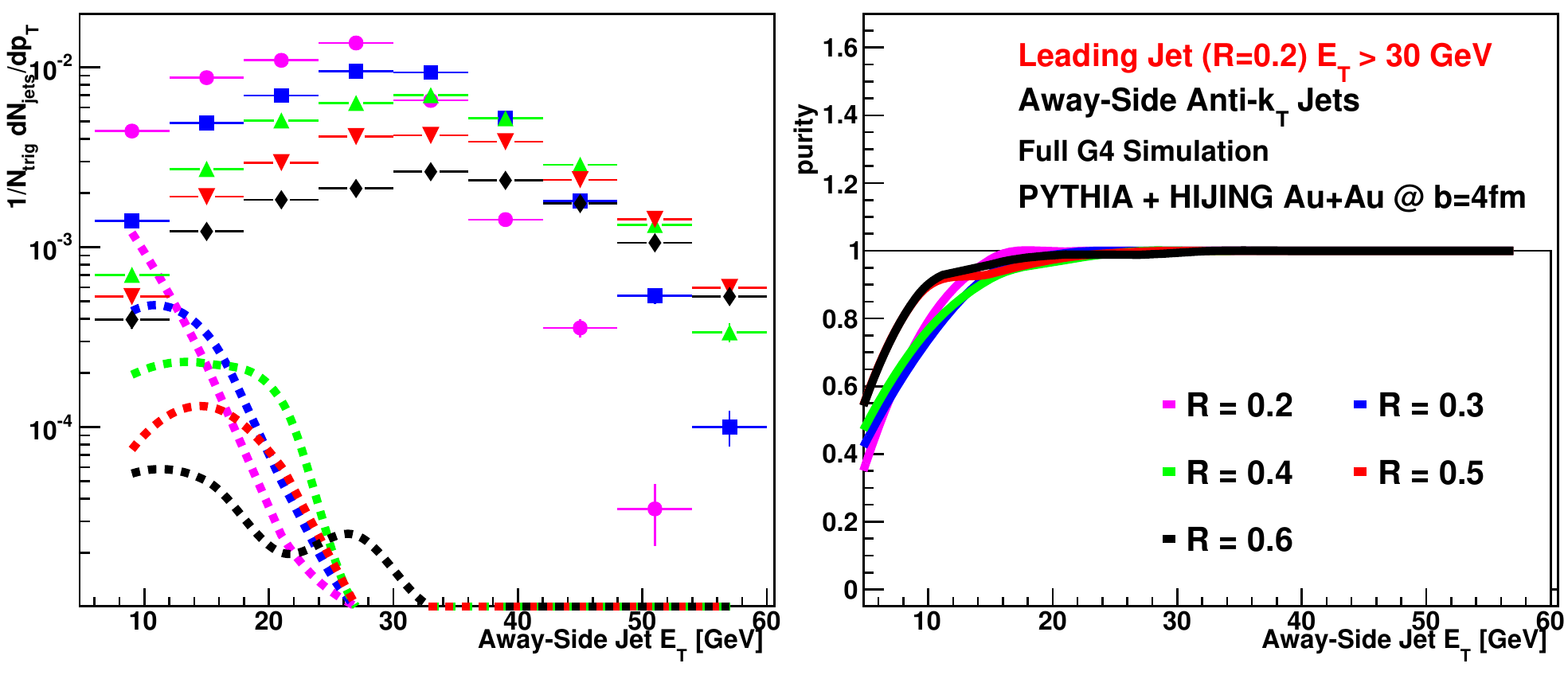}
  \hfill
  \includegraphics[width=0.9\linewidth]{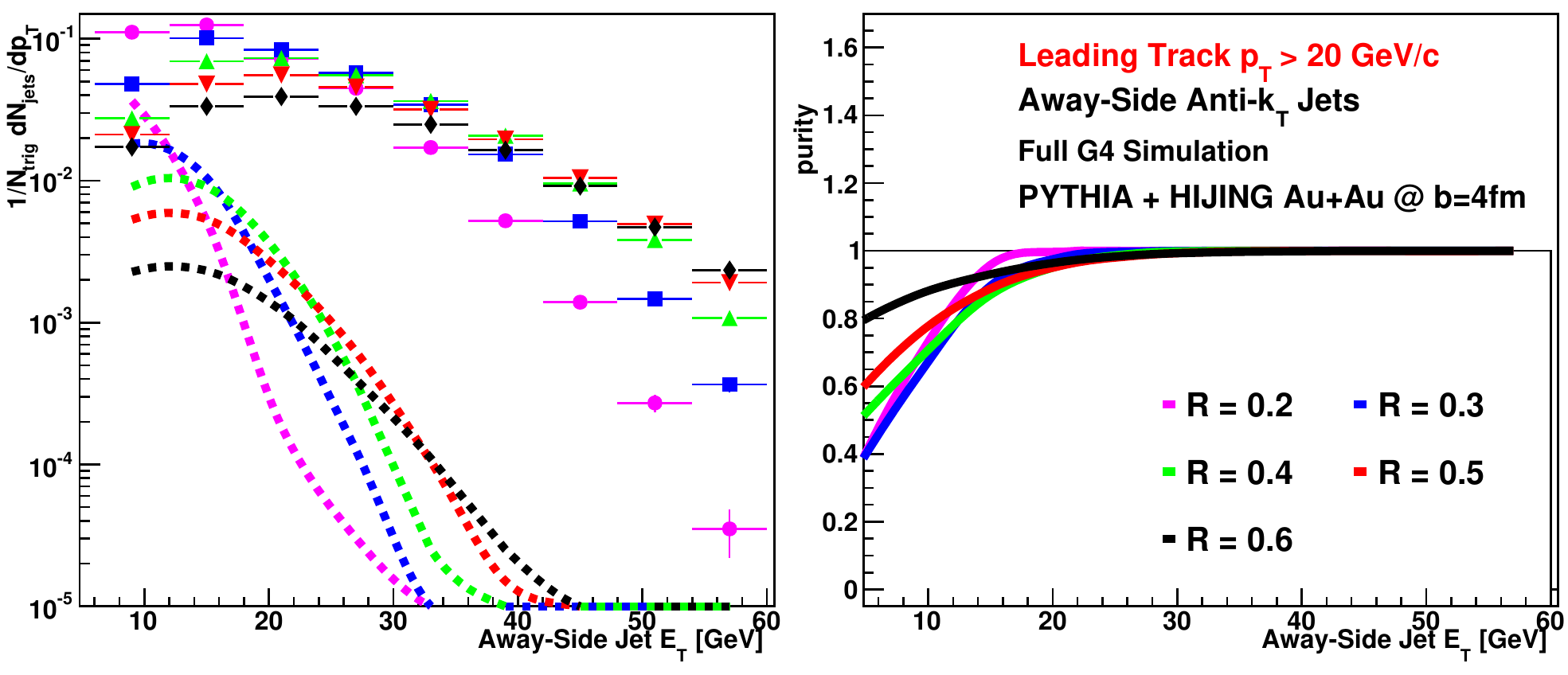}
  \caption[Full \geant simulations of \pythia dijets embedded in central
  \auau \hijing events with reconstruction of trigger and away-side
  jets, showing high purity even for large $R$]{(upper) Full \geant
    simulations with \pythia dijets embedded on central \auau \hijing
    events with reconstruction of trigger and away-side jets.  The left
    panel shows all reconstructed away-side jets for different $R$
    values opposite to a trigger jet with $R=0.2$ and $E_{T}>30$~GeV.
    The dashed lines indicate the \fake jet contributions.  The right
    panel shows the purity of away-side jets, which is quite high even
    for large radius away side jets.  (lower) Same quantity except now
    triggering on a charged hadron with $p_{T}>20$~GeV/$c$.  }
  \label{fig:g4_jetjetsim}
\end{figure}

Similar to the photon-jet case, the presence of a narrow cone jet or a
high-$p_\mathrm{T}$ track (both of which are indicative of a quark
jet) results in a purity for the away-side jet which is substantially
enhanced relative to that for inclusive jets at a comparable
$E_\mathrm{T}$. Figure~\ref{fig:g4_jetjetsim} shows the purity of
away-side jets as a function of away-side jet $E_\mathrm{T}$ for
different values of the jet radius $R$ in the presence of an $R=0.3$
$E_\mathrm{T} > 30$~GeV jet or a $p_\mathrm{T} > 20$~GeV/$c$ track on
the near side. It can be seen that the purity remains above $90$\% for
$> 20$~GeV jets, even for $R=0.6$ jets. 

\begin{figure}[hbt!]
  \centering
  \includegraphics[width=0.6\linewidth]{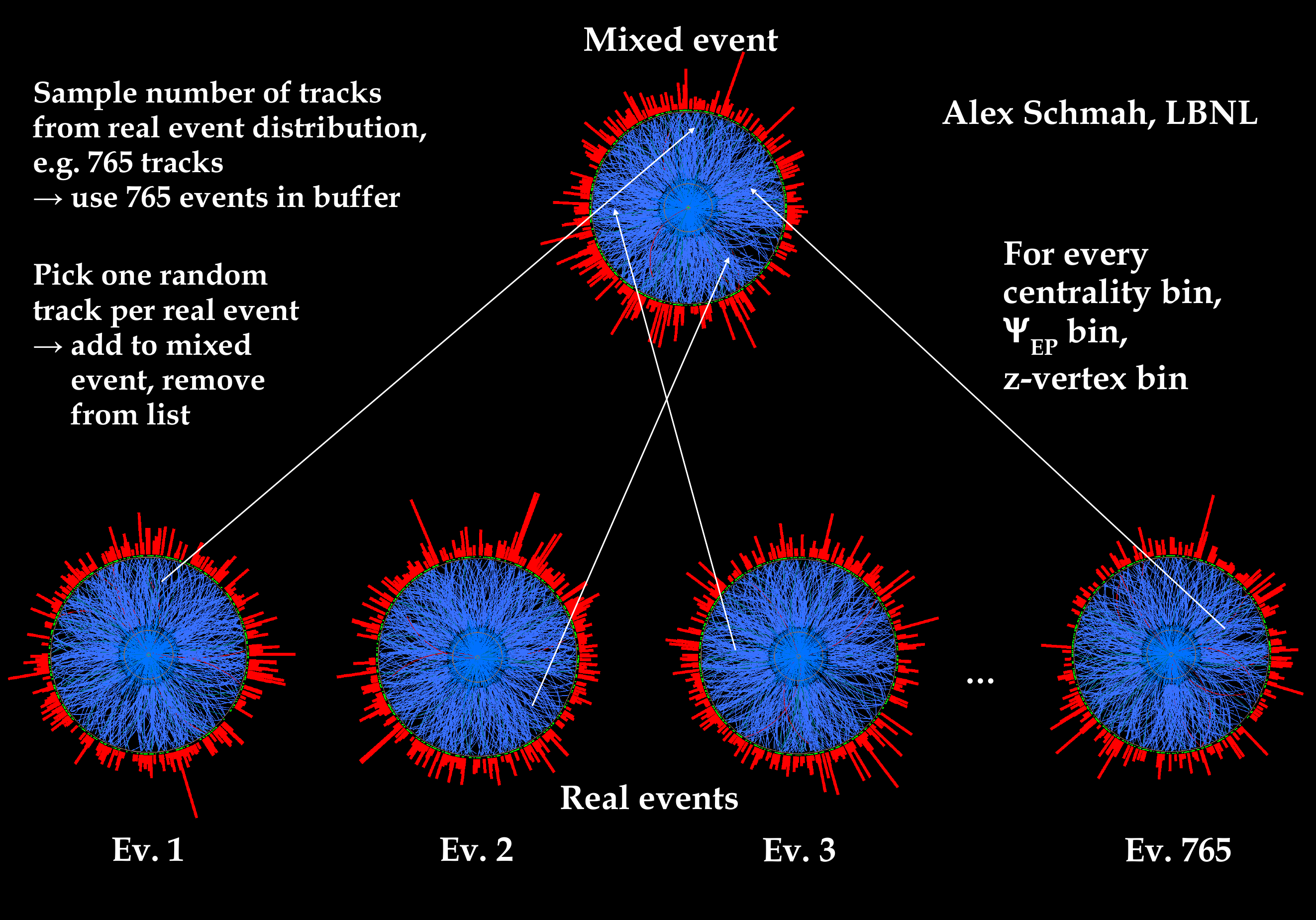}
  \caption{Illustration of a technique for ensemble mixing as
    developed by the STAR Collaboration.}
  \label{fig:ensemble_mixing}
\end{figure}

Additionally, there are other techniques which can be used to extend
the kinematic reach of the away-side jet to low $E_\mathrm{T}$ and
control the extent of the fake rate. An example of such a technique is
the statistical ensemble mixing method developed by the STAR
Collaboration (Figure~\ref{fig:ensemble_mixing}), which has been
successfully used in \auau collisions.

Thus, using small-$R$ jets or high-$p_\mathrm{T}$ single tracks as a
trigger object, a sample with an enhanced gluon content can be
reliably selected down to low $E_\mathrm{T}$. In tandem with the
quark-enhanced sample selected in photon-jet events, this will allows
for an experimental handle on the flavor-dependence of quenching in
sPHENIX. Of course, these selections on the trigger jet will surely
introduce a bias, such as in the distribution of path lengths for the
away-side jet. These issues were discussed previously in
Section~\ref{sec:jet_surface_emission_engineering}.

\begin{figure}[hbt!]
\centering
\includegraphics[width=0.47\linewidth]{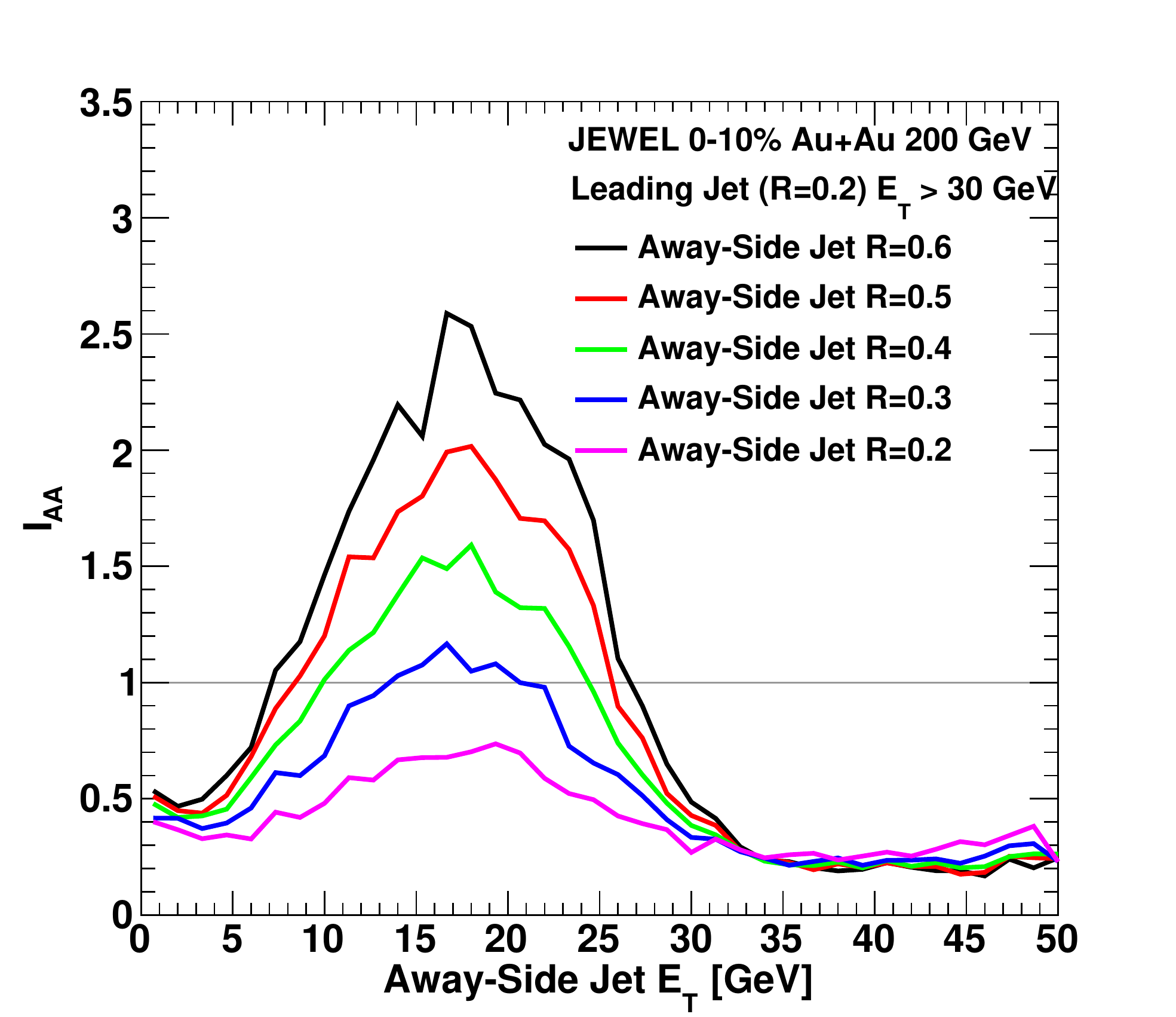}
\hfill
\includegraphics[width=0.47\linewidth]{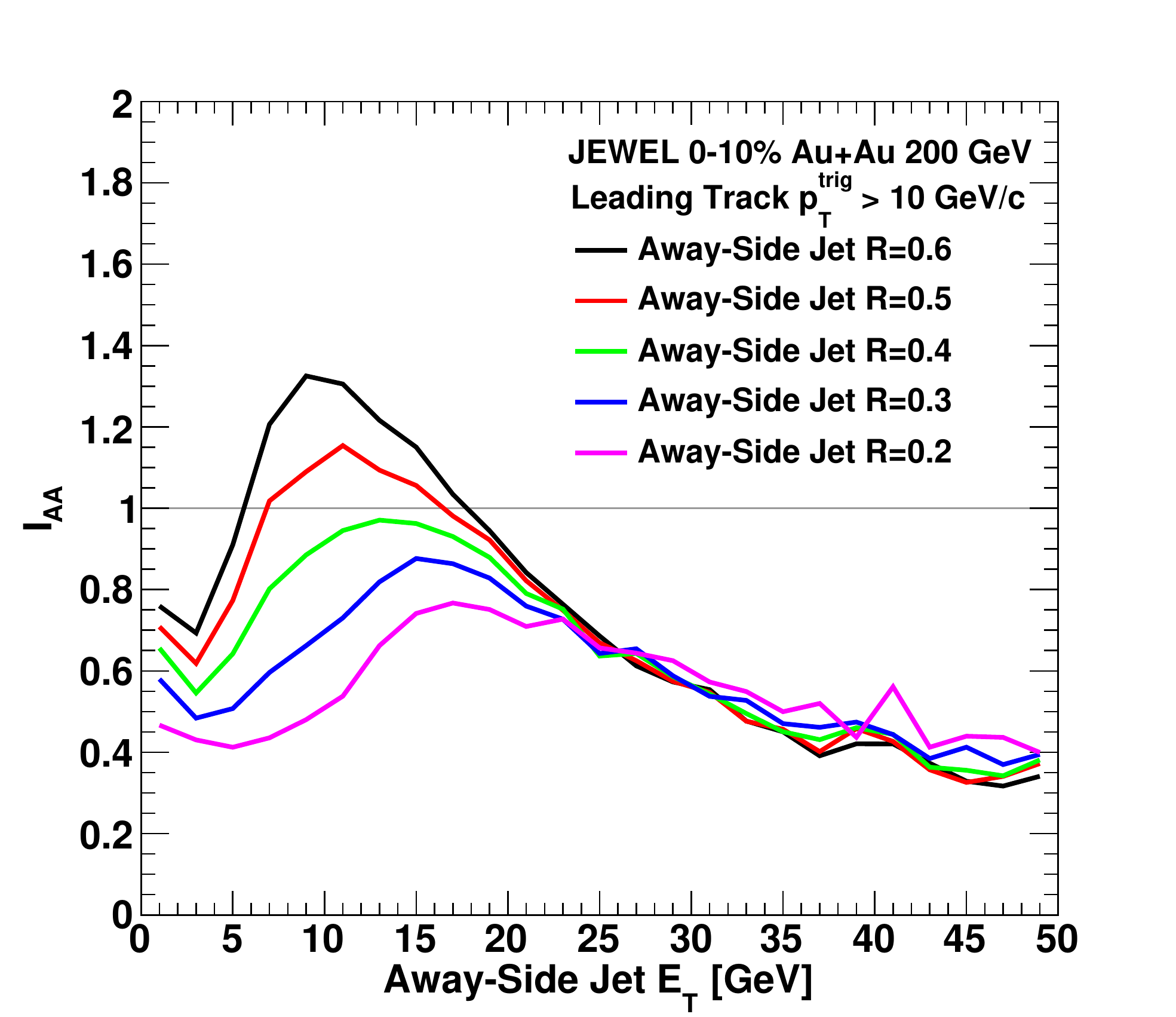}
\caption[JEWEL jet quenching Monte Carlo calculations in \auau 0-10\%
central collisions triggering on an anti-$k_{T}$ jet with $R=0.2$ with
$E_{T}>30$~GeV and measuring the modification of away-side correlated
jet yields ($I_{AA}$)]{(left) Shown are JEWEL jet quenching Monte Carlo
  calculations in \auau 0-10\% central collisions triggering on an
  anti-$k_{T}$ jet with $R=0.2$ with $E_{T}>30$~GeV and measuring the
  modification of away-side correlated jet yields ($I_{AA}$). The
  away-side jet yields are calculated for different jet sizes
  $R=0.2-0.6$. (right) Same quantity except now triggering on a charged
  hadron with $p_{T}>10$ ~GeV/$c$.  }
\label{fig:jewel_iaa}
\end{figure}

As an illustration of the benefit of reconstructing trigger-balancing
jets to low $E_\mathrm{T}$ and large $R$ values,
Figure~\ref{fig:jewel_iaa} shows an example of a measurement enabled
by this capability. The figure shows calculations from the JEWEL MC
generator of the modification of the away-side jet yield as a function
of jet $E_\mathrm{T}$ for events with a high-$E_\mathrm{T}$ narrow jet
or high-$p_\mathrm{T}$ charged hadron. In particular, the MC code
predicts a rich away-side jet cone size and $E_\mathrm{T}$ dependence
to the modifications, in a region which would be experimentally
accessible to sPHENIX given the purity studies above.

\clearpage

\section{Heavy Quark Jets}
\label{sec:hqjets}

Heavy quarks traversing the \qgp are an excellent test of our
understanding of the mechanisms of parton energy loss.  Due to the finite
velocity in medium there is an expected suppression of radiative
energy loss at small angles relative to the heavy quark. sPHENIX
has good tracking, including displaced vertex capabilities as detailed in
Section~\ref{sec:tracking}.

The rates for heavy quark jets shown in Figure~\ref{fig:heavyrates}
indicate a substantial number of accepted jets for sPHENIX in the
large data samples achievable.  With a jet tagged efficiency for
beauty jets of order 50\%, one would have thousands of beauty jets for
energy above 20~GeV and hundreds of thousands above 10~GeV.  The
tagging of heavy flavor dramatically reduces the ``fake'' jet
background and may allow one to push below the 20~GeV energy range
even in central \auau events.  Detailed \geant studies are at an early
stage and require detailed studies to document the full capabilities.

In this section, we investigate the feasibility of $b$-jet tagging
approaches through the requirement of charged tracks within the jet
with a large distance of closest (DCA) approach to the primary
vertex. This method, which exploits the long lifetime, displaced decay
and high multiplicity of the $B$ hadron, is sometimes called the
``Track Counting'' algorithm and has a wide use within the
literature. 

A description of the algorithm as it has been used in modern
experiments follows
below~\cite{ATLAS:2010sza,ATLAS:2010oba,CMS:2010hua,Chatrchyan:2012jua}. All
reconstructed charged tracks within the jet cone are considered, and
their distance of closest approach to the primary vertex is
determined, either for the full three dimensional distance or the two
dimensional distance in the transverse plane. The DCA value is signed
by comparing the vector from the primary vertex to the location of
closest approach along the track trajectory to the jet vector --- if
the location of closest approach is within the same hemisphere as the
jet vector (meaning the vectors have a positive dot product), the
DCA is taken to be positive. Otherwise, it is taken to be
negative. Finally, a signed DCA {\em significance} $S$ is defined by
dividing the reconstructed DCA value by its uncertainty, $S =
\mathrm{DCA}/\sigma_\mathrm{DCA}$. The $b$-tagging algorithm operates
by requiring some number of tracks to each have a significance $S$
above some minimum value $S_\mathrm{cut}$. As the quantity
$S_\mathrm{cut}$ and the number of tracks required change, so does the
performance of the algorithm as typically characterized by the
efficiency for tagging $b$-jets and the purity after applying this
cut. For a given sample of light, charm and bottom jets $N_l, N_c$ and
$N_b$ with cut efficiencies $\epsilon_l, \epsilon_c$ and $\epsilon_b$
respectively, the purity $P$ is defined as

\begin{equation}
P = \epsilon_b N_b / \left( \epsilon_l N_l + \epsilon_c N_c + \epsilon_b N_b \right)
\label{eq:bjet_tagging_purity}
\end{equation}

Alternately, one can also characterize the performance of the
algorithm in terms of the mis-identification probability or the
rejection factor for light and charm jets. Typically, one hopes to cut
in a way such that the purity of $b$-jets in the set of jets passing
the cut is high, while still maintaining a high enough efficiency to
give good statistics for the measurement.

In light quark and gluon jets, most charged hadrons originate very
near the primary vertex, resulting in a distribution of $S$ values
that is a Gaussian centered at zero with a width of $1$. However, some
small fraction of charged tracks within the jet, such as those from
the decay of strange hadrons, may have large positive significances
depending on the decay kinematics. In $b$-jets, many charged hadrons
originating from the $B$ hadron decay will generally have large DCA
significances. Thus, requiring a jet to have one or more tracks with a
large significance will preferentially select $b$-jets over light
jets. Since $B$ hadrons decays are accompanied by a large charged
particle multiplicity on average~\cite{Brandenburg:1999cs},
the requirement of a large number of tracks does not in and of itself
adversely affect the efficiency. The goal of the following study is to
quantify what tagging efficiency for $b$-jets can be achieved as a
function of the purity of $b$-jets in the tagged sample.

In this study, truth-level information with a parameterization of the
experimental DCA resolution was used to study the tagging
performance for $R=0.4$ light, charm and bottom jets with
$p_\mathrm{T} > 20$~GeV at the truth level. Separate samples of
$10^{6}$ \pythia events with $p_\mathrm{T} > 20$~GeV light, charm and
bottom jets were generated. To quantify the performance for each type
of jet, an unambiguous definition of jet flavor at the truth level is
needed. Following analogous studies in heavy flavor jet tagging at the
LHC~\cite{Aad:2012ma}, jet flavor is defined at the hadron (e.g. $B$ and $D$), and not
parton (e.g. $b$ and $c$ quark), level since this better corresponds
to the observed experimental signature. Bottom jets are defined as
those with a $p_\mathrm{T} > 5$~GeV $B$ hadron at any point in the
\pythia ancestry within $\Delta{R} < 0.4$ of the jet. Of the remaining
jets, those with a $p_\mathrm{T} > 5$~GeV $D$ hadron at any point in
the \pythia ancestry within $\Delta{R} < 0.4$ of the jet are defined
as charm jets. Jets which are not defined as either charm or bottom
jets are defined as light jets.

\begin{figure}[hbt!]
  \centering
  \includegraphics[width=0.45\textwidth]{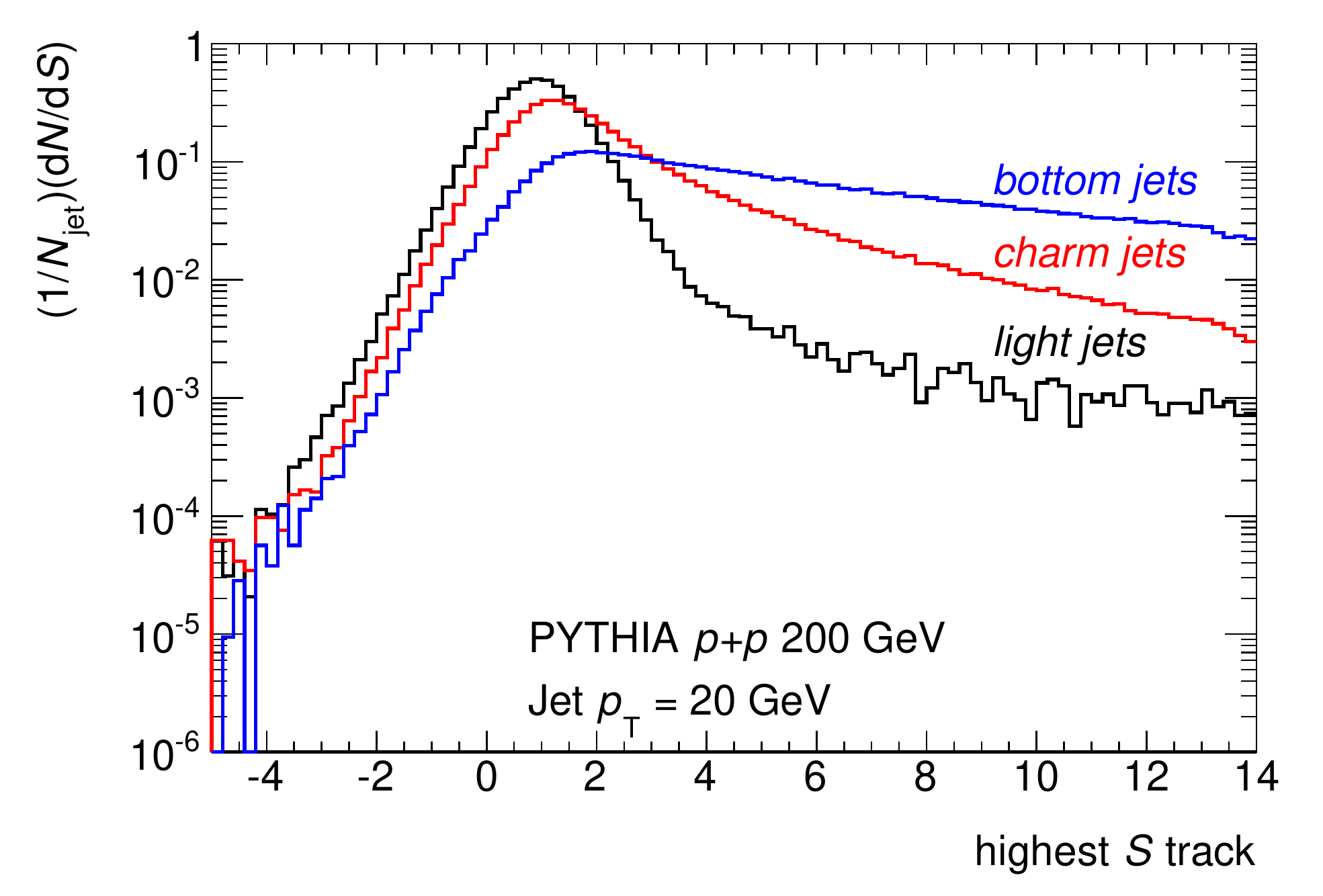}
  \includegraphics[width=0.45\textwidth]{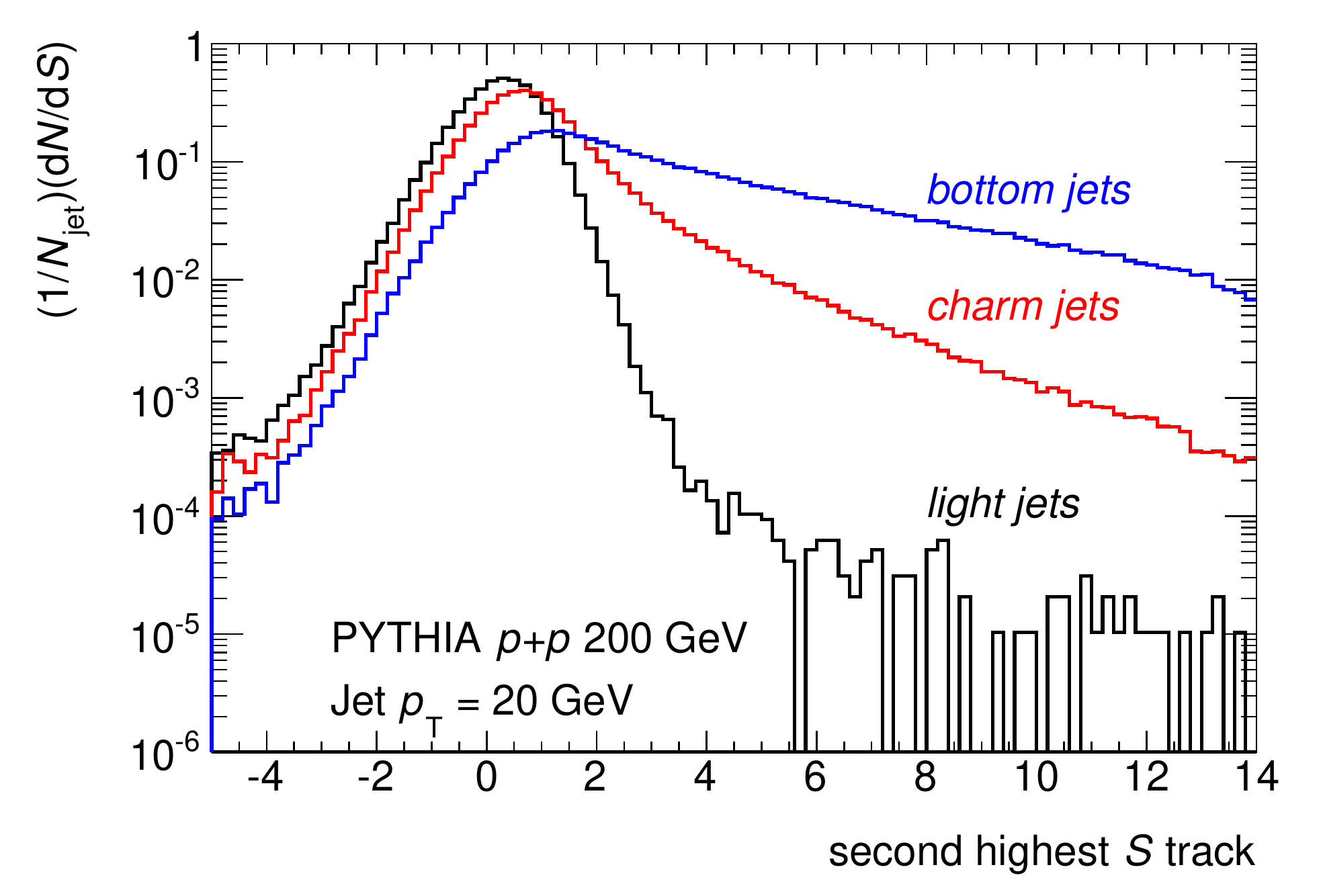}
  \caption[
Distribution of the DCA significance $S$ for the highest-$S$ track
(left panel) and second highest-$S$ track (right panel) in
$p_\mathrm{T} = 20$~GeV jets. Each panel shows the per-jet
distribution for light (black), charm (red) and bottom (blue) jets.
]{
Distribution of the DCA significance $S$ for the highest-$S$ track
(left panel) and second highest-$S$ track (right panel) in
$p_\mathrm{T} = 20$~GeV jets. Each panel shows the per-jet
distribution for light (black), charm (red) and bottom (blue) jets.
\label{fig:bjet_TrackCounting_highestS}}
\end{figure}

For each jet, the set of final state charged hadrons with
$p_\mathrm{T} > 0.5$~GeV that are within $\Delta{R}<0.4$ of the jet
axis are examined. Within this study, the 2-D DCA (e.g. the DCA in the
transverse plane) is used. To generate a reconstructed DCA for each
charged hadron, its trajectory is projected in a straight line in the
transverse plane to determine the truth DCA. Then, the DCA is smeared
according to the $p_\mathrm{T}$-dependent DCA resolution obtained
through full \geant studies of the tracking performance, as described
in Section~\ref{sec:tracking} and displayed in
Figure~\ref{fig:dca_hijing_3ptbins_svtxv1}. The significance $S$ is
calculated by dividing the smeared DCA by the nominal
$p_\mathrm{T}$-dependent resolution $\sigma_\mathrm{DCA}$, taken from
Gaussian fits to the core of the DCA resolutions, shown in
Figure~\ref{fig:dca_ptbins}. In this way, the study incorporates a
realistic description of the DCA performance in
sPHENIX. Figure~\ref{fig:bjet_TrackCounting_highestS} shows the
distribution of $S$ values for the first and second highest $S$ tracks
in jets of the three flavors.

In modern $b$-jet tagging approaches, efforts are made to exclude
hadrons which originate from strange decays by, for example, removing
pairs of tracks which reconstruct to a $\Lambda^0$ or $K_{s}^0$ mass
(called $V^0$'s), or by rejecting all tracks with a DCA so large that
they are dominated by strange decays instead of tracks from
$b$-jets. In this study, hadrons which originate from a $V^0$ or with
a DCA larger than $1$~mm are rejected. Such an approach will limit but
not remove the high-$S$ background in light jets which still enters
from, for example, decays of $\Sigma^\pm$ baryons which produce only a
single charged track associated with a neutral hadron and thus cannot
be identified through an invariant mass analysis.

\begin{figure}[hbt!]
  \centering
  \includegraphics[width=0.6\textwidth]{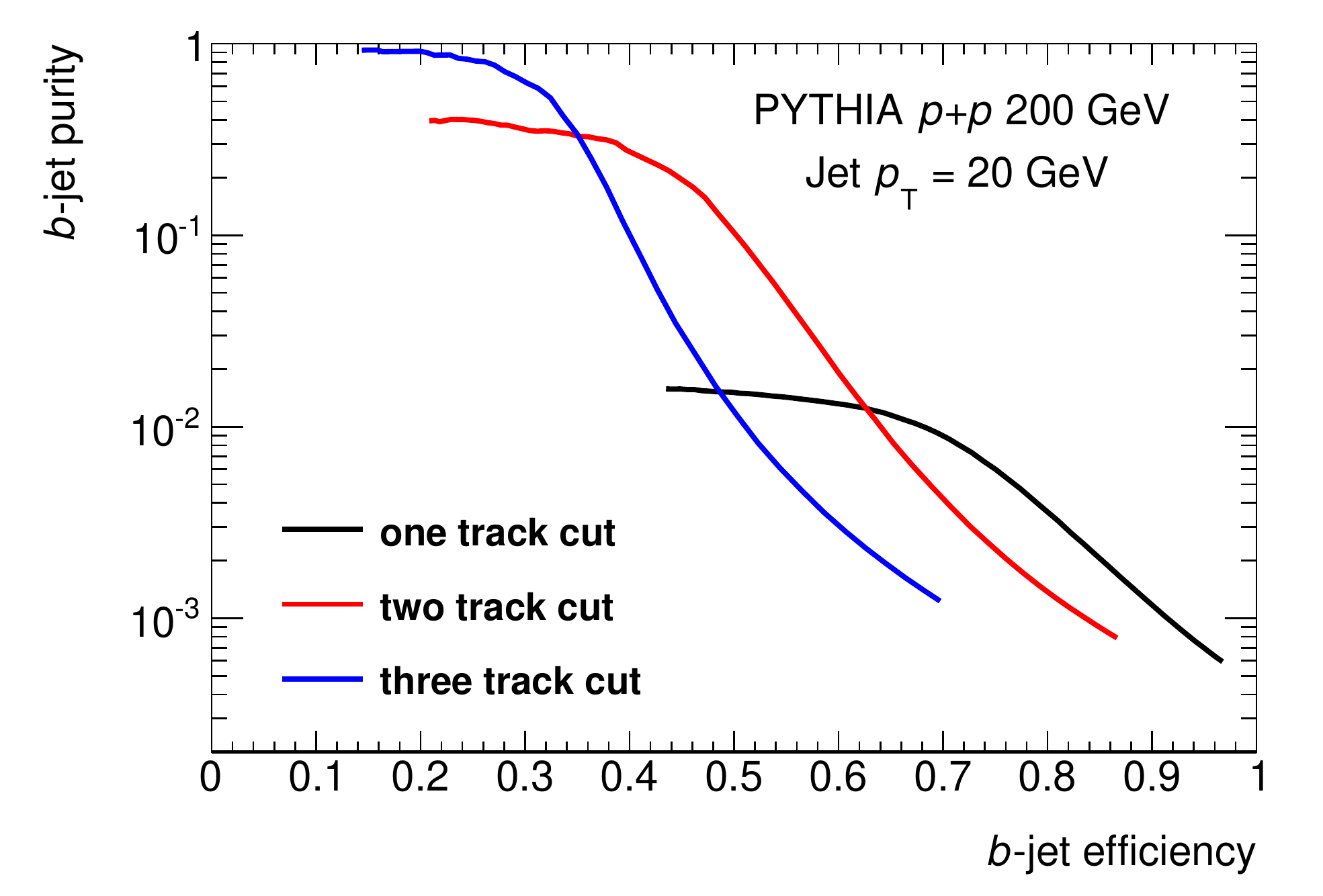}
  \caption[Performance of $b$-jet tagging algorithms for $p_\mathrm{T} =
  20$~GeV jets based on requiring at least one, two or
  three tracks in the jet to have a 2-D DCA significance above some
  minimum value]{Performance of 
    $b$-jet tagging algorithms for $p_\mathrm{T} = 20$~GeV jets based on
    requiring at least one (black), two (red) or three (blue) tracks in
    the jet to have a 2-D DCA
    significance above some minimum value. Purity vs. efficiency curves
    are generated by varying the minimum significance DCA
    \label{fig:bjet_TrackCounting_default}}
\end{figure}

For a given cut (specified by the $S_\mathrm{cut}$ and the number of
tracks required to have $S > S_\mathrm{cut}$), the efficiency for
light, charm and bottom jets is determined. Then, the purity $P$ of
$b$-jets is determined following
Equation~\ref{eq:bjet_tagging_purity}, with the initial mixture of
light, charm and bottom jets $N_l$, $N_c$ and $N_b$ at $p_\mathrm{T} =
20$~GeV/$c$ given by pQCD and FONLL calculations as is shown in
Figures~\ref{fig:nlo_jetrates}~and~\ref{fig:heavyrates}. The
performance is quantified by plotting the $b$-jet efficiency
$\epsilon_b$ against the $b$-jet purity $P$, which vary inversely with
one another as the details of the cut are changed.

The performance of the Track Counting algorithm in \pp\ collisions is
summarized in Figure~\ref{fig:bjet_TrackCounting_default}, showing the
behavior of the algorithm requiring one, two or three tracks in the
jet to all have $S > S_\mathrm{cut}$ as black, red and blue curves
respectively. The figure shows, as a function of the efficiency for
$b$-jets passing the cut, the purity of $b$-jets within the set of all
jets that pass the cut. The efficiency vs. purity curves are generated
by varying the value of $S_\mathrm{cut}$ between $0$ and $5$. It is
evident that less stringent requirements (fewer tracks, smaller
$S_\mathrm{cut}$ requirement) result in a high efficiency but a low
purity. On the other hand, stricter requirements (more tracks, each of
which has a large $S > S_\mathrm{cut}$ ) result in a low efficiency
but a high-purity.

The presence of high-DCA tails in light jets, either from hadrons
originating from strange decays or from hadrons originating from the
primary vertex but with a badly reconstructed DCA, are the limiting
factors in this approach. As can be seen in Figure~\ref{fig:bjet_TrackCounting_highestS}, 
examining only the highest-$S$ track
will reach a point of diminishing returns, as even large values of
$S_\mathrm{cut}$ will still leave a background of light jets. Thus,
the black curve in Figure~\ref{fig:bjet_TrackCounting_default}
saturates at a given maximum purity. Requiring more tracks results in
a higher maximum purity, since it is much rarer for a light jet to
have two tracks with a large $S$, but adversely affects the tagging
efficiency. Two- or three-track algorithms are able to reach a high
purity $P > 50$\% while still retaining a reasonable efficiency
$\epsilon_b \approx 40$--$50$\%, which is a promising indication of
the possible performance. When the cuts are chosen to give such a high
purity, the remaining non-$b$ jet background is composed predominantly
of $c$-jets.

\begin{figure}[hbt!]
  \centering
  \includegraphics[width=0.6\textwidth]{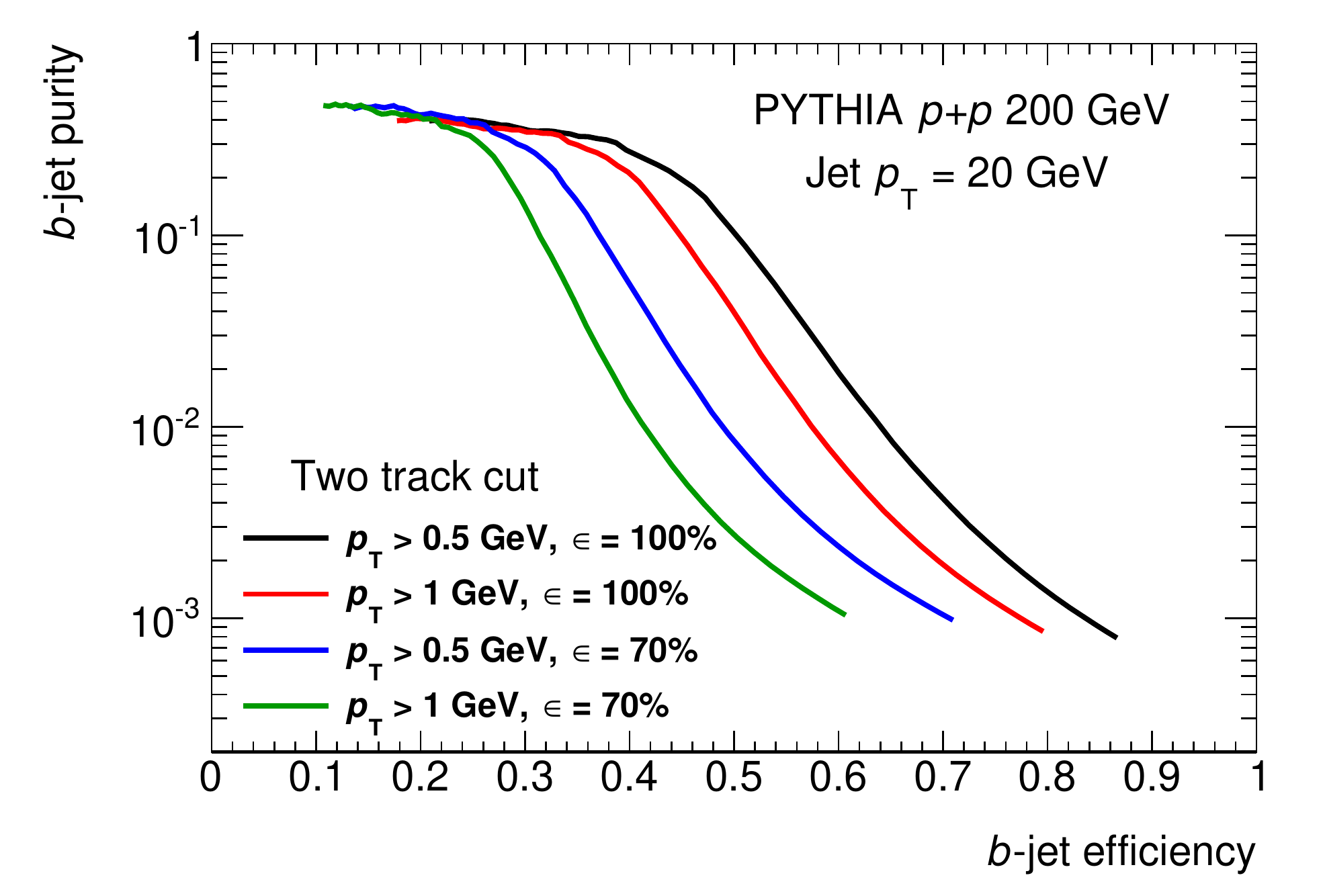}
  \caption[Performance of $b$-jet tagging algorithms for $p_\mathrm{T} =
  20$~GeV jets based on requiring at least two tracks in the jet to have
  a 2-D DCA significance above some
  minimum value]{Performance of $b$-jet tagging algorithms for
    $p_\mathrm{T} = 20$~GeV jets based on requiring at least two tracks
    in the jet to have a 2-D DCA
    significance above some minimum value. Curves of different colors
    show the performance under difference assumptions of the tracking
    performance and minimum track $p_\mathrm{T}$. Purity vs. efficiency
    curves are generated by varying the minimum significance DCA
 \label{fig:bjet_TrackCounting_efficiencyvariations}}
\end{figure}

While the first results shown above have been made assuming a perfect
tracking efficiency for charged hadrons with $p_\mathrm{T} > 0.5$~GeV,
the $b$-jet tagging algorithm has also been studied under different
assumptions of the tracking
performance. Figure~\ref{fig:bjet_TrackCounting_efficiencyvariations}
demonstrates how the two track-based cuts change under two
variations. In the first variation the overall tracking efficiency is
taken to be 70\%, while in the second variation only tracks with
$p_\mathrm{T} > 1$~GeV are considered. Both variations are also
considered together. Since the tracking design and offline tracking
reconstruction parameters are still being optimized for a
high-multiplicity environment, these variations are meant to bracket a
reasonable range of the possible tracking performance. As expected,
the performance systematically degrades with decreasing
efficiency. Additionally, restrictions to the track $p_\mathrm{T}$
decrease the multiplicity of tracks which are available to pass the
cuts and thus reduce the efficiency as well. This study highlights the
necessity of robust tracking capability in sPHENIX to carry out a
$b$-jet program. In particular, high efficiency in the inner pixel
layers, which determine the DCA, is required.

\begin{figure}[hbt!]
  \centering
  \includegraphics[width=0.6\textwidth]{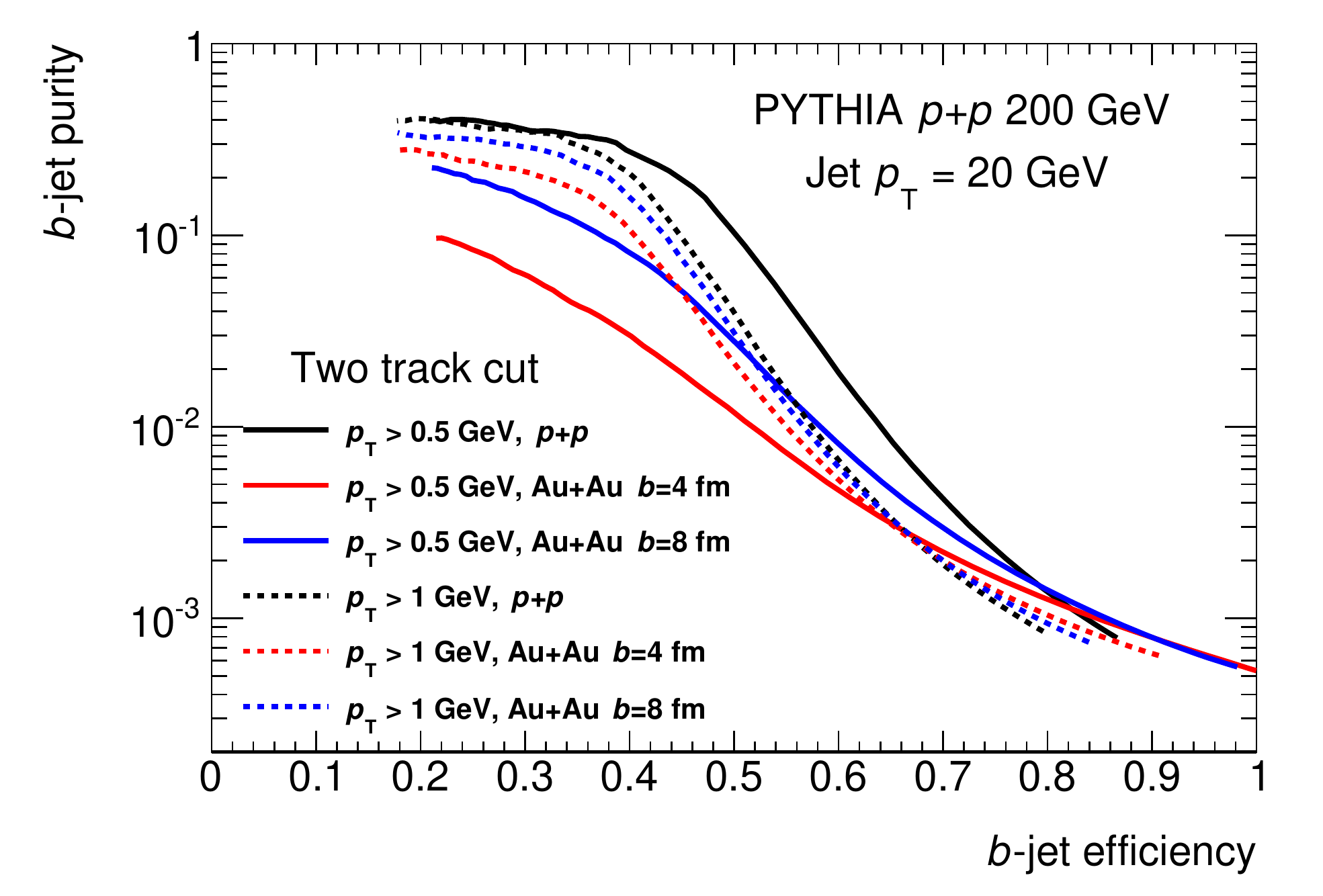}
  \caption[Performance of $b$-jet tagging algorithms for $p_\mathrm{T} =
  20$~GeV jets requiring at least two tracks in the jet to have
  a 2-D DCA significance above some
  minimum value]{Performance of $b$-jet tagging algorithms for
    $p_\mathrm{T} = 20$~GeV jets based on requiring at least two tracks
    in the jet to have a 2-D DCA
    significance above some minimum value. Curves of different colors
    show the performance in the presence of different heavy ion
    backgrounds, while different line styles correspond to different
    choices of the minimum track $p_\mathrm{T}$. Purity vs. efficiency
    curves are generated by varying the minimum significance DCA
    \label{fig:bjet_TrackCounting_UEeffects}}
\end{figure}

Additional studies have been performed to determine how the
performance of the algorithms may be affected by the presence of the
underlying event in \auau events. The large amount of UE charged
particles within the jet cone cannot be experimentally distinguished
from those arising from the jet fragmentation and $B$ hadron decay,
and will each have their own reconstructed
DCA significance. Therefore, for a given $b$-jet tagging cut, the
efficiency for all jet flavors will increase, as the addition of extra
tracks will give additional opportunities for a jet to have the
requisite number of tracks with $S > S_\mathrm{cut}$. 

To quantify the effect on the performance, jets were randomly embedded
into \hijing events with $b = 4$~fm and $b = 8$~fm, and the additional
charged particles from the underlying event particles were assumed to
originate from the primary vertex but otherwise have the same
DCA resolution as determined in \geant simulations in
Figure~\ref{fig:dca_hijing_3ptbins_svtxv1}. At fixed $b$-jet tagging
efficiency, the $b$-jet purity was found to be systematically worse in
central \hijing events than in peripheral \hijing events, and worse
there than in the \pythia only events. In particular, the two-track
cut with $p_\mathrm{T} > 0.5$~GeV track was no longer found to give
sufficiently good performance, necessitating the need to only examine
tracks with $p_\mathrm{T} > 1$~GeV (since the additional number of UE
tracks is smaller) or to require three
tracks. 

Figure~\ref{fig:bjet_TrackCounting_UEeffects} summarizes the
performance for the two-track cuts, showing how the same cuts perform
in the $p$+$p$ only case and after embedding into $b = 4$~fm and $b =
8$~fm \hijing events. Although the performance of the algorithm with
$p_\mathrm{T} > 0.5$~GeV tracks is very sensitive to the UE
background, we observe that even a mild track $p_\mathrm{T}$
requirement of $1$~GeV makes the algorithm much more resilient against
the effects of the UE. At the moment, this study focuses on the
effects of the UE on the tagging performance, and does not consider
the additional effects of the larger jet energy resolution introduced
by the presence of UE fluctuations.

\begin{figure}[hbt!]
  \centering
  \includegraphics[width=0.7\textwidth]{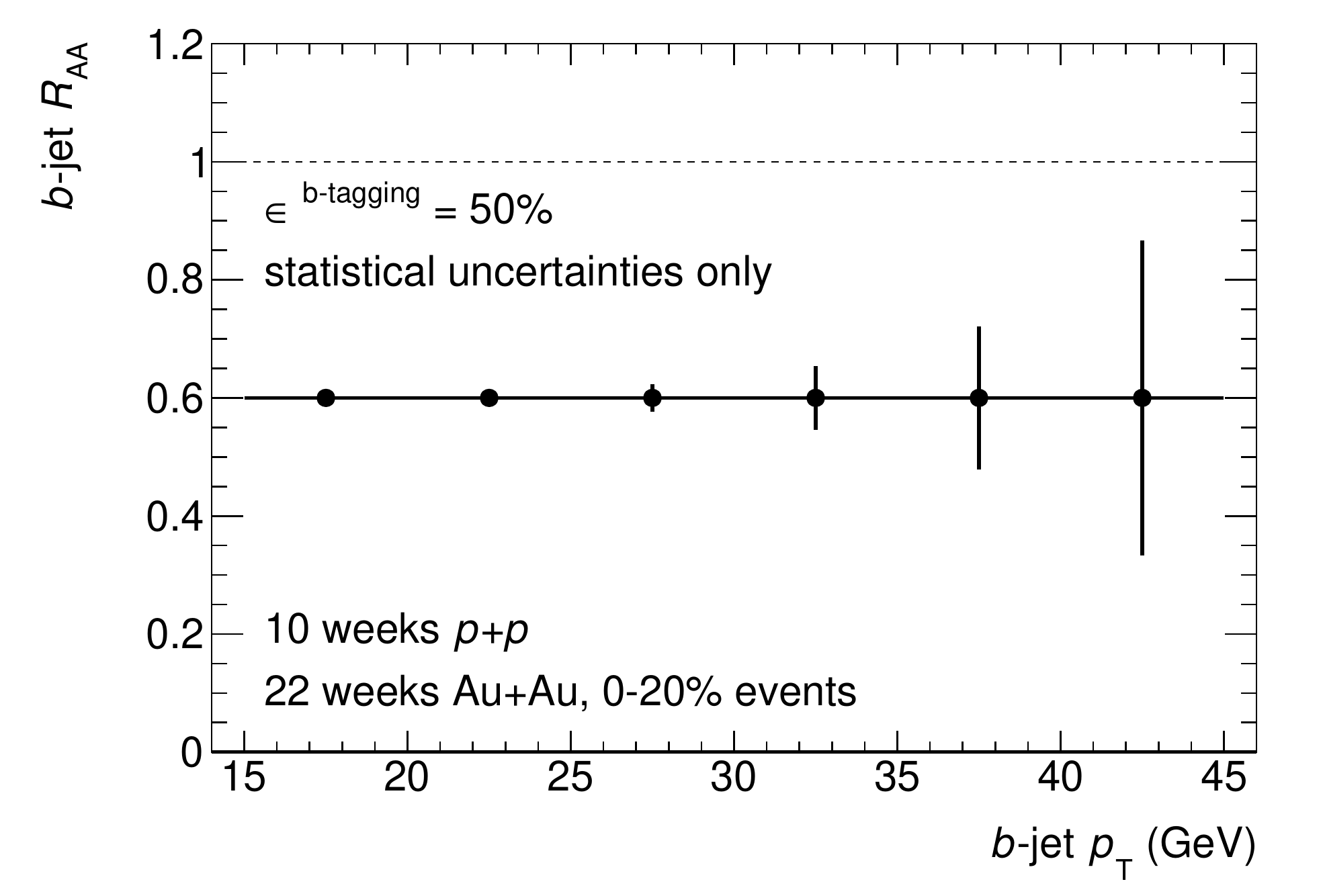}
  \caption[Projected statistical uncertainties on the $R_\mathrm{AA}$
  for $b$-jets in $0$--$20$\% \auau events, assuming an overall
  suppression of $R_\mathrm{AA} = 0.6$, a $b$-jet tagging efficiency of
  $50$\%, $\mathcal{L} = 630$~pb$^{-1}$ of $p$+$p$ events and
  $0.6\times10^{12}$ sampled minimum bias \auau events]{Projected
    statistical uncertainties on the $R_\mathrm{AA}$ for $b$-jets in
    $0$--$20$\% \auau events, assuming an overall suppression of
    $R_\mathrm{AA} = 0.6$, a $b$-jet tagging efficiency of $50$\%,
    $\mathcal{L} = 630$~pb$^{-1}$ of $p$+$p$ events and
    $0.6\times10^{12}$ sampled minimum bias \auau events.
    \label{fig:bjet_projectedRAA}}
\end{figure}

The studies detailed here have demonstrated that $b$-jet tagging
through the identification of associated tracks with a large distance
of closest approach with respect to the primary vertex can reach a
high $b$-jet purity while still maintaining good $b$-jet efficiency,
and is a plausible approach for $b$-jet tagging in sPHENIX. However,
they have also demonstrated the need for high-efficiency, precision
tracking within sPHENIX to achieve reasonable $b$-jet tagging
performance. These studies have been performed at the ``truth'' hadron
level using \pythia and \hijing events. However, they incorporate a
description of the DCA resolution (the main driver of the $b$-jet
identification and light jet rejection capability) as determined with
full \geant simulations of the tracking performance. Future studies
will be needed to determine the role of other detector effects, such
as the degree to which a realistic jet energy scale non-closure and
finite jet energy resolution affect the tagging performance for jets
at a given reconstructed (rather than truth) $p_\mathrm{T}$.

Additional studies of $b$-jet tagging through the identification of an
electron with large momentum transverse to the jet axis
($p_\mathrm{T}^\mathrm{rel}$), often called the ``Soft Lepton
Tagging'' (SLT) method~\cite{ATLAS:2010afa}, have resulted in the
similar conclusion that a high $b$-jet purity is achievable while
maintaining good efficiency. (In the case of the SLT method, the
overall efficiency is with respect to the fraction of $b$-jets which
have a semileptonic decay and thus produce an electron.) In this case,
the mis-identification of light jets as $b$-jets is due to the
possibility of pions being erroneously identified as electrons. Thus,
unlike the Track Counting method, the SLT performance is defined much
more by the electron identification than the DCA reconstruction
performance in sPHENIX. Having several distinct $b$-jet tagging
approaches is attractive, because it allows for independent
cross-calibration of the algorithms with respect to one
another. However studies of the SLT method, while promising, are still
at an early stage and are not detailed here. Early studies of $b$-jet
tagging through the direct reconstruction of the displaced secondary
vertex, a popular method within the LHC
experiments~\cite{Chatrchyan:2013exa,ATLAS:2010rza}, are also
underway. Finally, the performance of the Track Counting algorithm has
been benchmarked above using simulations. However, there are important
$b$-tagging methods which use data-driven approaches to boot-strap the
efficiency and purity of the method without the need for detailed MC
input. An example of these is the jet probability
algorithm~\cite{Buskulic:1993ka}, which uses the negative DCA-tails
of tracks in data as a reference against which to understand the light
jet contribution to the set of jets with tracks that have a large
positive DCA.

A statistical projection for the $b$-jet $R_\mathrm{AA}$ in
central \auau events is shown in
Figure~\ref{fig:bjet_projectedRAA}. This projection is constructed
through the FONLL-based predictions of the $b$-jet rates as a function
of $p_\mathrm{T}$ as shown in Figure~\ref{fig:heavyrates}, an
assumption of the total luminosity received in \auau and $p$+$p$
data-taking as detailed in Section~\ref{Section:Rates}, and a nominal
combined $b$-jet reconstruction and tagging efficiency of $50$\%. This
projection indicates that with the given $p_\mathrm{T}$ binning, the
$R_\mathrm{AA}$ will have good statistical power out to $40$~GeV,
enabling a measurement of inclusive $b$-jet suppression over a large
kinematic range. Furthermore, $b$-tagged jet-hadron correlation
analyses (such as those described in
Sections~\ref{sec:jet_surface_emission_engineering}~and~\ref{sec:fragmentation_function})
may offer the opportunity to probe how the internal structure of heavy
quark-initiated jets is modified by the medium.

\begin{figure}[hbt!]
  \centering
  \includegraphics[width=0.48\textwidth]{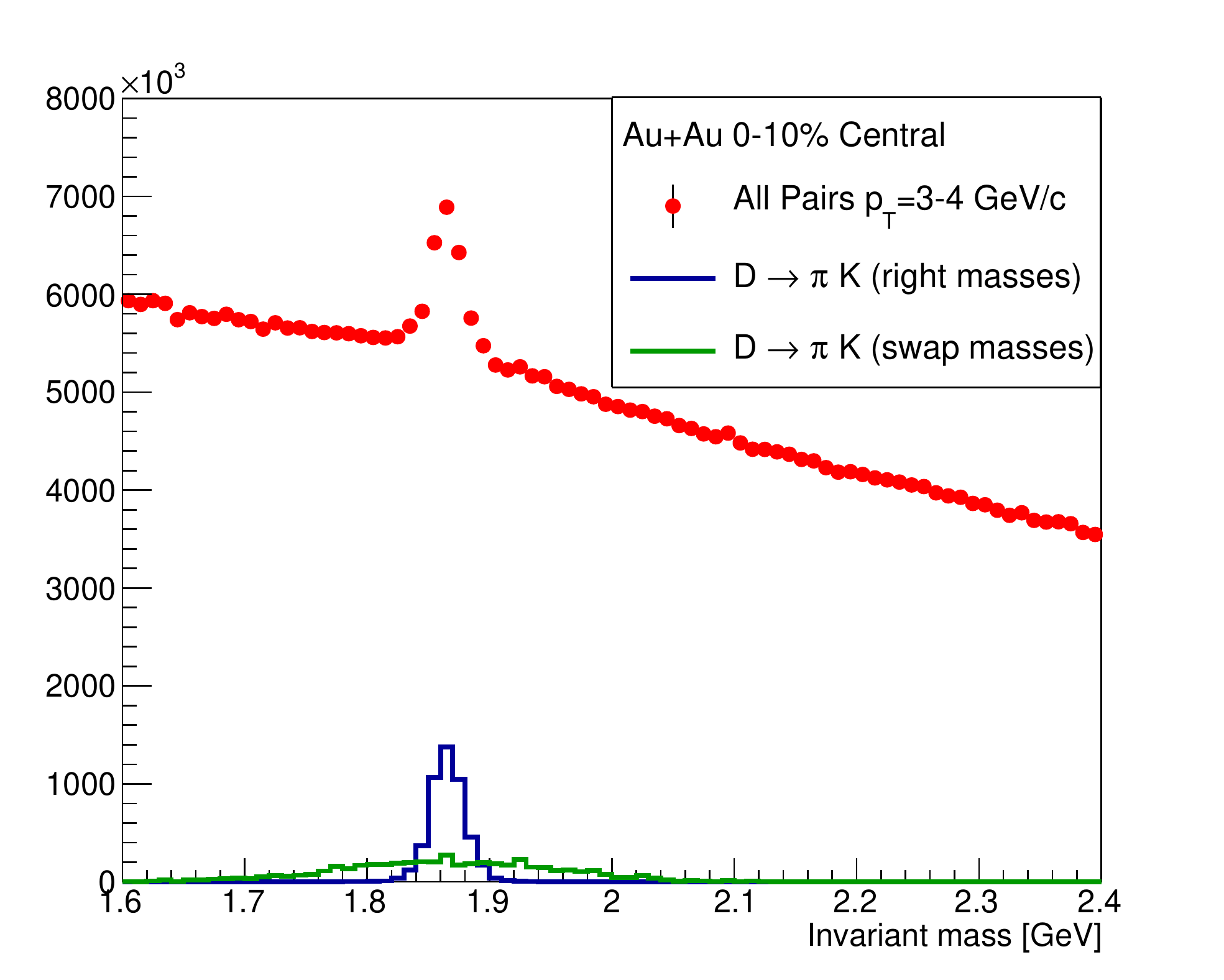}
  \hfill
  \includegraphics[width=0.48\textwidth]{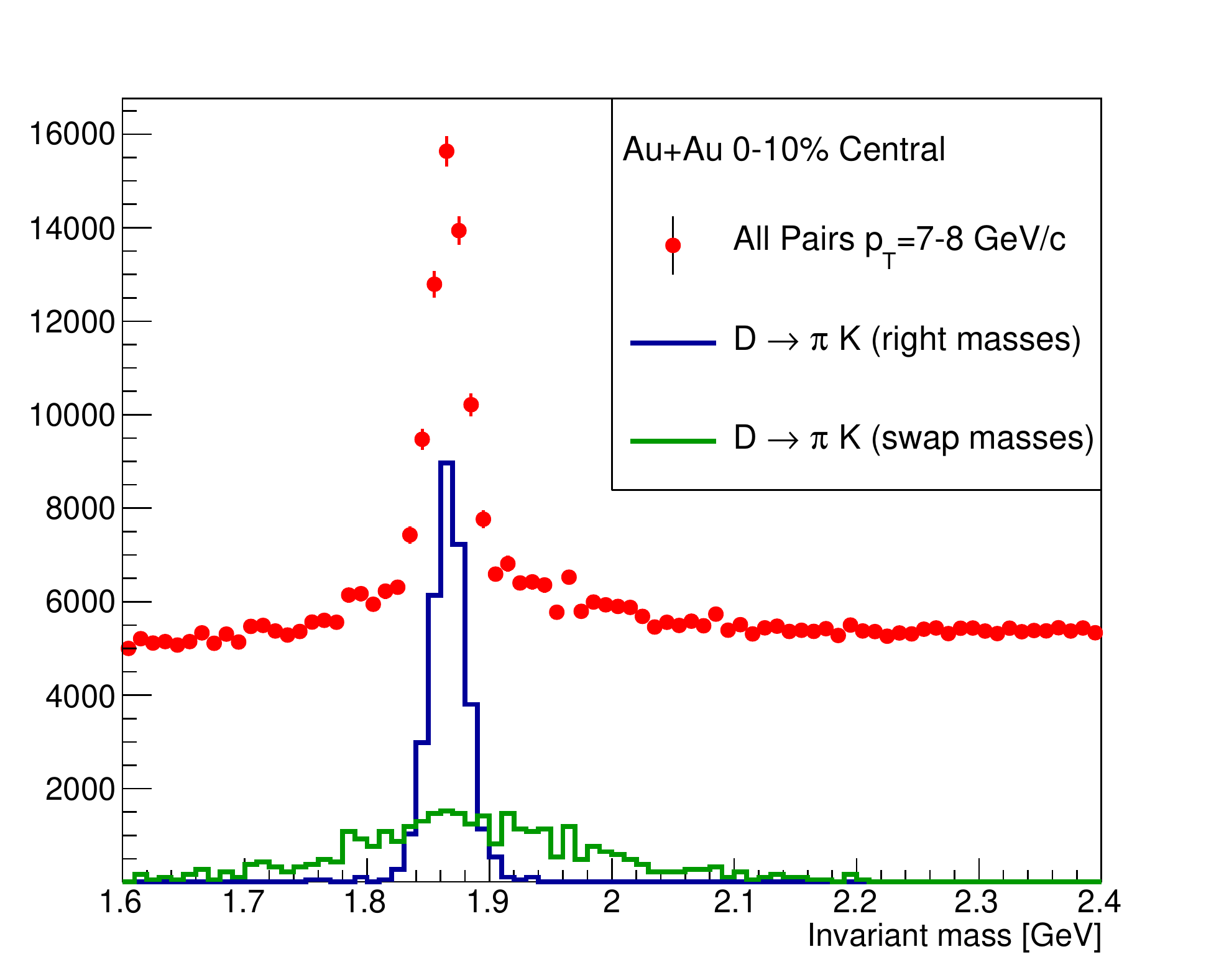}
  \caption[Reconstructed $D$ meson signal to background in 0-10\% central \auau collisions.]
	  {Shown are the invariant mass distributions of reconstructed charged particle track pairs in \pt bins
of 3-4 GeV/c (left) and 7-8 GeV/c (right).   The histograms indicate the contribution from correctly matched $D$
mesons and cases where the daughter mass assumption is swapped.
    \label{fig:Dmesons}}
\end{figure}

Reconstruction of charm jets with similar methods described above is very challenging.
However, with the ability to reconstruct secondary vertices, we can reconstruct $D$ mesons
via their $\pi+K$ decay channel for example and then associate them with reconstructed jets
via the calorimeter.   Shown in Figure~\ref{fig:Dmesons} are the simulated reconstructed invariant
mass for all charged particle pairs assuming one is a pion and one is a kaon in 0-10\% central \auau collisions.   
The histograms are the real $D$ meson contribution where the daughter product masses are assigned correctly
and where they are swapped.   The simulation includes \hijing generated uncorrelated and correlated backgrounds.
The signal to background is small at low \pt as expected and increases significantly at higher \pt.   
The precision secondary vertex reconstruction provides the important background suppression.   Associating
these $D$ mesons with reconstructed jets then allows a measure of the fragmentation functions which is
predicting to have large energy loss sensitivity.

\section{Proton-Nucleus Collision Jet Physics}
\label{sec:pajets}

In addition to a rich program of jet studies of hot nuclear matter,
sPHENIX will able to capitalize on its capabilities to study
high-$p_\mathrm{T}$ processes in ``cold'' nuclear systems, provided by
a high statistics $p$+Au 200~GeV run. Measurements of hadron, jet and
photon cross-sections in $p$+Au collisions are important for a number
of reasons. At their most basic level, they provide an overall test of
pQCD calculations of hard processes based on the standard collinear
factorization and parton distribution function formalism. They are
also sensitive to the so-called ``cold nuclear matter'' effects, which
may arise from a number of sources including the modification of the
parton densities in the nuclear environment, the initial state energy
loss of the hard scattering partons, and so forth. More generally,
$p$+Au collisions are a useful laboratory within which to understand
the interplay between hard processes and experimental handles on the
collision geometry.

While the above motivations address interesting physics questions in
and of themselves, $p$+Au collisions are also crucial experimental
context for the sPHENIX physics program, since they will provide the
reference against which to interpret the modifications of hard process
rates and jet shapes observed in \auau collisions.

\begin{figure}[p]
  \centering
  \includegraphics[width=0.9\textwidth]{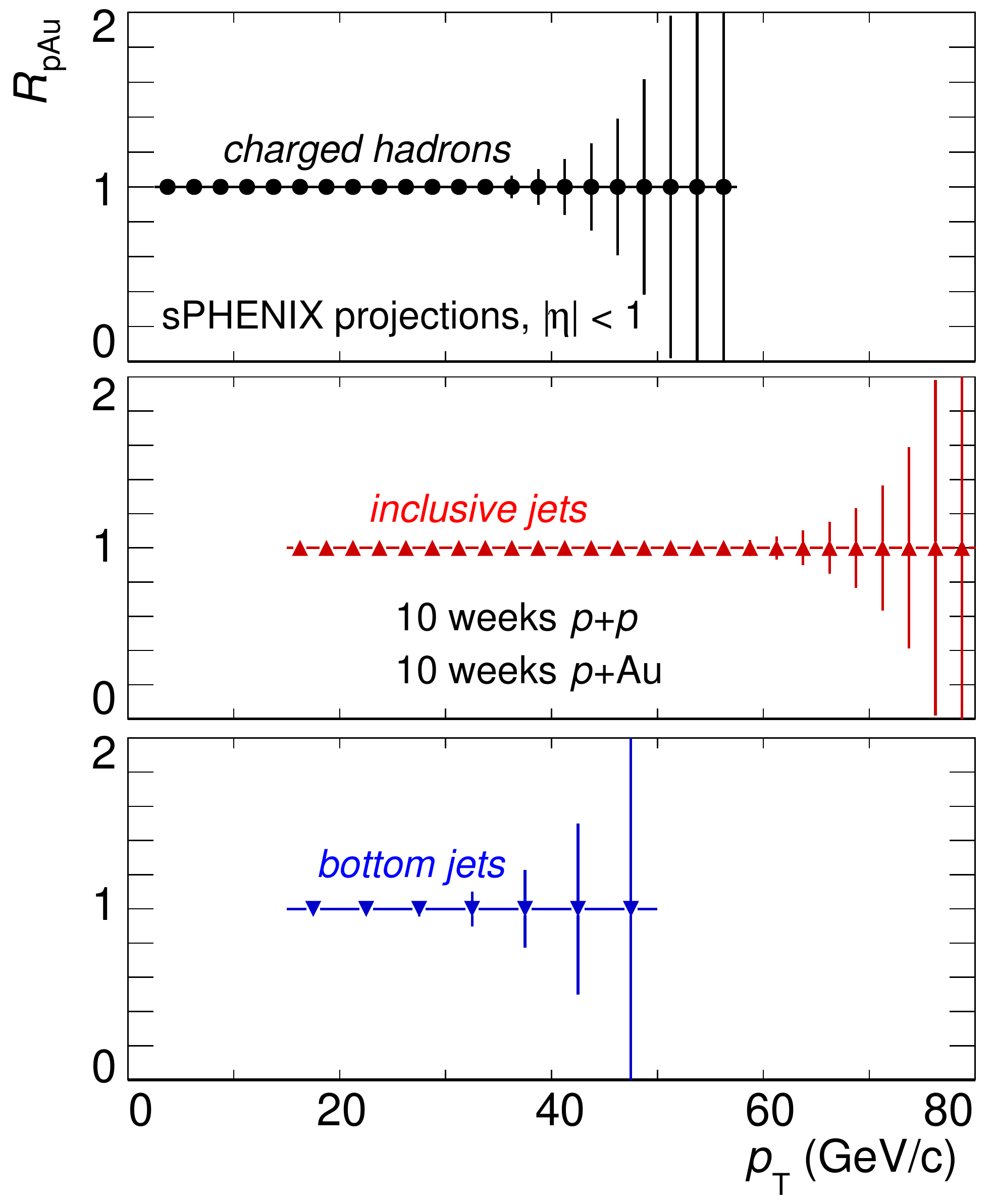}
  \caption[Projected statistical uncertainties on the $R_\mathrm{pAu}$ 
  for charged pions inclusive jets and $b$-jets for minimum bias $p$+Au
  events corresponding to 10 weeks of minimum bias $p$$+$Au 
  and $p$+$p$ running.]{Projected statistical uncertainties on the
    $R_\mathrm{pAu}$ for charged pions (top plot), inclusive jets
    (middle plot) and $b$-jets (bottom plot) for minimum bias $p$+Au
    events, assuming no overall modification ($R_\mathrm{pAu} = 1$), a
    $b$-jet tagging efficiency of $50$\%, corresponding to 10 weeks of
    minimum bias $p$+Au and $p$+$p$ running.
    \label{fig:pAphysics_projections}}
\end{figure}

Figure~\ref{fig:pAphysics_projections} shows the projected statistical
uncertainties for the nuclear modification factor $R_\mathrm{pAu}$ in
200~GeV $p$+Au collisions corresponding to the first two years of
sPHENIX data-taking. The $R_\mathrm{pAu}$ projections for unidentified
charged hadrons, inclusive fully reconstructed jets, and even
$b$-tagged jets are shown. (Measurements of direct photons, while not
shown in this compilation, would have modestly more statistics than
the $b$-jets, extending $\approx5$-$10$~GeV/$c$ farther in
$p_\mathrm{T}$.)

\begin{figure}[hbt!]
  \centering
  \raisebox{3.0ex}{\includegraphics[width=0.55\textwidth]{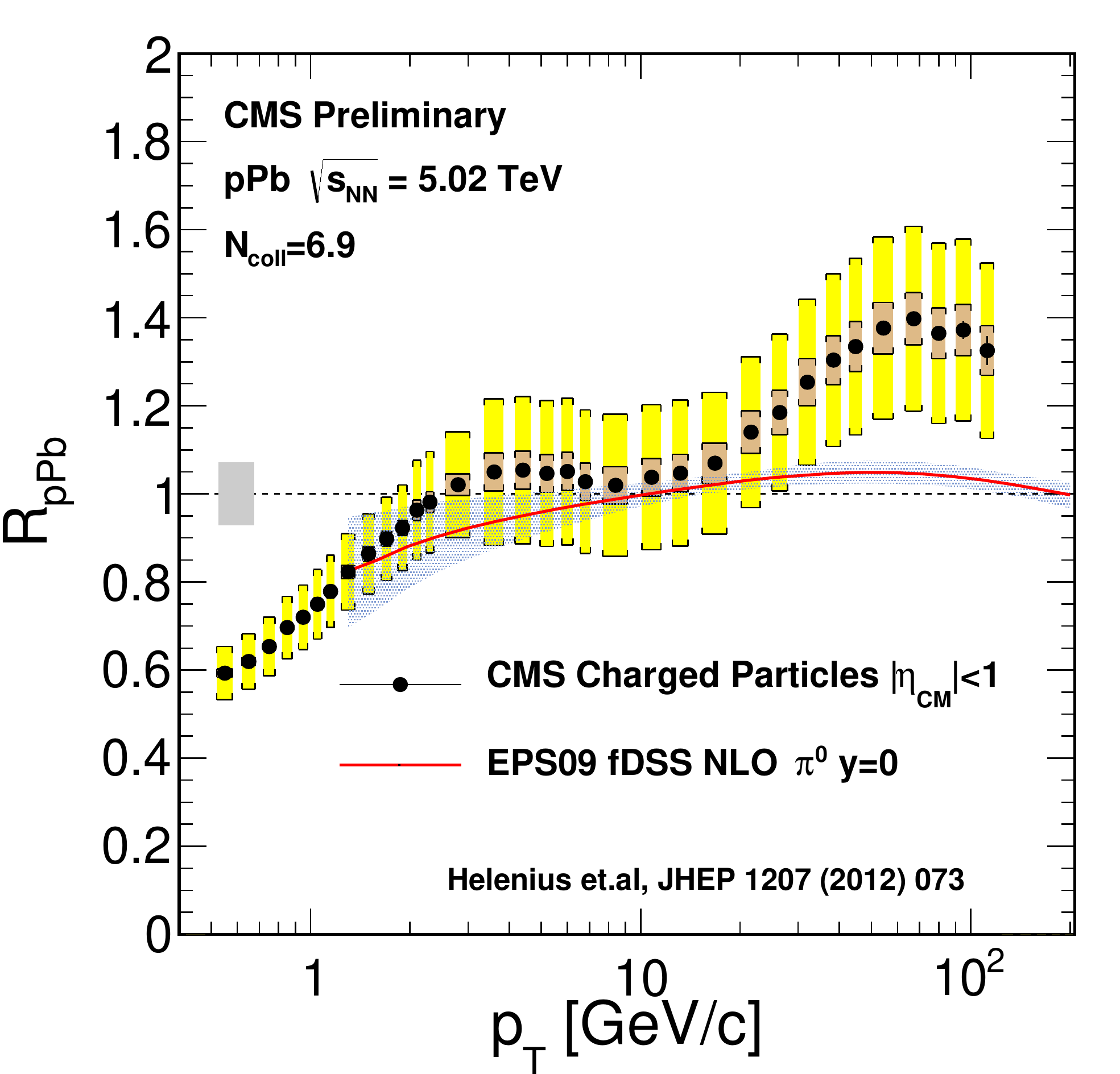}}
  \includegraphics[width=0.40\textwidth]{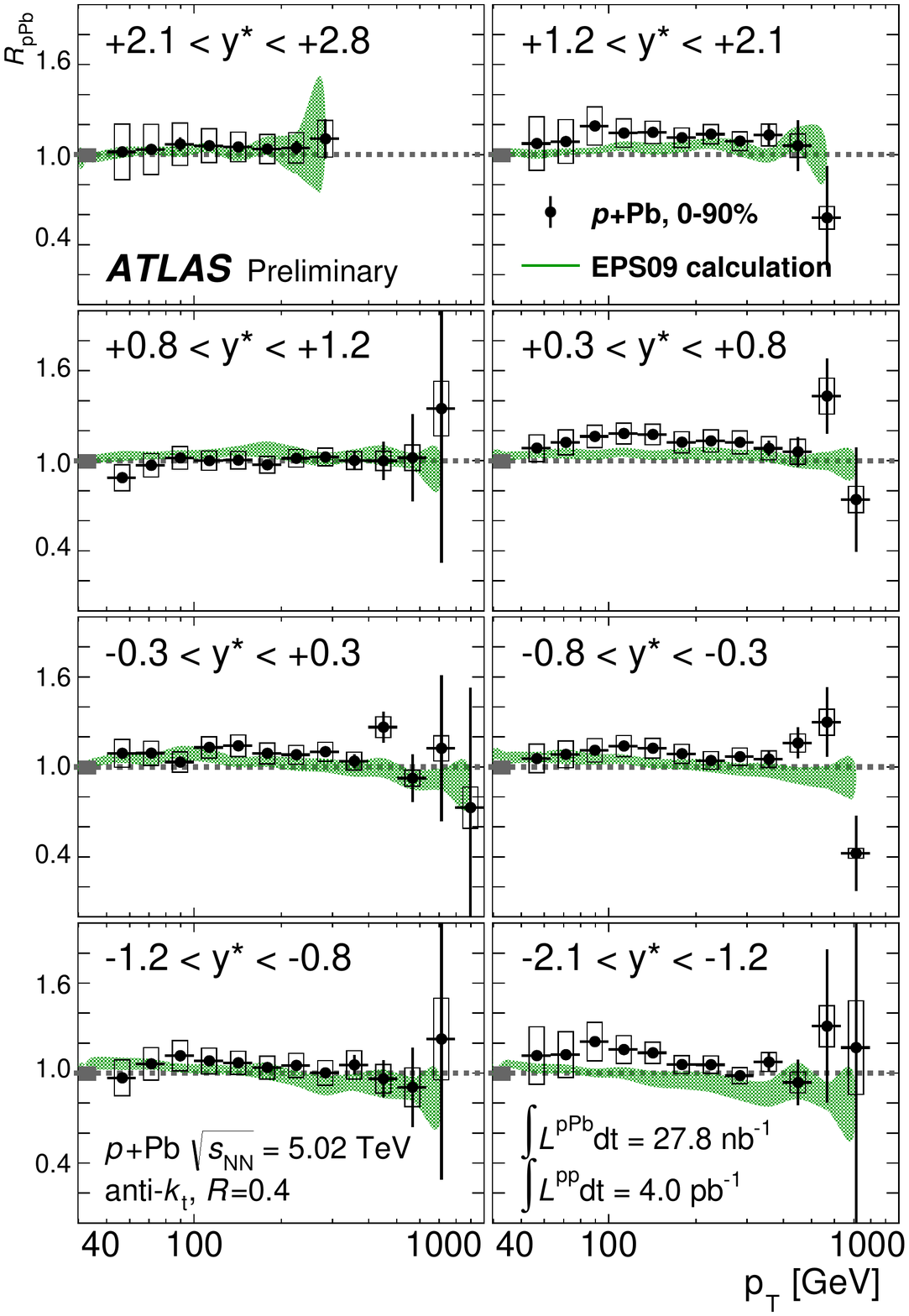}
  \caption[$R_\mathrm{pPb}$ for mid-rapidity charged particles vs
  $p_\mathrm{T}$ as measured by CMS, and for inclusive full jets vs jet
  $p_\mathrm{T}$ as measured by ATLAS]{left: 
    $R_\mathrm{pPb}$ for mid-rapidity charged particles as a function of
    $p_\mathrm{T}$ as measured by CMS~\cite{CMS:2013cka}. Right:
    $R_\mathrm{pPb}$ for inclusive full jets as a function of jet
    $p_\mathrm{T}$, with each panel at a fixed center of mass rapidity
    $y^*$, as measured by ATLAS~\cite{ATLAS-CONF-2014-024}.
    \label{fig:LHC_pPb_hadrons}}
\end{figure}

Recent LHC data on jet and hadron production in $p$+Pb collisions at
5.02~TeV has revealed several striking and unexpected
results. Figure~\ref{fig:LHC_pPb_hadrons} compares the
$R_\mathrm{pPb}$ for charged particles and jets at mid-rapidity in
minimum bias $p$+Pb collisions. While the overall jet rate is in line
with the geometric expectation~\cite{ATLAS-CONF-2014-024} (modulo some
small effects from the modification of the parton distribution
functions), the charged particle $R_\mathrm{pPb}$ shows an unexpected
rise above the geometric
expectation~\cite{CMS:2013cka,ATLAS-CONF-2014-029}. This anomalous
$R_\mathrm{pPb}$ presents a strong challenge to nPDF-based pictures of
$p$+A collisions, and its implications on the
$p_\mathrm{T}$-dependence of the charged hadron $R_\mathrm{AA}$
observed in Pb+Pb collisions is still an open question.

sPHENIX is situated in a prime kinematic region to investigate this
effect. As shown in Figure~\ref{fig:pAphysics_projections}, sPHENIX
will have excellent statistics for charged hadrons and jets in a large
overlapping $p_\mathrm{T}$ range, allowing for simultaneous
measurements of the jet and hadron $R_\mathrm{pAu}$. In particular,
sPHENIX will be able to measure the charged particle spectrum in
$p$+Au in the $20$--$40$~GeV/$c$ range, where the anomalous
$R_\mathrm{pPb}$ at the LHC has the strongest $p_\mathrm{T}$
dependence. Furthermore, unlike the LHC experiments, sPHENIX will be
able to benefit from $p$+$p$ reference data at the same collision
energy and center of mass frame as the $p$+Au data, resulting in
potentially smaller systematic uncertainties.

\begin{figure}[hbt!]
  \centering
  \includegraphics[width=0.63\textwidth]{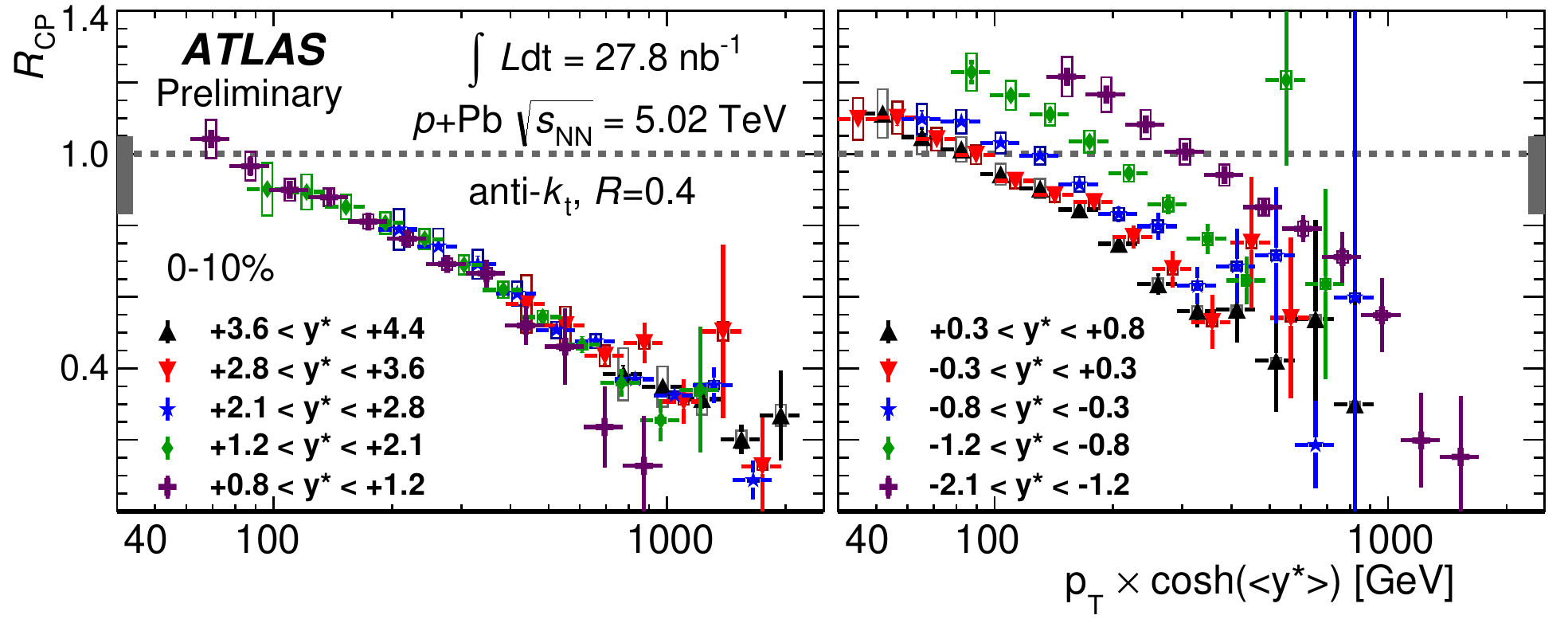}
  \includegraphics[width=0.36\textwidth]{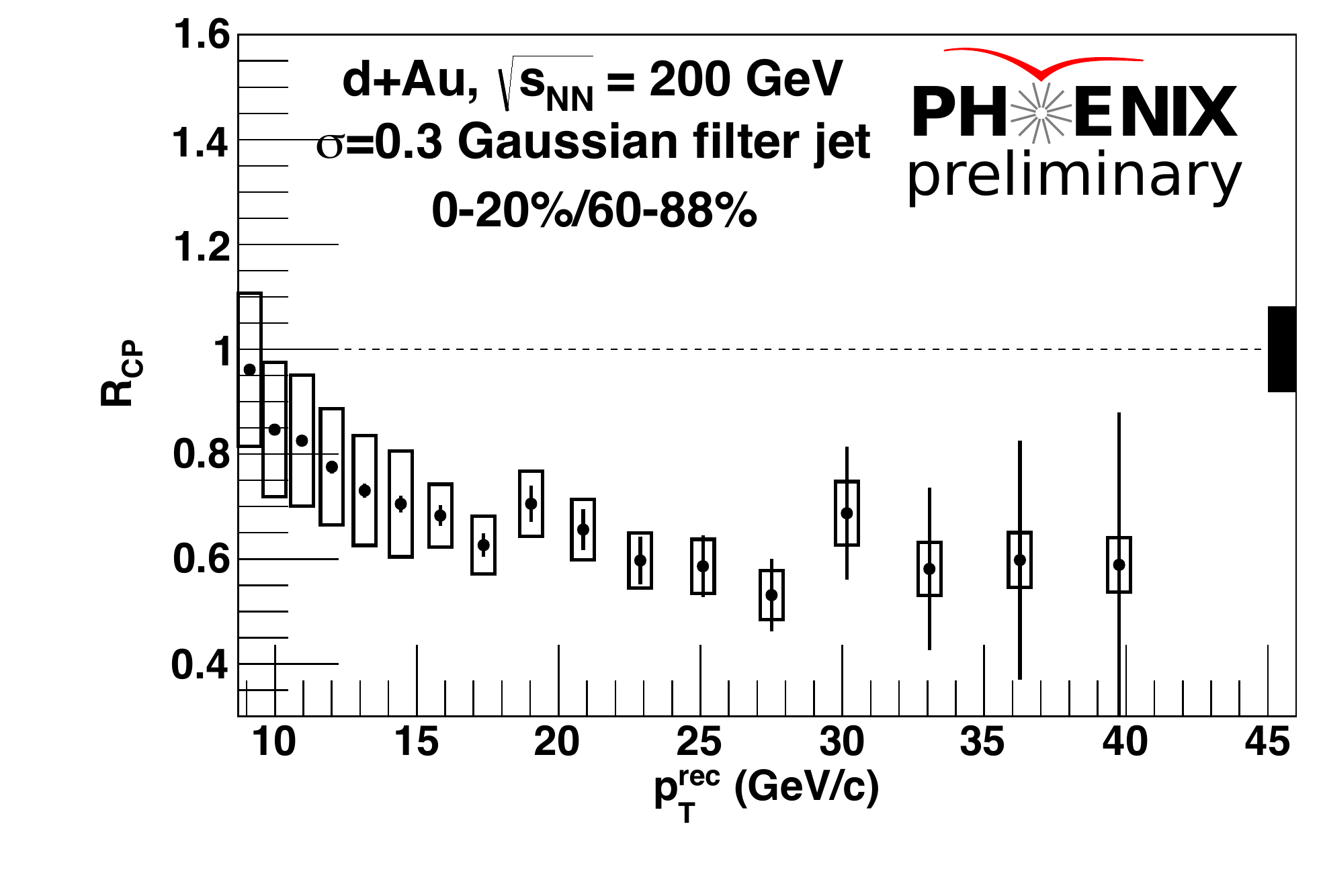}
  \caption[$R_\mathrm{CP}$ for inclusive
  jets vs approximate total jet energy $p =
  p_\mathrm{T} \times \cosh(y^*)$, for
 forward  and mid-rapidity  regions, as
  measured by ATLAS, and 
  for inclusive jets at mid-rapidity as a function of the detector-level
  jet $p_\mathrm{T}$ as measured by PHENIX]{left and center: Central to
    peripheral ratio ($R_\mathrm{CP}$) for inclusive jets shown as a
    function of the approximate total jet energy $p = p_\mathrm{T}
    \times \cosh(y^*)$, showing multiple $y^*$ selections in the forward
    (left plot) and mid-rapidity (center plot) regions, as measured by
    ATLAS~\cite{ATLAS-CONF-2014-024}. Right: $R_\mathrm{CP}$ for
    inclusive jets at mid-rapidity as a function of the detector-level
    jet $p_\mathrm{T}$ as measured by PHENIX~\cite{Perepelitsa:2013jua}.
    \label{fig:LHC_pPb_jets}}
\end{figure}

The LHC data have also revealed a striking pattern of jet modification
with respect to the $p$+Pb event centrality. Measurements of inclusive
jet production at ATLAS~\cite{ATLAS-CONF-2014-024} and dijet
production at CMS~\cite{Chatrchyan:2014hqa} have shown an unexpected
excess of jets and dijets in apparently peripheral collisions and a
suppression in apparently central ones. Furthermore, these effects were
found to be systematically larger in the forward (downstream proton)
direction, where any biases on the centrality determination (typically
characterized with detectors in the downstream Pb direction) from the
presence of the hard process are expected to be systematically
smaller. The left panels of Figure~\ref{fig:LHC_pPb_jets} shows a
kinematic pattern discovered by ATLAS in the suppression of the
central to peripheral ratio $R_\mathrm{CP}$ of these jet spectra. In
the figure, the $R_\mathrm{CP}$ over a wide rapidity range is seen to
only depend on the total jet energy $p$. An analogous effect had been
previously observed at mid-rapidity in a preliminary measurement of
jets in $d$+Au by PHENIX~\cite{Perepelitsa:2013jua}, shown in the
right panel of Figure~\ref{fig:LHC_pPb_jets}. 

sPHENIX is poised to investigate these effects using full jets over a
large kinematic range. Although the projected $R_\mathrm{pAu}$ in
Figure~\ref{fig:pAphysics_projections} is for minimum bias $p$+Au
events, the centrality-selected $R_\mathrm{pAu}$ will have only
modestly larger statistical uncertainties. This will be important for
investigating the $p_\mathrm{T}$ dependence of any jet modification,
which the LHC results imply systematically grows with the total jet
energy. Furthermore, due to the large dijet containment fraction
within the sPHENIX acceptance, measurements of dijet production, which
have a better connection to the hard scattering kinematics $(x_p,
x_\mathrm{Au}, Q^2$), will also be possible. More generally, the
nature of the correlations between hard processes and the soft
particle production which gives the centrality signal at forward
rapidities in $p$+$p$ and $p$+Au collisions can be systematically
studied with high statistics.

Finally, the $b$-jet tagging capabilities being developed at sPHENIX
will allow for a measurement of the $b$-jet $R_\mathrm{pAu}$ (shown in
Figure~\ref{fig:pAphysics_projections}) as a benchmark against which
to interpret the $b$-jet $R_\mathrm{AA}$, analogous to the recent
measurement by CMS~\cite{CMS:2014tca}.

\clearpage

\section{Jet physics at lower RHIC energies}
\label{sec:fake_jets_100GeV}

\begin{figure}
  \centering
  \includegraphics[width=0.48\linewidth]{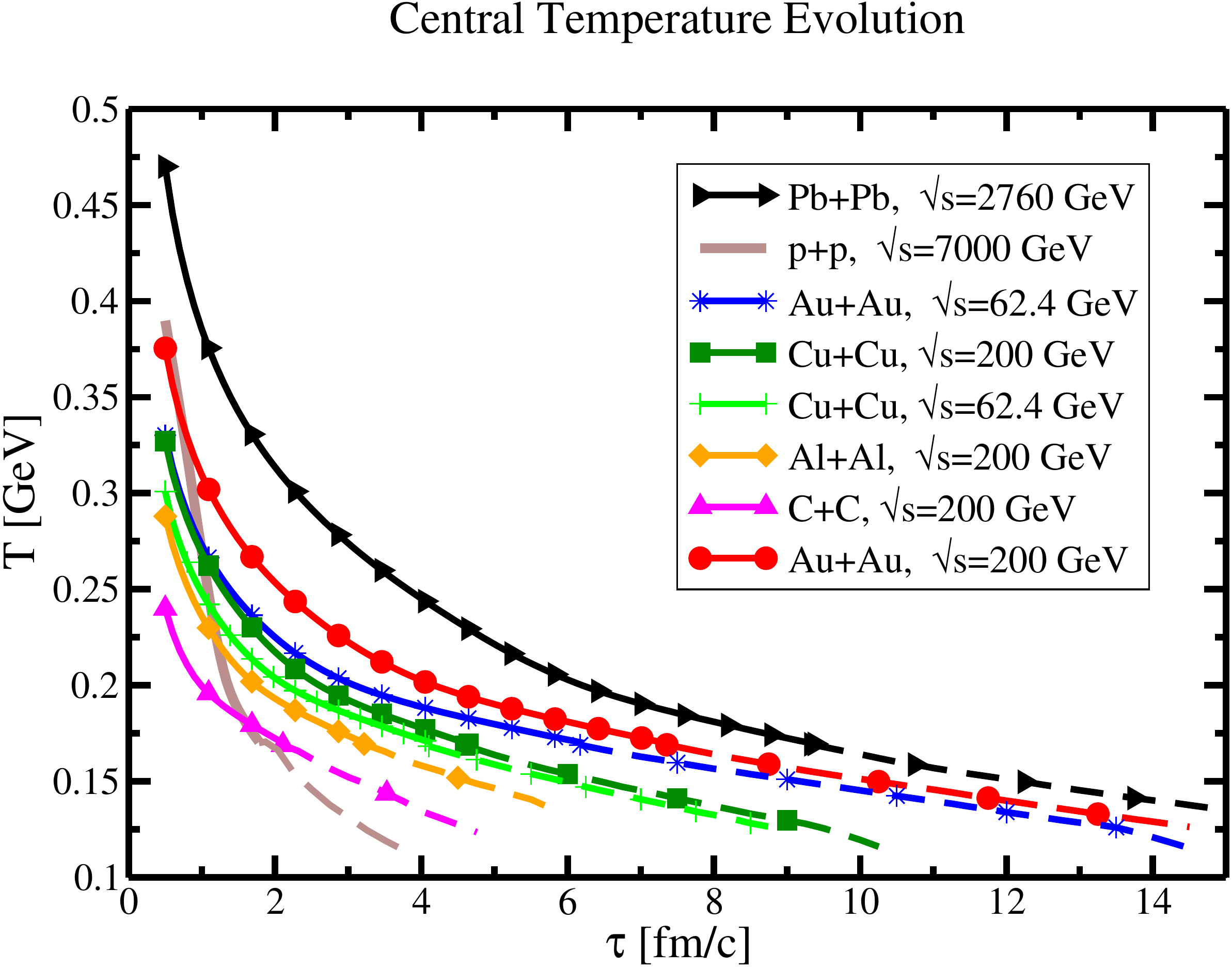}
  \hfill
  \includegraphics[width=0.48\linewidth]{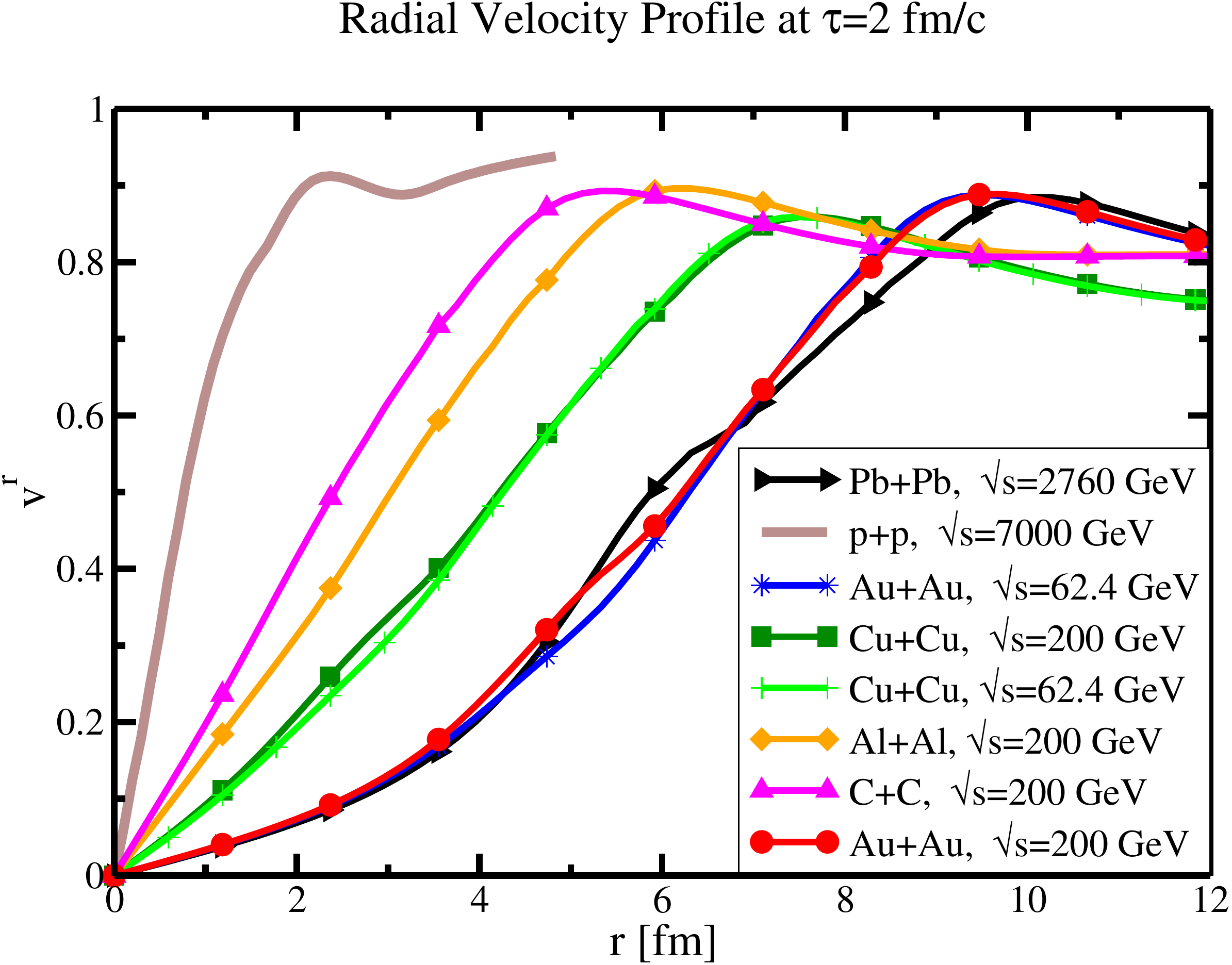}
  \caption[Temperature profile vs time from a full dynamical calculation
  for various collision species and energies, and results for the
  transverse flow velocity profile at time $\tau = 2$~fm/$c$.]{(left)
    Temperature profile for the central $r=0$ cell as a function of time
    from full dynamical calculations~\cite{Habich:2014jna} for various
    collision species and energies.  (right) The same calculation with
    results for the transverse flow velocity profile at time $\tau =
    2$~fm/$c$. }
  \label{fig:hydro_other}
\end{figure}

If additional running time becomes available and if physics
investigations indicate interest in this direction, there is the
potential for extending sPHENIX jet measurements to lower energies and with
lighter ions at RHIC.  Shown in Figure~\ref{fig:hydro_other} are the temperature
profile for the central cell (left) and the flow velocity profile at $\tau=$ 2 fm/c 
for various collision species and energies~\cite{Habich:2014jna}.
The calculations including pre-equilibrium dynamics, viscous hydrodynamics, and
hadronic cascade are for the central cell.  These results indicate that the
lever arm in temperature and medium dynamics that can be extended with lower
energy collisions and careful selection of light ion collisions.  Measurements of
jet observables in light ions at $\sqrt{s_{NN}} = 200$~GeV with statistics comparable to the \auau
results previously discussed can be obtained in 10 weeks of physics running, and
have significantly lower underlying event backgrounds.  

As an example, for lower energy running, in a 20 week running period, one can sample 10 billions \auau
events at $\sqrt{s_{NN}} = 100$~GeV.  Although the background
multiplicity in these events is lower than in corresponding collisions
at $\sqrt{s_{NN}} = 200$~GeV, the true jet spectrum at the lower
collision energy is steeper.  
NLO calculations have been performed and projected for the most central 20\% \auau collisions at 
$\sqrt{s_{NN}} = 100$~GeV.  The projected
luminosity delivered by the collider is lower than in the 200~GeV
case, and one still can obtain a substantial sample of jets
reaching out to 35~GeV.
We have performed simulations to
demonstrate that we can reconstruct jets in this environment.

\begin{figure}
  \centering
  \raisebox{0.09in}{\includegraphics[trim = 200 0 0 0, clip, width=0.48\linewidth]{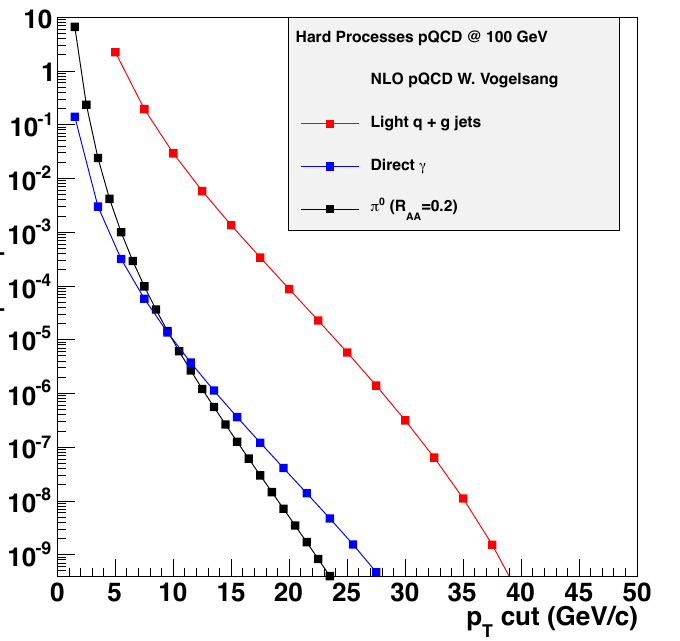}}
  \includegraphics[width=0.48\linewidth]{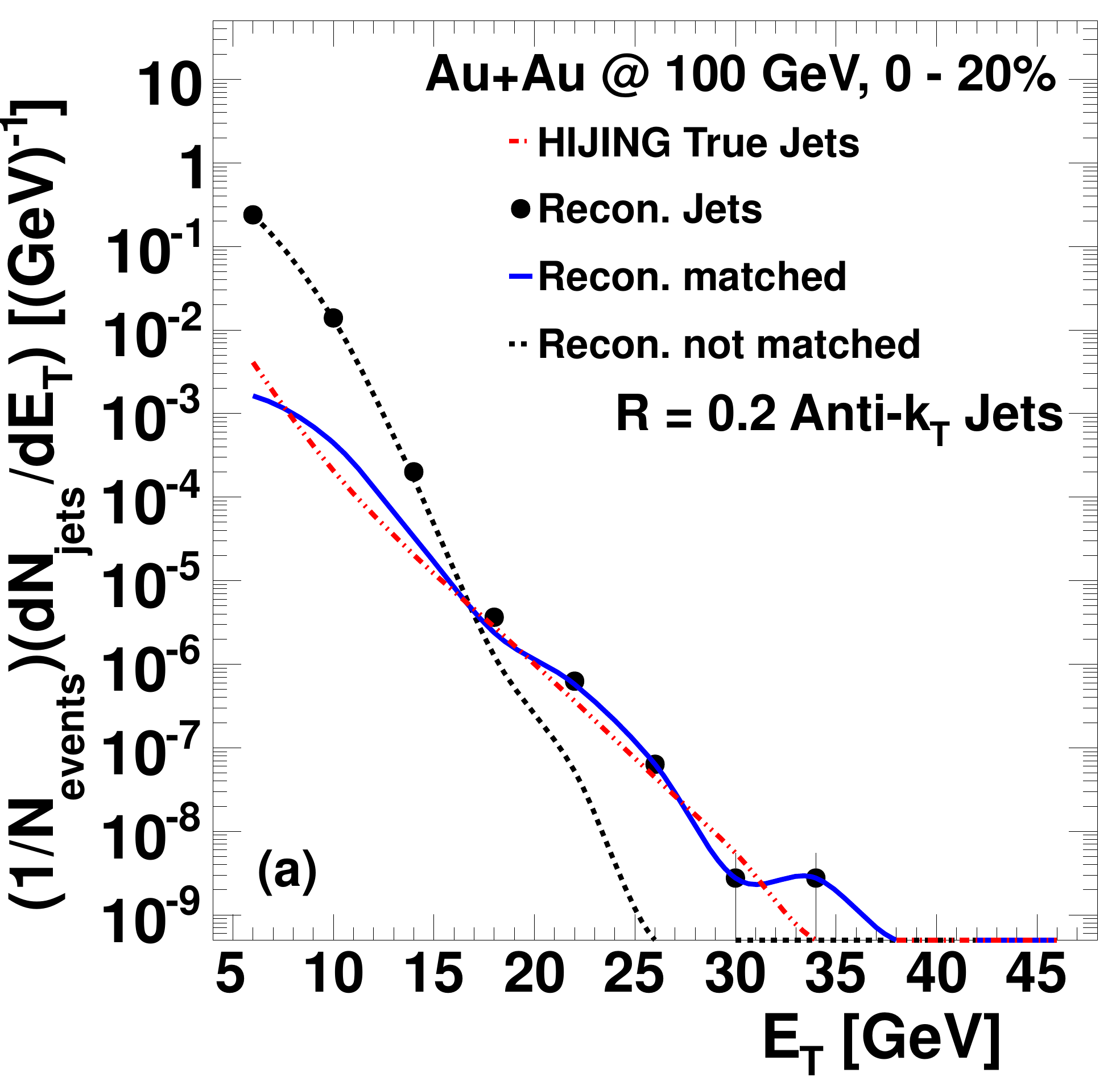}
  \caption[Jet, photon and $\pi^{0}$ rates with $|\eta|<1.0$ from NLO
  pQCD]{(left) Jet, photon and $\pi^{0}$ rates with $|\eta|<1.0$ from
    NLO pQCD~\protect\cite{Vogelsang:NLO}{}. A nominal 20 week RHIC run
    corresponds to 1.7 billion central \auau events at $\sqrt{s_{NN}} =
    100$~GeV.  (right) Results of a fake jet study at $\sqrt{s_{NN}} =
    100$~GeV for the most central 20\% of the cross section.  The
    \antikt algorithm with $R = 0.2$ was used to reconstruct jets.  True
    jets dominate over fake jets for $E_T > 20$~GeV.}
  \label{fig:fake_jets_100GeV}
\end{figure}

A procedure identical to that used for evaluating the jet finding
performance at the top RHIC energy was followed to evaluate the jet
finding performance for \auau collisions at $\sqrt{s_{NN}} =
100$~GeV.  A sample of 400 million \hijing events at the lower
collision energy was generated and the procedure of
Section~\ref{sec:fake_jets} was employed.  The results are shown in
Figure~\ref{fig:fake_jets_100GeV}.  The effects of the steeper jet
spectrum and of the reduced multiplicity at the lower collision energy
largely negate one another, and the true jet signal dominates
over the background at transverse energies quite similar to that seen
for $\sqrt{s_{NN}} = 200$~GeV.

\section{Jet Physics Summary}
\label{sec:jet_performance_summary}

Overall we conclude that a robust jet, dijet, $\gamma$-jet, fragmentation function program
with high statistics is achievable with the sPHENIX detector upgrade.
These observables indicate excellent discriminating ability between
scenarios with different medium coupling strengths and jet quenching
mechanisms.

\begin{figure}[hbt!]
  \centering
  \includegraphics[width=0.9\textwidth]{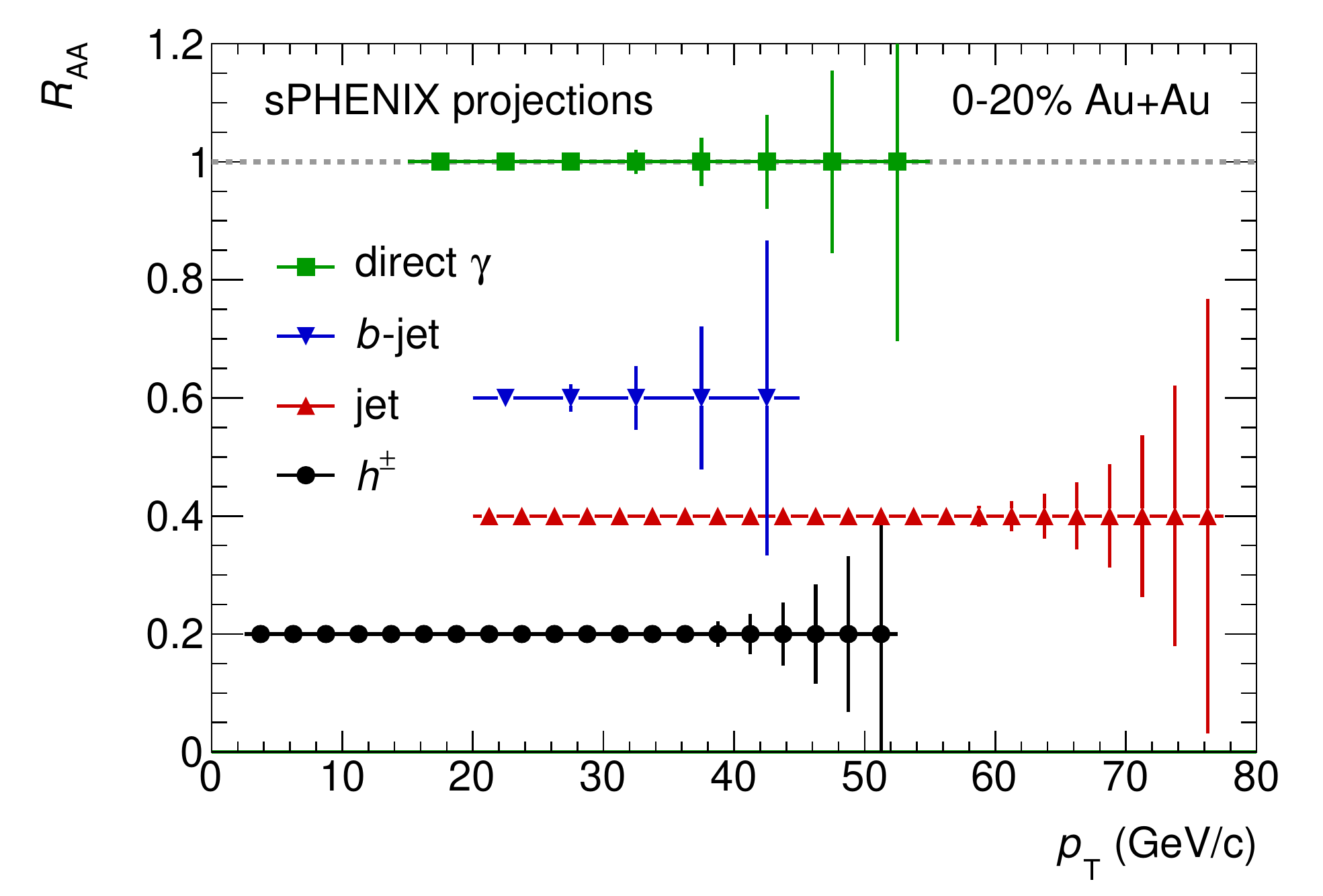}
  \caption[Projected statistical uncertainties on $R_\mathrm{AA}$
    for inclusive photons, $b$-jets,
    inclusive jets and
    charged hadrons  ]{Projected statistical uncertainties on the $R_\mathrm{AA}$
    for inclusive photons (green points, assuming $R_\mathrm{AA} =
    1$), $b$-jets (blue points, assuming $R_\mathrm{AA} = 0.6$),
    inclusive jets (red points, assuming $R_\mathrm{AA} = 0.4$) and
    charged hadrons (black points, assuming $R_\mathrm{AA} =
    0.2$). These projections are made with a $b$-jet tagging
    efficiency of $50$\%, 10 weeks of $p$+$p$ and 22 weeks of \auau
    data taking.
    \label{fig:AAphysics_projections}}
\end{figure}

Figure~\ref{fig:AAphysics_projections} shows the projected statistical
uncertainties for the $R_\mathrm{AA}$ of single inclusive hard probes
that will be possible within the first two years of sPHENIX
operations. This figure demonstrates that the large acceptance and
physics capabilities of the sPHENIX detector, along with the projected
performance of RHIC, would allow measurements of jet quenching over a
very large kinematic kinematic. sPHENIX would have enough statistics to
measure the overall suppression of inclusive jets out to
$\approx70$--80~GeV/$c$, single hadrons and photons out to
$\approx50$~GeV/$c$, and $b$-tagged jets to $40$~GeV/$c$ in 0--20\%
\auau events.

The statistical uncertainties in mid-central and mid-peripheral
centrality selections (such as $30$--$50$\% collisions) would only be
modestly larger, allowing for precise differential measurements such
as the $R_\mathrm{AA}$ as a function of the angle with respect to the
reaction plane. Finally, it should be noted that the statistical
uncertainty in the $R_\mathrm{AA}$ is limited by the $p$+$p$, not the
\auau, luminosity. Thus, any increase in the nominal $p$+$p$ dataset
of 10 weeks would only serve to improve the $p_\mathrm{T}$ reach of
all $R_\mathrm{AA}$ measurements, and measurements which rely only on
the \auau data would potentially have an even larger kinematic reach.

\clearpage
\pagebreak 

\section{Beauty Quarkonia Performance}
\label{sec:quarkonia_signal}
\makeatletter{}
Measurements of beauty quarkonia, the Upsilon states, provide critical
information by probing the \qgp at different length scales determined
by the size of the state.  Here we detail the sPHENIX measurement
capabilities to access this physics with precision.

We first report on the expected yield and line shape of the
$\Upsilon$(1s), $\Upsilon$(2s) and $\Upsilon$(3s) signal from decays
to dielectrons. The results were obtained with single simulated
$\Upsilon$ events in a \geant simulation with the reference sPHENIX
silicon tracking configuration inside the BaBar solenoid.

The baseline \pp cross section for $\Upsilon(1S+2S+3S)$ of $B_{ee}
d\sigma/dy_{|y=0} = 108 \pm 38(stat) \pm 15(sys) \pm 11$(global)~pb is
taken from a PHENIX central arm measurement~\cite{Adare:2014hje}. The
rapidity dependence was taken from \pythia. The relative yields of the
three $\Upsilon$ states were taken from CDF measurements at
1.8\,TeV~\cite{Acosta:2001gv}.  Estimates of the $\Upsilon$ yields for
the three states in sPHENIX are shown in
Table~\ref{tab:upsilon_yields} for a 10 week \pp run, a 22 week \auau
run and a 10 week \pAu run. These projections assume binary scaling
and no suppression of any of the $\Upsilon$ states in \auau or \pAu
collisions. The yields are calculated using the electron
identification efficiencies shown in the table, which are dependent on
the event multiplicity. The electron pair identification efficiency of
49\% in central \auau collisions corresponds to a hadron rejection of
90 in the EMCal and HCal, based on a full GEANT4 study of central
Hijing events.

The \auau data sample consists of 100~B minimum bias events. The \pp
and \pAu data for Upsilons will be obtained using a trigger.  The
trigger efficiency at the needed rejection is estimated to be $>$90\%
for \pp events. A trigger efficiency of 90\% is assumed for central
\pAu events.

\renewcommand{\arraystretch}{1.9}
\addtolength{\tabcolsep}{-0.5pt}

\begin{table}[hbt!]
  \centering
  \caption[The yields of the three $\Upsilon$ states obtained in 10 weeks of \pp,
  22 weeks of \auau and 10 weeks of \pAu  RHIC running]{The yields of
    the three $\Upsilon$ states obtained in 10 weeks of \pp, 22 weeks
    of \auau and 10 weeks of \pAu  RHIC running. All yields include
    the effect of electron identification efficiency. The numbers for
    \auau and \pAu are calculated assuming no suppression of any of
    the $\Upsilon$ state yields.}   
  \label{tab:upsilon_yields}
\begin{tabular}{cccccccc} 
\toprule
Species & $\mathbf{\displaystyle\int L\,dt (|Z|< 10 cm)}$  & Events & 
$\left<N_{\mathrm coll}\right>$ & eID eff. & $\Upsilon$(1S) & $\Upsilon$(2S) & $\Upsilon$(3S) \\
\midrule
\pp 	&   	175 $pb^{-1}$ &  7350 B	& 	1	& 0.9 &   8770  & 2205  & 1155 
\\
\auau (MB) & 	& 100 B		&    240.4  & 0.57    &   16240  &   4080 & 2140 
\\
\auau  (0--10\%) & 		& 10 B		&     962     & 0.49 &    5625 	&  1415     &  740
\\
$p$$+$Au (MB) &   960 $nb^{-1}$	&  1680 B       & 	   4.3     & 0.84 &    6560 	&  1650     &  860 
\\
$p$$+$Au (0--20\%) &   & 336 B	&   8.2     & 0.8 &   2360 &  592     &  311 
\\
\bottomrule
\end{tabular}
\end{table}

A critical question is whether the proposed tracking system is capable
of adequately resolving the $\Upsilon$(1S), $\Upsilon$(2S) and
$\Upsilon$(3S) states from each other. To answer this we have
performed \geant simulations with the reference tracking
configuration, containing seven silicon tracking layers covering
$2\pi$ in azimuth and two units in pseudorapidity, with specifications
given in Table~\ref{table:tracker}.

\begin{figure}
  \begin{center}
    \includegraphics[width=0.8\textwidth]{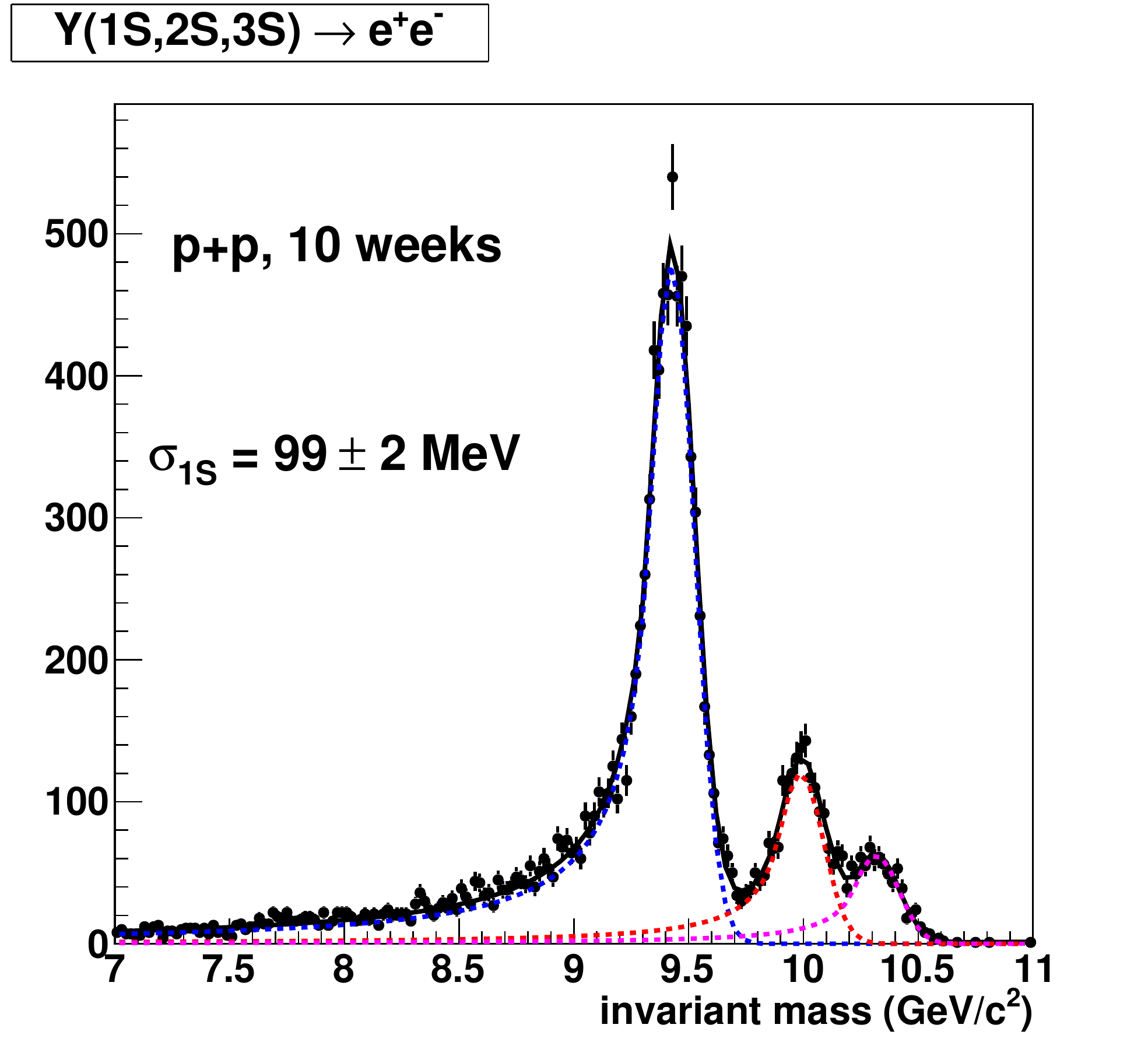}
  \end{center}
  \caption[The mass spectrum for reconstructed electron decay tracks
  from the three Upsilon states, corresponding to 10 weeks of \pp
  running, including the effects of electron identification efficiency
  and trigger efficiency. ]{\label{fig:quarkonia_upsilonlineshape} The
    mass spectrum (signal only) from reconstructed electron decay
    tracks for the three Upsilon states combined.  The yield
    corresponds to that for 10 weeks of \pp running, including the
    effects of electron identification efficiency and trigger
    efficiency.  }
\end{figure}

The reconstructed mass spectrum for dielectron decays (signal only) is
shown in Figure~\ref{fig:quarkonia_upsilonlineshape}, including a fit
using a Crystal Ball function that accounts for the radiative tail
contribution at low invariant mass~\cite{Gaiser:1982yw}.  This example
spectrum contains the number of Upsilons expected in 10 weeks of \pp
running.  There are significant low mass tails on the Upsilon mass
peaks due to radiative energy loss in the material of the silicon
tracker.  However at the mass resolution of 99 MeV obtained with the
reference design, and the relatively low thickness of the tracker
(about 10\% of a radiation length), the peaks are well defined and
easily obtained from the Crystal Ball fit.

In \pp, \pAu and \auau, the background under the Upsilon peaks
contains an irreducible physics background due to dileptons from
correlated charm, correlated bottom and Drell-Yan. There is also
combinatorial background from misidentified charged pions. The latter
can be estimated and removed by like sign or mixed event subtraction.
To study the physics background, correlated charm and bottom
di-electron invariant mass distributions predicted by \pythia were
normalized to the PHENIX measured charm and bottom cross-sections in
\auau collisions.  The \pythia Drell-Yan di-electron invariant mass
distribution was normalized to a theoretical prediction by
W. Vogelsang (private communication).

The combinatorial background was studied by generating events with
fake electrons due to misidentified pions, using input pion
distributions taken from PHENIX measured $\pi^0$ spectra in \auau collisions.
A $p_T$-independent rejection factor was applied to the $\pi^{+/-}$
spectra to imitate fake electron spectra. For the 0--10\% most central 
Au+Au collisions a rejection factor of 90 is assumed at a single electron track
efficiency of 70\% (giving a pair efficiency of 49\%). The pair efficiency is 
increased to 90\% as \auau collisions become more peripheral. The combinatorial 
background due to misidentified pions is assumed here to be zero in \pp collisions,
with an electron matching efficiency of greater than 90\%. 
The rejections in central \auau collisions are derived from \geant studies of
the electromagnetic calorimeter response to electrons and charged pions. The efficiencies
are obtained by embedding electrons in HIJING events.
The rejection and efficiency are still being optimized for the detector configuration 
relevant for electron identification.

\begin{figure}
  \begin{center}
    \includegraphics[width=0.50\textwidth]{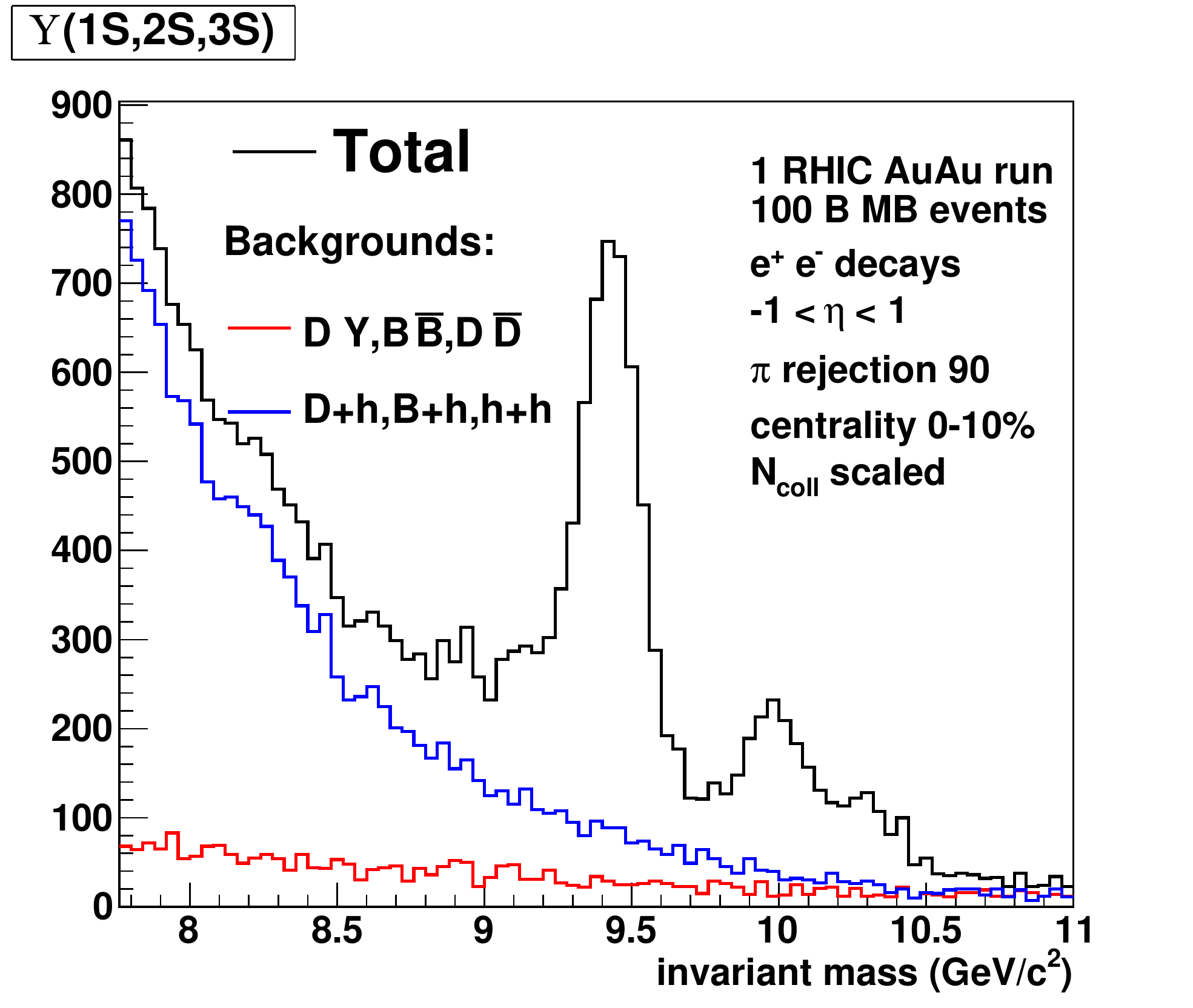}
    \includegraphics[width=0.46\textwidth]{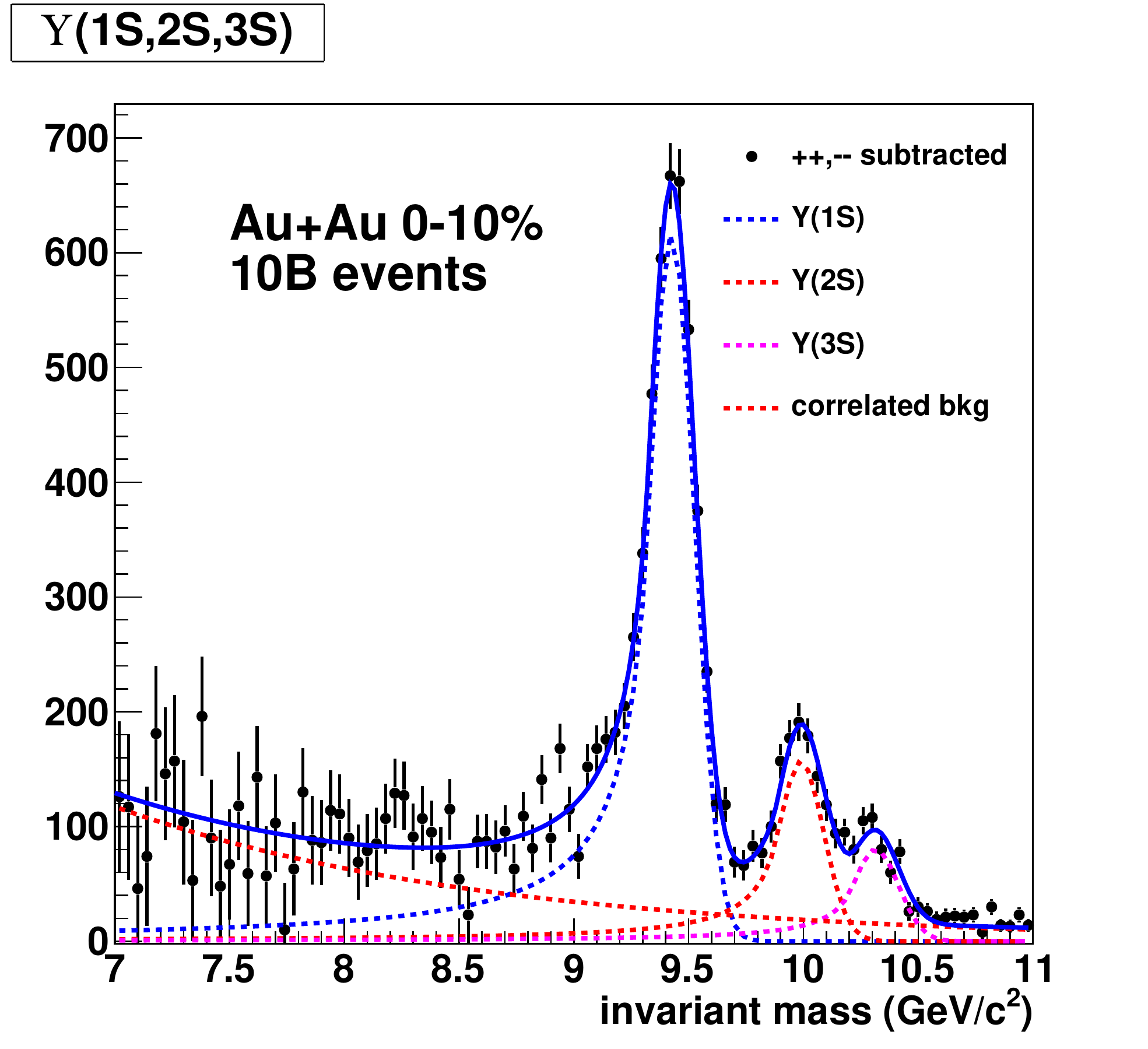}
  \end{center}
  \caption[The signal plus background in the Upsilon mass region for
  ten billion 0--10\% central \auau events, assuming a pion rejection
  of 90:1 and an electron pair identification efficiency of
  49\%]{\label{fig:quarkonia_bg} (Left) The signal plus background in
    the Upsilon mass region for ten billion 0--10\% central \auau
    events, assuming a pion rejection factor of 90, with the signal
    reduced by a pair identification efficiency of 49\%.  The combined
    backgrounds due to correlated bottom, correlated charm, and
    Drell-Yan are shown as the red curve. The combined backgrounds due
    to fake electrons combining with themselves, bottom, and charm are
    shown as the blue line.  (Right) The expected invariant mass
    distribution for ten billion 0--10\% central \auau events, after
    subtraction of combinatorial background using the like-sign
    method.  The remaining background from correlated bottom, charm
    and Drell-Yan is not removed by like sign subtraction. It must be
    estimated and subtracted.  }
\end{figure}

All combinations of fake electrons from misidentified pions were made
with each other, and with high $p_T$ electrons from physics sources.
The combinatorial background is found to be dominated by pairs
of misidentified pions, with only 30\% or so coming from combinations
of misidentified pions with electrons.
The results are summarized in Figure~\ref{fig:quarkonia_bg} (left), which
shows the signal + background in the $\Upsilon$ mass region for the ten billion
0--10\% most central events, along with our estimates of the total
correlated physics background and the total uncorrelated
combinatoric backgrounds.  In Figure~\ref{fig:quarkonia_bg} (right) we
show the di-electron invariant mass distribution for ten billion 0--10\%
central \auau events after the combinatorial background has been
removed by subtracting all like-sign pairs.

\begin{figure}
  \begin{center}
    \includegraphics[width=0.6\textwidth]{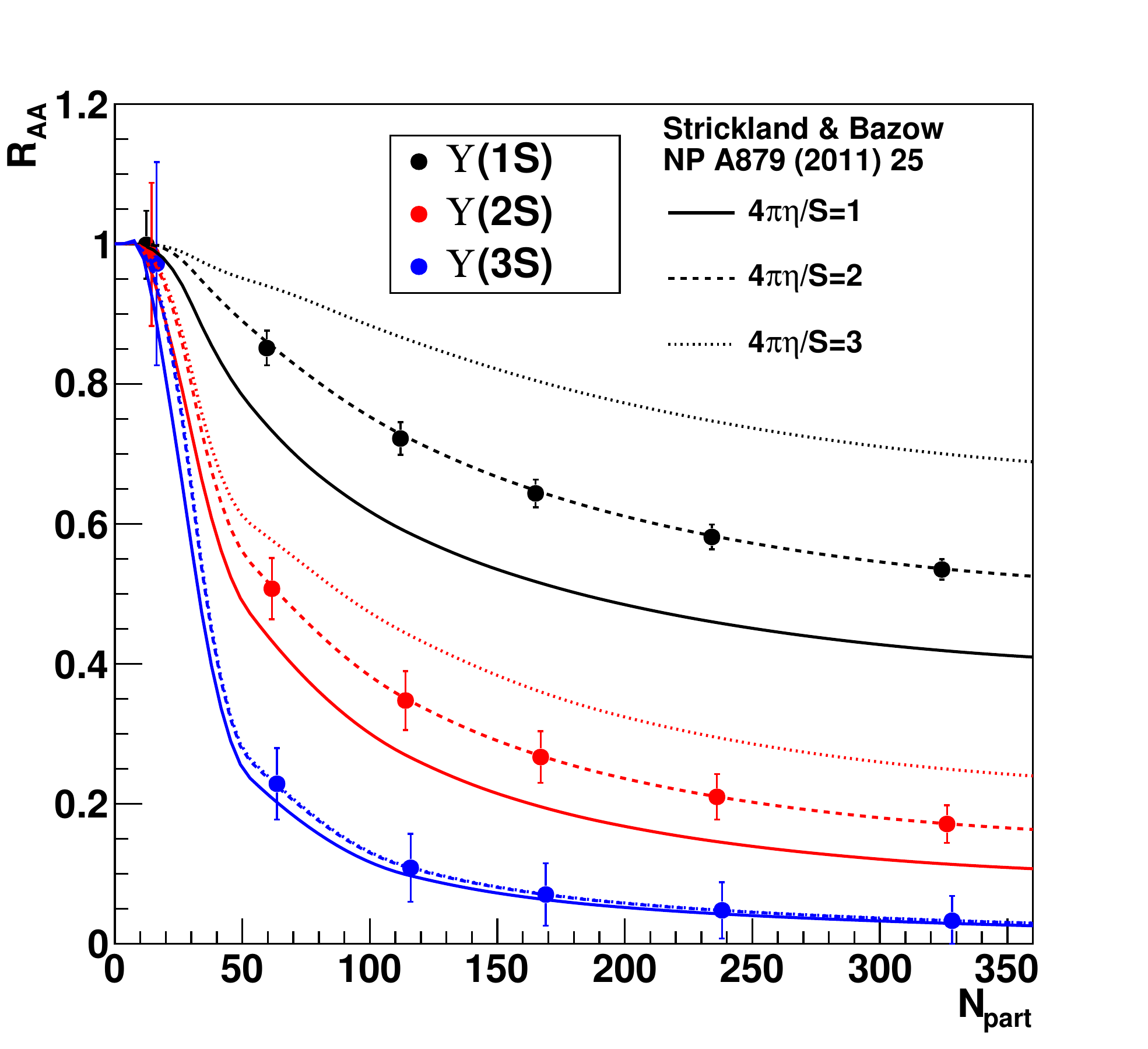}
  \end{center}
  \caption[Estimate of the statistical precision of a measurement of
  the $\Upsilon$ states in \auau collisions using sPHENIX, assuming
  that the measured $R_{AA}$ is equal to the results of a recent
  theory calculation]{\label{fig:quarkonia_upsilon_raa_statistics}
    Estimate of the statistical precision of a measurement of the
    $\Upsilon$ states in \auau collisions using sPHENIX, assuming that
    the measured $R_{AA}$ is equal to the results of a recent theory
    calculation~\cite{Strickland:2011aa}. The yields assume 100
    billion recorded \auau events.  }
\end{figure}

From Figure~\ref{fig:quarkonia_bg} (left) we estimate that without
$\Upsilon$ suppression the S/B ratios are $\Upsilon$(1S): 1.6,
$\Upsilon$(2S): 0.9, and $\Upsilon$(3S): 0.8 for central \auau collisions.  
Using our estimates of the signal and S/B ratio at each centrality as the unsuppressed 
baseline, we show in
Figure~\ref{fig:quarkonia_upsilon_raa_statistics} the expected
statistical precision of the measured $R_{AA}$ for 100 billion recorded \auau
events assuming that the suppression for each state
is equal to that from a theory 
calculation~\cite{Strickland:2011aa}.  For each state, at each value
of $N_\mathrm{part}$, both the $\Upsilon$ yield and the S/B ratio were
reduced together by the predicted suppression level.

\begin{figure}
  \begin{center}
    \includegraphics[width=0.6\textwidth]{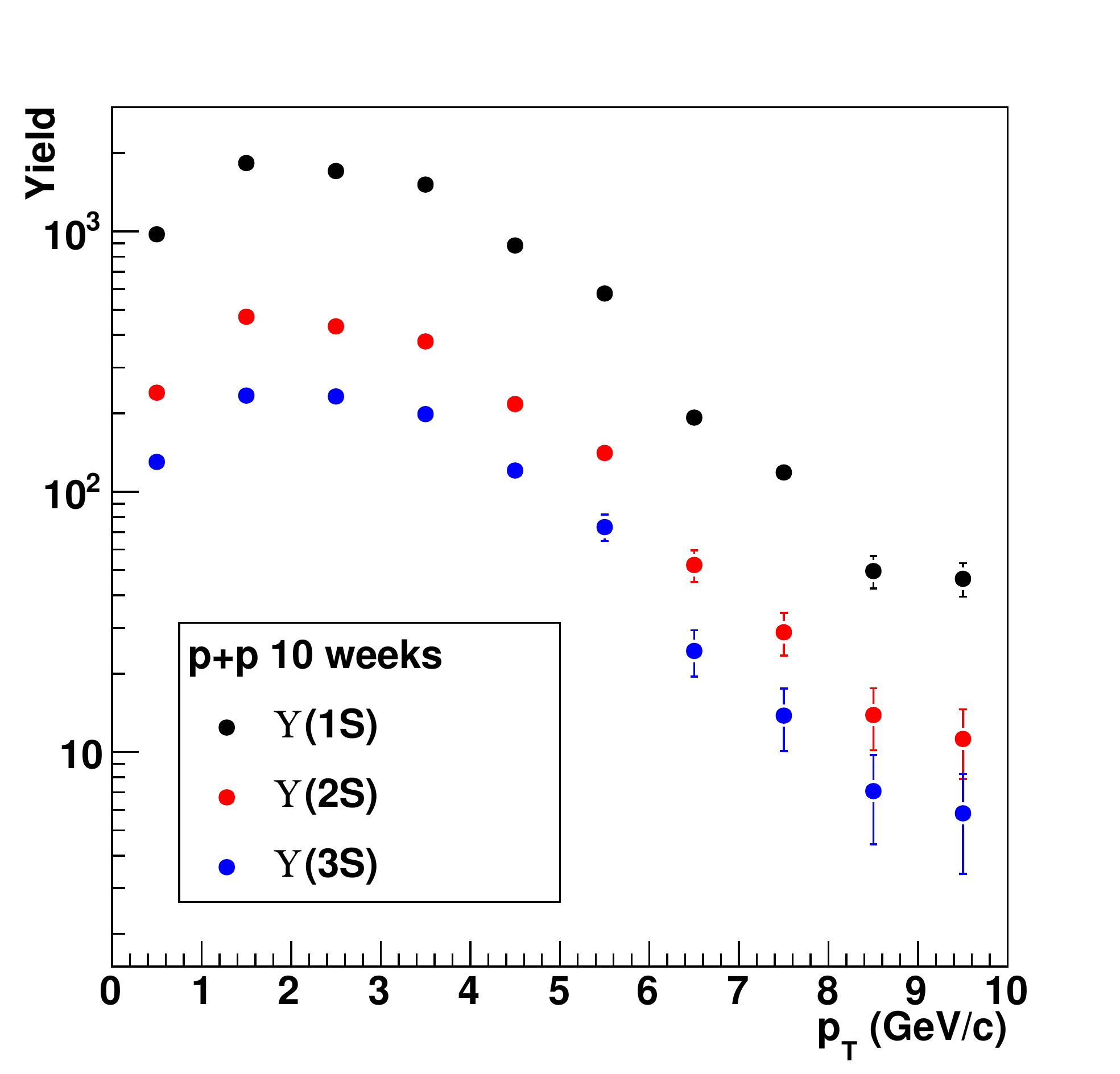}
  \end{center}
  \caption{\label{fig:upsilon_pp_pT_yields} Estimate of the yields
    expected for the three $\Upsilon$ states as a function of \pt from
    a 10 week \pp run.}
\end{figure}

\begin{figure}
  \begin{center}
    \includegraphics[width=0.45\textwidth]{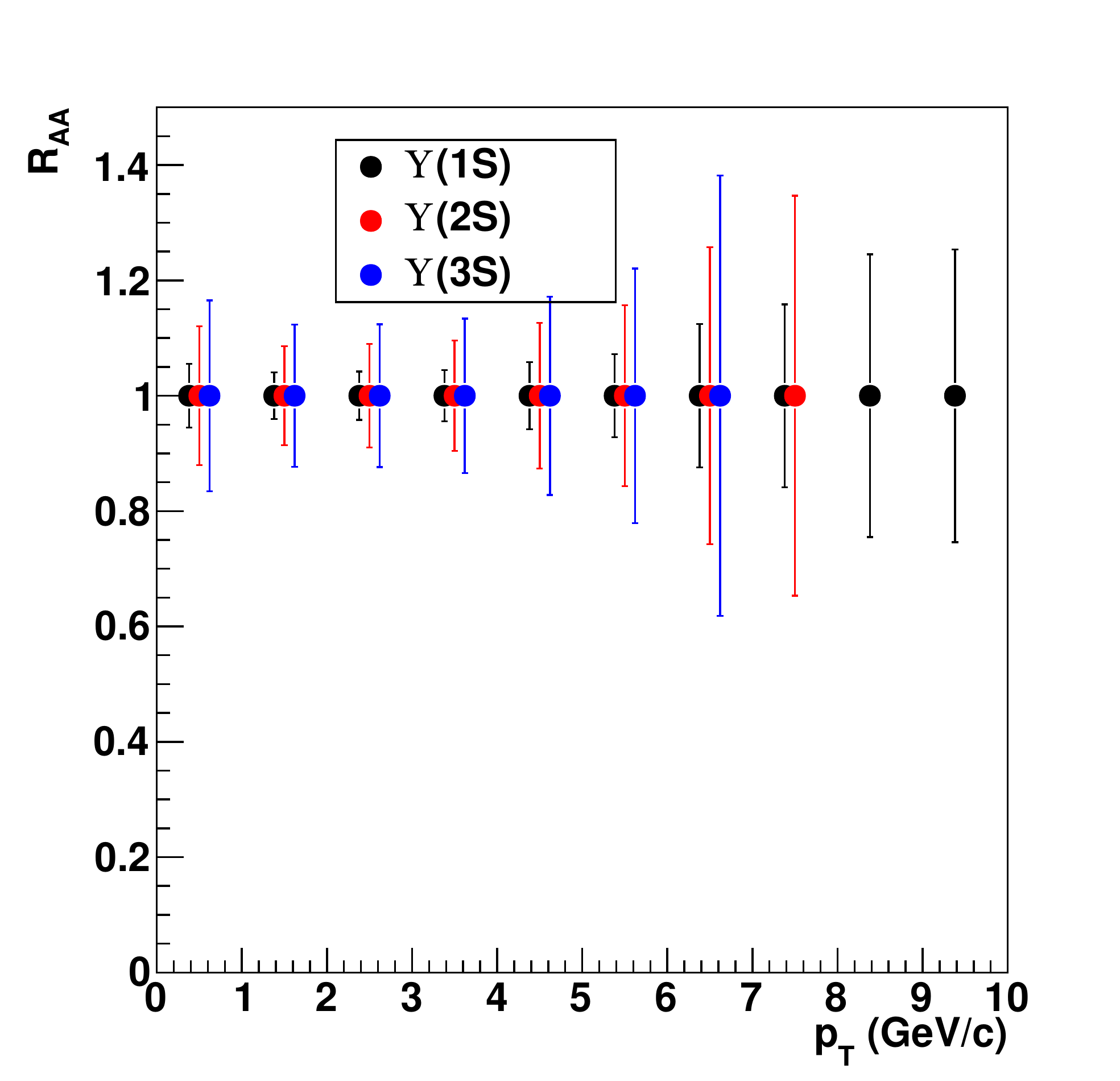}
    \includegraphics[width=0.45\textwidth]{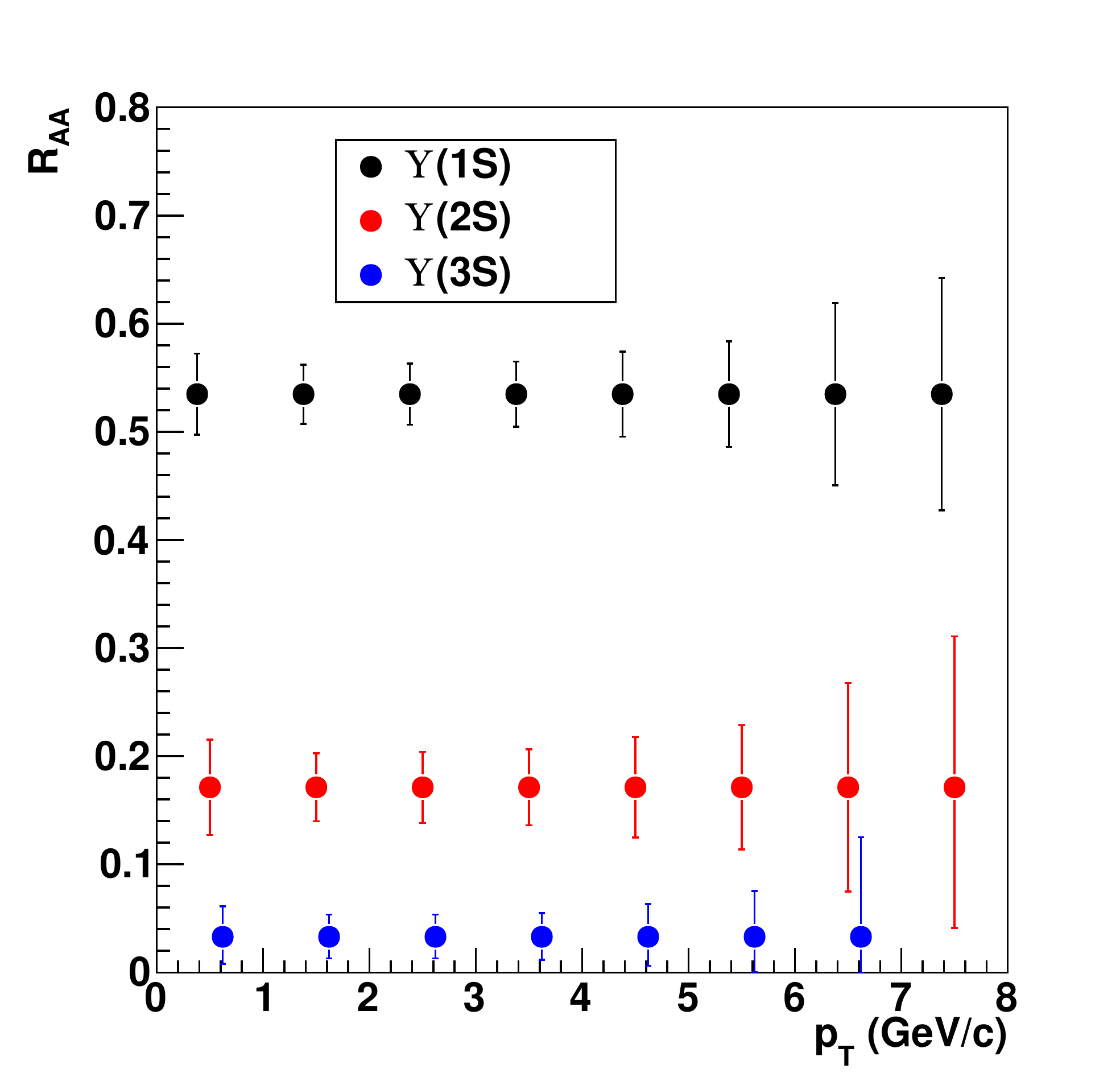}
  \end{center}
  \caption[Estimate of the statistical precision of a measurement of
  $R_{AA}$ versus \pt for the $\Upsilon$ states using sPHENIX, for the
  most central 0--10\% of events]{\label{fig:quarkonia_upsilon_raa_pt}
    Estimate of the statistical precision of a measurement of $R_{AA}$
    versus \pt for the $\Upsilon$ states using sPHENIX, for the most
    central 0--10\% of events.  The left panel shows the result if
    there is no suppression, the right panel shows the result
    assuming that the measured $R_{AA}$ is equal to the theory
    results in~\cite{Strickland:2011aa}. The yields assume 100 billion
    recorded \auau events.  }
\end{figure}

The \pt dependence of the $\Upsilon$ modification in nuclear
collisions places strong constraints on models, so we present here
some estimates of the statistical precision we expect from
measurements with sPHENIX.  Figure~\ref{fig:upsilon_pp_pT_yields}
shows the expected yields as a function of \pt for 10 weeks of \pp
running --- the baseline for the $R_{AA}$ measurement.  The expected
statistical precision of the measured \auau $R_{AA}$ versus \pt is
illustrated in Figure~\ref{fig:quarkonia_upsilon_raa_pt}.  These
estimates are made assuming that the signal to background ratio is
independent of \pt. Estimates are shown assuming no suppression of the
$\Upsilon$ states (left panel) and assuming the suppression predicted
in~\cite{Strickland:2011aa} (right panel).

\begin{figure}
  \begin{center}
    \includegraphics[width=0.6\textwidth]{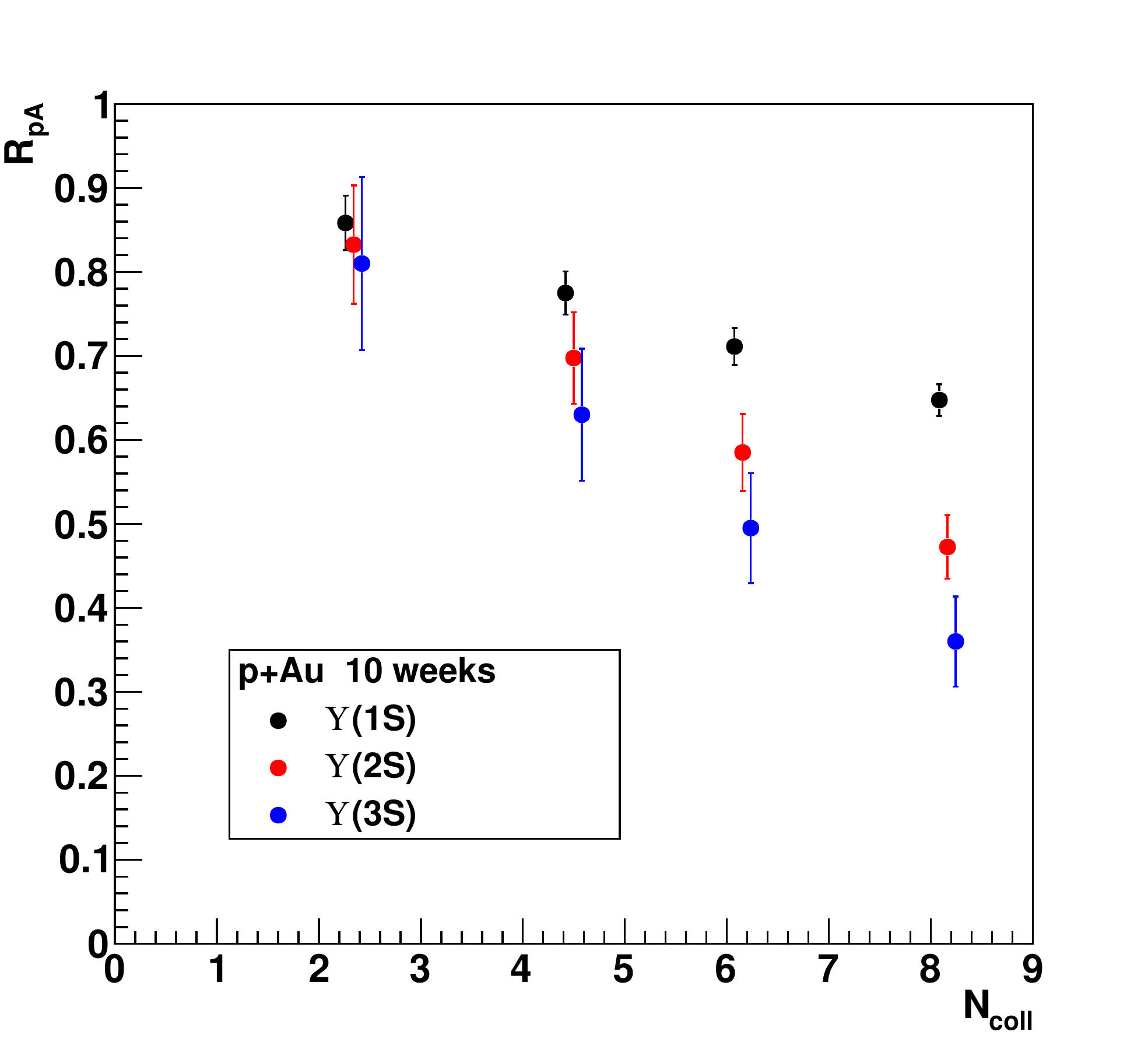}
  \end{center}
  \caption[Estimate of the statistical precision of a measurement of
  $R_{p\mathrm{Au}}$ for the $\Upsilon$ states using sPHENIX, in four
  centrality bins]{\label{fig:upsilon_RpAu} Estimate of the
    statistical precision of a measurement of $R_{p\mathrm{Au}}$ for
    the $\Upsilon$ states using sPHENIX, in four centrality bins.  The
    suppression is taken to be linear with centrality and the
    centrality integrated suppression for each state is set to match
    the double ratios of $\Upsilon(2S)/\Upsilon(1S)$ and
    $\Upsilon(3S)/\Upsilon(1S)$ measured by CMS at 5.02 TeV collision
    energy in \pPb and \pp collisions. The yields are for 10 weeks of
    \pAu running. }
\end{figure}

The expected statistical precision for $\Upsilon$ measurements with
sPHENIX in a 10 week \pAu run is illustrated in
Figure~\ref{fig:upsilon_RpAu}. The suppression values used in the plot
are set to match the double ratios of $\Upsilon(2S)/\Upsilon(1S)$ and
$\Upsilon(3S)/\Upsilon(1S)$ measured by CMS at 5.02 TeV collision
energy in \pPb and \pp collisions. The $\Upsilon(1S)$ is
taken to be unsuppressed except for the modified feed down from the
excited states, and the suppression of the $\Upsilon(2S)$ and
$\Upsilon(3S)$ states is arbitrarily taken to be linear with
centrality. The signal to background ratios in \pAu collisions are
taken to be the same as those in peripheral \auau collisions.

We conclude from these results that the proposed upgrade to the
sPHENIX detector would provide an excellent $\Upsilon$ measurement,
and would have the required mass resolution and
S/B to separate the $\Upsilon$(1S) state from the $\Upsilon$(2S) and
$\Upsilon$(3S) states. Further, we expect that by fitting a line
shape --- which could be determined very well from the $\Upsilon$(1S)
peak --- we could extract the $\Upsilon$(2S) and $\Upsilon$(3S) yields
separately with precision.

The expected higher statistics from the LHC experimental measurements of Upsilons over the next decade
underscore the need for measurements at RHIC with sufficient statistics to differentiate cold
and hot nuclear matters effects.   In particular the almost order of magnitude larger acceptance for sPHENIX
compared with the STAR MTD enables the precision required in \pp, \pAu and \auau to test models for the onset of suppression,
where CMS already observes differential suppression of the three states.

\cleardoublepage

\appendix

\makeatletter{}\chapter{Forward Hadronic Calorimeter and Barrel Preshower Options}
\label{chap:PreshowerFHCal}

The sPHENIX detector reference design presented in this proposal
consists primarily of a superconducting solenoid surrounded by
calorimetry and tracking covering $|\eta| < 1.1$.  This foundation
presents intriguing possibilities for physics opportunities which
would be enabled by instrumentation beyond the base detector.  
In particular, exciting physics in polarized \pp and \pau collisions enabled by a
forward detector has been documented in Ref.~\cite{forwardWhitePaper},
and is referred to
as fsPHENIX.    We believe these opportunities
are very interesting, and at the same time, we have segregated their
description from the main part of the proposal to convey clearly their
relationship to the base sPHENIX detector.  
We present two such additional detector systems and the physics they
would provide here in this Appendix.  

The first is a Forward Hadronic Calorimeter that would extend the jet
coverage significantly with particular physics emphasis in \pp and
\pau collisions.  The initial design explored here is with this
Forward Calorimeter acting as the endcap of the BaBar solenoid.  This
could be an initial configuration with later moving it back as part of
the integrated fsPHENIX design, again detailed in
Ref.~\cite{forwardWhitePaper}.  This simple configuration captures the
essential physics enabled by increased acceptance in $A$$+$$A$ collisions,
and could be considered a first step towards the fsPHENIX design if
the calorimeter were moved back and extended to accommodate the
fsPHENIX forward arm.  Such a calorimeter system would deliver physics
in itself and also represents a step towards the broader fsPHENIX
program.

The second is a barrel preshower detector in front of the Electromagnetic
Calorimeter, that would improve the signal to background for direct
photon measurements particularly in the range \pt $\approx 10$--20~GeV/c.

\section{Forward Hadronic Calorimeter}
\label{sec:fHCal}

\begin{figure}[hbt!]
\begin{center}
\includegraphics[width=0.9\linewidth]{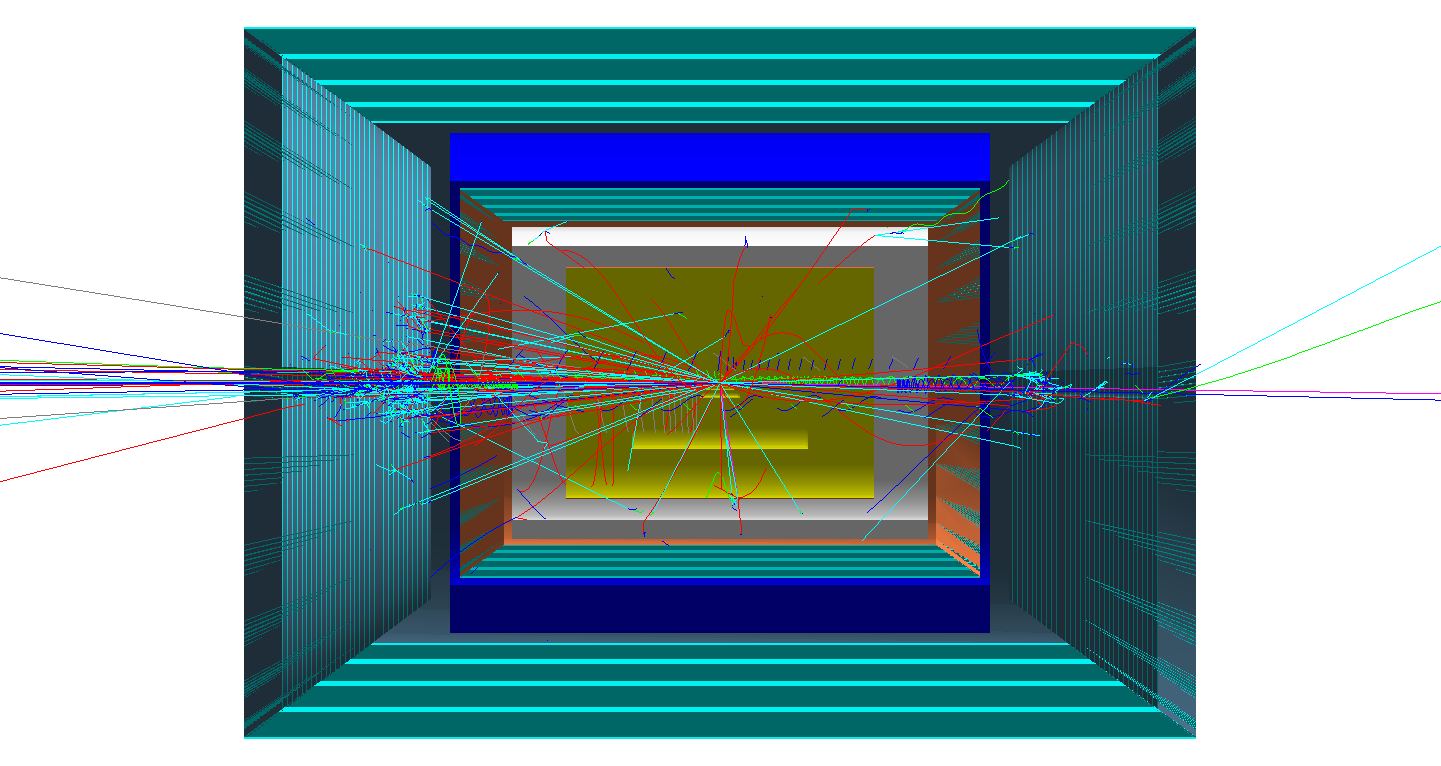}
\end{center}
\caption{Event display for a central HIJING \pA collision at $\sqrt{s}
  = 200$~GeV.  This detector concept adds calorimetrically
  instrumented endcaps to the sPHENIX barrel tracking and
  calorimetry.}
\label{fig:event}
\end{figure}

There is a great deal of enthusiasm in the Collaboration for the
physics enabled by instrumentation at forward rapidities.  For the
cases of \pp and \pA, the Collaboration has produced a white paper
detailing the many physics opportunities of forward
measurements~\cite{forwardWhitePaper}.  

For A$+$A collisions, forward calorimeter (fHCAL) coverage would
augment the base sPHENIX program in at least two key ways: one, it
provides event characterization (centrality, event plane) away from
observables measured at mid-rapidity; and, two, an fHCAL effectively
extends the acceptance of the sPHENIX barrel, enabling jet and dijet
measurements over wider rapidity range and potentially up to $\eta =
4$.

Extending the acceptance of hadronic calorimetry to forward rapidity
amplifies the already substantial complementarity of the RHIC and LHC
heavy-ion programs, with forward RHIC and mid-rapidity LHC
measurements enabling access to similar Bjorken $x$ values.  A global
analysis of experimental data at the LHC and RHIC including the
forward suppression region is crucial for a precision study and
understanding of the \pA physics.

There are various possibilities for instrumenting the forward
acceptance, with one option being to transform the steel
flux-returning endcaps of the base sPHENIX design into active steel
and scintillator calorimeters.  The fHCAL then works as an active flux
return yoke of the solenoidal magnet of the sPHENIX barrel. 

Figure \ref{fig:event} shows an event for a central HIJING \pA
collision at $\sqrt{s} = 200$~GeV and $b = 4$~fm.  There is no gap
between the barrel hadron calorimeter and the forward hadron
calorimeter (fHCAL) at $\eta = 1.1$.  The fHCALs are shown located on
both ends of sPHENIX in order to study what a very large $\eta$
acceptance can provide for sPHENIX, and to enable simulations of the
back-scattered secondary particles from flux return yoke.

For the current study of the detector performance, the fHCAL is
assumed to have the geometry of a truncated cone, and to be located at
$z = 2$~m.  The projective tower segmentation in polar and azimuthal
angles is defined to be roughly 10~cm$\times$10~cm per tower.  The
shower size defines the tower size, which naturally leads to large
$\Delta\eta \times \Delta\phi$ tower size.  The outermost ring of
towers has a segmentation that matches the azimuthal segmentation of
the central barrel.  Each tower consists of 30 layers of iron and
scintillator (= 4:1) and is one meter deep. The sampling fraction is
estimated to be 4.3\%.

\begin{figure}[hbt!]
  \centering
  \includegraphics[clip,width=1.0\linewidth]{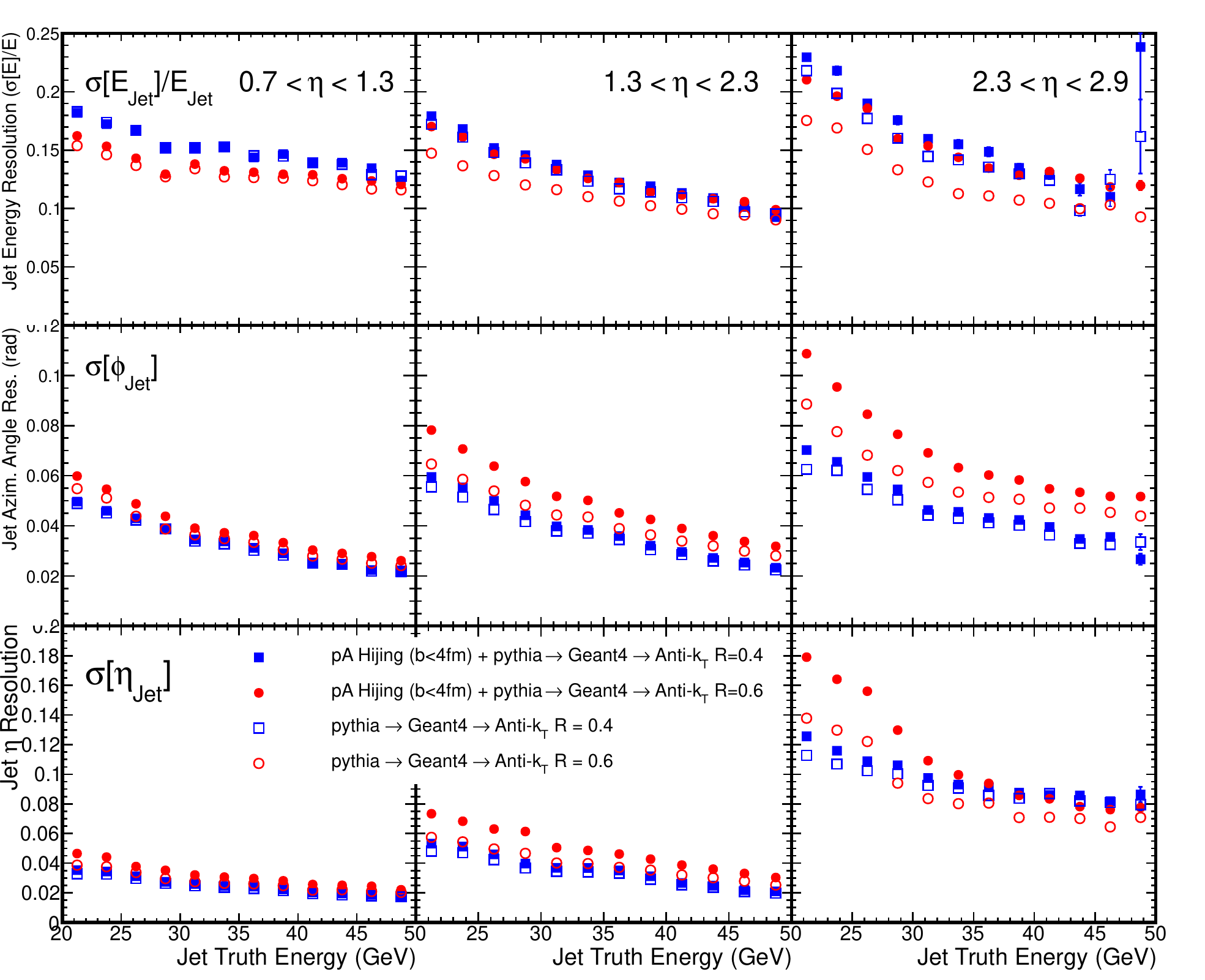}
  \caption{The \geant simulated jet resolution of single jets for
    energy (top row), $\phi$ (middle row) and $\eta$ (bottom row) in
    \pp (open markers) and \pA (closed markers) collisions
    reconstructed with the \fastjet \antikt algorithm with $R = 0.4$
    (blue) and $R = 0.6$ (red).}
 \label{fig:jet_fhcal}
\end{figure}

The jet reconstruction resolution of the fHCAL has been studied using
full \geant simulations for both \pp collisions and central \pAu
collisions. in Figure \ref{fig:jet_fhcal}. The acceptance was divided
in three regions: $0.7<\eta<1.3$ where the central and forward
calorimeter come together, $1.3<\eta<2.3$ where each forward calorimeter
blocks cover $d\phi$ of 0.1--0.2~rad, and $2.3<\eta<2.9$ where each
forward calorimeter blocks cover $d\phi$ of 0.2--0.4~rad.

In the first stage, full \pythia events were generated which contain at
least one forward-going single jet.  The energy and direction for the
truth jet was calculated with \fastjet \antikt
algorithm~\cite{Cacciari:2005hq} with $R = 0.4$ and 0.6 using the
\pythia generated particles.  Then the full event was simulated in the
\geant setup.  The tower energy was reconstructed from the energy
deposition in the active components in all four calorimeter systems
(forward hadron calorimeter, central EM calorimeter and two layers of
hadron calorimeters).  All calorimeter towers were analyzed using
\fastjet \antikt algorithm to again form the \geant simulated jets.
By comparing the truth and \geant jets, the resolution for energy and
angle for the jets were shown as the open markers in Figure
\ref{fig:jet_fhcal}.  The energy resolution in the forward direction
roughly matches that in the central direction (Figure
\ref{fig:jet_energy_resolution}).  For jets of $E_{jet}>20$~GeV, the
direction determination is better than 60~mrad for $\eta<2.3$, and
better than 0.12~rad for $2.3<\eta<2.9$.  This resolution presents a
minor effect in kinematic and azimuthal asymmetry smearing for jets
measurements in \pp.  In addition, no obvious energy leakage was
observed up to jet energy of 50~GeV.

In the second stage, the \pythia particles were embedded in central
\hijing \pA collisions with impact parameter $b$ varying between 0 and
4~fm.  The embedded event was simulated through the full \geant setup
again and formed jets in the \pA collisions. These jets were also
comparing to the truth jet based on the original \pythia particles,
and the resolutions are shown in closed circle markers as in
Figure~\ref{fig:jet_fhcal}.  As expected, the \antikt $R = 0.4$ jets
show less deterioration in performance when compared with the \antikt
$R = 0.6$ jets, and the effect from \pA background is stronger for the
more forward directions (e.g., $2.3<\eta<2.9$).  However, the change
of jet performance with \pA background is less than 20\%.

\clearpage
 
\makeatletter{}\section{Preshower}
\label{sec:preshower}

The measurement of direct photons in \pp, \pau, and \auau collisions represents
a golden channel for measuring the energy loss of the opposing quark and the medium response 
-- see Section~\ref{sec:direct_photons_and_ff} for details.  For \pt greater than 20 GeV/c, the 
direct photon signal to background is significantly greater than one in central \auau events and 
isolation cuts can be utilized effectively in \pp and \pau collisions as well.   There is a proposal
to include a preshower detector in front of the barrel electromagnetic calorimeter (EMCal) to enhance the
direct photon signal to background in order to increase
the kinematic range and give a larger statistical sample, particularly for\pt $\approx$ 10-20 GeV/c.    
In addition, this would extend the neutral pion measurements to significantly higher \pt than the EMCal alone.

The preshower concept is to provide discrimination between single photons and close-by pairs of photons, primarily
from $\pi^{0}$ decays.    The proposed preshower detector is to be placed
right in front of the electromagnetic calorimeter, at $\sim 88$~cm in radius from the beam line.
The total area is $\sim12$~m$^2$ if covering the entire sPHENIX acceptance of $| \eta | < 1.1$.
The current plan is to cover $\sim 25$\% of the acceptance to lower the construction cost while 
maintaining the physics capabilities.
The design consists of a tungsten converter, $\sim 2$ radiation lengths thick,
followed by a single-layer silicon pad detector.    We have implemented this geometry in 
the sPHENIX \geant simulations with two pad sizes of $5$ mm $\times 5$ mm and $10$ mm$\times 10$ mm.
We are currently working to optimize the pad size in terms of achieving a good $\pi^0$ identification
over a wide transverse momentum range, and keeping the number of readout channels at a reasonable level,
 30--120~k.   These pad sizes result in occupancies of $\sim 3$\% or less
in central \auau collisions at 200 GeV as determined from full \hijing events run through \geant.
A common readout scheme to the sPHENIX silicon tracking layers using the SVX4 chip is one option being explored.

\begin{figure}
\centering
\includegraphics[width=0.48\linewidth]{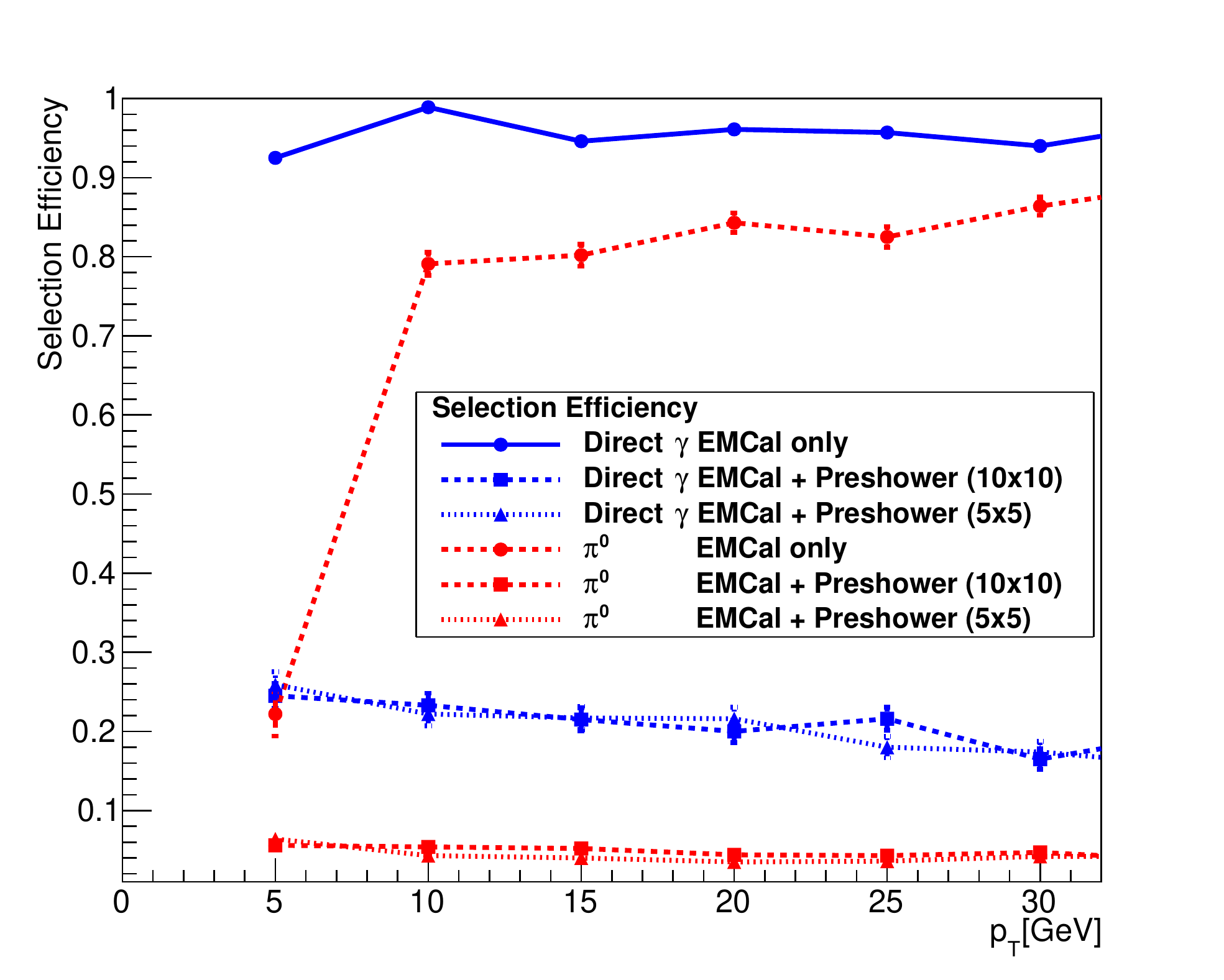}
\hfill
\includegraphics[width=0.48\linewidth]{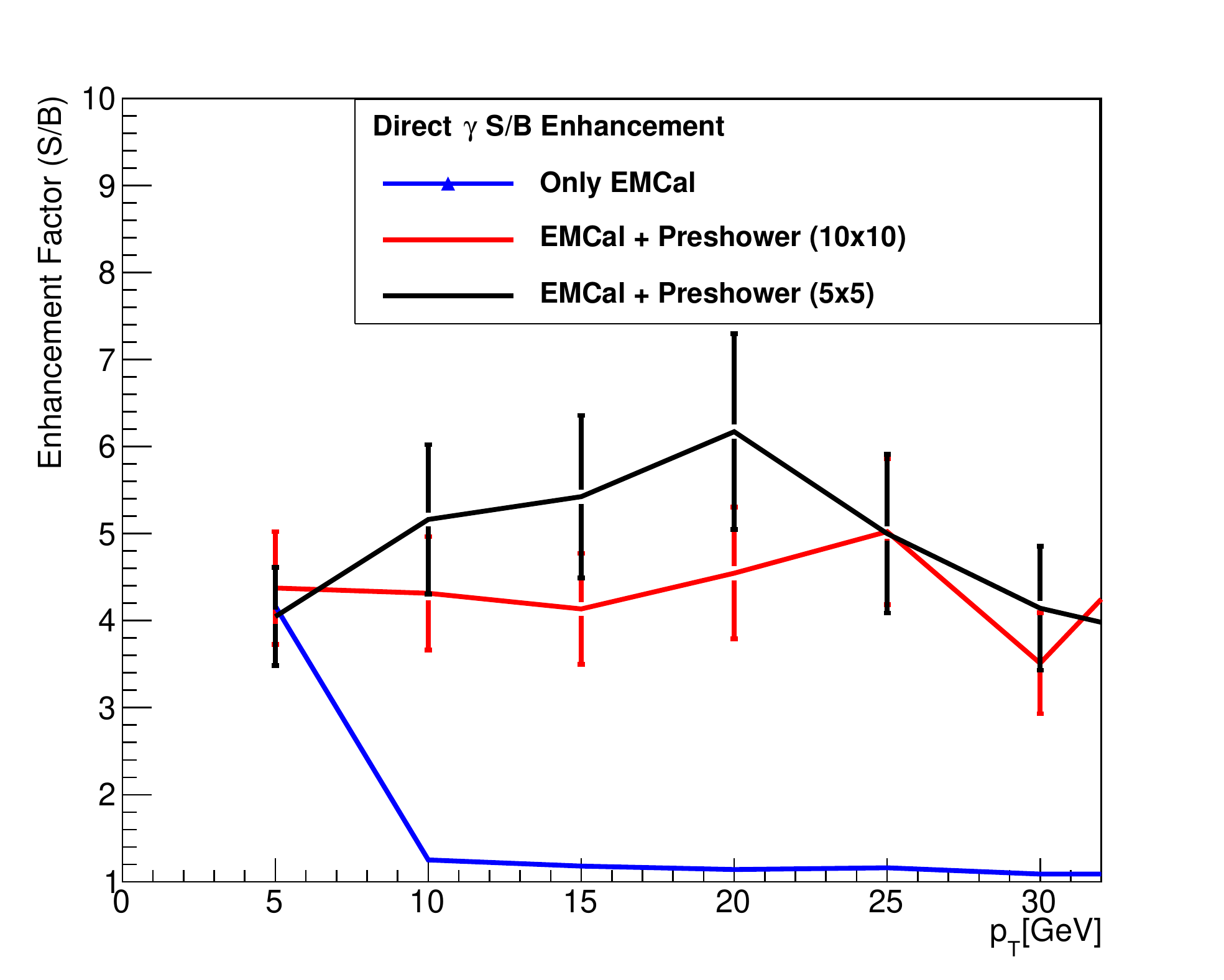}
\caption{(Left) Efficiency for direct photons and two-pairs from neutral pion decay as a function
of \pt for different EMCal and Preshower cuts.  (Right) The corresponding enhancement in the direct
photon S/B as a function of \pt.   Note that no isolation cuts are applied.
\label{fig:preshower}}
\end{figure}

A cluster algorithm applied to the EMCal, combined with a cluster shape metric, indicates good separation
of single photon from two photon showers up to 8-10 GeV, after which the discrimination degrades quickly.
Shown in the left panel of Figure~\ref{fig:preshower} is the efficiency for selecting direct photons
and two-photons from $\pi^{0}$ with EMCal selection only, EMCal selection with a Preshower with 10x10 pads, and 
then for 5x5 pads.  Note that Preshower and EMCal cuts are set to reject two-photon showers, thus enhancing
the signal to background for direct photons.   One can see that the EMCal alone works quite well at 5 GeV, and
already by 10 GeV provides limited discrimination.   The right panel of Figure~\ref{fig:preshower} shows
the enhancement in the signal to background (S/B) with the different selections.   Currently a factor of
five improvement in S/B is achieved and we believe with an optimized pad configuration can be increased.
We note that the EMCal resolution has a negligible change in resolution with the preshower included 
in the \geant simulation.   

The potential for extending the sPHENIX capabilities in this way has attracted
significant international interest, including from Japanese and Korean institutions.
Our goal is to have this upgrade installed and available for physics on day-one.
Funding through proposals to US-Japan and JSPS are being pursued toward the goal.

\makeatletter{}\chapter{Evolution to an EIC Detector}
\label{chap:ePHENIX}

\makeatletter{}The Nuclear Physics QCD community long term plan centers on the realization of 
an Electron Ion Collider (EIC).  
Brookhaven National Laboratory is working towards a specific realization of
an EIC with a potential turn-on date of 2025 with an electron beam
energy up to 21.2~GeV, hadron beam energies up to 255~GeV for protons
and 100~GeV/nucleon for gold ions, and design luminosities of
$10^{33}$ cm$^{-2}$s$^{-1}$ for 15.0 GeV on 255 GeV $e$$+$$p$
collisions.  The EIC detector proposed here, built on the foundation
of sPHENIX, will have excellent performance for a broad range of
exciting EIC physics measurements, providing powerful investigations
not currently available that will dramatically advance our
understanding of how quantum chromodynamics binds the proton and forms
nuclear matter.

From the beginning, it was realized that the sPHENIX detector design,
with its large bore superconducting solenoid, midrapidity calorimetry,
open geometry, coupled with the existing investment in infrastructure
in the PHENIX interaction region, provides an excellent foundation for
an EIC detector.  With this in mind, EIC design considerations for the
sPHENIX proposal have been incorporated from the
start~\cite{decadalPlan}.

\begin{figure}[p]
  \centering
  \includegraphics[trim = 280 150 240 140, clip, width=0.75\linewidth]{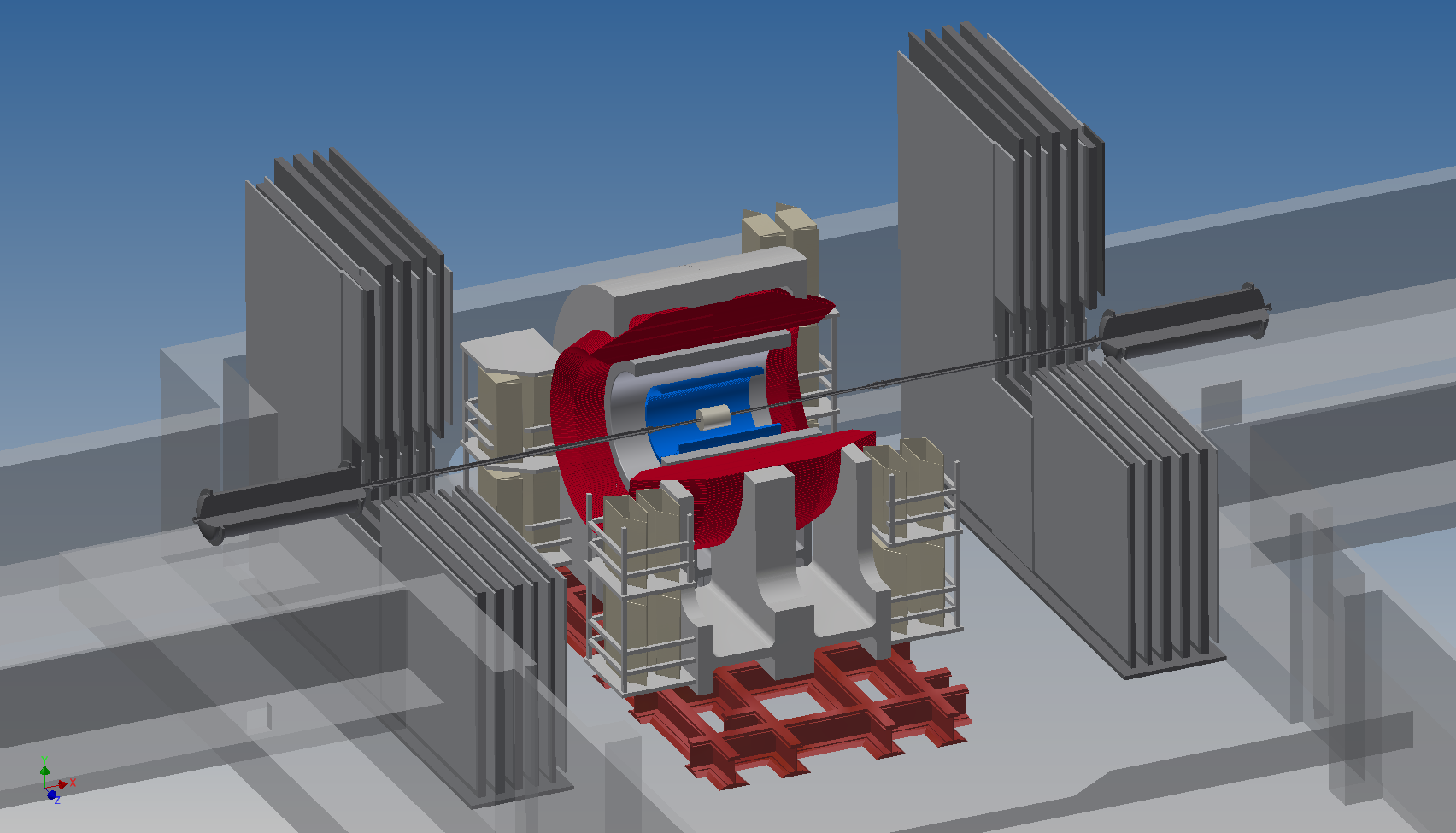}
  \vskip 0.1cm
  \includegraphics[width=0.75\linewidth]{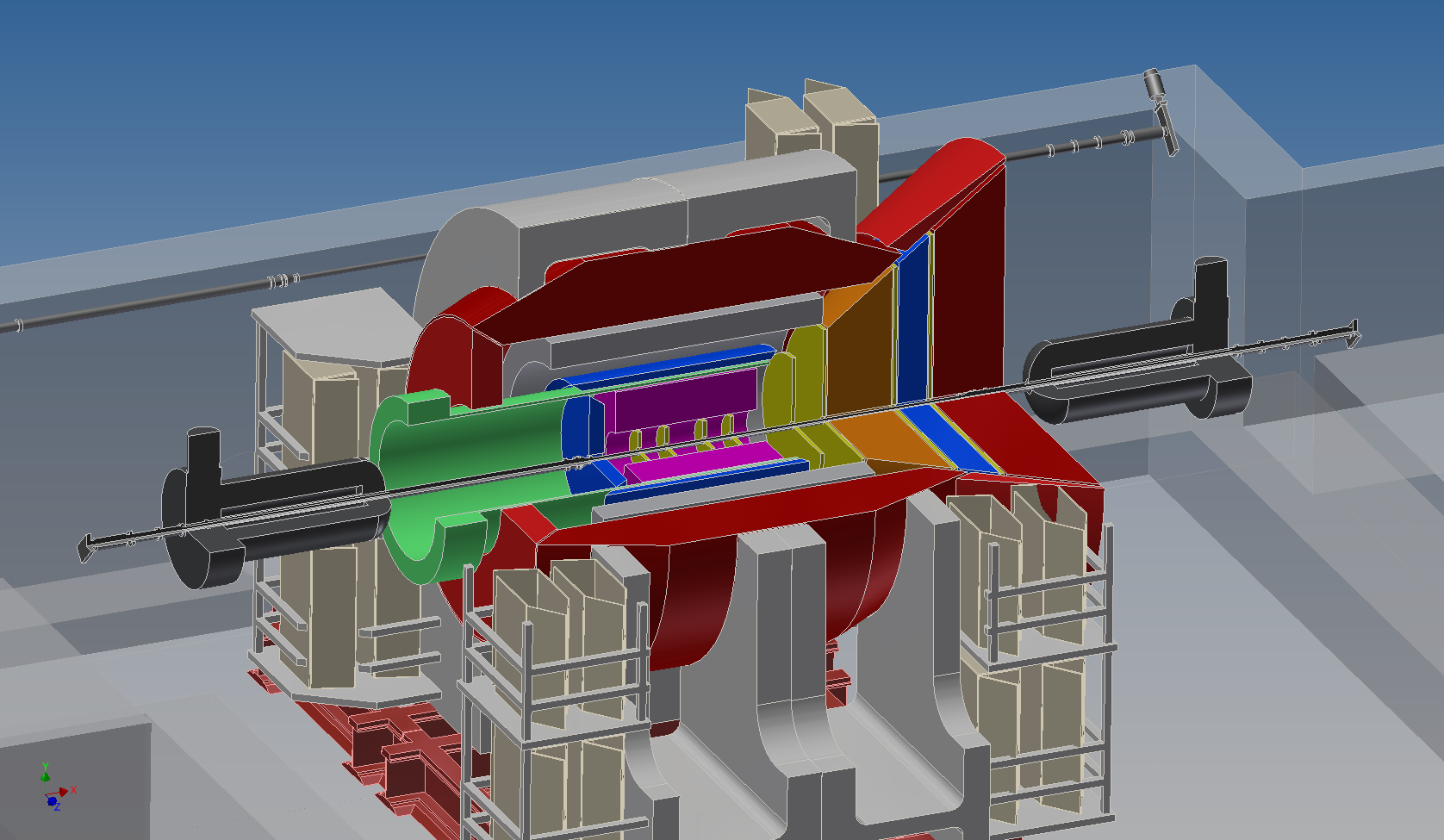}
  \caption[The evolution of the sPHENIX into a detector for eRHIC
  physics, with additional capabilities supporting its focus on
  $e$$+$$p$ and $e$$+$A collisions]{The evolution of the sPHENIX
    detector, with its focus on jets and hard probes in heavy-ion
    collisions, into ePHENIX, with additional capabilities supporting
    its focus on $e$$+$$p$ and $e$$+$A collisions. (top) The sPHENIX
    detector in the existing PHENIX experimental hall. (bottom) The
    EIC detector, in the same hall, showing the reuse of the
    superconducting solenoid and the electromagnetic and hadronic
    calorimeter system. The eRHIC focusing quadrupoles, each located
    4.5 m from the interaction point, and the height of the beam pipe
    above the concrete floor, set the dominant physical constraints on
    the allowable dimensions of the EIC detector.}
  \label{fig:ePHENIX}
\end{figure}

A full engineering rendering of the proposed detector --- showing how
it builds upon sPHENIX --- is shown in Figure~\ref{fig:ePHENIX}.  In
addition to fully utilizing the sPHENIX superconducting solenoid and
barrel calorimetry, new detectors are added in the barrel and \egoing
and \hgoing directions.  In the \egoing direction a crystal
calorimeter is added for electron identification and precision
resolution.  A compact time projection chamber, augmented by
additional forward and backward angle GEM detectors, provides full
tracking coverage.  In the \hgoing direction, behind the tracking is
electromagnetic and hadronic calorimetry.  Critical particle
identification capabilities are incorporated via a barrel DIRC, and in
the \hgoing direction, a gas RICH and an aerogel RICH.

The physics case for an EIC is documented in depth in the EIC White
Paper~\cite{Accardi:2012hwp}.  An EIC will enable major scientific advances in at least three main
areas: 1) Detailed imaging of the spin and momentum structure of the
nucleon; 2) Investigation of the onset of gluon saturation in heavy
nuclei; and 3) Study of hadronization in cold nuclear matter.
In this document we review each area with a focus on the connection to
detector acceptance and performance requirements.  
We consider each subsystem in sufficient detail to be able to map out
the performance using both parametrized and full \geant simulations.  
We find a broad suite of observables where this EIC detector realization has excellent capabilities.

The \ephenix capably addresses most all of the physics enabled at this
EIC machine.  We believe we have struck a strong balance between
capabilities and costs for this proposed detector, but there remain
clear targets for augmenting those capabilities---for instance, by
adding a silicon vertex detector to enable measurements of open charm
observables (e.g., $F_2^c$).  In addition, there is a possibility to
upgrade eRHIC to even higher energy electron beams at a future date,
and we believe \ephenix provides an excellent base upon which an
upgraded detector capable of exploiting the physics potential of those
collisions could be built. There is also the potential, if one can
realize appropriate instrumentation in the \hgoing direction while \pp
and $p$$+$A collisions are still available in RHIC, to pursue a rich
program of forward physics measurements.

The PHENIX collaboration itself has outstanding detector expertise and
technical support as a base for the construction of an EIC detector.
Nonetheless, we view \ephenix being by constructed and operated by a
fundamentally new collaboration that would require and welcome the
addition of new institutions bringing with them additional detector
expertise, physics insights, and scientific leadership.

The PHENIX collaboration was asked to produce a document detailing the capabilities
of an EIC detector utilizing electron beams of 5-10 GeV.   That document, as submitted 
in October 2013, is included as the remainder of this Appendix.  We note that now higher
initial electron beam energies are envisioned and the detector concept is readily being adapted to take full advantage of
this additional physics potential.

This Appendix is organized as follows.
Section~\ref{chap:physics_goals} illustrates the wide spectrum of EIC
physics that can be addressed.
Section~\ref{chap:ephenix_detector_requirements} describes the detector
requirements that follow from that physics and which drive this EIC detector
design.  Section~\ref{chap:ephenix_detector_concept} details the EIC
detector concept and shows its performance for key measurements.

\renewcommand{\arraystretch}{1.5}

\makeatletter{}\section[EIC Physics]{Physics at an Electron-Ion Collider}
\label{chap:physics_goals}

The 2007 Nuclear Physics Long Range Plan~\cite{NSAC_LRP:2007} states
that the Electron-Ion Collider (EIC) embodies ``the vision for
reaching the next QCD frontier.''  In this Chapter we review the
primary physics goals as detailed in the EIC White
Paper~\cite{Accardi:2012hwp} and the broad physics program that can be
carried out with the ePHENIX detector.

\subsection{Fundamental questions addressed by the EIC}
\label{sec:eic-quest}

The EIC is designed to address several important question that are described in detail in the recent EIC White Paper~\cite{Accardi:2012hwp}.   Quoting
from the White Paper, these questions are reproduced here:
\begin{itemize}
\item {\bf How are the sea quarks and gluons, and their spins,
distributed in space and momentum inside the nucleon?} How are these
quark and gluon distributions correlated with overall nucleon
properties, such as spin direction?  What is the role of the orbital
motion of sea quarks and gluons in building the nucleon spin?
\item {\bf Where does the saturation of gluon densities set in?} Is
there a simple boundary that separates this region from that of more
dilute quark-gluon matter? If so, how do the distributions of quarks
and gluons change as one crosses the boundary? Does this saturation
produce matter of universal properties in the nucleon and all nuclei
viewed at nearly the speed of light?
\item {\bf How does the nuclear environment affect the distribution of
quarks and gluons and their interactions in nuclei?}  How does the
transverse spatial distribution of gluons compare to that in the
nucleon? How does nuclear matter respond to a fast moving color charge
passing through it?   What drives the time scale for color
neutralization and eventual hadronization?
\end{itemize}

The White Paper describes in detail the ``golden'' measurements in inclusive
Deep Inelastic Scattering (DIS), Semi-Inclusive DIS (SIDIS), and exclusive 
scattering at a future $e$$+$$p$ and $e$$+$A collider which will address the above 
questions employing a perfect detector.

\section{eRHIC: realizing the Electron-Ion Collider}

The accelerator requirements for an EIC that can answer the questions listed
above are spelled out in the EIC White Paper~\cite{Accardi:2012hwp}.  Two possible designs are presented based on current facilities:  (1) the 
eRHIC design, which adds a Energy Recovery LINAC to the existing RHIC complex 
at Brookhaven National Laboratory (BNL) which can accelerate polarized protons 
up to 250~GeV and ions such as gold up to 100~GeV/nucleon, and (2) the 
ELectron-Ion Collider (ELIC) design, which uses the 12~GeV Upgrade of CEBAF at Jefferson Laboratory with a 
new electron and ion collider complex.  

As per the charge from the Brookhaven National
  Laboratory Associate Lab Director, we consider the following eRHIC design parameters:
\begin{itemize}
\item{A polarized electron beam with energy up to 10~GeV and polarization of
70\%,}
\item{A polarized proton beam with energy up to 250~GeV and polarization of 70\%,}
\item{An ion beam which can run a range of nuclei from deuteron to gold and 
uranium with energy up to 100~GeV/nucleon for gold,}
\item{Luminosity with a 10~GeV electron beam of $10^{33}$~cm$^{-2}$s$^{-1}$ 
for $e$$+$$p$ with 250~GeV proton beam energy, and 
 $6\times10^{32}$~cm$^{-2}$s$^{-1}$ for $e$$+$A with 100~GeV ion beams.} 
\end{itemize}

\section{Physics Deliverables}

The three fundamental and compelling questions in QCD to be addressed by the EIC discussed in
Section \ref{sec:eic-quest} can be broken down in to five golden measurements suggested in the 
EIC White Paper~\cite{Accardi:2012hwp}.

The first three relate to using the proton as a laboratory for fundamental QCD studies.
 
\begin{itemize}

\item {\bf The longitudinal spin of the proton:}   
With the good resolution calorimetry and tracking
in ePHENIX, Inclusive DIS measurements in polarized 
$e$$+$$p$ collisions will decisively determine the gluon and quark spin 
contributions to the proton spin.  
Further, planned particle identification capabilities will 
allow ePHENIX to pin down the spin contributions from the different 
quark flavors.

\item {\bf Transverse motion of quarks and gluons in the proton:} With the 
excellent particle identification capabilities of ePHENIX and the high luminosity of eRHIC, 
unparalleled SIDIS measurements will be possible, and enable
us to explore and understand how the intrinsic motion of partons in the 
nucleon is correlated with the nucleon or parton spin.

\item {\bf Tomographic imaging of the proton:}  The large acceptance of  
ePHENIX for tracking and calorimetry, far forward proton and neutron
detector capabilities,  the high luminosity of eRHIC and the phase space 
accessible in a collider geometry enables ePHENIX to significantly 
extend the kinematic coverage of exclusive measurements such as Deeply Virtual 
Compton Scattering (DVCS).  
With these, detailed images of how (sea) quarks and gluons are 
distributed in the proton will become possible for the first time.

\end{itemize}

The following two relate to extending these techniques to the heaviest stable nuclei.

\begin{itemize}

\item {\bf Hadronization and its modification in nuclear matter:}  With 
ePHENIX PID and the versatility of eRHIC to collide many different ions, 
measurements of identified hadrons in $e$$+$$p$ and $e$$+$A will allow 
precise study of how quarks hadronize in vacuum and in nuclear matter.

\item {\bf QCD matter at extreme gluon density:}  ePHENIX will enable   
measurements of diffractive and total DIS cross-sections 
in $e$$+$A and $e$$+$$p$.  Since the diffractive cross section is viewed as a double gluon
exchange process, the comparison of diffraction to total cross section in $e$$+$A and 
$e$$+$$p$ is a very sensitive indicator of the gluon saturation region. ePHENIX would
be an ideal detector to explore and study this with high precision. 

\end{itemize}

Below we discuss each of these points in more detail and with specific details on the
ePHENIX capabilities.

\subsubsection{The proton as a laboratory for QCD}

\begin{figure}
\centering
\includegraphics[width=0.8\textwidth]{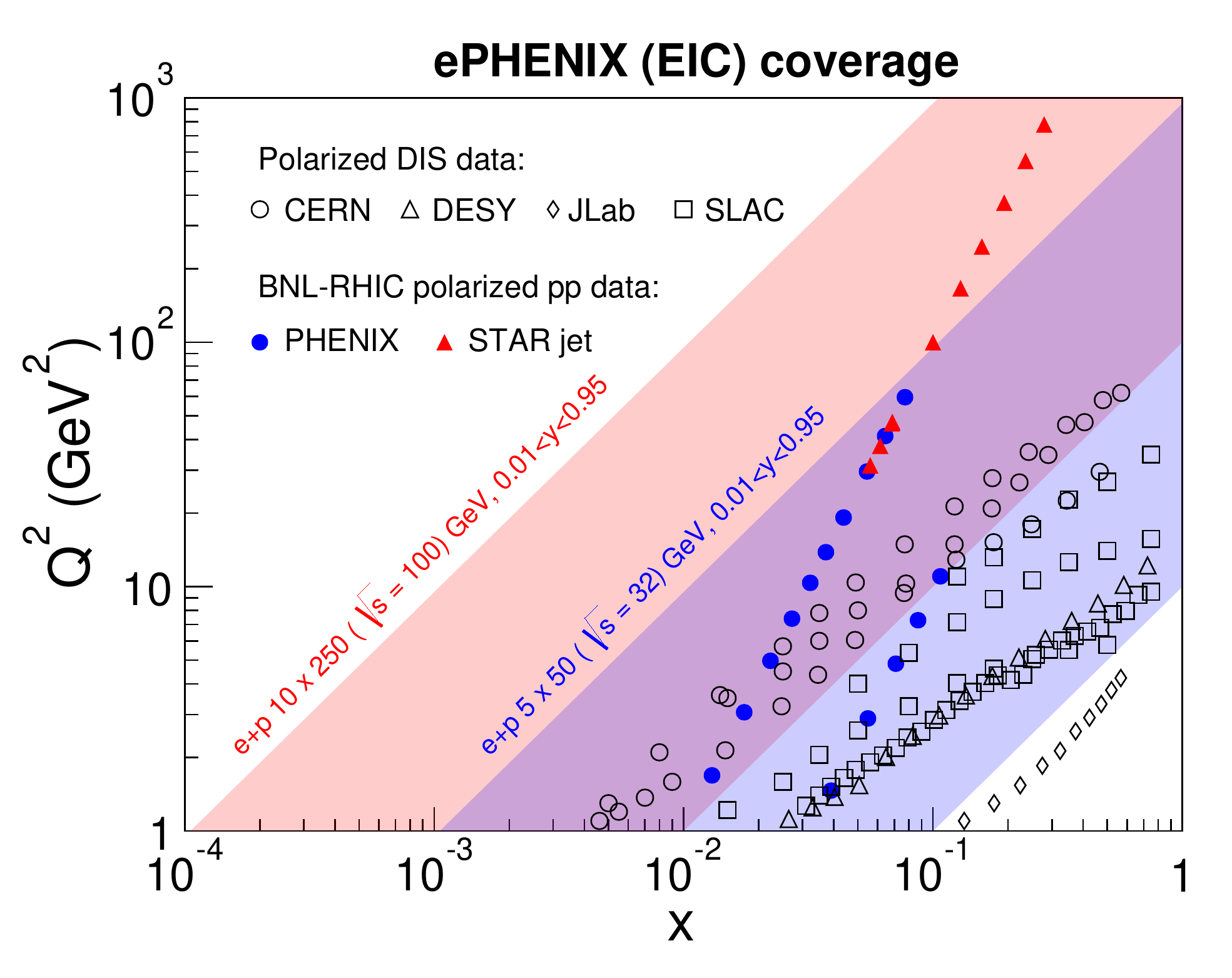}
\caption[Kinematic coverage of ePHENIX for two beam energy
configurations, 10$\times$250~GeV and 5$\times$50~GeV, which show the
range of eRHIC capabilities]{\label{fig:kin_helicity}{Kinematic
    coverage of ePHENIX for two beam energy configurations,
    10$\times$250~GeV and 5$\times$50~GeV, which show the range of
    eRHIC capabilities.  Also shown are data from current polarized
    fixed target DIS experiments and RHIC $p$$+$$p$ collisions.}}
\end{figure}

Deep Inelastic Scattering experiments over the last several decades 
have greatly enhanced our understanding of the proton substructure.  
Measurements with colliding beams at H1 and ZEUS at HERA have mapped out 
the momentum distributions of quarks and gluons, and shown that the 
gluons carry roughly half of the proton momentum.  Fixed target experiments,
with polarized nucleons and leptons at SLAC, CERN, DESY and JLab have 
revealed new surprises about proton structure, finding that only a small 
fraction of the proton spin comes from the quark spin and that there is 
significant 
correlation between the intrinsic motion of quarks and the nucleon spin.  
Measurements at both fixed target and colliders have started to image the 
proton through exclusive measurements.  

eRHIC will greatly enhance the kinematic coverage for DIS with polarized 
beams, as shown in Figure~\ref{fig:kin_helicity}.  With the capabilities of 
ePHENIX, we will significantly extend our understanding of the proton.  The 
gluon and flavor dependent sea quark spin contributions to the proton spin 
will be determined, as will the possible orbital angular momentum 
contributions.  The spatial and momentum distributions of (sea) quarks and 
gluons can be mapped, giving a multidimensional description of the proton.

\paragraph{Longitudinal spin of the proton}
\label{sec:longitudinal_spin}

\begin{figure}
  \centering
  \includegraphics[width=0.75\textwidth]{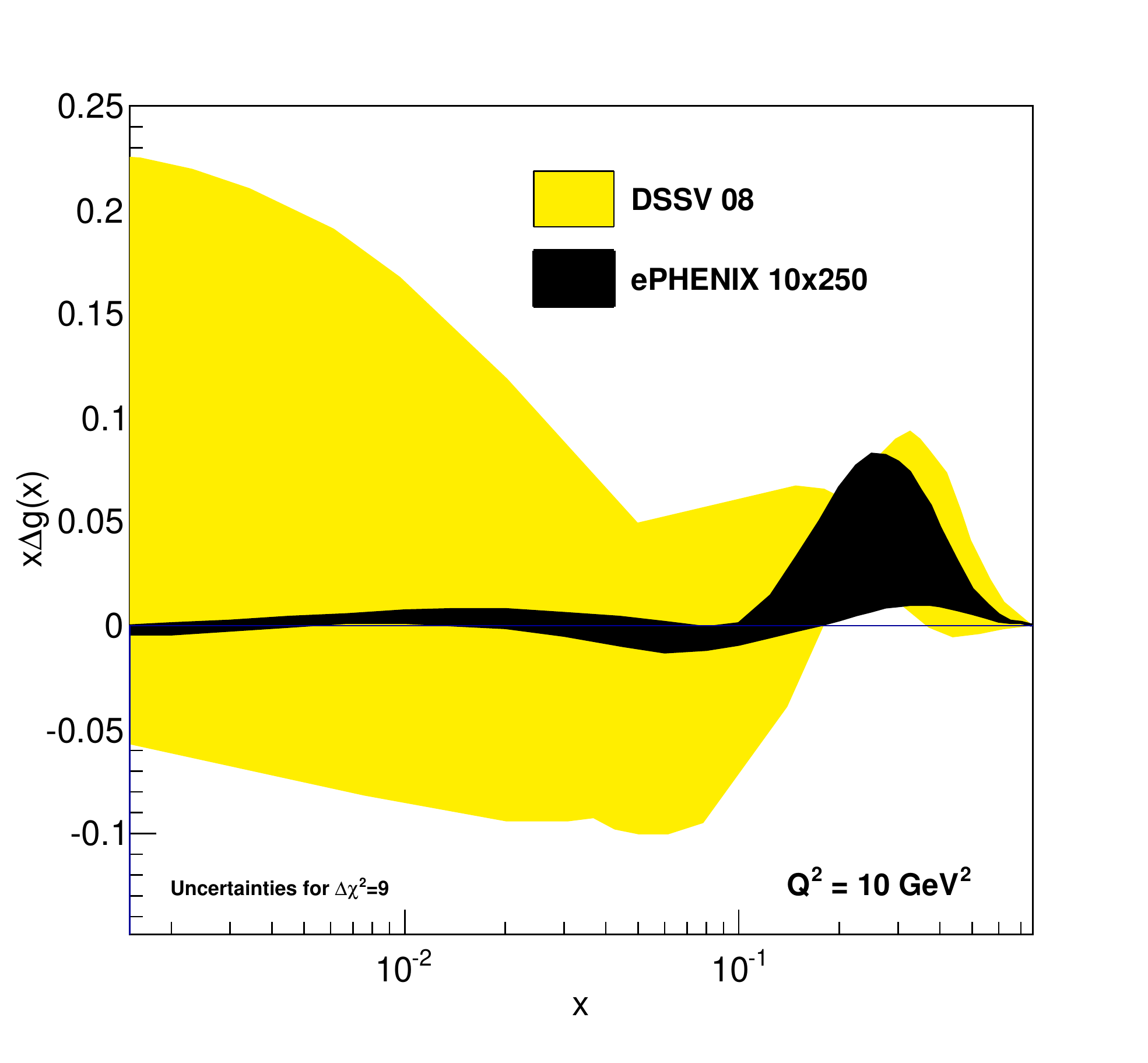}
  \caption[The projected reduction in the uncertainty on the gluon
  longitudinal spin distribution based on simulated \pythia events
  corresponding to an integrated luminosity of 10~fb$^{-1}$ at the
  10~GeV $\times$ 250~GeV beam energy configuration]{The projected
    reduction in the uncertainty (black) on the gluon longitudinal
    spin distribution based on simulated \pythia events corresponding
    to an integrated luminosity of 10~fb$^{-1}$ at the 10~GeV $\times$
    250~GeV beam energy configuration.  A $1\%$ systematic uncertainty
    in beam and target polarization is applied.  The yellow area shows
    the uncertainty from current data based on the analysis in
    Ref.~\cite{deFlorian:2009vb}.}
  \label{fig:deltag}
\end{figure}

The puzzle of the proton spin, to which the quark spin only contributes 
roughly a third, has spurred two decades of study.  Measurements from fixed
target polarized DIS have determined the quark contribution, but are less 
sensitive to the gluon due to the small kinematic coverage.  Current RHIC 
measurements indicate that the gluon spin contribution may be comparable 
or even larger than the quark spin contribution, but 
due to the limited coverage at low longitudinal momentum fraction, $x$, 
large uncertainty remains, as is shown in Figure~\ref{fig:deltag} (yellow band).

Determining the gluon longitudinal spin contribution is a primary goal of the 
EIC and of ePHENIX, and will be possible due to the large reach in $x$ and 
four-momentum transfer squared, $Q^2$.  Figure~\ref{fig:deltag} shows the 
expected 
impact from ePHENIX measurements of inclusive DIS on the uncertainty of the 
gluon helicity distribution as a function of $x$.  

With the ePHENIX particle identification (PID) detectors, measurements of pions and
kaons will greatly improve on the determination of the sea quark longitudinal
spin distribution as well, including that of the strange quark, $\Delta s$,
which has been of particular interest in the last few decades,
because of the contradictory results obtained from different data.
Current global analyses use hyperon beta decay to constrain $\Delta s$,
which indicates a negative value for the full integral over $x$.  Fixed target SIDIS
measurements of kaon asymmetries, which directly probe $\Delta s$, though
at low values of $Q^2$ and in a limited $x$ range, find a positive contribution for
$x>0.01$.  eRHIC provides data over a wide $x$ and $Q^2$ range. Further, 
ePHENIX will provide excellent particle ID capability
to identify kaons and allow direct measurements of strangeness spin
contribution to the nucleon down to $\sim2\times10^{-4}$.

\paragraph{Transverse motion of quarks and gluons in the proton}
\label{sec:transverse_motion}

\begin{figure}\begin{center}
\begin{minipage}[b]{0.38\textwidth} \centering
{\hskip -0.1in}
\includegraphics[width=0.96\textwidth]{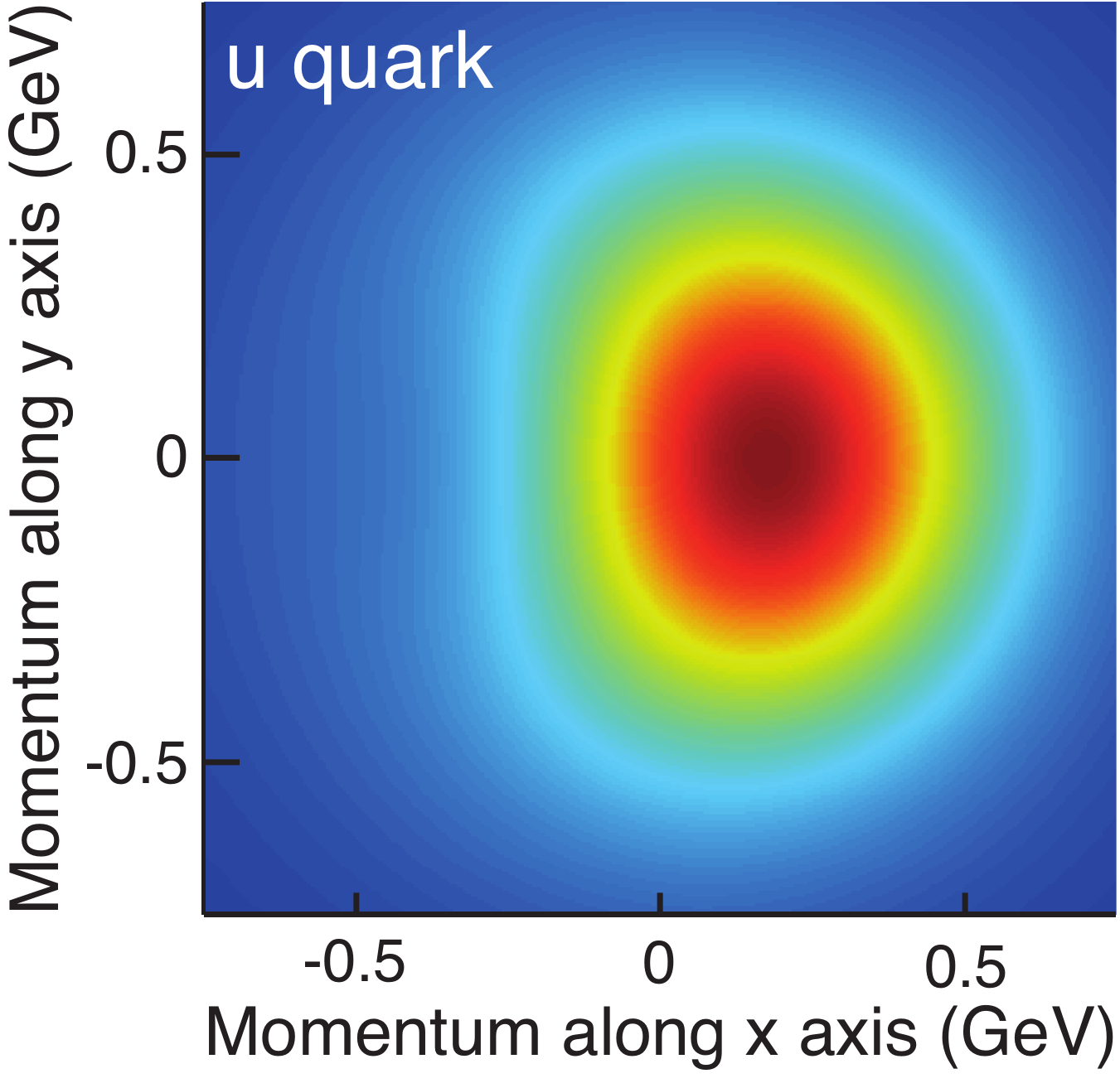}
\end{minipage} 
\hskip 0.05in
\begin{minipage}[b]{0.60\textwidth} \centering
\includegraphics[width=0.98\textwidth]{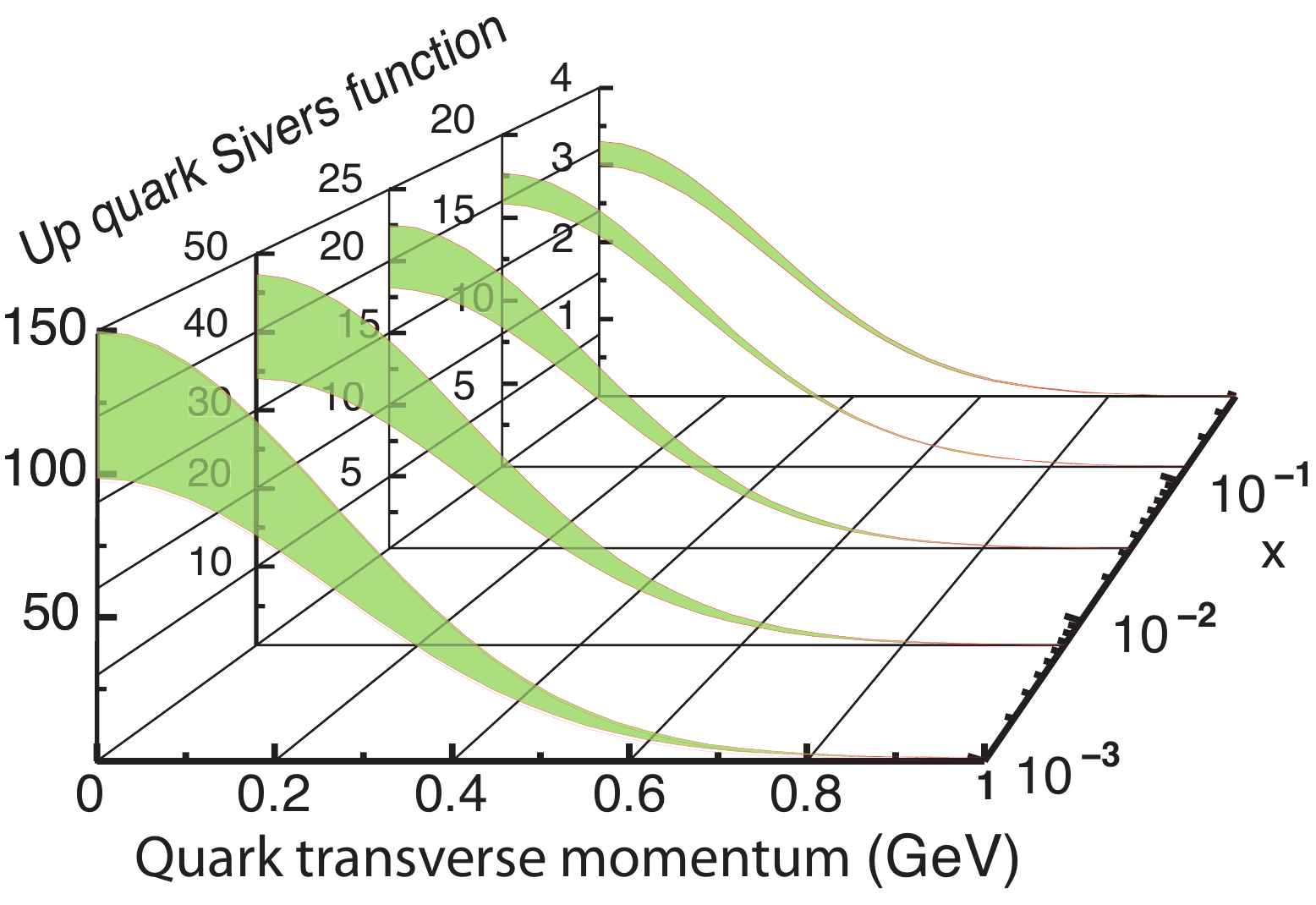}
\end{minipage}
\end{center} 
\vskip -0.05in
\caption[The transverse-momentum distribution of an up quark with $x =
0.1$ in a transversely polarized proton and the transverse-momentum
profile of the up quark Sivers function accessible at
eRHIC]{\label{fig:tmds}{ [Reproduced from
    Ref.~\cite{Accardi:2012hwp}.]  (left) The transverse-momentum
    distribution of an up quark with longitudinal momentum fraction
    $x=0.1$ in a transversely polarized proton moving in the
    z-direction, while polarized in the y-direction. The color code
    indicates the probability of finding the up quarks.  (right) The
    transverse-momentum profile of the up quark Sivers function at
    five $x$ values accessible with the kinematics avialable at eRHIC,
    and corresponding statistical uncertainties.}  }
\end{figure}

Large transverse spin asymmetries measured in fixed target 
SIDIS in the past decade have spurred significant 
theoretical work.   These asymmetries relate to the 
transversity distribution, the correlation between the transverse spin of the 
proton and a transversely polarized quark in it, and Transverse Momentum 
Distributions (TMDs), such as the Sivers or Boer-Mulders distributions, which 
describe correlations between either the proton or quark spin and the quark 
intrinsic motion, specifically the transverse momentum of the quark.  With 
measurements of identified pions and kaons, these asymmetries give a 2$+$1 dimensional 
description of the spin and momentum distributions of different quark flavors 
in the proton, such as is shown in Figure~\ref{fig:tmds}.

Current measurements, however, are only able to probe a small 
region in $x$ and $Q^2$, limiting the description to the valence quark region.
Understanding of how the sea quarks and gluons contribute requires a larger 
kinematic range, such as provided at eRHIC. With the PID capabilities of 
ePHENIX, asymmetry measurements with transversely polarized nucleons and 
electrons in SIDIS will enable the study of these TMDs over most of this 
range, significantly expanding our knowledge of the proton structure.  
The 
constraint on the Sivers distributions was discussed in the EIC White 
Paper~\cite{Accardi:2012hwp}, with the expectations shown in Figure~\ref{fig:tmds}.
For the first time, determination of the Sivers distribution over a 
wide range in $x$ will be possible, including the low $x$ region where gluons 
dominate.

The transversity distribution, when coupled with the Collins fragmentation asymmetry, 
would result in an azimuthal asymmetry in the hadron production. This
has been called the Collins effect, and is a measurement that goes to the
heart of establishing the transversity distribution in a proton \cite{Anselmino:2007fs}. 
Measurement over the wide kinematic region would not only allow us to
measure transversity, but the wide $x$-coverage possible at eRHIC would 
afford the first reliable measurement of the tensor charge of the proton
(the integral over $x$ of the transversity distribution). No other currently
operational or planned facility can do this.

\paragraph{Tomographic imaging of the proton}
\label{sec:tomographic_imaging}

Hard exclusive processes such as the Deeply Virtual Compton Scattering (DVCS) and
Deeply Virtual Vector Meson production (DVVM) involve interactions between the 
virtual photon and the partons in the proton without breaking the proton, resulting 
in the production of a real photon in DVCS or a vector meson in DVVM processes.
Just as elastic lepton-nucleon scattering gives information on the spatial 
distribution of the electric charge and magnetization in the nucleon, DVCS and DVVM 
processes probe the transverse distribution of quarks, anti-quarks and gluons. 
This information is encoded in generalized parton distributions (GPDs), 
which quantify the distributions of quarks and gluons in terms
of their positions in the transverse plane and longitudinal momentum fraction, 
providing 2$+$1 dimensional imaging of the nucleon. 
Measurements with polarized beams enable studies of spin-orbit correlations 
of quarks and gluons in the nucleon, by correlating the shift in the parton 
transverse distribution and proton transverse polarization. 
It is intuitively connected with orbital angular momentum carried 
by partons in the nucleon and hence of great interest in addressing the 
nucleon spin puzzle (nucleon spin decomposition)~\cite{Ji:1996ek}.

The existing data on GPDs from fixed target experiments cover only a 
limited kinematical range of $t$ (the squared momentum transfer to the 
proton), medium to high $x$ and low $Q^2$. 
The $t$ is connected through the Fourier transform with the impact parameter 
range probed.
While data from HERA collider experiments (ZEUS and H1) covered lower $x$ 
and a wide range in $Q^2$, they are statistically limited.  
Furthermore, the HERA proton beams were unpolarized, so ZEUS and H1 
were not able to study the proton-spin dependence in these measurements. 
With its large acceptance, excellent detection capabilities, 
high luminosity and broad range of energies of the polarized 
proton/helium beams available at eRHIC, ePHENIX will provide high precision 
data over a wide range of $x$, $Q^2$ and $t$. 
The wide range in $t$ possible at eRHIC is of crucial importance, 
and will be achieved by integrating Roman Pot detectors in the accelerator 
lattice from the outset. 
Similar measurements performed with ion beams will allow analogous imaging 
of nuclei, allowing the first look at the parton distributions inside 
the nuclei.

The EIC White Paper demonstrates the precision that can be achieved 
in such a program with Deeply Virtual Compton Scattering (DVCS) 
and exclusive $J/\psi$ production. 
The detector requirements for such measurements discussed in the
White Paper and what we propose as ePHENIX are similar. 
For such, we expect ePHENIX will be able to make high impact measurements 
of GPDs.

\subsubsection{Nucleus as a laboratory for QCD}

Electron scattering interactions from nuclei allow key tests of the modification of
parton distribution functions in nuclei of various sizes.  The EIC has the unprecedented energy
reach to probe deep into the low-$x$ quark and gluon region where there are predictions
of significant non-linear evolution effects and possibly the realization of a universal
state of the QCD vacuum at high gluon density.  
In addition, rather than looking at the modified number of deep inelastic scatterings, one can
study via SIDIS the changes in the process of a highly virtual struck quark to color neutralize and eventually
hadronize when in the presence of a nuclear medium.  

\paragraph{Hadronization and its modification in nuclear matter}
\label{sec:hadronization}

\begin{figure}[h]
  \centering
  \includegraphics[width=0.49\linewidth]{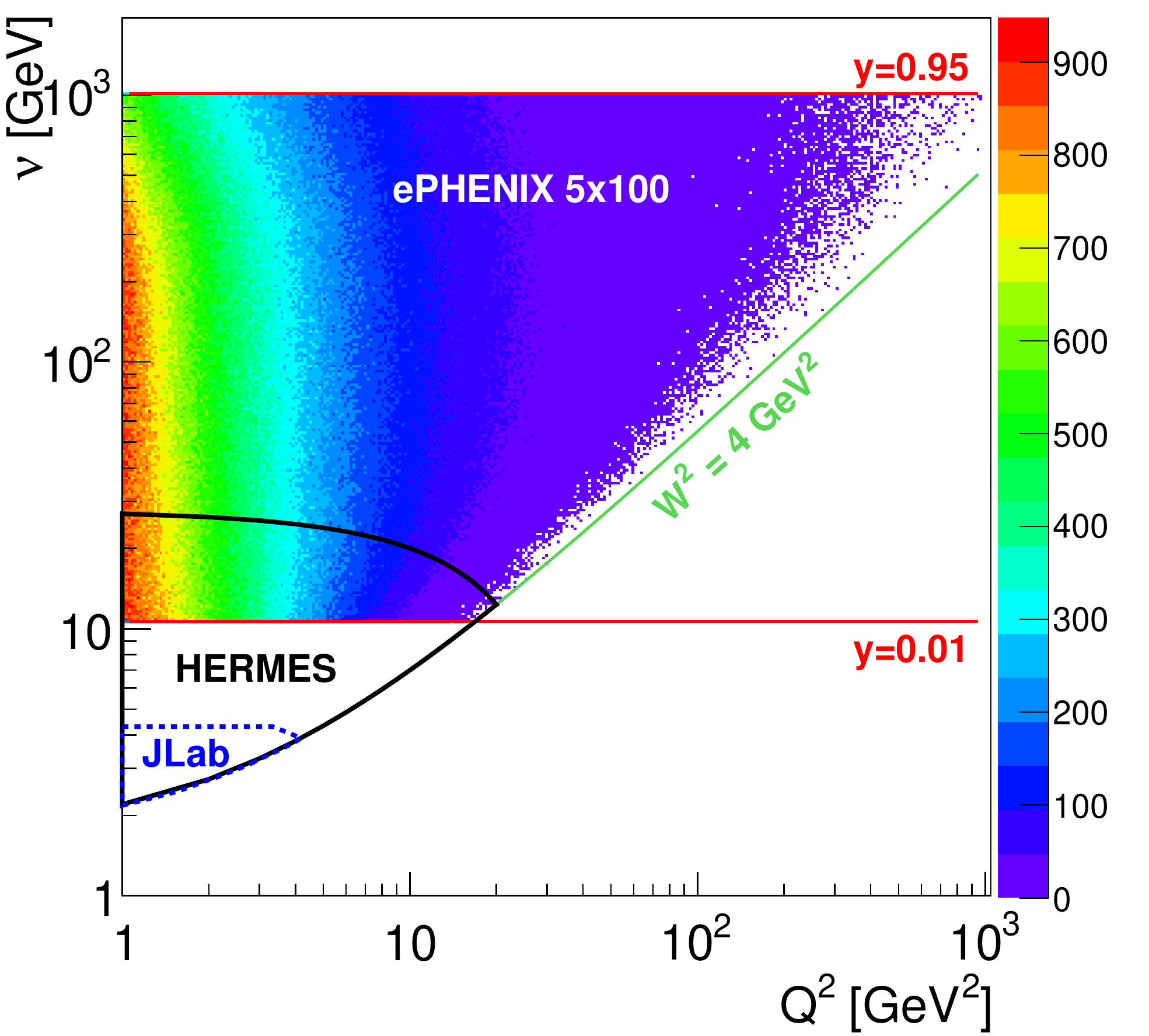}
  \includegraphics[width=0.49\linewidth]{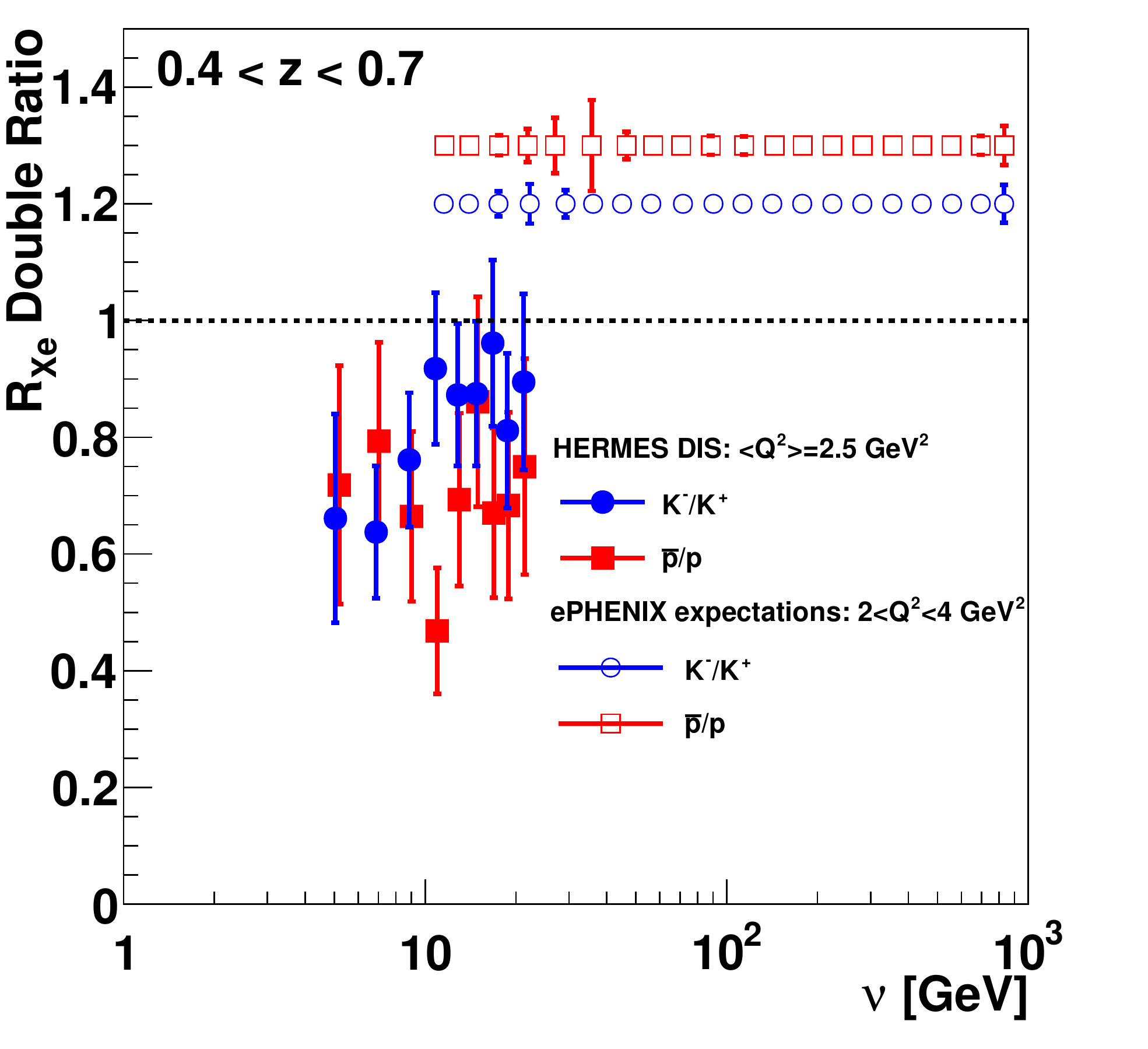}
  \caption[The virtuality $Q^{2}$ and $\nu$ coverage for ePHENIX (EIC)
  measurements with collisions of 5~GeV electrons on 100~GeV/nucleon
  heavy nuclei, data from HERMES on the modified fragmentation from
  xenon targets ($R_\mathrm{Xe}$)]{(left) Shown is the
    very large virtuality $Q^{2}$ and $\nu$ coverage for ePHENIX (EIC)
    measurements with collisions of 5~GeV electrons on 100~GeV/nucleon
    heavy nuclei.  The z-axis color scale shows the relative
    distribution of events from the \pythia event generator.  Also
    shown are the kinematic reach for the CLAS experiment at
    JLab~\cite{Daniel:2011nq} and for the HERMES
    results~\cite{Airapetian:2007vu}.  (right) Experimental data from
    HERMES~\cite{Airapetian:2007vu} on the modified fragmentation from
    xenon targets ($R_{Xe}$) in the range $0.4 < z < 0.7$ and with
    average $\left<Q^{2}\right> = 2.5$~GeV$^{2}$.  The filled points
    are the double ratio for antiprotons relative to protons (red) and
    for $K^{-}$ relative to $K^{+}$ (blue).  ePHENIX will measure with
    precision the modified fragmentation distribution with excellent
    $\pi, K, p$ particle identification over a very broad range of
    $Q^{2}$ and $\nu$.  The open symbols show the expected statistical
    precision for ePHENIX with its particle identification
    capabilities for one bin in $Q^2$, $2<Q^2<4$~GeV$^2$ based on
    2~fb$^{-1}$ at the 5~GeV $\times$ 100~GeV beam energy
    configuration.}
  \label{fig:hadronization}
\end{figure}

Deep inelastic scattering with heavy nuclear targets provides an
effective stop watch and meter stick with which one can measure the
color neutralization and hadronization times, and understand important
details of partonic interactions with the nucleus.  By varying the
size of the nuclear target (at eRHIC all the way up to uranium) and
changing key DIS parameters ($Q^{2}, \nu, z, p_{T}^{2}, \phi$) one can
calibrate this watch and meter stick.  Figure~\ref{fig:hadronization}
shows the kinematic reach for 5~GeV electrons scattering from
100~GeV/nucleon heavy nuclei in terms of the initial virtuality
$Q^{2}$ and the energy of the struck quark in the nuclear rest frame
$\nu$.  Earlier experiments with fixed targets have measured very
interesting modifications in apparent fragmentation functions, and yet
those results are limited to small values of $Q^{2}$ and $\nu$.  In
the case of the published HERMES results~\cite{Airapetian:2007vu} in 
Fig.~\ref{fig:hadronization}, one
observes a dramatic decrease in the number of high-$z$ hadrons (those
with a large fraction of the struck quark momentum) in scattering from
nuclear targets.  There are many possible explanations of the
experimental results, including parton energy loss due to multiple
scattering in the nucleus and induced gluon radiation --- a similar
mechanism has been used to explain the ``jet quenching'' phenomena
discovered in heavy ion collisions at RHIC.  Other theoretical
frameworks predict a strong correlation between a short color
neutralization timescale and high-$z$ resulting processes.  An
excellent review of the various theoretical approaches is given in
Ref.~\cite{Boer:2011fh}.  Figure~\ref{fig:hadronization} also shows the 
expected statistical precision with the ePHENIX PID capabilities 
over the full $\nu$ range in one $Q^2$ bin.

If the struck quark remains an undressed color charge while it
traverses the nucleus, one might expect that the ratio of final state
hadrons ($\pi^{+}, K^{+}, p$ and their anti-particles) would show the
same degree of nuclear modification.  Shown in the right panel of
Figure~\ref{fig:hadronization} are the double ratios of modifications
$R_{Xe}$ with a xenon target for antiprotons to protons and $K^{-}$ to
$K^{+}$.  It is notable that there is a larger suppression for the
hadrons with a larger cross section with nucleons (e.g.
$\sigma_{\overline{p}+N} > \sigma_{p+N}$ and $\sigma_{K^{-}+N} >
\sigma_{K^{+}+N}$).  If this is due to hadronization occurring within
the nucleus, then inelastic collisions can result in the differential
attenuation.  How does this attenuation vary with the energy of the
struck quark?  The EIC realization has the enormous reach in the
energy of the struck quark $\nu$ at fixed $Q^{2}$ to measure the full
evolution with high statistics. As demonstrated in this document,
ePHENIX will have excellent $\pi, K, p$ particle identification to
make exactly this measurement with high statistics.  In addition, one
can vary the virtuality which is also expected to play a significant
role in the length scale probed in the nucleus and thus rate of
initial radiation.

Tests with charm mesons via displaced vertex measurements are not in the initial suite of ePHENIX capabilities,
and could be added with a later inner thin silicon detector.  Measurements of the interactions of charm quarks with the nucleus
would be quite interesting in the context of suppressed radiation due to the ``dead-cone'' effect.  However, the relation
to kinematic variables $z$ and $\nu$ may depend on the balance of DIS events from intrinsic charm as opposed to photon-gluon fusion 
reactions resulting in $c\overline{c}$ pair production.

\paragraph{QCD matter at extreme gluon density}
\label{sec:saturation}

A key goal of any future EIC is to explore the gluonic matter at low $x$, 
where it is anticipated that the density of gluons will saturate as the rate 
of gluon recombination balances that of gluon splitting.  In fact, there are
well known modifications to the quark distribution functions in nuclei that
have  significant $x$ dependence:  high $x$ Fermi motion effects, then the EMC suppression,
anti-shadowing enhancement, and finally nuclear shadowing at the lowest $x$.  
The ePHENIX detector, combined with the large kinematic reach of an $e$$+$A collider, 
is in an excellent position to map this physics out in the gluon sector.

\begin{figure}[h!] 
  \begin{center}
    \includegraphics[trim = 0 0 -20 0, width=0.6\textwidth]{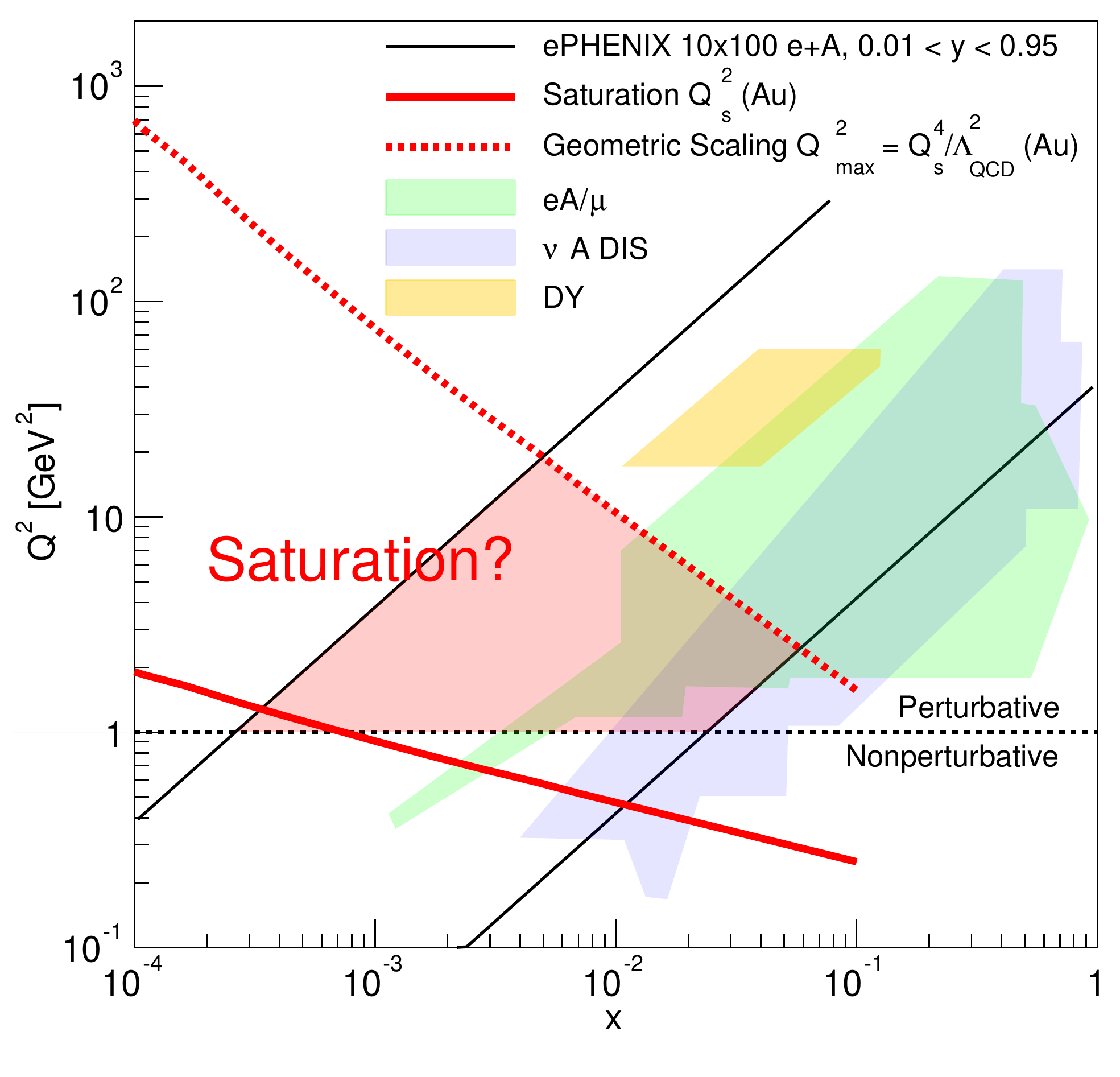}
  \end{center} 
  \caption[Coverage in $x$ and $Q^{2}$ for the EIC and the ePHENIX
  detector for 10~GeV electrons on 100~GeV/nucleon nuclei, and the
  kinematic coverage by previous experiments in $e$$+$A and $\nu$$+$A
  DIS and Drell-Yan measurements]{\label{fig:kin-wf} Shown is the
    coverage in $x$ and $Q^{2}$ for the EIC and the ePHENIX detector
    for 10 GeV electrons on 100 GeV/nucleon heavy nuclei.  The two
    black lines indicate the kinematic coverage with selections on the
    inelasticity $0.01 < y < 0.95$ (which might be slightly reduced
    depending on the final electron purity at low momentum).  Also
    shown are the kinematic coverage by previous experiments in
    $e$$+$A and $\nu$$+$A DIS and also Drell-Yan measurements.  The
    red solid line is an estimate of the $x$ dependence for the
    saturation scale $Q_{s}^{2}$.  The region where this universal
    saturated matter dictates the physics is estimated to extend over
    the geometric scaling region up to $Q^{2}_{max} =
    Q^{4}_{s}/\Lambda^{2}_{QCD}$ shown by the red dashed
    line~\cite{Iancu:2002tr}.  }
\end{figure} 

The lowest $x$ regime with saturated gluon densities is unique to QCD,
as gluons carry the QCD charge, ``color'', and so interact with
themselves.  In order to explore this saturation region, one must
probe nuclear matter at high center-of-mass energy, so as to reach as
low in $x$ as possible while still in the perturbative QCD regime
(i.e., $Q^2>1$~GeV$^2$).  Generally, a saturation scale, $Q_s$, is
defined to indicate the onset of saturation (where the gluon splitting
and recombination balance each other), with $Q_s$ falling as $x$
increases.  In reality the point at which recombination starts to
balance the gluon splitting is a range in $x$ and $Q^2$ and so making
measurements over a wide range in $x$ and $Q^2$ is necessary to fully
understand these effects.

eRHIC will have a significantly lower center-of-mass energy than 
HERA, and so cannot improve upon the minimum $x$ 
probed with measurements in $e$$+$$p$.  However, eRHIC will also be 
capable of accelerating heavy ions in $e$$+$A collisions.  As 
the $x$ probed is related to the resolution of the probe, collisions 
at the same $Q^2$ can resolve significantly lower $x$ due to the 
larger extent of the nucleus: the partons in the highly accelerated
nucleus are probed coherently. This effectively reduces the $x$ 
probed in $e$$+$A collisions by a factor of $A^{\frac{1}{3}}$, with 
$A$ the atomic weight, as this is proportional to the size of the 
nucleus.  At the energies planned for eRHIC, based on measurements 
in $p(d)$$+$A, one expects saturation effects in inclusive DIS in $e$$+$A.

\begin{figure}[t]
  \begin{center}
    \includegraphics[width=0.6\textwidth]{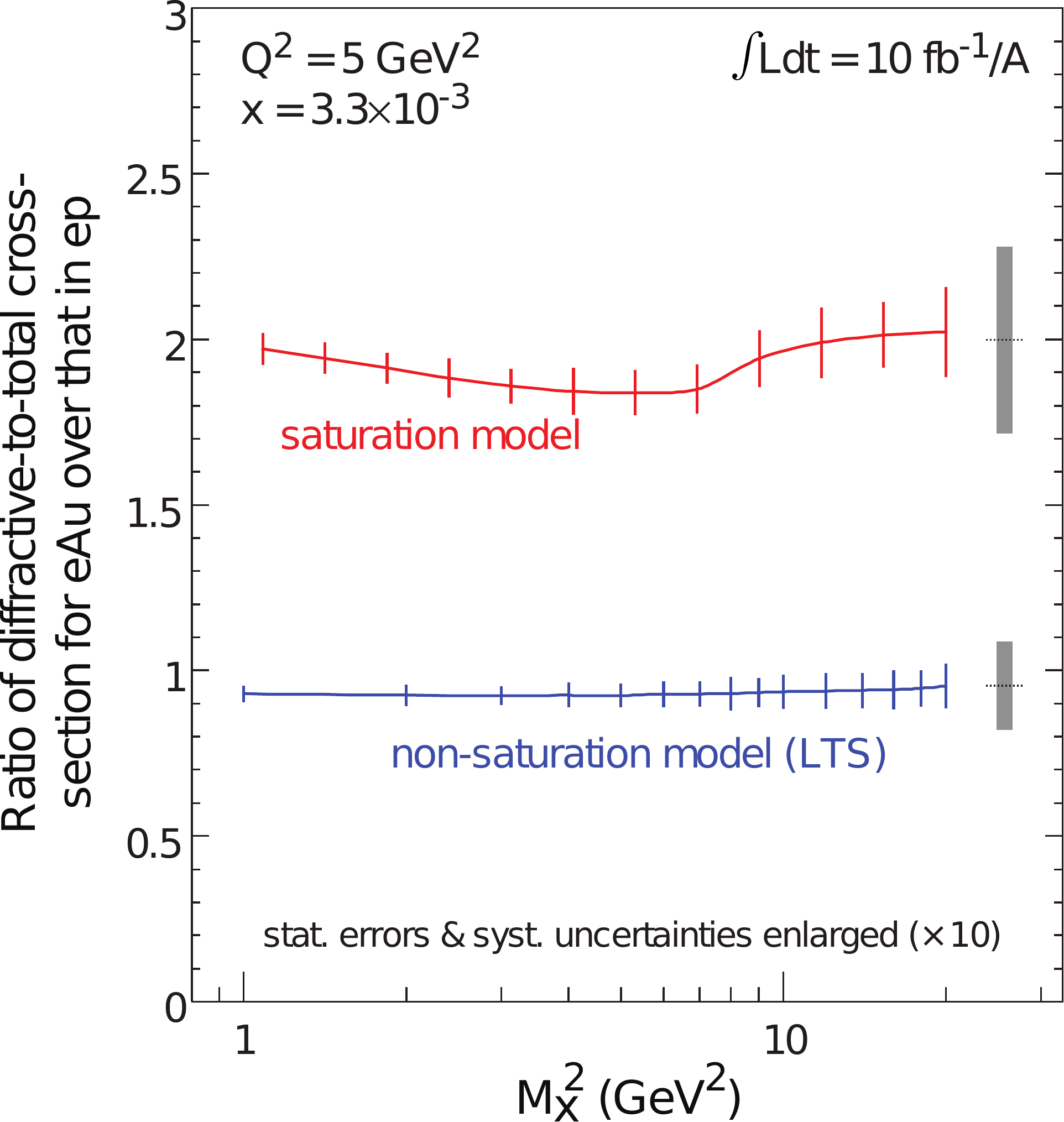}
  \end{center} 
  \caption[Ratio of diffractive-to-total cross-section for $e$$+$Au
  normalized to $e$$+$$p$ plotted as a function of the squared mass of
  the hadronic final state, $M^2_X$]{\label{fig:sat_diff}{[Reproduced
      from Ref.~\cite{Accardi:2012hwp}.]  Ratio of
      diffractive-to-total cross-section for $e$$+$Au normalized to
      $e$$+$$p$ plotted as a function of the squared mass of the
      hadronic final state, $M^2_X$.  The expected uncertainties for
      10~fb$^{-1}$ are scaled by a factor of 10 to be visible.  The
      ePHENIX detector will have similar capabilities as was assumed
      for this plot, and will achieve similar precision.  }  }
\end{figure} 

Figure~\ref{fig:kin-wf} 
shows the $x$ and $Q^2$ coverage of ePHENIX
for the 10~GeV~$\times$~100~GeV/nucleon configuration compared 
with the current fixed target data.  Two red lines are drawn, 
one (solid) showing expectations of $Q^2_s$ in $e$$+$Au and the other (dashed) 
showing the expected turn on of geometric scaling, which relates to the 
saturation scale by $Q^2_{max} = Q^4_s/\Lambda^2_{QCD}$.  The shaded red region
is where ePHENIX can search for saturation effects.

As described in the EIC White Paper~\cite{Accardi:2012hwp}, it can 
be even more effective to explore this region of dense
gluonic matter with diffractive physics, where at least two gluons are
exchanged in the interaction.  Therefore, a primary measurement to 
probe saturation effects at eRHIC will be comparing the diffractive-to-total 
cross-section from $e$$+$$p$ and $e$$+$A.  The ratio of these cross-sections 
will directly relate to the size of any saturation effects.  
Figure~\ref{fig:sat_diff}, taken from the EIC white paper~\cite{Accardi:2012hwp}, 
shows the prediction of one saturation model for this cross-section ratio 
with and without saturation, indicating large possible 
effects.  Note that the statistical and systematic uncertainties in this plot 
are scaled up by a factor of 10 in order to be visible.  This measurement 
relies on measuring events with a large rapidity gap, which is the 
signature of diffractive events due to the fact that the hadron remains 
intact after the scattering (though in the case of ions, the nucleus 
may still break up).  The ePHENIX detector will have wide calorimetric coverage, and so will be 
able to make a measurement of the ratio of diffractive-to-total 
cross-sections with 
comparable precision as shown in Figure~\ref{fig:sat_diff}.

\makeatletter{}\section{Detector Requirements}
\label{chap:ephenix_detector_requirements}

The detector requirements for Deep Inelastic Scattering measurements 
are well established by
previous DIS experiments (H1, ZEUS, HERMES, COMPASS, etc.) and by EIC group
studies~\cite{Accardi:2012hwp,Boer:2011fh}. Table~\ref{table:dreq} summarizes these 
basic requirements and how ePHENIX would meet them. 
After a brief overview of the relevant 
kinematic variables, detailed studies are presented in this chapter.

\begin{table}[hbtp]
\begin{center}
\caption{Detector requirements}
\label{table:dreq}
\begin{tabular}{p{0.49\linewidth}|p{0.49\linewidth}}
\toprule
Detector requirements & Detector solution \\
\midrule
{\bf Electron-ID:}  \newline
High purity (~99\%) identification of the scattered lepton over hadron 
and photon background \newline {\it Important for \egodir and barrel acceptance}& 
Electromagnetic Calorimetry and charged particle tracking \newline
Minimum material budget before EMCal \newline
Good energy and tracking resolution for $E/p$ matching \\
\midrule

{\bf Resolution in $x$ and $Q^2$:}  \newline
Excellent momentum and angle resolution of the scattered lepton to provide high 
survival probability (~80\%) in each ($x$,$Q^2$) bin (important for unfolding) 
\newline {\it Important for \egodir and barrel acceptance} &
High resolution EMCal and tracking in \egodir \newline
Good (tracking) momentum resolution for $E_e'<10$~GeV in barrel \newline
Good (EMCal) energy resolution for $E_e'>10$~GeV in barrel  \\
\midrule

{\bf Hadron identification:}  \newline
$>90\%$ efficiency and $>95\%$ purity 
&
In barrel acceptance: DIRC for $p_h<4$ GeV/c  \newline
In \hgodir:  Aerogel for lower momentum and gas RICH for higher momentum \\
\midrule

{\bf Wide acceptance for leptons and photons in DVCS:}   \newline
Ability to measure DVCS lepton and photon within $-4<\eta<4$ &
EMCal and tracking with good resolution over for lepton and photon measurements covering $-4<\eta<4$ \\
\midrule

{\bf Electron/Photon separation:}   \newline
Separate DVCS photon and electron in \egoing direction &
High granularity EMCal in \egoing direction \\
\midrule

{\bf Measurement of scattered proton in exclusive processes} &
Roman pots in \hgodir \\
\midrule

{\bf "Rapidity gap" measurement capabilities:}   \newline
Measure particles in $-2<\eta<4$ for diffractive event identification &
Hadronic calorimetry covering $-1<\eta<5$, and EMCal covering $-4<\eta<4$\\
\midrule
{\bf Forward Zero-Degree calorimetry:}  \newline
Measure neutrons from nucleus breakup in diffractive $e$$+$A events &
Zero-Degree calorimeter in \hgoing direction planned, in coordination with CAD \\
\bottomrule
\end{tabular}
\end{center}
\end{table}

The suggested ePHENIX detector configuration is shown in Figure~\ref{fig:ePHENIX}.
It is built around the sPHENIX detector,
which is a superconducting solenoid and electromagnetic
and hadronic calorimeter in the central region ($-1<\eta<1$ for pseudorapidity $\eta$).
This proposal would add to that detector the following detector subsystems:

\begin{description}
\item[\egodir ($-4<\eta<-1$):] High resolution Crystal EMCal with GEM tracking.
\item[Barrel ($-1<\eta<1$):] Compact-TPC for low mass tracking and PID for momentum $p<4$ GeV/c with DIRC
\item[\hgodir ($1<\eta<4$):] Hadronic and Electromagentic calorimeters, GEM trackers, and Aerogel-based ($1<\eta<2$) and gas-based RICH for PID up to momentum $p\sim50$~GeV. 
\item[Far-Forward in \hgodir:] Roman Pots and Zero-Degree Calorimeter.
\end{description}

\subsection{Kinematics}
\label{s:kinematics}

In DIS, a lepton is scattered off a target hadron
via the exchange of a virtual boson, which for electron beam energy $E_e<10$~GeV
can always be taken as a virtual photon.  Defining the four-momenta of the incoming and 
scattered electron and the incoming proton as $k$, $k'$ and $p$ respectively, we can
define the following Lorentz invariant quantities:
\begin{alignat}{3}
s   &\equiv& (k + p)^2& &=& 4E_eE_p \label{eq:rts}  \\
Q^2 &\equiv& -q^2 = -(k-k')^2& &=& 2 E_{e}E_{e}'\left(1 - cos\theta\right) \label{eq:Q2}\\
  y &\equiv& \frac{p \cdot q}{k \cdot p}& &=& 1 - \frac{E_{e}'}{E_e} + \frac{Q^2}{4E_e^2} \label{eq:y}\\
  x &\equiv& \frac{Q^2}{2p\cdot q}& &=& \frac{Q^2}{ys} \label{eq:x} \\
\nu &\equiv& \frac{p \cdot q}{M}&  &=& \frac{Q^2}{2Mx} \label{eq:nu}
\end{alignat}
where $s$ is the center-of-mass energy squared, $q$ is the 4-momentum 
transferred from scattered electron and $Q^2$ is the virtuality of the photon 
which gives the resolution scale of the scattering, $y$ is the inelasticity 
of the scattering and $x$ is Bjorken $x$, the fractional momentum carried by 
the struck parton.  Here, we have also written these in the lab frame in 
terms of the measured scattering angle, $\theta$ and the energies of the 
proton and incoming and scattered electron, $E_p$, $E_e$ and $E_e'$, 
respectively, under the approximation that the electron and proton mass are 
small compared to the beam energies.

For inclusive DIS, where only the kinematics of the scattered lepton are measured, 
Eq.~\ref{eq:rts}--\ref{eq:nu} fully describe the event.  For SIDIS, in which a final state
hadron is also measured, additional variables are needed.  The fraction of the scattered parton's
momentum carried by the hadron is defined as
\begin{equation}
\label{eq:z}
z \equiv \frac{p_h \cdot p}{q\cdot p}
\end{equation}
where $p_h$ is the four-momentum of the measured hadron.  Further, we can define 
$p_{h\perp}$ as the transverse momentum of the hadron w.r.t. the virtual photon, in the 
center-of-mass frame of the proton (or ion) and virtual photon.  

For exclusive processes, in addition to the scattered lepton, 
the final state photon in DVCS or meson in Deeply Virtual Meson Production 
as well as the scattered proton are measured.  
In this case, another kinematic variable is introduced -- 
the squared momentum transfer to the proton, $t$, 
defined as
\begin{equation}
\label{eq:t}
t \equiv (p' - p)^2
\end{equation}
where $p'$ is the four-momentum of the scattered proton.

\subsection{Inclusive DIS and scattered electron measurements}

In inclusive DIS, where only the kinematics of the scattered electron are 
necessary, the primary requirements of any detector are electron 
identification and sufficient resolution in $x$ and $Q^2$, 
which in turn mandates good energy and angle resolution for the scattered 
electron measurements (Eq.~\ref{eq:Q2}--\ref{eq:x}).  

\subsubsection{Electron Identification}

\begin{figure}[ht]
\begin{center}
\includegraphics[trim=0 0 0 370,clip,width=0.75\textwidth]{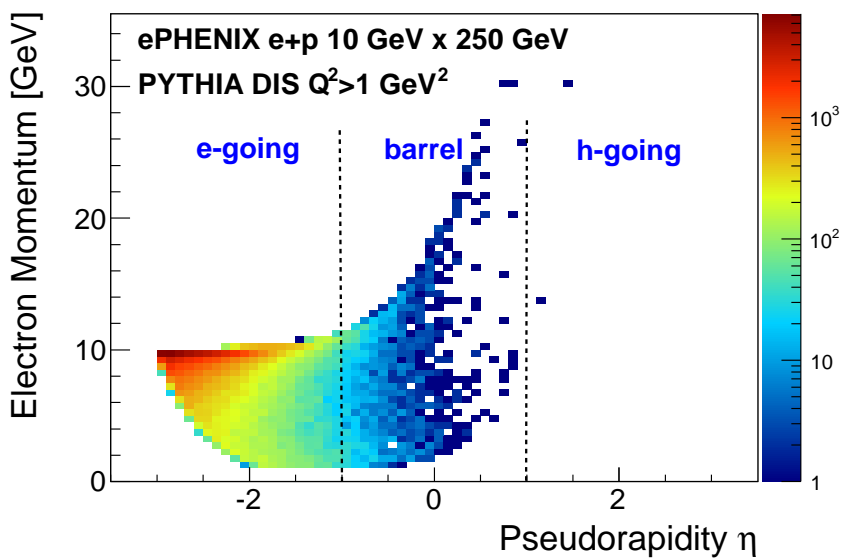}
\end{center}
\caption[The distribution of scattered electrons in pseudorapidity and
energy from \pythia DIS simulations for $e$$+$$p$ collisions with
$10~\mathrm{GeV} \times 250~\mathrm{GeV}$ beam energies]{Shown is the
  distribution of scattered electrons in pseudorapidity and energy.
  The results are from \pythia DIS simulations for $e$$+$$p$
  collisions with $10~\mathrm{GeV} \times 250~\mathrm{GeV}$ beam
  energies.  The events are selected as DIS with $Q^{2} >
  1$~GeV$^{2}$.  }
\label{fig:eeta}
\end{figure}

In collider geometry, the DIS electrons are scattered mainly in the \egodir
and central rapidities (barrel acceptance), see Figure~\ref{fig:eeta}. 
Central rapidity selects scatterings with higher
$Q^2$ and higher $x$ (due to its correlation with $Q^2$).
The higher the electron beam energy, the more scattering there is in 
the \egodir. The energy of the scattered electron varies
in the range from zero up to the electron beam energy and even to higher values
for electrons detected in the barrel acceptance, see Figure~\ref{fig:eeta}.

\begin{figure}[ht]
\begin{center}
\includegraphics[width=0.95\textwidth]{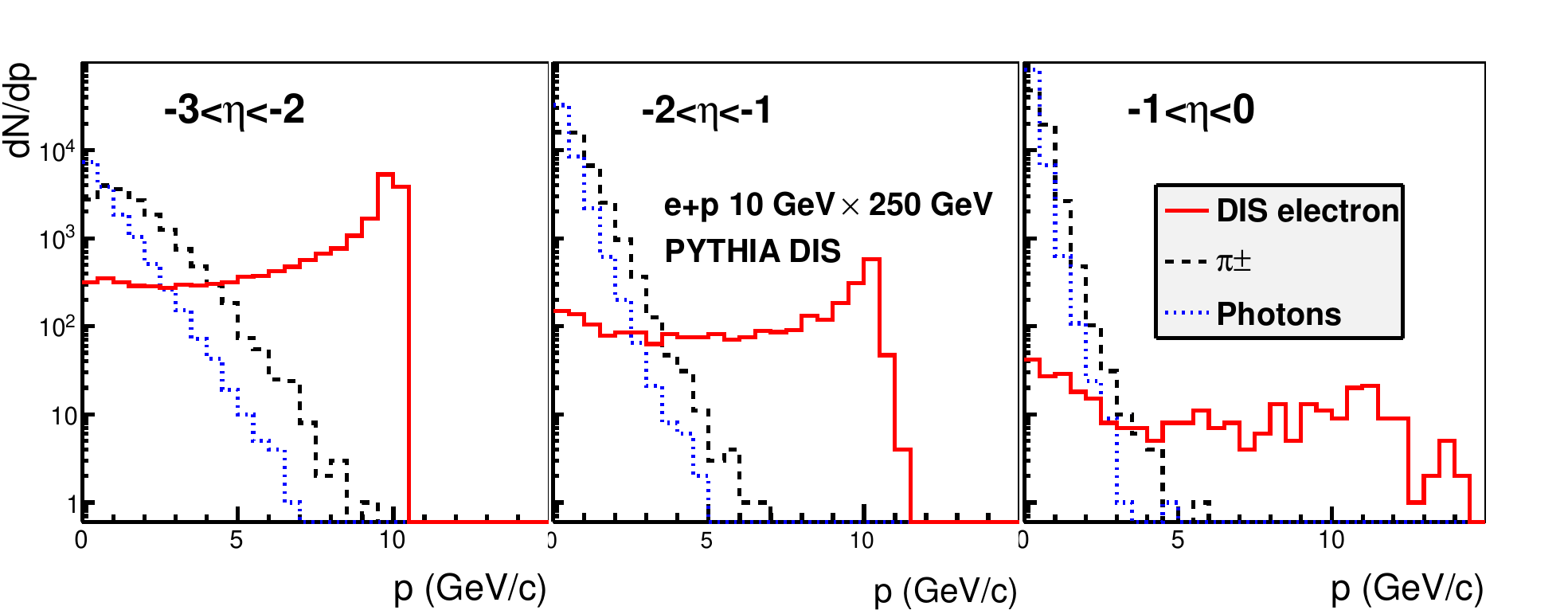}
\end{center}
\caption[Momentum spectra for scattered electrons, charged pions and
photons for the $10~\mathrm{GeV} \times 250~\mathrm{GeV}$ beam energy
configuration]{For $10~\mathrm{GeV} \times 250~\mathrm{GeV}$ beam
  energy configuration: Momentum spectra for scattered electron (red),
  charged pions (black) and photons (blue).  }
\label{fig:spectra}
\end{figure}

Collider kinematics allow clear separation of the scattered electrons from
other DIS fragments --- hadrons and their decay products --- which are detected
preferably in the \hgodir, leaving much softer spectra in the central region and
the \egodir. Figure~\ref{fig:spectra}
shows scattered electron momentum spectra along
with photon (mainly from hadron decays) and charged pion spectra.  For the 10~GeV
electron beam, hadronic and photonic backgrounds are small above $\sim5$~GeV/c, but 
increase rapidly at lower momenta.

\begin{figure}[ht]
\begin{center}
\includegraphics[width=0.95\textwidth]{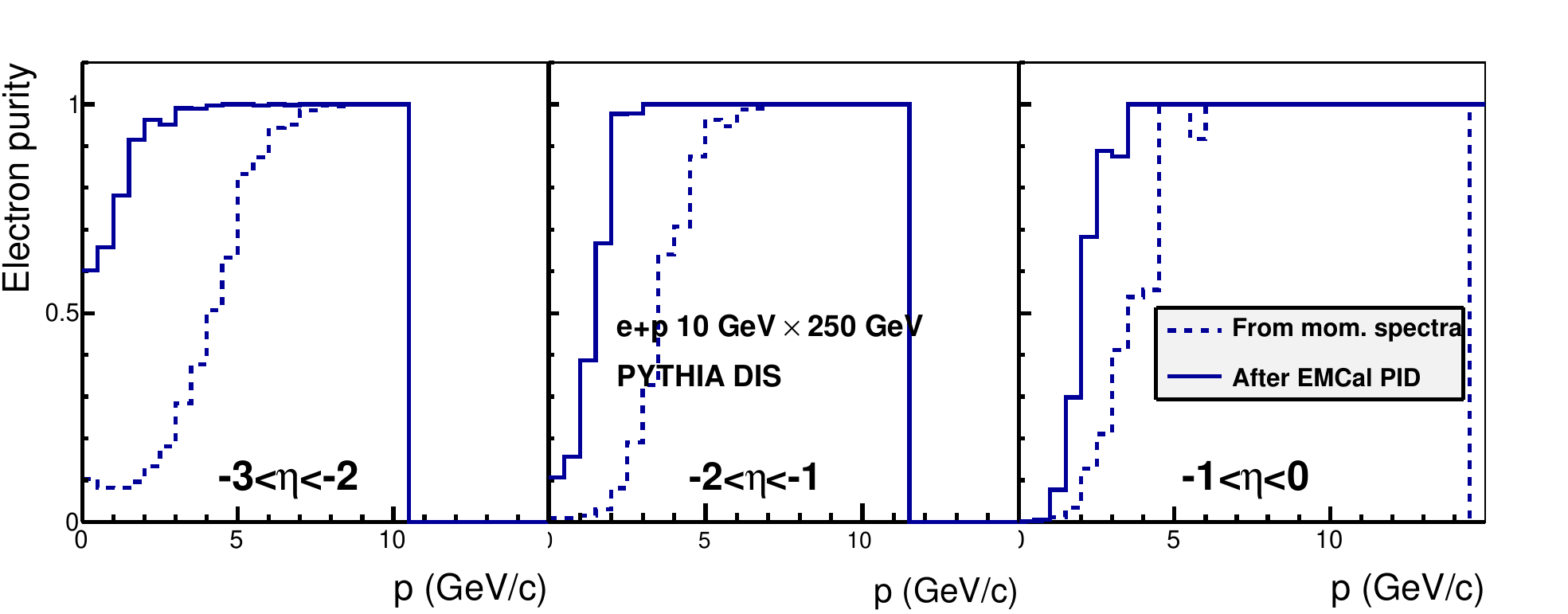}
\end{center}
\caption[For 10 GeV $\times$ 250 GeV beam energy configuration:
fraction of charged particles from DIS electrons before and after
electron identification with the EMCal response and E/p matching ]{For
  10 GeV $\times$ 250 GeV beam energy configuration: The fraction of
  charged particles from DIS electrons   before electron identification (dotted) and after identification
  with the EMCal response and E/p matching (solid).  }
\label{fig:epurity}
\end{figure}

\begin{figure}[ht]
\begin{center}
\includegraphics[width=0.75\textwidth]{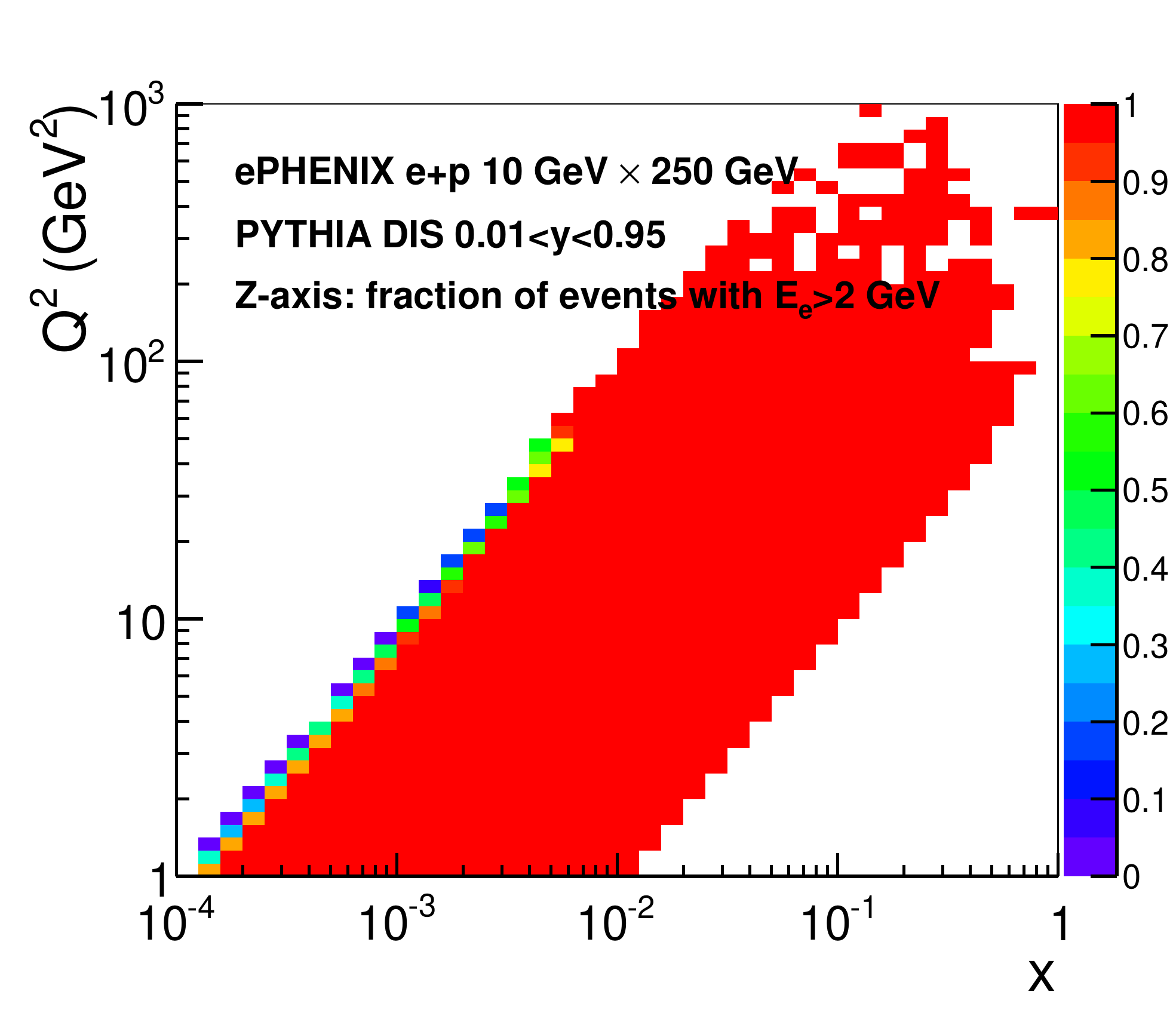}
\end{center}
\caption[For 10 GeV $\times$ 250 GeV beam energy configuration: the
fraction of events in $(x,Q^2)$ space surviving after a $>2$~GeV
energy cut on the DIS scattered electron ]{For 10 GeV $\times$ 250 GeV
  beam energy configuration: The color axis indicates the fraction of
  events in $(x,Q^2)$ space surviving after a $>2$~GeV energy cut on
  the DIS scattered electron.  }
\label{fig:kin_2gev}
\end{figure}

The different response of the EMCal to hadrons and electrons,
along with a direct comparison of energy deposited in the EMCal
and momentum measured in the tracking system (i.e., $E/p$ matching)
provides a significant suppression of hadronic background
in DIS scattered electron measurements:
from a factor of 20--30 at momenta near 1 GeV/c to a factor of greater than
100 for momenta above 3~GeV/$c$.
Figure~\ref{fig:epurity} shows the effectiveness of electron identification 
with the EMCal and tracking, 
providing high purity for DIS scattered electron measurements
at momenta $>$3 GeV/$c$ for the 10~GeV electron beam 
(and $>$1.5 GeV/$c$ for the 5~GeV electron beam).
The evaluations above are done with a parametrized
response of the EMCal to hadrons and electrons, and EMCal and tracking
resolutions described in Sections~\ref{sec:EMCAL}
and \ref{sec:VertexTracking}.
Further enhanced electron identification is expected from the use of the 
transverse shower profile.  
We are also studying possible electron identification improvement with 
longitudinal segmentation in the crystal calorimeter in the \egodir.
These are expected to move the detector capabilities for high purity electron 
identification down to 2 GeV/c (1 GeV/c) for 10 GeV (5 GeV) electron beam, 
which only marginally limits the $(x,Q^2)$ space probed in our measurements, 
see Figure~\ref{fig:kin_2gev}.

Photon conversion in material between the collision point and the tracker 
(mainly beam pipe, with thickness as small as $0.3\%$ of radiation length) 
is not expected to contribute sizable background. 
Moreover, conversion electron-positron pairs will be well identified by our 
tracking system in the magnetic field and additionally suppressed by E/p 
matching cut. 
A detailed GEANT simulation study is ongoing to quantify this effect.

\subsubsection{Resolution in $x$ and $Q^2$  and bin survival probability}
\label{sec:resolution}

Measurements of the scattered electron energy and polar angle impact
the DIS kinematic reconstruction, Eq.~\ref{eq:Q2}--\ref{eq:x}.
Unfolding techniques are generally used to correct for smearing in
$(x,Q^2)$ due to detector effects, and the effectiveness of this
technique depends on the degree to which events migrate from their
true $(x,Q^2)$ bin to another.  This migration can be characterized by
the likelihood of an event remaining in its true $(x,Q^2)$ bin --- the
bin survival probability.

The energy resolution $\sigma_E$ is directly propagated to $\sigma_{Q^2}$, 
so that $\sigma_{Q^2}/Q^2=\sigma_E/E$. The EMCal energy and tracking
momentum resolutions will provide excellent precision for $Q^2$
measurements.  Conversely, the $\sigma_x$ resolution is magnified by a
factor of $1/y$ as $\sigma_x/x=1/y \cdot \sigma_E/E$, and so the energy
resolution in this approach effectively defines the limit of our
kinematic reach at low y.

\begin{figure}[ht]
  \begin{center}
    \includegraphics[width=0.95\textwidth]{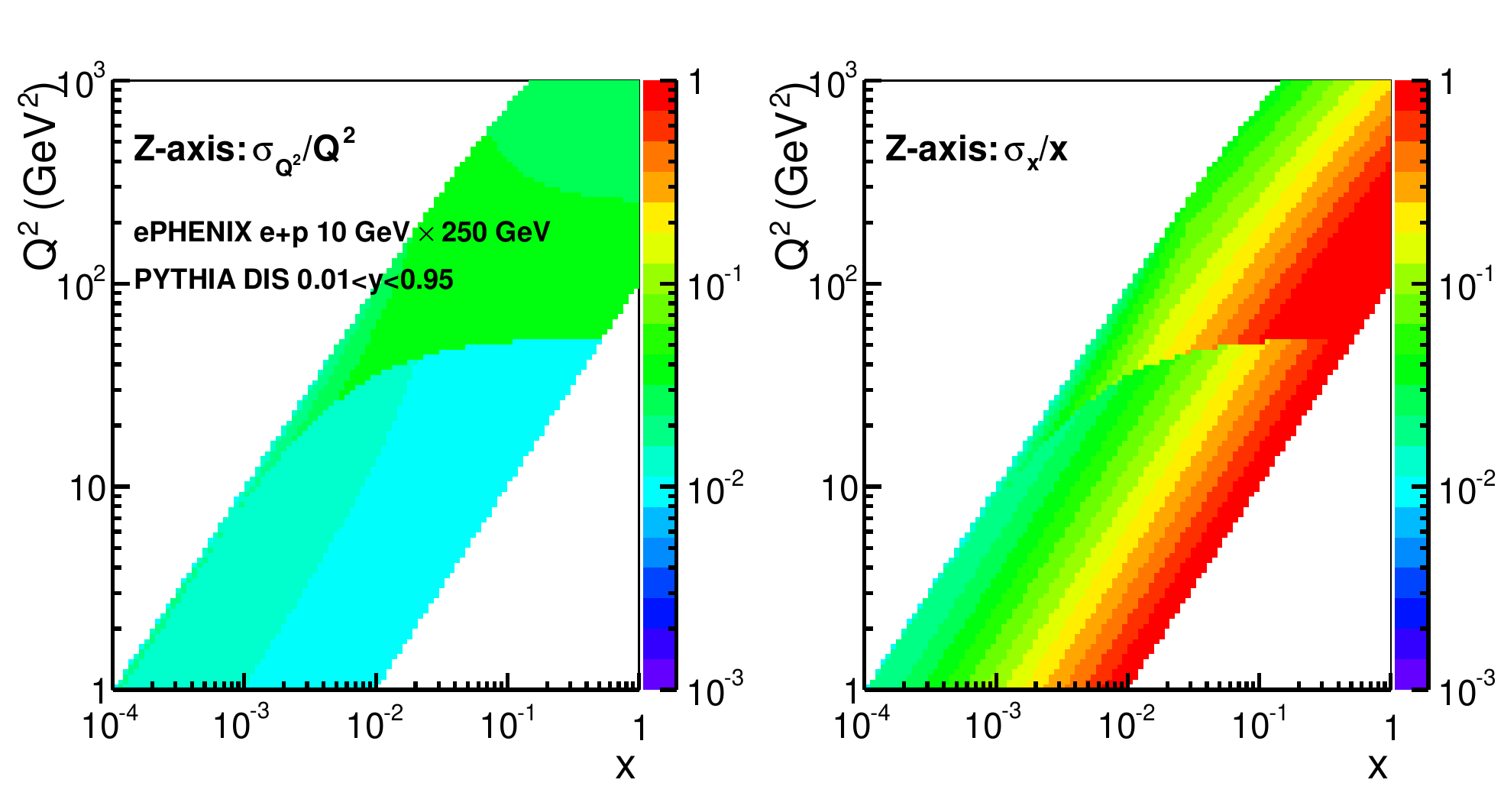}
  \end{center}
  \caption[For $10~\mathrm{GeV} \times 250~\mathrm{GeV}$ beam energy
  configuration: the relative resolution for $Q^2$ and $x$ as a
  function of ($x,$$Q^2$)]{For $10~\mathrm{GeV} \times
    250~\mathrm{GeV}$ beam energy configuration: the relative
    resolution for $Q^2$ (left) and $x$ (right) as a function of
    ($x,$$Q^2$).  }
\label{fig:res}
\end{figure}

\begin{figure}[ht]
\begin{center}
\includegraphics[width=0.75\textwidth]{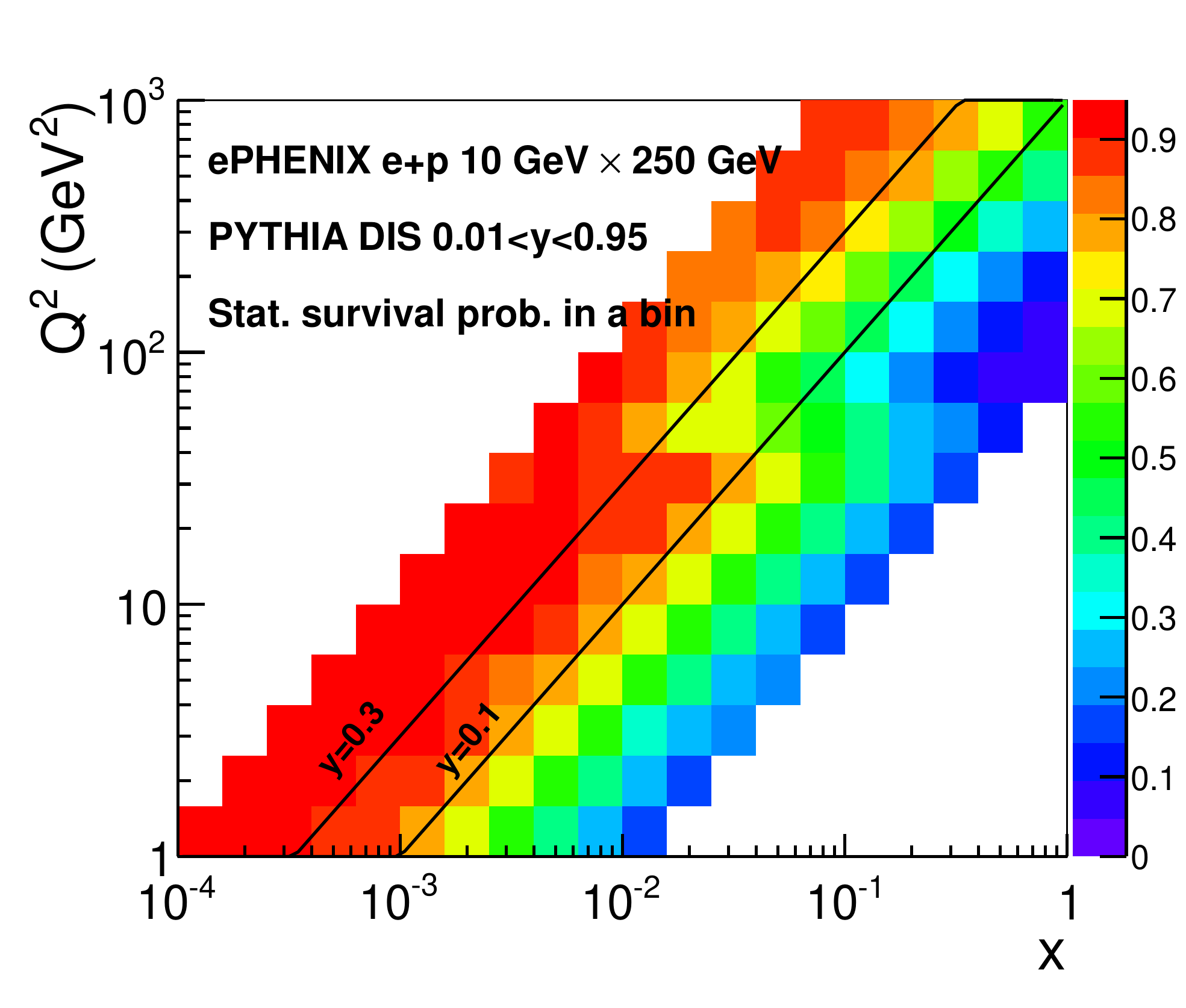}
\end{center}
\caption{For $10~\mathrm{GeV} \times 250~\mathrm{GeV}$ beam energy
  configuration: Statistics survivability in ($x$,$Q^2$) bins.}
\label{fig:res2}
\end{figure}

Figure~\ref{fig:res} shows the relative resolution in $Q^2$ and $x$
measurements using the standard ``electron'' method, in which the 
scattered electron is measured.  While the $Q^2$ relative
uncertainty, $\sigma_{Q^2}/Q^2$, is better than 10\% over whole
$x$-$Q^2$ acceptance, the relative uncertainty on $x$, $\sigma_{x}/x$,
clearly demonstrates its $y$-dependence (the same $y$ points are on
the diagonal, as from Eq.~\ref{eq:x}, $Q^2=syx$). The step in resolution
around $Q^2=50$~GeV$^2$ in
these plots corresponds to the transition from the \egodir to the barrel
acceptance, which differ mainly in the resolution of the different
electromagnetic calorimeters covering those two regions of the
acceptance.  All of this translates to the statistics survival probability
in a bin shown in Figure~\ref{fig:res2}, which is calculated for five bins per
decade in each of $x$ and $Q^2$.  The survival probability is
$>80\%$ for $y>0.1$ in the \egodir and for $y>0.3$ in the barrel acceptance.

The effect of the polar angle resolution $\theta$ in 
Eq.~\ref{eq:Q2}--\ref{eq:x}, is the biggest for forward scattering 
(small $\theta$). It was found that crystal EMCal position resolution 
(better than 3~mm for $>1$~GeV electrons, see Section~\ref{sec:CrystalEMCAL}) 
provides enough precision for scattered electron angle measurements, 
so that it affects the statistics migration in bins on 
Figure~\ref{fig:res2} only marginally.

The Jacquet-Blondel method using the hadronic final state is an
alternative approach to reconstruct DIS kinematics.  Its resolution
for inelasticity $y$, and hence for $x$, is nearly flat, so it
provides much better precision for $x$ determination than the
``electron'' method, in the region with small $y$. It is also
better in the higher $Q^2$ region corresponding to the barrel
acceptance, where the resolution of the ``electron'' method is
limited by the EMCal resolution.

The Jacquet-Blondel method requires the measurement of all final state
hadrons produced in $e$$+$$p$ or $e$$+$A scattering.  A study with
the \pythia generator shows that the precision of this approach does
not deteriorate if the hadron detection capabilities are limited to
$|\eta| < 4$. This method provides relative precision for the
measurement of $x$ of better than 20\%, which satisfies the bin
statistics migration criteria discussed above.  It was found that for
$y<0.3$ the precision of this approach deteriorates only slightly when
hadron measurements are limited to the barrel and forward acceptance
$-1<\eta<4$ (the acceptances we plan to equip with hadron
identification capabilities, see
Section~\ref{sec:HadronPIDDetectors}). As was shown above,
measurements at higher $y$ are well provided by the ``electron'' method.

Therefore, combining the electron and hadronic final state measurements provides
precise determination of basic kinematic variable $x$, $y$ and $Q^2$
in the whole kinematical space.

QED radiative effects (radiation of real or virtual photons) are
another source of smearing which is usually corrected with unfolding
techniques. Unlike energy-momentum resolutions which introduces
Gaussian-like smearing, radiative corrections are tail-like. 
They can be responsible for as much as 10--20\% of statistics migrating away 
from a bin, and dominate over energy-momentum smearing at higher $y$
(compare to Figure~\ref{fig:res2}).

\subsection{Semi-inclusive DIS and hadron ID}

As was discussed in
Chapter~\ref{chap:physics_goals},
measurements of hadrons in SIDIS events are necessary to determine
both the (sea)quark separated helicity distributions and TMDs.  It is
also important for understanding the hadronization process in nuclear
matter.  For these measurements, one needs to identify the hadron,
particularly in the case of pions and kaons.  In this section, we
discuss the kinematic ranges of interest for pions, kaons and protons,
and in
Chapter~\ref{chap:ephenix_detector_concept},
we discuss technology choices which can effectively make these
measurements.

\begin{figure}[ht]
  \centering
  \includegraphics[trim=0 0 20 370, clip, width=0.8\textwidth]{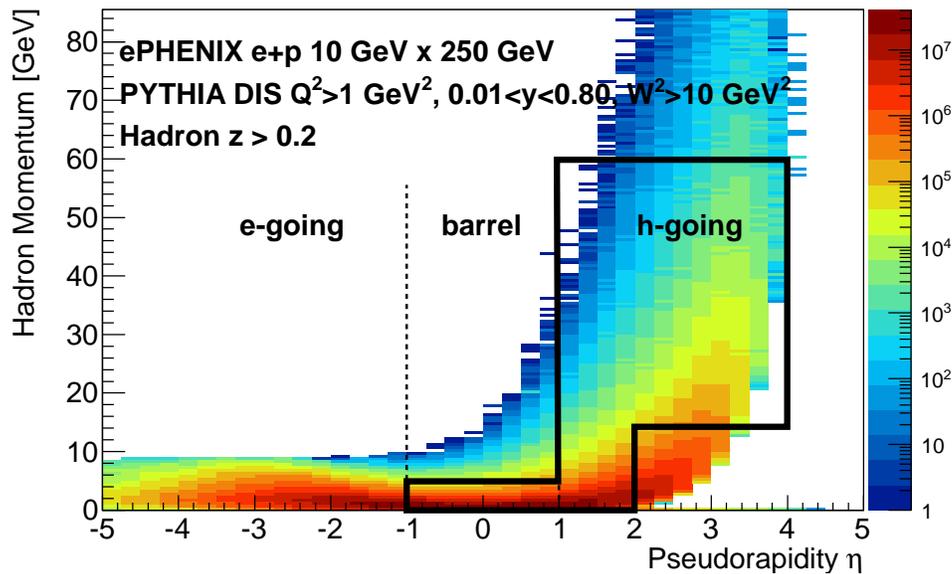}
  \caption[ Distribution of hadrons from DIS events in $e$$+$$p$ as a
  function of momentum and pseudorapidity, based on \pythia
  simulations of the $10~\mathrm{GeV} \times 250~\mathrm{GeV}$ beam
  energy configuration]{Shown is the distribution of hadrons from DIS
    events in $e$$+$$p$ as a function of momentum and pseudorapidity,
    based on \pythia simulations of the $10~\mathrm{GeV} \times
    250~\mathrm{GeV}$ beam energy configuration.  The black outline
    indicates the pseudorapidity and momentum range covered for kaons
    by the planned PID detectors in ePHENIX.}
  \label{fig:250x10_mometa}
\end{figure}

Figure~\ref{fig:250x10_mometa} shows the yields of positively charged
hadrons as a function of momentum and pseudorapidity for the 10~GeV 
$\times$ 250~GeV beam configuration.  A minimum $z$
cut of $z > 0.2$ to remove soft physics effects and beam remnant is
applied.  For $\eta<0$, the hadron momenta are limited by the electron 
beam momentum, while in the \hgodir, the hadron
momenta extend almost to the full proton beam energy.
The results are similar for other beam energy configurations.

\begin{figure}[ht]
\centering
\includegraphics[width=0.48\textwidth]{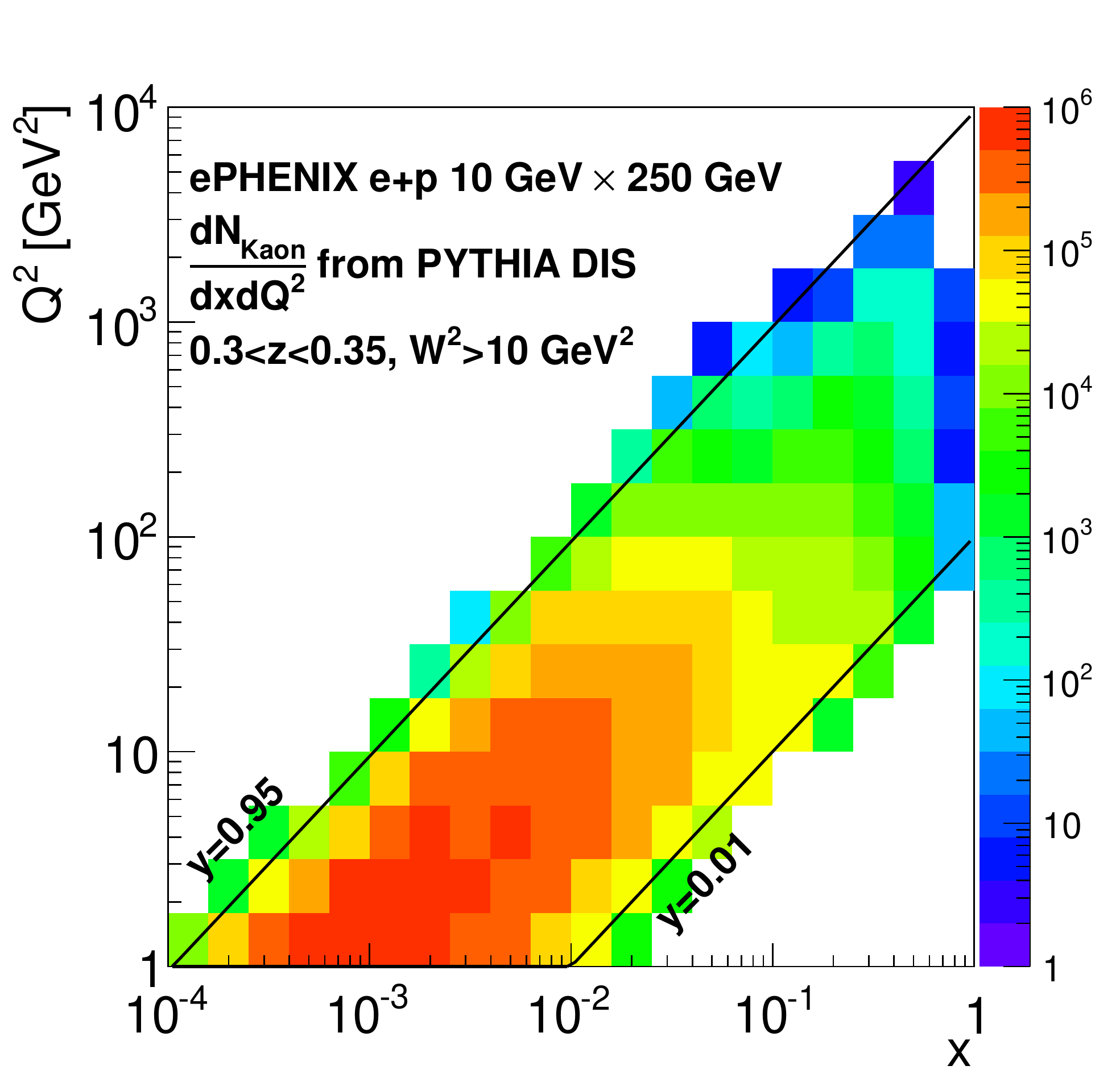}
\includegraphics[width=0.48\textwidth]{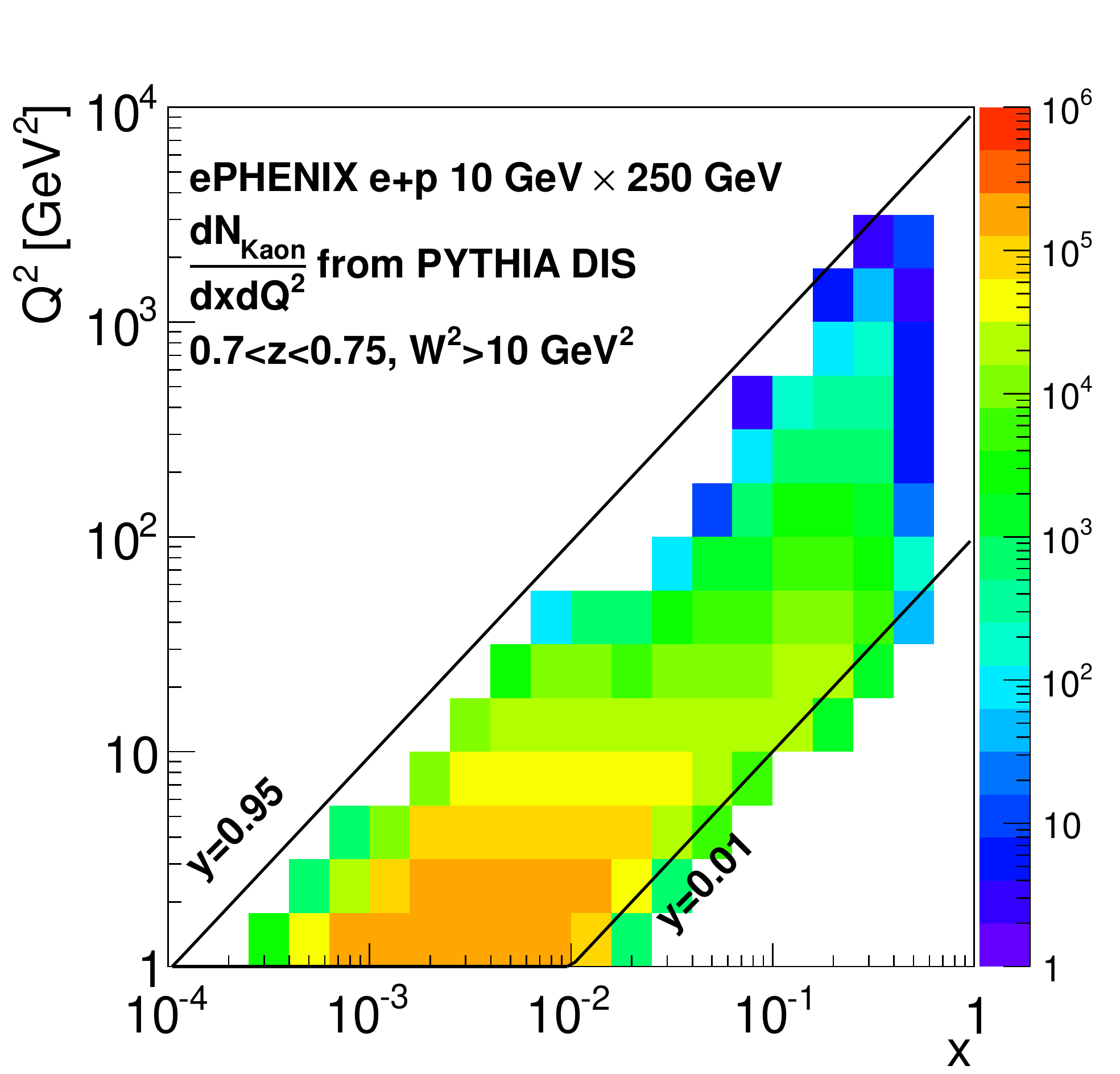}
\caption{$x$ and $Q^2$ distribution of events with kaons which can be
  identified with the ePHENIX PID detectors in expected binning at 
  (left) low and (right) high $z$. }
\label{fig:SIDISxQ2}
\end{figure}

As was stated above, ePHENIX will have three PID systems: (1) a DIRC
covering $|\eta| < 1$ providing $\pi$-$K$ separation below
3.5--4~GeV/$c$ (depending on purity and efficiency requirements), (2)
an aerogel based RICH covering $1<\eta<2$ providing $\pi$-$K$
($K$-$p$) separation below 6 (10)~GeV/$c$ and (3) a gas based RICH
covering $1<\eta<4$ providing $\pi$-$K$ separation for
$3<p<50$~GeV/$c$ and $K-p$ separation for $15<p<60$~GeV/$c$ (depending
on the balance between efficiency and purity chosen).  Based on these
numbers, the PID for kaons would cover the $\eta$ and $p$ region
outlined in black in Figure~\ref{fig:250x10_mometa}.  The resulting
ePHENIX $x$ and $Q^2$ coverage for SIDIS events with an identified 
kaon is shown in Figure~\ref{fig:SIDISxQ2},
for low ($0.30 < z <0.35$) and high ($0.70<z<0.75$) $z$ bins, along
with lines indicating the accessible DIS $y$ range ($0.01<y<0.95$).

\begin{figure}[ht]
  \centering
  \includegraphics[width=0.95\textwidth]{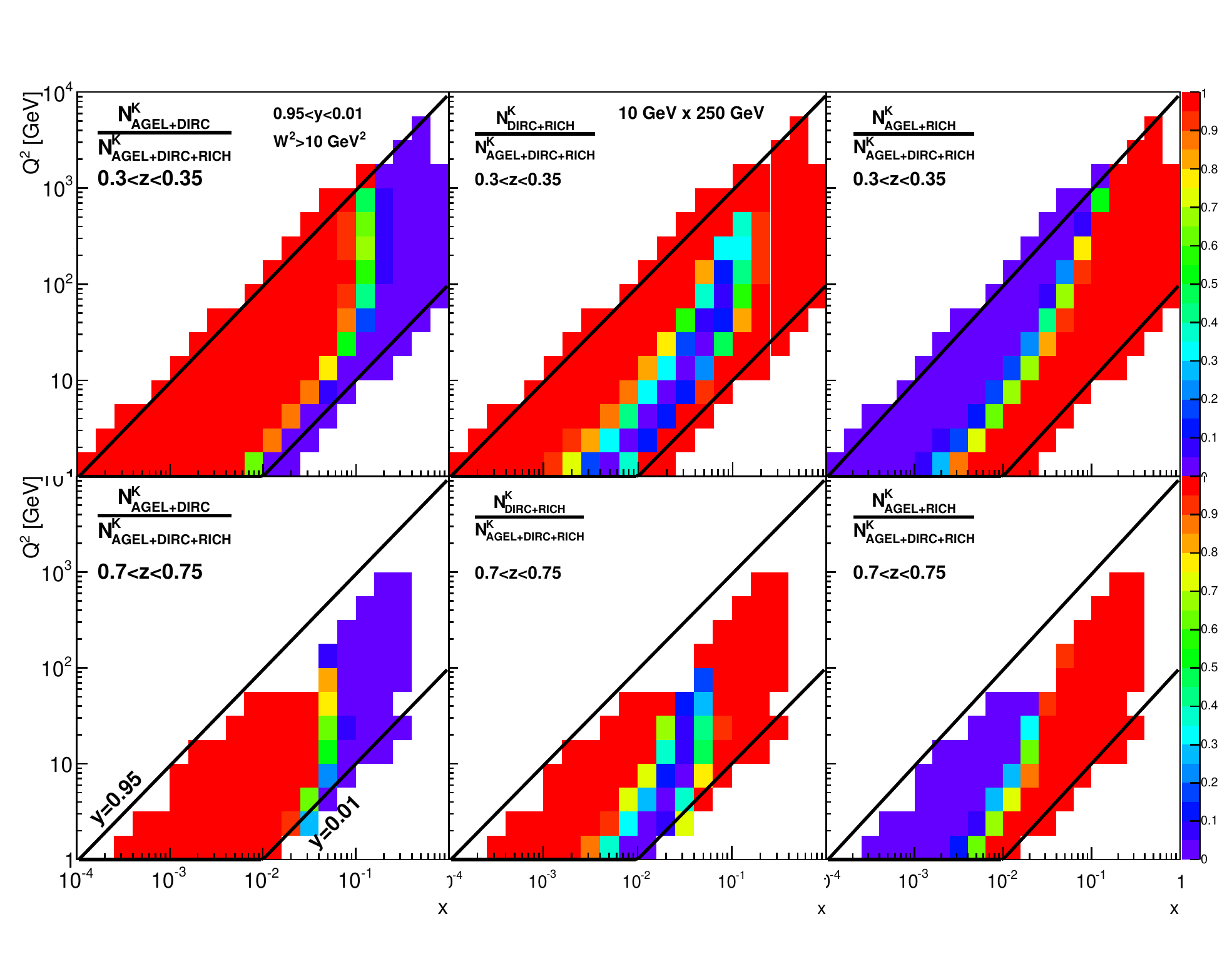}
  \caption{Efficiency as a function of $x$ and $Q^2$ of kaon
    identification when comparing to baseline ePHENIX design with a
    DIRC, RICH and Aerogel when one of these subsystems is removed.
    The top three plots are for low $z$ ($0.3<z<0.35$) and the bottom
    three are for high $z$ ($0.7<z<0.75$).  Also shown are lines
    indicating different values of $y$.}
  \label{fig:kaon_pid_impact}
\end{figure}

Figure~\ref{fig:kaon_pid_impact} shows the impact on the $x$ and $Q^2$
coverage of removing one of the three PID detectors planned for
ePHENIX at low and high $z$.  The plots show the ratio of kaon yields
when using only two PID detectors to those with all three detectors
(i.e., standard ePHENIX).  If the gas-based RICH detector is removed
(left), the high $x$ reach, particularly at high $Q^2$, is lost.  If
the aerogel-based RICH is removed (middle), sensitivity to the region
of moderate $x$, $Q^2$ and $y$ is lost.  Finally, if the DIRC is
removed, significant kinematic coverage at low $x$, as well as
moderate $x$ and high $Q^2$ is lost.  To achieve a wide $x$ and $Q^2$
coverage, all three detectors are necessary. Extending the
aerogel-based RICH to $\eta>2$ does not extend the kinematic coverage;
the momentum range covered by such a detector corresponds to very low
values of $y$.

\subsection{Exclusive DIS}

\begin{figure}[ht]
  \begin{center}
    \includegraphics[width=0.75\textwidth]{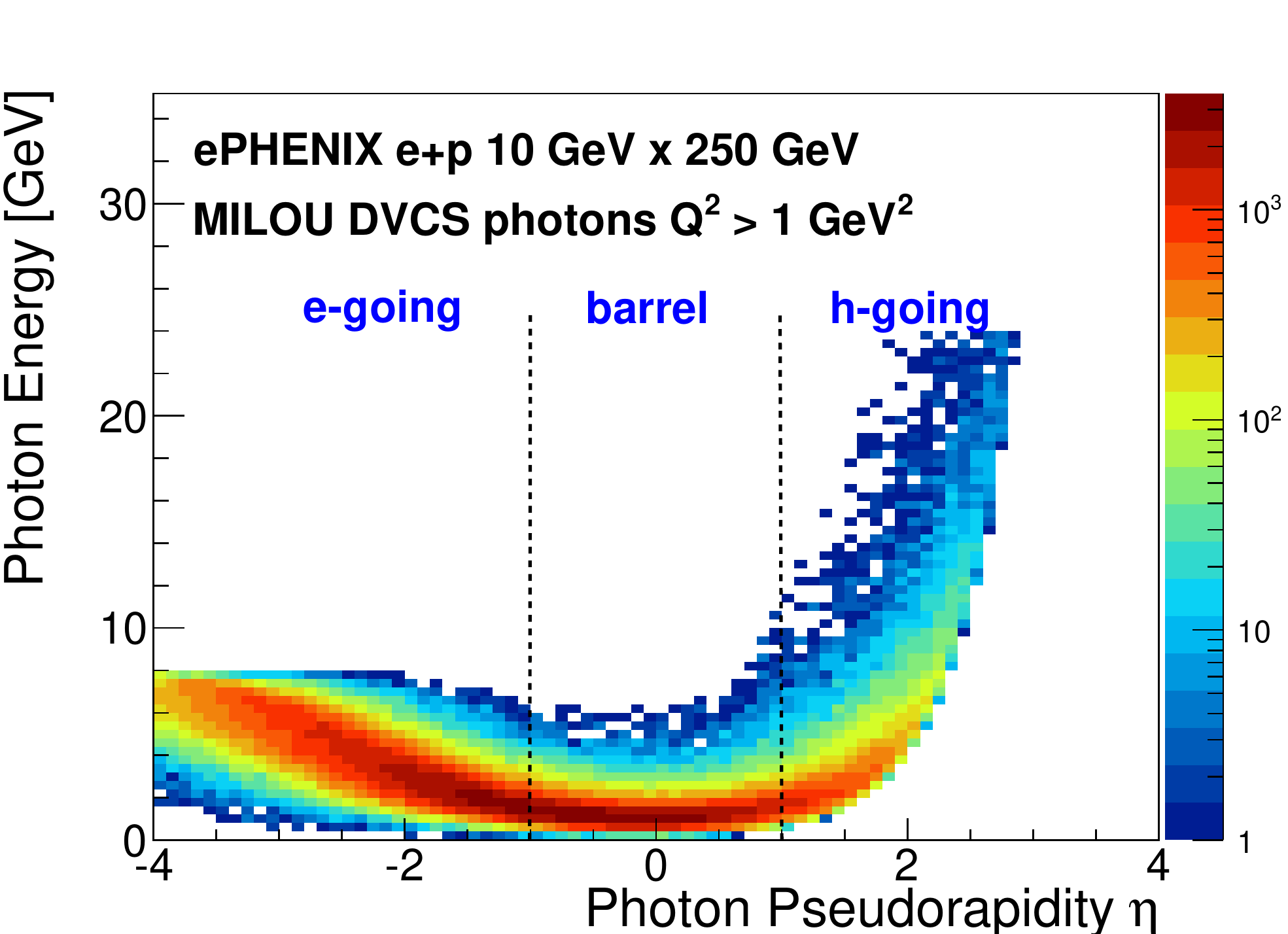}
  \end{center}
  \caption{For the $10~\mathrm{GeV} \times 250~\mathrm{GeV}$ beam
    energy configuration: DVCS photon energy vs pseudorapidity
    distribution; the $z$-axis scale shows the relative distribution
    of events from the \milou event generator.  }
  \label{fig:dvcs_eeta}
\end{figure}

Among exclusive processes, Deeply Virtual Compton Scattering
(DVCS) is of special interest (see
Section~\ref{sec:tomographic_imaging}). The produced DVCS photon
energy versus pseudorapidity distribution is shown in
Figure~\ref{fig:dvcs_eeta}.  Most of the photons fall in the \egodir and
the barrel (central rapidity) acceptance.  The photon energy for $-1<\eta<1$ 
varies in the range $\sim1$--4~GeV/$c$ and is nearly independent of the 
beam energy in the range considered for eRHIC.  Photons in the \egodir are
more correlated with the electron beam and have energy from
1~GeV up to electron beam energy.

\begin{figure}[ht]
  \begin{center}
    \includegraphics[width=0.95\textwidth]{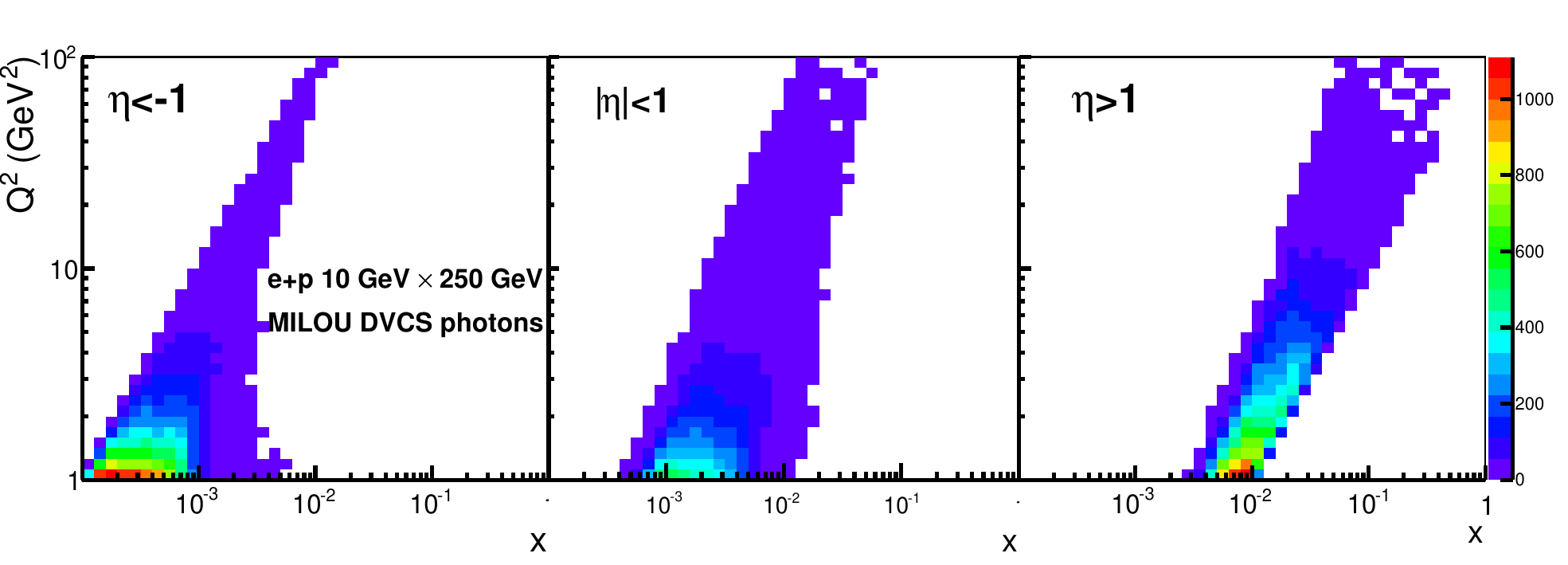}
  \end{center}
  \caption{For $10~\mathrm{GeV} \times 250~\mathrm{GeV}$ beam energy
    configuration: $x$-$Q^2$ coverage for DVCS events with photon
    detected in the \egodir, $\eta<-1$ (left), or central rapidities,
    $|\eta|<1$ (middle) and \hgodir, $\eta>1$ (right). The $z$-axis scale
    shows relative distribution of events from the \milou event
    generator.}
  \label{fig:dvcs_q2x}
\end{figure}

Figure~\ref{fig:dvcs_q2x} shows the $x$-$Q^2$ range covered by
DVCS measurements for different rapidity ranges, emphasizing the
importance of measurements over a wide rapidity range. Wide
kinematical coverage is also important for separating DVCS events from
Bethe-Heitler (BH) events (when a photon is radiated from the initial
or final state lepton), which share the same final state.  This can be done by 
utilizing the different kinematic distributions of DVCS and BH
photons (e.g., in rapidity and inelasticity $y$).  The planned EMCal
and tracking cover $|\eta| < 4$ (Section~\ref{sec:EMCAL} and
~\ref{sec:VertexTracking}) and will provide excellent capabilities for
DVCS measurements.

To ensure the reliable separation of electromagnetic showers in the
EMCal from the scattered electron and the DVCS photon, sufficient EMCal
granularity is necessary. The minimal angle separation between the 
electron and the photon is reached for electrons with the smallest scattering 
angle (i.e., the smallest $Q^2$) and is inversely proportional to
electron beam energy.  For a 10~GeV electron beam and $Q^2>1$~GeV$^2$,
the minimum angle is $\sim 0.1$ rad.  The proposed crystal 
EMCal in the \egodir, with granularity $\sim 0.02$ rad 
(see Section~\ref{sec:CrystalEMCAL}), will provide the necessary 
electron and photon shower separation.

It is also important to ensure the exclusiveness of the DVCS measurements, 
and so it is highly
desirable to reconstruct the scattered beam proton. The proton
scattering angle is inversely proportional to proton beam energy and
varies from 0 to 5~mrad for 250~GeV proton beam and four-momentum
transfer $-t<1$~GeV$^2$.  It can be detected with the planned "Roman Pots"
detectors located along the beam line (See Section~\ref{sec:beamline}).  

\subsection{Diffractive measurements}

\begin{figure}[ht]
\begin{center}
\includegraphics[width=0.95\textwidth]{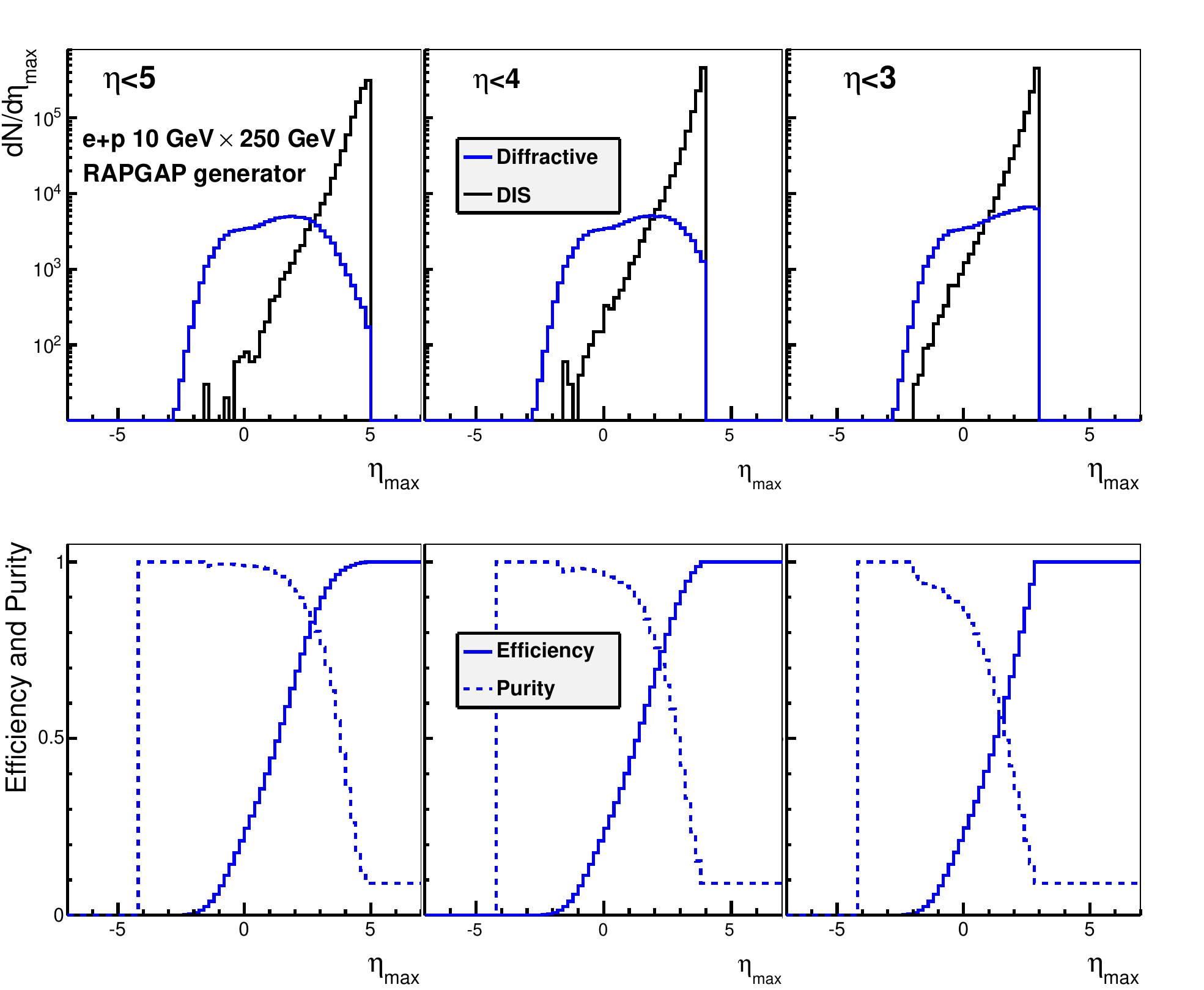}
\end{center}
\caption{For the $10~\mathrm{GeV} \times 100~\mathrm{GeV}$ beam energy
  configuration: Top: Pseudorapidity distribution for the most forward
  going particle in DIS events (black) and in diffractive events
  (blue); Bottom: Efficiency (dashed) and purity (solid) for
  diffractive event identification as a function of pseudorapidity cut
  defining the rapidity gap, for different detector acceptance:
  $|\eta|<5$ (left), $|\eta|<4$ (middle), $|\eta|<3$ (right).
  Obtained using the \rapgap generator developed at HERA and tuned to
  H1 and ZEUS data.}
\label{fig:diff}
\end{figure}

Diffractive event measurements play an important role in nucleon and
nucleus imaging. They are particularly sensitive to the gluon distribution
in nuclei and hence to gluon saturation phenomena.  Diffractive events
are characterized by a rapidity gap, i.e. an angular region in the
direction of the scattered proton or nucleus devoid of other
particles.  Figure~\ref{fig:diff} shows the pseudorapidity
distribution for the most forward going particle in DIS events and in
diffractive events.  Extending the forward acceptance of the detector
to $\eta = 4$ and beyond is important if one is to have good
capability using the rapidity gap method for detecting diffractive
events and to separate them from DIS processes.

The planned ePHENIX EMCal and tracking coverage of $|\eta| < 4$ and
hadronic calorimetry coverage of $-1 <\eta < 5$ are expected to provide
excellent identification capabilities for diffractive events.  In addition,
to separate coherent (the nucleus remains intact) and incoherent (the
nucleus excites and breaks up) diffractive events, we plan to place a
zero degree calorimeter after the first RHIC dipole magnet 
(see Section~\ref{sec:beamline}), which is
expected to be very efficient at detecting nuclear break-up by
measuring the emitted neutrons.

\makeatletter{}\section{Detector Concept}
\label{chap:ephenix_detector_concept}

\begin{figure}[h]
  \centering
  \includegraphics[width=0.8\linewidth]{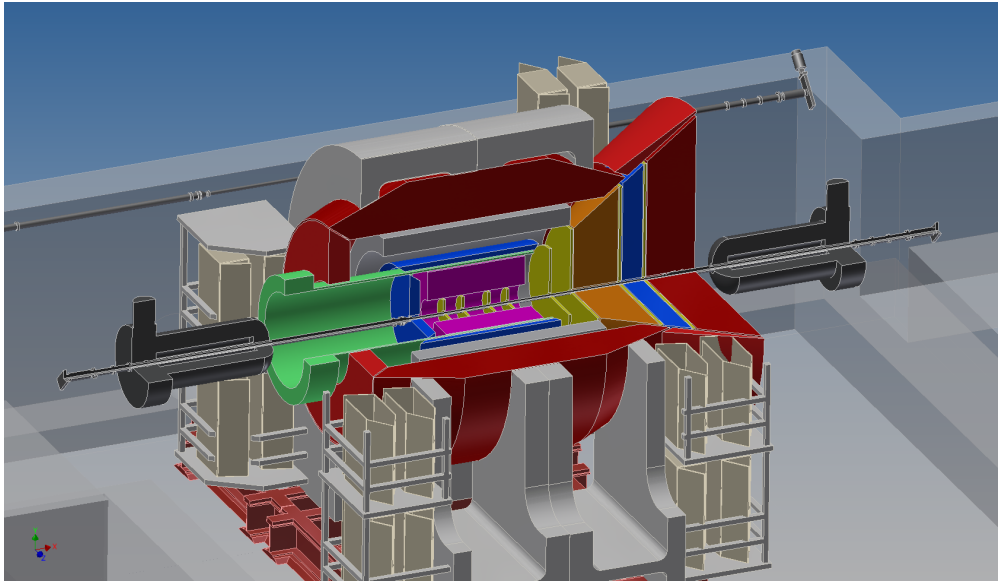}
  \caption{Engineering rendering of ePHENIX in the PHENIX experimental
    hall.  The drawing shows the location of the final eRHIC focusing
    quadrupoles as well as the electron bypass beamline behind the
    detector.}
  \label{fig:ephenix_drawing}
\end{figure}

A full engineering rendering of ePHENIX is shown in
Figure~\ref{fig:ephenix_drawing}.  The drawing shows the ePHENIX
detector in the existing PHENIX experimental hall and illustrates the
reuse of the superconducting solenoid and the electromagnetic and
hadronic calorimeter system of sPHENIX.  The rendering also shows the final eRHIC
focusing quadrupoles, each located 4.5 m from the interaction point (IP).
Those magnets and the height of the beam pipe above the concrete
floor, set the dominant physical constraints on the allowable
dimensions of ePHENIX. This Section will describe the ePHENIX detector
concept in terms of its component subdetectors and their expected
performance.

\begin{figure}[ht]
  \begin{center}
    \includegraphics[trim = 0 90 0 0, clip, width=1.0\textwidth]{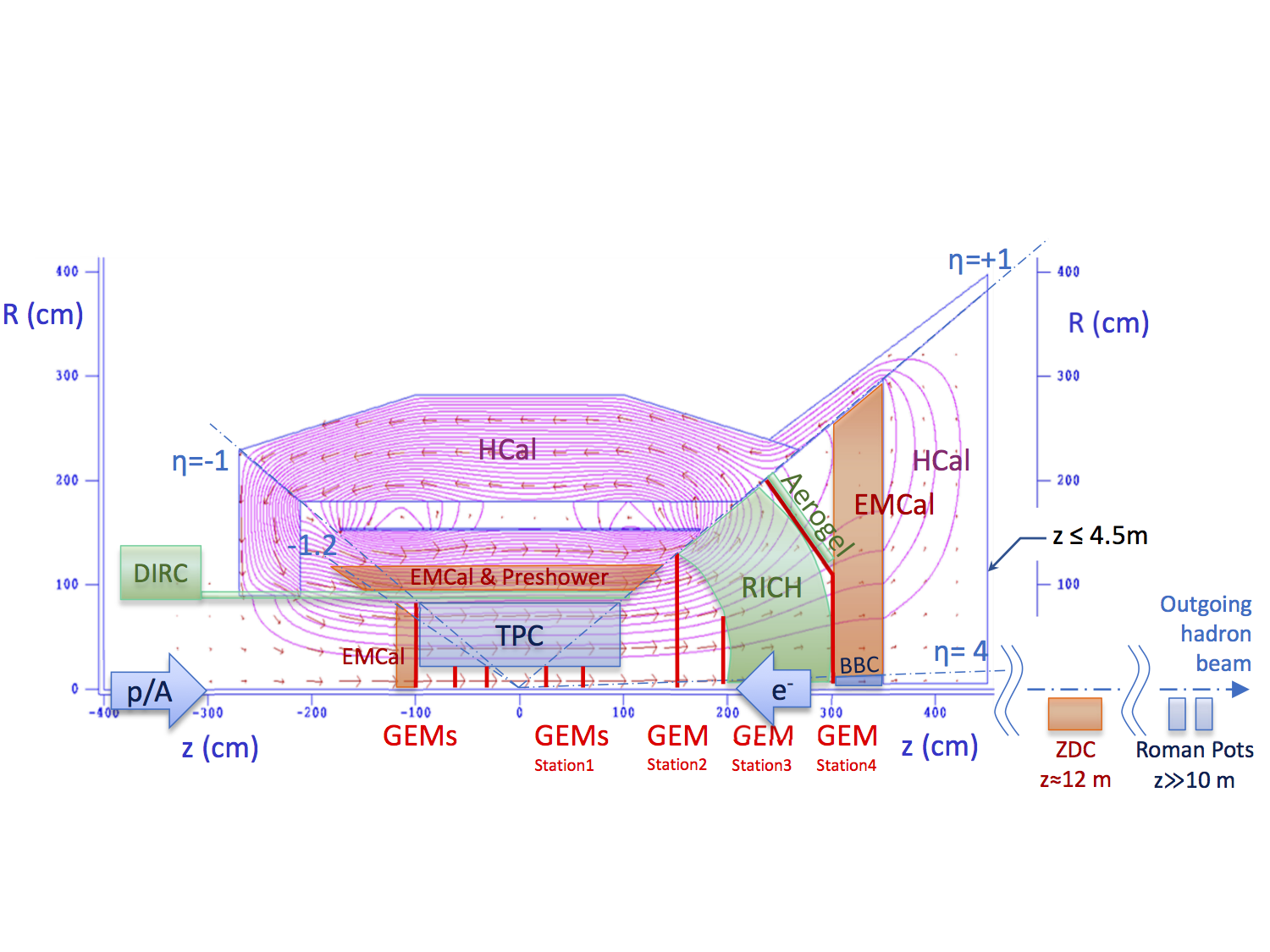}
  \end{center}
  \caption{A cross section through the top-half of the ePHENIX
    detector concept, showing the location of the superconducting
    solenoid, the barrel calorimeter system, the EMCal in the \egoing
    direction and the system of tracking, particle identification
    detectors and calorimeters in the \hgoing direction.  Forward
    detectors are also shown along the outgoing hadron beamline.
    The magenta curves are contour
    lines of magnetic field potential as determined using the 2D
    magnetic field solver, POISSON.}
  \label{fig:straw-man}
\end{figure}

The ePHENIX detector consists of a superconducting solenoid with
excellent tracking and particle identification capabilities covering a
large pseudorapidity range, as shown in Figure~\ref{fig:straw-man}.  It
builds upon an excellent foundation provided by the proposed sPHENIX
upgrade~\cite{Aidala:2012nz} detailed in the MIE proposal submitted to
the DOE Office of Nuclear Physics by Brookhaven National Laboratory in
April 2013.  The strong sPHENIX focus on jets for studying the
strongly-coupled quark-gluon plasma in $p$$+$$p$, $p$/$d$$+$A and
A$+$A is enabled by excellent electromagnetic and hadronic calorimetry
in the central region ($|\eta|<1$).

The C-AD Interaction Region (IR) design at the time the Letter of Intent charge was issued
had the final focusing quadrupoles of the accelerator positioned
$\pm4.5$~m from the IP and employed a ``crab crossing''
to maintain high luminosity while allowing the electron and hadron
beams to intersect at an angle of 10~mr (see Figure~\ref{fig:ZDC}).
The ePHENIX detector concept shown in Figure~\ref{fig:ephenix_drawing}
and Figure~\ref{fig:straw-man} respects these constraints.  For
instance, the hadronic calorimeter in the \hgoing direction fits
within the 4.5~m constraint imposed by the accelerator magnets, and
the detector is aligned so that the electron beam travels along the
symmetry axis of the magnetic field.  Clearly, the progress of ePHENIX
from concept to final design will be done in close consultation with
C-AD to ensure that the design of IR and the design of the detector
remain synchronized.

\begin{figure}[ht]
  \begin{center}
    \includegraphics[clip, width=0.7\linewidth]{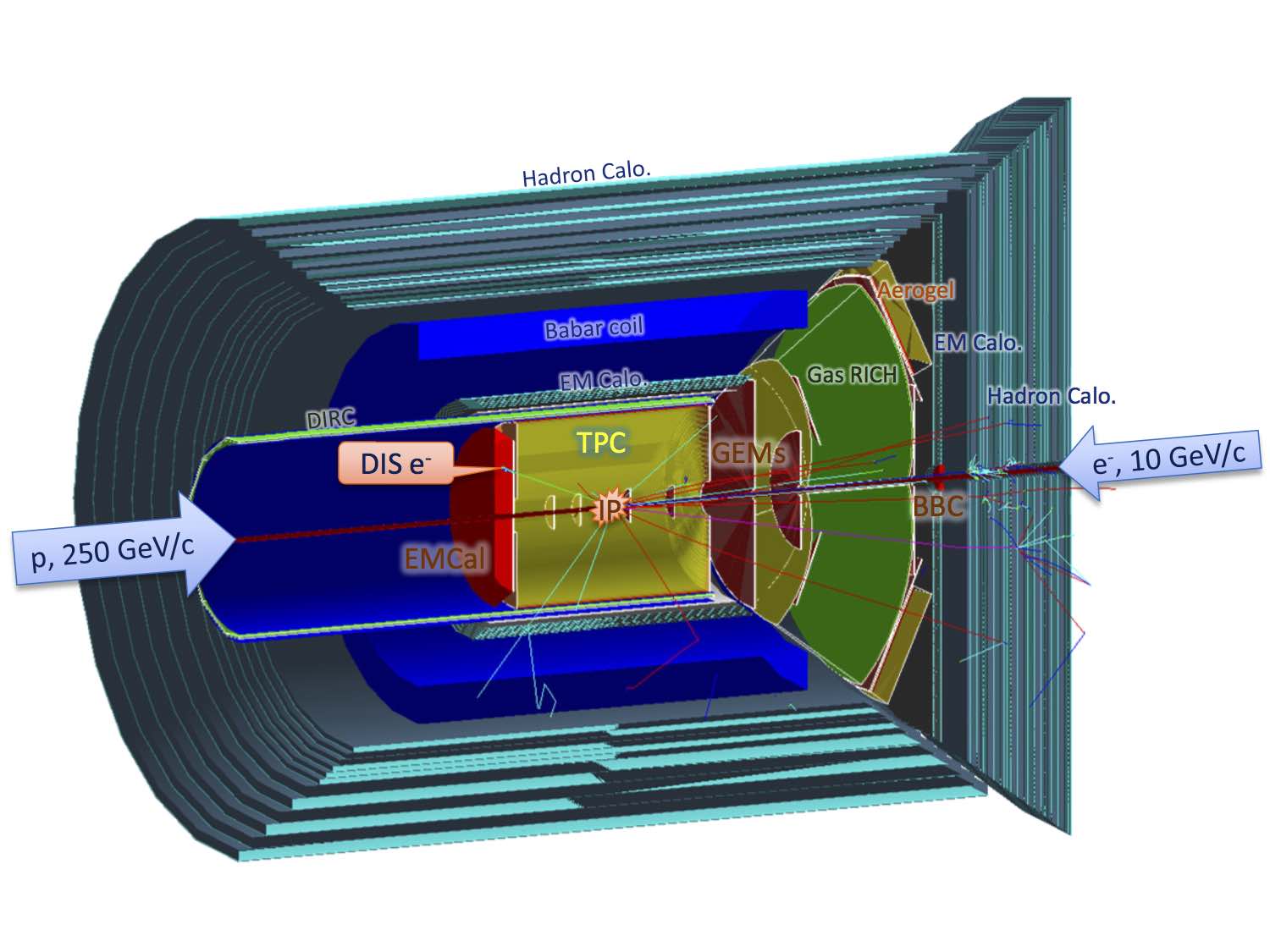}
  \end{center}
  \caption{The response of the ePHENIX detector to a single event, as
    determined using \geant.  The field map in this simulation was
    determined using the 2D magnetic field solver OPERA.  These same
    OPERA calculations were used to verify and validate the
    calculations underlying the magnetic field lines shown in
    Figure~\ref{fig:straw-man}.}
  \label{fig:geant4}
\end{figure}

We have an extensive \geant description of the ePHENIX detector, based
on the same software framework as used in PHENIX and sPHENIX, which
enables ready use of many existing PHENIX software analysis tools.  An
example of running a DIS event through the \geant detector description
is shown in Figure~\ref{fig:geant4}.

The DOE funded sPHENIX subsystems which will be reused in ePHENIX are:
\begin{description}
\item[Superconducting solenoid:] The sPHENIX detector concept reuses
  the BaBar superconducting solenoid to provide a 1.5~Tesla
  longitudinal tracking magnetic field. Its field is shaped in the
  forward directions with an updated yoke design in the ePHENIX
  detector as discussed in Section~\ref{sec:MagneticSystem}.

\item[Electromagnetic calorimeter:] A tungsten-scintillator sampling
  electromagnetic calorimeter with silicon photomultipliers (SiPMs)
  enables a compact barrel calorimeter positioned inside the bore of
  the superconducting solenoid.  The calorimeter system provides full
  azimuthal coverage for $|\eta|<1$ with an energy resolution of
  $\sim12\%/\sqrt{E}$. The readout is segmented into towers measuring
  roughly $\Delta\eta\times\Delta\phi\sim0.024\times0.024$.

\item[Hadronic calorimeter:] A $5\lambda_{\mbox{int}}$-depth hadron
  calorimeter surrounds the solenoid. An iron-plate and scintillator
  sampling design provides an energy resolution of better than
  $\sim100\%/\sqrt{E}$ with full azimuthal coverage. It also serves as
  part of the magnetic flux return for the solenoid.
\end{description}

In addition, new subsystems will be added to the ePHENIX detector,
which will be further discussed in this Section. These subsystems
include:
\begin{description}
\item[Electron going direction:] GEM detectors~\cite{Sauli:1997qp,
    Abbon:2007pq} and lead-tungstate crystal electromagnetic
  calorimeters
  \item[Central barrel:] Fast, compact TPC tracker and DIRC
  \item[Hadron going direction:] GEM tracking system, gas-based RICH, aerogel-based RICH, beam-beam counter (BBC),
    electromagnetic and hadron calorimeter
  \item[Beam line of hadron-going direction:] Roman pot detectors and
    a zero-degree calorimeter
\end{description}

\subsection{Magnet system}
\label{sec:MagneticSystem}

\renewcommand{\arraystretch}{1.9}
\addtolength{\tabcolsep}{-0.5pt}
\begin{table}
\caption{Main characteristics of BaBar solenoid~\cite{Bell:1999vf}}
\centering
\begin{tabular}{ll}
\toprule
Central Induction & 1.5~T\tabularnewline
Winding structure & 2~layers, $2/3$ higher current density at both ends\tabularnewline
Winding axial length & 3512~mm\tabularnewline
Winding mean radius & 1530~mm\tabularnewline
BaBar operation current & 4596~A, 33\% of critical current\tabularnewline
Total turns & 1067\tabularnewline
\bottomrule
\end{tabular}
\label{tab:Main-characteristics-of-BaBar}
\end{table}

As with sPHENIX, ePHENIX is based around the BaBar superconducting
solenoid~\cite{Bell:1999vf} with no modifications to its inner
structure. The major specifications for its coil are listed in
Table~\ref{tab:Main-characteristics-of-BaBar}.  A notable feature of
the BaBar magnet is that the current density of the solenoid can be
varied along its length, i.e., lower current density in the central
region and higher current density at both ends. This is accomplished
by using narrower windings (5~mm) for the last 1~m at both ends. The
central winding uses 8.4~mm-width coils~\cite{Bell:1999vf}. The main
purpose of the graded current density is to maintain a high field
uniformity in the bore of the solenoid, which is also a benefit for
ePHENIX.  This design feature enhances the momentum analyzing power in
both the \egoing and \hgoing directions.

A magnetic flux return system, consisting of the forward
steel/scintillator hadron calorimeter, a flaring steel lampshade, and
a steel endcap not only returns the flux generated by the solenoid,
but shapes the field in order to aid the momentum determination for
particles in the \hgoing and \egoing directions.  As shown
in~Figure~\ref{fig:straw-man}, the flux return system consists of the
following major components:

\begin{itemize}
\item Forward steel/scintillator hadron calorimeter, at $z=3.5$ to $4.5$~m
  \item Steel flux shaping lampshade, along the $\eta\sim1$ line
  \item Barrel steel/scintillator hadron calorimeter, from $r=1.8$ to $2.8$~m
  \item Steel end cap, at $z=-2.1$ to
    $-2.7$~m and $r>90$~cm
\end{itemize}

The magnetic field lines were calculated and cross checked using three
different 2D magnetic field solvers (POISSION, FEM, and OPERA) and are
shown in Figure~\ref{fig:straw-man}. In the central region, a 1.5~Tesla
central field along the electron beam direction is produced. The field
strength variation within the central tracking volume is less than
$\pm3\%$.

\subsection{Vertex and Tracking}
\label{sec:VertexTracking}

\begin{figure}
  \centering
  \includegraphics[trim = 33 0 24 19, clip, width=0.95\textwidth]{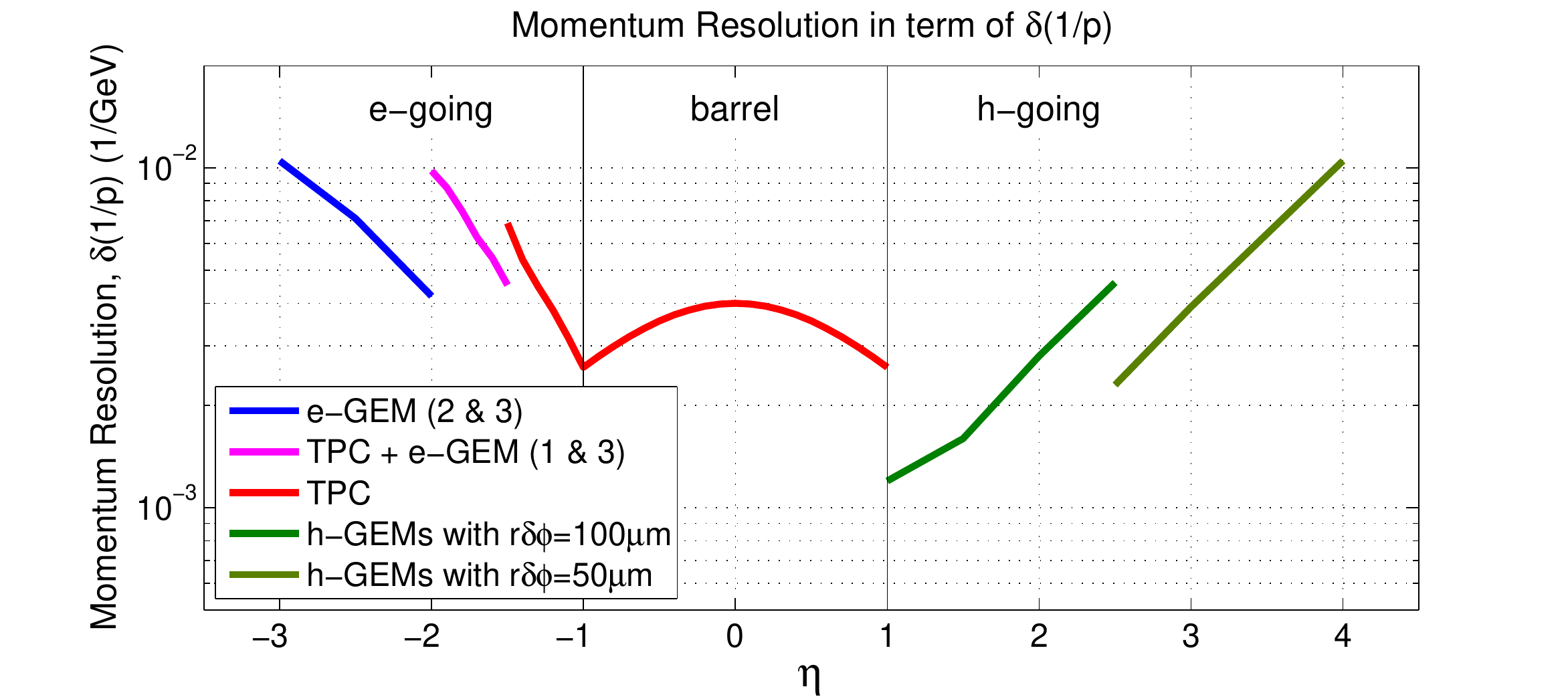}

  \caption{Momentum resolution over the full pseudorapidity coverage
    of the planned tracking system in the high momentum limit.
    Multiple scattering contribution to the relative momentum resolution 
    (not shown on the plot) was studied with GEANT4 simulation, 
    and found to vary from below 1\% at low pseudorapidity 
    to $\sim$3\% at $|\eta|$=3.}
  \label{fig:momentum_resolution}
\end{figure}

The $z$-location of the primary event vertex will be determined using
a timing system enabling a precision of $\Delta z \leq 5$~mm. The
ePHENIX tracking system utilizes a combination of GEM and TPC trackers
to cover the pseudorapidity range of $-3<\eta<4$. The momentum
resolution for the full device is summarized in
Figure~\ref{fig:momentum_resolution}.

\subsubsection {Event vertex measurement}

The vertex information is used for the determination of photon
kinematics and for assisting the track fitting.  Precise vertex
information is important for momentum determination in the \egoing
direction, where tight space constraints limit the possible number of
tracking planes.  The location of the vertex will be measured by:

\begin{itemize}
\item For non-exclusive processes, we propose to identify the
  $z$\nobreakdash-location for the vertex using timing information
  from a BBC detector in the \hgodir in
  coincidence with the electron beam RF timing. The BBC detector
  covers $\eta = 4$--5 at $z = 3.0$~m. A timing resolution of
  30~ps or better enables the measurement of the vertex with
  resolution of $\Delta z=5$mm.  It leads to a sub-dominant error for
  the momentum determination for the \egoing direction ($\delta p/p =
  2$\%).  This timing resolution can be provided by the existing
  technology of Multigap Resistive Plate Chamber
  (MRPC)~\cite{An:2008zzc} or by microchannel plate detectors (MCP)
  photomultiplier~\cite{Adams:2013} with a thin quartz \v{C}erenkov
  radiator, a technology which is under active current development.

\item We plan to measure the average transverse beam position by
  accumulating tracking information over the course of a one hour run.
  The statistical precision for the beam center determination is
  expected to be much smaller than the distribution of the transverse
  collision profile ($\sigma_{x,y}\sim 80$~$\mu$m), and therefore
  a negligible contribution to the uncertainty for event-by-event
  vertex determination.

\end{itemize}

\subsubsection {Tracking in the central region, $-1<\eta<1$}

A fast, compact Time Projection Chamber (TPC) will be used for
tracking in the central region, occupying the central tracking volume
of $r = 15$--80~cm and $|z| < 95$~cm and covering $-1 < \eta < 1$. A
TPC will provide multiple high resolution space point measurements
with a minimal amount of mass and multiple scattering.  The design is
based on a GEM readout TPC, similar to a number of TPCs that have
either already been built or are currently under design. For example,
the LEGS TPC~\cite{Geronimo:2005} utilized a fine chevron-type
readout pattern with a pad
size of 2~mm $\times$ 5 mm and achieved a spatial resolution $\sim$
200~$\mu$m. The use of such a readout pattern
helps minimize the total channel count for the electronics and hence
the total cost. The GEM TPC upgrade for
ALICE~\cite{ALICE_TPC:2012,Ketzer:2013laa} and the large GEM readout
TPC for ILC~\cite{Abe:2010aa,Schade:2011zz} are other examples of
large GEM TPCs that have recently been studied.

It is assumed that the TPC will have a single high voltage plane at $z
= 0$~cm and be read out on both ends, resulting in a maximum drift
distance $\sim 95$~cm. It will use a gas mixture with a fast drift
time, such as 80\% argon, 10\% CF$_4$ and 10\% CO$_2$, which, at an
electrical field of 650~V/m, achieves a drift speed $\sim
10$~cm/$\mu$s, and would result in a maximum drift time of 10~$\mu$s.
With a position resolution of $\sigma(r\Delta \phi) =300$~$\mu$m and
65 readout rows, the expected transverse momentum resolution would be $\delta
(1/p_{T}) = 0.4\%/($GeV$/c)$ for high momentum tracks.

\subsubsection {Tracking in \hgodir, $\eta>1$}

The design of the magnetic flux return enables tracking in the \hgoing
direction in the main and fringe fields of the BaBar magnet. Compared
to a compact solenoid with no current density gradient, the BaBar
magnet system improves the momentum analyzing power for forward tracks
by about a factor of four due to two main factors: 1) the BaBar magnet
has a length of 3.5~m, which provides a longer path length for
magnetic bending; 2) the higher current density at the ends of the
solenoid improves the magnetic field component transverse to forward
tracks, and therefore provides higher analyzing power.

The tracking system at high $\eta$ in the \hgoing direction utilizes four stations of GEMs.

\begin{itemize}
\item Station~1 consists of two planes with complementary $\eta$
  coverages. They are located at $z = 17$ and 60~cm, respectively,
  covering a radius of $r = 2$--15~cm.
\item Stations~2--4 are at $z = 150$, 200, 300~cm, respectively,
  covering $\eta = 1$--4.
\end{itemize}

The readout planes for these devices are optimized to preserve high
position resolution in the azimuthal direction ($\sim 200$~$\mu$m in
$r \delta \phi$ using a chevron-type readout with a pad size similar to the
central TPC) and $\sim$10--100 mm in $\delta r$, while minimizing the
readout channel cost. However, the $r$-$\phi$ resolution can be improved
to be better than 100~$\mu$m, even for tracks at larger angles (up to
45~degrees), by the use of mini-drift GEM detectors, in which a small
track segment, or vector, is measured for each track at each measuring
station.  These detectors, which are currently under development
~\cite{EIC_Tracking_RD:2013}, would provide improved position
resolution with less material and lower cost than multiple stations of
planar GEM detectors. For this letter, we assumed that a high resolution
GEM readout pattern (1~mm wide chevron-type readout) with a  $r \delta \phi \sim 50$~$\mu$m
for the inner tracking region ($\eta>2.5$).
For the outer tracking region ($1<\eta<2.5$), mini-drift GEM
with 2~mm chevron-type readout provide $r \delta \phi \sim 100$~$\mu$m.
The momentum resolution is estimated in Figure~\ref{fig:momentum_resolution}.

It should be noted that the size of the GEM trackers for Stations 2--4
are quite large ($\sim 5$--20~m$^2$).  It is currently challenging to
produce such large GEM foils and to do so at an affordable cost.
However, there has been substantial progress in this area in recent
years at CERN due to the need for large area GEM detectors for the CMS
Forward Upgrade~\cite{CMS_GEM:2012}. CERN has developed a single mask
etching technology which allows fabrication of very large area GEMs
(up to 2~m $\times$ 0.5~m), and they plan to transfer this technology
to various commercial partners (such as Tech Etch in the US, which
supplied the GEM foils for the STAR Forward GEM Detector).  We
anticipate being able to procure such large area GEM detectors by the
time they are needed for EIC.

\subsubsection {Tracking in the \egodir, $\eta<-1$}

The electron direction tracking is designed to fit in the space
limited by the DIRC ($R < 80$~cm) and the electromagnetic calorimeter
($z > -100$~cm). Three GEM tracking stations, located at $z = 30$, 55
and 98~cm, are used in combination with the TPC and vertex information
to determine the momentum vector.

\begin{itemize}
  \item For $\eta = -1.5$ to $-1$, TPC track segment and vertex are used
  \item For $\eta = -2.0$ to $-1.5$, vertex, TPC track segment, GEM station 1 and 3 are used.
  \item For $\eta = -3.0$ to $-2.0$, vertex, GEM station 2 and 3 are used.
\end{itemize}

Similar to the \hgoing direction,
the position resolution for these detectors is $r\Delta\phi$ 50~$\mu$m for $-3<\eta<-2$
using 1~mm wide chevron-type readout. For $-2<\eta<-1$, the mini-drift GEM technology~\cite{EIC_Tracking_RD:2013}
and 2~mm wide chevron-type readout provide $r \delta \phi \sim 100$~$\mu$m.
The radial resolution is $\delta r = 1$~cm (stations 1 and 2) and
$\delta r = 10$~cm (station 3). As shown in
Figure~\ref{fig:momentum_resolution}, a momentum resolution of $\Delta
p/p < 5\%$ can be achieved for tracks of $p<4$~GeV$/c$ and
$-1<\eta<-3$, which is sufficient for the calorimeter
$E$\nobreakdash-$p$ matching cut for the electron
identification.
For DIS kinematics reconstruction the tracking radial resolution is not
crucial as enough precision for scattered electron polar angle measurements
will be provided by the EMCal, see Section~\ref{sec:resolution}.

\subsection{Electromagnetic calorimeters}
\label{sec:EMCAL}

The ePHENIX detector will have full electromagnetic calorimeter
coverage over $-4<\eta<4$.
The sPHENIX barrel electromagnetic calorimeters will also be used in
ePHENIX, covering $-1<\eta<1$ with an energy resolution of
$\sim12\%/\sqrt{E}$.
In addition, crystal and lead-scintillator
electromagnetic calorimeter are planned for the \egoing and \hgoing
direction, respectively.
Optimization of the design of the barrel and endcap calorimeters
will aim for uniform response in the overlap region between $-1.2<\eta<-1$.

\subsubsection {Crystal Electromagnetic calorimeter}
\label{sec:CrystalEMCAL}

\begin{figure}
\centering
\includegraphics[width=0.4\textwidth]{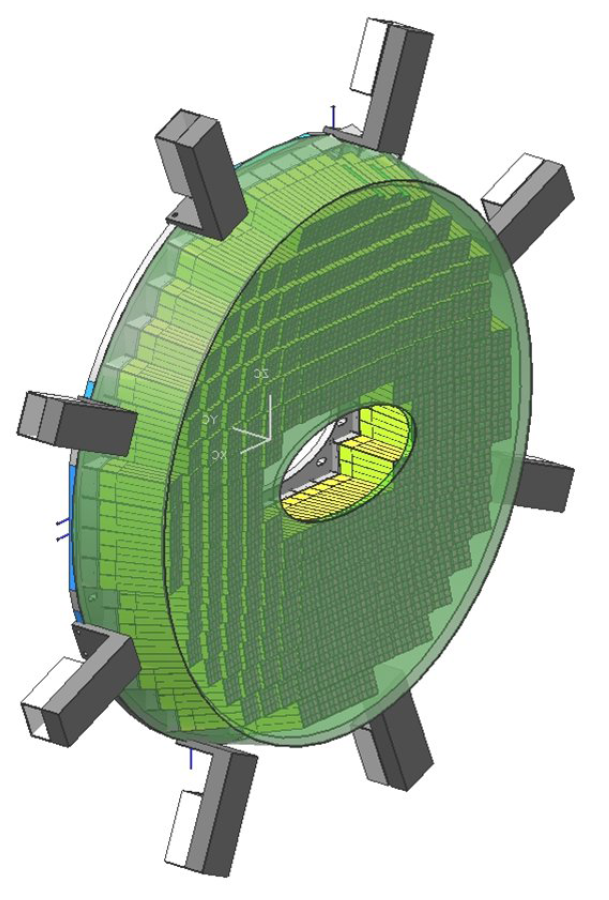}
\caption{PANDA Crystal Endcap Calorimeter~\cite{PANDA:2008uqa}. The
  PWO crystal modules are shown in green color, which is projective
  towards the target.}
\label{fig:PANDA_Endcap}
\end{figure}

The calorimeter on the electron\nobreakdash-going side consists of an
array of lead tungstate (PbWO$_4$) crystals (commonly known as PWO),
similar to the PANDA endcap crystal calorimeter shown in
Figure~\ref{fig:PANDA_Endcap} ~\cite{PANDA:2008uqa}. An enhanced light
output version of lead tungstate (PWO-II) was chosen to provide high
light yield ($\sim 20$~p.e./MeV at room temperature) at a moderate
cost ($\sim$ \EUR{5}/cm$^3$). It will provide an energy resolution
$\sim 1.5$\%/$\sqrt{E}$ and position resolution better than
$3~\mathrm{mm}/\sqrt{E}$ in order to measure the scattered electron energy
and angle in the \egoing direction down to low momentum with high precision.

The ePHENIX PWO calorimeter will consist of $\sim 5000$ crystals,
compared with 4400 crystals for the PANDA endcap, and will have a
similar size and shape to the PANDA crystals. They will be $\sim 2~\mathrm{cm}
\times 2~\mathrm{cm}$ (corresponding to one $R_M^2$) and will be read out with
four SiPMs. This is different than the PANDA readout, which uses large
area ($\sim 1$~cm$^2$) APDs. The SiPMs will provide higher gain, thus
simplifying the readout electronics, and will utilize the same readout
electronics as the other calorimeter systems in sPHENIX. It is also
expected that the cost of SiPMs will be less than that of APDs
covering the same area by the time they are needed for ePHENIX.

\subsubsection {Lead-scintillator electromagnetic calorimeter}

The electromagnetic calorimeter in the \hgoing direction consists of a
lead-scintillating fiber sampling configuration, similar to the
tungsten-scintillating fiber calorimeter in the central sPHENIX
detector. Lead is used instead of tungsten in order to reduce the
cost, but it is otherwise assumed to be of a similar geometry. It will
cover the rapidity range from $1<\eta<4$ and have 0.3~$X_0$ sampling
(2~mm lead plates) with 1~mm scintillating fibers, which will give an
energy resolution $\sim 12$\%/$\sqrt{E}$.  The segmentation and
readout will also be similar to the central tungsten calorimeter,
with $\sim 3~\mathrm{cm} \times 3~\mathrm{cm}$ towers (roughly 1~$R_M^2$) that are read
out with SiPMs. This segmentation leads to $\sim 26$K towers. By using
the same type of readout as the central calorimeter, the front end
electronics and readout system will also be similar, resulting in an
overall cost savings for the combined calorimeter systems.

\subsection{Hadron calorimeter}

The hadron calorimeter in the \hgoing direction consists of a
steel-scintillating tile design with wavelength shifting fiber
readout, similar to the central sPHENIX hadron calorimeter. It will be
$\sim 5$ $L_{abs}$ thick and cover a rapidity range from $1<\eta<5$.
The steel in the absorber will also serve as part of the flux return
for the solenoid magnet. The segmentation will be $\sim 10~\mathrm{cm}
\times 10~\mathrm{cm}$, resulting in $\sim 3000$ towers. The readout will also be
with SiPMs, similar to the central sPHENIX HCAL, which will again
provide an advantage in being able to use a common readout for all of
the calorimeter systems.

\subsection{Hadron PID detectors}
\label{sec:HadronPIDDetectors}

\begin{figure}
\centering
\subfloat[Aerogel and RICH gas radiators for hadron-going
direction]{\includegraphics[width=.4\textwidth]{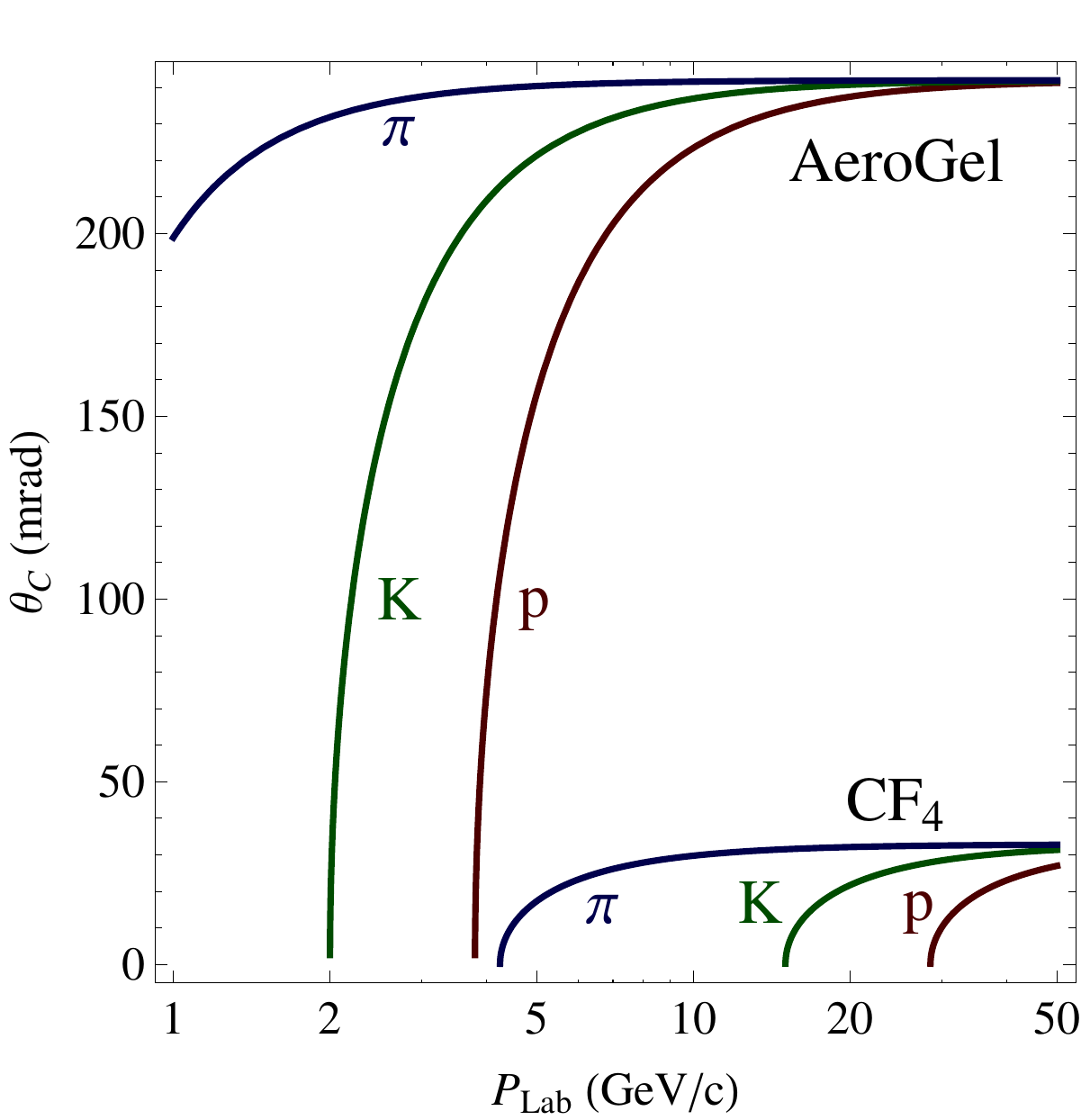}}
\subfloat[Fused silica radiator for barrel DIRC detector. Data are
measured by the BaBar DIRC~\cite{Adam:2004fq}]{\includegraphics[bb=0bp
  7bp 227bp 227bp,clip,width=0.45\textwidth]{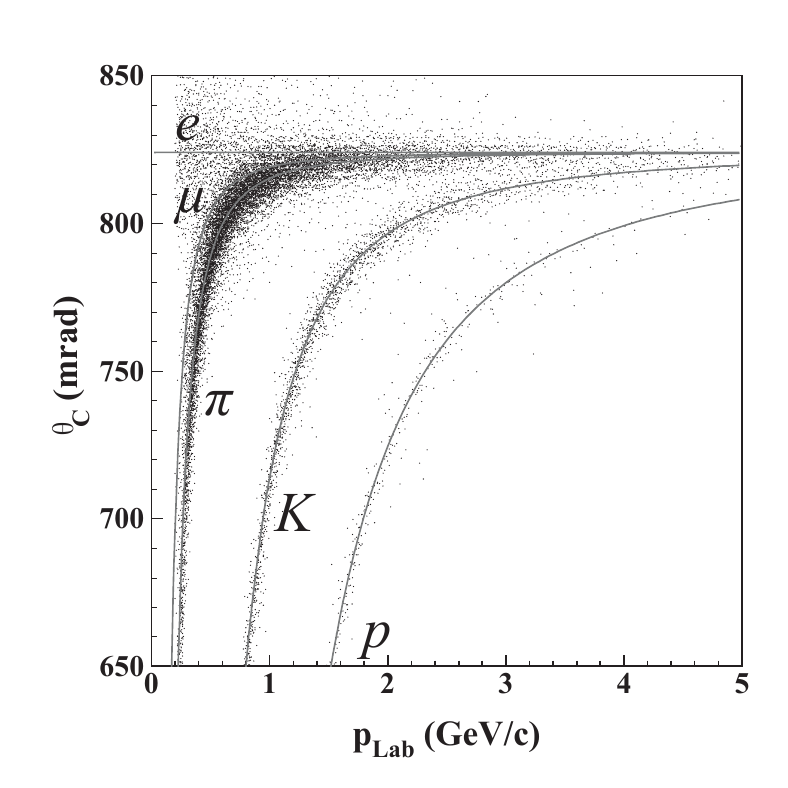}}
\caption{\v{C}erenkov angle versus momentum for various particle
  species.}
\label{fig:RICH_Angles}
\end{figure}

Hadron PID is planned for the \hgoing and barrel regions, covering
$-1.2 < \eta < 4$. In the \hgoing direction, two PID detectors cover
complementary momentum range: a gas-based RICH detector for the higher momentum
tracks and an aerogel-based RICH detector for the lower momentum region.  As in
the BaBar experiment~\cite{Adam:2004fq}, a DIRC detector identifies
hadron species in the central barrel.  In addition, the TPC detector assists
with PID by providing $dE/dx$ information for the low momentum region.

\subsubsection {\label{sec:RICH}Gas RICH detector}

\begin{figure}
\centering
\includegraphics[width=0.65\textwidth]{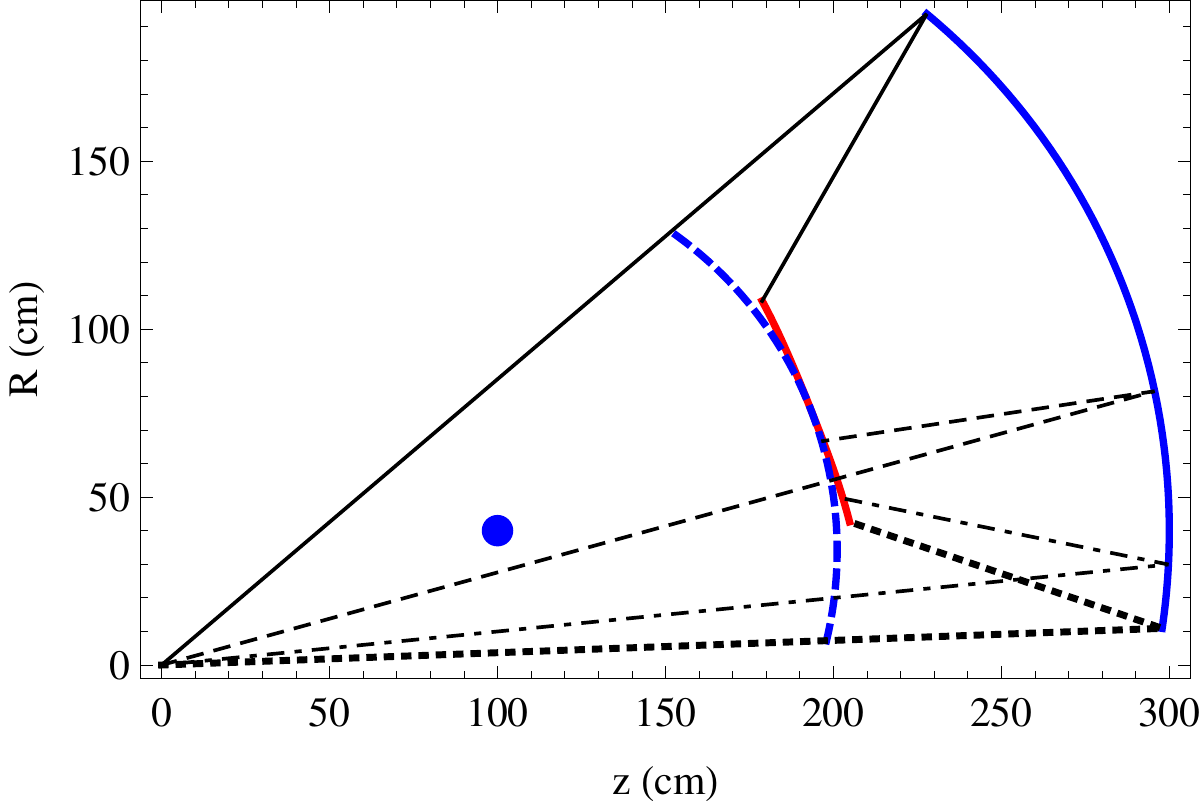}
\caption{The cross-section of the gas-based RICH detector in the $r$-$z$ plane
  that crosses the mirror center. The interaction point is
  centered at $(0,0)$. The geometric center of the mirror
  is shown as the blue dot at $(r, z) = (40~\mathrm{cm}, 100~\mathrm{cm})$. The mirror and
  RICH entrance window are shown by the solid and dashed blue curves,
  respectively.  Several example tracks and the central axis of their
  \v{C}erenkov light cone are illustrated by the black lines. The
  \v{C}erenkov photons are reflected by the mirror to the focal plane,
  shown in red. }
\label{fig:RICH_setup}
\end{figure}

High momentum hadron PID is provided by an optically focused RICH
detector using a gas radiator. The main features for this RICH setup
are
\begin{itemize}
\item One meter of CF$_4$ gas is used as the \v{C}erenkov radiator.
  The pion, kaon and proton thresholds are 4, 15 and 29 GeV,
  respectively.
\item \v{C}erenkov photons are focused to an approximately flat focal
  plane using spherical mirrors of 2~m radius, as shown in
  Figure~\ref{fig:RICH_setup}. The geometric center of the mirror is
  at $(r,z) = (40~\mathrm{cm}, 100~\mathrm{cm})$, as highlighted by the blue dot.
\item There are six azimuthal segmented RICH sextants.
\item The photon detector consists of CsI-coated GEM
  detectors~\cite{Anderson:2011jw}, which are installed on the focal
  plane. The CsI coating converts the \v{C}erenkov photons into electrons
  which are then amplified by the GEM layers and readout through
  mini-pads. The photon detector for each RICH sextant assumes a
  roughly triangle shape and covers an area of 0.3~m$^2$.
\end{itemize}

\begin{figure}
\centering
\includegraphics[width=0.65\textwidth]{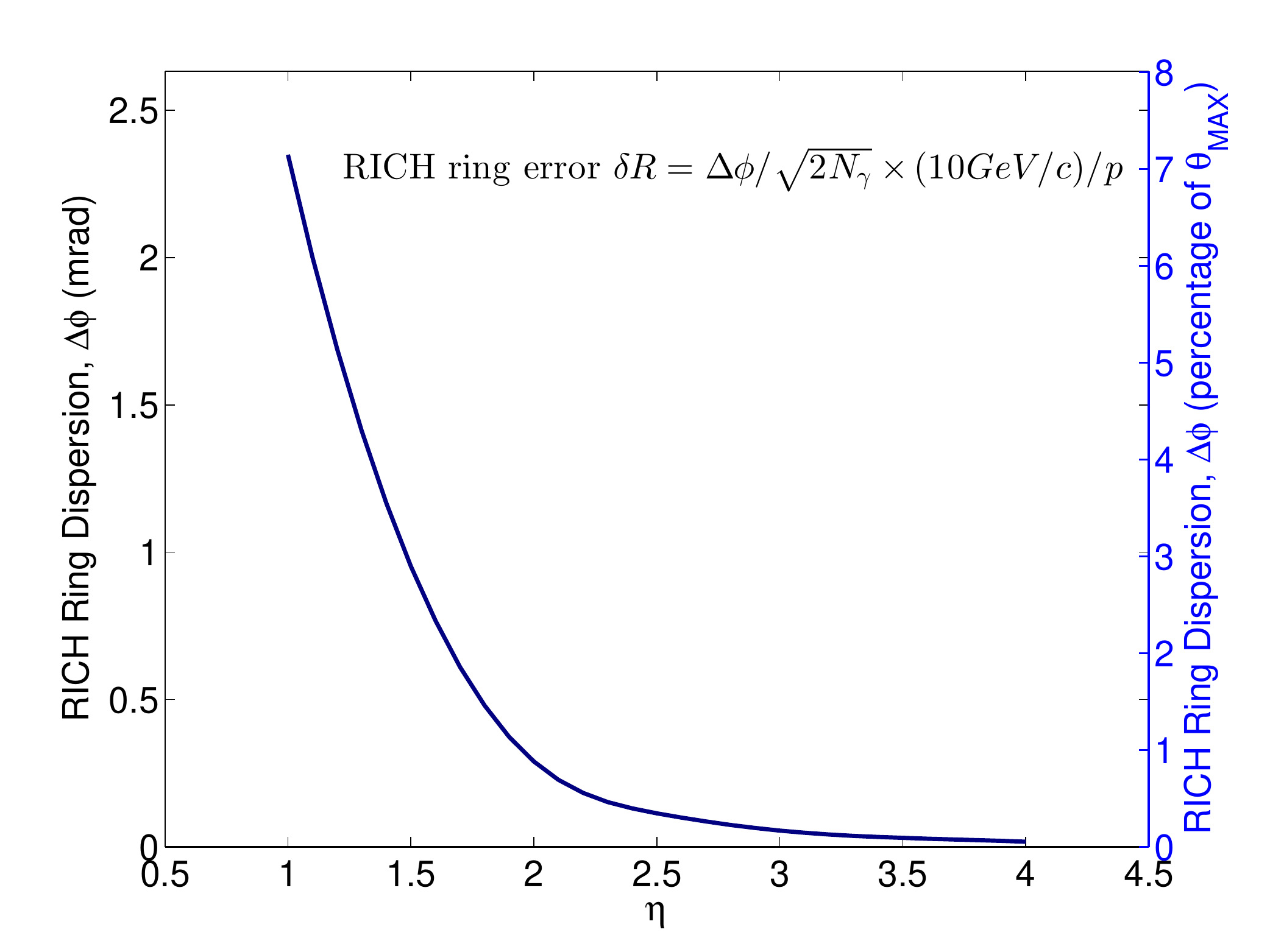}
\caption{Azimuthal angular dispersion of gas-based RICH ring due to fringe magnetic field for a $p=10$~GeV$/c$ track. It is compared to the maximum RICH ring angle as shown on the right vertical axis.}
\label{fig:RICH_field_effect}
\end{figure}
Two distortion effects were estimated to be sub-dominant in error contributions for most cases:
\begin{itemize}
\item Strong residual magnetic field ($\sim0.5$~Tesla) are present in the
  RICH volume. This field will bend the tracks as they radiate
  photons, and therefore smear the \v{C}erenkov ring in the azimuthal
  direction. However, the field design ensures that the field
  component is mostly parallel to the track inside RICH and therefore
  this smearing effect is minimized. The RMS size of the smearing,
  $\Delta\phi$, is evaluated as in Figure~\ref{fig:RICH_field_effect}.
  The uncertainty contribution to the RICH ring angular radius is
  $\delta R=\Delta\phi/\sqrt{2N_{\gamma}}(10\mbox{ GeV}/c)/p$, which
  is sub-dominant comparing to the photon measurement error for
  $\eta>1.5$. The field contribution was included in the RICH
  performance estimation.
\item For tracks that originate from an off-center vertex, their focal
  point may be offset from the nominal focal plane as shown in
  Figure~\ref{fig:RICH_setup}. The effect is $\eta$ dependent. 
  For the most extreme case, that a track of $\eta = 1$ originates from
  the vertex of $z = 40$~cm (1.5 sigma of expected vertex width), an
  additional relative error of $5\%/\sqrt{N_{\gamma}}$ 
  is contributed to the ring radius measurement, which averages over 
  all vertices to below $2\%/\sqrt{N_{\gamma}}$ contribution.
  For high $\eta$ tracks, the
  difference is negligible comparing to the nominal RICH error. 
\end{itemize}

We simulated the RICH performance with a radiator gas CF$_{4}$
(index of refraction 1.00062). We use \pythia to generate 
the momentum distributions for pions, kaons, and protons.  
For each particle species, we use the momentum resolution 
and RICH angular resolution, to calculate the particle 
mass $m(p,\theta_{Crk})$ distribution. For higher momentum tracks 
the combined information from tracking system and energy deposit in HCal 
helps to improve momentum resolution particularly at higher rapidities, 
where momentum resolution from tracking degrades. 
For example, at pseudorapidity $\eta$=4, the tracking momentum 
resolution for 50~GeV/c tracks is $\sim$50\% 
(see Figure~\ref{fig:momentum_resolution}), while HCal can provide 
energy measurements with precision $100\%/\sqrt{50\mathrm{[GeV]}} \sim 14\%$. 
Our simulation showed that the HCal is very effective in improving 
the resolution for high momentum track measurements even when 
this and other tracks (usually with lower momenta) are merged in a single 
cluster of deposited energy in HCal. 
In such a case, the contribution of lower energy tracks in HCal can be 
evaluated and subtracted based on momentum measurements in tracking system. 

\begin{figure}
  \centering
  \includegraphics[bb=0bp 0bp 550bp 262bp,clip,width=1.0\textwidth]{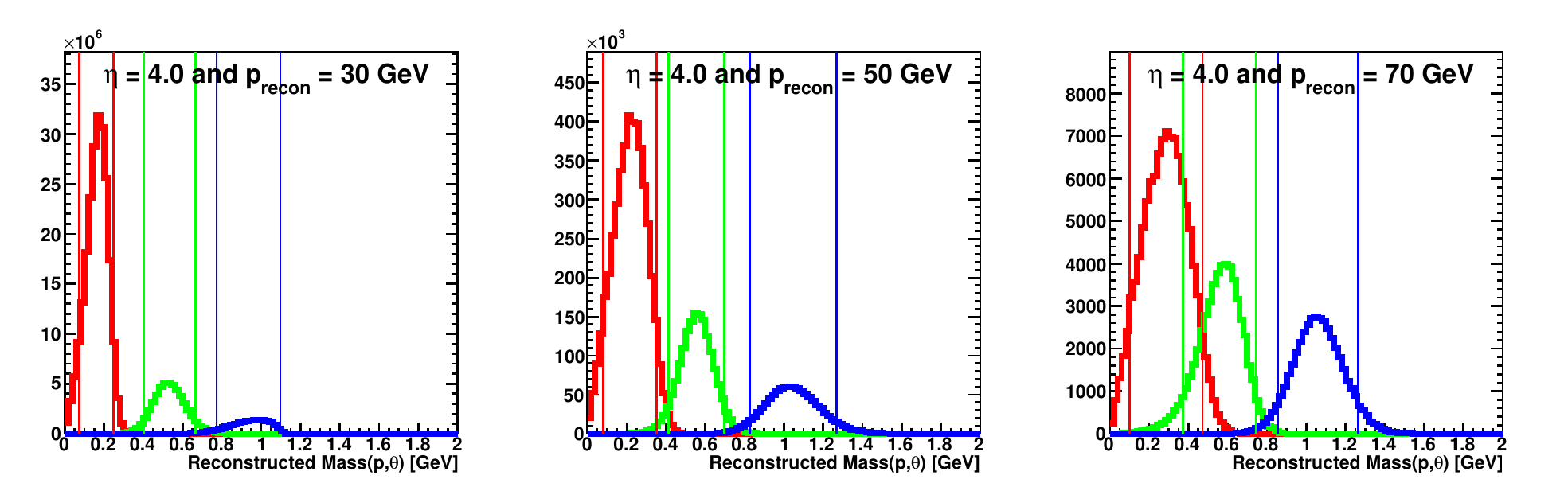}
  \caption{Reconstructed mass distribution via m(p,$\theta_{Crk}$) at $\eta=4$
    for reconstructed momenta 30 GeV/c (left), 50 GeV/c (middle) and 70 GeV/c 
    (right), for pions (red), kaons (green) and protons (blue), 
    with the parent momentum and particle
    abundances from the \pythia generator.  
    Vertical lines indicate the symmetric mass cuts corresponding 
    to 90\% efficiency. 
    Note that particle true momentum is on the average smaller than 
    reconstructed momentum, see Figure~\ref{fig:rich_purity}. 
  }
  \label{fig:rich_mreco}
\end{figure}

\begin{figure}
  \centering
  \includegraphics[width=1.0\textwidth]{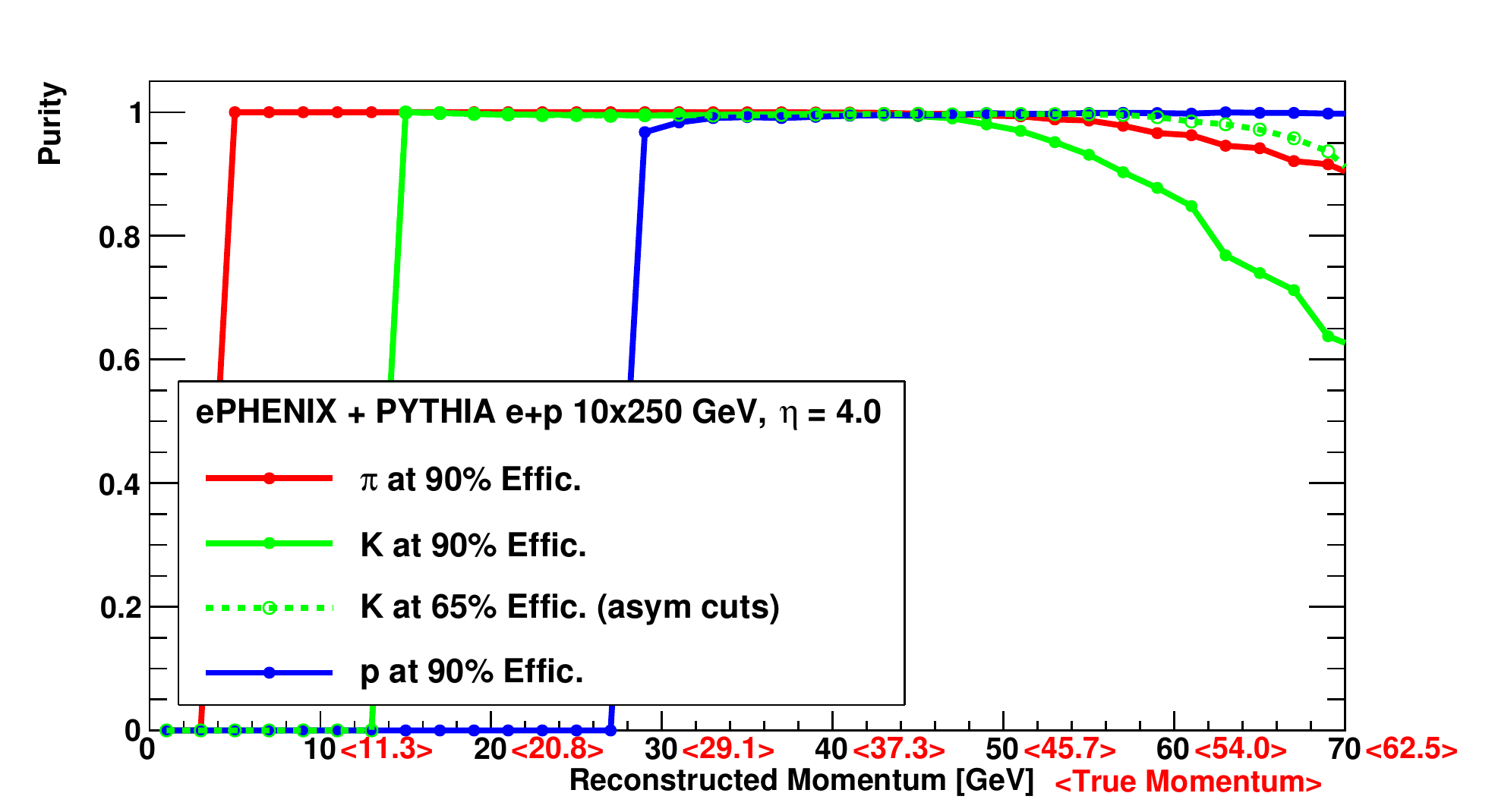}
  \caption{ $\pi, K, p$ purities at
    pseudorapidity 4.0 as a function of reconstructed momentum, 
    based on symmetric cut on reconstructed mass corresponding to 90\% 
    efficiency (solid lines), and asymmetric cut with stricter selection 
    on the kaons with efficiency 65\% (dashed line);
    Also indicated in angle brackets are the values of the average 
    true momentum at each reconstructed momentum, which are different due 
    to momentum smearing and sharply falling momentum spectra.
    }
  \label{fig:rich_purity}
\end{figure}

Figure~\ref{fig:rich_mreco} shows mass distributions for the 
most challenging high rapidity region $\eta$=4 for different reconstructed 
track momenta. We make a symmetric 90\% efficiency cut on the mass 
distributions, and calculate the purity for $\pi, K, p$, 
shown in Figure~\ref{fig:rich_purity}. One can see high purity for all 
particle species up to momenta $\sim$50 GeV/c. 
Introducing asymmetric cuts on the mass distributions (and sacrificing some 
efficiency) extends further our capabilities for high purity hadron 
identification. 

It is notable that the limitation on the mass resolution comes 
from the estimated 2.5\%
radius resolution per photon for the RICH from the EIC R\&D RICH
group.  Our calculation includes the effect of the magnetic field
distortion mentioned above, which is sub-dominant.  This is a somewhat
conservative estimate and LHCb and COMPASS have quoted values near 1\%
per photon.  The R\&D effort is working towards the best radius
resolution, though there are challenges in having the light focus and
readout within the gas volume in this configuration compared with LHCb
or COMPASS.

\subsubsection {Aerogel RICH detector}

The aerogel detector will provide additional particle ID for kaons in
the momentum range $\sim 3$--15~GeV/c when used in conjunction with
the gas RICH. Pions can be identified by the signal they produce in
the gas RICH starting at a threshold of $\sim$ 4 GeV/c,\ and kaons
will begin producing a signal in the aerogel at a threshold $\sim$ 3
GeV. Reconstructing a \v{C}erenkov ring in the aerogel enables one to
separate kaons from protons up to a momentum $\sim 10$~GeV/c with
reduced efficiency above that.

\begin{figure}
\centering
\includegraphics[width=0.6\textwidth]{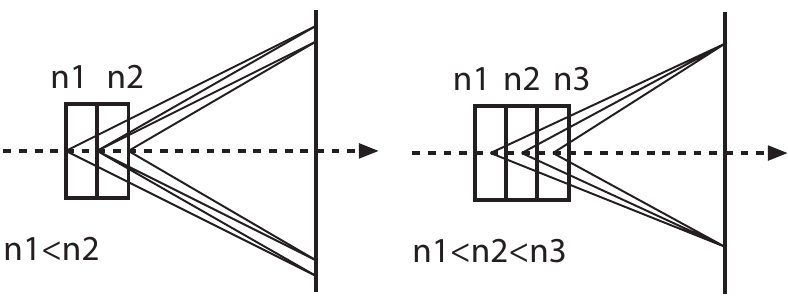}
\caption{Approximate focusing method using two (left) and three
  (right) layers of aerogel with slightly different indicies of
  refraction proposed by Belle~II~\cite{Iijima:2005qu}}
 \label{fig:Aerogel_setup}
\end{figure}

Measuring a ring in the aerogel detector is a challenging technical
problem for a number of reasons. Due to its relatively low light
output, it will require detecting single photons in the visible
wavelength range with high efficiency inside the rather strong fringe
field of the superconducting solenoid. Also, due to the limited space
available, it is difficult to have a strong focusing element in the
RICH to focus the light into a ring in a short distance. One
possibility for how this might be accomplished has been proposed by
the Belle II experiment~\cite{Iijima:2005qu} and is shown in
Figure~\ref{fig:Aerogel_setup}. It uses several layers of aerogel with
slightly different indices of refraction to achieve and approximate
focusing of the light onto an image plane located behind the radiator.
It should be possible to add additional layers of aerogel and optimize
their thickness for producing the best quality ring for kaons using
this technique, and therefore achieve good kaon-proton separation up
to the highest momentum.
One possibility for the photon detector would be large area
Microchannel Plate detectors (MCPs), such as those being developed by
the Large Area Picosecond Photodetector (LAPPD)
Collaboration~\cite{Adams:2013}. This effort is based on utilizing flat
panel screen technology to produce large area MCPs at very low cost,
while also preserving their excellent timing resolution
(typically $\sim$ 20-30 ps). These devices would use multi-alkali
photocathodes, which would be suitable for detecting the Cherenov light
from aerogel with high efficiency, and also provide high gain for detecting
single photoelectons. The excellent time resolution would also provide
additional time of flight capability when used in conjunction with the
BBC to further enhance hadron particle ID. While this is still an R\&D effort,
it has already produced very encouraging results and has substantial support
within the high energy physics community, and we feel that this would offer
an attractive low cost, high performance readout for the aerogel detector.

\subsubsection {DIRC}

The main form of particle ID in the central region will be provided by
a DIRC (Detection of Internally Reflected \v{C}erenkov Light). The
DIRC will be located at a radius of $\sim 80$~cm and extend $\sim
8$--10~cm in the radial direction. As we will be using the BaBar
magnet for ePHENIX, it would be a major benefit to also acquire the
BaBar DIRC, which was specifically designed to fit inside this magnet,
and would completely satisfy the physics requirements for ePHENIX.
However, since it is not certain at this time that the BaBar DIRC will
be available for ePHENIX, we consider it more as a model for the type
of DIRC that would be required in terms of its construction and
performance.

\begin{figure}
  \centering
  \subfloat[Nominal elevation view]{
    \includegraphics[width=0.64\textwidth]{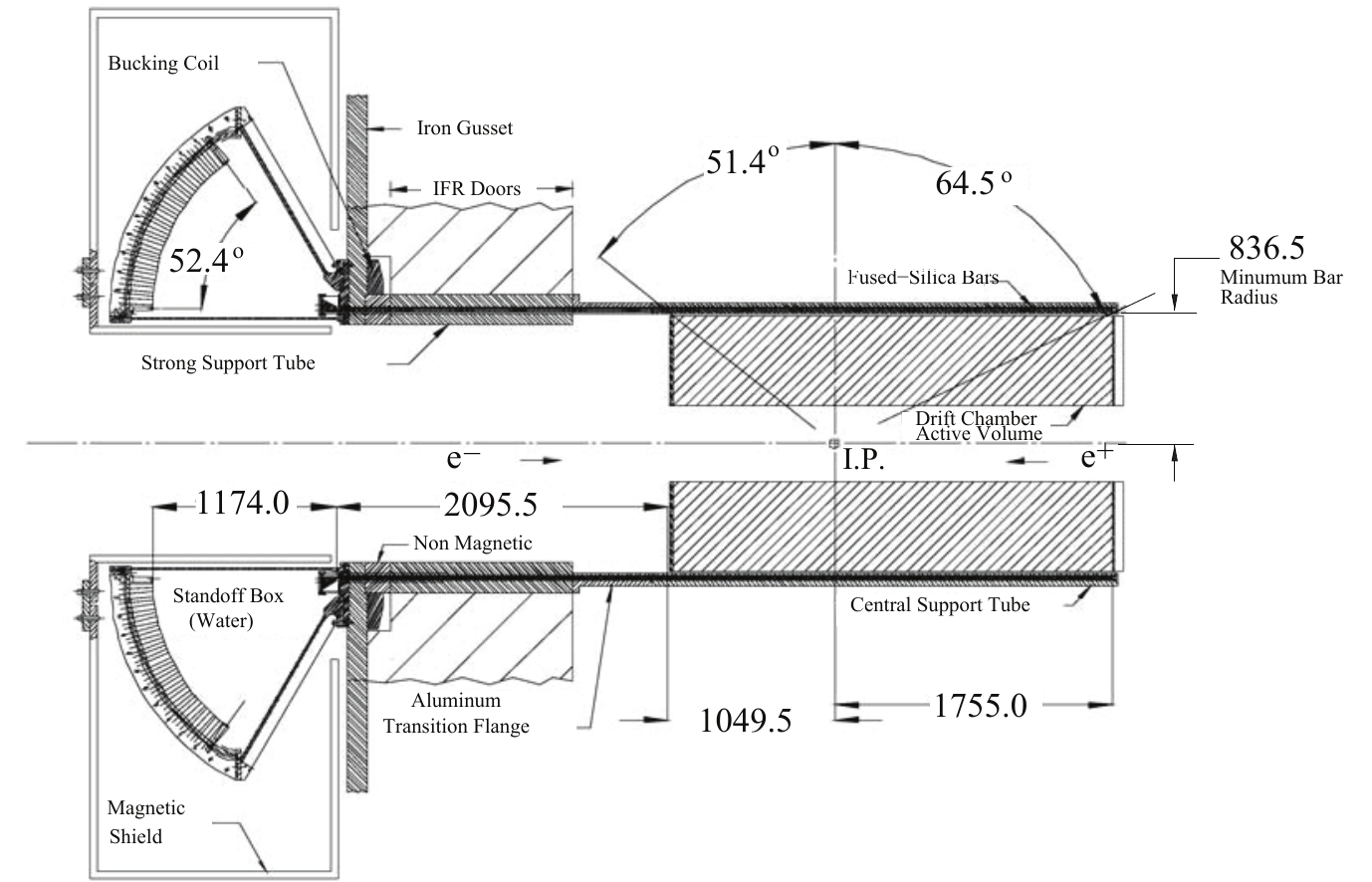}
  }\subfloat[Cross section through the central tube]{
    \includegraphics[width=0.35\textwidth]{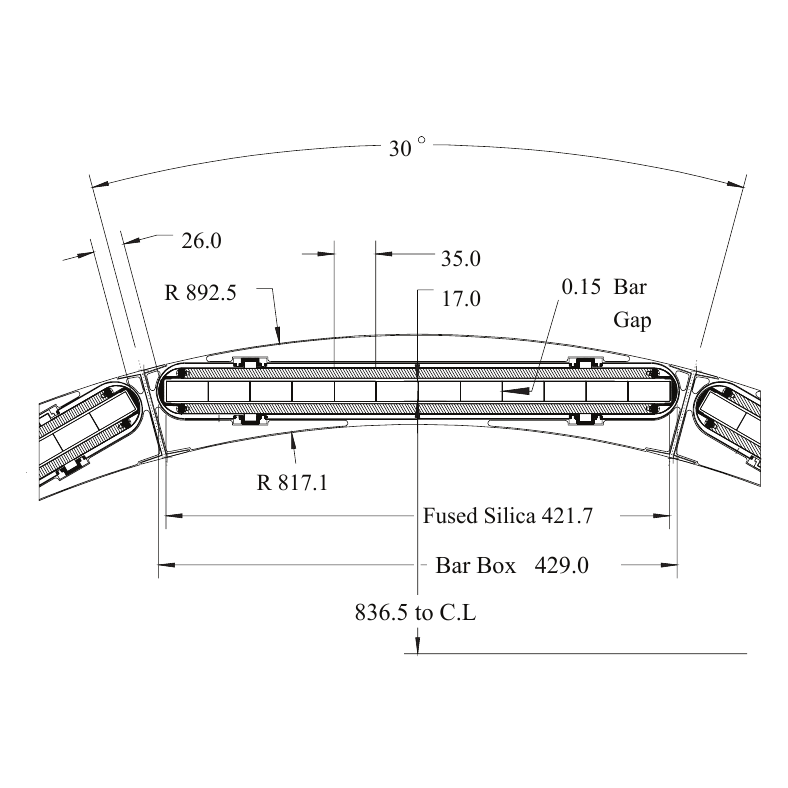}
  }
  \caption{BaBar DIRC geometry~\cite{Adam:2004fq}. All dimensions are given in mm.}
  \label{fig:BaBar_DIRC}
\end{figure}

The BaBar DIRC, shown in Figure~\ref{fig:BaBar_DIRC}, consists of 144
precision fabricated quartz radiator bars that collect \v{C}erenkov light
produced by charged particles traversing the bars. In the BaBar DIRC,
the quartz bars were read out on one end utilizing a large water
filled expansion volume to allow the light to spread out and be read
out using a large number (over 10,000) 28~mm diameter photomultiplier
tubes.

The BaBar design, while allowing for a conventional PMT readout
without the use of any focusing elements, requires a large expansion
volume and this places stringent demands on the mechanical
specifications for the detector. After the shutdown of BaBar at SLAC,
it was proposed to use the DIRC in the SuperB Experiment in Italy. In
doing so, it was also proposed to convert the original DIRC into a
Focusing DIRC (FDIRC)~\cite{Grauges:2010fi}, which would utilize
mirrors at the end of the radiator bars, allowing for a considerable
reduction in the size of the expansion region,
more highly pixellated PMTs, and an
overall expected improvement in performance. We would therefore
propose the same modification of the BaBar DIRC for ePHENIX, or would
construct a similar FDIRC if the BaBar DIRC were not available.

\begin{figure}
\centering
\includegraphics[width=.7\textwidth]{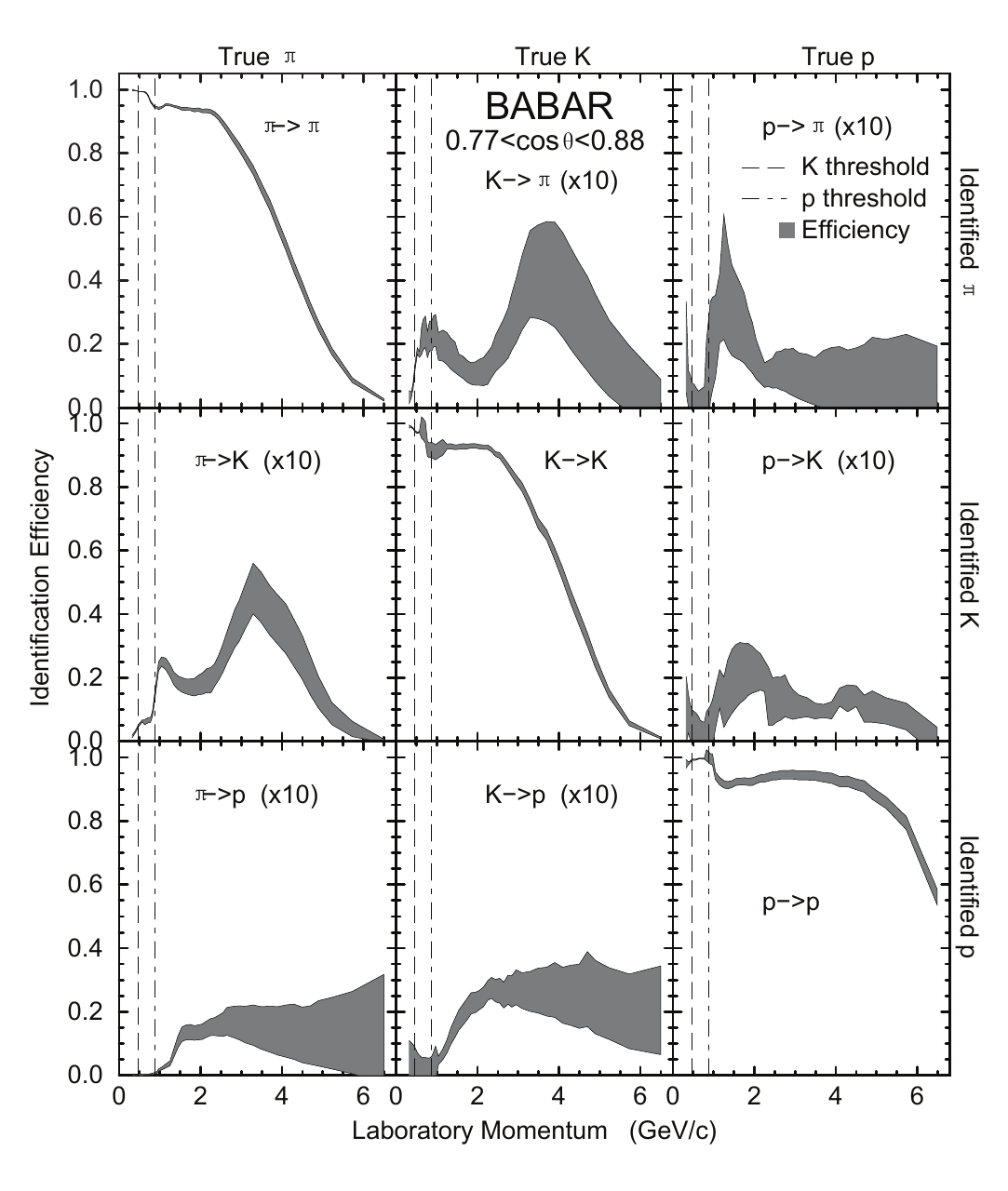}
\caption{Simulated PID Efficiency matrix and its uncertainty for
  $1.0<\eta<1.4$ region, utilizing combined information of the BaBar
  DIRC and $dE/dx$ measured in the tracking
  detector~\cite{Adam:2004fq}.  Note that the off-diagonal efficiency
  values are scaled by a factor of 10.}
\label{fig:FDIRC_PID}
\end{figure}

Similar to the BaBar technique~\cite{Adam:2004fq}, the hadron PID in
the barrel will be analyzed using an event likelihood analysis with
the DIRC and TPC $dE/dx$ information simultaneously.  A $dE/dx$
measurement in the TPC gives very good hadron separation for very low
momentum particles.  But the ability of that technique to separate
K-$\pi$ and p-K drops off quickly around 0.5~GeV$/c$ and 0.8~GeV$/c$,
respectively.  Meanwhile, the pions and kaons exceed their respective
DIRC \v{C}erenkov thresholds in this momentum region, as shown in
Figure~\ref{fig:RICH_Angles}. Therefore, the DIRC sensitivity for
K-$\pi$ and p-K turns on sharply. A combined analysis of both pieces
of information can give high PID purity up to a few GeV$/c$, as shown
by the BaBar experiment~\cite{Adam:2004fq}.  At higher momenta, the
DIRC ring resolution limits the separation capability. As shown in
Figure~\ref{fig:FDIRC_PID}, the K-$\pi$ and p-K separation gradually
drops below plateau above momentum of 2 and 5~GeV$/c$, respectively. A
$\sim20\%$ pion and kaon efficiency can still be maintained at
5~GeV$/c$.  The vast majority of hadron kinematics in SIDIS can be
covered in the $5\times100$~GeV$/c$ collisions. In the
$10\times250$~GeV$/c$ collisions, the low to intermediate-$z$ region
in SIDIS are still well covered by this design.

\subsection{\label{sec:beamline}Beamline detectors}

\begin{figure}
\centering
\includegraphics[bb=0bp 100bp 720bp 400bp,clip,width=1.0\textwidth]{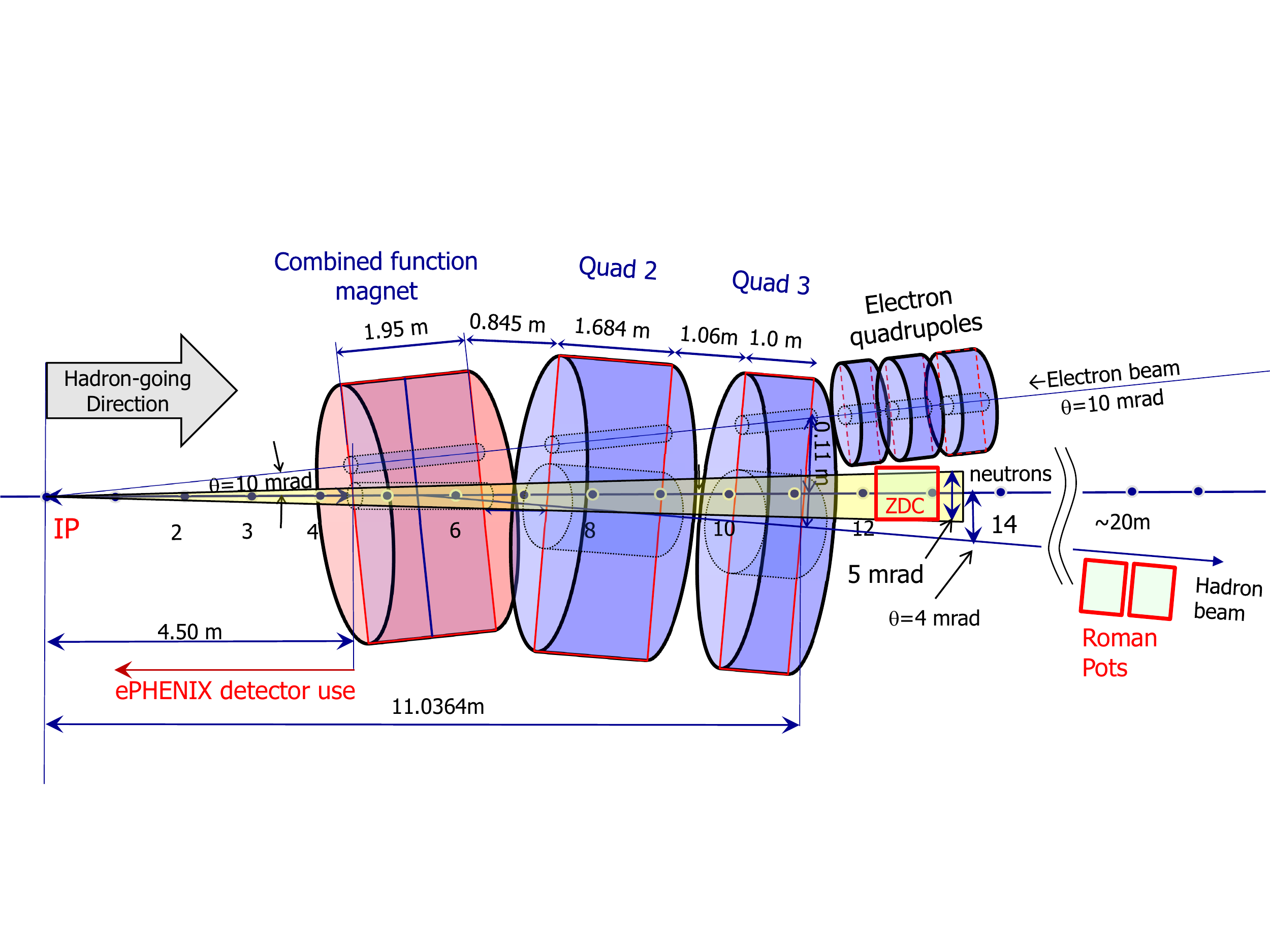}
\caption[Floor plan showing the locations of ZDC, Roman Pots, and IR
magnets]{Floor plan showing the locations of ZDC and Roman Pots
  relative to the ePHENIX interaction point. One layout of the
  interaction point magnets is also shown~\cite{Trbojevic:2013}.}
\label{fig:ZDC}
\end{figure}

Two detectors will be installed near the outgoing hadron beam,
downstream of the ePHENIX detector. They will be included in the eRHIC
machine lattice design~\cite{Accardi:2012hwp}.

\begin{description}
\item[Zero Degree Calorimeter:] A Zero Degree Calorimeter (ZDC) is
  planned for the \hgoing direction for the ePHENIX IP. Consistent with the eRHIC IR design (Figure~\ref{fig:ZDC}),
  the ZDC will be installed about 12 meters downstream of the
  IP centered on the hadron direction at the IP. A  5~mrad cone
  opening of the IP is guaranteed by the ePHENIX detector and beam
  line magnets. The ZDC for the current PHENIX
  experiment~\cite{Adler:2000bd} and its design can be reused for this
  device.
\item[Roman Pots:] In exclusive deep inelastic $e$$+$$p$ scattering,
  the final state proton will have a small scattering angle and escape
  the main ePHENIX detector. Two silicon tracking stations (also
  called the Roman Pot spectrometer) will be installed close to the beam,
  inside the beam pipe,
  downstream in the \hgoing direction to capture such protons. Each of
  the ePHENIX Roman Pot stations utilizes four tracking modules to
  cover the full azimuthal angles. Each of the tracking modules can use
  the design of the existing STAR Roman Pots~\cite{Adamczyk:2012kn}.
  Depending on the eRHIC lattice and magnet design, their location
  will be around 20~meters from the IP. This Roman Pot spectrometer will
  provide high efficiency for the exclusive DIS events in the
  $e$$+$$p$ collisions.

  \end{description}

\cleardoublepage

\backmatter

\cleardoublepage
\phantomsection
\addcontentsline{toc}{chapter}{List of Tables}
\listoftables

\cleardoublepage
\phantomsection
\addcontentsline{toc}{chapter}{List of Figures}
\listoffigures

\cleardoublepage
\phantomsection
\addcontentsline{toc}{chapter}{References}

\bibliographystyle{unsrturl}

\end{document}